\documentclass[a4paper,11pt]{book}
\usepackage[latin1]{inputenc}
\usepackage{amsmath}
\usepackage{amsfonts}
\usepackage{amssymb}
\usepackage{graphicx}
\usepackage{fancyhdr}
\usepackage{psfig}
\usepackage{graphicx}
\usepackage{color}
\usepackage{lscape}
\usepackage[scriptsize]{caption}
\usepackage{longtable}
\usepackage[english]{babel}
\newenvironment{changemargin}[2]{%
 \begin{list}{}{%
  \setlength{\topsep}{0pt}%
  \setlength{\leftmargin}{#1}%
  \setlength{\rightmargin}{#2}%
  \setlength{\listparindent}{\parindent}%
  \setlength{\itemindent}{\parindent}%
  \setlength{\parsep}{\parskip}%
 }%
\item[]}{\end{list}}
\begin{document}

\begin{titlepage}
  \begin{changemargin}{-2cm}{-4cm}
  
\begin{center}
\begin{Large}
\textbf{UNIVERSIT\`A DEGLI STUDI DELL'INSUBRIA} \\
\end{Large}

\vspace{10pt}
\begin{large}
\textsc{Facolt\`a di Scienze Matematiche, Fisiche e Naturali}\\
\end{large}
\vspace{10pt}
\begin{normalsize}
\textsc{Dottorato di Ricerca in Astronomia e Astrofisica} \\
\end{normalsize}
%\end{center}
\vspace{1cm}

\begin{figure}[htbp]
% \begin{center}
\hspace{5.5cm} \includegraphics[angle=0, height=3.5cm]{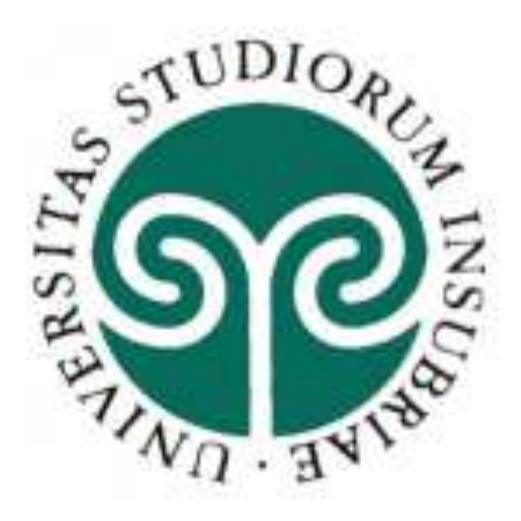} 
%\includegraphics[angle=0, height=3.5cm]{./ps_files/logo_inaf_little2}
% \end{center}
\end{figure}

\vspace{1cm}

\begin{LARGE}
\begin{center}
\textbf{The X-ray behaviour of {\it Fermi}/LAT pulsars}
\end{center}
\end{LARGE}

\vspace{2.0cm}

\begin{normalsize}
\begin{table}[htbp]
\begin{center}
\begin{tabular}{ll}
\textit{\bf Supervisor:}     \hspace{140pt} &    \\
{\bf Dott.ssa Patrizia A. Caraveo}   {\bf (INAF-IASF Milano)}  &     \\
                                        &     \\
                                        &     \\
                                        &  \textit{\bf Tesi di Dottorato di:}  \\  
                                        &  {\bf Martino Marelli}                 \\  
\end{tabular}
\end{center}
\end{table}
\end{normalsize}

\vspace{1.5cm}

%\line(1, 0){338} \\
\line(1, 0){430} \\
\begin{normalsize}
\textsc{XXIII ciclo}
\end{normalsize}

\end{center}

  \end{changemargin}
\end{titlepage}

\title{The X-ray behaviour of {\it Fermi}/LAT pulsars}

\author{Marelli Martino}

{\bf Structure of the Thesis}

More than 40 years after the discovery of Isolated Neutron Stars, the comprehension of their physics is
still rather poor. This thesis is based on a program of multiwavelength observations of pulsars
which yielded new and important pieces of information about the overall proprieties of this
class of sources.

The thesis is organized as follows:\\
- In chapter 1 we give a very brief overview of the current status of the understanding of Isolated Neutron Stars.
We also talk about the {\it Fermi} revolution that occurred in the last three years, focusing on the {\it Fermi} contribution
to the knowledge of neutron stars. Then, we describe the results led by the synergy between X-ray and $\gamma$-ray bands.\\
- In chapter 2 we give a detailed description of X-ray analyses we did and surprising results we obtained for two different radio-quiet pulsars. Such neutron stars,
J0007+7303 and J0357+3205, can be considered $"$extreme$"$ in the {\it Fermi} pulsars' zoo due to their energetics and ages. Both the X-ray observations
and analyses are very different so that they can be considered as the standing-up examples of all the following analyses.\\
- In chapter 3 we describe the analysis we done in the X-ray band and briefly report
the obtained spectra of each pulsar and its nebula, if present. Then, we study the X-ray and $\gamma$-ray pulsars' luminosities
as a function of their rotational energies and ages in order to find any relationship between these values and any difference
between the two populations of radio-quiet and radio-loud pulsars.\\
- In chapter 4 we report the $"$identity card$"$ of all {\it Fermi} pulsars, the detailed description of the analyses done and results obtained for each pulsar.

Finally, in appendix we report our accepted proposals of the most significative X-ray observations used in this thesis plus the
article on the X-ray behaviour of {\it Fermi}/LAT pulsars we published on the Astrophyisical Journal.

\clearpage

\tableofcontents

\clearpage

\chapter{Introduction}

\section{An overview of Neutron Stars' proprieties}

Neutron Stars (NS) host some of the most extreme physical phenomena that we can observe in the universe.
Their immense gravitational and magnetic fields will never be duplicated in terrestrial laboratories; therefore, the study of NSs
provides a unique opportunity to understand the physics of matter and radiation under extreme conditions. Despite the
time elapsed since their discovery, as of now no theory can fully explain
their emission, spanning all the electromagnetic spectrum. In this chapter, we will review the basic information on the current
status of theory and observable parameters of Isolated Neutron Stars (INS) which is required as a background for the reading of the
thesis work.

\subsection{The discovery}

Speculations about the existence of NSs were made for the first time right after Chandwick's discovery of the neutron, by Lev Landau (1932).
Using the newly-established Fermi-Dirac statistics and basic quantum mechanics, he estimated that such stars composed by neutrons
would form some sort of giant nuclei, with a radius of $\sim$ 3 $\times$ 10$^5$ cm (Landau, 1932).
Two years later Badee \& Zwicky (1934) proposed that NSs could form during
supernova (SN) explosions while in 1939 Oppenheimer \& Volkoff proposed a model to describe the structure and equilibrium of NSs.
Such objects were expected to be unobservable because of their small size and low optical luminosity so that
all the theories about NSs were not seriously considered by the scientific community
for more than 30 years, when the emission of these objects were instead detected.\\
The discovery of NSs was a great success of Radio astronomers. In 1967 Jocelyn Bell and Anthony Hewish, using a radio telescope designed to study
the scintillation in the radio signal from distant quasars, detected a pulsating radio signal coming from an unknown source in the sky.
The pulses, which repeated extremely regularly every 1.33 seconds, were something new to the
point that the emitting source was called LGM1, Little Green Man, as a facetious
reference to an extraterrestrial intelligence.
After the detection of few more similar sources with different periodicity,
it became immediately clear that a new class of celestial objects had been discovered. The
discovery of pulsars (contraction of pulsating stars) was published in a letter to Nature in 1968 (Hewish et al., 1968).
The identification of these pulsating
radio sources with fast spinning NSs was proposed by Pacini, 1967 and Gold, 1968, who introduced the concept of rotation-powered pulsars
which radiate electromagnetic energy at the expense of their rotational energy.

\subsection{What is a Neutron Star?}

As theorized in 1934, NSs are supposed to be born in Type II supernova explosions, from the collapse of massive stars progenitors
(M = 8-15 M$_{\odot}$, Baym, 1991). They are fast spinning objects (P $\sim$ 10ms) and have very intense magnetic fields (B $\sim$ 10$^{12}$ G)
as expected from the momentum conservation and the compression of the magnetic field of the progenitor star.
NSs are expected to have a radius of order 8-16 km, depending on the poorly known equation of state, and masses of order 0.3-2 M$_{\odot}$
(Arnett \& Bowers, 1977); canonically, we assume a radius of 10 km and a mass of 1.4 M$_{\odot}$. The surface temperature of young NSs is expected to be
of order 10$^7$ K (Chiu \& Salpeter, 1964) decreasing with time. Their surfaces are possibly covered by atmospheres composed mainly by H and He,
but there is no agreement on its actual structure. Very high spatial velocities (of order 100-1000 km/s, Lyne \&
Lorimer 1994) have been observed, as a consequence of $"$kicks$"$ due to asymmetries in supernovae explosions (Woosley, 1987).

The basic model that describes a neutron star is the rotating dipole (originally formulated by Pacini et al., 1967). In this picture a fast
spinning, strongly magnetized NS looses rotational energy (and therefore spins down) via electromagnetic radiation of the rotating magnetic
dipole (not aligned with the spin axis) and emission of relativistic particles. The electromagnetic emission is collimated in beams aligned
with the magnetic axis: when the beam intercept our line of sight a radio-pulse is observed.
The basic model of the rotating dipole (see e.g. Lyne \& Graham-Smith 1998; Becker \& Pavlov, 2002) makes it possible to obtain information on NS's parameters
simply measuring the pulsar's spin period and its derivative. Although the slow-down rate of a pulsar is small ($\dot{P}$ $\sim$ 10$^{-13}$ s/s),
the corresponding rate of rotational energy loss is huge ($\sim$ 10$^{38}$ erg/s for the Crab pulsar):\\
$\dot{E} = -4\pi^2I\dot{P}P^{-3}$\\
where the moment of inertia I is canonically set to 10$^{45}$ g cm$^2$.
Both the surface magnetic field and the pulsar age (under the hypothesis of a constant slow-down) can be obtained as a function of P and $\dot{P}$:\\
$B=3.2\times10^{19}(P\dot{P})^{1/2}$\\
$\tau_c=P/(2\dot{P})$\\
where $\tau_c$, called characteristic age, is canonically assumed as a measure of the age of a pulsar.
The rotating dipole does not provide any information about the physical processes which operate in the NS magnetosphere and which are
responsible for the observed broad band emission spectrum, from radio to high energy wavelengths.

\subsection{Emission theories}

The electromagnetic emission from Isolated Neutron Stars (INS) may be basically divided into two different components: (i) non-thermal emission, the rotation-powered
component, due to the fast rotation of the strongly magnetized neutron star; (ii) thermal emission from the hot surface of
the star.\\
Only $\sim$ 10\% of the rotational energy is converted in electromagnetic radiation: the rest leaves the star as a wind of relativistic particles,
which can interact with the surrounding interstellar medium (ISM) and emit synchrotron and Inverse Compton (IC) radiation in the whole electromagnetic
spectrum. Such a structure is called Pulsar Wind Nebula (PWN).

{\bf Non-thermal emission}

Non-thermal emission is believed to originate within the NS magnetosphere. This is the region between the star surface and the so-called
light cylinder (the virtual surface whose radius is defined by the condition that the azimuthal velocity corresponding to co-rotation with the
NS is equal to the speed of light), where the magnetic field, coupled to an high-energy plasma, is supposed to be co-rotating with the NS
(see e.g. Goldreich \& Julian, 1969). Somehow peculiar charge distributions are maintained within the magnetosphere, with
charge-depleted regions (gaps) where charged particles may be accelerated to ultrarelativistic energies in the very strong electric fields
generated by the rotating magnetic field. Such relativistic particles emit collimated radiation (responsible for the pulsed emission)
with a typical power-law spectral distribution, from radio to $\gamma$-rays. A number of magnetospheric emission models exist, but
there is no generally accepted theory.

The radio emission mechanisms are poorly understood, in spite of the wealth of data collected since the discovery of pulsars. The very high
temperature brightness has suggested that radio emission is due to some coherent process. The most promising mechanism involves bunches
of charged particles moving in synchronism along the magnetic field lines near the star poles and emitting coherent curvature radiation
(see e.g. Michel, 1991). The complex phenomenology of the radio pulses still lacks a complete modelization.

Conversely, the high-energy emission must be incoherent. There is still no consensus about the actual emission mechanism. There are different
main types of models, based on the location of the particles' acceleration zone. An accelerating field may develop along open field
lines near the NS surface, giving rise to polar cap models, or in vacuum gaps in the outer magnetosphere, giving rise to outer gap models. 
In the original versions of these models, the polar cap accelerator was confined to much less than a stellar
radius above the polar cap surface and the outer gap accelerator was confined to the region above the null charge surface.
Recently, both models have been undergoing revisions that have changed
the geometry of the accelerators and expanded them into overlapping territories.
Anyway, in both models the radiation is due to particles generated in cascade processes in charge-free gaps and accelerated to ultrarelativistic
energies, emitting via synchrotron/curvature radiation and Inverse Compton scattering of thermal soft X-ray photons.

A gamma-ray pulsar may be radio-quiet if its gamma-ray beam is wider than
its radio beam or if these beams have different orientations.
Thus the pulsar's high energy beam can sweep us while the radio one does not,
depending both on the magnetic angle and the angle of view of the pulsar.
{\em Anyway, as of now, no empirical evidence has been found to
associate the radio-quiet pulsars' population with a particular geometry.}

{\bf Thermal emission}

The existence of NSs' magnetospheric emission, as described in the previous chapter, came as an observational surprise to neutron-star
theorists. By contrast, not long after the existence of NSs was proposed (Baade \& Zwicky, 1934), it was also pointed out that such a star
would be hot, with surface temperatures over 10$^6$ K, and that surface emission might be detectable (Zwicky, 1938).
The first detections of thermal emission from the NSs' surfaces came much later, following the launch of the Einstein (Helfand et al., 1980)
and EXOSAT (e.g. Brinkmann \& Ogelman, 1987) X-ray observatories in the 1970's and 1980's.

Neutron stars are expected to be very hot objects, born with temperature of order $\sim$10$^{11}$ K (Chiu \& Saltpeter, 1964) in the cores of supernova
explosions. The physics of surface thermal emission from NSs is very complex and may be divided into two main components:\\
- Neutron Stars cooling mechanism: NSs loose thermal energy by neutrino emission from their interiors and by photon emission from their surfaces,
the first mechanism being perhaps dominating for the first $\sim$10$^5$ years of the NS life. The actual thermal evolution as a function of
time is very sensitive to the composition and structure of the star interior, in particular to the equation of state at supernuclear densities.
Different scenarios have been proposed (see e.g. Tsuruta, 1998), but the problem is far from being understood.\\
- Neutron Stars re-heating mechanism: NSs may experience heating at later stages of their life. Such heating are expected to occur both in the
star interior, owing to dissipative processes (see e.g. Alpar et al., 1984), and on the star surface, owing to bombardment
of the polar caps by relativistic particles from magnetosphere (see e.g. Cheng \& Ruderman, 1980).

The resulting surface temperature distribution is expected to be highly anisotropic, owing to anisotropic heat conduction from the NS interior
and to non-uniform re-heating of the surface. Moreover, the effects of radiative transfer in the NS atmosphere may significantly distort the
emerging spectrum of the thermal radiation. The resulting spectrum strongly depends on the atmospheric chemical composition, density, magnetic
field, temperature distribution and ionization state (see e.g. Zavlin \& Pavlov, 2002). Such an anisotropic temperature distribution
has been mapped using phase-resolved spectroscopy for bright pulsars such as the three musketeers (De Luca et al. 2005) and Vela (Manzali et al. 2005).

{\bf Nebular emission}

It is generally accepted that most of the energy leaves the pulsars' magnetosphere in the form of magnetized wind (Michael 1969).
In the ideal Golderich \& Julian's approximation of the aligned rotating magnetic dipole, it appears that the poloidal field goes
from a nearly dipolar structure within the light cylinder to a split monopole structure outside the light cylinder. The toroidal
component, which is small out of the light cylinder, grows rapidly outside it. Charged particles flow outward following the magnetic field
forming a magnetized wind which is ultimately accelerated to very high energies. This highly relativistic magnetized wind eventually
interacts with the surrounding medium and emits synchrotron radiation from Radio to TeV wavelengths.\\
The detailed structure and luminosity of the PWN should depend on the pulsar's spin-down energy history and space velocity
as well as on the density profile of the surrounding medium.

\section{Pulsars in the X and $\gamma$-ray bands}

\subsection{The {\it Fermi} Revolution}

{\it Fermi} Gamma-ray Space Telescope spacecraft was launched
on June 11, 2008. {\it Fermi} carries two instruments: the Large Area Telescope (LAT)
and the Gamma-ray Burst Monitor (GBM). LAT is the main instrument and
can detect gamma-rays with unprecedented sensitivity in the high-energy range
from 20 MeV to more than 300 GeV. GBM complements LAT in its observations
of transient sources and it is sensitive to X-rays and gamma-rays with
energies between 8 keV and 40 MeV.\\
LAT design allows for a great improvement with the respect to its
predecessor EGRET. The field of view is of $\sim$ 2.4 sr and the angular
resolution, which improves with increasing energy, is $\sim$ 0.6$^{\circ}$ at 1 GeV. The
sensitivity for a point-like source to be detected in a given time span is
50 times better than the EGRET one, and the precision in source localization is
$<$ 0.5'. These amazing characteristics are coupled to a highly improved
cosmic ray rejection.\\
Nearly the entire first year in orbit was dedicated to an all-sky survey,
imaging the entire sky every two orbits, i. e. every 3 hours. This pointing
strategy allowed 1452 sources to be detected in the first 11 months of the mission
(Abdo et al.(2010b)).\\
One of the main goal of the instrument is the gamma-ray study of neutron
stars in order to better understand their emission mechanisms and features.
To detect such kind of sources two strategies have been applied:
multiwavelenght campaigns and blind searches. Monitoring programmes
were arranged for hundreds of promising radio pulsars, selected on the basis
of their rotational energy loss weighted by the distance squared.
Meanwhile, the expectation of a significant contribution from radio quiet,
Geminga-like, gamma-ray pulsars led to the development of new codes in
order to search for pulsation in the sparse photon harvest typical of $\gamma$-ray
observations (Atwood et al., 2006).\\
ATNF database (Manchester et al. 2005) lists 1826 pulsars, and more have
been discovered and await publication LAT observes them continuously
during its all-sky survey. The best candidates for gamma-ray emission
are the pulsars with high $\dot{E}$, which often have substantial timing noise,
thus making their ephemerides validity range very short. {\it Fermi} collaboration
have obtained 726 contemporaneous pulsar ephemerides from radio observatories
and 5 from X-ray telescopes.\\
Applying the blind search techniques to all unidentified gamma-ray
bright sources, most of them already detected by EGRET and even by COS-B,
several gamma-ray pulsars were found, establishing the radio quiet pulsars
as a major fraction of the pulsar family (Abdo et al. (2009a); Saz Parkinson et al. 2010). Many of the newly found
pulsars were indeed unidentified EGRET sources which had been already
studied at X-ray wavelengths thus providing promising candidates whose
positions could be injected in the data analysis, establishing a new synergy
between X and gamma-ray astronomies (Caraveo et al. 2009). For all these pulsars the timing
ephemerides were determined directly from LAT data. In addition, LAT
data provided the best timing solutions for two others pulsars: the radio
quiet pulsar Geminga and PSR J1124-5916, which is extremely faint in radio.
Both techniques, therefore, have provided a major increase in the
known gamma-ray pulsar population, including pulsars discovered first in
gamma-rays and millisecond pulsars.\\
The current number of LAT observed gamma-ray pulsar is 88 (46 of
which were detected in the first six months of observation (Abdo et al., 2010a)):\\
- 34 radio-selected pulsars;\\
- 27 gamma-ray-selected pulsars;\\
- 27 millisecond pulsars.\\
Three gamma-ray-selected pulsars were afterwards discovered in radio wavelengths.
Moreover, despite the detection of pulsations from RL pulsars J1513-5908 and J1531-5610, the presence
of nearby sources makes their $\gamma$-ray fluxes uncertain.\\
By firmly establishing the gamma-ray-selected (radio-quiet, Geminga-like)
and millisecond gamma-ray pulsars populations, {\it Fermi} transformed GeV
pulsar astronomy into a major probe of the energetic pulsar population and its
magnetosphere physics.

{\bf {\it Fermi} Results}

The Large Area Telescope on {\it Fermi} has provided a major increase (nearly
an order of magnitude) in the known gamma-ray pulsar population, thus
allowing astronomers to solve some astrophysical puzzles:\\
- Are all gamma-ray pulsars consistent with one type of spectrum?\\
- How does the gamma-ray pulsar population compare with the radio
one?\\
- Are LAT pulsars associated with supernova remnants, pulsar wind
nebulae, unidentified EGRET sources or TeV sources?\\
- Can we constrain emission models (outer-gap, polar-cap and slot-gap)?\\
We will now briefly summarize the results reported in the first LAT pulsar
catalogue paper (Abdo et al. 2010a).\\
The spectral analysis performed on the whole LAT pulsar sample gave a
unique, consistent result: the pulsed energy spectrum can be described by a
power law with an exponential cutoff.\\
$dN/dE= KE(GeV)^{-\Gamma} exp(-E/E_{cutoff})$\\
where $\Gamma$ is the photon index at low energy, E$_{cutoff}$ is the cutoff energy, and
K is a normalization factor. The cutoff
energies range is $\sim$ 1 - 5 GeV. In all cases, the gamma-ray emission seen by
LAT is dominated by the pulsed emission.\\
The rotational energy loss rate $\dot{E}$ of the neutron stars detected by {\it Fermi}
spans 5 decades, from 3 $\times$ 10$^{33}$ erg/s to 5 $\times$ 10$^{38}$ erg/s.
Spatial associations imply that many of the pulsars detected by LAT are
young with ages less than 20 kyr: indeed, at least 19 of the 46 pulsars
of the first catalogue are associated with a pulsar wind nebula and/or
a supernova remnant, and at least 12 of the pulsars are associated with TeV
sources, 9 of which are also associated with pulsar wind nebulae.
Population studies suggested that LAT would detect a comparable number
of radio loud and radio quiet pulsars in the first year (Zhang et al. 2007). The
ratio of radio-selected to gamma-ray-selected pulsars is a sensitive discriminator
of emission models, since the outer magnetosphere models predict
much smaller ratios than inner gap models. LAT data suggest that $\gamma$ray-
selected young pulsars are born at a rate comparable to that of their
radio-selected cousins and that the birthrate of all gamma-ray-detected pulsars
is a substantial fraction of the expected Galactic supernova rate. Thus
LAT detections are consistent with the predicted range, pointing towards the
outer magnetosphere models.\\
Pulse shapes can also help to probe the geometry and physics of the emission
region. The light curves are substantially different, but roughly 75\% of
them shows two dominant, relatively sharp, peaks, separated by $>$ 0.2 of
rotational phase, suggesting that we are seeing caustics from the edge of a
hollow cone. When a single peak is seen, it tends to be broader, suggesting
a tangential cut through an emission cone. Thus, for most of the pulsars,
gamma-ray emission appears to come mainly from the outer magnetosphere.
Another constraint on the emission model comes from the study of other
observables. Comparison of the spectral cutoff energy with the surface magnetic
field shows no significant correlation, thus arguing against polar-cap
models. The values of E$_{cutoff}$ , as already seen, have a small range and
this strongly implies that the gamma-ray emission originates in similar locations
in the magnetosphere. The observed correlation between E$_{cutoff}$ and
the light cylinder magnetic field is also expected in all outer magnetosphere
models.

{\bf The blind search technique}

The improved sensitivity of the LAT has resulted in
the detection of an order of magnitude more $\gamma$-ray pulsars than were previously known.
In addition to detecting $\gamma$-ray pulsations from known
radio pulsars, the LAT
is the first $\gamma$-ray telescope to independently discover pulsars through blind searches of
$\gamma$-ray data. Searching for pulsars in $\gamma$-ray data poses significant challenges, the main
one being a scarcity of photons. Despite its huge improvement in sensitivity, the LAT
still only detects a relatively small number of $\gamma$-ray photons from a given source. For
example, the LAT detects fewer than 1000 ($>$30 MeV) photons per day (fewer than
1 per 1000 rotations) from the Vela pulsar, the brightest steady $\gamma$-ray source in the
sky. Typical $\gamma$-ray pulsars result in tens or at most hundreds of photons per day.
The detection of $\gamma$-ray pulsations therefore requires observations spanning long periods
of time (up to years), during which the pulsars not only slow down, but often also
experience significant timing irregularities, such as timing noise or glitches.

In order to lessen the impact of the long integrations
required for blind searches of $\gamma$-ray pulsars, a new technique, known as $"$time-differencing$"$,
was developed, in which FFTs are computed on the time differences
of events, rather than the times themselves. By limiting the maximum time window up
to which differences are computed to $\sim$days, rather than months or years, the required
number of FFT bins is greatly reduced. The reduced frequency resolution results in a
larger step size required in frequency derivative, $\dot{f}$ , thus greatly reducing the number of
$\dot{f}$ trials needed to cover the requisite parameter space, with the added bonus of making
such searches less sensitive to timing irregularities than a traditional coherent search.
The net result is a significant reduction in the computational and memory costs, relative
to the standard FFT methods, with only a modest effect on the overall sensitivity.

\subsection{The synergy between X and $\gamma$-ray bands}

Our work focus on the connection of pulsar's emissions in X-ray and $\gamma$-ray bands
linking it to the geometry and efficiency of each pulsar. Moreover, we are interested
in pinpointing differences (or lack thereof) between RQ and RL pulsars. Could a different geometry
translate into different X- and $\gamma$-ray emissions?\\

The physical origin of the non-thermal X-ray emission
from radio pulsars is still uncertain. Polar cap models give
predictions on the pulsed X-ray luminosity, which is attributed
to inverse Compton scattering of higher order
generation pairs on soft photons emitted by the surface
of the neutron star and/or by hot polar caps (Zhang \&
Harding 2000). The soft tail in the inverse Compton scattering
spectrum can explain the non-thermal X-ray component
observed in many pulsars.\\
Outer gap models attribute the pulsed non-thermal
X-ray emission to synchrotron radiation of downward cascades
from the outer gap particles, and include (as in the
inner gap models) a thermal component from the hot polar
caps heated by impinging particles.\\
{\em One interesting question is whether the non-thermal X-ray spectrum
is part of the same emission as that detected at $\gamma$-ray energies}. For the youngest
of these pulsars (Crab, PSR B1509-58), the total power in pulsed emission peaks
in the hard X-ray band. There is strong emission through the entire
X-ray and $\gamma$-ray bands and a smooth connection between the two. However, in the
case of the Crab the smoothness of the spectrum is misleading as the situation may
be more complicated. Kuiper et al. (2001) have argued for several separate emission
components in the Crab optical to $\gamma$-ray spectrum, evidenced by the strong frequency
dependence of the interpeak emission and the peak 2 to peak 1 ratio, both of which
have a maximum around 1 MeV. The middle-aged pulsars seem to have comparatively weak
non-thermal emission in the X-ray band since their power peaks at GeV energies,
and there is a gap in the detected spectrum between the X-ray and $\gamma$-ray bands.
In several cases (Vela, PSR B1055-52) an extrapolation between the
two is plausible, but in others (Geminga, PSR B1706-44; Gotthelf et al., 2002) a
connection is not clear.

The X-ray connection plays a very important role for RQ pulsars.
The limited accuracy of gamma-ray positions (tens of arcmins) hampers {\it Fermi} periodicity searches.
While the timing signature of a source can be easily found using a week time span,
it is difficult to keep track of the pulsar phase over a much longer time.
To preserve the phase over a time span of months, photon arrival times must be precisely barycentrized,
a standard procedure whose accuracy depends on the knowledge of the position of the spacecraft
as well as on the source positioning. A rough source position translates into a rough 
barycentric correction which could weaken or even destroy the phase coherence of the source photons.
Pinpointing likely X-ray counterparts of {\it Fermi} gamma-ray selected pulsars, i.e. 
isolated neutron stars (INSs) not detected at radio wavelengths, has proven very helpful (Abdo et al. 2009a).
We asked for {\it SWIFT} observations of all the uncovered {\it Fermi} INS error boxes, unveiling one or more counterparts for $\sim$30\% of the $\gamma$ INSs.
When the {\it SWIFT} coverage yielded more than one candidate, the timing analysis was used to select the best one, or to discard 
all candidates if none yielded a significant improvement to the overall timing solution, yielding 4
confirmed counterparts.
We also searched for X-ray observations performed of all the {\it Fermi} INS error boxes, finding a confirmed counterpart
for PSRs J2032+4127, J1023-5746 and recently for J1135-6055.\\
In these years, we asked for long {\it XMM-Newton} and {\it Chandra} observations of some interesting RQ {\it Fermi} pulsars,
such as the first one detected by {\it Fermi} (J0007+7303) and the least energetic RQ $\gamma$-ray pulsar detected
in the first year of {\it Fermi} operations, J0357+32.

In the following chapters, we report the results of the X-ray observations asked for J0007+7303 and J0357+32. Next we revise the current status of the
X-ray coverage of {\it Fermi} pulsars, providing all the X- and $\gamma$-ray spectral characteristics.
The study of the correlation between pulsars' X- and $\gamma$-ray emissions will follow, focusing on the search for
differences between RQ and RL pulsars.

\clearpage

\chapter{Study of two $"$extreme$"$ radio-quiet pulsars}

In these years, we asked for long {\it XMM-Newton} and {\it Chandra} observations of two interesting RQ {\it Fermi} pulsars,
such as the first one detected by {\it Fermi} (J0007+7303) and the least energetic RQ $\gamma$-ray pulsar detected
in the first year of {\it Fermi} operations, J0357+32.

CTA-1 {\it XMM-Newton} observation provides a good example of what we can obtain from the X-ray band:\\
- X-ray data revealed the presence of a bright nebula surrounding the pulsar: analysis of the morphology and spectrum
of extended emission is helpful in order to establish the PWN symmetry axis, presumably reflecting the pulsar spin axis (see e.g. Romani et al. 2005);\\
- Caraveo et al. 2011 also revealed a thermal hot-spot component in the spectrum. As previously said, this is strictly dependant from the equation of
state that characterize pulsars, in case of thermal cooling emission, or from the magnetosphere structure, in case of hot spots emission;\\
- Using the $\gamma$-ray ephemerides, it was possible to find the X-ray pulsations of the thermal component and make a phase-resolved spectrum.
Pulsations in the thermal emission can show the position of the pulsar's poles, providing a powerful instrument to find the pulsar spin axis;\\
- Phase-integrated spectra of both the pulsar and the nebula provided a new measurement of pulsar's distance through the column density value.
Even if analyses of the SNR already provided such a value in the case of CTA-1, the $\gamma$-ray analysis alone
isn't able to give a clear estimation of pulsars' distances.

In order to identify the X-ray counterpart of one of the least energetic FERMI pulsars J0357+32, we also proposed
a multiwavelength project within the frame of {\it Chandra}/NOAO joint observations. Due to its high angular
resolution, {\it Chandra} results to be the best in-orbit telescope for an in-depth search for X-ray counterpart.
We found only one source inside the {\it Fermi} error box with a very high X-to-optical flux ratio, typical of pulsars only.
After a spectral analysis of this confirmed counterpart, we analyze the peculiar (and unexpected) trail of J0357+32, studying its morphology
and spectrum.

\section{Energetic and young ones: the pulsar in CTA1}
\label{cta1}

The {\it Fermi} LAT discovery (Abdo et al. 2008) of a pulsed gamma-ray signal from the
position of the candidate neutron star  (NS) RXJ0007.0+7303 (Halpern 2004)
 inside CTA 1, a 5,000 to 15,000 y old supernova remnant (SNR) at a
distance of 1.4$\pm$0.3 kpc (Pineault 1993), heralded a new era in pulsar astronomy.
The discovery prompted us to ask for an orbit-long {\it XMM-Newton} observation to
study the NS X-ray behaviour.

Previous X-ray studies of CTA 1 central regions unveiled a central
filled SNR (ASCA and ROSAT observations, Seward 1995) and a point source 
({\it Chandra} and a short XMM observations, Slane 2004 and Halpern 2004, with a jet-like feature, embedded in a
compact nebula. Standard FFT searches on the XMM data failed to detect pulsation, mainly owing
to the source faintness.  

\subsection{Observations and data reduction}
We used the deep {\it XMM-Newton} observation of the CTA 1 system started on 2009, March 7 
at 15:11:10 UT and lasted 130.1 ks. The PN camera (Strueder 2001) 
of the EPIC instrument was operated in Small Window mode 
(time resolution of $\sim5.6$ ms over a $4'\times4'$ field of view), 
while the MOS detectors (Turner 2001) were set in Full frame mode 
(2.6 s time resolution on a 15$'$ radius field of view). 
We used the {\it XMM-Newton} Science Analysis Software v8.0. After standard 
data processing (using the {\tt epproc} and {\tt emproc} tasks) and screening 
of high particle background time intervals (following De Luca 2004), 
the good, dead-time corrected exposure time is 66.5 ks for the PN and 93.5 ks for 
the two MOS. The resulting 0.3-10 keV MOS image 
is shown in Figure \ref{cta1image}. 
In order to get a sharp view of the diffuse emission
in the CTA 1 system,
we also used a {\it Chandra}/ACIS (Garmire 2003) observation of the field, 
performed on 2003, April 13 (50.8 ks observing time - such dataset was  
included in the investigation by Halpern et al. 2004). 
We retrieved  ``level 2'' data from the {\it Chandra} Science Archive and 
used the {\it Chandra} Interactive Analysis of Observation (CIAO) software v3.2.

{\bf Spatial-spectral analysis}
\label{deconv}
The angular resolution of
{\it XMM-Newton}  telescopes' is not sufficient to resolve the pulsar (PSR) from the
surrounding pulsar wind nebula (PWN). 
Thus, we used  the spatial-spectral deconvolution method  
developed by Manzali 2007
to disentangle the point source from the diffuse emission,
taking advantage of their different spectra and angular distribution.

\begin{enumerate}
\item for each EPIC instrument, we extracted spectra from three concentric 
regions of increasing radii (0-5$''$, 5$''$-10$''$, 10$''$-15$''$). 
\item based on the well 
known angular dependence of the EPIC Point Spread Function (PSF), we 
estimated the PSR encircled fraction in each region. 
Since the target is on-axis and most of the counts are below 1 keV, 
we used PSF model parameters 
for an energy of 0.7 keV
and null off axis angle. 
\item we used 
{\it Chandra} data to compute the PWN encircled fraction in each region. 
To this aim, we subtracted 
the point-like PSR ($1.5''$ radius)
from the ACIS 0.3-10 keV image using the 
XIMAGE task and
we replaced its counts with a poissonian distribution having 
a mean value evaluated in a surrounding annulus $5''$ wide. 
Then, as in Manzali 2007, we degraded the 
angular resolution to match the EPIC PSF, obtaining a map of the PWN surface 
brightness as seen by EPIC. 
\item we fit a two component (PSR+PWN) model 
to all spectra, freezing the PSR and PWN normalization ratios to the results 
of the previous steps.  Uncertainty in best fit parameters induced by errors
in the encircled fractions is estimated to be
negligible with respect to statistical errors.

\end{enumerate}

Disentangling the PSR from the PWN, 
such an approach yields best fit parameters for both the PSR and the PWN spectral
models. A more detailed description of the method can be found in Manzali 2007.
Together with the XMM spectra, we fitted the spectra obtained from the pulsar
(1.5$''$ circle radius) and the nebula (15$''$ circle radius)in the
{\it Chandra} observation. We used CIAO 4.1.2 software {\tt acisspec} to
generate the spectrum and the response and effective areas.

We focus here into the spectral analysis (step 4). 
Background spectra for each EPIC camera were extracted from source-free
regions within the same chip. Ad-hoc response and effective area 
files were generated using the SAS tasks {\tt rmfgen} and {\tt arfgen}.
Since in our approach encircled energy fractions for the PSR and PWN 
are computed a priori and then used in the spectral analysis, effective
area files are generated with the prescription for extended sources, 
without modeling the PSF distribution of the source counts.
Spectra from the three regions were included in a simultaneous fit using
the combination: \\

(interstellar absorption)$\times$(\,$\rho_i$(PSR model)\,+\,$\epsilon_i$(PWN model)\,) \\

where $\rho_i$ and $\epsilon_i$
are the PSR and PWN encircled fractions within the $i^{th}$ extraction region.

The interstellar absorption coefficient does not depend
on $i$. For the PWN, we used a power law model. Although
the PWN spectrum is expected to vary as a function of
the position, we fitted a single photon index $\Gamma_{PWN}$ 
to all regions, due to the relatively small photon statistic.
For the PSR emission, we tried both a power law 
and the combination of a power law and a blackbody
(of course, PSR parameters do not vary in the different annuli).
 
The resulting parameters are summarized in Table 1. 
Both the purely non-thermal and the composite thermal+non thermal
models for the PSR emission yield acceptable fits
(power law: $\chi^2_{\nu}$ =91.5, 124 d.o.f.; 
blackbody+power law: $\chi^2_{\nu}$=85.8, 121 d.o.f.). 

Using the best fit blackbody+power law model, within a 15$''$ 
circle in 0.3-10 keV, we estimate that 
47\%  of the PN counts come from the PSR, 
32\% from the PWN and 21\% are background (instrumental as well as cosmic). 

In order to discriminate between the two descriptions of the
pulsar emission, the high resolution
temporal information provided by the PN instrument
is crucial.

\subsection{Timing Analysis}
4989 PN events in the 0.15-10 keV energy range were extracted from a 15$''$ circle, 
centered on the gamma-ray pulsar. PATTERN selection was performed as by 
Pellizzoni 2008. X-ray photons' times of arrival were
barycentered 
according to the PSR {\it Chandra} position (RA 00:07:01.56, Dec 73:03:08.3) and
then folded according to an accurate {\it Fermi}-LAT timing solution (Abdo 2010a)
that overlap our XMM dataset (the pulsar period at the start of our {\it XMM-Newton}
observation is P=0.3158714977(3) s).
Such exercise
was repeated selecting photons in different energy ranges.

A 4.7$\sigma$ pulsation is seen in the 0.15-2 keV energy range 
(null hypothesis probability of 
$1.1\times10^{-6}$, according to a $\chi^2$ test), 
characterized by a single peak, 
which is out of phase with respect to the gamma-ray emission.  Light curves 
computed for different energy ranges are shown in Figure \ref{cta1lc}, 
phase aligned with the gamma-ray one.   

Results from the previous section make it possible to compute a net
(background and PWN - subtracted) pulsed fraction of $85\pm15\%$  
in the 0.15-0.75 keV  energy range. No pulsation is seen in the 2-10 keV
energy range. Assuming a sinusoidal pulse profile, we evaluated
a $3\sigma$ upper limit of 57\% on the net pulsed fraction. 
Such a difference in the overall source pulsation as a function of 
the photon energy does not support the single-component model for the PSR
emission, pointing to the composite blackbody plus power law model. 

\subsection{Phase-resolved spectroscopy}
Phase-resolved spectral analysis  was performed by  selecting on- and
off-pulse portions of the light curve. We selected PN events 
(15$''$ extraction radius, PATTERN 0) from the phase intervals
corresponding 
to the peak and to the minimum of the folded light curve.  These spectra  are 
plotted in Figure \ref{cta1phaseres} where they are seen to differ only at 
low energy,  
while they appear superimposed for E$>1.2$ keV.  We adopted the  
best fit model computed in sect.~\ref{deconv}, featuring
the composite 
blackbody + powerlaw spectrum for the PSR, 
as a template to describe the phase-resolved spectra.
Following Caraveo 2004 and De Luca 2005, we fixed
all spectral parameters (including the PWN component)
at their phase-averaged best fit values and we used the 
PSR blackbody and power law normalization to describe
the pulse-phase modulation.
The spectral variation may be well
described as a simple modulation of the emitting radius of the blackbody 
component, keeping the power law component fixed
($\chi^2=26.2$, 31 d.o.f.). 
The blackbody emitting radius
varies from $242_{-242}^{+111}$ m to $600_{-75}^{+68}$ m as a  function of the star rotation phase.
Such a variation could easily account for the totality of the X-ray pulsation. 

Fitting the on-off spectra using a single power law component does not yield
acceptable 
results ($\chi^2=48.9$, 31 d.o.f.),
while the paucity of the counts does not allow to test a model where both
thermal and 
non-thermal components are allowed to vary
nor to use atmosphere models to account for the thermal emission.

\subsection{Extended emission spectral analysis}
\begin{figure}
\centering
\includegraphics[angle=0,scale=.50]{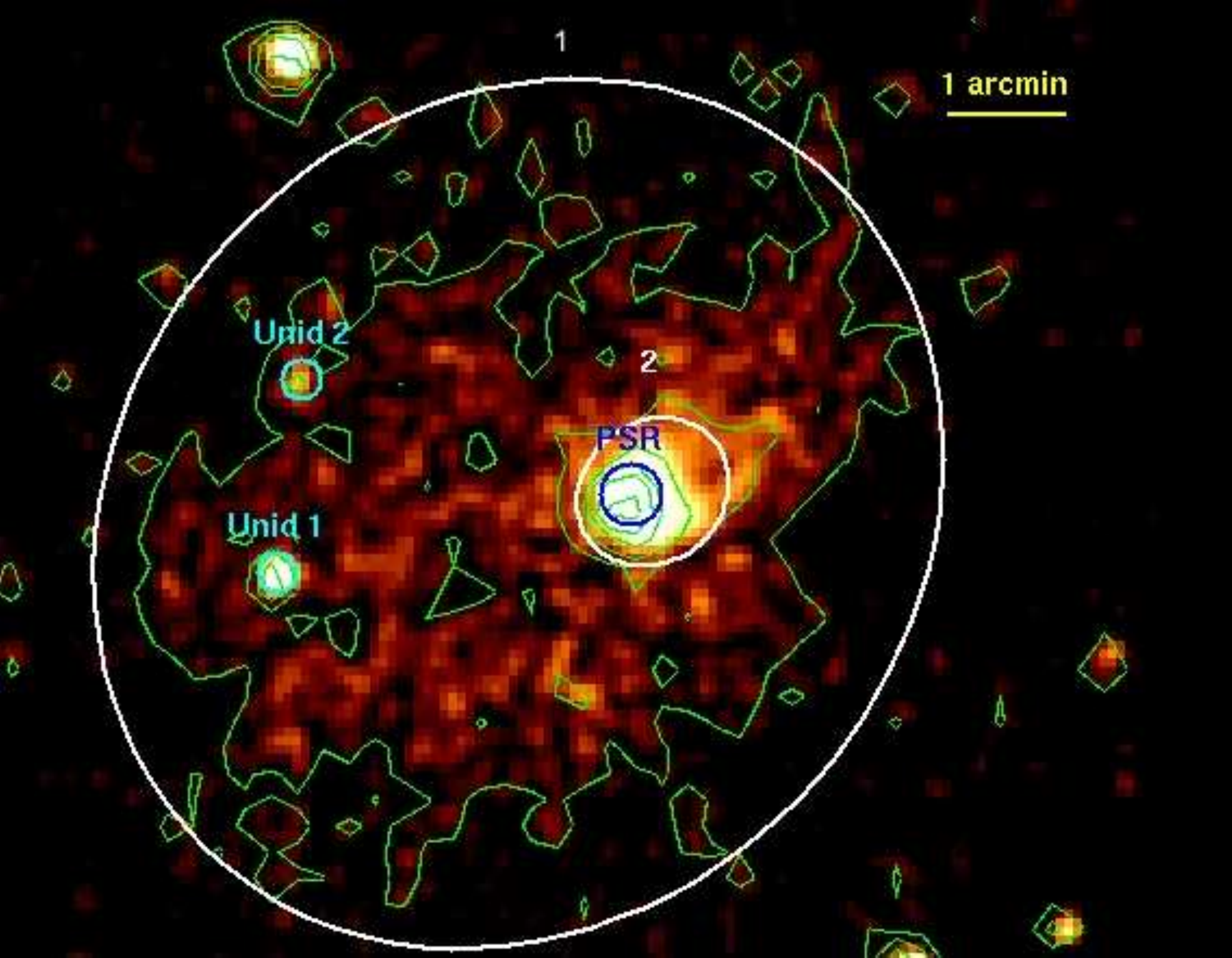}
\caption{0.35-10 keV MOS Imaging. The two MOS exposure-corrected images have been added 
and smoothed with a Gaussian with a Kernel Radius of $13''$. 
The two white ellipses indicate the extended source and the compact PWN.
A $15"$ blue circle indicates the PSR position while two unidentified sources are marked with 
cyan circles.\label{cta1image}}
\end{figure}

\begin{figure}
\centering
\includegraphics[angle=0,scale=.50]{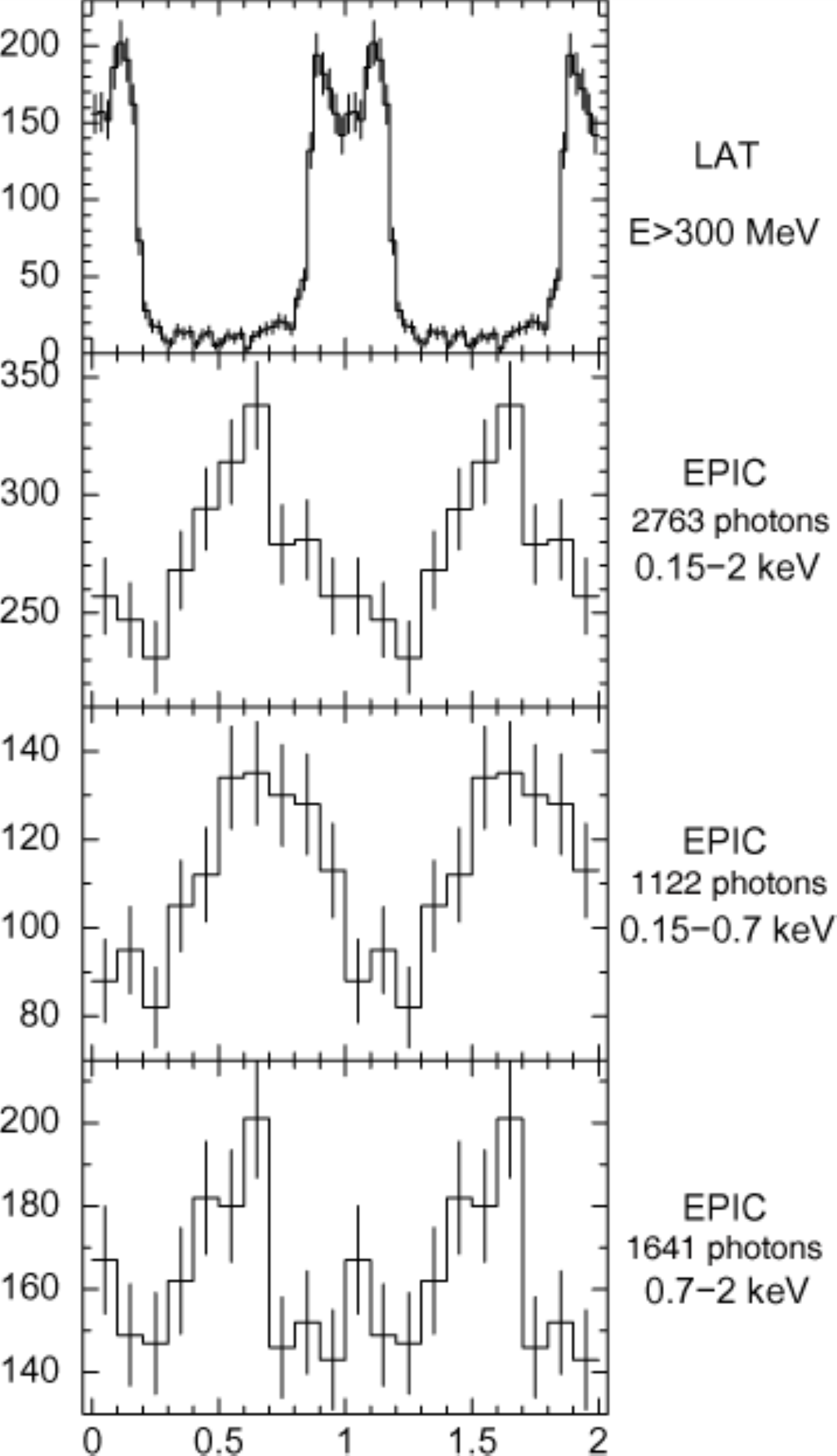}
\caption{EPIC/PN folded light curves in different energy ranges using
photons within a $15"$
radius from the {\it Chandra} position. X-ray photons' phases were computed according 
to an accurate {\it Fermi}-LAT ephemeris overlapping
with the XMM dataset: the pulsar period at the start of the XMM observation is P=0.3158714977(3) s
and the $\dot{P}$ contribution was taken in account.
PATTERN 0 events have been selected in the 
0.15-0.35 keV energy range while PATTERN $\leq$4 have been used in the 0.35-2 keV range.
The upper panel shows the LAT light curve
of the CTA 1 pulsar from Abdo 2010a to which the XMM
light curves have been aligned in phase.\label{cta1lc}}
\end{figure}

Diffuse emission, already
observed by ROSAT and ASCA (Halpern 2004, Slane 2004),
pervades the entire EPIC/MOS field
of view. A thorough analysis of such emission,
requiring ad-hoc background subtraction/modeling techniques, 
is beyond the scope of this thesis, focused the pulsar phenomenology.
For completeness, we include a simple study of the inner and brighter portion
of the diffuse emission (within $\sim150''$ from the 
pulsar). Using the brightness profile along different radial 
directions, we selected two elliptical regions (ellipse 1 and 2,
see \ref{cta1image}) -- excluding the inner 15$''$ radius circle --
and we extracted the corresponding spectra from the MOS data. Background spectra
were extracted from a region outside ellipse 2.
Since ellipse 1 lies within the PN field of view, we extracted also a
PN spectrum for such a region. Owing to the dimension of the PN field of view,
the PN background spectrum was extracted 
from a region within ellipse 2. However, the 
difference in surface brightness between the two ellipses is large enough to
induce a negligible distortion to the PN ellipse 1 spectrum.
The extended emission is well described by a power law spectrum
with an index of $1.59\pm0.18$ in the inner portion (ellipse 1, 
$\chi^2$=123.9, 90 dof) and of $1.80\pm0.09$ in the outer portion 
(ellipse 2, $\chi^2$=189.3, 137 dof). The observed flux is of
1.60$\pm$0.09 $\times10^{-13}$ erg cm$^{-2}$ s$^{-1}$ and of 1.98$\pm$0.06 $\times10^{-12}$ erg cm$^{-2}$ s$^{-1}$
for the inner and outer portions, respectively.
Owing to smaller
collecting area as well as larger background per unit solid angle, {\it Chandra}/ACIS data yield consistent, although less constrained, results for such an extended 
emission.

\subsection{Discussion and Conclusions}

After Geminga, PSR J0007+7303 became the second example of a
radio-quiet gamma-ray pulsar also seen to pulsate in X-rays. 

Our X-ray analyses characterize the system emission
as follows.

The PWN X-ray spectrum at the position of the pulsar can be described by a power law 
with index $\Gamma_{PWN}=1.5\pm0.3$. Diffuse, non-thermal emission 
with a decreasing surface brightness is seen across the EPIC field of view,
with a photon index steepening as a function of the distance from the pulsar
($\Gamma=1.80\pm0.09$ at $1\div2.5'$ distance).

The X-ray spectrum of the pulsar is a combination of thermal emission,  
with T=$1.2\times10^6$  K  from an emitting surface of $640^{+880}_{-220}$ m 
radius,  superimposed to a non-thermal power law component with index 
$\Gamma_{PSR}=1.3\pm0.2$. The best fit absorbing column is 
N$_H=(1.66_{-0.76}^{+0.89})\times10^{21}$.  
The hot spot is larger than the polar cap computed for a dipole model for 
PSR J0007+7303 (about 100 m radius) but far smaller than the entire surface 
of any reasonable NS. At variance with the majority of X-ray emitting isolated  
pulsars (e.g. Kaspi 2006), no thermal component from the whole NS surface 
is discernible from 
the XMM spectrum. The $3\sigma$ upper limit  on the temperature of a 
10 km radius NS is $5.3\times10^5$ K. This makes PSR J0007+7303 by far 
the coldest NS  for its age interval,  
suggesting a rapid cooling for this young gamma-ray pulsar.   

The detection of X-ray pulsation makes it possible to directly compare the PSR J0007+7303  
multiwavelength phenomenology with that of other prototypical pulsars.

With a rotation energy loss of 4.52 $\times 10^{35} erg s^{-1}$ and a kinematic age of 13 ky, PSRJ0007+7303
is 50 times younger and 10 times more energetic than Geminga, for many years
the only known radio-quiet gamma-ray pulsar (for a review, see Bignami \&
Caraveo, 1996).  While PSR J0007+7303 is a relatively young,  Vela-like pulsar, its
rotational energy loss is intermediate between Geminga and Vela. Thus, it makes sense to
compare PSR J0007+7303 with Vela and Geminga, two NSs with a well
established  multiwavelength phenomenology
(Sanwal 2002, Manzali 2007, Caraveo 2004, De Luca 2005 and Jackson 2005).

Starting from the source flux values, if we consider the ratio between the
gamma-ray and non-thermal X-ray fluxes, we find a value of $(5.6 \pm 1.1) \times10^3$ for
CTA 1, to be compared with $(6.8 \pm 0.4) \times 10^3$ for Geminga  and
$(1.3\pm0.3) \times 10^3$ for
Vela. Thus, the young and energetic gamma-ray pulsar PSR J0007+7303 is
somewhat under-luminous in X-rays, joining  Geminga  and PSR J1836+5925, another
radio-quiet pulsar also known as Next Geminga (Halpern 2007, Abdo 2010b),
in the extreme region of the gamma-X-ray flux ratio distribution. 

Turning now to the phase resolved spectral analysis, we note that the peak
emission of the newly measured  single-peak X-ray light curve can be
ascribed to a hot spot,  apparently varying
throughout the pulsar rotation. Although the hot spot dimension seems too
big to be reconciled with the NS polar cap, but far too small to account for
the entire NS surface, the varying thermal contribution  is indeed
reminiscent of the behaviour of middle-aged pulsars such as Geminga and PSR
B1055-52 (De Luca 2005).

Our CTA-1 observation provides a good example of the breath of valuable information which
can be gained through deep X-ray observations.

\begin{figure}
\centering
\includegraphics[angle=0,scale=.40]{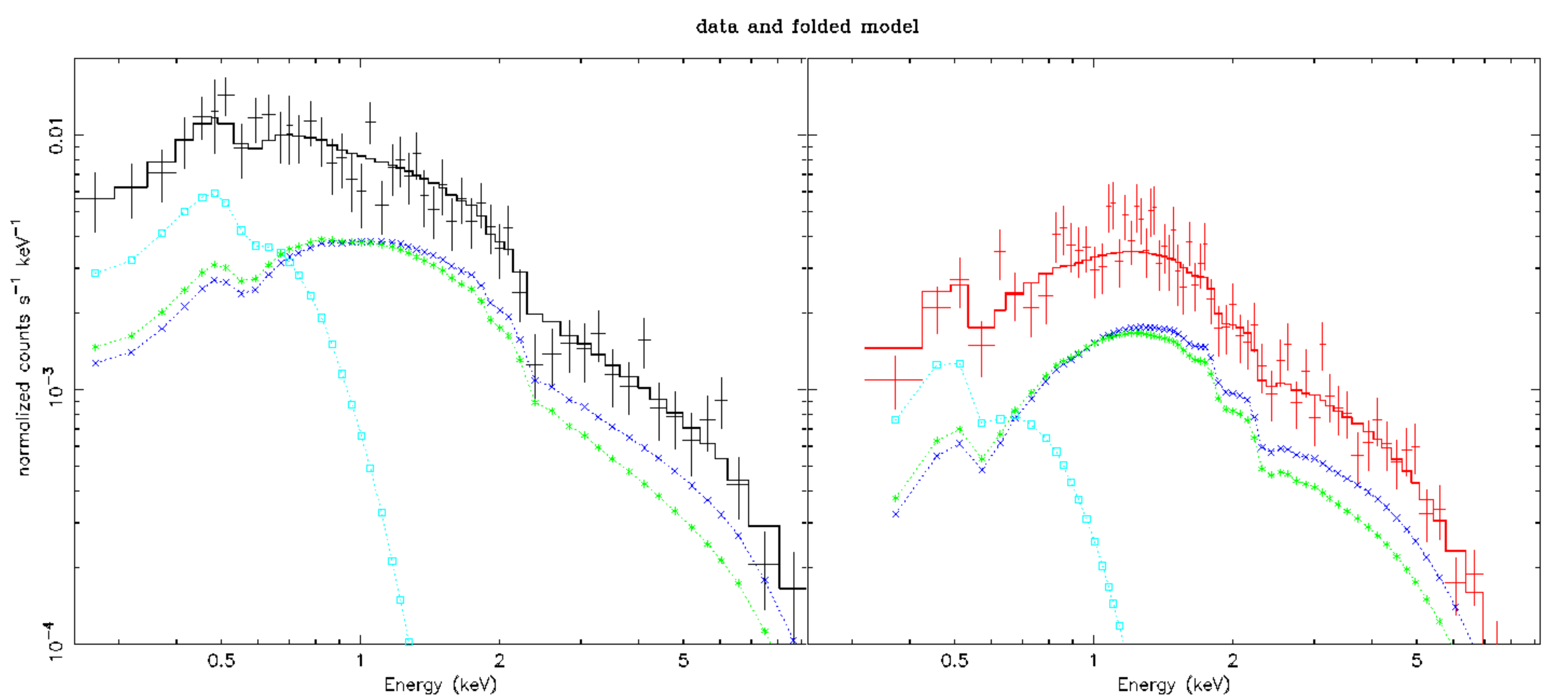}
\caption{PN and MOS spectra of PSR J0007+7303.
PATTERN 0 PN events  and PATTERN $\leq$12 MOS events 
have been selected among photons within 15$''$ from the target position.
The spectra are rebinned in order to have at least 25 counts
per bin and no more than 3 spectral bins per energy resolution
interval.
The black and the red curves show respectively the PN and MOS data and spectral 
fits.  
Cyan square-marked curve shows the blackbody component, 
while pulsar power law is shown with blue curves and PWN one with green asterisks.
\label{cta1spec}}
\end{figure}

\begin{figure}
\centering
\includegraphics[angle=0,scale=.50]{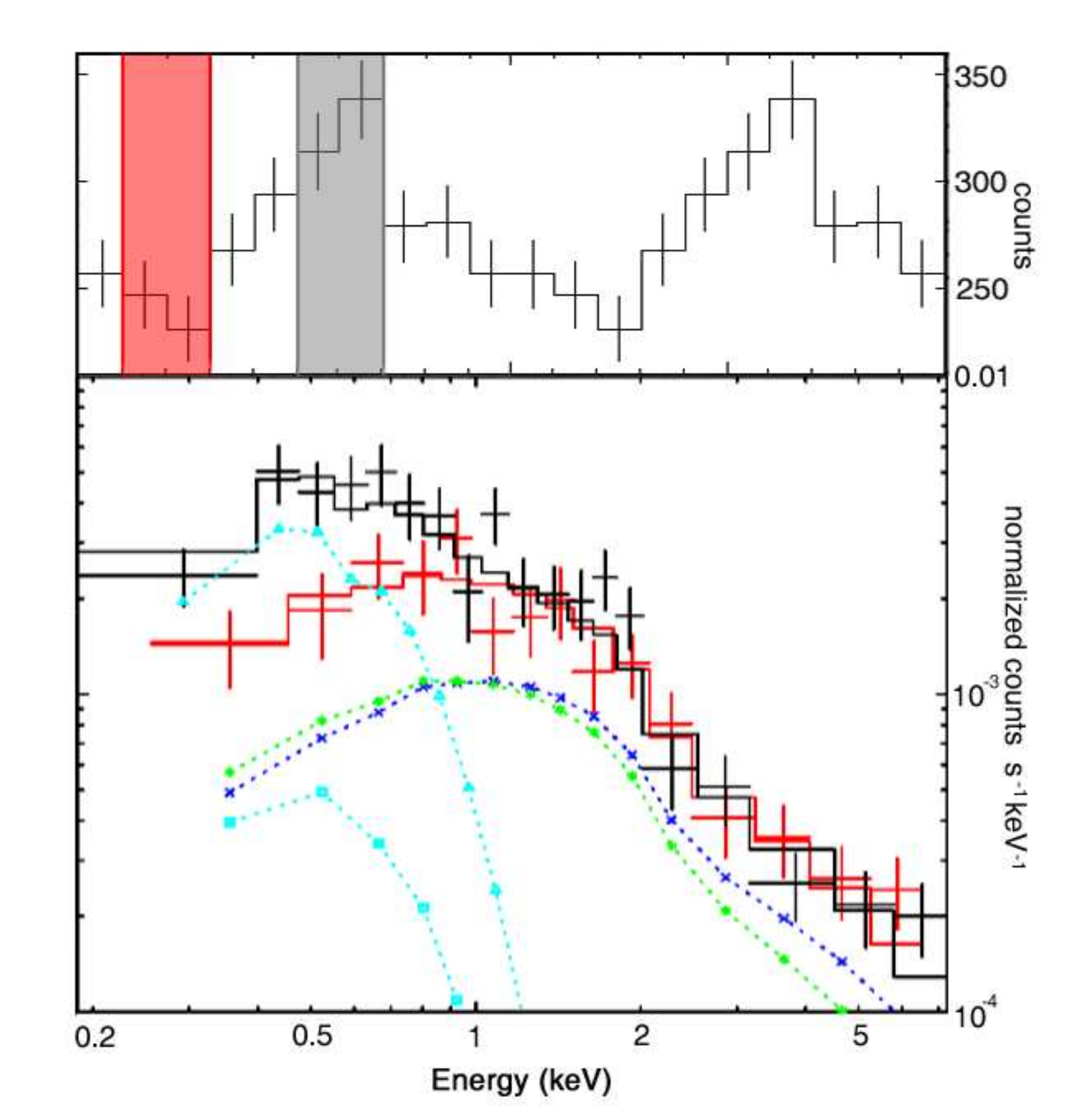}
\caption{
Upper panel: X-ray folded light curve 
of PSR J0007+7303 (0.15-2 keV).
Lower panel: X-ray spectra relative to the phase intervals shaded 
in gray (on-pulse, black line) and red (off-pulse, red line). 
Both spectra were fitted with a 3-component model to account for 
the NS thermal emission (cyan symbols) and power law (blue dotted line) 
as well as the PWN power law (green dotted line). With the power law 
contributions unchanged in the two spectra, a significant thermal component 
is present only in the on-pulse spectrum (cyan triangles) 
while it appears suppressed in the off-pulse one (cyan squares).
\label{cta1phaseres}}
\end{figure}

\clearpage

\begin{table}
\begin{center}
\caption{CTA 1 Pulsar and nebula spectra.\label{tab-1}}
\begin{tabular}{l|rr|r|r}
 &Pulsar&J0007+7303&Inner PWN&Outer PWN\\
Parameter & PL & PL+BB & & \\
 N$_{H}$($10^{21}$)                                & 0.63$_{-0.23}^{+0.25}$ & 1.66$_{-0.76}^{+0.89}$ & 1.67$_{-1.22}^{+1.38}$ & 1.83$_{-0.27}^{+0.30}$\\
 $\Gamma_{PWN}$                                     & 1.25$_{-0.15}^{+0.17}$ & 1.53$_{-0.27}^{+0.33}$ & 1.59$\pm$0.18 & 1.80$\pm$0.09\\
 $\Gamma_{PSR}$                                     & 1.36$_{-0.14}^{+0.16}$ & 1.30$\pm$0.18 & - & -\\
 kT(keV)                                          & - & 0.102$_{-0.018}^{+0.032}$ & - & -\\
 r$_{1.4kpc}$(km)                                  & - & $0.64_{-0.20}^{+0.88}$ & - & -\\
 $\chi^{2}$                                         & 91.56 & 85.81 & 123.89 & 189.30\\
 d.o.f.                                            & 124 & 121 & 90 & 137\\
 Total Flux$_{0.3-10 keV}$($10^{-14}$erg/cm$^2$s)  & 12.00$\pm$0.10 & 13.90$\pm$0.36 & 16.0$\pm$0.9 & 198$\pm$6\\
 Total Flux$_{2-10 keV}$($10^{-14}$erg/cm$^2$s)    & 8.83$_{-0.28}^{+0.37}$ & 8.69$_{-0.86}^{+0.97}$ & 10.1$\pm$0.6 & 105$\pm$5\\
 PSR Flux$_{0.3-10 keV}$($10^{-14}$erg/cm$^2$s)    & 6.54$\pm$0.53 & 8.41$\pm$0.98 & - & -\\
 PSR Flux$_{2-10 keV}$($10^{-14}$erg/cm$^2$s)      & 3.98$\pm$0.72 & 4.30$_{-0.61}^{+1.62}$ & - & -\\
 Thermal Flux$_{0.3-10 keV}$($10^{-14}$erg/cm$^2$s)& - & 1.55$\pm$1.01 & - & -\\
\end{tabular}
\caption{X-ray spectrum of the pulsar and the nebula.  
Inner and Outer PWN correspond to emission from ellipse 1 and
ellipse 2, respectively (see text and Figure 1).  
For the pulsar we provide both the power law 
and the blackbody + power law spectral fits.
For the pulsar and inner nebula we used {\it Chandra}, MOS and PN data,
while for the outer PWN spectra we used
only data from MOS1+2 instruments owing to the small FOV of PN.}
\end{center}
\end{table}

\clearpage

\section{The oldest and less energetic RQ pulsars: MORLA}
\label{morla}

Also appeared on the {\it Chandra X-ray Observatory} image release on 2011, July, 13th.

\subsection{Introduction}

The most important objects to constrain
pulsar models are the ``extreme'' ones, accounting
for the tails of the population distribution in energetics,
age, magnetic field.
In this respect, PSR J0357+3205 is one of
the most interesting pulsars discovered by LAT. It is listed
in the catalogue of the 205 brightest sources compiled after
3 months of sky scanning
(Abdo et al. 2009b), with a flux of $\sim1.1\times10^{-7}$
ph cm$^{-2}$ s$^{-1}$ above 100 MeV.
The source is located off the Galactic plane,
at a latitude $\sim-16^{\circ}$. A blind search allowed to
unambiguously detect the timing signature of a pulsar,
with P$\sim0.444$ s and $\dot{P}\sim1.3\times10^{-14}$ s s$^{-1}$
(Abdo et al. 2009b; see Ray et al. 2010 for the updated timing parameters).
The characteristic age of PSR J0357+3205
($\tau_C=5.4\times10^5$ yr) is not outstanding 
among $\gamma$-ray pulsars. The ``Three Musketeers'' 
(Geminga, PSR B0656+14, PSR B1055-52, see e.g. De Luca et al. 2005),
have ages in the 115-550 kyr range and are prominent $\gamma$-ray sources
(Geminga and PSR B1055-52 are known to pulsate in $\gamma$-rays since
the EGRET era).
However, the spin-down luminosity of PSR J0357+3205
is as low as $\dot{E}_{rot}\, = \, 5.8\times10^{33}$
erg s$^{-1}$, which is almost an order of magnitude lower than that 
of the Three Musketeers. Indeed, PSR J0357+3205
is the non-recycled $\gamma$-ray pulsar with the smallest rotational energy loss
detected so far.
This suggests PSR J0357+3205 to be rather close to us: by scaling
its $\gamma-$ray flux using the so-called $\gamma$-ray ``pseudo-distance'' relation
(e.g. Saz Parkinson 2010),
a distance of $\sim500$ pc is inferred.
PSR J0357+3205  shows
that even mature pulsars with a rather low spin-down luminosity
can sustain copious, energetic particle acceleration in their magnetosphere,
channeling a large fraction of their rotational energy loss in gamma rays. 
Thus, it stands out as a powerful test bed for pulsar models.

In view of its proximity, PSR J0357+3205 is a natural 
target for X-ray observations.
The lack of any plausible counterpart in
7 ks archival Swift/XRT data coupled with the quite large uncertainty in the position
of the $\gamma$-ray pulsar available in an earlier phase of the {\em Fermi} mission
(Abdo et al. 2009b)
called for deep X-ray and optical observations
in order to identify the pulsar counterpart 
as an X-ray source with a high X-ray to optical flux ratio. This requires
(i) a deep X-ray observation with sharp angular resolution
to nail down the position of faint X-ray sources with sub-arcsec
accuracy and (ii) a sensitive multicolor optical coverage
of the X-ray sources detected inside the $\gamma-$ray error circle.
This second step is
crucial to reject unrelated field sources such as stars or
extragalactic objects.
To this aim, in 2009, we were granted a joint
{\it Chandra} (80 ks) and NOAO program (4 hr in the V band and 3 hr in the Ks band 
at the Kitt Peak North Mayall 4m Telescope).
We also made use of optical images in the B, R and I bands collected at the 2.5
m Isaac Newton Telescope (INT) at the La Palma Observatory (Canary Islands) in
2010 as a part of an International Time Programme aimed at a first follow-up of
{\em Fermi} $\gamma$-ray pulsars (Shearer et al., in preparation).

\subsection{Observations}

{\bf X-ray observations and data reduction}

The {\it Chandra} observation of PSR J0357+3205 
was split between two consecutive
satellite revolutions. The first observation started on 25 October 2009 at
00:56 UT
and lasted  29.5 ks;
the second observation started on 26 October 2009 and lasted 47.1 ks. The two observations
are almost co-aligned, with very similar pointing directions and 
satellite roll angles. The target position
was placed on the back-illuminated ACIS S3 chip. The time resolution 
of the observation is 3.2 s. The VFAINT exposure mode was adopted.
We retrieved ``level 1'' data from the {\it Chandra} X-ray Center Science Archive
and we generated ``level 2'' event files using the {\it Chandra} Interactive
Analysis of Observations (CIAO v.4.2)
software. 
We also produced a combined event file using the {\em merge\_all} 
script.

{\bf Optical observations and data reduction}
Deep optical and near infrared images of the field of PSR J0357+3205 were collected 
at the 4m Mayall Telescope
at Kitt Peak North National Observatory as a part of our joint {\it Chandra}-NOAO program.
Optical observations in the V band (``V Harris'' filter, $\lambda=5375$ \AA,
$\Delta \lambda=$945.2 \AA) were performed using the large-field
($36'\times36'$) MOSAIC CCD Imager (Jacoby 1998) on 2009, November 10$^{th}$. 
Sky was mostly clear, with a few thin cirrus. Seeing was always better than $1.1''$.
We obtained a first set of 5 exposures of 10 min each and a second set of 18 exposures
of 12 min each, for a total integration time 
of 4 hr 26 min. 55\% of the observations were performed in dark conditions, 
45\% had partially ($\sim43\%$) illuminated Moon,
about 85$^{\circ}$ away from the target position. We used a standard 5-point dithering pattern. 
We performed standard data reduction (bias subtraction and flat fielding), CCD
mosaic, and image co-addition using the package {\tt mscred}
available in IRAF.

In our resulting co-added image, point sources have a full width at half
maximum of $\sim1.0''$ close to the expected target position. An astrometric solution
was derived using more than 1000 stars from the Guide Star Catalogue 2
(GSC2.3 Lasker 2008)
with a r.m.s. deviation of $\sim0.25''$ across the whole field of view. 
Following Lattanzi et al. 1997, and taking into account the mean positional error 
in the GSC2 source coordinates as well as the uncertainty on the alignment
of GSC2 
with respect to the International Celestial Reference Frame (Lasker 2008),
our absolute astrometric accuracy is $0.29''$.
In view of the non-optimal sky conditions, 
photometric calibration of the image was performed using a set 
of more than 400 unsaturated sources, also listed in the
GSC2.3 catalogue, taking into account the transformation
from the photographic band to the Johnson band (Russell 1990) assuming a 
flat spectrum as a function of frequency. The r.m.s. of 
the fit is $\sim0.12$ mag.

Near Infrared observations were performed at Mayall on 2010,
February 2, using 
the Florida Multi-Object Imaging Near-infrared 
Grism Observational Spectrometer (FLAMINGOS Elston 2003),
having a field of view of $10'\times10'$, using the Ks filter 
($\lambda=2.16\,\mu$m, $\Delta \lambda=0.31\,\mu$m).
Sky conditions were not optimal, with passing thin to moderate cirrus clouds.
Seeing was good, always better than 0.9$''$.
To allow for subtraction of the variable IR sky background,
observations were split in 15 sequences (stacks) of short dithered exposures 
with integration time of 30 s.
Data reduction has been performed using 
MIDAS and SciSoft/ECLIPSE packages.
The near IR raw science images have
been linearised, dark subtracted and flat fielded.  The flat fields
have been generated from science frames via median stacking to ensure
that the flat field was stable over the two hours observing time.
Data consist of fifteen stacks, each one composed of 16
or 25 jittered images on a $4\times4$ or $5\times5$ grid.  
In a first step, the reduced raw frames of a
stack have been co-added using the SciSoft {\em jitter} command.  The sky was
subtracted as a moving average from the frames before the co-addition step.
A bad pixel map, derived from the flat field, was used for masking these
pixels.  For some of the stacks the jitter offset values required manual
adjustments in order to improve the image alignment.  As the ambient
conditions became less stable in the second half of the observing run
a brighter correlation star had to be used  further away
from the center of the field of view.
As a last step, the fifteen intermediate products have been co-added to
generate the final deep image. This image is composed of 254 raw frames,
and corresponds to a total integration time of 2h 7min.
An astrometric solution was computed ($\sim0.2''$ accuracy), based on a set of
stars from 
the Two-Micron All-Sky Survey (2MASS Skrutskie 2006),
catalogue Photometric calibration, owing to poor
sky conditions, was performed on the image using a set of 35 stars also 
listed in the 2MASS catalogue, with a r.m.s. of 0.12 mag.

Additional optical observations of PSR J0357+3205 in the B
($\lambda=4298$ \AA, $\Delta \lambda=$1065 \AA), R ($\lambda=$6380 \AA, $\Delta
\lambda=$1520 \AA),
and I ($\lambda=8063$ \AA,$\Delta \lambda=$1500 \AA) bands were
obtained in dark time with the Wide Field Camera (WFC) at the 2.5
m Isaac Newton Telescope (INT) at the La Palma Observatory (Canary Islands)
on the
nights of January 16-17 2010, with seeing in the $1.1''-1.3''$ range 
(Shearer et al., in preparation), for a total
integration time of 6000 s in each band. The WFC is a mosaic of four thinned
2048$\times$2048 pixel CCDs, with a pixel size of 0.33$"$ and a full field of
view of $34 \times 34'$. To compensate for the 1$'$ gaps
between the CCDs and for the fringing in the I band, observations were split in
sequence of 600 s exposures with a 5-point dithering pattern. Data reduction was
also performed with IRAF.
 Our astrometric solution was computed using 13 
USNOB stars\footnote{The stars used in the astrometric solution were 1220-0055300,
1220-0055302,1220-0055314, 1220-0055341, 1220-0055345, 1220-0055354,
1220-0055361, 1220-0055377, 1220-0055381, 1221-0061652, 1221-0061691,
1221-0061693, 1221-0061695.}
with $0.26''$ accuracy. For photometric calibration 45 USNO-B1 stars
(Monet 2003)
were used for I band images, 30 for B and 16 for R.

\subsection{Results}

{\bf The X-ray counterpart of PSR J0357+3205}
In order to identify the X-ray counterpart of the $\gamma-$ray pulsar, 
we  searched 
the most recent {\em Fermi}-LAT timing error circle 
for X-ray sources
showing 
the expected signature of isolated neutron stars, i.e.
a very high X-ray to optical flux ratio.

\begin{figure}
\centering
\includegraphics[angle=0,scale=0.3]{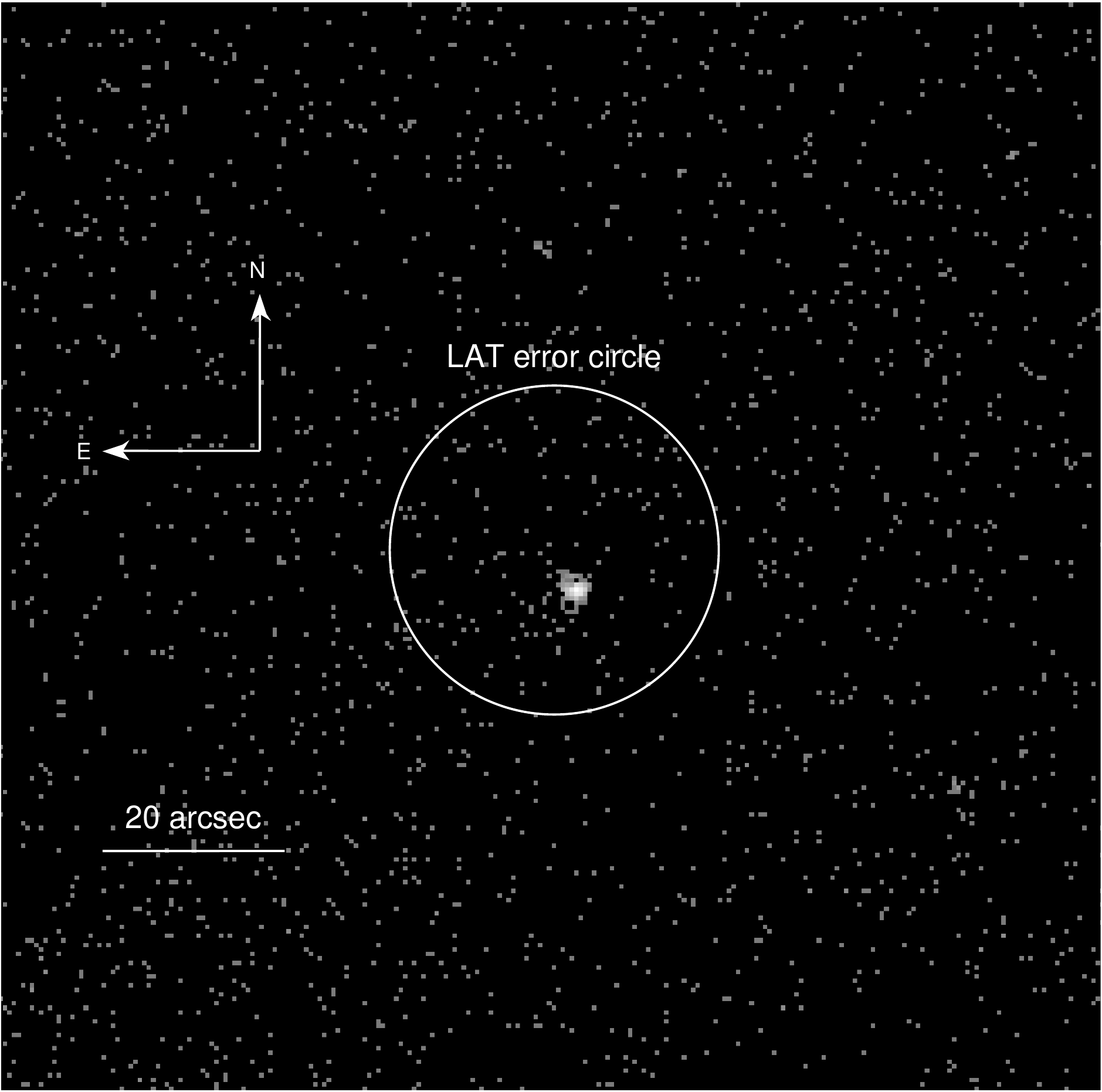}
\caption{Inner portion ($2'\times2'$) of the {\it Chandra}/ACIS 
image (0.5-6 keV) of the field of PSR J0357+3205. The {\em Fermi}-LAT 
timing error ellipse for the pulsar is superimposed. \label{chandrapsr}}
\end{figure}

\begin{figure}
\centering
\includegraphics[angle=0,scale=0.3]{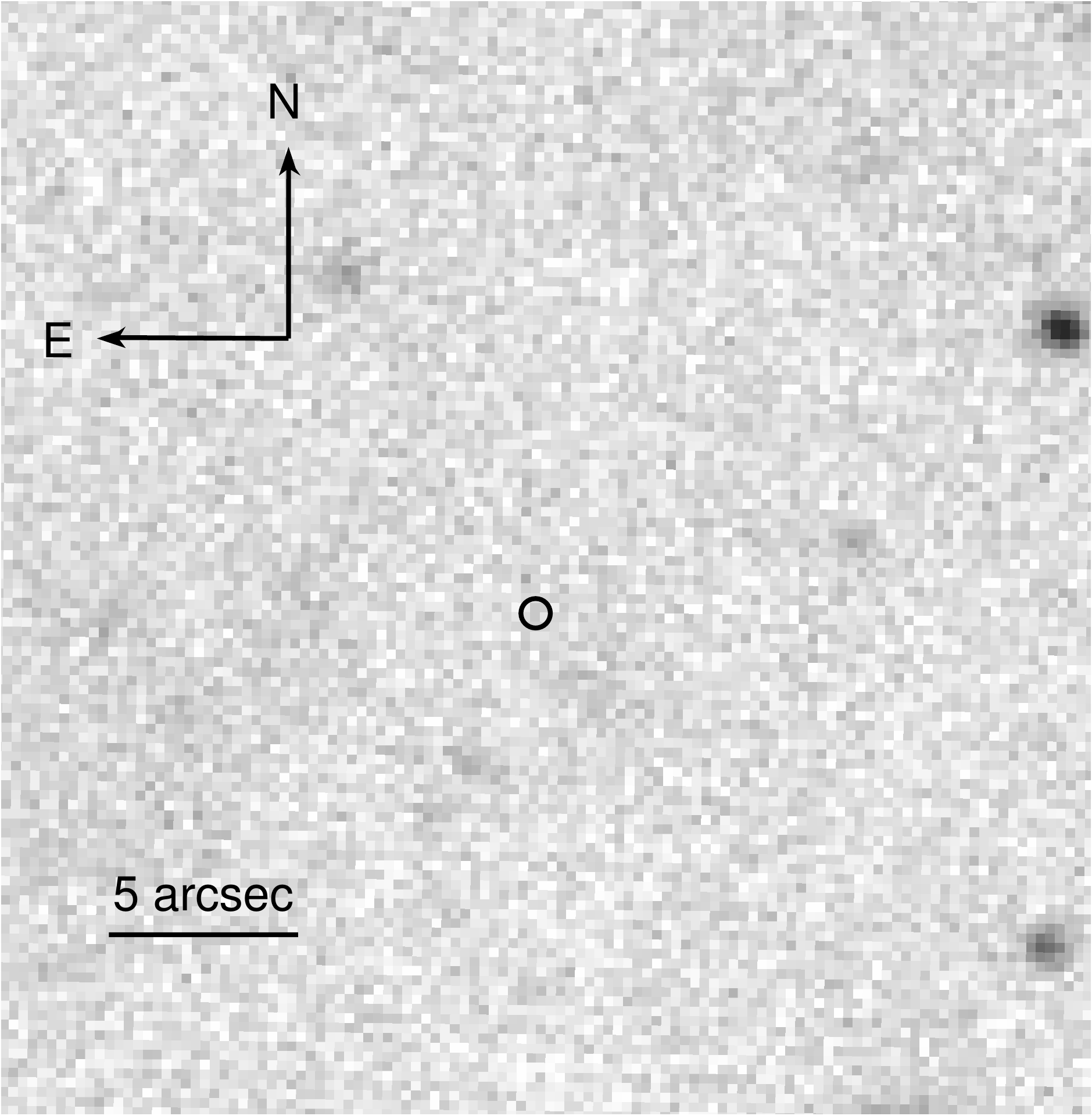}
\caption{
Inner region ($30''\times30''$) of the field as seen by the CCD Mosaic Imager
at the KPNO 4m telescope in the V band. Integration time is $\sim4.3$ hr. 
The $1\sigma$ error circle ($0.4''$ radius) for the X-ray source
consistent with the position PSR J0357+3205 is shown. Positional error accounts for 
the uncertainty in the absolute astrometry of both X-ray and optical images.
No sources are seen at the position of the {\it Chandra} source (nor within
$\sim2.5''$ from it), down to V$>26.7$.  
\label{kpnopsr}}
\end{figure}

We generated an X-ray image in the 0.5-6 keV energy range 
using the ACIS original pixel size (0.492$"$).  
We ran a source detection using the 
{\it wavdetect} task, with wavelet
scales ranging from 1 to 16 pixels, spaced by a factor $\sqrt{2}$. A
detection threshold of $10^{-5}$ was selected in order not to miss
faint sources. 
The {\em Fermi}-LAT
timing error circle for PSR J0357+3205 is centered at R.A.=03:57:52.5,
Dec.=$+$32:05:25 and has a radius
of $18''$ (Ray 2010). Only one X-ray source,
positioned at  R.A.(J2000)= 03:57:52.32, Dec=+32:05:20.6,
is detected within such region, with a background-subtracted 
count rate of $(6.3\pm0.3)\times10^{-3}$ cts s$^{-1}$ in the 0.5-6 keV
energy range (see Fig.\ref{chandrapsr}).
In order to check the accuracy of the {\it Chandra}/ACIS absolute astrometry,
we cross-correlated positions 
of ACIS sources detected at $>4.5\sigma$ within 3$'$ from the aimpoint 
with astrometric 
catalogues. We found two coincidences in the GSC2.3, 
with offsets of $0.15''-0.3''$. One of such two sources is also
listed in 
2MASS,
with a $0.15''$ 
offset with respect to the {\it Chandra} position. 
Although we could not derive an improved astrometric solution, 
such an exercise suggests that the {\it Chandra} 
astrometry is not affected by any systematics in our observations. 
Thus, we attach to the coordinates of our candidate counterpart
a nominal positional error of $0.25''$ (at $68\%$ confidence 
level). 
No coincident optical/infrared sources were found 
in our deep images
collected at Kitt Peak,
down to $5\sigma$ upper limits V$>26.7$, Ks$>19.9$ (the inner portion of the 
field, as seen in the V band, is shown in Fig.\ref{kpnopsr}). 
The INT telescope observation allows us to set 5 $\sigma$ upper limits 
of B$>25.86$, R$>25.75$ and I$>23.80$ (see Fig.~\ref{int}).
Assuming the best fit spectral model for the X-ray source (see below),
the corresponding X-ray to optical (V band) flux ratio is F$_X$/F$_V>520$,
while the X-ray to near infrared (Ks band) flux ratio is F$_X$/F$_{Ks}>30$.
Thus, positional coincidence coupled to very high X-ray to optical flux ratio prompt us to
conclude that our X-ray source is the 
counterpart of PSR J0357+3205.

To evaluate the source spectrum, we extracted photons within a $1.5"$ radius 
(561 counts in the 0.2-6 keV range, with 
a background contribution $<$0.004)
and we generated an ad-hoc
response matrix and effective area file using the CIAO script 
{\em  psextract}.
We used the C-statistic
approach (e.g. Humphrey 2009) implemented in 
XSPEC 
(requiring neither spectral grouping, nor background subtraction),
well suited to study sources with low photon statistics. Errors are at
$90\%$ confidence level for a single parameter.
The pulsar emission is well described
(the p-value, i.e. probability of obtaining the data if the model is correct, is 0.62) 
by a simple power law model, with a steep 
photon index ($\Gamma=2.53\pm0.25$),
absorbed by a hydrogen column density $N_H=(8\pm4)\times10^{20}$ cm$^{-2}$. A blackbody model
yields a poor fit (p-value $<0.00005$).
Assuming the best fit power law model, the 0.5-10 keV observed flux is 
$(3.9^{+0.7}_{-0.6})\times10^{-14}$ erg cm$^{-2}$ s$^{-1}$. The unabsorbed 0.5-10 keV flux 
is $4.7\times10^{-14}$ erg cm$^{-2}$ s$^{-1}$.

The limited statistics prevent us from constraining a more complex composite, thermal
plus non-thermal model, due to spectral parameter degeneracy
(e.g. $N_H$ vs. the normalization of the pulsar emission components).
To ease the problem, we can set an independent upper limit to the $N_H$.
The total Galactic absorption in the direction of the target is $(7-10)
\times 10^{20}$ cm$^{-2}$ (Dickey 1990, Karberla 2005). 
Since such values, based on HI surveys, could differ significantly
with respect to the actual X-ray absorption,
we used our X-ray data to get an independent $N_H$ estimate.
Our brightest point source 
(source ``A'',
see fig.~\ref{trail})
is a {\it bona fide} AGN,
with a power law spectrum ($\Gamma=1.75\pm0.15$), a flux 
of $\sim1.1\times10^{-13}$ erg cm$^{-2}$ s$^{-1}$ and a F$_X$/F$_{opt}$ ratio
of $\sim11$. 
Its absorbing column is
N$_H=(1.0\pm0.3)\times10^{21}$ cm$^{-2}$. Thus, we can assume conservatively
N$_H=1.3\times10^{21}$ cm$^{-2}$ as the maximum
possible value for the absorption towards the target. 
Such a constraint on N$_H$ allowed us to estimate upper limit temperatures
for any thermal emission from PSR J0357+3205
originating (i) from a hot polar cap  and (ii) from the whole 
neutron star surface. Assuming standard blackbody emission and
the standard polar cap radius 
($r_{PC}=(2 \pi R^3 / c P)^{1/2}=320$ m), 
we obtain kT$<122$ eV (at 3$\sigma$ confidence level) 
for a 500 pc distance.
Similarly, for a NS radius of 13 km, we obtain 
kT$<$35 eV as a limit to the temperature of 
the whole surface of the star (at 3$\sigma$ confidence level).
Blackbody radii and temperature reported above are the values 
as seen by a distant observer.

{\bf The extended tail of X-ray emission}

Our {\it Chandra} data unveil the existence of a peculiar X-ray feature 
in the field of PSR J0357+3205. An extended structure of diffuse
X-ray emission, apparently protruding from the pulsar position,
is seen in the ACIS image, extending $>9'$ in length 
and $\sim1.5'$ in width (see Fig.~\ref{trail}). 
A total of $1550\pm75$ background-subtracted counts
in the 0.5-6 keV band are collected from
such feature (the ``tail'', thereafter). 

We studied the morphology of the tail, 
extracting surface brightness profiles on different regions (see Fig.~\ref{regions}).
First, we searched for diffuse emission in the pulsar surroundings, 
by comparing the source intensity profile to the expected ACIS
Point Spread Function (PSF). Assuming the pulsar best fit spectral model,
we simulated a PSF using the
ChaRT 
and MARX - We set
  the {\em DitherBlur} parameter to the value of $0.25''$ (smaller than
the default value of $0.35''$) in order to obtain a better reproduction of the
shape of the PSF in the inner core, as discussed by (Misanovic 2008).
Results in the 0.5-6 keV energy range are shown in
Fig.~\ref{psf}, where the lack of any significant diffuse emission 
within 20$''$ of the pulsar position is apparent.
Then, we extracted exposure-corrected, background-subtracted 
surface brightness profiles on a larger angular scale.
Along the main (North-West to South-East) axis, 
the tail emerges from background $\sim20"$ 
away from PSR J0357+3205, shows a broad
maximum after $\sim 4'$  
and then fades away at more than $9'$ from the pulsar.
(see Fig.~\ref{length}). 
A possible local minimum in the
surface brightness is also seen at $\sim2'$ from the pulsar position.
In the direction orthogonal to the main axis, 
the profile shows a sharper edge towards North-East
(rising to the maximum within 15$''$) and a shallower decay to the South-West
(fading to background in $\sim70''$), as shown in Fig.~\ref{width}. 
We also extracted energy-resolved images in
``soft'' (0.5-1.5 keV) and ``hard'' (1.5-6 keV) energy bands. However, no significant
differences in the brightness profiles are observed (see Fig.~\ref{length} and
Fig.~\ref{width}).

Spectral analysis of the tail emission is hampered by the low signal-to-noise
ratio. In the extraction region, background accounts for $\sim57\%$ of the
total counts in 0.5-6 keV. A background spectrum was extracted from a source-free
region north-east of the tail. Response and effective area files were
generated using the CIAO {\em specextract}
script.
The spectrum of the tail is described well ($\chi^2_{\nu}=1.0$, 70 dofs) by 
a non-thermal emission model (power law photon index $\Gamma=1.8\pm0.2$),
absorbed by a column $N_H=(2.0\pm0.7)\times10^{21}$
cm$^{-2}$. 
Confidence contours for $N_H$ and photon index of the diffuse 
feature, compared to the ones of the pulsar, are shown in Fig.~\ref{contours}.
Fixing $N_H$ to
 $8\times10^{20}$ cm$^{-2}$ (best fit value for the pulsar counterpart)
yields a statistically
acceptable fit ($\chi^2_{\nu}=1.1$, 71 dofs), with a photon index 
$\Gamma=1.55\pm0.15$.
Adopting the latter model, 
the total observed flux of the tail
in the 0.5-10 keV energy range
is $(2.4\pm0.4)\times10^{-13}$ erg cm$^{-2}$ s$^{-1}$, corresponding to an
average surface brightness of $2.5\times10^{-14}$ erg cm$^{-2}$ s$^{-1}$
arcmin$^{-2}$. The unabsorbed flux in the same energy range
is $2.9\times10^{-13}$ erg cm$^{-2}$ s$^{-1}$.
We note that a thermal bremsstrahlung model also fits well the data
($\chi^2_{\nu}=1.0$, 70 dofs)
with an absorbing column $N_H=(1.4\pm0.7)\times10^{21}$ cm$^{-2}$, 
but requiring an unrealistically high temperature, kT=5.4$\pm$1.7 keV.

Spatially-resolved spectroscopy was also performed, 
using two separate extraction
regions, both along the tail and across it, assuming
a power law model. 
However, no significant spectral differences were found,
which is consistent with the results of our energy-resolved imaging reported
above. 
For instance, dividing the tail in two sections, for $N_H=8\times10^{20}$ cm$^{-2}$, we found 
photon indexes 
$\Gamma=1.45\pm0.15$ in the first half of the tail 
and $\Gamma=1.60\pm0.15$ in the second half. 

No point sources are detected superimposed to the tail,
with the exception of two objects located close to the SE end 
(``source 1'' and ``source ``2'' in Fig.~\ref{trail}). 
X-ray spectroscopy points to non-thermal 
emission spectra for such sources. 
Both sources have very likely optical
counterparts in our ground-based images. 
The resulting X-ray to optical flux ratio
is F$_X$/F$_V\sim13$ and F$_X$/F$_V\sim4.5$
for source 1 and source 2, respectively. 
Such results allow us to conclude that they are unrelated extragalactic
sources. 
Our ground-based images do not show any bright optical
source possibly associated to the tail, nor hints of correlated, diffuse
emission. 
We also retrieved and analyzed public data at radio wavelengths
from the NRAO VLA Sky Survey 
(NVSS Condon 1998). The images at 1.4 GHz do not show any counterpart
for the tail. 
We could set upper limits of 6.1 mJy to the tail radio emission 
over the whole extension of the X-ray feature (T. Cheung, private communication). 
The tail has also been detected by {\it Suzaku}
(Y. Kanai, private communication), in a 40 ks long observation,
although such data could not resolve its shape, nor
yield a better characterization of its spectrum.

\subsection{Discussion} 

{\bf The X-ray counterpart of PSR J0357+3205}

Our multiwavelength campaign allowed us to identify the faint X-ray counterpart
of the $\gamma$-ray only pulsar PSR J0357+3205. 
Bright in $\gamma$-rays (Abdo et al 2009a, Abdo et al 2009b), with a $\gamma$-ray
to X-ray flux ratio of $F_{\gamma}/F_X\sim1,300$, PSR J0357+3205
is an unremarkable X-ray source. Although the small photon statistics
does not allow us to draw firm conclusions, the non-negligible interstellar absorption
points to a distance of a few hundred parsecs for the source, in broad agreement
with the value of $\sim500$ pc estimated by scaling its $\gamma$-ray flux,
using the $\gamma$-ray pseudo-luminosity relation by (Saz Parkinson 2010).

The ACIS spectrum is consistent 
with a purely non-thermal origin of the X-ray emission.
The 0.5-10 keV luminosity (at 500 pc) is $L_X=1.4\times10^{30}$ erg s$^{-1}$, 
accounting 
for $\sim2.4\times10^{-4}$ of the pulsar rotational energy loss
$\dot{E}_{rot}$, in broad agreement with the dependence of the 
X-ray luminosity of rotation-powered pulsars on the spin-down luminosity 
(Becker 1997, Possenti 2002, Kargaltsev 2008). The photon index 
is significantly steeper than the typical value of $\sim1.8$ 
observed for middle-aged pulsars (De Luca 2005).

No thermal emission from the neutron star surface was detected. The
$3\sigma$ upper limit to the bolometric luminosity is
$\sim5\times10^{31}$ erg $s^{-1}$. Such limit can
be compared to
the bolometric luminosity of the well studied surface thermal
emission of the Three Musketeers, which have a characteristic age similar to
that of our
target. The upper limit to the thermal emission from PSR J0357+3205 
is a factor of 10 lower than the bolometric luminosity
of PSR B0656+14 and 
PSR B1055-52\footnote{The revision of the distance to PSR B1055-52 suggested
  by Mignani 2010 would translate to a factor $\sim4$ smaller luminosity.}
(De Luca 2005), but it is
comparable to the luminosity of Geminga (Caraveo 2004). 
Although PSR J0357+3205 turns out to be
the coldest neutron star in its age range (0.1-1 Myr), the upper limit
to its thermal emission  
is consistent with the expectations of several
cooling models (e.g. Tsuruta 2009, Page 2009). 
On the other hand, the apparent lack of
emission from the polar caps is also interesting, 
since PSR J0357+3205 is a bright $\gamma$-ray
pulsar, channeling about 40\% of its spin-down
luminosity in $\gamma$-rays of magnetospheric origin and thus
polar cap re-heating by ``return currents'' in the magnetosphere 
would be expected. 
Our limit to the temperature
of a hot polar cap points to a bolometric luminosity
$L_{PC}<5\times10^{30}$ erg s$^{-1}$, 
a factor $>5$ lower than the polar cap luminosity 
for PSR B0656+14 and PSR B1055-52 (De Luca 2005), but
a factor $\sim10$ larger than the polar cap luminosity of Geminga (Caraveo 2004).
The upper limit is a factor of a few lower than the luminosity 
expected by heating models based on return currents
of e$^{+}$/e$^{-}$ generated above the polar caps
by curvature radiation photons,
but it is consistent with expectations for polar cap heating
due to bombardment by particles created only by inverse Compton
scattered photons (Harding 2001, Harding 2002). 
PSR J0357+3205 is close to the death line for production of
e$^{+}$/e$^{-}$ by curvature radiation photons (Harding 2002),
which could explain the reduced polar cap heating. As a further 
possibility, the system's viewing geometry could play some role,
as in the case of Geminga, where the emitting area  and luminosity
of the thermally emitting polar cap suggested an almost aligned
rotator, seen at high inclination angle (Caraveo 2004, De Luca 2005).

When compared to other well-known
middle-aged rotation-powered pulsars,
the X-ray spectrum of PSR J0357+3205 is
remarkably different. 
Indeed, it is reminiscent of a number of {\em older} ($\tau_C \sim 10^{6}-10^{7}$ yr) pulsars, such as, e.g.,
PSR B1929+10 (Becker 2006), B1133+16 (Kargaltsev 2006), B0943+10
(Zhang 2005), B0628+28 (Becker 2005). 
A non thermal origin for the bulk of the X-ray emission from such
pulsars was proposed by (Becker 2004, Becker 2006), although such a picture
was questioned, e.g., by Zavlin 2004 and Misanovic 2008, 
who preferred a composite, thermal plus
non-thermal spectral model.

{\bf The X-ray tail}

The morphology of the tail, apparently protruding from
PSR J0357+3205 and smoothly connected to the pulsar counterpart
strongly argues for a physical association of the two systems.
This is also supported by the lack of any other source 
possibly related to the extended feature. 
 Sources ``1'' and ``2'' are extragalactic objects. An interpretation 
of the feature as an AGN jet, associated e.g. to Source 2, can be safely
discarded, owing to
the lack of radio emission for both the point source
and the diffuse feature, at variance with all known AGN jets (Harris 2006).
Furthermore, the angular extent of the feature would imply an unrealistic 
physical size, unless the source is quite local (a huge 200 kpc-long 
jet would 
imply an angular scale distance 
of order 80 Mpc, assuming standard cosmological parameters), which would call for a rich
multiwavelength phenomenology (the host galaxy itself -- with an angular 
scale well in excess of 1$'$ -- should be clearly
resolved in our ground-based optical images).

Assuming an association of the feature  to PSR J0357+3205, the observed extension of the tail, at 
a distance of 500 pc, would correspond to a physical length of $\sim1.3$ pc
(assuming no inclination with respect to the plane of the sky).

A few elongated ``tails'' of X-ray emission associated to  rotation-powered
pulsars have been discovered by
Gaensler 2004, McGowan 2006, Becker 2006, Kargaltsev 2008.
Such features are interpreted within
the framework of bow-shock, ram-pressure dominated, pulsar wind nebulae (Gaensler 2006).
If the pulsar moves supersonically, shocked pulsar
wind is expected to flow in an
elongated region downstream of the termination shock (basically, the cavity in the interstellar medium
created by the moving neutron star and its wind),
confined by ram pressure. X-ray emission is due to synchrotron emission
from the wind particles accelerated at the termination shock, which 
is typically
seen (if angular resolution permits) as the brightest portion of the extended
structure (e.g. Kargaltsev 2008), 
as expected from MHD simulations (Bucciantini 2002, Van der Swaluw 2003, Bucciantini 2005).

Although for our radio-quiet pulsar we have no information about the proper
motion, the bow-shock PWN scenario would seem the most natural
explanation. Of course, such a picture would suggest for PSR
J0357+3205  a large space velocity aligned with the tail, in the direction opposite to the tail extension. 
If this is the case, the pulsar would be moving almost 
parallel to the plane of the Galaxy, which would suggest 
that it was born out of the Galactic plane, at an height 
of order 140$d_{500}$ pc (where $d_{500}$ is the distance to the pulsar 
in units of 500 pc), possibly from a
``runaway'' high mass star (Mason 1998).

The luminosity of the tail in the 0.5-10 keV energy range
(assuming d=500 pc) is $8.8\times10^{30}$ erg s$^{-1}$, corresponding to 
a fraction $1.5\times10^{-3}$ of the pulsar spin-down luminosity. Indeed, such value
is fully compatible with that measured for other pulsars, which channel
into their tails $10^{-2}-10^{-4}$ of their rotational energy loss.
Synchrotron cooling of the particles  
injected at the termination shock induces
a significant softening of the emission
spectrum as a function of the distance from the pulsar in Bow-shock PWNe. 
For the tail 
of PSR J0357+3205 we do not have firm evidence for such a spectral variation.

However, explaining the tail 
of PSR J0357+3205 within the bow-shock PWN frame is not straightforward.
A first difficulty arises from energetic requirements for 
the emitting particles --
indeed, the hypothesis that the observed X-rays
from the tail are due to synchrotron emission
is somewhat challenging for a pulsar with
such a low $\dot{E}_{rot}$. As
in the case of PSR B1929+10, discussed by Becker 2006 and
De Jager 2008, the problem lies with the maximum 
energy of the particles injected in the PWN.
Particle acceleration mechanisms in PWNe are not yet fully
understood. The maximum energy to which electrons can be
accelerated (via acceleration of the pulsar wind 
and then re-acceleration at the termination shock)
is expected to be a fraction of the polar cap potential
($\sim0.1$ for the Crab, see de Jager et al.1996; see also de Jager \&
 Djannati-Ata\"i 2008, Bandiera 2008).
According to Goldreich 1969, 
the maximum potential drop between the pole and the light cylinder 
(in an aligned pulsar) is $\Delta \Phi = (3\dot{E}_{rot}/2c)^{1/2}$. 
For PSR J0357+3205, this would correspond to electron acceleration 
in the pulsar magnetosphere up to a maximum Lorentz factor
$\gamma_{max}\sim10^8$, which can be considered as an upper limit
for the electrons injected in the PWN. 
The characteristic energy of synchrotron
photons is $\sim0.5 B_{-5} \gamma_8^2$ keV,
where $B_{-5}$ is the ambient magnetic field in units of 10 $\mu G$ and 
$\gamma_8$ is the Lorentz factor of the radiating electrons
in units of $10^8$. It is clear that, in order to produce bright emission
at few keV, the typical Lorentz factor of the electrons 
in the tail has to
be of the same order of $\gamma_{max}$, 
implying the presence of e$^+$/e$^-$ accelerated 
at the highest possible energy, as well as
an ambient magnetic field 
as high as $\sim50$ $\mu G$. 
 If this is the case, it is possible to estimate 
the synchrotron cooling time of the emitting electrons
as $\tau_{sync}\,\sim100\,(B/50 \mu G)^{-3/2}\,(E/1\,keV)^{-1/2}$ yr.
Coupling such value with the estimated physical length of the feature
yields an estimate of the bulk flow speed of the emitting particles
of $\sim15,000$ km s$^{-1}$, assuming no inclination w.r.t. 
the plane of the sky. Such a value is consistent with results 
for other bow-shock PWNe (Kargaltsev 2008).

A second difficulty for the bow-shock interpretation arises owing to the 
lack of diffuse emission surrounding the pulsar.
Bright emission from
the wind termination shock should be seen there as 
the maximum surface brightness portion of the diffuse feature,
as observed in all other known cases 
(e.g. Gaensler 2004, McGowan 2006, Kargaltsev 2008).
As a possible way out, we evaluate under what conditions the termination shock could be unresolved by {\it Chandra}.
Assuming standard relations (Gaensler 2006), the distance between the 
pulsar and the head of the termination shock is expected to be
$r_S=(\dot{E}_{rot}/4 \pi c
\rho_{ISM} v^2_{PSR})^{1/2}$, where $\rho_{ISM}$ is the ambient density
and $ v_{PSR}$ is the pulsar space velocity.
For PSR J0357+3205,
$r_S\sim10^{16}v_{PSR,100}^{-1}n_{ISM,1}^{-1/2}$ cm, 
where $v_{PSR,100}$ is the 
pulsar space velocity in units of 100 km s$^{-1}$ and $n_{ISM,1}$ is the 
ambient number density in units of 1/cm$^3$. At a distance of 500 pc, 
this corresponds to $\sim1.3''v_{PSR,100}^{-1}n_{ISM,1}^{-1/2}$.
The surface of the 
termination shock (in the hypothesis of an isotropic pulsar wind) 
should assume an elongated
shape, extending $\sim6 r_S$ ($\sim8''v_{PSR,100}^{-1}n_{ISM,1}^{-1/2}$ at 500
pc) behind the pulsar. The termination shock could
hide within the point spread function of the pulsar if $6r_s<0.5''$, which would require
an unrealistically large ambient number density (of order several hundred per
cm$^3$), and/or a pulsar space velocity of at least 1000 km
s$^{-1}$. Anisotropies in the pulsar wind could also play some role.
Such a picture would suggest that a significant fraction
of the flux of the point source is due to emission from the wind 
termination shock.

\begin{figure}
\centering
\includegraphics[angle=0,scale=0.3]{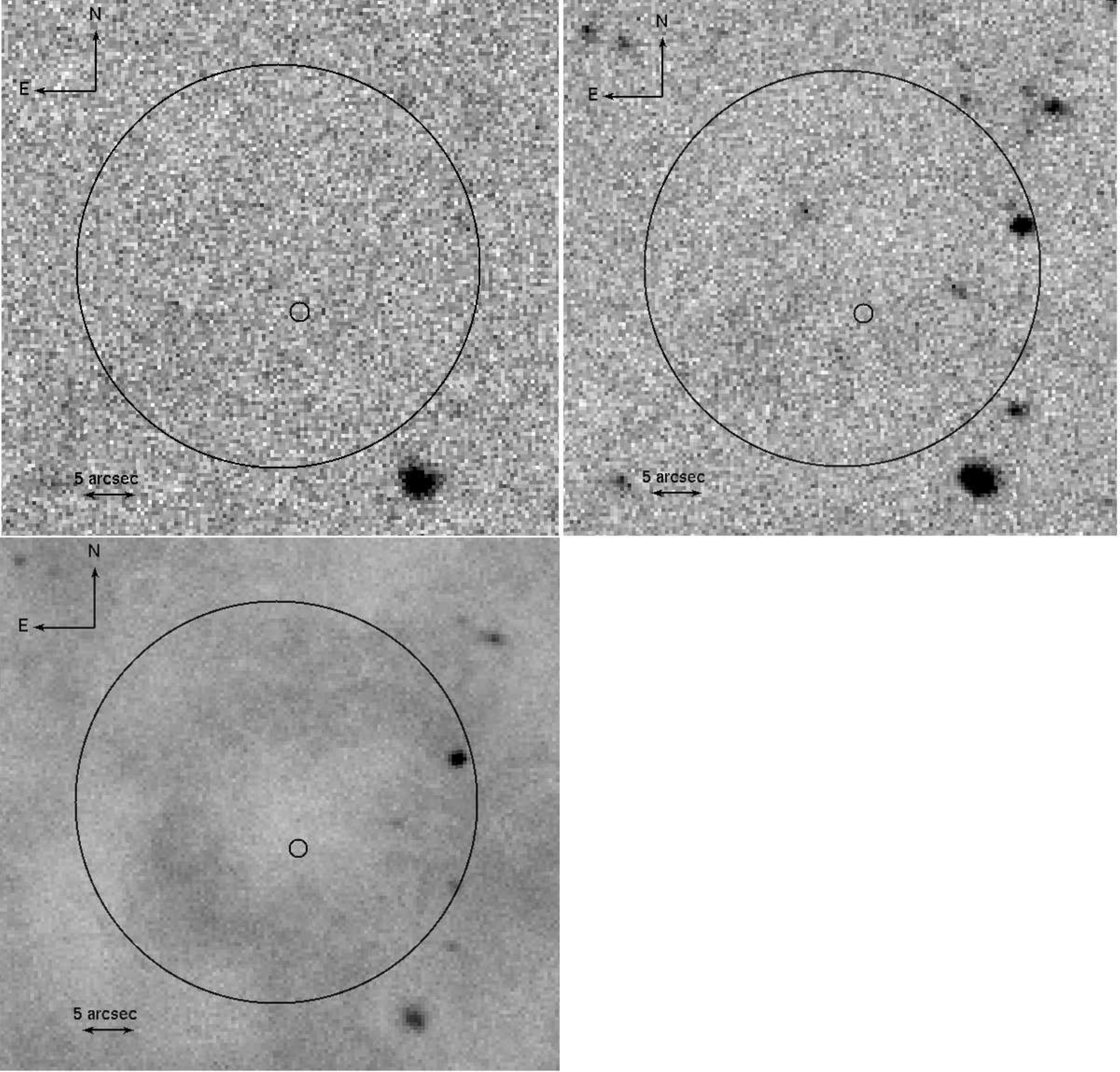}
\caption{Inner region ($40''\times40''$) of the field as seen by the WFC
on the INT
2.5m telescope in the B,R (top) and I (bottom) bands. Integration time is 6000
seconds in all cases. The smaller circle (0.8$''$ radius) marks the
$2\sigma$ error circle of the X-ray source consistent with the
position of PSR J0357+32. The larger circle shows the {\it Fermi} LAT 18$''$
error circle.
The positional error accounts for the uncertainty in the absolute
astrometry of both X-ray and
optical images. No sources are seen at the position of the {\it Chandra}
source (nor within $\sim2.5''$
from it), down to a 5 $\sigma$ limit of B$>$25.86, R$>$25.75 and I$>$23.80.
\label{int}
}
\end{figure}

\begin{figure}
\centering
\includegraphics[angle=0,scale=0.3]{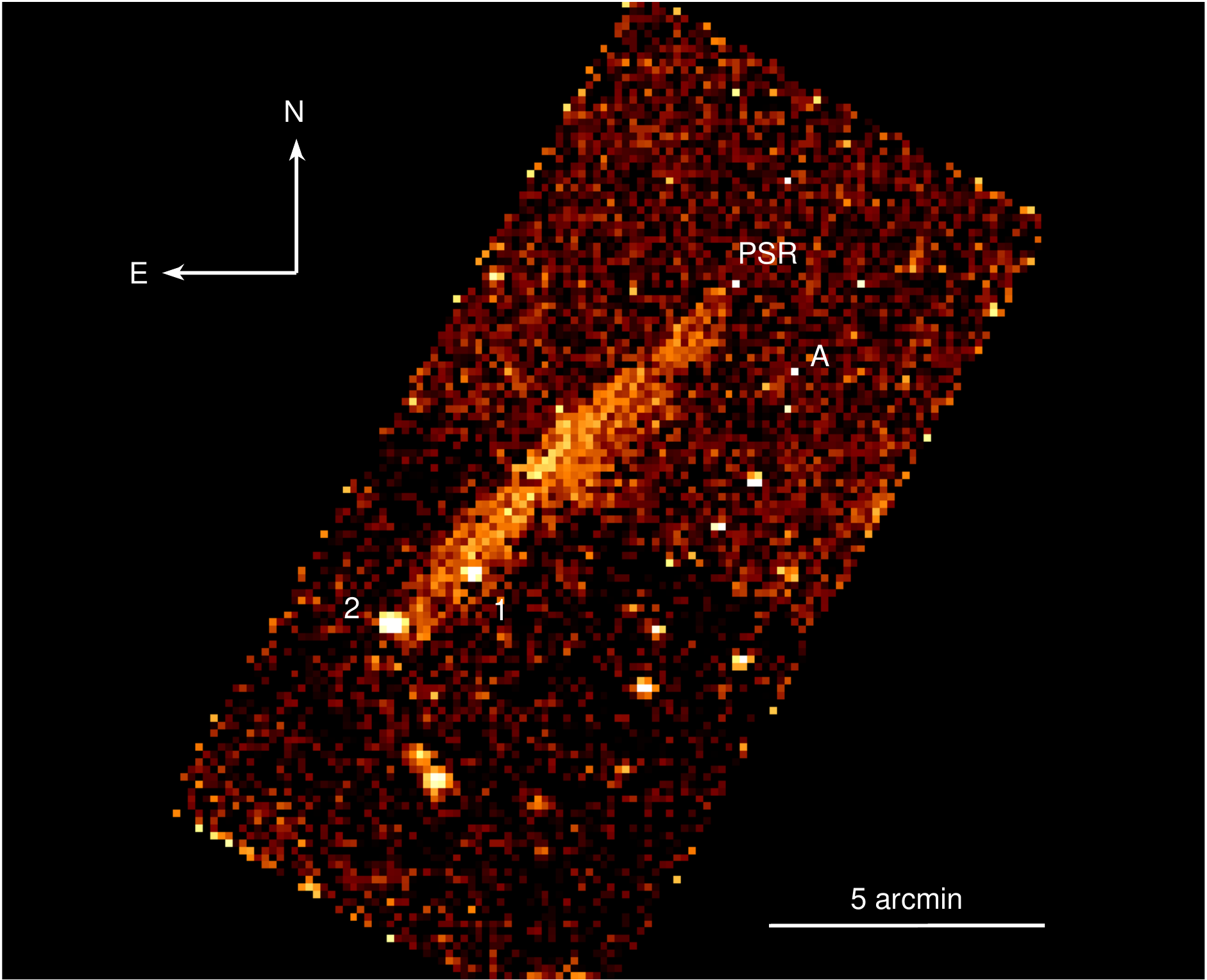}
\caption{Exposure corrected {\it Chandra}/ACIS image of the field of PSR J0357+3205
in the 0.5-6 keV energy range. The image has been rebinned to a pixel scale
of $8''$. No smoothing has been applied. A large tail of diffuse X-ray
emission is apparent, with a length (North-West to South-East) of $\sim9'$
and a width of $\sim1.5'$ in its central portion. The pulsar emission
is enclosed in a single pixel. The same is true for the brightest point
source in the field (marked as ``A''), an AGN which allowed us to estimate
the overall Galactic absorption (see text). Two
point sources are seen close to the southern end of the tail (marked as 
``1'' and ``2''). Multiwavelength data suggest they are unrelated
extragalactic objects.
\label{trail}
}
\end{figure}

\begin{figure}
\centering
\includegraphics[angle=0,scale=0.3]{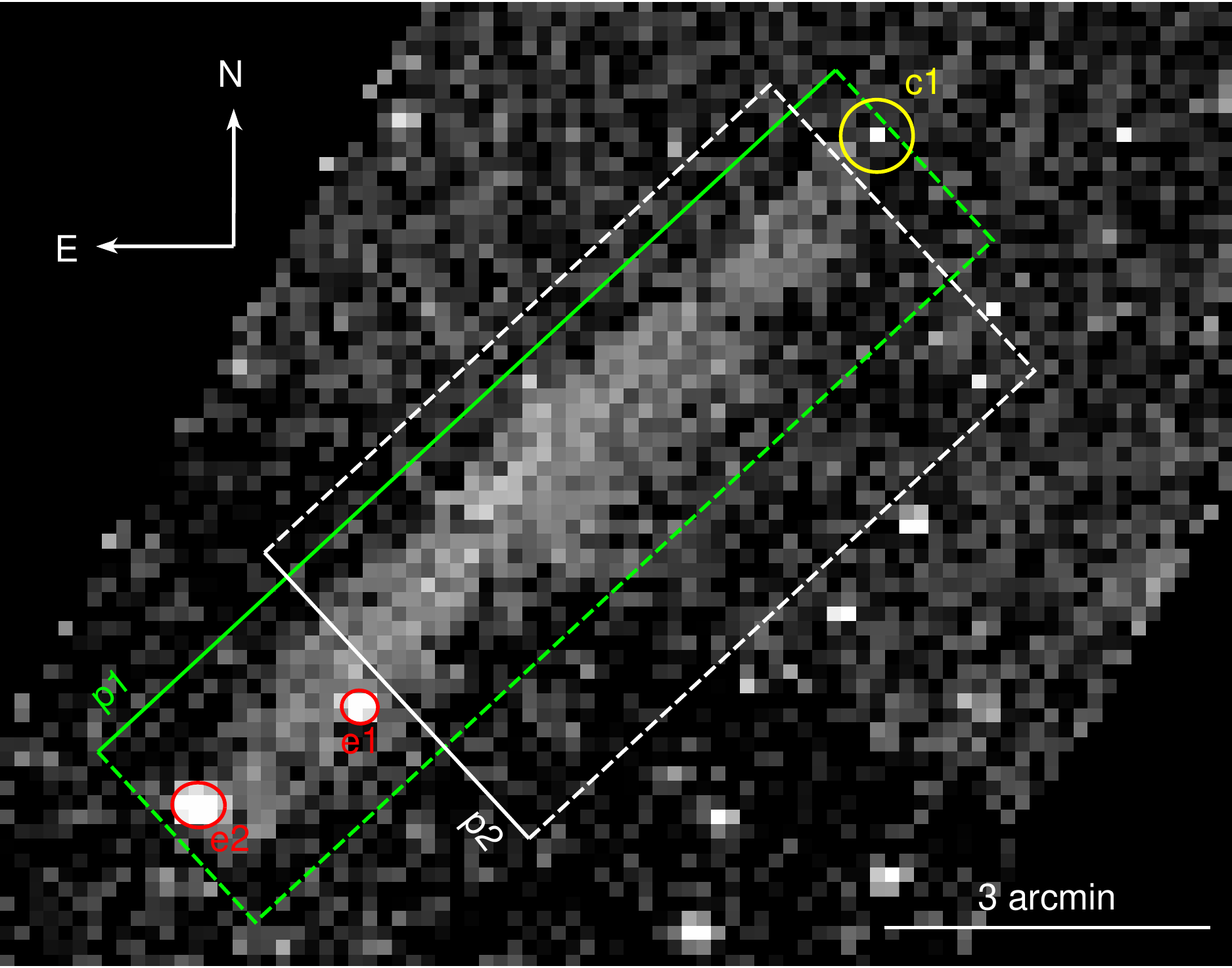}
\caption{Same as Fig.\ref{trail}. Regions used to generate surface 
brightness profiles for PSR J0357+3205 and its tail.
Circle c1 marks the 20$''$ radius region
from which we extracted
the radial profile of the pulsar counterpart shown in Fig.~\ref{psf}.
The regions from which
the brightness profiles of the tail (shown in Fig.~\ref{length} and Fig.~\ref{width}) were extracted are 
marked as p1 (along the main axis) and p2 (in the orthogonal direction).
Ellipses e1 and e2, computed using the {\em wavdetect} tool, were excluded
from the analysis to remove the counts from the
point-like sources ``1'' and ``2'' (see Fig.~\ref{trail}).  
\label{regions}
}
\end{figure}

\begin{figure}
\centering
\includegraphics[angle=0,scale=0.3]{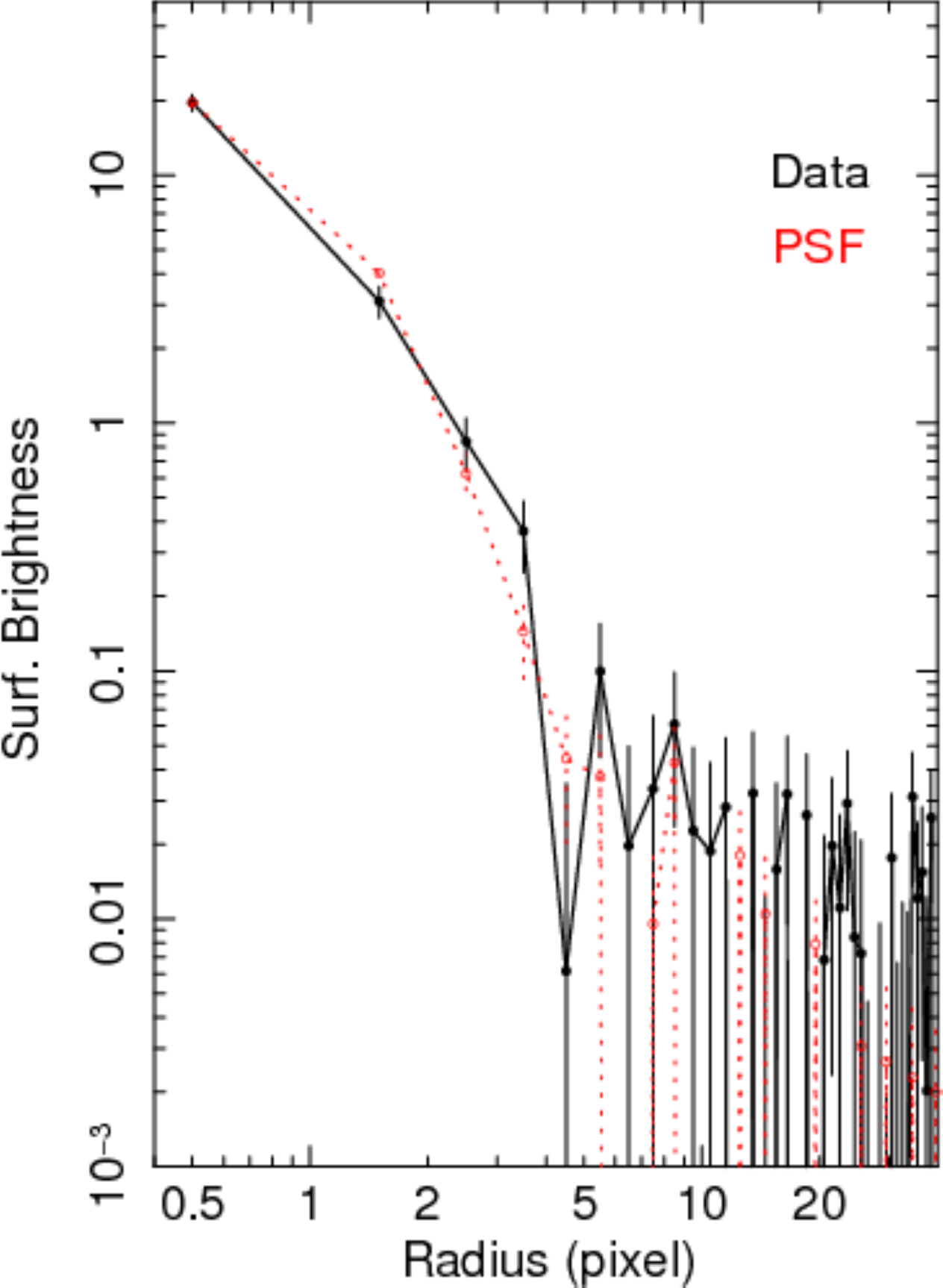}
\caption{Radial profiles (0.5-6 keV energy range) for the X-ray counterpart of PSR J0357+3205 
(background-subtracted, black points) and for a simulated point source 
(red points) having flux, spectrum
and detector coordinates coincident with the ones of the pulsar counterpart
(see text for details). The two profiles agree very well and there is no
evidence for significant diffuse emission in the surroundings of the pulsar
up to $20''$. 
\label{psf}
}
\end{figure}

\begin{figure}
\centering
\includegraphics[angle=0,scale=0.3]{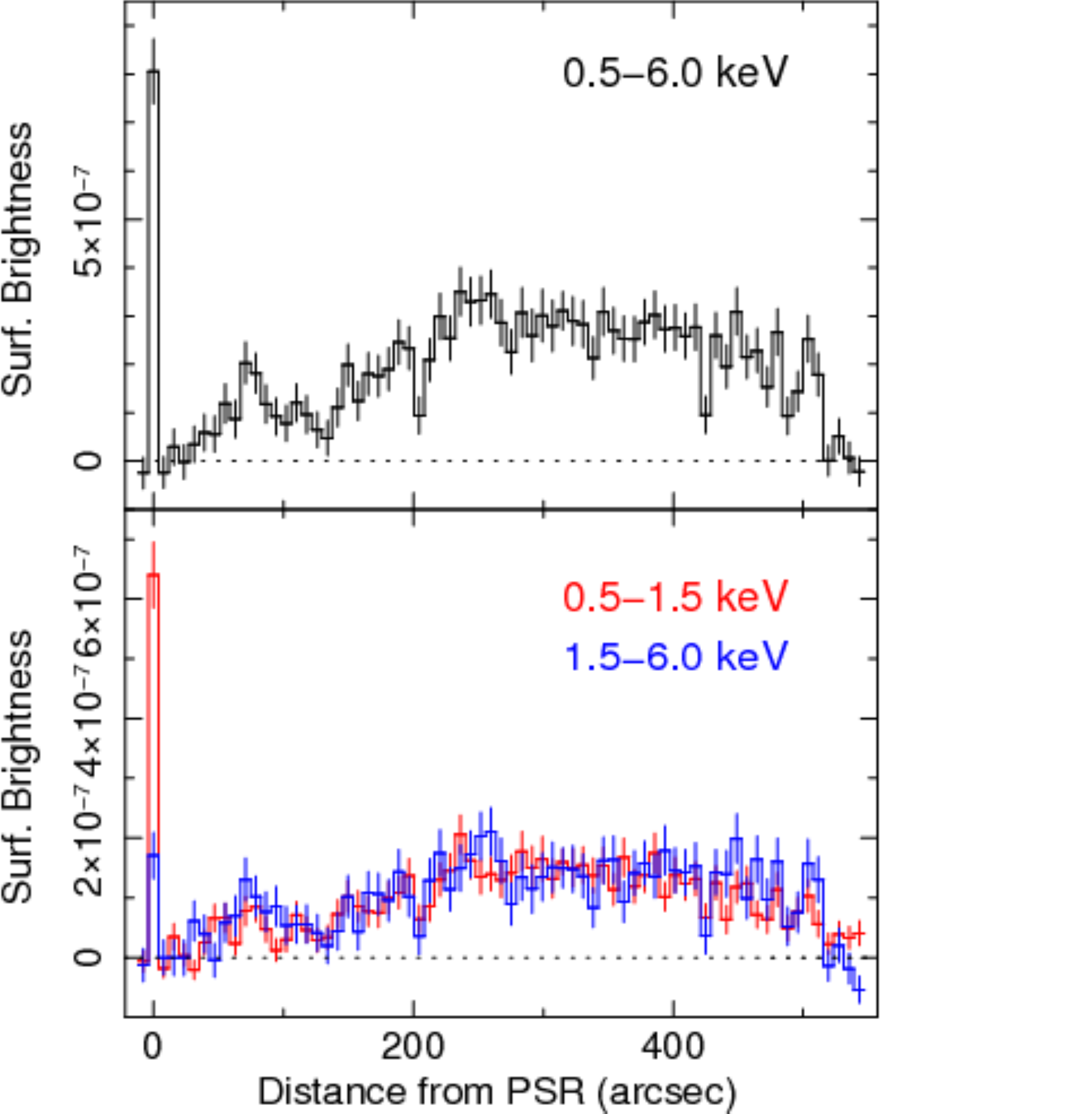}
\caption{Exposure-corrected, background subtracted surface
  brightness profiles of the tail along its 
main (North-West to South-East) axis (see also Fig.~\ref{regions}).  The upper panel 
shows the 0.5-6.0 keV energy range; the lower panel 
shows the 0.5-1.5 keV and 1.5-6.0 keV energy ranges.
The peak corresponding to PSR J0357+3205 is easily seen. Source ``1'' and ``2''
(see Fig.\ref{trail}) have been removed.
The rather smooth profile of the tail
as well as its broad maximum $\sim4.5'$ away from the pulsar is apparent. 
A possible local minimum in the
surface brightness is also seen at $\sim2'$ from the pulsar position.
The profiles in the soft (0.5-15.5 keV) and hard (1.5-6 keV) energy range
are almost indistinguishable.
\label{length}}
\end{figure}

\begin{figure}
\centering
\includegraphics[angle=0,scale=0.3]{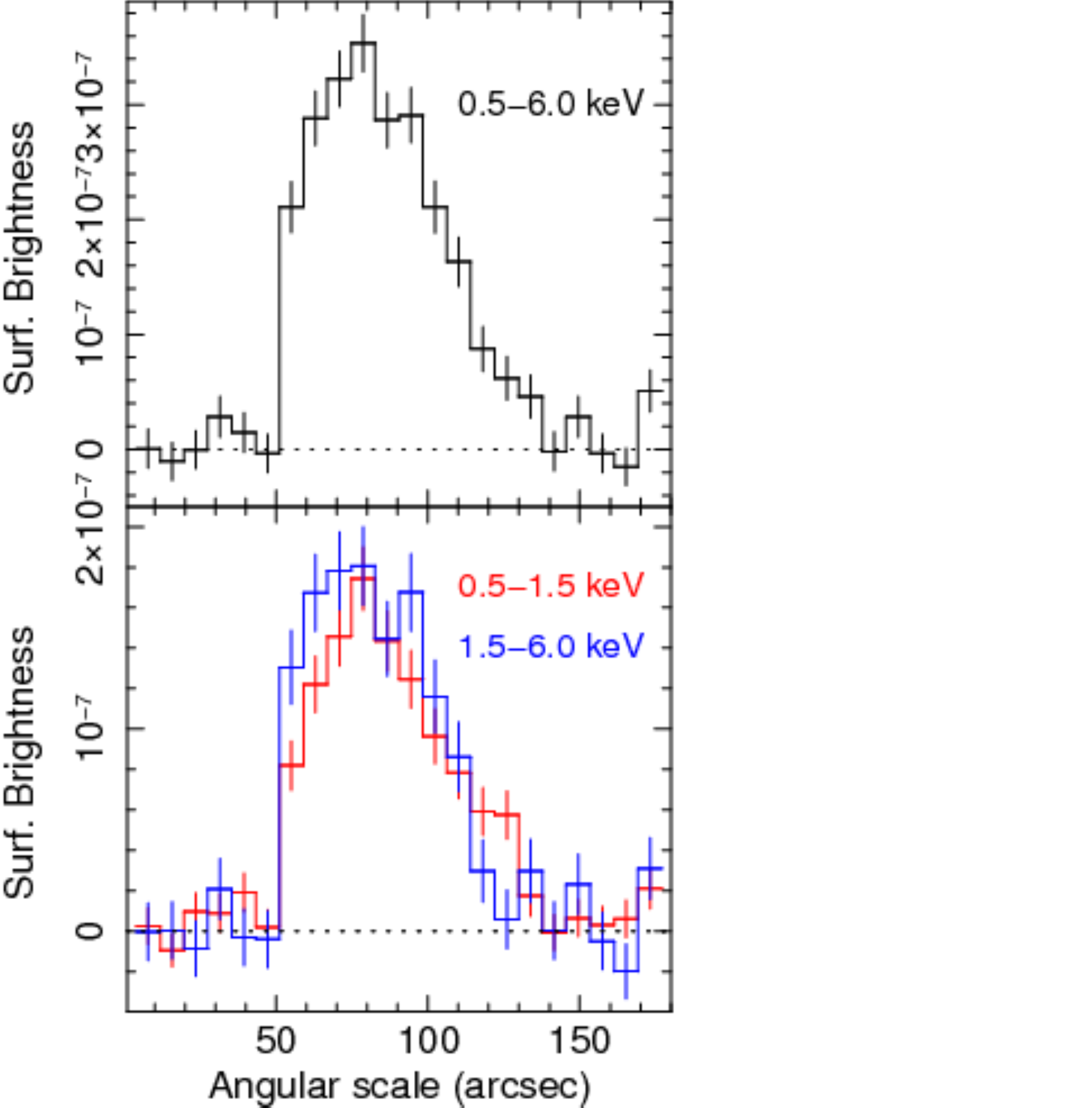}
\caption{Exposure-corrected, background subtracted surface
  brightness profiles of the tail along its width. 
The angular scale refers to the North-East to South-West direction
marked as ``p2'' in Fig.~\ref{regions}.
The sharp edge on the Northeastern
side is apparent, as well as the shallower decay on the opposite side. The
profiles in the soft (0.5-1.5 keV) and hard (1.5-6 keV) energy ranges
are very similar, with a slightly sharper edge in the hard band.
\label{width}}
\end{figure}

\begin{figure}
\centering
\includegraphics[angle=0,scale=0.3]{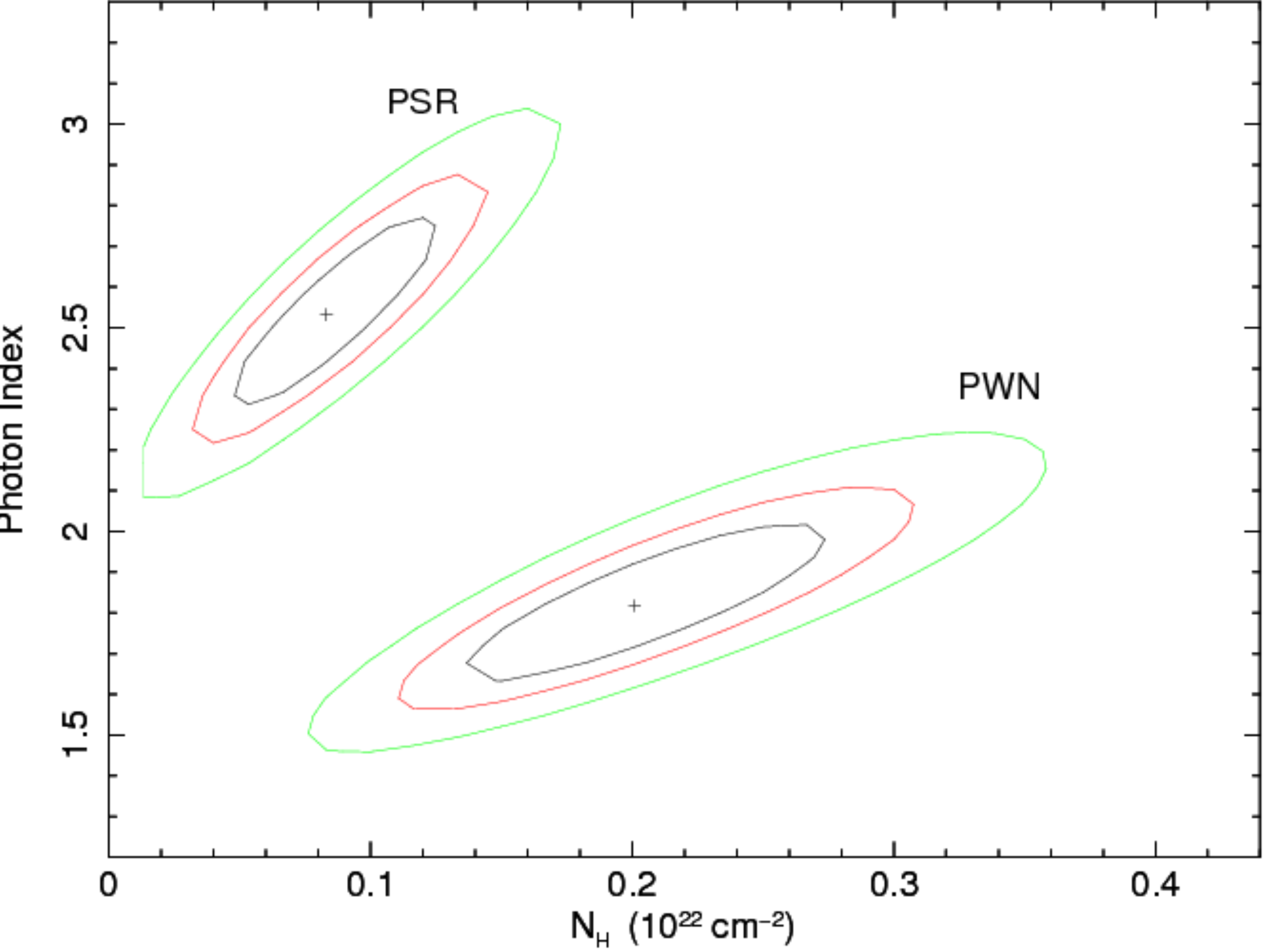}
\caption{Results of spectroscopy on the pulsar counterpart
as well as on the diffuse emission feature. Using an absorbed
power law model (see text), error ellipses (at 68\%, 90\% and 99\%
confidence level) for the
absorbing column $N_H$ and the photon index $\Gamma$ are shown.
\label{contours}}
\end{figure}

A further problem with the bow-shock picture is related to the brightness
profile of the tail, which  is remarkably different from what is observed for all
other diffuse structures interpreted as ram-pressure dominated PWN. Figures \ref{int}
and \ref{trail} clearly show that
the surface brightness grows as a function of angular distance from PSR J0357+3205 
and reaches a broad maximum $\sim4'$ away from the pulsar, while 
all the elongated structures imaged so far have their peak 
close to their parent pulsar position
(although localized, bright ``blobs''
along the tails have been observed, see e.g. Kargaltsev \& Pavlov 2008, and 
interpreted as due to kink instabilities in the particle flow). 
Invoking geometric effects, such as bending of the tail along the line of sight,
producing limb brightening (higher column density of emitting particles) 
would require ad-hoc assumptions for the tail 3-D structure. Since no
plausible explanations for the origin of such bending are apparent, we discard
such a possibility. Lack of any significant spectral evolution along the tail 
ultimately prevents us from drawing conclusions on the physical nature of its peculiar profile.

The ``asymmetric'' brightness profile of the tail 
in the direction perpendicular
to its main axis (with its sharp north-eastern edge and its shallower
decay towards South-West)
is also remarkably different from what is observed for any
other diffuse structure interpreted as ram-pressure dominated PWN, but
is reminiscent of the case of the peculiar diffuse X-ray feature
associated to PSR B2224+65, powering the ``guitar'' nebula
(Hui 2007, Johnson 2010). 
The extended feature seen there is
remarkably misaligned (by $118^{\circ}$) wrt. the direction of the pulsar proper motion
(Hui 2007) and therefore it has been interpreted in a different frame, either as a 
``magnetically-confined'' PWN (Bandiera 2008), or as a jet
 from the pulsar (Johnson 2010). Both pictures naturally predict  
the feature to be brighter in the leading edge
(the one in the direction
of the proper motion), where ``fresh'' electrons are injected. 
The profile in the trailing edge is expected to fade smoothly,
dominated by cooling of the electrons deposited by the moving source
(i.e. the feature is a ``synchrotron wake'' along its minor axis). 
Thus, if this is the case, 
the proper motion of the pulsar 
should not be aligned with the tail main axis
and the tail itself should display a proper motion.
In view of the lack of information about the 
pulsar proper motion, it is premature to discuss 
such scenarii any more. We note, however, that
both pictures are not free from 
difficulties. For instance, the jet explanation cannot 
easily explain the lack of any appreciable bending of 
the structure due to ram pressure from the ISM. On the other  
hand, the magnetically confined PWN would require a very 
intense (50 $\mu$G), ordered ambient magnetic field 
(for further details on such pictures for the case of PSR B2224+65 Bandiera 2008, Johnson 2010).
Moreover, the broad maximum at a large distance from the pulsar 
would not be accounted for easily in these pictures (which, similarly to the
Bow-shock picture, predict
a brightness peak close to the pulsar). 

\subsection{Discussion}

We have detected the faint X-ray counterpart of the middle-aged, $\gamma$-ray only, pulsar
PSR J0357+3205, together with an associated, elongated feature of diffuse
X-ray emission. The pulsar emission is consistent with a purely
magnetospheric, non-thermal origin. Future deep X-ray observations will allow
to better constrain the interstellar absorption (consistent with a distance 
of a few hundred parsecs) and possibly to detect pulsations. As for the case of the
$\gamma$-ray only pulsar in the CTA-1 supernova remnant (Caraveo 2010),
this could unveil the presence of thermal emission from rotating hot spots,
possibly associated to polar cap reheating by magnetospheric currents.
The diffuse feature is $\sim9'$ long (to our knowledge, 
considering its angular extension, this
is the largest ``tail'' of X-ray emission so far associated to 
a rotation-powered pulsar) and displays a hard, non-thermal spectrum. The nature 
of such feature cannot be firmly established. A crucial piece of information
could come from  the pulsar proper motion. In this respect, if 
the lack of a discernible pulsar wind termination shock is indeed due to 
a very high pulsar velocity ($\sim1000$ km s$^{-1}$), at a distance of 500 pc
this would translate to a proper motion of
$\sim0.42''$ yr$^{-1}$, a value which is within the reach of {\it Chandra},
even with a short time baseline ($\sim2$ yr). We note that precise timing of LAT 
photons is not expected to be sensitive to the proper motion of PSR J0357+3205 
(Ray et al. 2010 estimated that timing based on 5 years of LAT data
will yield a positional accuracy of $\sim2''$). 
A proper motion aligned with the tail
main axis would point to a bow-shock PWN interpretation, which will have, in
any case, to
face difficulties related to the energetics of the emitting particles as 
well as to the peculiar brightness profile.  
Conversely, a proper motion misaligned with respect to the tail axis
would point to a ``Guitar''-like system, to be interpreted as a 
magnetically confined PWN or as a pulsar jet. In such a case, 
proper motion of the tail itself could be detected. 
A further check on the tail nature could come from deep  X-ray observations,
which could allow to detect spectral steepening in its emission,
possibly shedding light on the geometry of the injection of particles
in the nebula and pointing either to the bow-shock scenario, or 
to the Guitar-like picture.
A long XMM observation, recently granted,  will clarify the spectral
behaviour 
of the pulsar as well as of its record long tail.

\clearpage

\clearpage

\chapter{Study of the pulsar population in the X-ray band}

A previous study, based on the pulsars published in the first {\it Fermi} pulsar catalogue (Abdo et al. 2010a)
has been published in Astrophysical Journal (and is reported in appendix).
Here, we apply the same method to a larger pulsar sample. As will be seen, using a larger sample
makes it possible to strengthen the somewhat qualitative conclusions by Marelli et al. 2011.

\section{Pulsars' analysis}

\subsection{Description of Data}
\label{data}

We consider all $\gamma$-ray pulsars discovered by {\it Fermi} until May 2011.
Our sample comprehends 88 pulsars:\\
- 61 detected using radio ephemerides, 27 of which are millisecond pulsars;\\
- 24 found through blind searches; of these 3 (J1741-2054, J1907+06 and J2032+4127) were later found to have also a radio emission
and, as such, they were added to the radio emitting ones.\\
Moreover, despite the detection of pulsations from RL pulsars J1513-5908 and J1531-5610, the presence
of nearby sources makes their $\gamma$-ray fluxes uncertain so that we decided not to use them in all the analyses.\\
Thus, our sample of $\gamma$-ray emitting neutron stars consists of 35 radio-loud pulsars (RLP), 28 radio-loud millisecond pulsars (MSP) and 24
radio-quiet pulsars (RQP).
All the $\gamma$-ray fluxes and proprieties are taken from Saz Parkinson et al.(2010), Abdo et al.(2009) or, where necessary,
are directly provided from the {\it Fermi}/LAT collaboration (see Abdo et al. 2012 in preparation).
While for pulsars listed in Abdo et al. (2009), Saz Parkinson et al. (2010) an exponential cutoff was used for fitting the spectrum,
some the newest (and not jet published) ones are fitted with a simple powerlaw (see Table \ref{tottab-3} for details).

Differently from the $\gamma$-ray band, the X-ray coverage of the {\it Fermi} LAT pulsars is uneven since the majority of the newly discovered radio-quiet PSRs 
have never been the target of a deep X-ray observation,
while for other well-known $\gamma$-ray pulsars - such as Crab, Vela and Geminga - one can rely on a lot of observations.
To account for such an uneven coverage, we classify the X-ray spectra on the basis of the public X-ray data available, thus assigning:\\
- label $"$0$"$ to pulsars with no confirmed X-ray counterparts (or without a non-thermal spectral component);\\
- label $"$1$"$ to pulsars with a confirmed counterpart but too few photons to assess its spectral shape;\\
- label $"$2$"$ to pulsars with a confirmed counterpart for which the data quality allows for the analysis of both the pulsar and the nebula (if present).\\
An $"$ad hoc$"$ analysis was performed for seven pulsars for which the standard analysis couldn't be applied (e.g. owing to the very high thermal component
of Vela or to the closeness of J1418-6058 to an AGN). Table 2 provides details on such pulsars.\\
We searched for X-ray counterparts' positions and 3$\sigma$-confidence errors
into {\it XMM-Newton} observations by using the standard {\em edetect\_chain} task,
following the prescriptions reported in the
$"$Users Guide to the XMM-Newton Science Analysis System$"$, Issue 8.0, 2011 (ESA: XMM-Newton SOC).
Similarly, for {\it Chandra} analyses we used the {\em celldetect} tool's prescriptions
from the $"$CIAO Science Threads Guide$"$ \footnote{http://cxc.harvard.edu/ciao/threads/all.html}
and the {\em detect} tool inside the XIMAGE distribution for {\it SWIFT/XRT} analyses
\footnote{http://heasarc.gsfc.nasa.gov/docs/xanadu/ximage/manual/ximage.html}.\\
We consider an X-ray counterpart to be confirmed if:\\
- X-ray pulsation has been detected;\\
- X and Radio coordinates coincide;\\
- X-ray source position has been validated through the blind-search algorithm developed by the {\it Fermi} collaboration (Abdo et al. 2008, Ray et al. 2011).\\
If none of these conditions apply, $\gamma$-ray pulsar is labeled as $"$0$"$.\\
According to our classification scheme we have 33 type-0, 5 type-1 and 48 type-2 pulsars. In total
53 $\gamma$-ray neutron stars, 25 RLP, 13 MSP and 15 RQP have an X-ray counterpart.

Since the X-ray observation database is continuously growing, the results available in literature encompass only fractions of the X-ray data now available.
Moreover, they have been obtained with different versions of the standard analysis software or using different
techniques to account for the PWN contribution.
Thus, with the exception of the well-known and bright X-ray pulsars, 
such as Crab or Vela, we re-analyzed all the X-ray data publicly available
following an homogeneous procedure.
If only a small fraction of the data are publicly available, we quoted results from a literature search.

In order to assess the X-ray spectra of {\it Fermi} pulsars, we used photons with energy 0.3$<$E$<$10 keV collected
by all the on-flight telescopes operating in the soft X-ray band:
{\it Chandra}/ACIS (Garmire et al. 2003), {\it XMM-Newton} (Struder
et al. 2001; Turner et al. 2001) and {\it SWIFT}/XRT (Burrows
et al. 2005).  
We selected all the public observations (as of April 2011) that overlap the error box 
of {\it Fermi} pulsars or the Radio coordinates.
We chose not to use Suzaku data due to its poor spatial resolution ($\sim$1') that makes it impossible
to disentangle pulsars from their nebulae nor from nearby sources. We neglected also all 
{\it Chandra}/HRC observations owing to the lack of energy resolution of the instrument.\\
To analyze {\it Chandra} data, we used the {\it Chandra} Interactive Analysis of Observation software (CIAO version 4.1.2).
The {\it Chandra} point spread function depends on the off-axis angle: we used for all the point sources
an extraction area around the pulsar that contains 90\% of the events. For instance, for on-axis sources we
selected all the photons inside a 2$"$ radius circle, while we extracted photons
from the inner part of PWNs (excluding the 2$"$ radius circle of the point source) in order to assess the nebular spectra: such
extended regions vary for pulsar to pulsar as a function of the nebula dimension and flux (see the following
subsections for details).\\
We analyzed all the {\it XMM-Newton} data (both from PN and MOS1/2 detectors)
with the {\it XMM-Newton} Science Analysis Software (SASv8.0). 
The raw observation data files (ODFs) were processed using standard pipeline tasks (epproc for PN, emproc for MOS data); we used only photons
with event pattern 0-4 for the PN detector and 0-12 for the MOS1/2 detectors.
When necessary, an accurate screening for soft proton flare events was done, following the prescription by De Luca \&
Molendi (2004).\\
If, in addition to XMM data, deep {\it Chandra} data were also available, we made an XMM spectrum of the entire PSR+PWN and used the {\it Chandra} higher resolution
in order to disentangle the two contributions. When only {\it XMM-Newton} data were available, the point source 
was analyzed by selecting all the photons inside a 20$"$ radius circle while 
the whole PWN (with the exception of the 20$"$ radius circle of the point source) was used in order to assess the nebular spectrum.\\
Where deeper observations were not available,
we analyzed all the {\it SWIFT}/XRT data with HEASOFT version 6.5 selecting all the photons inside a 20$"$ radius circle.
If multiple data sets collected by the same instruments were found, spectra, response and effective area files
for each dataset were added by using the mathpha, addarf and addrmf HEASOFT tools.

All the spectra have been studied with XSPEC v.12 (Arnaud 1996) choosing, whenever possible, the same
background regions for all the different observations of each pulsar. All the data were rebinned in order to 
have at least 25 counts per channel, as requested for the validity of $\chi^2$ statistic.
When needed, due to the limited number of counts, we used the C-statistic
approach implemented in XSPEC which requires no spectral grouping, nor background
subtraction; owing to these requirements, such approach can be used only when the background
contribution is negligible.\\
The XMM-{\it Chandra} cross calibration studies (Stuhlinger
et al. 2008) report only minor changes in flux
($<$10\%) between the two instruments. When both XMM and {\it Chandra} data were available, a constant has been introduced to
account for such uncertainty. Conversely, when the data were collected only by one instrument, a systematic error
was introduced.\\
All the PSRs and PWNs have been fitted with absorbed powerlaws; when statistically needed,
a blackbody component has been added to the pulsar spectrum. Since PWNs typically show a 
powerlaw spectrum with a photon index which steepens moderately as a function 
of the distance from the PSR (Gaensler \& Slane
2006), we used only the inner part
of each PWN. 
Absorption along the line of sight has been obtained through the fitting procedure
but for pulsars with very low statistics we used values derived from
observations taken in different bands.

\subsection{X-ray Analysis}
\label{analysis}

For pulsars with a good X-ray coverage we proceed as follows.\\
If only {\it XMM-Newton} public observations were available, we tried to take into account the PWN contribution.
First we searched the literature for evidence, if any, of the presence of a PWN and, if nothing was found, we analyzed
the data to search for extended emission. To perform such an analysis, we derived a radial brilliance
profile and we fitted it with the predicted Point Spread Function (PSF) of a pointlike source (XMM calibration document CAL-TN-0052).
If no evidence for the presence of a PWN was found, 
we used PN and MOS1/2 data in a simultaneous spectral fit.
On the other hand, if a PWN was present, its contribution was evaluated on a case by case basis. If the statistic was good enough, we studied
simultaneously the inner region, containing both PSR and PWN, and the extended region surrounding it.
The inner region data were described by two absorbed (PWN and PSR) powerlaws, while the outer one by a single (PWN) powerlaw.
The N$_H$ and the PWN photon index values were the same in the two (inner and outer) datasets.\\
When public {\it Chandra} data were available, we evaluated separately PSR and PWN (if any) in a similar way.
In order to search for extended emission in {\it Chandra} datasets, we made a radial brilliance profile and we compared it
with the simulated profile of a pointlike source obtained by using CHART and MARX tools.

If both {\it Chandra} and XMM public data were available, we exploited {\it Chandra} space resolution to evaluate the PWN contribution by:\\
- obtaining two different spectra of the inner region (a), encompassing both PSR and PWN and of the outer region (b) encompassing only the PWN;\\
- extracting a total XMM spectrum (c) containing both PSR and PWN: this is the only way to take into account the XMM's larger PSF;\\
- fitting simultaneously a,b,c with two absorbed powerlaws and eventually (if statistically significant) an absorbed blackbody, using the same N$_H$; 
a multiplicative constant was also introduced in order to account for a possible discrepancy between 
{\it Chandra} and XMM calibrations;\\
- forcing to zero the normalization(s) of the PSR model(s) in the {\it Chandra} outer region and freeing the other normalizations in the {\it Chandra} datasets; 
fixing the XMM PSR normalization(s) at the inner {\it Chandra} dataset one and the XMM PWN normalization at the inner+outer normalizations of the {\it Chandra} PWN.\\
Only for few well-known pulsars, or pulsars for which the dataset is not yet entirely public, we used results taken from 
the literature (see Table \ref{tottab-2}). Where necessary,
we used XSPEC in order to obtain the flux in the 0.3-10 keV energy range and to evaluate the unabsorbed flux.\\

For pulsars with a confirmed counterpart but too few photons to discriminate the spectral shape, we evaluated an hypothetical unabsorbed
flux by assuming that a single powerlaw spectrum with a photon index of 2 to describe PSR+PWN. We also assumed that 
the PWN and PSR thermal contributions are 30\% of the entire source flux (a sort of mean value of all
type 2 pulsars considered).
To evaluate the absorbing column, we need a distance value which can come either from
the radio dispersion or - for radio-quiet pulsars - from the pseudo-distance defined in Saz Parkinson et al. (2010) as:
$d=0.51\dot{E_{34}}^{1/4}/F_{\gamma,10}^{1/2}$ kpc \\
where $\dot{E}=\dot{E}_{34}\times10^{34}erg/s$ and $F_{\gamma}=F_{\gamma,10}\times10^{-10}erg/cm^2s$
for a beam correction factor f$_{\gamma}$ set to be 1 (Watters et al. 2009) for all pulsars.\\
Then, the HEASARC WebTools were used to find the galactic column density (N$_H$) 
in the pulsar direction; with the distance information,
we could rescale the column density value of the pulsar.
We found the source count rate by using the XIMAGE task (Giommi et al. 1992).
Then, we used the WebPimms tool inside the WebTools package to evaluate the source unabsorbed flux.
Such a value has been corrected to account for the PWN and PSR thermal contributions.
We are aware that each pulsar can have a different
photon index, as well as thermal and PWN contributions so that we used these mean values only as a first
approximation. Pulsars well fitted by both a simple thermal and non-thermal models will also be listed
as type 1, using the best fit non-thermal spectrum to obtain their luminosity.\\
All the low-quality pulsars (type 1) will be treated separately. Indeed, all the results of this paper
will be based only on high-quality objects (type 2).

For pulsars without a confirmed counterpart we evaluated the X-ray unabsorbed flux upper limit assuming
a single powerlaw spectrum with a photon index of 2 to describe PSR+PWN  and using a signal to noise of 3.\\
The column density has been evaluated as above.
Under the previous hypotheses, we used the signal-to-noise definition in order to compute the upper limit to the absorbed flux of the X-ray
counterpart. Next we used XSPEC to find the unabsorbed upper limit flux.

On the basis of our X-ray analysis we define a subsample of {\it Fermi} $\gamma$-ray pulsars for which we have, at once, 
reliable X-ray data (type 2 pulsars) and satisfactory distance estimates such as parallax, radio dispersion
measurement, column density estimate, SNR association.
Such a subsample contains 23 RLP, 8 MSP and 6 RQP. The low number of radio quiet pulsars is to be
ascribed to lack of high quality X-ray data.
Moreover, we have 7 additional radio-quiet pulsars and 4 millisecond pulsars with reliable X-ray data but without a satisfactory distance estimate.

In Tables \ref{tottab-2} and \ref{tottab-3} we reported the gamma-ray and X-ray parameters of the selected pulsars. 
Using the P and $\dot{P}$ values taken from Abdo et al. 2010a and Saz Parkinson et al. 2010, we computed
the values reported in Table \ref{tottab-1}. Most of the distance values are taken from Abdo et al. 2010a and Saz Parkinson et al. 2010 (see Table \ref{tottab-1}).

All the pulsars' detailed analyses are reported in Chapter ~\ref{detail}.

\clearpage

\setlength{\LTleft}{-1pt}
\begin{footnotesize}
\begin{landscape}
\begin{center}
\begin{longtable}{cccccccccc}
PSR Name & P$^a$ & $\dot{P}^a$ & $\tau_c$ & $\tau_{snr}^b$ & B$_{lc}$ & d$^a$ & $\dot{E}$ & Type$^e$ & PWN$^f$ \\
 & (ms) & $(10^{-15})$ & (ky) & (ky) & (kG) & (kpc) & ($10^{34}$erg/s) & & \\
\endhead
J0007+7303 & 316 & 361 & 14 & 13$\pm$3 & 3.1 & 1.4$\pm$0.3 & 45.2 & g & Y\\
J0030+0451 & 4.9 & $10^{-5}$ & 7.7$\times10^{6}$ & - & 17.8 & 0.30$\pm$0.09 & 0.3 & m & N\\
J0034-0534 & 1.9 & 5$\times10^{-6}$ & 6$\times10^{6}$ & - & 136.0 & 0.53$\pm$0.21 & 5.6 & m & N\\
J0101-6422 & 2.6 & 4.8$\times10^{-6}$ & 8.4$\times10^{6}$ & - & 60.8 & $\sim$0.55 & 1.1 & m & ?\\
J0205+6449 & 65.7 & 194 & 5 & 4.25$\pm$0.85 & 115.9 & 2.9$\pm$0.3 & 2700 & r & Y\\
J0218+4232 & 2.3 & 7.7$\times10^{-5}$ & 5$\times10^{5}$ & - & 313.1 & 3.25$\pm$0.75 & 24 & m & N\\
J0248+6021 & 217 & 55.1 & 63 & - & 3.1 & 5.5$\pm$3.5 & 21 & r & ?\\
J0340+4130 & 3.3 & 7$\times10^{-6}$ & 7.5$\times10^{6}$ & - & 40.5 & 1.8 & 0.768 & m & ?\\
J0357+32  & 444 & 12 & 590 & - & 0.2 & 0.5$^c$ & 0.5 & g & Y\\
J0437-4715 & 5.8 & 1.4$\times10^{-5}$ & 6.6$\times10^{6}$ & - & 13.7 & 0.1563$\pm$0.0013 & 0.3 & m & N\\
J0534+2200 & 33.1 & 423 & 1.0 & 0.955 & 950 & 2.0$\pm$0.5 & 46100 & r & Y\\
J0610-2100 & 3.9 & 1.2$\times10^{-5}$ & 5.1$\times10^{6}$ & - & 34.9 & 3.5 & 0.798 & m & ?\\
J0613-0200 & 3.1 & 9$\times10^{-6}$ & 5.3$\times10^{6}$ & - & 54.3 & 0.48$^{+0.19}_{-0.11}$ & 1.3 & m & N\\
J0614-3330 & 3.1 & 1.78$\times10^{-5}$ & 2.8$\times10^{6}$ & - & 75.5 & 1.9 & 2.36 & m & ?\\
J0631+1036 & 288 & 105 & 44 & - & 2.1 & 2.185$\pm$1.440 & 17.3 & r & ?\\
J0633+0632 & 297 & 79.5 & 59 & - & 1.7 & 1.1$^c$ & 11.9 & g & Y\\
J0633+1746 & 237 & 11 & 340 & - & 1.1 & 0.250$_{-0.062}^{+0.12}$ & 3.3 & g & N\\
J0659+1414 & 385 & 55 & 110 & 86$\pm$8 & 0.7 & 0.288$_{-0.027}^{+0.033}$ & 3.8 & r & N\\
J0734-1559 & 155 & 12.1 & 203 & - & 3.5 & 1.3 & 12.8 & g & ?\\
J0742-2822 & 167 & 16.8 & 160 & - & 3.3 & $2.07_{-1.07}^{+1.38}$ & 14.3 & r & ?\\
J0751+1807 & 3.5 & 6.2$\times10^{-6}$ & 8$\times10^{6}$ & - & 32.3 & 0.6$_{-0.2}^{+0.6}$ & 0.6 & m & N\\
J0835-4510 & 89.3 & 124 & 11 & 13$\pm$1 & 43.4 & 0.287$_{-0.017}^{+0.019}$ & 688 & r & Y\\
J0908-4913 & 107 & 15.1 & 112 & - & 9.9 & 6.57$_{-0.90}^{+1.30}$ & 48.6 & r & Y\\
J0940-5428 & 87.5 & 32.9 & 42 & - & 24.2 & 2.95 & 193 & r & ?\\
J1016-5857 & 107 & 80.8 & 21 & 10 & 23.0 & 9$_{-2}^{+3}$ & 260 & r & Y\\
J1023-5746 & 111 & 384 & 4.6 & - & 44 & 2.4$^c$ & 1095 & g & Y\\
J1024-0719 & 5.2 & 1.85$\times10^{-5}$ & 4.4$\times10^{6}$ & - & 21.1 & 0.53$\pm$0.12 & 0.519 & m & N\\
J1028-5819 & 91.4 & 16.1 & 90 & - & 14.6 & 2.33$\pm$0.70 & 83.2 & r & ?\\
J1044-5737 & 139 & 54.6 & 40.3 & - & 9.5 & 1.5$^c$ & 80.3 & g & ?\\
J1048-5832 & 124 & 96.3 & 20 & - & 16.8 & 2.71$\pm$0.81 & 201 & r & Y\\
J1057-5226 & 197 & 5.8 & 540 & - & 1.3 & 0.72$\pm$0.20 & 3.0 & r & N\\
J1119-6127 & 408 & 4022 & 1.61 & - & 5.7 & 8.4$\pm$0.4 & 234 & r & Y\\
J1124-5916 & 135 & 747 & 3 & 2.99$\pm$0.06 & 37.3 & 4.8$_{-1.2}^{+0.7}$ & 1190 & r & Y\\
J1135-6055 & 114 & 78.5 & 23 & - & 19.3 & 2.88 & 209 & g & Y\\
J1231-1411 & 3.7 & 1.86$\times10^{-5}$ & 3.1$\times10^{6}$ & - & 50.3 & 0.4 & 1.47 & m & N\\
J1357-6429 & 166 & 360 & 7.3 & - & 16.2 & 2.4$\pm$0.6 & 310 & r & Y\\
J1410-6132 & 50 & 31.8 & 25 & - & 96.7 & 15.6 & 1000 & r & ?\\
J1413-6205 & 110 & 27.7 & 62.9 & - & 12.3 & 1.4$^c$ & 82.7 & g & ?\\
J1418-6058 & 111 & 170 & 10 & - & 29.4 & 3.5$\pm$1.5 & 495 & g & Y\\
J1420-6048 & 68.2 & 83.2 & 13 & - & 69.1 & 5.6$\pm$1.7 & 1000 & r & Y\\
J1429-5911 & 116 & 30.5 & 60.2 & - & 11.3 & 1.6$^c$ & 77.5 & g & ?\\
J1459-60 & 103 & 25.5 & 64 & - & 13.6 & 1.5$^c$ & 91.9 & g & ?\\
J1509-5850 & 88.9 & 9.2 & 150 & - & 11.8 & 2.6$\pm$0.8 & 51.5 & r & Y\\
J1513-5908 & 150.6 & 1537 & 1.55 & 13$\pm$7 & 42.6 & 5.2$\pm$1.4 & 1774 & r$^{g}$ & Y\\
J1531-5610 & 84.2 & 13.7 & 97.4 & - & 17.2 & 2.1 & 90.5 & r$^{g}$ & ?\\
J1600-3053 & 3.6 & 9.5$\times10^{-6}$ & 6$\times10^{6}$ & - & 37.9 & 1.53 & 0.8 & m & ?\\
J1614-2230 & 3.2 & 4$\times10^{-6}$ &  1.2$\times10^{6}$ & - & 33.7 & 1.27$\pm$0.39 & 0.5 & m & N\\
J1648-4611 & 165 & 23.7 & 110 & - & 4.2 & 5.71 & 20.8 & r & ?\\
J1658-5324 & 2.44 & 1.17$\times10^{-5}$ & 3.3$\times10^{6}$ & - & 111 & 0.9 & 3.2 & m & ?\\
J1709-4429 & 102 & 93 & 18 & 5.5$\pm$0.5 & 26.4 & 2.5$\pm$1.1 & 341 & r & Y\\
J1713+0747 & 4.57 & 8.53$\times10^{-5}$ & 8.5$\times10^{5}$ & - & 62.6 & 1.05 & 3.5 & m & ?\\
J1718-3825 & 74.7 & 13.2 & 90 & - & 21.9 & 3.82$\pm$1.15 & 125 & r & Y\\
J1730-3350 & 139.5 & 14.8 & 149 & - & 5.1 & 4.24 & 21.5 & r & ?\\
J1732-31 & 197 & 26.1 & 120 & - & 2.7 & 0.6$^c$ & 13.6 & g & ?\\
J1741-2054 & 414 & 16.9 & 390 & - & 0.3 & 0.38$\pm$0.11 & 0.9 & r & ?\\
J1744-1134 & 4.1 & 7$\times10^{-6}$ & 9$\times10^{6}$ & - & 24 & 0.357$_{-0.035}^{+0.043}$ & 0.4 & m & N\\
J1747-2958 & 98.8 & 61.3 & 26 & 163$_{-39}^{+60}$ & 23.5 & 2.0$\pm$0.6 & 251 & r & Y\\
J1801-2451 & 125 & 128 & 15.5 & - & 19.6 & 5.2$\pm$0.5 & 258 & r & Y\\
J1809-2332 & 147 & 34.4 & 68 & 50$\pm$5 & 6.5 & 1.7$\pm$1.0 & 43 & g & Y\\
J1813-1246 & 48.1 & 17.6 & 43 & - & 76.2 & 2.0$^c$ & 626 & g & ?\\
J1823-3021A& 5.44 & 0.0034 & 2.55$\times10^{4}$ & - & 255 & 7.9 & 82.8 & m & ?\\
J1826-1256 & 110 & 121 & 14 & - & 25.2 & 1.2$^c$ & 358 & g & Y\\
J1833-1034 & 61.9 & 202 & 5 & 0.87$_{-0.15}^{+0.20}$ & 137.3 & 4.7$\pm$0.4 & 3370 & r & Y\\
J1836+5925 & 173 & 1.5 & 1800 & - & 0.9 & 0.4$_{-0.15}^{+0.4 d}$ & 1.2 & g & N\\
J1846+0919 & 226 & 9.93 & 360 & - & 1.2 & 1.2$^c$ & 3.4 & g & ?\\
J1902-5105 & 1.74 & 9$\times10^{-6}$ & 3.1$\times10^{6}$ & - & 227 & 1.2 & 6.7 & m & ?\\
J1907+06 & 107 & 87.3 & 19 & - & 23.2 & 1.3$^c$ & 284 & r & ?\\
J1939+2134 & 1.56 & 1.1$\times10^{-4}$ & 2.4$\times10^{5}$ & - & 1020 & 7.7 & 109 & m & N\\
J1952+3252 & 39.5 & 5.8 & 110 & 64.0$\pm$18 & 71.6 & 2.0$\pm$0.5 & 374 & r & Y\\
J1954+2836 & 92.7 & 21.2 & 69.5 & - & 16.4 & 1.7$^c$ & 105 & g & ?\\
J1957+5036 & 375 & 7.08 & 838 & - & 0.3 & 0.9$^c$ & 0.5 & g & ?\\
J1958+2841 & 290 & 222 & 21 & - & 3.0 & 1.4$^c$ & 35.8 & g & ?\\
J1959+2048 & 1.6 & 1.7$\times10^{-5}$ & 1.5$\times10^{6}$ & - & 385 & 2.5$\pm$1.0 & 16.4 & m & Y\\
J2017+0603 & 2.9 & 6.9$\times10^{-6}$ & 6.7$\times10^{6}$ & - & 55.5 & 1.6 & 1.1 & m & ?\\
J2021+3651 & 104 & 95.6 & 17 & - & 26 & 2.1$_{-1.0}^{+2.1}$ & 338 & r & Y\\
J2021+4026 & 265 & 54.8 & 77 & - & 1.9 & 1.5$\pm$0.45 & 11.6 & g & N\\
J2030+3641 & 200 & 6.5 & 488 & - & 1.4 & 8.1 & 3.2 & r & ?\\
J2032+4127 & 143 & 19.6 & 120 & - & 5.3 & 3.60$\pm$1.08 & 26.3 & r & ?\\
J2043+1710 & 2.38 & 5.6$\times10^{-6}$ & 6.7$\times10^{6}$ & - & 82 & 1.8 & 1.6 & m & ?\\
J2043+2740 & 96.1 & 1.3 & 1200 & - & 3.6 & 1.80$\pm$0.54 & 5.6 & r & N\\
J2055+25 & 320 & 4.08 & 1227 & - & 0.3 & 0.4$^c$ & 0.5 & g & ?\\
J2124-3358 & 4.9 & 1.2$\times10^{-5}$ & 6$\times10^{5}$ & - & 18.8 & 0.25$_{-0.08}^{+0.25}$ & 0.4 & m & N\\
J2214+3002 & 3.12 & 1.4$\times10^{-5}$ & 3.5$\times10^{6}$ & - & 65.9 & 1.5 & 1.8 & m & ?\\
J2229+6114 & 51.6 & 78.3 & 11 & 3.90$\pm$0.39 & 134.5 & 3.65$\pm$2.85 & 2250 & r & Y\\
J2238+59 & 163 & 98.6 & 26 & - & 8.6 & 2.1$^c$ & 90.3 & g & ?\\
J2240+5832 & 139 & 15.2 & 145 & - & 5.2 & 7.7 & 22.3 & r & ?\\
J2241-5236 & 2.2 & 6.1$\times10^{-6}$ & 5.7$\times10^{6}$ & - & 106 & 0.5 & 2.3 & m & ?\\
J2302+4442 & 5.2 & 1.3$\times10^{-5}$ & 6.2$\times10^{6}$ & - & 18 & 1.2 & 0.38 & m & N\\
\caption{\\
a : P, $\dot{P}$ and most of the values of the distance are taken from Abdo et al. 2010a, Saz Parkinson 2010.\\
b : Age derived from the associated SNR. Respectively taken from Slane 2004, Gotthelf 2007, Rudie 2008, Hwang 2001, Thompson 2003, Gorenstein 1974, Winkler 2009, Bock 2002, Halpern 2009, Roberts 2008, Tam 2008, Brogan 2005, Bietenholz 2008, Mignani 2008, Kothes 2006.\\
c : These distances are taken from Saz Parkinson 2010 and are obtained under the assumption of a beam correction factor f$_{\gamma}$ = 1 for the gamma-ray emission cone of all pulsars. In this way one obtains:\\
$d=0.51\dot{E_{34}}^{1/4}/F_{\gamma,10}^{1/2}$ kpc\\
where $\dot{E}=\dot{E}_{34}\times10^{34}erg/s$ and $F_{\gamma}=F_{\gamma,10}\times10^{-10}erg/cm^2s$.
See also Saz Parkinson 2010.\\
e : g = radio-quiet pulsars ; r = radio-loud pulsars ; m = millisecond pulsars.\\
f : Only bright PWNs have been considered (with F$^{pwn}_x$ $>$ 1/10 F$^{psr}_x$). The presence or the absence of a bright PWN has been valued by re-adapting the X-ray data (except for the X-ray analyses taken from literature, see the following table).\\
g : PSRs J1513-5908 and J1531-5610 were excluded from the analyses due to their unreliable $\gamma$-ray spectra.
\label{tottab-1}}
\end{longtable}

\begin{longtable}{cccccccccc}
PSR Name & X$^a$ & Inst$^b$ & F$_X^{nt}$ & F$_X^{tot}$ & N$_H$ & $\Gamma_X$ & kT & R$_{BB}$ & Eff$_X^d$\\
 & & & ($10^{-13}erg/cm^{2}s$) & ($10^{-13}erg/cm^{2}s$) & ($10^{20}cm^{-3}$) & & (10$^6$K) & (km) & \\
\endhead
J0007+7303 & 2 & X/C & 0.984$\pm$0.007 & 1.032$\pm$0.008 & 16.6$_{-7.6}^{+8.9}$ & 1.30$\pm$0.18 & 1.18$_{-0.21}^{+0.37}$ & 0.64$_{-0.20}^{+0.88}$ & 5.13e-05\\
J0030+0451 & 2 & X & 2.55$\pm$0.29 & 4.17$\pm$0.47 & 6.4$_{-2.4}^{+3.4}$ & 3.44$\pm$0.26 & 2.17$_{-0.13}^{+0.09}$ & 0.111$_{-0.023}^{+0.040}$ & 8.21e-4\\
J0034-0534 & 0 & X & $<$0.058 & 0.058$\pm$0.011 & $<$56.3 & - & 2.23$_{-1.22}^{+0.62}$  & 0.036$_{-0.025}^{+0.525}$ & -\\
J0101-6422 & 0 & S & $<$2.31 & $<$2.31 & 1$^c$ & 2 & - & - & -\\
J0205+6449 & 2 & C & 19.7$\pm$0.7 & 2.18$\pm$0.08 & 45.0$_{-1.1}^{+1.3}$ & 1.77$\pm$0.03 & 1.88$_{-0.13}^{+0.14}$ & 1.67$_{-0.36}^{+0.43}$ & 7.38e-05\\
J0218+4232 & 2 & X & 4.62$_{-0.63}^{+0.43}$ & 4.62$_{-0.63}^{+0.43}$ & 2.70$_{-2.70}^{+3.76}$ & 1.11$_{-0.11}^{+0.08}$ & - & - & 2.34e-3\\
J0248+6021 & 0 & S & $<$9.00 & $<$9.00 & 80$^c$ & 2 & - & - & -\\
J0340+4130 & 0 & X & $<$0.20 & $<$0.20 & 5$^c$ & 2 & - & - & -\\
J0357+32 & 2 & C & 0.64$_{-0.06}^{+0.09}$ & 0.64$_{-0.06}^{+0.09}$ & 8.0$\pm$4.0 & 2.53$\pm$0.25 & - & - & -\\
J0437-4715 & 2 & X/C & 7.91$_{-0.60}^{+0.50}$ & 12.1$_{-0.8}^{+0.7}$ & 1.58$_{-1.09}^{+0.93}$ & 2.98$_{-0.11}^{+0.09}$ & 2.60$\pm$0.05 & 0.0639$_{-0.0057}^{+0.0065}$ & 8.20e-4\\
J0534+2200 & 2 & L$^f$ & 44300$\pm$1000 & 44300$\pm$1000 & 34.5$\pm$0.2 & 1.63$\pm$0.09 & - & - & 4.62e-3\\
J0610-2100 & 0 & S & $<$3.46 & $<$3.46 & 3$^c$ & 2 & - & - & -\\
J0613-0200 & 2* & X & 0.959$_{-0.443}^{+0.685}$ & 0.959$_{-0.443}^{+0.685}$ & $<$3.30 & 2.05$_{-0.49}^{+0.22}$ & - & - & 2.22e-4\\
J0614-3330 & 1 & S & 0.518$\pm$0.201 & 0.518$\pm$0.201 & 42.1$_{-23.7}^{+87.7}$ & 4.91$_{-3.37}^{+4.94}$ & - & - & 9.52e-4\\
J0631+1036 & 0 & X & $<$0.225 & $<$0.225 & 20$^c$ & 2 & - & - & -\\
J0633+0632 & 2 & C & 0.625$\pm$0.050 & 1.71$\pm$0.14 & 6.08$_{-6.08}^{+21.91}$ & 1.45$_{-0.82}^{+0.76}$ & 1.46$_{-0.39}^{+0.28}$ & 0.817$_{-0.611}^{+8.944}$ & -\\
J0633+1746 & 2 & L$^f$ & 4.97$_{-0.27}^{+0.09}$ & 12.6$_{-0.7}^{+0.2}$ & 1.07$^e$ & 1.7$\pm$0.1 & 0.190$\pm$0.030 & 0.04$\pm$0.01 & 1.14e-4\\
J0659+1414 & 2 & L$^f$ & 4.06$_{-0.59}^{+0.03}$ & 168$_{-24}^{+1}$ & 4.3$\pm$0.2 & 2.1$\pm$0.3 & 0.125$\pm$0.003 & 1.80$\pm$0.15 & 1.06e-4\\
J0734-1559 & 0 & S & $<$2.36 & $<$2.36 & 20$^c$ & 2 & - & - & -\\
J0742-2822 & 0 & X & $<$0.225 & $<$0.225 & 20$^c$ & 2 & - & - & -\\
J0751+1807 & 2 & X & 0.592$\pm$0.094 & 0.901$\pm$0.142 & 8.74$_{-8.74}^{+2.10}$ & 1.31$_{-0.60}^{+0.52}$ & 1.69$_{-0.54}^{+0.68}$ & 0.167$_{-0.145}^{+5.689}$ & 4.48e-4\\
J0835-4510 & 2* & L$^f$ & 65.1$\pm$15.7 & 281$\pm$67 & 2.2$\pm$0.5 & 2.7$\pm$0.6 & 0.129$\pm$0.007 & 2.5$\pm$0.3 & 9.36e-06\\
J0908-4913 & 0 & X & $<$0.393 & $<$0.393 & 80$^c$ & 2 & - & - & -\\
J0940-5428 & 0 & C & $<$0.129 & $<$0.129 & 50$^c$ & 2 & - & - & -\\
J1016-5857 & 2 & C & 1.47$_{-1.31}^{0.40}$ & 1.47$_{-1.31}^{0.40}$ & 57.5$_{-19.5}^{+23.5}$ & 1.52$_{-0.20}^{+0.40}$ & - & - & 5.49e-4\\
J1023-5746 & 2* & C & 0.942$_{-0.595}^{+0.190}$ & 0.942$_{-0.595}^{+0.190}$ & 117$_{-33}^{+37}$ & 1.54$\pm$0.33 & - & - & -\\
J1024-0719 & 0 & X & $<$0.111 & 0.111$\pm$0.051 & $<$3.58 & - & 2.83$_{-0.37}^{+0.44}$ & 0.0295$_{-0.0059}^{+0.0258}$ & -\\
J1028-5819 & 1 & S & 1.5$\pm$0.5 & 1.5$\pm$0.5 & 50$^c$ & 2 & - & - & 1.17e-4\\
J1044-5737 & 0 & S & $<$3.93 & $<$3.93 & 50$^c$ & 2 & - & - & -\\
J1048-5832 & 2* & C+X & 0.490$_{-0.342}^{+0.181}$ & 0.490$_{-0.342}^{+0.181}$ & 46.0$\pm$2.3 & 1.22$\pm$0.46 & - & - & 2.17e-5\\
J1057-5226 & 2 & C+X & 1.51$_{-0.13}^{+0.02}$ & 24.5$_{-2.5}^{+0.3}$ & 2.7$\pm$0.2 & 1.7$\pm$0.1 & 0.179$\pm$0.006 & 0.46$\pm$0.06 & 3.17e-4\\
J1119-6127 & 2 & C+X & 1.48$\pm$0.21 & 5.89$\pm$0.84 & 185$_{-38}^{+42}$ & 1.74$_{-0.24}^{+0.45}$ & 2.15$\pm$0.33 & 4.96$_{-0.61}^{+5.08}$ & 5.36e-4\\
J1124-5916 & 2 & C & 9.78$_{-1.08}^{1.18}$ & 10.90$_{-1.26}^{+1.32}$ & 30.0$_{-4.8}^{+2.8}$ & 1.54$_{-0.17}^{+0.09}$ & 0.426$_{-0.018}^{+0.034}$ & 0.274$_{-0.077}^{+0.089}$ & 2.26e-4\\
J1135-6055 & 2 & C & 0.370$_{-0.321}^{+0.145}$ & 0.370$_{-0.321}^{+0.145}$ & 41.9$_{-15.2}^{+18.9}$ & 1.15$_{-0.50}^{+0.52}$ & - & - & -\\
J1231-1411 & 2 & X & 4.12$_{-1.74}^{+0.88}$ & 4.49$_{-1.75}^{+0.90}$ & 11.3$\pm$5.1 & 3.84$_{-0.50}^{+0.40}$ & 2.05$_{-0.56}^{+0.68}$ & 0.0797$_{-0.0094}^{+0.0078}$ & -\\
J1357-6429 & 2 & X & 0.419$\pm$0.156 & 0.985$\pm$0.365 & 18.9$_{-4.5}^{+4.8}$ & 1.48$\pm$0.30 & 2.27$_{-0.30}^{+0.32}$ & 0.478$_{-0.045}^{+0.914}$ & 9.32e-6\\
J1410-6132 & 0 & - & - & - & - & - & - & - & -\\
J1413-6205 & 0 & S & $<$4.9 & $<$4.9 & 40$^c$ & 2 & - & - & -\\
J1418-6058 & 2 & C+X & 0.359$\pm$0.144 & 0.359$\pm$0.144 & 225$_{-45}^{+52}$ & 1.80$_{-0.30}^{+0.47}$ & - & - & 1.06e-5\\
J1420-6048 & 2* & X & 1.6$\pm$0.7 & 1.6$\pm$0.7 & 202$_{-106}^{+161}$ & 0.84$_{-0.37}^{+0.55}$ & - & - & 5.82e-5\\
J1429-5911 & 0 & S & $<$5.76 & $<$5.76 & 80$^c$ & 2 & - & - & -\\
J1459-60 & 0 & S & $<$3.93 & $<$3.93 & 100$^c$ & 2 & - & - & -\\
J1509-5850 & 2 & C+X & 0.534$_{-0.180}^{+0.197}$ & 0.534$_{-0.180}^{+0.197}$ & 79.5$_{-16.5}^{+22.1}$ & 1.36$\pm$0.20 & - & - & 8.38e-5\\
J1513-5908 & 2* & C+L$^{f}$ & 520$\pm$180 & 520$\pm$180 & 91.8$\pm$0.2 & 2.05$\pm$0.04 & - & - & -\\
J1531-5610 & 1 & C & 2.31$\pm$0.75 & 2.31$\pm$0.75 & 40$^c$ & 2 & - & - & -\\
J1600-3053 & 0 & X & $<$0.0685 & 0.0685$\pm$0.0214 & 10$^e$ & - & 4.10$_{-1.31}^{+4.10}$ & 0.099$_{-0.038}^{+0.394}$ & -\\
J1614-2230 & 0 & C+X & $<$0.286 & 0.286$_{-0.086}^{+0.015}$ & 2.9$_{-2.9}^{+4.3}$ & 2 & 0.236$\pm$0.024 & 0.92$_{-0.35}^{+0.73}$ & -\\
J1648-4611 & 0 & C & $<$0.216 & $<$0.216 & 100$^c$ & 2 & - & - & -\\
J1658-5324 & 0 & S & $<$2.81 & $<$2.81 & 20$^c$ & 2 & - & - & -\\
J1709-4429 & 2 & C+X & 3.78$_{-0.94}^{+0.37}$ & 9.04$_{-2.25}^{+0.87}$ & 45.6$_{-2.9}^{+4.4}$ & 1.88$\pm$0.21 & 0.166$\pm$0.012 & 4.3$_{-0.86}^{+1.72}$ & 8.20e-5\\
J1713+0747 & - & - & - & - & - & - & - & - & -\\
J1718-3825 & 2 & X & 1.18$_{-0.97}^{+0.58}$ & 1.18$_{-0.97}^{+0.58}$ & 40.7$_{-15.5}^{+14.6}$ & 1.55$\pm$0.48 & - & - & 1.65e-4\\
J1730-3350 & 0 & C+X & $<$0.0562 & $<$0.0562 & 100$^c$ & 2 & - & - & -\\
J1732-31 & 2 & C & 0.369$\pm$0.130 & 0.369$\pm$0.130 & 9.39$_{-9.39}^{+28.58}$ & 2.01$_{-0.62}^{+0.90}$ & - & - & -\\
J1741-2054$^{g}$ & 2 & C & 6.24$_{-1.14}^{+0.34}$ & 11.07$_{-2.02}^{+0.58}$ & 15.3$_{-3.6}^{+5.1}$ & 2.71$_{-0.11}^{+0.14}$ & 0.829$_{-0.084}^{+0.080}$ & 2.60$_{-1.64}^{+4.27}$ & 1.15e-3\\
J1744-1134 & 0 & C & $<$0.256 & 0.256$_{-0.114}^{+0.020}$ & 9.40$_{-9.40}^{+11.50}$ & - & 3.31$_{-0.46}^{+0.50}$ & 0.0220$_{-0.0116}^{+0.0342}$ & -\\
J1747-2958 & 2* & C+X & 48.7$_{-6.0}^{+21.3}$ & 48.7$_{-6.0}^{+21.3}$ & 256$_{-6}^{+9}$ & 1.51$_{-0.44}^{+0.12}$ & - & - & 9.32e-4\\
J1801-2451 & 2 & C+X & 9.97$\pm$2.02 & 9.97$\pm$2.02 & 374$_{-108}^{+120}$ & 1.54$_{-0.44}^{+0.28}$ & - & - & 1.25e-3\\
J1809-2332 & 2 & C+X & 1.40$_{-0.23}^{+0.25}$ & 3.14$_{-0.53}^{+0.57}$ & 61$_{-8}^{+9}$ & 1.85$_{-0.36}^{+1.89}$ & 0.190$\pm$0.025 & 1.54$_{-0.44}^{+1.26}$ & 1.14e-4\\
J1813-1246 & 1 & S & 15.1$\pm$3.8 & 15.1$\pm$3.8 & 396$_{-195}^{+379}$ & 1.72$_{-0.87}^{+1.56}$ & - & - & -\\
J1823-3021A& 0 & X+C & - & - & - & - & - & - & -\\
J1826-1256 & 2 & C & 1.12$\pm$0.25 & 1.12$\pm$0.25 & 126$_{-46}^{+53}$ & 0.79$\pm$0.39 & - & - & -\\
J1833-1034 & 2 & X+C & 66.3$\pm$1.5 & 66.3$\pm$1.5 & 210$\pm$1 & 1.52$\pm$0.02 & - & - & 5.23e-4\\
J1836+5925 & 2 & X+C & 0.311$_{-0.214}^{+0.036}$ & 0.417$_{-0.283}^{+0.061}$ & 0.7$_{-0.7}^{+10.6}$ & 2.05$_{-0.32}^{+0.54}$ & 0.631$_{-0.260}^{+0.235}$ & 0.982$\pm$0.868 & 5.22e-5\\
J1846+0919 & 0 & S & $<$2.92 & $<$2.92 & 20$^c$ & 2 & - & - & -\\
J1902-5105 & 0 & S & $<$1.54 & $<$1.54 & 3$^c$ & 2 & - & - & -\\
J1907+06 & 1 & C & 2.75$\pm$1.01 & 2.75$\pm$1.01 & 398$_{-375}^{+468}$ & 3.16$_{-2.28}^{+2.76}$ & - & - & -\\
J1939+2134 & 2 & C & 3.95$\pm$0.71 & 3.95$\pm$0.71 & 109$_{-44}^{+63}$ & 1.15$_{-0.32}^{+0.46}$ & - & - & 2.57e-3\\
J1952+3252 & 2 & X+C & 40.7$\pm$1.5 & 44.0$\pm$1.5 & 33.3$\pm$0.9 & 1.71$\pm$0.03 & 1.61$_{-0.15}^{+0.16}$ & 4.18$_{-1.85}^{+2.95}$ & 5.26e-4\\
J1954+2836 & 0 & S & $<$3.65 & $<$3.65 & 50$^c$ & 2 & - & - & -\\
J1957+5036 & 0 & S & $<$2.98 & $<$2.98 & 10$^c$ & 2 & - & - & -\\
J1958+2841 & 1 & S & 1.11$\pm$0.71 & 1.11$\pm$0.71 & 40$^c$ & 1.14$_{-0.84}^{+0.79}$ & - & - & -\\
J1959+2048 & 2 & X+C & 0.549$_{-0.440}^{+0.100}$ & 0.654$_{-0.524}^{+0.120}$ & 3.72$_{-3.72}^{+3.79}$ & 1.37$_{-0.48}^{+0.43}$ & 3.11$_{-0.78}^{+0.70}$ & 1.56$_{-1.01}^{+8.50}$ & 2.51e-4\\
J2017+0603 & 0 & S & $<$0.826 & $<$0.826 & 10$^c$ & 2 & - & - & -\\
J2021+3651 & 2 & C+X & 2.15$_{-0.49}^{+0.24}$ & 5.98$_{-1.38}^{+0.66}$ & 63.8$_{-0.39}^{+0.50}$ & 1.68$_{-0.13}^{+0.18}$ & 1.66$\pm$0.13 & 4.60$_{-1.83}^{+5.95}$ & 3.39e-5\\
J2021+4026 & 2 & C & 0.148$\pm$0.009 & 0.965$\pm$0.057 & 65.2$_{-37.3}^{+30.5}$ & 0.86$_{-0.86}^{+1.87}$ & 2.82$_{-0.62}^{+1.00}$ & 0.230$_{-0.114}^{+0.516}$ & 3.44e-5\\
J2030+3641 & 0 & S & $<$4.52 & $<$4.52 & 80$^c$ & 2 & - & - & -\\
J2032+4127 & 2 & C+X & 0.273$_{-0.156}^{+0.139}$ & 0.273$_{-0.156}^{+0.139}$ & 47.8$_{-14.9}^{+13.1}$ & 2.00$\pm$0.33 & - & - & 1.61e-4\\
J2043+1710 & 0 & S & $<$0.980 & $<$0.980 & 6$^c$ & 2 & - & - & -\\
J2043+2740 & 2 & X & 0.218$_{-0.113}^{+0.030}$ & 0.218$_{-0.113}^{+0.030}$ & $<$3.62 & 2.98$_{-0.29}^{+0.44}$ & - & - & 1.47e-4\\
J2055+25 & 2 & X & 0.433$_{-0.277}^{+0.121}$ & 0.433$_{-0.277}^{+0.121}$ & 15.1$_{-11.7}^{+14.2}$ & 2.04$_{-0.52}^{+0.66}$ & - & - & -\\
J2124-3358 & 2 & X & 0.668$_{-0.344}^{+0.150}$ & 0.959$_{-0.494}^{+0.216}$ & 2.76$_{-2.76}^{+4.87}$ & 2.89$_{-0.35}^{+0.45}$ & 3.11$_{-0.35}^{+0.37}$ & 0.019$_{-0.009}^{+0.012}$ & 1.24e-4\\
J2214+3002 & 2 & C & 0.743$\pm$0.025 & 0.743$\pm$0.025 & $<$21.3 & 3.32$_{-0.46}^{+1.01}$ & - & - & -\\
J2229+6114 & 2 & C+X & 51.3$_{-5.8}^{+9.3}$ & 51.3$_{-5.8}^{+9.3}$ & 30$_{-4}^{+9}$ & 1.01$_{-0.12}^{+0.06}$ & - & - & 3.65e-4\\
J2238+59 & 0 & S & $<$4.49 & $<$4.49 & 70$^c$ & 2 & - & - & -\\
J2240+5832 & 0 & S & $<$4.60 & $<$4.60 & 70$^c$ & 2 & - & - & -\\
J2241-5236 & 2 & C & 0.522$\pm$0.072 & 0.522$\pm$0.072 & $<$24.8 & 2.59$_{-0.42}^{+0.95}$ & - & - & -\\
J2302+4442 & 2 & X & 0.681$_{-0.383}^{+0.140}$ & 0.681$_{-0.383}^{+0.140}$ & 13.0$_{-5.2}^{+9.1}$ & 2.91$_{-0.33}^{+0.46}$ & - & - & -\\
\caption{X-ray spectra of the pulsars. The fluxes are unabsorbed and here the non-thermal and total fluxes are shown. The model used is an absorbed powerlaw plus blackbody, where statistically necessary. The only exceptions are PSR J0437-4715 (double PC plus powerlaw), J0633+1746 and J0659+1414 (double BB plus powerlaw): here only the most relevant thermal component is reported. All the errors are at a 90\% confidence level.\\
a : This parameter shows the confidence of the X-ray spectrum of each pulsar, based on the available X-ray data. An asterisk mark the pulsars for which ad ad-hoc analysis was necessary. See section ~\ref{tot-xdata}.\\
b : C = {\it Chandra}/ACIS ; X = {\it XMM-Newton}/PN+MOS ; S = {\it SWIFT}/XRT ; L = literature. Only public data have been used.\\
c : here, the column density has been fixed by using the galactic value in the pulsar direction obtained by Webtools and scaling it for the distance (see Table \ref{tab-1}).\\
d : the beam correction factor f$_X$ is assumed to be 1, which can result in an efficiency(=L/$\dot{E}$) $>$ 1. See Watters et al.(2009). Here the errors are not reported.\\
e : The statistic is very low so that it was necessary to freeze the column density parameter; the values have been evaluated by using WebTools.\\
f : Respectively taken from Kargaltsev 2008, De Luca 2005, De Luca 2005, Webb 2004, Mori 2004, Abdo et al. 2010c.\\
g : The spectrum is well fitted also by a single blackbody.
\label{tottab-2}}
\end{longtable}

\begin{longtable}{cccccccc}
PSR Name & F$_R$ & F$_{\gamma}$ & $\Gamma_{\gamma}$ & Cutoff$_G$ & Eff$_{\gamma}^a$ & F$_{\gamma}$/F$_{X nt}$ & F$_{\gamma}$/F$_{X th}$\\
 & (mJy) & ($10^{-10}erg/cm^{2}s$)& & (GeV) & & & \\ 
\endhead
J0007+7303 & $<$0.006$^c$ & 3.82$\pm$0.11 & 1.38$\pm$0.04 & 4.6$\pm$0.4 & 0.2 & 3882$\pm$115 & 79583$\pm$2828\\
J0030+0451 & 0.6 & 0.527$\pm$0.035 & 1.22$\pm$0.16 & 1.8$\pm$0.4 & 0.17 & 207$\pm$27 & 325$\pm$42\\
J0034-0534 & - & 0.19$\pm$0.03 & 1.5$\pm$0.3 & 1.7$\pm$0.7 & 0.022 & $>$3276 & -\\
J0101-6422 & - & 0.108$\pm$0.012 & 0.63$\pm$0.38 & 1.5$\pm$0.5 & 0.036 & $>$46.8 & -\\
J0205+6449 & 0.04 & 0.665$\pm$0.054 & 2.09$\pm$0.14 & 3.5$\pm$1.4 & 0.0025 & 33.8$\pm$3.0 & 317$\pm$30\\
J0218+4232 & 0.9 & 0.362$\pm$0.053 & 2.02$\pm$0.23 & 5.1$\pm$4.2 & 0.2 & 78.4$_{-15.7}^{+13.6}$ & -\\
J0248+6021 & 13.7$\pm$2.7 & 0.308$\pm$0.058 & 1.15$\pm$0.49 & 1.4$\pm$0.6 & 0.735 & $>$34.2 & -\\
J0340+4130 & - & 0.208$\pm$0.018 & 0.96$\pm$0.22 & 2.3$\pm$0.5 & 1.05 & $>$1040 & -\\
J0357+32 & $<$0.043$^c$ & 0.639$\pm$0.037 & 1.29$\pm$0.18 & 0.9$\pm$0.2 & 5.23 & 998$_{-110}^{+152}$ & -\\
J0437-4715 & 140 & 0.186$\pm$0.022 & 1.74$\pm$0.32 & 1.3$\pm$0.7 & 0.02 & 23.5$_{-3.3}^{+3.2}$ & 44.4$\pm$5.7\\
J0534+2200 & 14 & 13.07$\pm$1.12 & 1.97$\pm$0.06 & 5.8$\pm$1.2 & 0.001 & 0.295$\pm$0.026 & -\\
J0610-2100 & 0.4 & 0.107$\pm$0.016 & 2.23$\pm$0.10 & - & 1.97 & $>$30.9 & -\\
J0613-0200 & 1.4 & 0.324$\pm$0.035 & 1.38$\pm$0.24 & 2.7$\pm$1.0 & 0.07 & 338$_{-160}^{+244}$ & -\\
J0614-3330 & - & 1.09$\pm$0.14 & 1.44$\pm$0.12 & 4.49$\pm$1.44 & 2.00 & 2097$\pm$858 & -\\
J0631+1036 & 0.8 & 0.304$\pm$0.051 & 1.38$\pm$0.35 & 3.6$\pm$1.8 & 0.14 & $>$1350 & -\\
J0633+0632 & $<$0.003$^c$ & 0.801$\pm$0.064 & 1.29$\pm$0.18 & 2.2$\pm$0.6 & 1.4 & 1282$\pm$145 & 738$\pm$85\\
J0633+1746 & $<$1 & 33.85$\pm$0.29 & 1.08$\pm$0.02 & 1.90$\pm$0.05 & 0.78 & 6812$_{-375}^{+136}$ & 4436$_{-253}^{+74}$\\
J0659+1414 & 3.7 & 0.317$\pm$0.030 & 2.37$\pm$0.42 & 0.7$\pm$0.5 & 0.01 & 78.1$_{-13.6}^{+7.5}$ & 1.93$_{-0.33}^{+0.18}$\\
J0734-1559 & - & 0.537$\pm$0.027 & 1.79$\pm$0.10 & 2.1$\pm$0.5 & 0.08 & $>$228 & -\\
J0742-2822 & 15 & 0.183$\pm$0.035 & 1.76$\pm$0.40 & 2.0$\pm$1.4 & 0.07 & $>$812 & -\\
J0751+1807 & 3.2 & 0.109$\pm$0.032 & 1.56$\pm$0.58 & 3.0$\pm$4.3 & 0.08 & 184$\pm$61 & 353$\pm$117\\
J0835-4510 & 1100 & 88.06$\pm$0.45 & 1.57$\pm$0.01 & 3.2$\pm$0.06 & 0.01 & 1353$\pm$326 & 408$\pm$97\\
J0908-4913 & 10 & 0.446$\pm$0.047 & 1.44$\pm$0.37 & 0.7$\pm$0.3 & 0.47 & $>$1134.86\\
J0940-5428 & - & - & - & - & - & - & -\\
J1016-5857 & 0.46 & 0.713$\pm$0.081 & 2.26$\pm$0.06 & - & 0.27 & 485.034$_{-436}^{+143}$ & -\\ 
J1023-5746 & $<$0.031 & 1.55$\pm$0.10 & 1.47$\pm$0.14 & 1.6$\pm$0.3 & 0.12 & 1645$_{-1045}^{+349}$ & -\\
J1024-0719 & 0.66 & 0.057$\pm$0.013 & 2.19$\pm$0.18 & - & 0.037 & $>$487 & 487$\pm$250\\
J1028-5819 & 0.36 & 1.77$\pm$0.12 & 1.25$\pm$0.17 & 1.9$\pm$0.5 & 0.14 & 1182$\pm$403 & -\\
J1044-5737 & $<$0.021 & 1.03$\pm$0.07 & 1.60$\pm$0.12 & 2.5$\pm$0.5 & 0.45 & $>$262 & -\\
J1048-5832 & 6.5 & 1.73$\pm$0.11 & 1.31$\pm$0.15 & 2.0$\pm$0.4 & 0.08 & 3531$_{-2474}^{+1323}$ & -\\
J1057-5226 & 11 & 2.72$\pm$0.08 & 1.06$\pm$0.08 & 1.3$\pm$0.1 & 0.56 & 1804$_{-164}^{+59}$ & 118$_{-13}^{+4}$\\
J1119-6127 & - & 0.37$\pm$0.05 & 1.6$\pm$0.4 & 2.4$\pm$1.8 & 0.13 & 250$\pm$49 & 83.9$\pm$16.5\\
J1124-5916 & 0.08 & 0.380$\pm$0.058 & 1.43$\pm$0.33 & 1.7$\pm$0.7 & 0.01 & 38.9$\pm$7.4 & 339$_{-75}^{+67}$\\
J1135-6055 & - & 0.444$\pm$0.060 & 1.53$\pm$0.18 & 2.00$\pm$0.49 & 1200$_{-1054}^{+497}$ & -\\
J1231-1411 & - & 1.033$\pm$0.122 & 1.40$\pm$0.12 & 2.98$\pm$0.68 & 0.13 & 251$_{-110}^{+61}$ & 2792$_{-338}^{+373}$\\
J1357-6429 & - & 0.338$\pm$0.035 & 1.65$\pm$0.30 & 0.9$\pm$0.4 & 0.0075 & 807$\pm$312 & 597$\pm$229\\
J1410-6132 & - & 0.604$\pm$0.093 & 2.11$\pm$0.09 & - & 0.176 & - & -\\
J1413-6205 & $<$0.025 & 1.29$\pm$0.10 & 1.32$\pm$0.16 & 2.6$\pm$0.6 & 0.43 & $>$263 & -\\
J1418-6058 & $<$0.03$^c$ & 2.36$\pm$0.32 & 1.32$\pm$0.20 & 1.9$\pm$0.4 & 0.08 & 6672$\pm$3049 & -\\
J1420-6048 & 0.9 & 1.59$\pm$0.29 & 1.73$\pm$0.20 & 2.7$\pm$1.0 & 0.06 & 426$\pm$112 & -\\
J1429-5911 & $<$0.022 & 0.926$\pm$0.081 & 1.93$\pm$0.14 & 3.3$\pm$1.0 & 0.45 & $>$160.7639 & -\\
J1459-60 & $<$0.038$^c$ & 1.06$\pm$0.10 & 1.83$\pm$0.20 & 2.7$\pm$1.1 & 0.52 & $>$269 & -\\
J1509-5850 & 0.15 & 0.969$\pm$0.101 & 1.36$\pm$0.23 & 3.5$\pm$1.1 & 0.15 & 1815$_{-640}^{+696}$ & -\\
J1513-5908 & 0.94 & 0.293$\pm$0.074 & 2.50$\pm$0.16 & - & 0.0054 & 0.563$\pm$0.241 & -\\ % da controllare in un nuovo catalogo
J1531-5610 & 0.6 & - & - & - & - & - & -\\ % da controllare in un nuovo catalogo
J1600-3053 & 3.2 & 0.055$\pm$0.011 & 1.00$\pm$0.06 & 2.0$\pm$0.7 & $>$803 & 803$\pm$298\\
J1614-2230 & - & 0.274$\pm$0.042 & 1.34$\pm$0.36 & 2.4$\pm$1.0 & 1.03 & $>$958 & -\\
J1648-4611 & 0.58 & 0.559$\pm$0.087 & 1.58$\pm$0.22 & 5.8$\pm$2.6 & 1.05 & $>$2587.963 & -\\
J1658-5324 & - & 0.293$\pm$0.025 & 1.60$\pm$0.21 & 1.2$\pm$0.3 & 0.090 & $>$104 & -\\
J1709-4429 & 7.3 & 12.42$\pm$0.22 & 1.70$\pm$0.03 & 4.9$\pm$0.4 & 0.33 & 3285$_{-819}^{+327}$ & 2361$_{-590}^{+228}$\\
J1713+0747 & 0.8 & 0.127$\pm$0.019 & 2.21$\pm$0.10 & - & 0.048 & - & -\\
J1718-3825 & 1.3 & 0.673$\pm$0.160 & 1.26$\pm$0.62 & 1.3$\pm$0.6 & 0.09 & 570$_{-488}^{+311}$ & -\\
J1730-3350 & 3.2 & 0.364$\pm$0.056 & 2.34$\pm$0.34 & - & 0.36 & $>$6477 & -\\
J1732-31 & $<$0.008$^c$ & 2.42$\pm$0.12 & 1.27$\pm$0.12 & 2.2$\pm$0.3 & 1.33 & 6558$\pm$2333 & -\\
J1741-2054 & 0.156$^c$ & 1.28$\pm$0.07 & 1.39$\pm$0.14 & 1.2$\pm$0.2 & 0.24 & 205$_{-39}^{+16}$ & 265$_{-50}^{+20}$\\
J1744-1134 & 3 & 0.280$\pm$0.046 & 1.02$\pm$0.59 & 0.7$\pm$0.4 & 0.1 & $>$1094 & -\\
J1747-2958 & 0.25 & 1.31$\pm$0.14 & 1.11$\pm$0.28 & 1.0$\pm$0.2 & 0.02 & 26.9$_{-4.3}^{+12.1}$ & -\\
J1801-2451 & - & 0.714$\pm$0.097 & 1.90$\pm$0.08 & - & 0.09 & 13.96$\pm$3.41 & -\\
J1809-2332 & $<$0.026$^c$ & 4.13$\pm$0.13 & 1.52$\pm$0.06 & 2.9$\pm$0.3 & 0.33 & 2951$_{-494}^{+535}$ & 2374$\pm$416\\
J1813-1246 & $<$0.028$^c$ & 1.69$\pm$0.11 & 1.83$\pm$0.12 & 2.9$\pm$0.8 & 0.20 & 112$\pm$29 & -\\
J1823-3021A& 0.72 & 0.158$\pm$0.026 & 2.16$\pm$0.11 & - & 0.14 & - & -\\
J1826-1256 & $<$0.044$^c$ & 3.34$\pm$0.15 & 1.49$\pm$0.09 & 2.4$\pm$0.3 & 20.7 & 2982$\pm$679 & -\\
J1833-1034 & 0.07 & 1.02$\pm$0.12 & 2.24$\pm$0.15 & 7.7$\pm$4.8 & 0.01 & 15.4$\pm$1.8 & -\\
J1836+5925 & $<$0.01$^c$ & 6.00$\pm$0.11 & 1.35$\pm$0.03 & 2.3$\pm$0.1 & 1.01 & 19293$_{-13280}^{+2261}$ & 56604$_{-36860}^{+13390}$\\
J1846+0919 & $<$0.004 & 0.358$\pm$0.035 & 1.60$\pm$0.19 & 4.1$\pm$1.5 & 2.1 & $>$123 & -\\
J1902-5105 & - & 0.218$\pm$0.017 & 1.62$\pm$0.15 & 2.7$\pm$0.8 & 0.056 & $>$141.6 & -\\
J1907+06 & 0.0034$^b$ & 2.75$\pm$0.13 & 1.84$\pm$0.08 & 4.6$\pm$1.0 & 0.30 & 1000$\pm$370 & -\\
J1939+2134 & - & 0.198$\pm$0.036 & 1.43$\pm$1.27 & 1.15$\pm$1.15 & 0.13 & 50.1$\pm$12.8 & -\\
J1952+3252 & 1 & 1.34$\pm$0.07 & 1.75$\pm$0.10 & 4.5$\pm$1.2 & 0.02 & 32.9$\pm$2.1 & 406$\pm$21\\
J1954+2836 & $<$0.004 & 0.975$\pm$0.068 & 1.55$\pm$0.14 & 2.9$\pm$0.7 & 0.39 & $>$267 & -\\
J1957+5036 & $<$0.025 & 0.227$\pm$0.020 & 1.12$\pm$0.28 & 0.9$\pm$0.2 & 5.6 & $>$76.2 & -\\
J1958+2841 & $<$0.005$^c$ & 0.846$\pm$0.069 & 0.77$\pm$0.26 & 1.2$\pm$0.2 & 1.6 & 762$\pm$491 & -\\
J1959+2048 & - & 0.134$\pm$0.016 & 1.33$\pm$0.66 & 1.30$\pm$0.69 & 0.061 & 244$_{-198}^{+53}$ & 1276$_{-1032}^{+287}$\\
J2017+0603 & - & 0.371$\pm$0.043 & 1.00$\pm$0.32 & 3.12$\pm$1.32 & 1.02 & $>$449 & -\\
J2021+3651 & 0.1 & 4.70$\pm$0.15 & 1.65$\pm$0.06 & 2.6$\pm$0.3 & 0.07 & 2186$_{-503}^{+254}$ & 1227$_{-288}^{+140}$\\
J2021+4026 & $<$0.011$^c$ & 9.77$\pm$0.18 & 1.79$\pm$0.03 & 3.0$\pm$0.2 & 2.2 & 66014$\pm$4195 & 11958$\pm$736\\
J2030+3641 & - & 0.370$\pm$0.035 & 0.68$\pm$0.28 & 1.5$\pm$0.3 & 9.09 & $>$81.9 & -\\
J2032+4127 & 0.05$^c$ & 1.12$\pm$0.12 & 0.68$\pm$0.38 & 2.1$\pm$0.6 & 0.64 & 4103$_{-2385}^{+2135}$ & -\\
J2043+1710 & - & 0.278$\pm$0.021 & 1.47$\pm$0.11 & 3.00$\pm$0.15 & 0.66 & $>$284 & -\\
J2043+2740 & 7 & 0.155$\pm$0.027 & 1.07$\pm$0.55 & 0.8$\pm$0.3 & 0.09 & 711$_{-389}^{+158}$ & -\\
J2055+25 & $<$0.106 & 1.15$\pm$0.07 & 0.71$\pm$0.19 & 1.0$\pm$0.2 & 5.4 & 2656$_{-1707}^{+760}$ & -\\
J2124-3358 & 1.6 & 0.276$\pm$0.035 & 1.05$\pm$0.28 & 2.7$\pm$1.0 & 0.05 & 413$_{-219}^{+107}$ & 948$_{-503}^{+246}$\\
J2214+3002 & - & 0.332$\pm$0.045 & 1.44$\pm$0.24 & 2.53$\pm$0.85 & 0.49 & 447$\pm$62 & -\\
J2229+6114 & 0.25 & 2.20$\pm$0.08 & 1.74$\pm$0.07 & 3.0$\pm$0.5 & 0.025 & 42.9$_{-5.1}^{+7.9}$ & -\\
J2238+59 & $<$0.007$^c$ & 0.545$\pm$0.059 & 1.00$\pm$0.36 & 1.0$\pm$0.3 & 0.52 & $>$121 & -\\
J2240+5832 & 2.7$\pm$0.7 & 0.10$\pm$0.06 & 1.8$\pm$0.7 & 5.7$\pm$5.4 & 0.32 & $>$21.7 & -\\
J2241-5236 & 4.1$\pm$0.1 & 0.341$\pm$0.019 & 1.35$\pm$0.11 & 3.2$\pm$0.6 & 0.044 & 653$\pm$97 & -\\
J2302+4442 & - & 0.394$\pm$0.032 & 1.25$\pm$0.27 & 2.97$\pm$1.05 & 1.81 & 579$_{-329}^{+128}$ & -\\
\caption{$\gamma$-ray spectra of the pulsars. A broken powerlaw spectral shape is assumed for all the pulsars and the values are taken from Abdo et al.(2009b), Saz Parkinson et al.(2010). The gamma-ray flux is above 100 GeV. The radio flux densities (at 1400MHz) are taken from Abdo 2009, Saz Parkinson 2010. All the errors are at a 90\% confidence level.\\
a : f${_\gamma}$ is assumed to be 1, which can result in an efficiency $>$ 1. See Watters et al.(2009). Here the errors are not reported.\\
b : taken from Abdo 2010.\\
c : taken from Ray 2010.
\label{tottab-3}}
\end{longtable}

\begin{longtable}{ccccccc}
PSR Name & PWNx & F$_{X}^{pwn}$ & $\Gamma_{X}^{pwn}$ & PWN$\gamma$ & F$_{\gamma}^{pwn}$ & $\Gamma_{\gamma}^{pwn}$\\
 & & (10$^{-13}erg/cm^{2}s$) & & & (10$^{-12}erg/cm^{2}s$) & \\ 
\endhead
J0007+7303 & Y & 21.40$_{-0.17}^{+0.14}$ & 1.78$\pm$0.08 & N & $<$69.94 & -\\
J0030+0451 & N & - & - & N & $<$7.83 & -\\
J0034-0534 & N & - & - & Y & 11.09$\pm$2.68 & 2.27$\pm$0.17\\
J0101-6422 & ? & - & - & ? & - & -\\ 
J0205+6449 & Y & 24.0$\pm$0.5 & 2.00$\pm$0.03 & N & $<$12.88 & -\\
J0218+4232 & N & - & - & N & $<$13.65 & -\\
J0248+6021 & ? & - & - & N & $<$8.59 & -\\
J0340+4130 & ? & - & - & ? & - & -\\
J0357+32 & Y & 3.72$_{-1.36}^{+0.62}$ & 2.20$_{-0.32}^{+0.37}$ & N & $<$4.36 & -\\
J0437-4715 & N & - & - & N & $<$9.41 & -\\
J0534+2200 & Y & 396000$\pm$1000 & 2.12$\pm$0.01 & Y & 540.92$\pm$46.73 & 2.15$\pm$0.03\\
J0610-2100 & ? & - & - & ? & - & -\\
J0613-0200 & N & - & - & N & $<$7.46 & -\\
J0614-3330 & ? & - & - & ? & - & -\\
J0631+1036 & ? & - & - & N & $<$20.72 & -\\
J0633+0632 & Y & 2.92$_{-0.81}^{+0.79}$ & 1.19$_{-0.22}^{+0.59}$ & N & $<$35.97 & -\\
J0633+1746 & Y & 0.172$\pm$0.001 & 1.70$\pm$0.06 & Y & 749.44$\pm$22.24 & 2.24$\pm$0.02\\
J0659+1414 & N & - & - & N & $<$5.58 & -\\
J0734-1559 & ? & - & - & ? & - & -\\
J0742-2822 & ? & - & - & N & $<$6.63 & -\\
J0751+1807 & N & - & - & N & $<$10.53 & -\\
J0835-4510 & Y & 128$\pm$1 & 1.4$\pm$0.1 & Y & 210.25$\pm$13.87 & 2.3$\pm$0.1\\
J0908-4913 & ? & - & - & ? & - & -\\
J0940-5428 & ? & - & - & ? & - & -\\
J1016-5857 & Y & 3.53$_{-2.77}^{+0.26}$ & 1.61$_{-0.36}^{+0.41}$ & ? & - & -\\
J1023-5746 & Y & 0.853$_{-0.593}^{+0.193}$ & 1.54$\pm$0.33 & Y & 27.58$\pm$13.73 & 1.05$\pm$0.36\\
J1024-0719 & N & - & - & ? & - & -\\
J1028-5819 & ? & - & - & N & $<$98.27 & -\\
J1044-5737 & ? & - & - & N & $<$13.2 & -\\
J1048-5832 & Y & 0.608$_{-0.426}^{+0.224}$ & 1.22$\pm$0.46 & N & $<$16.96 & -\\
J1057-5226 & N & - & - & N & $<$11.35 & -\\
J1119-6127 & Y & 0.601$\pm$0.194 & 1.55$\pm$0.54 & ? & - & -\\
J1124-5916 & Y & 5.17$_{-0.30}^{+0.24}$ & 1.78$_{-0.05}^{+0.03}$ & N & $<$13.38 & -\\
J1135-6055 & Y & 1.87$_{-1.05}^{+0.39}$ & 1.84$_{-0.36}^{+0.40}$ & ? & - & -\\
J1231-1411 & N & - & - & ? & - & -\\
J1357-6429 & Y & 3.81$_{-0.52}^{+0.38}$ & 1.30$_{-0.11}^{+0.21}$ & ? & - & -\\
J1410-6132 & ? & - & - & ? & - & -\\
J1413-6205 & ? & - & - & N & $<$5.34 & -\\
J1418-6058 & Y & ? & ? & N & $<$86.03 & -\\
J1420-6048 & ? & - & - & N & $<$139.42 & -\\
J1429-5911 & ? & - & - & N & $<$23.53 & -\\
J1459-60 & ? & - & - & N & $<$25.64 & -\\
J1509-5850 & Y & 2.47$_{-0.54}^{+0.32}$ & 1.17$\pm$0.13 & N & $<$28.19 & -\\
J1513-5908 & Y & 1459.5$_{-12.7}^{+12.9}$ & 1.87$\pm$0.01 & ? & - & -\\
J1531-5610 & ? & - & - & ? & - & -\\
J1600-3053 & ? & - & - & ? & - & -\\
J1614-2230 & N & - & - & N & $<$24.36 & -\\
J1648-4611 & ? & - & - & ? & - & -\\
J1658-5324 & ? & - & - & ? & - & -\\
J1709-4429 & Y & 8.36$_{-0.67}^{+0.52}$ & 1.44$\pm$0.05 & N & $<$39.39 & -\\
J1713+0747 & ? & - & - & ? & - & -\\
J1718-3825 & Y & 1.33$_{-0.95}^{+0.55}$ & 1.18$\pm$0.40 & N & $<$9.44 & -\\
J1730-3350 & ? & - & - & ? & - & -\\
J1732-31 & ? & - & - & N & $<$8.19 & -\\
J1741-2054 & Y & 1.68$_{-0.34}^{+0.28}$ & 1.67$_{-0.14}^{+0.22}$ & N & $<$11.64 & -\\
J1744-1134 & N & - & - & N & $<$30.64 & -\\
J1747-2958 & Y & 84.5$_{-4.0}^{+10.2}$ & 2.04$\pm$0.06 & ? & - & -\\
J1801-2451 & Y & 3.27$\pm$1.24 & 1.94$_{-0.81}^{+0.87}$ & ? & - & -\\
J1809-2332 & Y & 11.1$\pm$1.2 & 1.43$\pm$0.10 & N & $<$21.25 & -\\
J1813-1246 & ? & - & - & Y & 119.03$\pm$9.29 & 2.65$\pm$0.14\\
J1823-3021A& ? & - & - & ? & - & -\\
J1826-1256 & Y & 1.52$\pm$0.33 & 0.86$\pm$0.39 & N & $<$160.67 & -\\
J1833-1034 & Y & 721$\pm$5 & 1.85$\pm$0.01 & N & $<$10.38 & -\\
J1836+5925 & N & - & - & Y & 524.16$\pm$34.03 & 2.07$\pm$0.03\\
J1846+0919 & ? & - & - & N & $<$5.3 & -\\
J1902-5105 & ? & - & - & ? & - & -\\
J1907+06 & ? & - & - & N & $<$18.83 & -\\
J1939+2134 & N & - & - & ? & - & -\\
J1952+3252 & Y & 77.7$\pm$1.5 & 1.81$\pm$0.02 & N & $<$18.68 & -\\
J1954+2836 & ? & - & - & N & $<$23.78 & -\\
J1957+5036 & ? & - & - & N & $<$6.04 & -\\
J1958+2841 & ? & - & - & N & $<$17.07 & -\\
J1959+2048 & Y & 0.168$_{-0.071}^{+0.061}$ & 1.79$_{-0.36}^{+0.49}$ & ? & - & -\\
J2017+0603 & ? & - & - & ? & - & -\\
J2021+3651 & Y & 10.4$\pm$0.6 & 1.46$_{-0.04}^{+0.06}$ & N & $<$101.24 & -\\
J2021+4026 & N & - & - & Y & 888.12$\pm$8.56 & 2.36$\pm$0.02\\
J2030+3641 & ? & - & - & ? & - & -\\
J2032+4127 & ? & - & - & N & $<$171.45 & -\\
J2043+1710 & ? & - & - & ? & - & -\\
J2043+2740 & ? & - & - & N & $<$2.99 & -\\
J2055+25 & ? & - & - & Y & 17.59$\pm$3.34 & 2.51$\pm$0.15\\
J2124-3358 & Y & 0.140$_{-0.069}^{+0.094}$ & 1.90$_{-0.43}^{+0.47}$ & Y & 21.81$\pm$4.44 & 2.06$\pm$0.14\\
J2214+3002 & ? & - & - & ? & - & -\\
J2229+6114 & Y & 11.4$_{-1.0}^{+0.8}$ & 1.31$_{-0.04}^{+0.06}$ & N & $<$15.55 & -\\
J2238+59 & ? & - & - & N & $<$165.1 & -\\
J2240+5832 & ? & - & - & ? & - & -\\
J2241-5236 & ? & - & - & ? & - & -\\
J2302+4442 & N & - & - & ? & - & -\\
\caption{$\gamma$ and X-ray spectra of PNWe associated with {\it Fermi} pulsars. A powerlaw spectral shape is assumed for all the nebulae. The gamma-ray flux is above 100 GeV while the X-ray one in the 0.3-10 keV energy range. All the errors are at a 90\% confidence level.\\
\label{tottab-4}}
\end{longtable}
\end{center}
\end{landscape}
\end{footnotesize}

\section{Discussion}

\subsection{Study of the X-ray luminosity}
\label{tot-xdata}

The X-ray luminosity, L$_X$, 
is correlated with the pulsar spin-down luminosity $\dot{E}$. The scaling was firstly noted by Seward\&Wang(1988) who used Einstein data of 22 pulsar
- most of them just upper limits - to derive a linear relation between log$F_{0.2-4 keV}^{X}$ and log$\dot{E}$. Later, Becker\&Trumper(1997) investigated 
a sample of 27 pulsars by using ROSAT, yielding the simple scaling L$_X^{0.1-2.4keV}\simeq10^{-3}\dot{E}$. The uncertainty due to soft X-ray absorption
translates into very high flux errors; moreover it was very hard to discriminate between the thermal and powerlaw spectral components.
A re-analysis was performed by Possenti et al.(2002), who studied in the 2-10 keV band a sample of
39 pulsars observed by several X-ray telescopes. However, they could not separate
the PWN from the pulsar contribution. Moreover, they conservatively adopted, for most of the pulsars, an uncertainty of $40\%$ on the distance values.
A better comparison with our data can be done with the results by Kargaltsev\&Pavlov(2008), who recently used high-resolution {\it Chandra}
data in order to disentangle the PWN and pulsar fluxes. Focusing just on {\it Chandra} data, and rejecting XMM observations, they obtain a
poor spectral characterization which translates in high errors on fluxes. They also adopted an uncertainty of $40\%$ on the distance values for most pulsars.
Despite the big uncertainties, mainly due to poor distance estimates, all these datasets show that the L$_X$ versus $\dot{E}$ relation
is quite scattered. The high values of the $\chi^2_{red}$ seem to exclude a simple statistical effect.

We are now facing a different panorama, since our ability to evaluate pulsars' distances has improved (Abdo et al.2009b, Saz Parkinson et al.2010) and we are now much better in
discriminating pulsar emission from its nebula.
The use of XMM data makes it possible to build good quality spectra allowing to disentangle the non-thermal from the thermal 
contribution, when present. In particular, we can study the newly discovered radio-quiet
pulsar population and compare them with the "classical" radio-loud pulsars.
We investigate the relations between the X and $\gamma$ luminosities and pulsar parameters, making use of the data collected in Tables \ref{tottab-1}-\ref{tottab-2}-\ref{tottab-3}-\ref{tottab-4}.\\

Using the 37 {\it Fermi} type 2 pulsars with a clear distance estimate and with a well-constrained X-ray spectrum,
the weighted least square fit yields:
\begin{equation}
log_{10}L^X_{30}=(0.71\pm0.06)+(0.88\pm0.04)log_{10}\dot{E}_{34}
\end{equation}
where $\dot{E}=\dot{E}_{34}\times10^{34}erg/s$ and $L_X=L^X_{30}\times10^{30}erg/s$. All the uncertainties are at 90\% confidence level.
We can evaluate the badness of this fit using the reduced chisquare value $\chi^2_{red}=13$;
a double linear fit does not significantly change the value of $\chi^2_{red}$.
A more precise way to evaluate the dispersion of the dataset around the fitted curve is the parameter:\\
$W^2=(1/n)\sum_{i=1->n}(y_{oss}^i-y_{fit}^i)^2$\\
where $y_{oss}^i$ is the actual i$^{th}$ value of the dataset (in our case $log_{10}L^X_{30}$) and $y_{fit}^i$ the expected one. A lesser spread in the dataset
translate into a lower value of $W^2$. We obtain $W^2=0.512$ for the $L_x-\dot{E}$ relationship.
Such high values of both $W^2$ and $\chi_{red}^2$ are an indication of an important scattering of the $L_X$ values around the fitted relation.\\
Our results are in agreement with Possenti et al.(2002), Kargaltsev\&Pavlov(2008).

\subsection{Study of the $\gamma$-ray luminosity}
\label{tot-gdata}

The gamma-ray luminosity, L$_{\gamma}$, is correlated with the pulsar spin-down luminosity $\dot{E}$.
Such a trend is expected in many theoretical models (see e.g. Zhang et al.2004, Muslimov\&Harding 2003)
and it's shortly discussed in the {\it Fermi} LAT catalogue of gamma-ray pulsars (Abdo et al. 2009c).

Selecting the same subsample of {\it Fermi} pulsar used in the previous chapter to assess the relation
between $L_{\gamma}$ and $\dot{E}$, we found that a linear fit
yields an unacceptable value of $\chi^2_{red}$.\\
Inspection of the distribution of residuals lead us to try a double-linear relationship, 
which yields:
\begin{eqnarray}
log_{10}L_{30}^{\gamma}=(2.63\pm0.10)+(1.48\pm0.14)log_{10}\dot{E}_{34} & , & \dot{E}<E_{crit}\\
log_{10}L_{30}^{\gamma}=(4.84\pm0.40)+(0.10\pm0.12)log_{10}\dot{E}_{34} & , & \dot{E}>E_{crit}
\end{eqnarray}
with $E_{crit}=4.04\pm2.02\times10^{35}erg/s$ and $\chi^2_{red}=7$. An f-test shows
that the probability for a chance $\chi^2$ improvement is less than $10^{-3}$.
Neglecting PSR J0659+1414, the $\chi^2_{red}$ value would half. This NS
presents singular features, such as the soft $\gamma$-ray spectrum and an extremely low efficiency,
that makes any interpretation of its emission difficult and unsatisfactory (Weltevrede et al. 2009).
This double-linear relation is in agreement with the data reported in Abdo et al. (2009c) for the entire dataset
of {\it Fermi} $\gamma$-ray pulsars.
Indeed, the $\chi^2_{red}$ obtained for the double linear fit is better than that obtained for the 
$L_X$-$\dot{E}$ relationship. We obtain $W^2=0.328$ for the double linear $L_{\gamma}-\dot{E}$ relationship. 
Both the $\chi^2_{red}$ and $W^2$ are in agreement with a higher scatter in the $L_X-\dot{E}$ graph.
A difference between the X-ray and $\gamma$-ray emission geometries - that translates in different values of 
f$_{\gamma}$ and f$_X$ - could explain such a behaviour.

The existence of an $\dot{E}_{crit}$ has been posited from the theoretical point for different
pulsar emission models.
Revisiting the outer-gap model for pulsars with $\tau<10^7$ yrs and assuming initial conditions as well as
pulsars' birth rates, Zhang et al.(2004) found a sharp boundary, due to the saturation of the gap size, for $L_{\gamma}=\dot{E}$.
They obtain the following distribution of pulsars' $\gamma$-ray luminosities:
\begin{eqnarray}
log_{10}L_{\gamma}=log_{10}\dot{E}+const. & , &  \dot{E}<\dot{E}_{crit}\\
log_{10}L_{\gamma}\sim0.30log_{10}\dot{E}+const. & , & \dot{E}>\dot{E}_{crit}
\end{eqnarray}
By assuming the fractional gap size from Zhang et al.(1997), they obtain $\dot{E}_{crit}=1.5\times10^{34}P^{1/3}erg/s$.
While Equation 4 is similar to our double linear fit (Equation 3), the $\dot{E}_{crit}$ they obtain
seems to be lower than our best fit value.\\
On the other hand, in slot-gap models (Muslimov\&Harding 2003), the break occurs
at about $10^{35}erg/s$, when the gap is limited by screening of the acceleration field by pairs.\\
We can see from Figure \ref{fig-2} that radio-quiet pulsars have higher luminosities than the radio-loud ones, for similar values of
$\dot{E}$. As in the $L_X-\dot{E}$ fit, we can't however discriminate between the two population due to the big errors stemming from distance estimate.

\subsection{Study of the $\gamma$-to-X ray luminosity ratio}
\label{sec-GX}

At variance with the X-ray and gamma-ray luminosities, the ratio between the X-ray and gamma-ray luminosities is independent
from pulsars' distances. This makes it possible to significatively reduce the error bars leading to more 
precise indications on the pulsars' emission mechanisms.\\
Figure \ref{fig-3} reports the histogram of the F$_{\gamma}$/F$_X$ values using only type 2 (high quality X-ray data) pulsars. 
The radio-loud pulsars have
 $<F_{\gamma}/F_X>\sim970$, the radio-quiet population has $<F_{\gamma}/F_X>\sim9460$
while the millisecond radio-loud pulsars have $<F_{\gamma}/F_X>\sim290$ (see \ref{tottab-5} for the main statistical parameters of the logs of F$_{\gamma}$/F$_X$).
Applying the Kolmogoroff-Smirnov test to type 2 pulsars' log($F_{\gamma}/F_X$) values we obtained that the chance for the RQ and RL
datasets belong to the same population is 0.00026. Similarly, the KS test applied on MS and RQ type-2 pulsars give a
probability of 1.35 $\times$ 10$^{-5}$.
We can conclude, with a 3$\sigma$ confidence level, that the radio-quiet and radio-loud datasets we used are somewhat different;
similarly our radio-quiet and millisecond pulsars' populations are different with a 5$\sigma$ confidence level.
On average, MS pulsars have the smallest $F_{\gamma}/F_X$ values. In particular all MS pulsars have lesser values of $F_{\gamma}/F_X$
than all the RQ ones.
These values are similar to those of some of our high-$\dot{E}$, young RL pulsars, such as the Crab.
Recently, it has been argued that some MSPs would have co-located radio and $\gamma$ emitting regions,
similar to some high-$\dot{E}$, young $\gamma$-ray pulsars (Abdo et al. 2010 J0034, Ravi et al. 2010).
However, other millisecond pulsars anyway don't show such a co-location of the two emitting regions so that we cannot easily use this
to explain the low $F_{\gamma}/F_X$ values measured for out entire MSPs' sample.\\
Anyway, the apparent inconsistency we find between RQ and MS pulsars makes more and more interesting future studies
on radio-quiet MS pulsars {\bf (if any)}.

{\bf A distance independent spread in F$_{\gamma}$/F$_X$}

\begin{figure}
\centering
\includegraphics[angle=0,scale=.40]{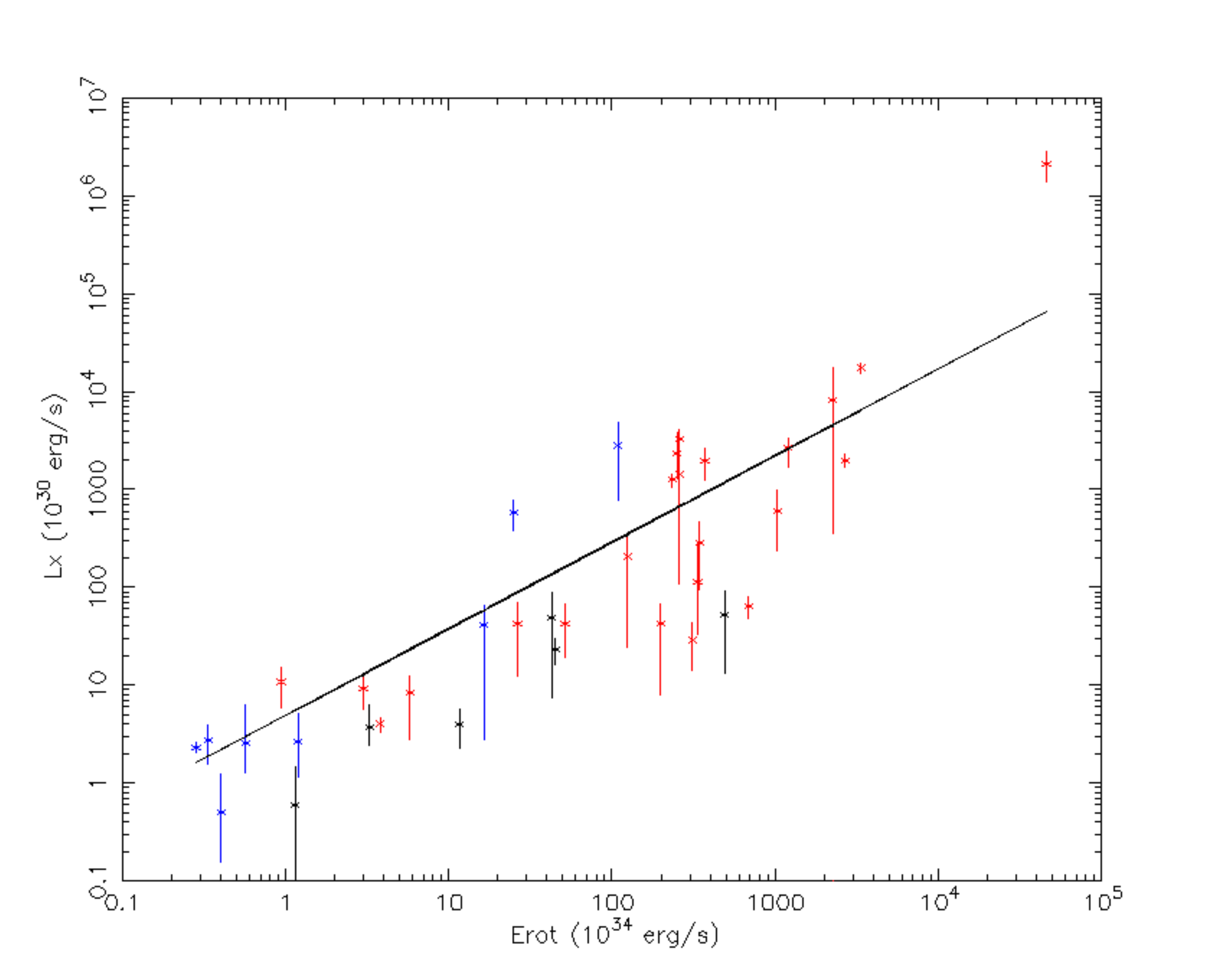}
\caption{$\dot{E}$-L$_X$ diagram for all pulsars classified as type 2 and with a clear distance estimation, assuming f$_X$=1 (see Equation 5). Black: radio-quiet pulsars; red: radio-loud pulsars; blue: millisecond pulsars. The linear best fit of the logs of the two quantities is shown. \label{fig-1}}
\end{figure}

\begin{figure}
\centering
\includegraphics[angle=0,scale=.40]{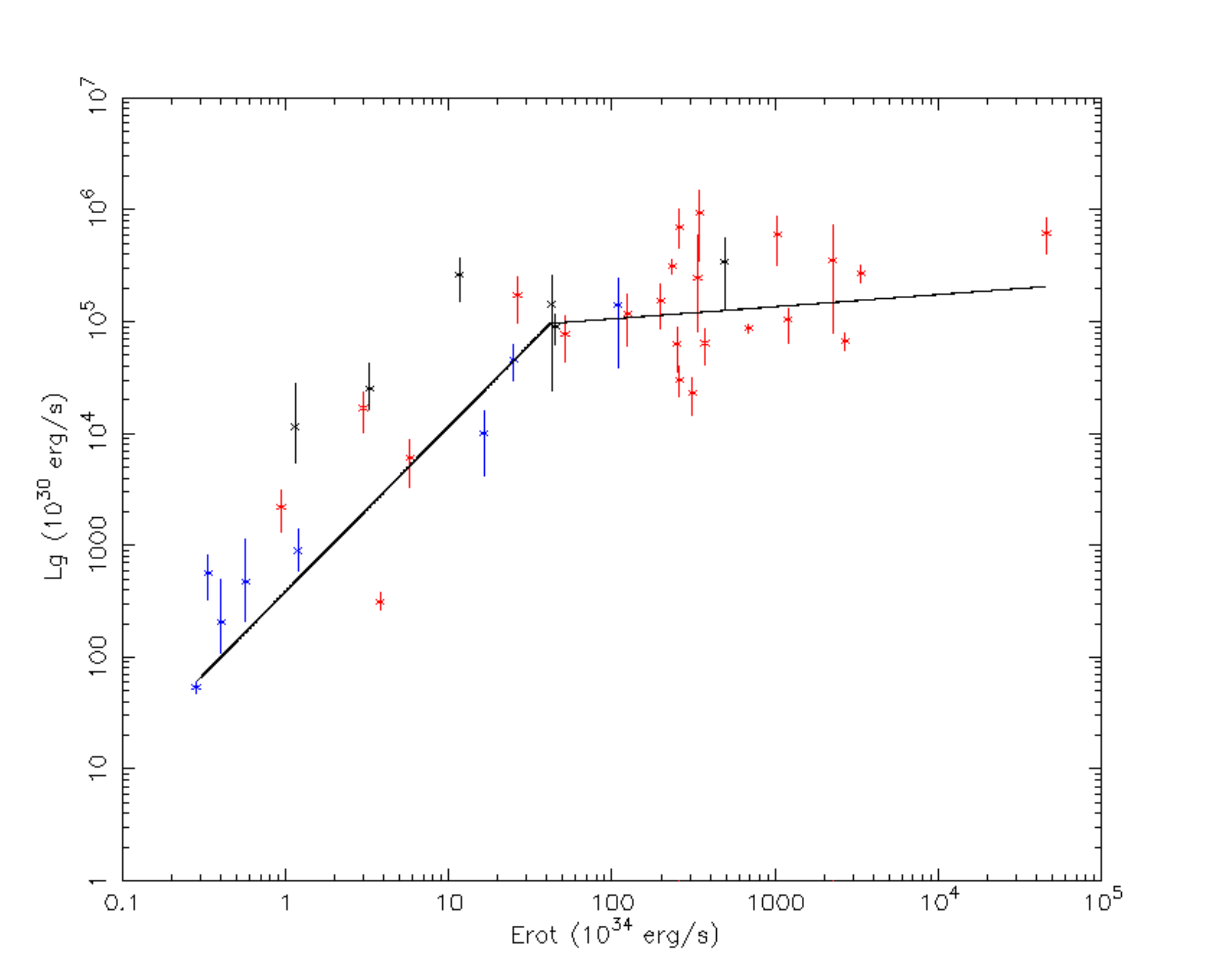}
\caption{$\dot{E}$-L$_{\gamma}$ diagram for all pulsars classified as type 2 and with a clear distance estimation, assuming f$_{\gamma}$=1 (see Equation 5). Black: radio-quiet pulsars; red: radio-loud pulsars; blue: millisecond pulsars. The double linear best fit of the logs of the two quantities is shown. \label{fig-2}}
\end{figure}

Figure \ref{fig-4} shows F$_{\gamma}$/F$_X$ as a function of $\dot{E}$ for our entire sample of $\gamma$-ray emitting NSs
and also only the pulsar with "high quality" X-ray data have been selected.
Even neglecting the upper and lower limits (shown as triangles) as well as the low quality points (see Figure \ref{fig-4}), one immediately notes
the scatter on the F$_{\gamma}$/F$_X$ parameter values for a given value of $\dot{E}$. Such an apparent spread
cannot obviously be ascribed to a low statistic. An inspection of Figure \ref{fig-4} makes it clear
that a linear fit cannot satisfactory describe the data. In a sense, this finding should not come as a surprise since Figure \ref{fig-4} is a combination
of Figures \ref{fig-1} and \ref{fig-2} and we have seen that figure \ref{fig-2} requires a double linear fit.
However, combining the results of our previous fits (Equations 1 and 3) we obtain the
dashed line in Figure \ref{fig-4}, clearly a very poor description of the data.
For $\dot{E}\sim<5\times10^{36}$ the F$_{\gamma}$/F$_X$ values scatter around a mean value of $\sim$1000 with a spread
of a factor about 100. For higher $\dot{E}$ the values of F$_{\gamma}$/F$_X$ seem to decrease drastically to an average value of $\sim$50,
reaching the Crab with F$_{\gamma}$/F$_X$$\sim0.1$.

The spread in the F$_{\gamma}$/F$_X$ values for pulsars with similar $\dot{E}$ is obviously unrelated to distance uncertainties. 
Such a scatter can be due to geometrical effects. For both X-ray and $\gamma$-ray energy bands:
\begin{equation}
L_{\gamma,X}=4\pi f_{\gamma,X}F_{obs}D^2
\end{equation}
where f$_X$ and f$_{\gamma}$ account for the X and $\gamma$ beaming geometries (which may or may not be related).
If the pulse profile observed along the line-of-sight at $\zeta$ (where $\zeta_E$ is the Earth line-of-sight) for a pulsar with
magnetic inclination $\alpha$ is $F(\alpha,\zeta,\phi)$, where $\phi$ is the pulse phase, than we can write:
\begin{equation}
f=f(\alpha,\zeta_E)=\frac{\int\int F(\alpha,\zeta,\phi)sin(\zeta)d\zeta d\phi}{2\int F(\alpha,\zeta_E,\phi)d\phi}
\end{equation}
where $f$ depends only from the viewing angle and the magnetic inclination of the pulsar. With an high
value of this correction coefficient, the emission is disfavored. Obviously
F$_{\gamma}$/F$_X$=L$_{\gamma}$/L$_X\times$f$_X$/f$_{\gamma}$. Different $f_{\gamma}/f_X$ values
for different pulsars can explain the scattering seen in the F$_{\gamma}$/F$_X$-$\dot{E}$ relationship.\\
Watters et al. (2009) assume a nearly uniform emission efficiency while
Zhang et al. (2004) compute a significant variation in the emission efficiency as a function of the geometry of pulsars.
In both cases, geometry plays an important role through magnetic field inclination as well as through viewing angle.\\
The very important scatter found for F$_{\gamma}$/F$_X$ values is obviously due to the different geometrical
configurations which determine the emission at different wavelength of each pulsar. While geometry is clearly playing
an equally important role in determining pulsar luminosities, the F$_{\gamma}$/F$_X$ plot makes its effect
easier to appreciate.

The line in Figures \ref{fig-4} is the combination of the best fits of L$_{\gamma}$-$\dot{E}$ and 
L$_X$-$\dot{E}$ relationship, considering f$_{\gamma}$=1 and f$_X$=1 so that represent the hypothetical
value of F$_{\gamma}$/F$_X$ that each pulsar would have if f$_{\gamma}$=f$_X$: all the pulsars with a value of F$_{\gamma}$/F$_X$
below the line have f$_X<$f$_{\gamma}$.
We have seen in Section \ref{sec-GX} that the radio-quiet dataset shows an higher mean value of F$_{\gamma}$/F$_X$.
This is clearly visible in Figure \ref{fig-4} where all the radio-quiet points are above the expected values (dashed line)
so that all the radio-quiet pulsars should have f$_X>$f$_{\gamma}$. Moreover, the radio-quiet 
dataset shows a lower scatter with respect to the radio-loud one pointing to more uniform values of f$_{\gamma}$/f$_X$ 
for the radio-quiet pulsars. A similar viewing angle or a similar
magnetic inclination for all the radio-quiet pulsars could explain such a behaviour (see Equation 6).\\
Also millisecond pulsars show a quite uniform distribution of $F_{\gamma}/F_X$ values. As now, we aren't able
to explain such a behaviour.

Figure \ref{fig-6} shows the $F_{\gamma}/F_X$ behaviour as a function of the characteristic pulsar age. 
In view of the uncertainty of this parameter, we have also built a similar plot using "real" pulsar
age, as derived from the associated supernova remnants (see Figure 6).
Similarly to the $\dot{E}$ relationship, for $\tau<10^4$ years, $F_{\gamma}/F_X$ values increase with
age (both the characteristic and real ones), while for $t>10^4$years the behaviour becomes more complex.\\

\subsection{Study of the selection effects}

There are two main selections we have done in order to obtain our sample of pulsars with both good $\gamma$ and X-ray spectra (type 2).
First, the two populations of radio-quiet and radio-loud pulsars are unveiled with different techniques: using the same dataset, 
pulsars with known rotational ephemerides have a detection threshold
lower than pulsars found through blind period searches. In the First {\it Fermi} LAT pulsar catalogue (Abdo et al. 2009c) the
faintest gamma-ray-selected pulsar has a flux $\sim$ 3$\times$ higher than the faintest radio-selected one.
Second, we chose only pulsars with a good X-ray coverage. Such a coverage depends on many factors (including the policy of
X-ray observatories) that cannot be modeled\\
Our aim is to understand if these two selections influenced in different ways the two populations of pulsars we are studying:
if this was the case, the results obtained would have been distorted
The $\gamma$-ray selection is discussed at length in the {\it Fermi} LAT pulsar catalogue (Abdo et al. 2009c). Since the radio-quiet population
has obviously a detection threshold higher than the radio-loud one, we could avoid such bias by selecting all the pulsars with a flux higher than
the radio-quiet detection threshold (6$\times10^{-8}$ph/cm$^2$s).
Some radio-loud type 2 pulsars are excluded
with F$_{\gamma}$/F$_X$ values ranging from $\sim$ 50 to $\sim$ 2000. We performed our analysis on such a reduced sample
and the results didn't change significatively.\\
We can, therefore, exclude the presence of an important bias due to the $\gamma$-ray selection on type 2 pulsars.

\begin{figure}
\centering
\includegraphics[angle=0,scale=.20]{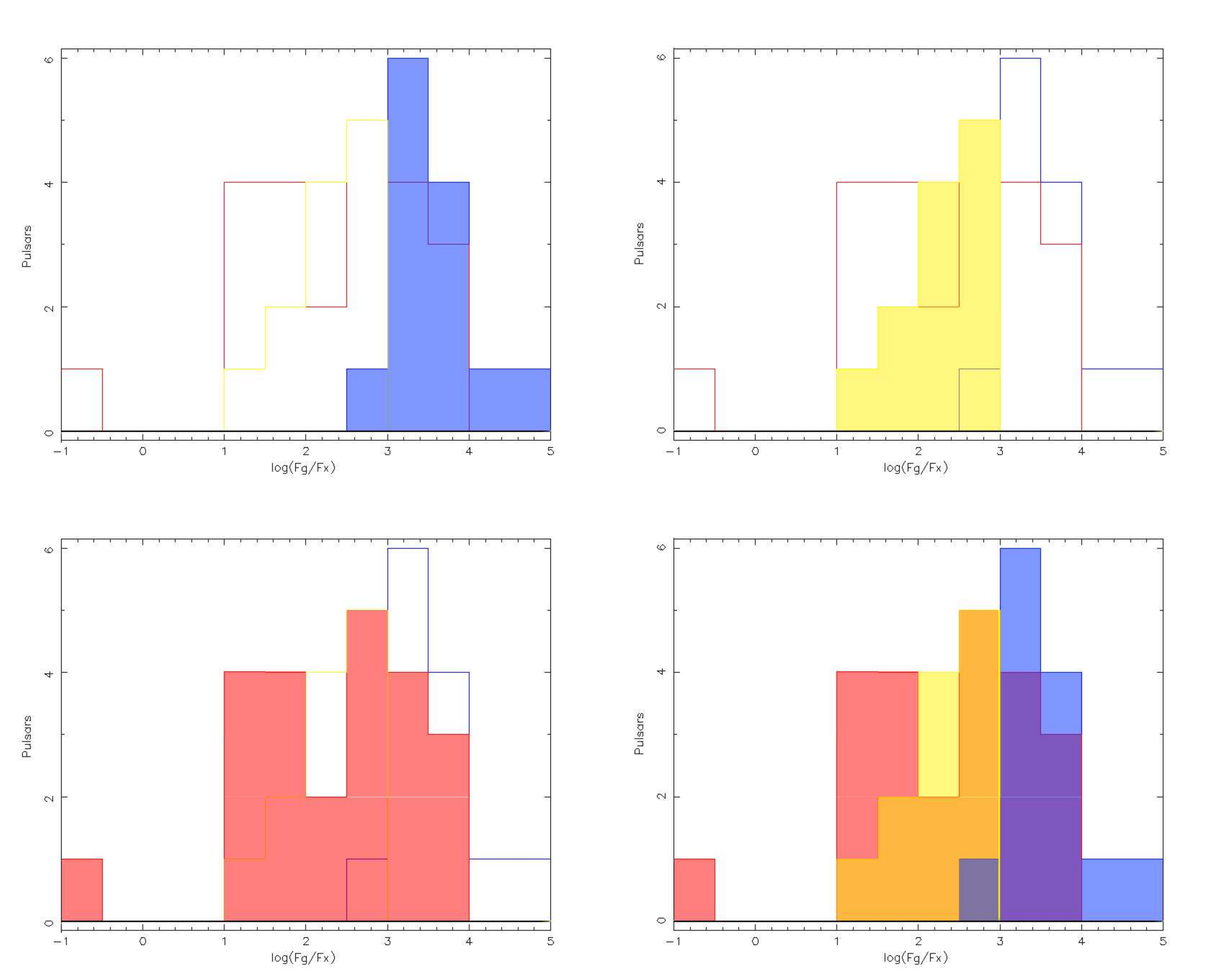}
\caption{log(F$_{\gamma}$/F$_X$) histogram. The step is 0.5; Red: radio-loud pulsars; yellow: millisecond radio-loud pulsars; blue: radio quiet pulsars. Only high confidence pulsars (type 2) have been used. \label{fig-3}}
\end{figure}

\begin{figure}
\centering
\includegraphics[angle=0,scale=.20]{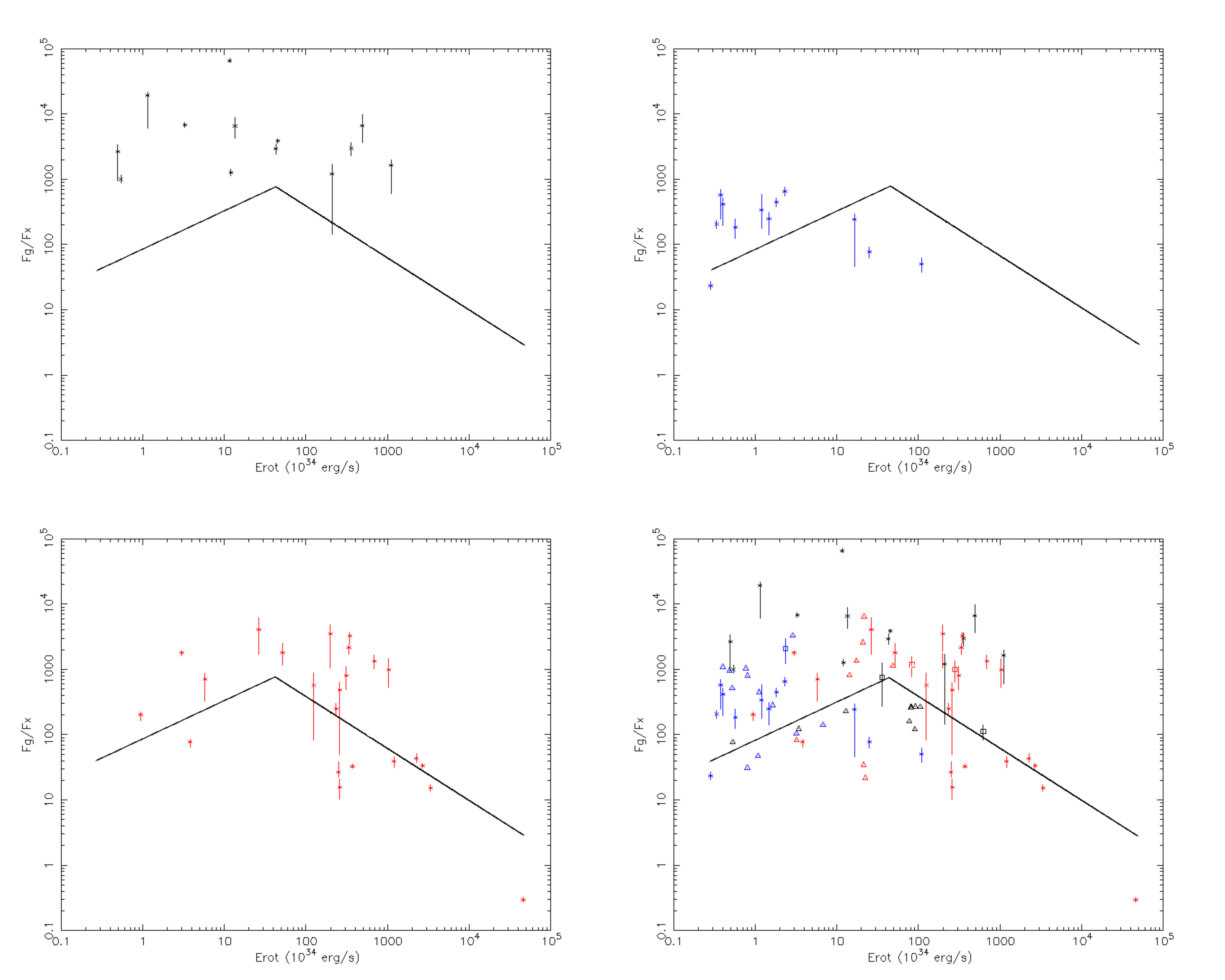}
\caption{$\dot{E}$-F$_{\gamma}$/F$_X$ diagram. Black: radio-quiet pulsars; red: radio-loud pulsars; blue: millisecond pulsars. The triangles are upper and lower limits, the squares indicate pulsars with a type 1 X-ray spectrum (see Table \ref{tottab-2}) and the stars pulsars with a high quality X-ray spectrum. The line is the combination of the best fitting functions obtained for Figure \ref{fig-1} and \ref{fig-2} with the geometrical correction factor set to 1 for both the X and $\gamma$-ray bands. In panels upper-left, upper-light and lower-left only high confidence pulsars are reported. \label{fig-4}}
\end{figure}

\begin{figure}
\centering
\includegraphics[angle=0,scale=.40]{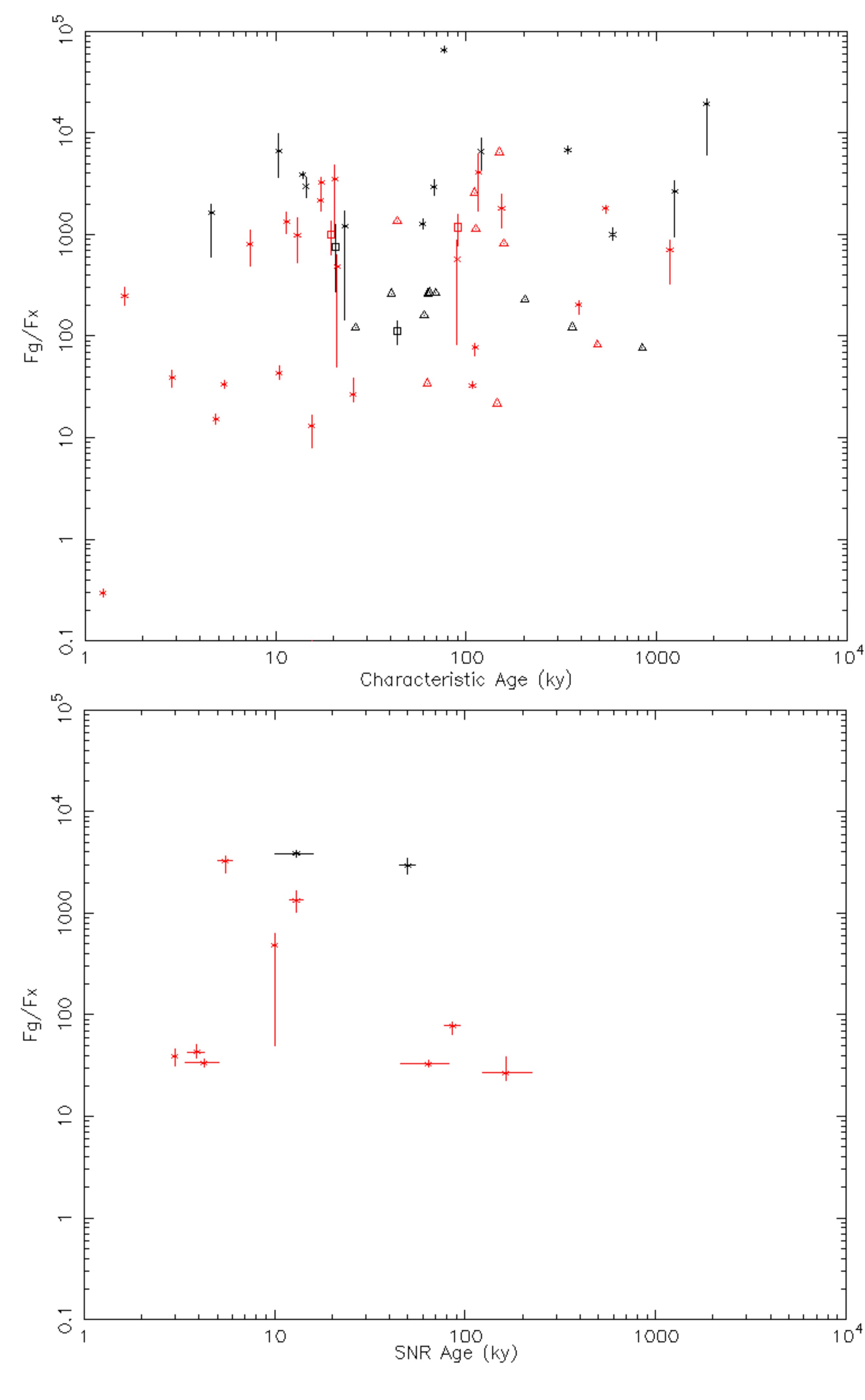}
\caption{Up: Characteristic Age-F$_{\gamma}$/F$_X$ diagram. Down: SNR Age-F$_{\gamma}$/F$_X$ diagram. Black: radio-quiet pulsars; red: radio-loud pulsars. Triangles are upper limits, squares are pulsars with a type 1 X-ray spectrum while stars are pulsars with a type 2 X-ray spectrum. \label{fig-6}}
\end{figure}

In order to roughly evaluate the selection affecting the X-ray observations, we used the method developed by Schmidt(1968) to compare the current
radio-quiet and -loud samples' spatial distributions, following the method also used in Abdo et al. (2009c).
For each object with an available distance estimate, we computed the maximum distance
still allowing detection from $D_{max}=D_{est}(F_{\gamma}/F_{min})^{1/2}$, where $D_{est}$ comes from Table 1, 
the photon flux and $F_{min}$ are taken from Abdo et al. (2009c), Saz Parkinson et al. 2010. 
We limited $D_{max}$ to 15 kpc, and compared $V$, the volume enclosed within the estimated source distance, to that
enclosed within the maximum distance, $V_{max}$, for a galactic disk with radius 10 kpc and thickness 1 kpc (as in Abdo et al. (2009c)).
The inferred values of $<V/V_{max}>$
are 0.375, 0.233, 0.657 and 0.343 for the entire gamma-ray pulsars' dataset, the radio-quiet pulsars, millisecond pulsars and
the radio-loud pulsars. These are quite close to the expected value of 0.5 even if $<V/V_{max}>^{rq}$ is lower than $<V/V_{max}>^{rl}$.
If we use only type 2 pulsars we obtain 0.314, 0.181, 0.530 and 0.298.
These lower values of $<V/V_{max}>$ indicate that we have a good X-ray coverage only for close-by - or very bright - pulsars,
not a surprising result. Millisecond pulsars show a different selection: we have detected an uniform
space distribution of MS RL pulsars. This translates in a possible selection effect acting when we compare the millisecond population
with the other two classes, lowering X-ray fluxes with respect of the other two populations. Such a selection effect
anyway would enhance the $\gamma$-to-X fluxes ratio, strengthening our results on the MS population more strong.
By using the X-ray-counterpart dataset, all the populations' $<V/V_{max}>$ values appear lower of about 0.1:
this seems to indicate that we used the same selection criteria for all the populations and we minimized 
the selection effects in the histogram of Figure \ref{fig-3}.\\
We can conclude that the $\gamma$-ray selection introduced no changes in the two populations,
while the X-ray selection excluded objects both faint and/or far away; any distortion, if present,
is not overwhelming.

\subsection{Conclusions}

The discovery of a number of radio-quiet pulsars comparable to that of radio-loud ones
together with the study of their X-ray counterparts made it possible, for the first time,
to address their behaviour using a distance independent parameters such as the ratio of their fluxes at X and gamma ray wavelengths.\\
First, we reproduced the well known relationship between the neutron stars luminosities and their rotational energy losses. Next, selecting only the {\it Fermi} pulsars with good X-ray data,  we computed  the ratio between the gamma and X-ray fluxes and studied its dependence on the overall rotational energy loss as well as on the neutron star age.\\
Much to our surprise, the distance independent F$_{\gamma}$/F$_X$ values computed for pulsars of similar age and energetic differ by up to 3 orders of magnitude, pointing to important (yet poorly understood) differences both in position and height of the regions emitting at X and $\gamma$-ray wavelengths within the pulsars' magnetospheres. Selection effects cannot account for the spread in the F$_{\gamma}$/F$_X$ relationship and any 
further distortion, if present, is not overwhelming.\\
In spite of the highly scattered values, a decreasing trend is seen when considering young and energetic pulsars.
Moreover, radio quiet pulsars are characterized by higher values of the F$_{\gamma}$/F$_X$ parameter  ($<F_{\gamma}/F_X>_{rl}\sim800$ and $<F_{\gamma}/F_X>_{rq}\sim4800$) so that a KS test points to a chance of 0.0016 for them to belong to the same population as the radio loud  ones. While it would be hard to believe that radio loud and radio quiet pulsars belong to two different neutron star populations, the KS test probably points to different geometrical configurations (possibly coupled with viewing angles) that characterize radio loud and radio quiet pulsars. Indeed 
the radio-quiet population we analyzed is less scattered than the radio-loud one, pointing to a more uniform viewing or 
magnetic geometry of radio-quiet pulsars.\\
Millisecond pulsars have lesser values of $F_{\gamma}/F_X$ than all the RQ population and, on mean, lesser than the RL ones.
Moreover, they show a quite uniform distribution of $F_{\gamma}/F_X$ values. As now, we aren't able
to explain such a behaviour.

Our work is just a starting point, based on the first harvest of gamma-ray pulsars. The observational panorama will quickly evolve. The gamma-ray pulsar list is continuously growing and this triggers more X-ray observations, improving both in quantity and in quality the database of the neutron stars detected in X and $\gamma$-rays to be used to compute our multiwavelength, distance independent parameter. Indeed this thesis contains 88 pulsars and represents a step forward from the 1st Fermi catalogue and from Marelli et al. 2011. Adding 34 $\gamma$-ray pulsars, as well as new X-ray observations, we confirmed and improved the tentative results of Marelli et al. 2011.
The difference between RQ and non-milliseconds RL is more robust and the different behaviour of millisecond pulsars is starting to clearly appear.
However, to fully exploit the information packed in the F$_{\gamma}$/F$_X$ a complete 3D modeling of pulsar magnetosphere is needed to account for the different locations and heights of the emitting regions at work at different energies. Such modeling could provide the clue to account for the spread we have observed for the ratios between $\gamma$ and X-ray fluxes as well as for the systematically higher values measured for radio-quiet pulsars.   

\begin{longtable}{ccccc}
Population & MinValue & MaxValue & Mean & St.Dev.\\
\endhead
Radio-Quiet & 2.999 & 4.820 & 3.611 & 0.516\\
Millisecond & 1.371 & 2.815 & 2.309 & 0.442\\
Radio-Loud & -0.530 & 3.613 & 2.333 & 1.050\\
\caption{Main statistical parameters of log(F$_{\gamma}$/F$_X$) values for the three pulsars' populations. Here are reported their minimum and maximum values, the mean and the standard deviation.\\
\label{tottab-5}}
\end{longtable}

\clearpage

\chapter{Detailed X-ray analyses of Fermi pulsars}
\label{detail}

{\bf J0007+7303 (CTA1) - type 2 RQP}

See sect.~\ref{cta1}.

{\bf J0030+0451 - Type 2 Radio-loud Millisecond Pulsar}

% da bogdanov et al. 2009
PSR J0030+0451 is one of the nearest rotation-powered,
recycled, millisecond pulsars (MSPs) (D = 300 $\pm$ 90 pc; Lommen et al. 2006),
with a spin period P = 4.87 ms and intrinsic spin-down rate
$\dot{P}$ = 10$^{-20}$, implying a surface dipole
magnetic field strength B $\sim$ 2.7 $\times$ 10$^8$ G, a characteristic age
$\tau$ $\sim$ 7.7 Gyr, and spin-down luminosity $\dot{E}$ $\sim$ 3 $\times$ 10$^{33}$ erg/s.
This solitary field MSP was discovered at radio frequencies in the
Arecibo drift scan survey (Lommen et al. 2000) and
was detected in $\gamma$-rays by the {\it Fermi}/LAT
collaboration using the first 6 months of data (Abdo et al. 2009).
No pulsar wind nebula emission was detected in the first 16 months of {\it Fermi} data
down to 7.83 $\times$ 10$^{-12}$ erg/cm$^2$s (Ackermann et al. 2010). 
PSR J0030+0451 X-ray counterpart has been firstly detected with
ROSAT PSPC (Becker et al. 2000), and subsequently by {\it XMM-Newton} (Becker
\& Aschenbach 2002, Bogdanov et al. 2009).
Its light curve shows two broad pulses with a $\sim$60\%-70\% pulsed fraction
in the 0.3-2 keV band, consistent with a mostly thermal origin.

\begin{figure}
\centering
\includegraphics[angle=0,scale=.40]{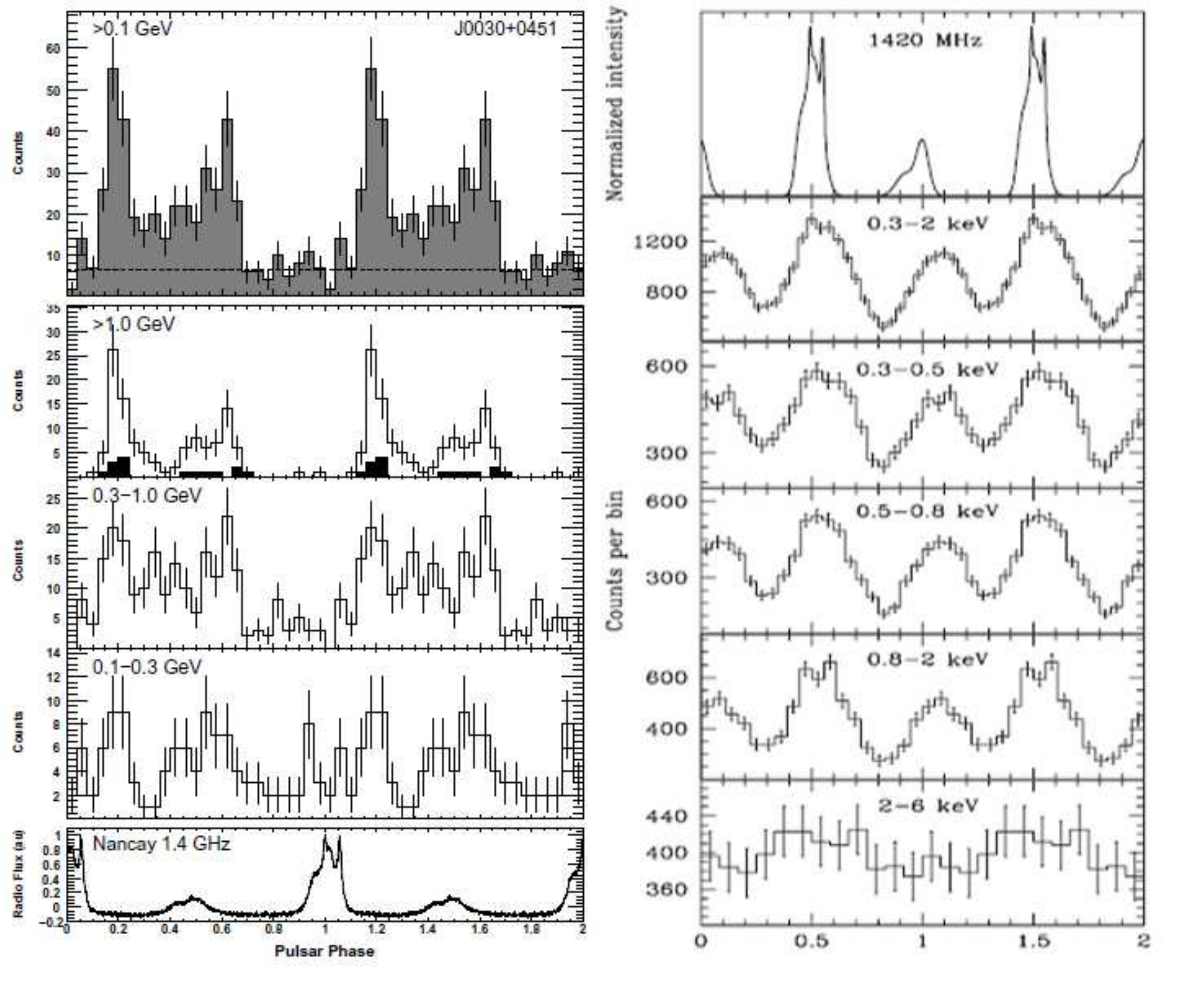}
\caption{PSR J0030+0451 Lightcurve.{\it Left: Fermi} $\gamma$-ray lightcurve folded with Radio
(Abdo et al. 2009c). {\it Right:} {\it XMM-Newton} EPIC PN X-ray pulse profiles
for different energy bands, folded with Radio. The choice of phase zero
and the alignment of the X-ray and radio profiles are arbitrary (Bogdanov et al. 2009).
\label{J0030-lc}}
\end{figure}

We used two {\it XMM-Newton} observations centered on the pulsar: the first (obs. id 0112320101) started on
2001, June 19 at 20:18:34 UT and lasted 30.6 ks, the second (obs. id 0502290101) started on
2007, December 12 at 21:04:04 UT and lasted 126.6 ks.
In both the observations the PN camera (Strueder et al. 2001) of the EPIC
instrument was operated in Fast Timing mode (time resolution of 30 $\mu$s at the expense of one 
imaging dimension), while the MOS detectors (Turner et al. 2001) were set in Full frame mode (2.6
s time resolution on a 15$'$ radius field of view). For
all three instruments, the thin optical filter was used.
The PN camera dataset wasn't used in order to perform the spectral analysis
due to the lack of spatial resolution.
After standard data processing (using the epproc and emproc tasks) and
screening of high particle background time intervals (following De Luca \& Molendi 2004),
the good, dead-time corrected exposure time are respectively 22.0 and 95.9 ks for the two
observations.\\
Using the XIMAGE and SAS dedicated tools, we detected the source at 
R.A.(J2000) = 00:30:27.259, decl. = +04:51:41.69 (5$"$ error radius).
We extracted the source spectrum from a circle of 20$"$ radius centered on the source position,
while we extracted the background from a source-free annulus between 40$"$ and 80$"$.
In the first observation we obtained 746 and 707 source counts in the 0.3-10
keV range for the two MOS cameras;
the background accounts for 3.1\% and 2.4\% of such values.
In the second observation we obtained 3119 and 3018 source counts in the 0.3-10
keV range, with a background contribution of 3.7\% and 3.4\%.
In the MOS1 and MOS2 images, the source count distribution
around the radio position of the pulsar is fully consistent
with that of a point source. Thus, no indication of
any diffuse extended emission that could arise due to a bow
shock was found (see also Bogdanov et al. 2009).\\
The pulsar emission is well described (probability of
obtaining the data if the model is correct - p-value - of 0.22) by a
combination of a blackbody and a power law model.
The powerlaw component has a steep photon index ($\Gamma$ = 3.44 $\pm$ 0.26), 
absorbed by a column N$_H$ = 6.4$_{-2.4}^{+3.4}$ $\times$ 10$^{20}$ cm$^{-2}$.
The best fit temperature of the blackbody component is 2.17$_{-0.13}^{+0.09}$ $\times$ 10$^6$ K
while the emitting radius is R$_{300pc} = $111$_{-23}^{+40}$ m. Such a small emitting radius
suggest an hot spot thermal emission, quite typical for such millisecond pulsars.
A simple blackbody model yields a poor fit (p-value $<$ 10$^{-14}$) as well as a simple
powerlaw model (p-value $<$ 10$^{-5}$).
Assuming the best fit model, the 0.3-10 keV unabsorbed thermal flux is 
(1.62 $\pm$ 0.18) $\times$ 10$^{-13}$ and the non-thermal flux is
(2.55 $\pm$ 0.29) $\times$ 10$^{-13}$ erg/cm$^2$ s. Under the hypothesis of a distance
of 400 pc, the two luminosities are L$_{300pc}^{bol}$ = (1.75 $\pm$ 0.19) $\times$ 10$^{30}$,
L$_{300pc}^{nt}$ = (2.75 $\pm$ 0.31) $\times$ 10$^{30}$ erg/s.

\begin{figure}
\centering
\includegraphics[angle=0,scale=.50]{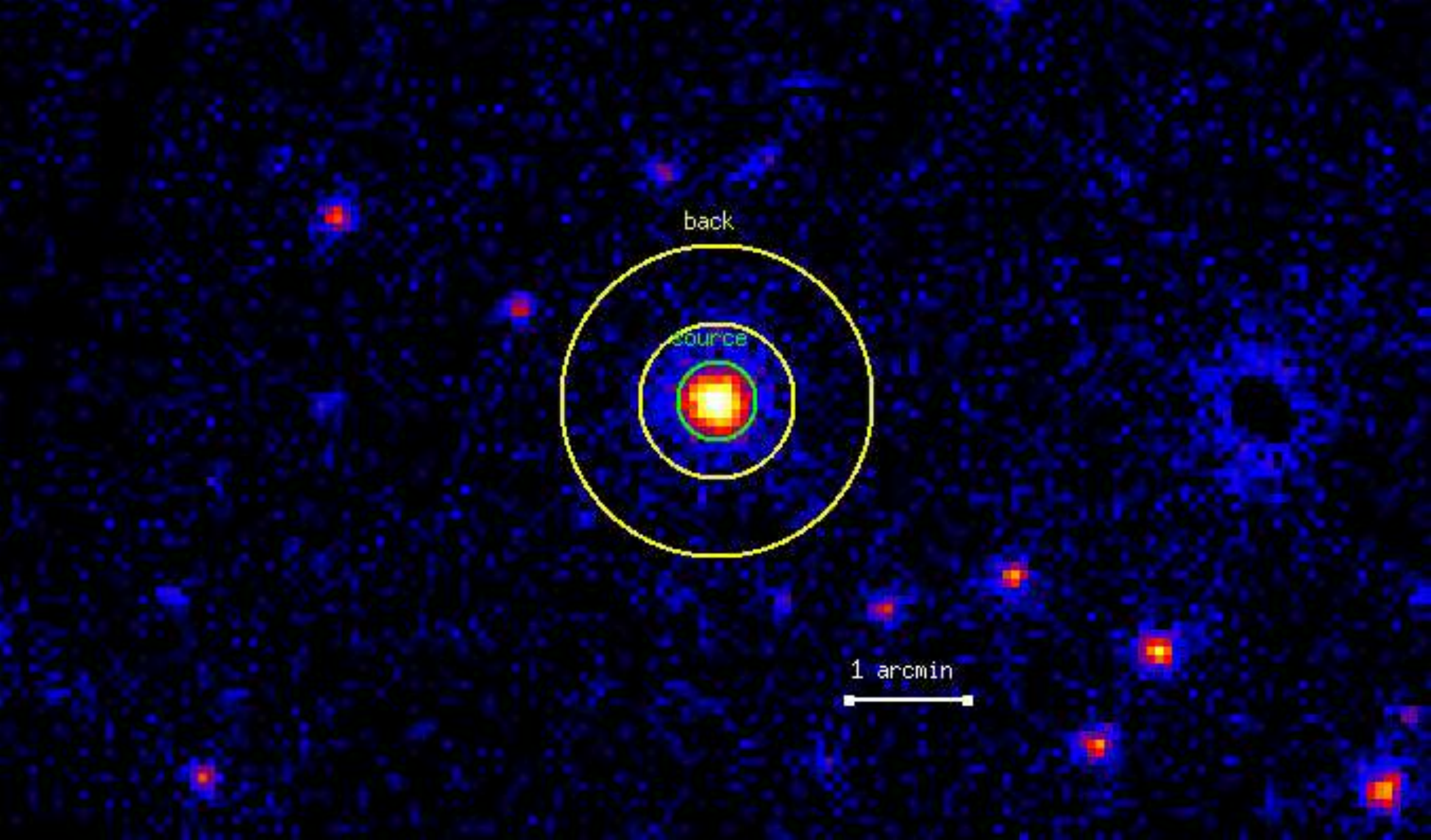}
\caption{PSR J0030+0451 0.3-10 keV MOS Imaging. The two MOS images have been added
by using the {\em sum} tool inside the XIMAGE package. 
The green circle marks the pulsar while the yellow annulus the background region used in the analysis.
\label{J0030-im}}
\end{figure}

\begin{figure}
\centering
\includegraphics[angle=0,scale=.50]{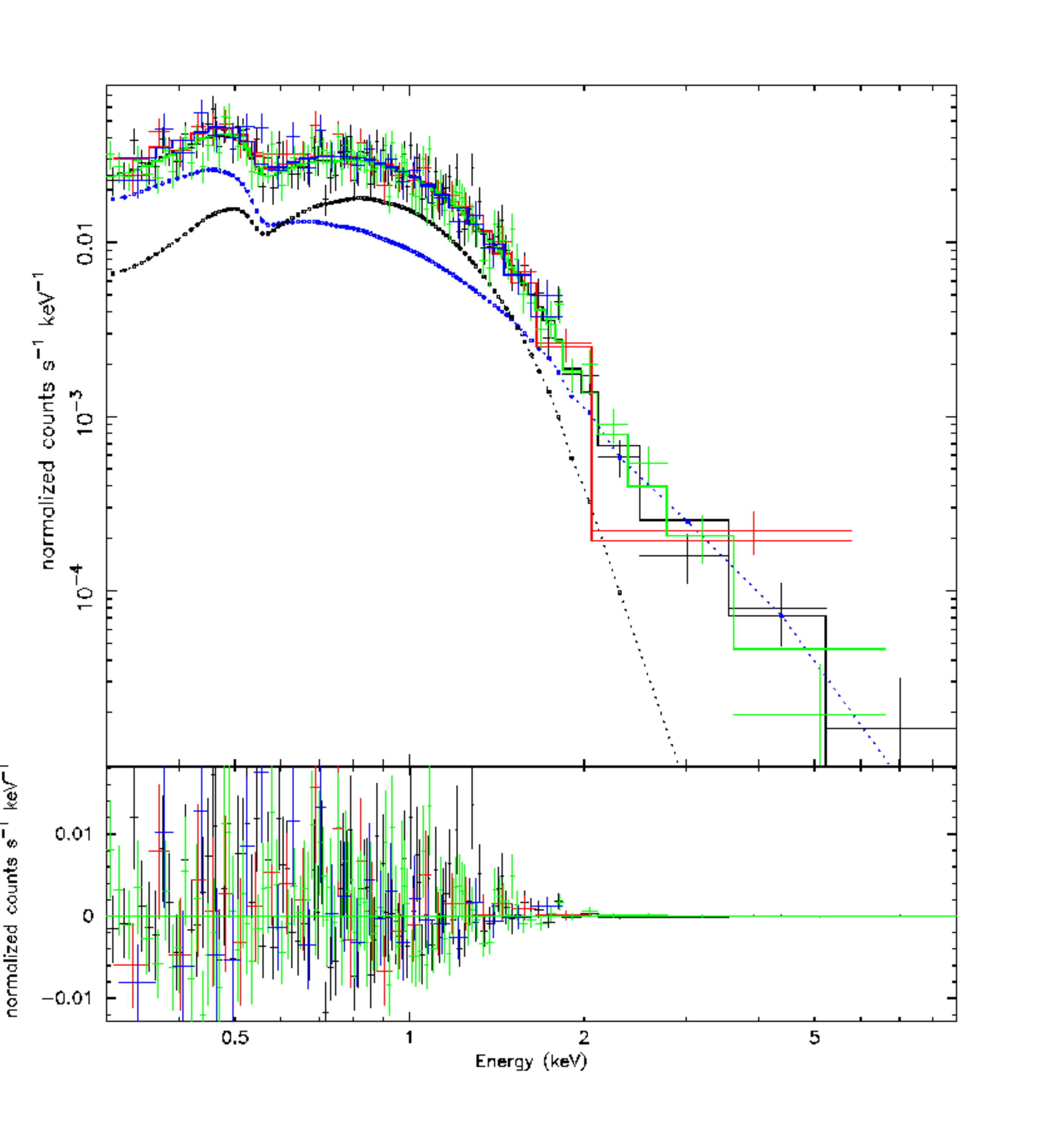}
\caption{PSR J0030+0451 Spectrum. Different colors mark all the different dataset used (see text for details).
Blue points mark the powerlaw component while black points the thermal component of the pulsar spectrum.
Residuals are shown in the lower panel.
\label{J0030-sp}}
\end{figure}

\clearpage

{\bf J0034-0534 - Type 0 RL MSP} % Nuova!

%Zavlin 2005
It is one of the fastest pulsar known, with a spin period P $\sim$ 1.9 ms. This pulsar is in a
circular binary system with a low-mass companion and 1.6 d orbital period (Zavlin 2006). The standard
estimates give its characteristic age $\tau_c$ $\sim$ 6 Gyr, rotational energy
loss $\dot{E}$ $\sim$ 5.56 $\times$ 10$^{34}$ erg s$^{-1}$,
and a distance to the pulsar d $\sim$ 0.53 $\pm$ 0.21 kpc, obtained from dispersion measurements (Brisken et al. 2002).
Before the {\it XMM-Newton} era, X-ray counterpart wasn't detected.

Only one X-ray observation of this source is available - {\it XMM-Newton} obs. id 0031740101,
start time on 2002, June 19 at 09:24:21 UT, exposure 35.4ks.
The PN camera of the EPIC
instrument was operated in Fast Timing mode, while the MOS detectors were set in Full frame mode.
For all three instruments, the medium optical filter was used.
The PN camera wasn't used to perform the spectral analysis
owing to the lack of spatial resolution.
After standard data processing (using the epproc and emproc tasks) and
screening of high particle background time intervals,
the good, dead-time corrected exposure time is 33.7 ks.\\
The X-ray source best fit position found using the XIMAGE and the SAS tools is 00 34 21.62 -05 34 36.42 (5$"$ error radius).
No search for a nebular emission is possible due to the faintness of the source; anyway,
we don't expect MSPs to have bright nebulae (Cheng et al. 2006, Hui\&Becker 2006).
We extracted the source spectrum from a circle of 20$"$ radius centered on the source position while we extracted the background
from a source-free annulus with radii 40$"$ and 80$"$.
Due to the very low number of counts in the XMM observation, we added the two MOS pulsar spectra using
mathpha tool and, similarly, the response
matrix and effective area files using addarf and addrmf.
We obtained a total of 204 source counts (background contribution of 27.5\%).
Only a simple blackbody spectrum can describe the pulsar emission
(a powerlaw model cannot fit the data while the counts are too few to study
a composite model).
The thermal spectrum requires a temperature T = 2.23$_{-1.22}^{+0.62}$ $\times$ 10$^6$ K
and an emitting radius of R$_{0.54kpc}$ = 35.8$_{-25.4}^{+525.3}$ m, typical of an hot spot emission,
absorbed by a column N$_H$ = $<$ 5.63 $\times$ 10$^{21}$ cm$^{-2}$ (90\% confidence level).
Assuming the best fit model, the 0.3-10 keV unabsorbed flux is
5.83 $\pm$ 1.11 $\times$ 10$^{-15}$ erg/cm$^2$ s.
Under the hypothesis of a distance
of 0.54 kpc, the pulsar luminosity is L$_{0.54kpc}^{bol}$ = 2.04 $\pm$ 0.39 $\times$ 10$^{29}$ erg/s.

\begin{figure}
\centering
\includegraphics[angle=0,scale=.30]{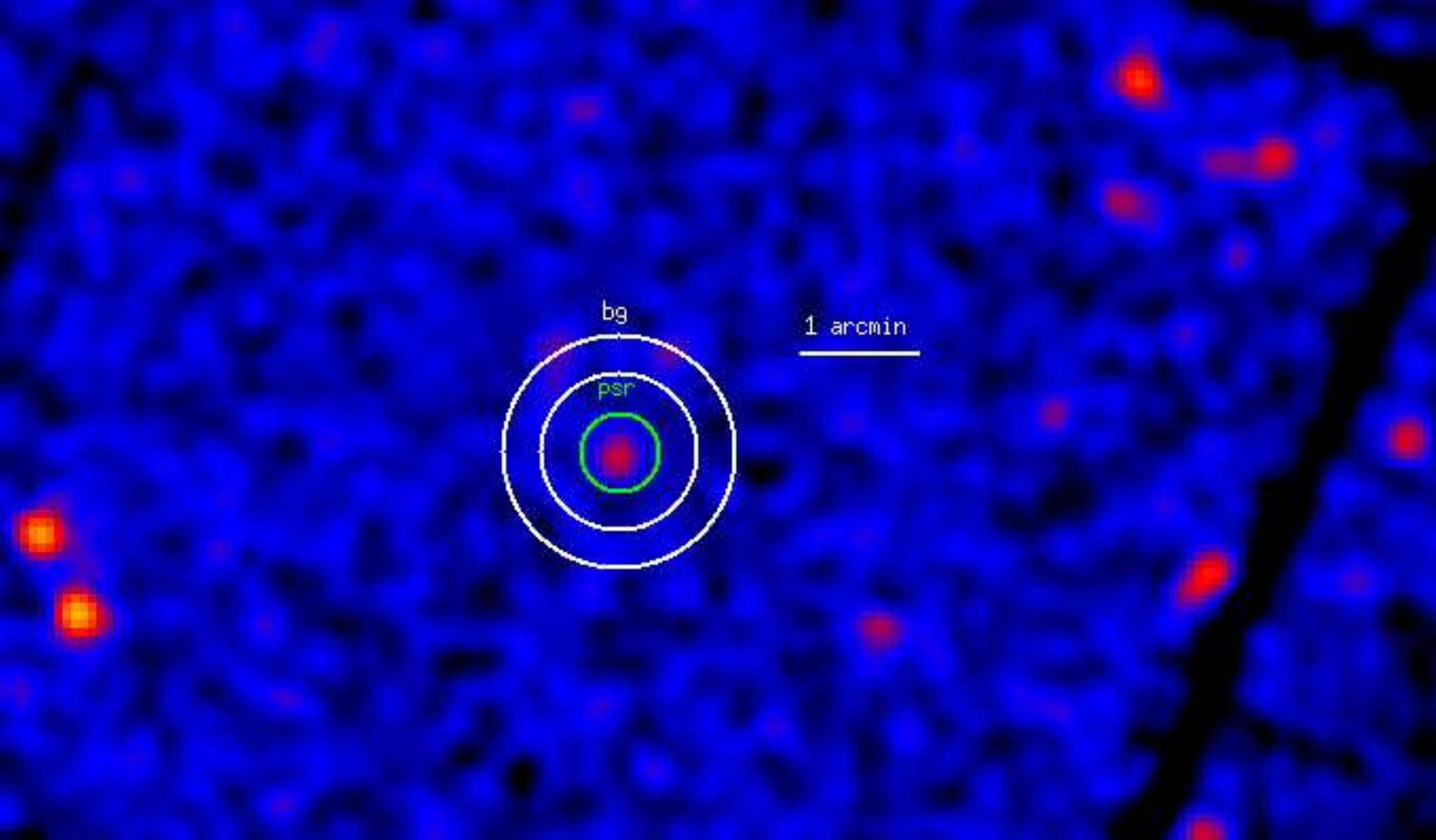}
\caption{J0034-0534 0.3-10 keV {\it Chandra} Imaging. The image has been smoothed with a Gaussian
with Kernel radius of $2"$. The green circle marks the pulsar while the yellow annulus the background region used in the analysis.
\label{J0034-im}}
\end{figure}

\begin{figure}
\centering
\includegraphics[angle=0,scale=.30]{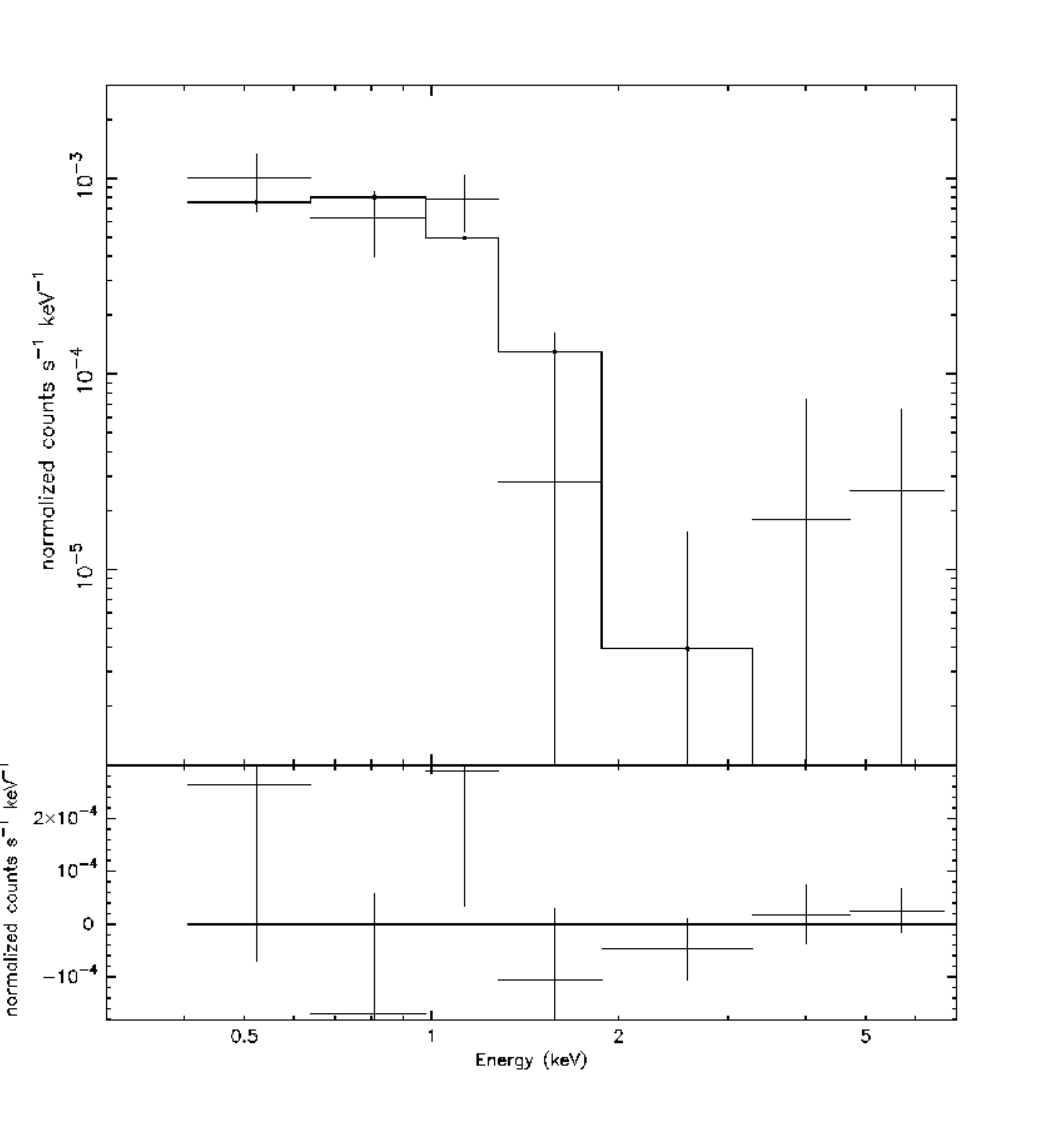}
\caption{J0034-0534 Chandra Spectrum (see text for details).
Residuals are shown in the lower panel.
\label{J0034-sp}}
\end{figure}

\clearpage

{\bf J0101-6422 - type 0 RL MSP} % Nuova!

Pulsations from J0101-6422 were detected by {\it Fermi} in the last months and
it was announced during the 3rd {\it Fermi} Symposium in Rome.
The pseudo-distance obtained from Radio dispersion measurements is
$\sim$ 0.55 kpc.

A {\it SWIFT} observation was performed on the position of the {\it Fermi} source
(obs id. 00031541001, 3.28 ks exposure).
No X-ray source was found at the Radio position
(01:01:11.04, -64:22:30.00).
Under the hypothesis of a distance of 0.55 kpc we found a
rough absorption column value of 1 $\times$ 10$^{20}$ cm$^{-2}$.
Using a simple powerlaw spectrum
for PSR+PWN with $\Gamma$ = 2 and a signal-to-noise of 3,
we obtained an upper limit
unabsorbed non-thermal flux of 2.31 $\times$ 10$^{-13}$ erg/cm$^2$ s
that translates in an upper limit luminosity L$^{nt}_{0.55kpc}$ = 8.40 $\times$ 10$^{30}$ erg/s.

{\bf J0205+6449 (3C58) - type 2 RLP} % cambiamento rispetto all'articolo : aggiunto il blackbody

% da abdo et al. 2009 su J0205
The radio source 3C 58 was recognized early to be a supernova
remnant (SNR G130.7+3.1; Caswell 1970), and
later classified as a pulsar wind nebula by Weiler \& Panagia 1978. Becker et al.
(1982) identified an X-ray point source in the heart of
3C 58 as a likely pulsar, and subsequent studies yielded
a distance of 3.2 kpc (Roberts et al. 1993). The distance evaluation
from kinematic models yields a value 2.9 $\pm$ 0.3 kpc (Green \& Gull 1982, Roberts et al. 1993). The pulsar
J0205+6449 was discovered in {\it Chandra} X-ray
Observatory data, with a period of 65.7 ms, while Rossi
X-ray Timing Explorer archives allowed a measurement
of the spindown rate of $\dot{P}$ = 1.94 $\times$ 10$^{-13}$ (Murray et al.
2002). The characteristic age of the pulsar is $\tau_c$ $\sim$ 5 ky and the age of the associated SNR is 
$\tau_{SNR}$ = 4.25 $\pm$ 0.85 ky (Gotthelf et al. 2007a).
The pulsar has a very high $\dot{E}$,
2.7 $\times$ 10$^{37}$ ergs s$^{-1}$ (one of the most energetic of the
Galactic pulsars), and a surface magnetic field strength of
3.6 $\times$ 10$^{12}$ G.
Both telescopes observed an X-ray profile with
two narrow peaks separated by 0.5 in phase. This was
followed by the detection of radio pulsations with
a pulse averaged flux density of $\sim$ 45 $\mu$Jy at 1.4GHz
and a sharp pulse of width 2ms (Camilo et al. 2002).
The $\gamma$-ray PWN wasn't detected down to
11.63 $\times$ 10$^{-12}$ erg/cm$^2$s (Ackermann et al. 2010).

\begin{figure}
\centering
\includegraphics[angle=0,scale=.30]{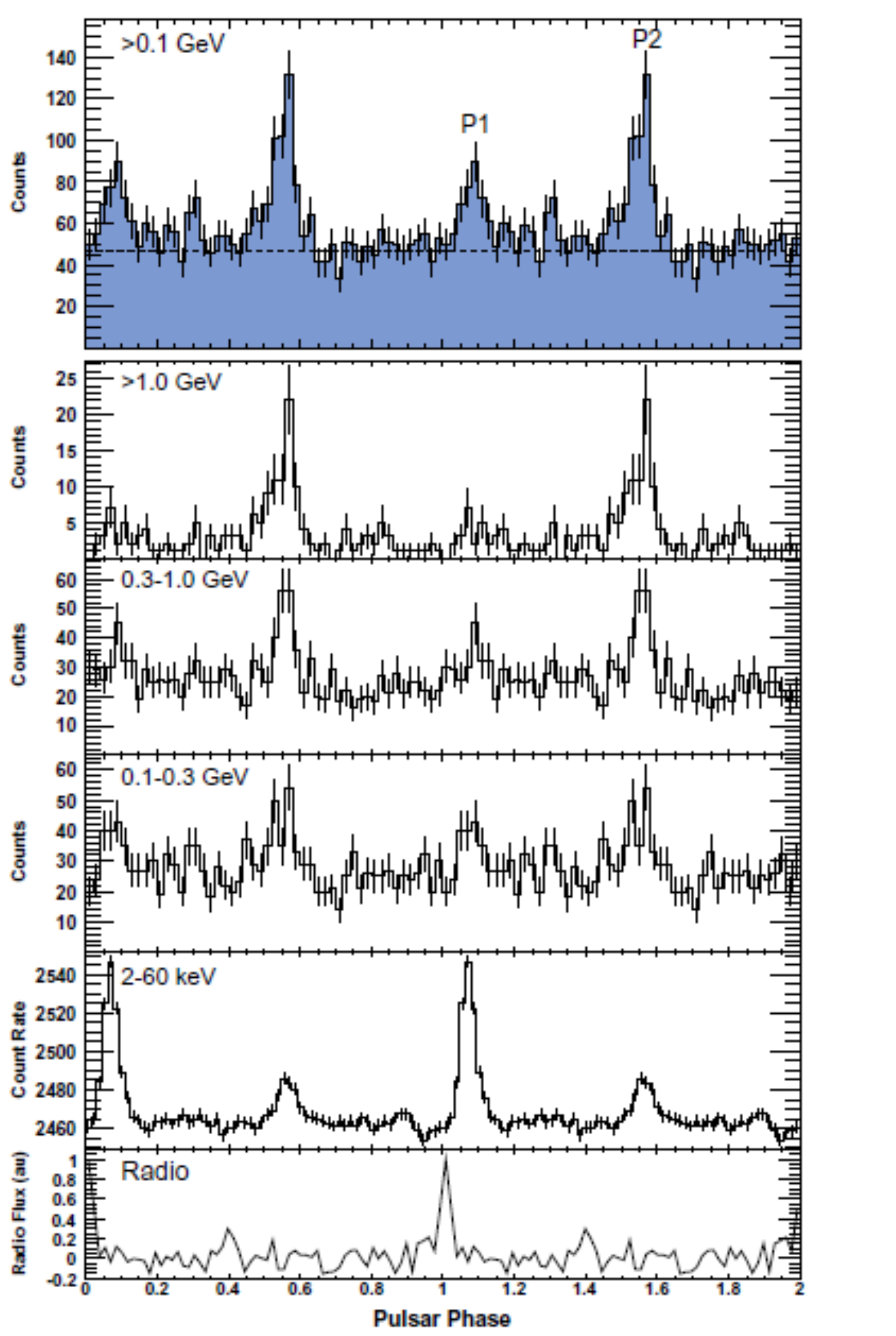}
\caption{PSR J0205+6449 Lightcurve. {\it Fermi} $\gamma$-ray lightcurve folded with Radio and X-ray (Abdo et al. 2009c). 
{\it Second panel from bottom:} Count rate in the energy band 2-60
keV from RXTE data (Livingstone et al. 2009). {\it Bottom panel:} Radio pulse profile based on 3.8 hours of GBT observations at a center
frequency of 2GHz with 64 phase bins.
\label{J0205-lc}}
\end{figure}

\begin{figure}
\centering
\includegraphics[angle=0,scale=.50]{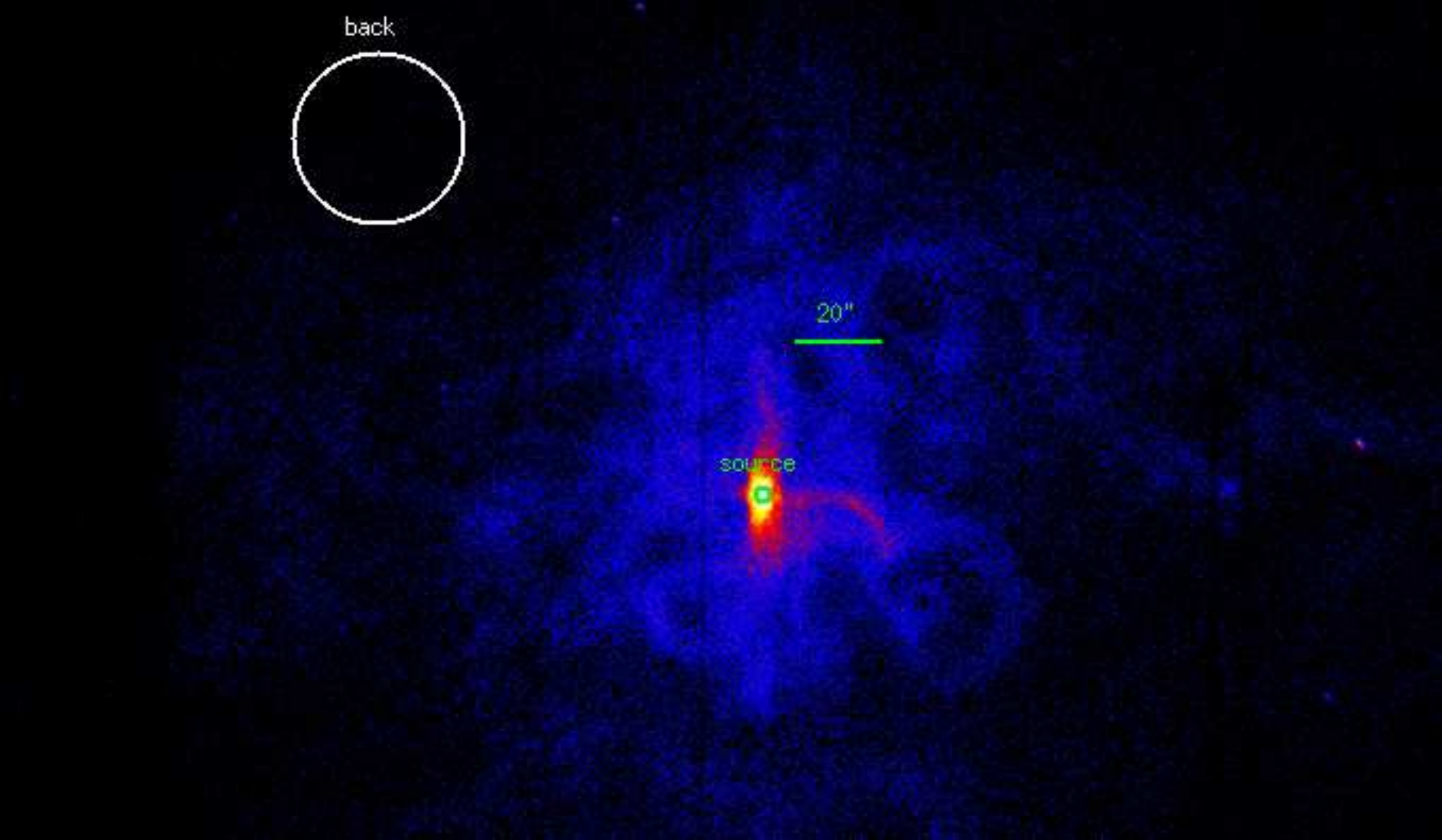}
\caption{PSR J0205+6449 0.3-10 keV {\it Chandra Imaging}. The green circle marks the pulsar while the white one the background region used in the analysis.
\label{J0205-im}}
\end{figure}

This pulsar presents a low scale ($\sim$ 5$"$) bright PWN and
a fainter pulsar wind nebula is evident few arcmins from the point source.
The low spatial resolution of XMM cameras makes it very difficult to
disentangle the different spectral components.
Due to the numerous X-ray observations available, we used only the {\it Chandra} ACIS-S observations of this source:\\
- obs. id 728, start time 2000, September 04 at 09:27:41 UT, exposure 50.0 ks;\\
- obs. id 3832, start time 2003, April 26 at 11:17:50 UT, exposure 139.3 ks;\\
- obs. id 4382, start time 2003, April 23 at 20:00:38 UT, exposure 167.4 ks;\\
- obs. id 4383, start time 2003, April 22 at 15:26:43 UT, exposure 37.4 ks.\\
In all the observations the pulsar position was imaged on the back-illuminated ACIS
S3 chip and the VFAINT exposure mode was adopted. The off-axis angle is negligible.
The X-ray source best fit position is 02:05:37.91 +64:49:41.40 (1$"$ error radius),
obtained by using the celldetect tool inside the CIAO distribution.
We extracted the source spectrum from a circle of 1.5$"$ radius centered on the source position in order to reduce the contamination from the bright PWN.
The nebular spectrum has not been analyzed due to the important thermal contribution from the SNR.
The background was extracted from a region far away from the nebula (see figure \ref{J0205-im})
in order to avoid any contamination from the thermal shell found in Gotthelf et al. 2007
between 2 and 5 $"$ from the source.
We obtained respectively 5915,17672,22023 and 4647 pulsar counts in the four observations
inside the extraction region (where the background contributions are $<$ 0.1\%).
Due to the presence of an important SNR around the pulsar, the nebular spectrum was
analyzed in the 2-10 keV energy band.
We obtained respectively 18973, 23603, 4862 and 7621 nebular counts in the four observations
inside the extraction region (where the background contributions are less than 0.01).
The pulsar emission is well described (probability of
obtaining the data if the model is correct - p-value - of 0.999, 882 dof) by a
combination of a blackbody and a power law model.
The powerlaw component has a photon index of 1.77 $\pm$ 0.03,
absorbed by a column N$_H$ = 4.50$_{-0.11}^{+0.13}$ $\times$ 10$^{21}$ cm$^{-2}$.
The best fit temperature of the blackbody component is 1.88$_{-0.13}^{+0.14}$ $\times$ 10$^6$ K
while the emitting radius is R$_{2.9kpc}$ = 1.67$_{-0.36}^{+0.43}$ km.
The nebular photon index is $\Gamma^{pwn}$ = 2.00 $\pm$ 0.03.
A simple powerlaw model yields an acceptable fit but the very low chance probability ($<10^{-20}$)
obtained with an f-test seems to exclude such a model.
Assuming the best fit model, the 0.3-10 keV unabsorbed non-thermal flux is 
1.97 $\pm$ 0.07 $\times$ 10$^{-12}$, the thermal flux is
2.10 $\pm$ 0.10 $\times$ 10$^{-13}$ and the nebular flux is
2.40 $\pm$ 0.05 $\times$ 10$^{-12}$ erg/cm$^2$ s. Using a distance
of 2.9 kpc, the luminosities are L$_{2.9kpc}^{bol}$ = 2.12 $\pm$ 0.10 $\times$ 10$^{32}$,
L$_{2.9kpc}^{nt}$ = 1.99 $\pm$ 0.07 $\times$ 10$^{33}$ and L$_{2.9kpc}^{pwn}$ = 2.42 $\pm$ 0.05 $\times$ 10$^{33}$ erg/s.

\begin{figure}
\centering
\includegraphics[angle=0,scale=.50]{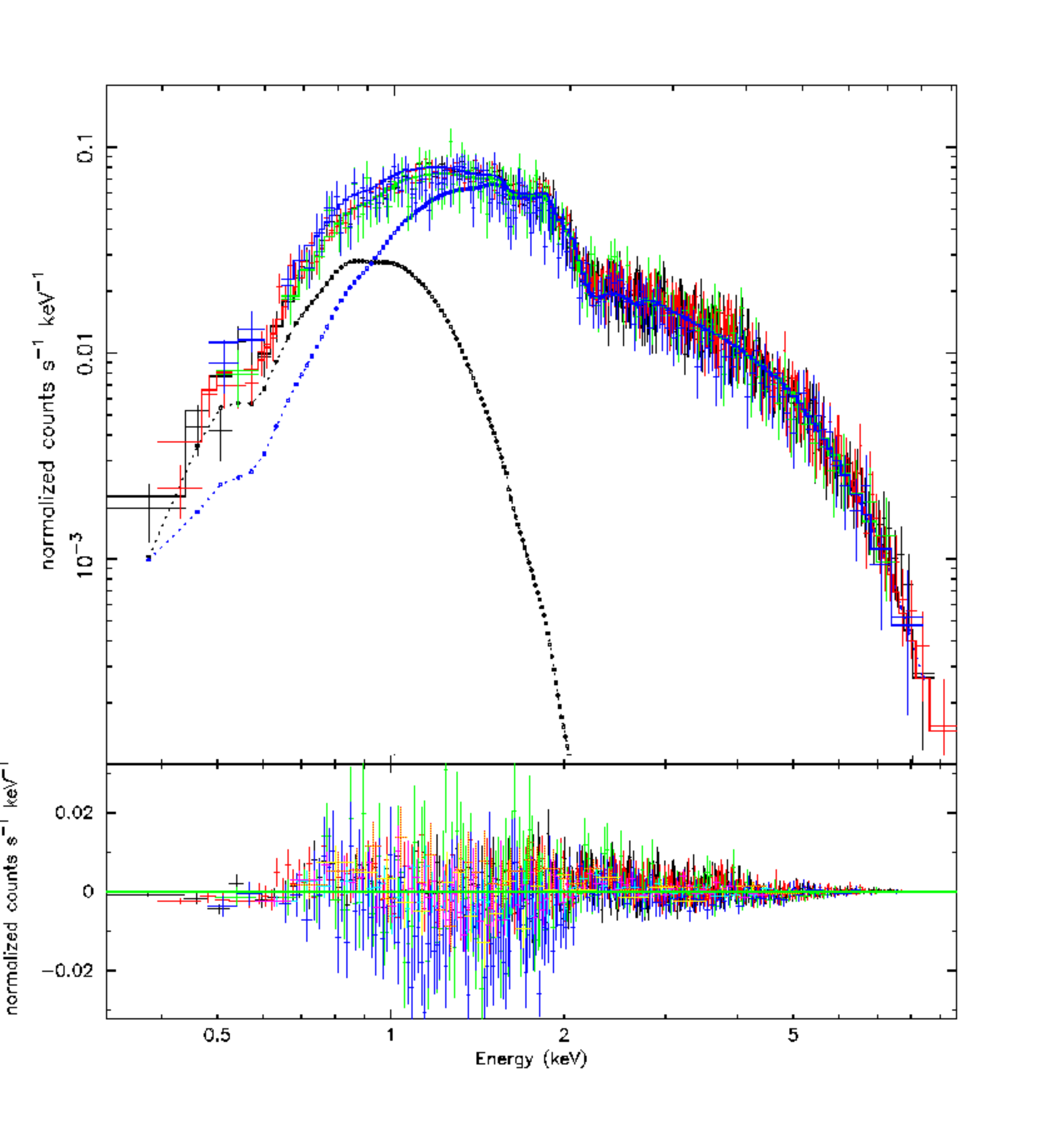}
\caption{PSR J0205+6449 Spectrum. Different colors mark all the different dataset used (see text for details).
Blue points mark the powerlaw component while black points the thermal component of the pulsar spectrum.
Residuals are shown in the lower panel.
\label{J0205-sp}}
\end{figure}

\clearpage

{\bf J0218+4232 - type 2 RL MSP}

% da webb et al. 2004
The radio source J0218+4232 has been known for many
years (Dwarakanath \& Shankar 1990; Hales et al. 1993) although
it was first confirmed to be a millisecond pulsar by
Navarro et al. (1995), using radio observations made with the
Lovell telescope in 1993. A pulse period of 2.3 ms was determined.
Navarro et al. (1995) also showed that this luminous millisecond
pulsar orbits a low mass white
dwarf (0.2 M$_s$), with an orbital period of about two days. From
the dispersion measure a lower limit on the distance of 5.7 kpc
was derived. Considerations on the companion star imply a 
distance of 2.5 to 4 kpc to the system (Bassa et al. 2003).
No PWN was detected in $\gamma$-rays down to
13.65 $\times$ 10$^{-12}$ erg/cm$^2$s (Ackermann et al. 2010).

\begin{figure}
\centering
\includegraphics[angle=0,scale=.30]{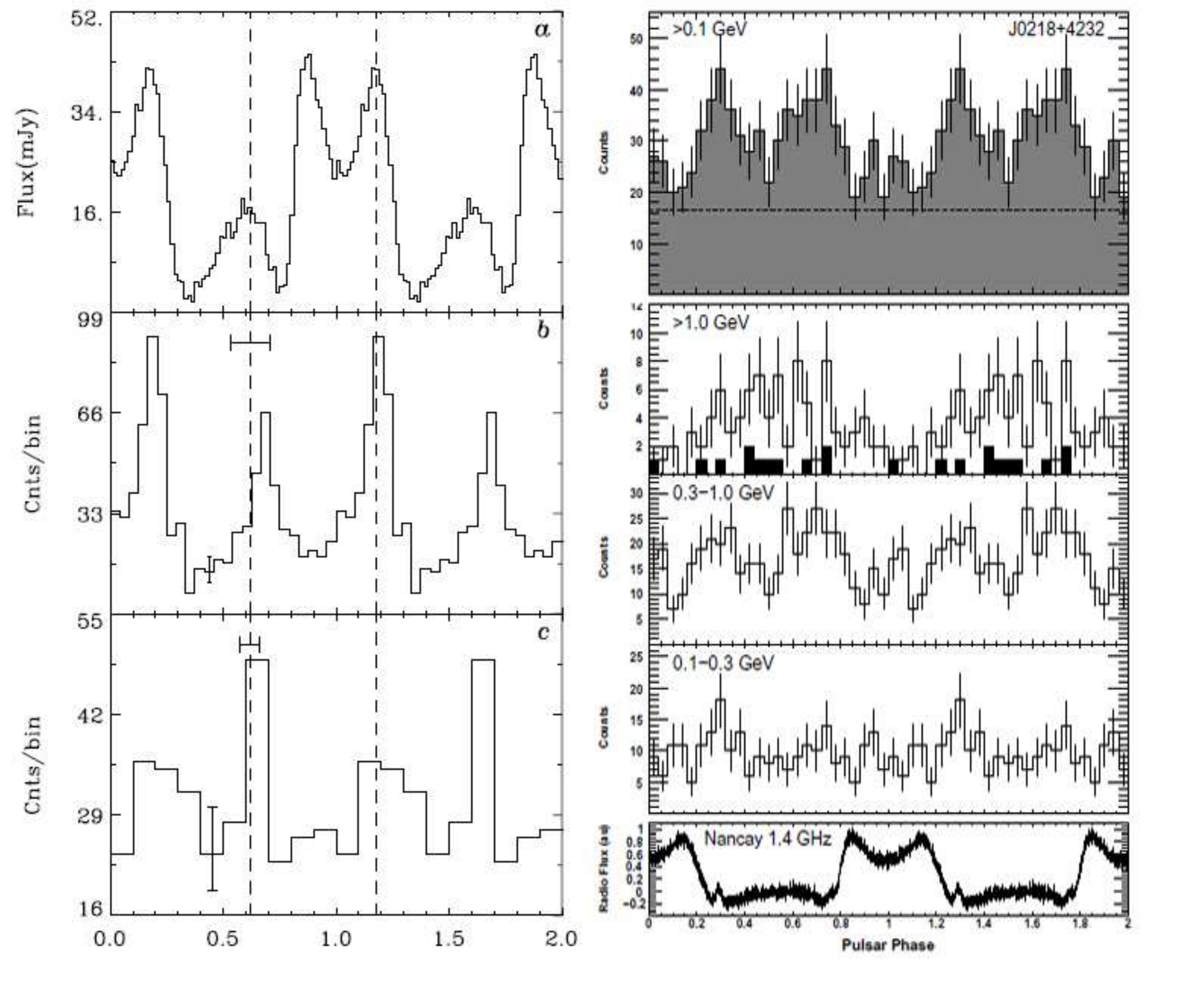}
\caption{PSR J0218+4232 Lightcurve.{\it Right: Fermi} $\gamma$-ray lightcurve folded with Radio
(Abdo et al. catalogue). {\it Left:} X-ray pulse profiles
for different energy bands, folded with Radio.  (Kuiper et al. 2002).
(a) Radio pulse profile at 610 MHz. (b) {\it Chandra} HRC-S X-ray pulse
profile (0.1-10 keV; timing accuracy 200 $\mu$s or 0.085 in phase). (c) {\it EGRET}
$\gamma$-ray pulse profile (0.1-1 GeV; timing accuracy 100 $\mu$s or 0.043 in phase).
Taken from Kuiper et al. 2002.
\label{J0218-lc}}
\end{figure}

We analyzed the only {\it XMM-Newton} observation centered on the pulsar
(observation id 0111100101 started on 2002, February 11 at 16:44:21
and lasted 35.6 ks.
The PN camera of the EPIC
instrument was operated in Fast Timing mode, while the MOS detectors were set in Full frame mode. For
all three instruments, the thin optical filter was used.
After standard data processing (using the epproc and emproc tasks) and
screening of high particle background time intervals,
the good, dead-time corrected exposure time is 26.7 ks.\\
The X-ray source best fit position found using the {\it Chandra} observations
is 2:18:6.305, +42:32:17.23 (1$"$ error radius).
The subarcsecond resolution image of the
HRC-I showed that the X-ray emission from PSR J0218+4232 is consistent with that of a
point source, excluding the presence of a compact nebula (Kuiper et al. 2004).
We extracted the source spectrum from a circle of 20$"$ radius centered on the source position while we extracted the background
from a source-free annulus between 40$"$ and 80$"$.
We excluded the low-energy counts from PN camera (E $<$ 0.5 keV)
due to the contribution of proton micro-flares.
We obtained 5019, 614 and 615 source counts in the 0.3-10
keV range for the PN and two MOS cameras, 
with background contributions of 63.6\%, 4.2\% and 4.6\% of the total.
The pulsar emission is well described (probability of
obtaining the data if the model is correct - p-value - of 0.72, 94 dof) by a
simple power law model with a photon index
$\Gamma$ = 1.11$_{-0.11}^{+0.08}$
absorbed by a column N$_H$ = 2.70$_{-2.70}^{+3.76}$ $\times$ 10$^{20}$ cm$^{-2}$.
An f-test analysis excluded a significative statistical improvement
by adding a blackbody component while a simple blackbody is not statistically allowed.
Assuming the best fit model, the 0.3-10 keV unabsorbed flux is
4.62$_{-0.63}^{+0.43}$ $\times$ 10$^{-13}$ erg/cm$^2$ s.
Using a distance
of 5.7 kpc, the pulsar luminosity is L$_{5.7kpc}$ = 1.80$_{-0.25}^{+0.17}$ $\times$ 10$^{33}$ erg/s.

\begin{figure}
\centering
\includegraphics[angle=0,scale=.50]{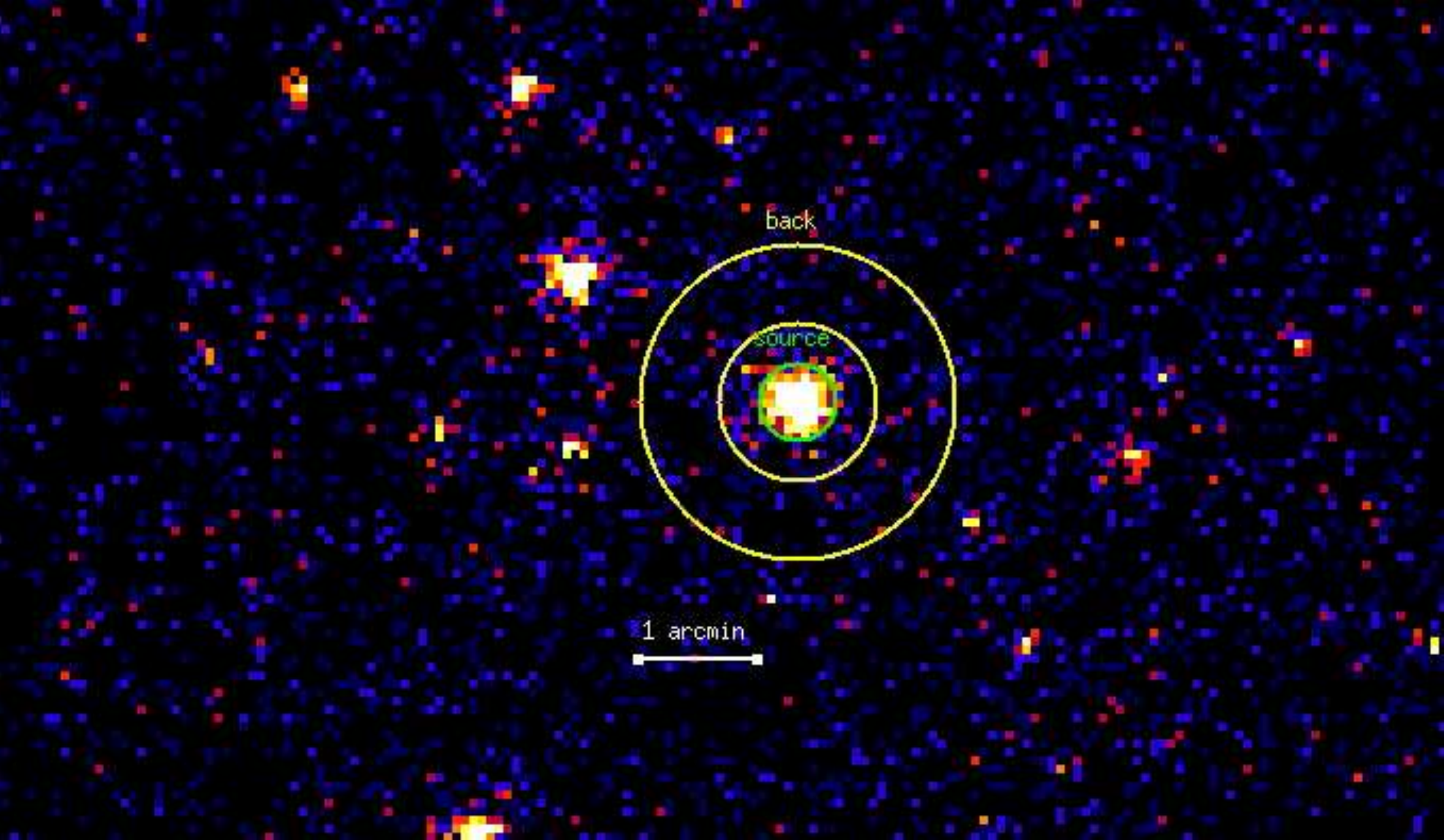}
\caption{PSR J0218+4232 0.3-10 keV MOS Imaging. The two MOS images have been added. 
The green circle marks the pulsar while the yellow annulus the background region used in the analysis.
\label{J0218-im}}
\end{figure}

\begin{figure}
\centering
\includegraphics[angle=0,scale=.50]{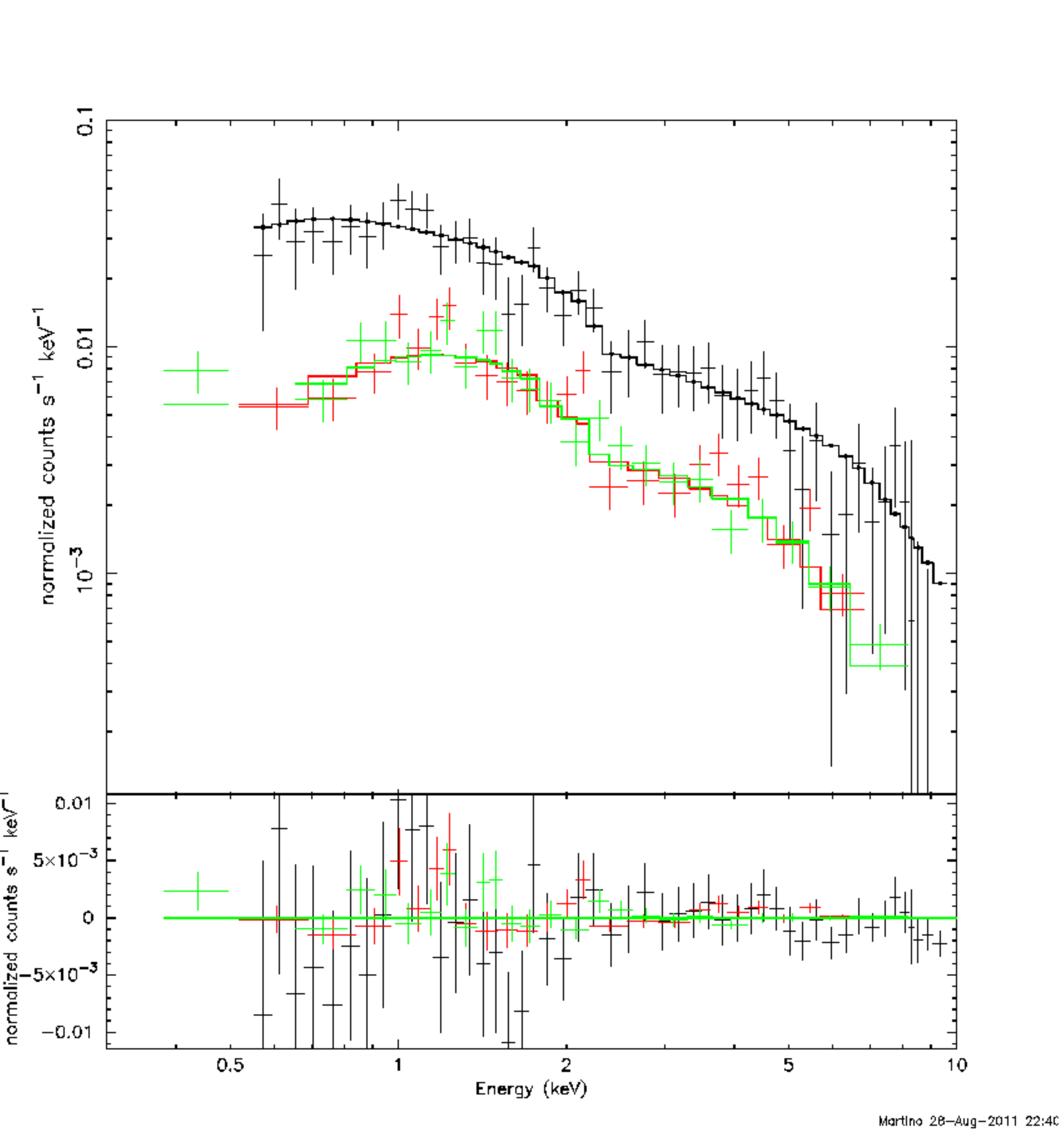}
\caption{PSR J0218+4232  Spectrum. Black points mark the PN spectrum while red and green ones the two MOS' spectra.
Residuals are shown in the lower panel.
\label{J0218-sp}}
\end{figure}

\clearpage

{\bf J0248+6021 - type 0 RLP} % verrà osservato da {\it Chandra}

The young pulsar J0248+6021 (with a rotation period P = 217 
ms and spin-down power $\dot{E}$ = 2.13 $\times$ 10$^{35}$ erg/s) was discovered
with the Nancay radio telescope (NRT) in a survey
of the northern Galactic plane (Foster et al. 1997).
The dispersion measurement places PSR J0248+6021 
at 5.5 $\pm$ 3.5 kpc.
Recently it was measured the proper motion
of the pulsar to be 53 $\pm$ 11 mas yr$^{-1}$.
Moreover, they suggest an association between the pulsar and
the giant HII region W5, the $"$Soul$"$ nebula in the Perseus arm.
This would place the pulsar at a distance of 2.0 $\pm$ 0.2 kpc.
Abdo et al. (2009c) places the pulsar between 2 and 9 kpc.
No PWN was detected in $\gamma$-rays down to
8.59 $\times$ 10$^{-12}$ erg/cm$^2$s (Ackermann et al. 2010).

The only X-ray observations that cover the Radio position
of the pulsar were performed by {\it SWIFT} (obs. id 00031559001 and
00031559002) for a total exposure time of 7.2 ks.
After the data reduction, no X-ray source was found at the
Radio position of the pulsar.
For a distance of 5.5 kpc, we computed a
rough absorption column value of 8 $\times$ 10$^{21}$ cm$^{-2}$
and using a simple powerlaw spectrum
for PSR+PWN with $\Gamma$ = 2 and a signal-to-noise of 3.
We obtained an upper limit non-thermal unabsorbed flux of 9 $\times$ 10$^{-13}$ erg/cm$^2$s,
that translates in an upper limit luminosity of L$^{nt}_{5.5kpc}$ = 3.27 $\times$ 10$^{33}$ erg/s.

{\bf J0340+4130 - type 0 RL MSP} % Nuova!

J0340+4130 was one of the last millisecond radio
pulsars discovered by the {\it Fermi} collaboration.
The pseudo-distance obtained from Radio dispersion measurements is
$\sim$ 1.80 kpc.

Two XMM observations were performed in order to find the counterpart
of the three-months {\it Fermi} source:\\
- obs. id 0605470101, start time 2009, August 05 at 10:01:58 UT, exposure of 19.2ks;\\
- obs. id 0605470801, start time 2009, August 07 at 09:49:22 UT, exposure of 17.2 ks.\\
Both the PN and MOS cameras were operating in Full Frame mode, with a thin optical filter
for the PN camera and a medium for the MOS ones.
After standard data processing (using the epproc and emproc tasks) and
screening of high particle background time intervals,
the good, dead-time corrected exposure time is 12.7 ks.
Four X-ray sources were found inside the {\it Fermi} Error Box
using both the XIMAGE and the SAS dedicated tools.
None of them is coincident with the Radio position
(03:40:29.76 +41:29:49.20).
For a distance of 1.8 kpc, we found a
rough absorption column value of 5 $\times$ 10$^{20}$ cm$^{-2}$ and
using a simple powerlaw spectrum
for PSR+PWN with $\Gamma$ = 2 and a signal-to-noise of 3,
we obtained an upper limit non-thermal unabsorbed flux of 2.00 $\times$ 10$^{-14}$ erg/cm$^2$ s,
that translates in an upper limit luminosity L$_{1.8kpc}^{nt}$ = 7.78 $\times$ 10$^{30}$ erg/s.

{\bf J0357+32 (Morla) - type 2 RQP}

This pulsar was discovered by the {\it Fermi/LAT} collaboration in the first three
months' dataset. J0357+32 was informally named $"$Morla$"$ (from the ancient turtle
appeared in $"$The Neverending Story$"$, by M. Ende) due to its very high age and long
spin period (it's the slowest pulsar discovered by Fermi so far).
See sect.~\ref{morla}.

{\bf J0437-4715 - type 2 RL MSP} % nel paper l'avevo presa da letteratura

% Zavlin et al. 2002
At a parallattic distance d = 156.3 $\pm$ 1.3 pc (Deller et al. 2008),
PSR J0437-4715 is the nearest and brightest X-ray millisecond
pulsar known at both radio and X-ray wavelengths. It was
discovered by Johnston et al. (1993) during the Parkes Southern
radio pulsar survey. It has a spin period P = 5.76 ms, and
a characteristic age $\tau_c$ = 6.6
Gyr and rotation energy loss rate
$\dot{E}$ $\sim$ 3 $\times$ 10$^{33}$ erg/s. It is in a 5.74 d binary orbit with a white
dwarf companion of a low mass M $\simeq$ 0.2M$_s$. Optical observations
in H$\alpha$ have revealed a bow-shock pulsar-wind nebula
(PWN), caused by the supersonic motion of the pulsar through
the interstellar medium, with the bow-shock apex at about 10$"$
south-east of the pulsar, in the direction of the pulsar's proper
motion (Bell, Bailes, \& Bessell 1993; Bell et al. 1995).
PSR J0437-4715 was the first millisecond pulsar detected in
X-rays - Becker \& Trumper 1993 observed it with the PSPC in 1992 and discovered X-ray
pulsations with a single broad pulse and a pulsed fraction
f$_p$ $\sim$ 30\% in the 0.1-2.4 keV energy range.
No PWN was instead detected in $\gamma$-rays down to
9.41 $\times$ 10$^{-12}$ erg/cm$^2$s (Ackermann et al. 2010).

\begin{figure}
\centering
\includegraphics[angle=0,scale=.30]{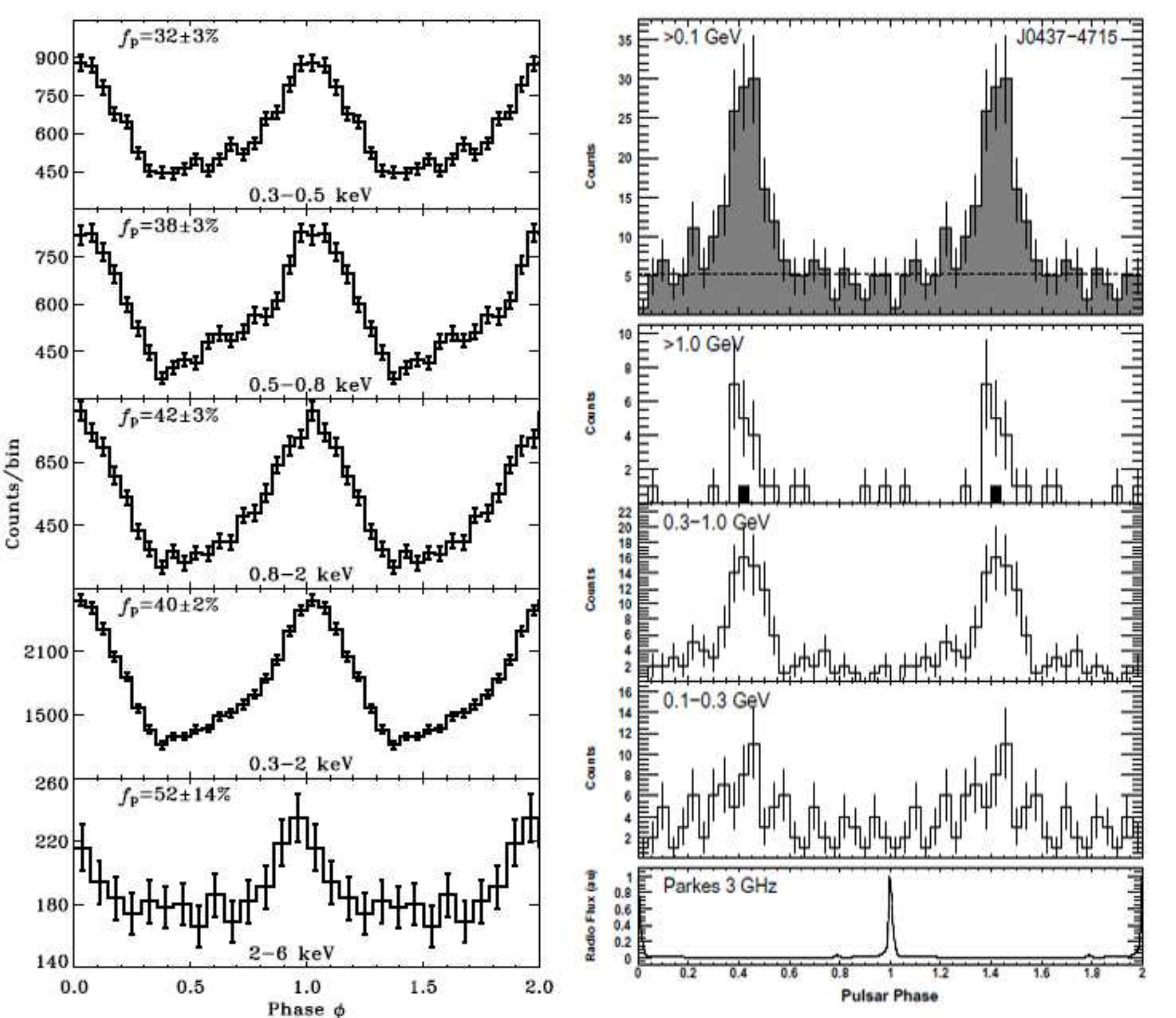}
\caption{PSR J0437-4715 Lightcurve.{\it Right: Fermi} $\gamma$-ray lightcurve folded with Radio
(Abdo et al. 2009c). {\it Left: XMM-Newton} EPIC PN X-ray pulse profiles
for different energy bands, folded with Radio. Here are reported estimated values of the
intrinsic source pulsed fraction f$_p$. The choice of phase zero
and the alignment of the X-ray and radio profiles are arbitrary (Zavlin 2006).
\label{J0437-lc}}
\end{figure}

J0437-4715 was the target of many X-rays observations,
both by {\it Chandra} and XMM. We used the following XMM observation
centered on the pulsar:\\
- obs. id 0112320201, started on 2002, October 09 at 03:33:16 UT and lasted 68.4 ks;\\
- obs. id 0603460101, started on 2009, December 15 at 20:00:24 UT and lasted 129.4 ks.\\
The PN camera of the EPIC
instrument was operated in Fast Timing mode, while the MOS detectors were set in Full frame mode. For
all three instruments, the thin optical filter was used.
We excluded the low-energy counts from PN camera (0.4 $<$ E $<$ 0.5 keV)
due to the contribution of proton micro-flares.
The screening of high particle background time intervals was
not necessary due to the goodness of the first observation.
After standard data processing (using the epproc and emproc tasks) and
screening of high particle background time intervals,
the good, dead-time corrected exposure time of the second observation is 113.5 ks.
We also used all the {\it Chandra} ACIS-S observations of this source:\\
- obs. id 741, start time 2000, May 29 at 23:46:35 UT, exposure 28.2 ks;\\
- obs. id 6154, start time 2005, February 18 at 10:28:31 UT, exposure 25.2 ks;\\
- obs. id 6155, start time 2005, March 03 at 10:06:30 UT, exposure 25.1 ks;\\
- obs. id 6156, start time 2005, April 28 at 06:02:33 UT, exposure 25.2 ks;\\
- obs. id 6157, start time 2005, October 10 at 15:38:30 UT, exposure 9.5 ks;\\
- obs. id 6158, start time 2005, November 28 at 03:54:01 UT, exposure 20.9 ks;\\
- obs. id 6767, start time 2006, February 10 at 04:10:21 UT, exposure 23.2 ks;\\
- obs. id 6768, start time 2006, May 09 at 01:42:31 UT, exposure 24.9 ks;\\
- obs. id 7216, start time 2006, February 10 at 10:52:31 UT, exposure 18.9 ks.\\
In all the {\it Chandra} observations the pulsar position was imaged on the back-illuminated ACIS
S3 chip and the VFAINT exposure mode was adopted. The off-axis angle is negligible.
The X-ray source best fit position is 04:37:15.86 -47:15:09.01 (0.8$"$ error radius).
Zavlin et al. 2002 found no structures in {\it Chandra} observations
which could be interpreted as an X-ray PWN associated with 
the bow shock observed in H$\alpha$.
In the {\it Chandra} observations we extracted the source spectrum from a circle of 2$"$ radius centered on the source while
the background was extracted from a source-free annulus with radii 10$"$ and 15$"$.
In the XMM observation we extracted the source spectrum from a circle of 20$"$ radius centered on the source while we extracted the background
from a source-free annulus with radii 40$"$ and 80$"$.
We obtained 52450, 10308 and 10852 source counts (background contributions of 30.6\%, 3.0\% and 2.7\%)
in the PN and two MOS camera during the first observation and
91039, 16527 and 16227 source counts (background contributions of 31.6\%, 3.3\% and 3.3\%)
in the second {\it XMM-Newton} observation. We also obtained
a total of 5002 source counts 
(where the background contribution is $<$ 0.001) in the {\it Chandra} observations.
The pulsar emission is well described (probability of
obtaining the data if the model is correct - p-value - of 0.007, 2268 dof) by a
combination of a blackbody and a power law model.
The powerlaw component has a steep photon index ($\Gamma$ = 2.98$_{-0.11}^{+0.09}$),
absorbed by a column N$_H$ = 1.58$_{-1.09}^{+0.93}$ $\times$ 10$^{20}$ cm$^{-2}$.
The best fit temperature of the blackbody component is (2.60 $\pm$ 0.05) $\times$ 10$^6$ K
while the emitting radius is R$_{156pc}$ = 63.9$_{-5.7}^{+6.5}$ m. Such a small emitting radius
suggest an hot spot thermal emission, quite typical for such millisecond pulsars.
A simple blackbody model yields a poor fit (p-value $<$ 10$^{-100}$) as well as a simple
powerlaw model (p-value $<$ 10$^{-10}$).
Assuming the best fit model, the 0.3-10 keV unabsorbed thermal flux is 
4.19 $\pm$ 0.20 $\times$ 10$^{-13}$ and the non-thermal flux is
7.91$_{-0.60}^{+0.50}$ $\times$ 10$^{-13}$ erg/cm$^2$ s. Using a distance
of 156 pc, the two luminosities are L$_{156pc}^{bol}$ = (1.23 $\pm$ 0.06) $\times$ 10$^{30}$,
L$_{156pc}^{nt}$ = 2.32$_{-0.18}^{+0.15}$ $\times$ 10$^{30}$ erg/s.

\begin{figure}
\centering
\includegraphics[angle=0,scale=.40]{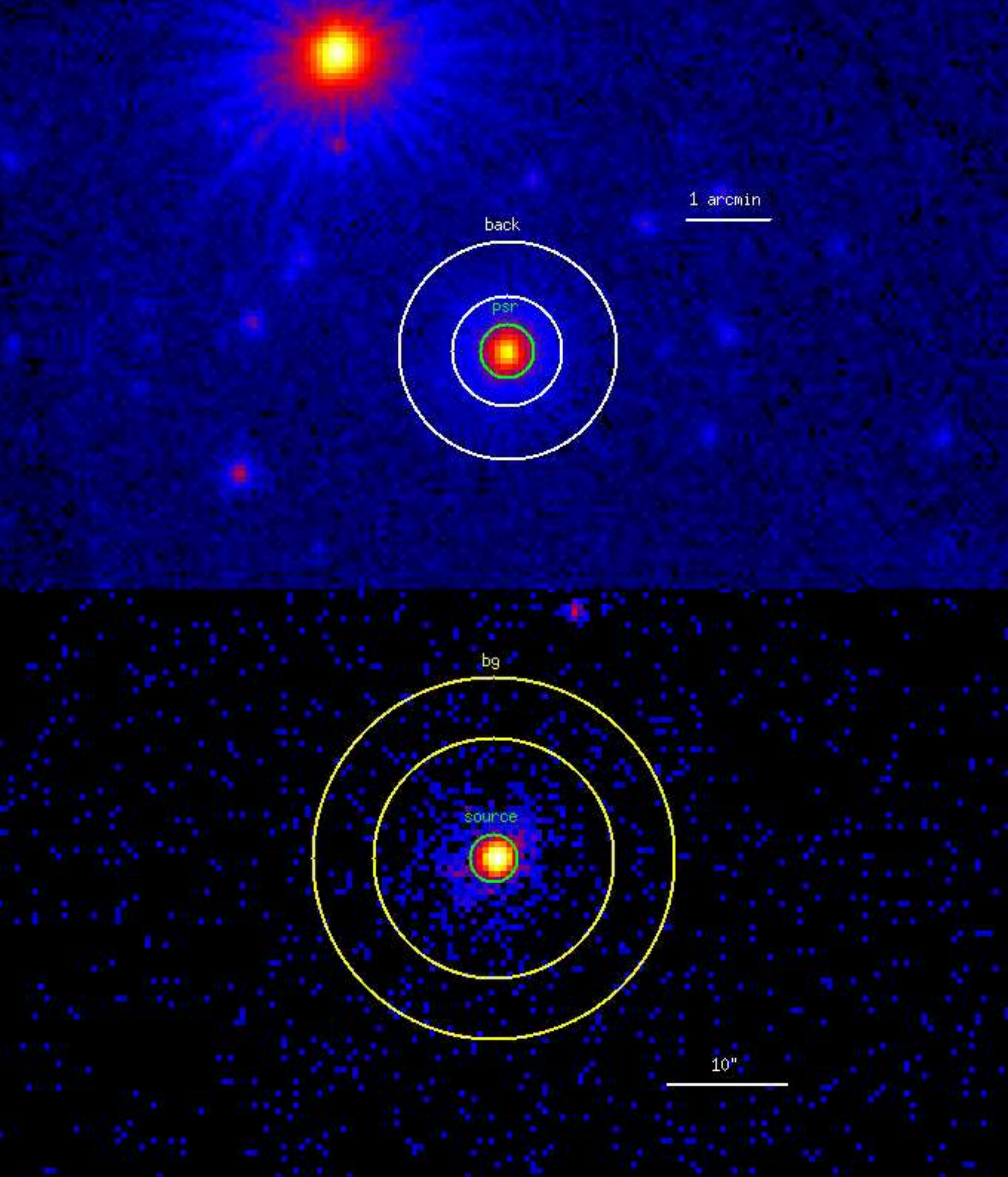}
\caption{PSR J0437-4715 Imaging. {\it Up:} 0.3-10 keV MOS Imaging. The two MOS images have been added. 
The green circle marks the pulsar while the yellow annulus the background region used in the {\it XMM-Newton} analysis.
{\it Down:} 0.3-10 {\it Chandra} Imaging. The green circle marks the pulsar while the 
yellow annulus the background region used in the {\it Chandra} analysis.
\label{J0437-im}}
\end{figure}

\begin{figure}
\centering
\includegraphics[angle=0,scale=.50]{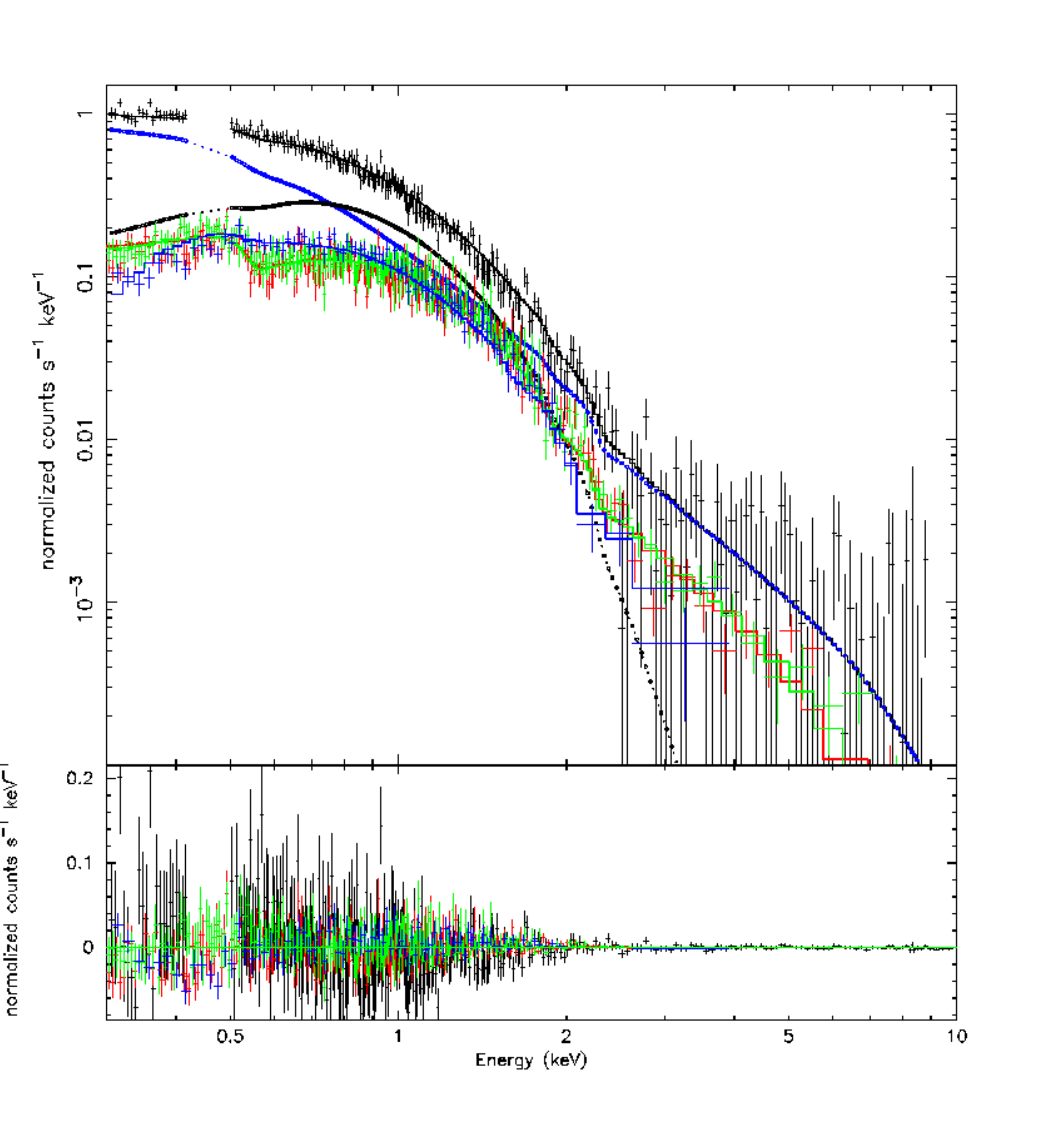}
\caption{PSR J0437-4715 Spectrum. Different colors mark all the different dataset used (see text for details).
Blue points mark the powerlaw component while black points the thermal component of the pulsar spectrum.
Residuals are shown in the lower panel.
\label{J0437-sp}}
\end{figure}

\clearpage

{\bf J0534+2200 (Crab) - type 2 RLP}

The Crab Nebula and Pulsar constitute an intricate system, observed throughout the
electromagnetic spectrum. 
PSR J0534+2200 spins at a rotation frequency of 30 Hz and is the most deeply studied
of all isolated pulsars. The Crab pulsar was first discovered in 1968 and has been to subject
to many long term observation programs at optical and radio wavelengths (Groth 1975a;
Gullahorn et al. 1977; Lohsen 1981; Lyne, Pritchard \& Smith 1988). There have been
several prior analyses of the long term properties of the pulse phase of the Crab pulsar.

\begin{figure}
\centering
\includegraphics[angle=0,scale=.30]{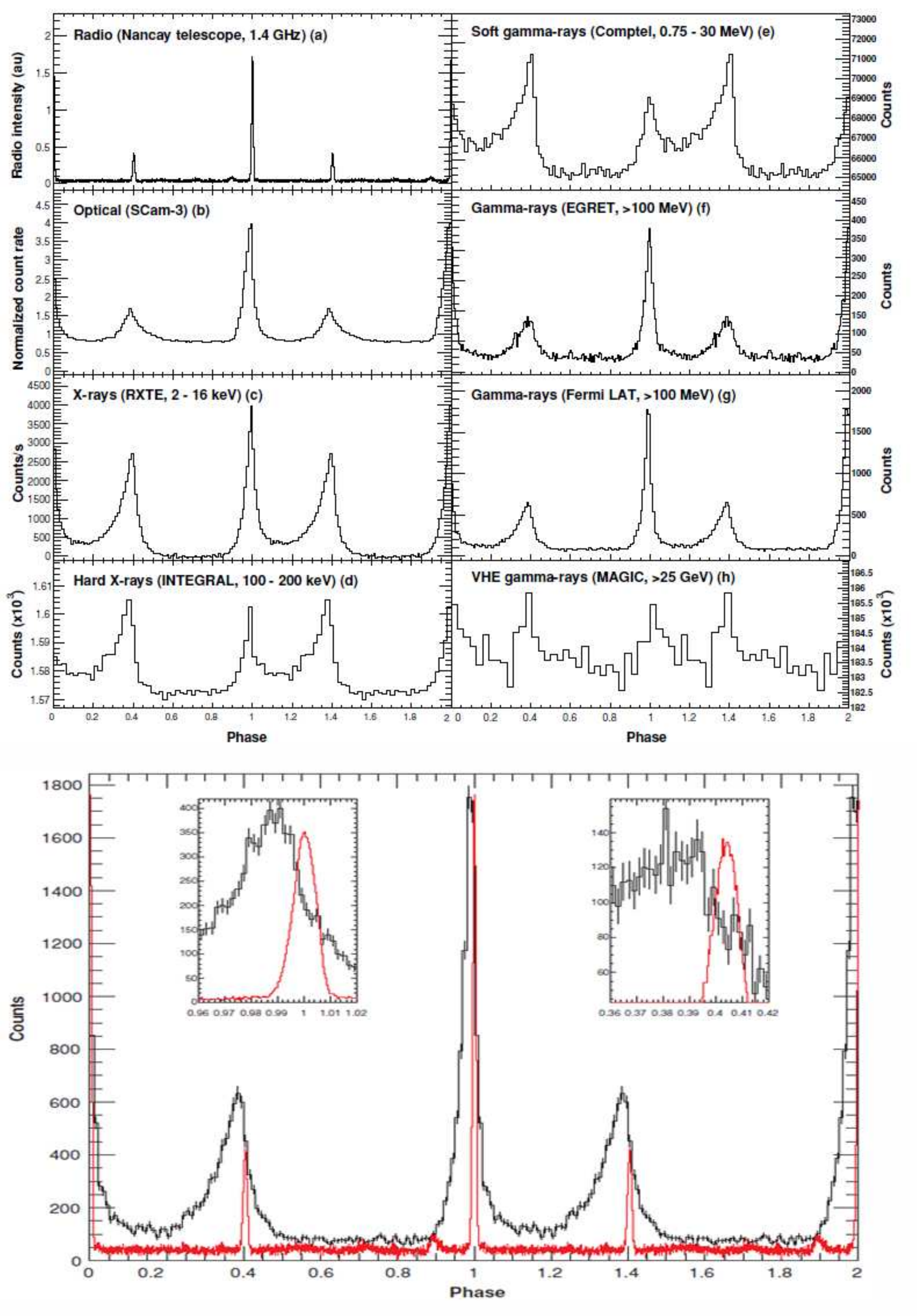}
\caption{PSR J0534+2200 Lightcurve. {\it Upper panel:} Light curves at different wavelengths.
Two cycles are shown. References: (a) from the Nancay radio telescope; (b) Oosterbroek et al. 2008; (c) Rots et al.
2004; (d) Mineo et al. 2006; (e) Kuiper et al. 2001; (f) EGRET, Kuiper et al. 2001; (g) This paper; (h) Aliu et al. 2008.
{\it Lower panel: Fermi}  light curve obtained with photons above 100 MeV. The radio light curve (red line) is overlaid.
See Abdo et al. (2010b) for details.
\label{crab-lc}}
\end{figure}

Due to the complex X-ray structure of the inner nebula and
pulsar, the unprecedented angular resolution of the {\it Chandra} X-ray Observatory has proven
invaluable in probing the nature of this region.
All the results are obtained using {\it Chandra} observations and are taken from Kargaltsev \& Pavlov 2008.\\
Recently the nebula was found to be variable but the X-ray and $\gamma$-ray behaviour of the
pulsar appears unchanged.

\begin{figure}
\centering
\includegraphics[angle=0,scale=.50]{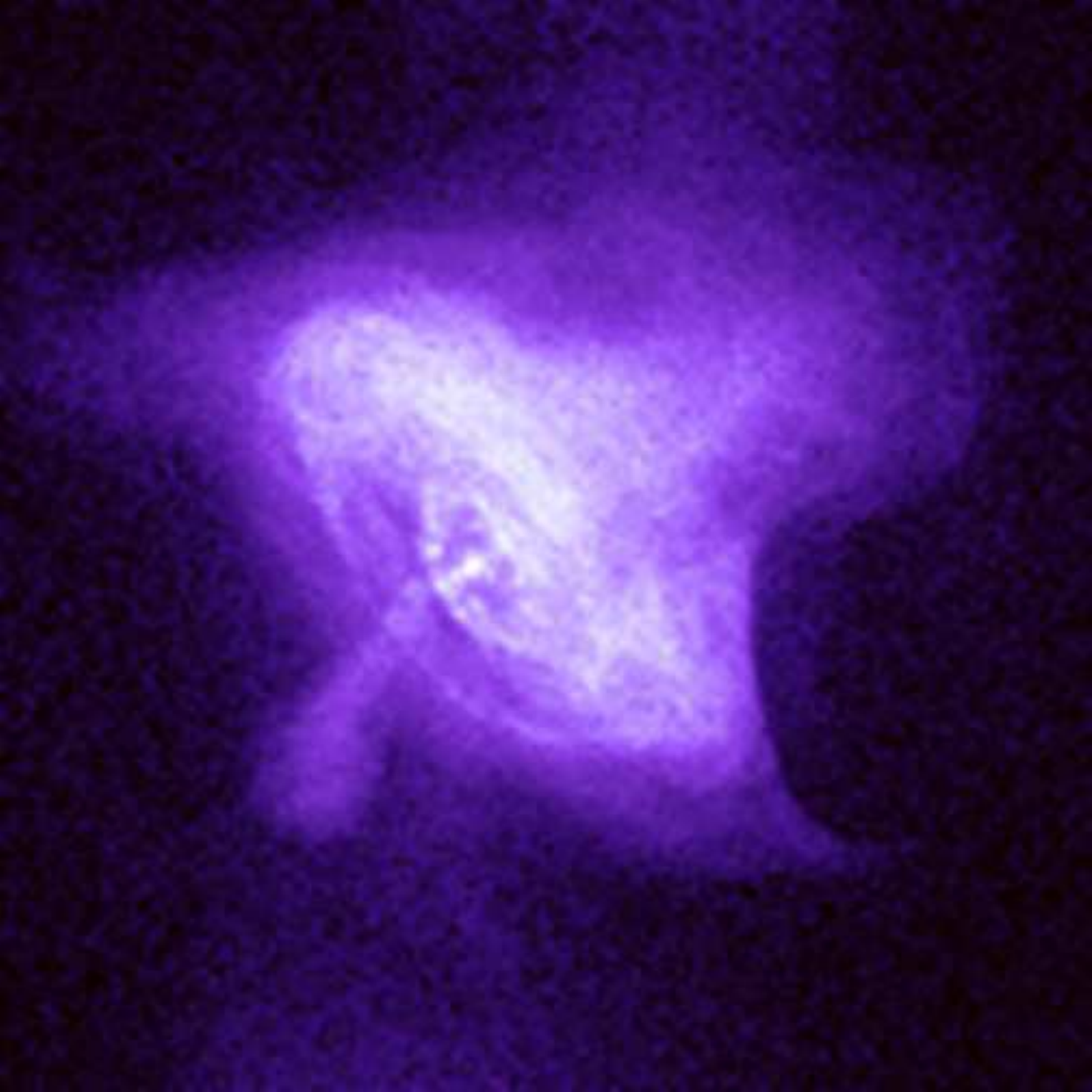}
\caption{PSR J0534+2200 0.3-10 keV {\it Chandra} Imaging. \label{crab-im}}
\end{figure}

\begin{figure}
\centering
\includegraphics[angle=0,scale=.40]{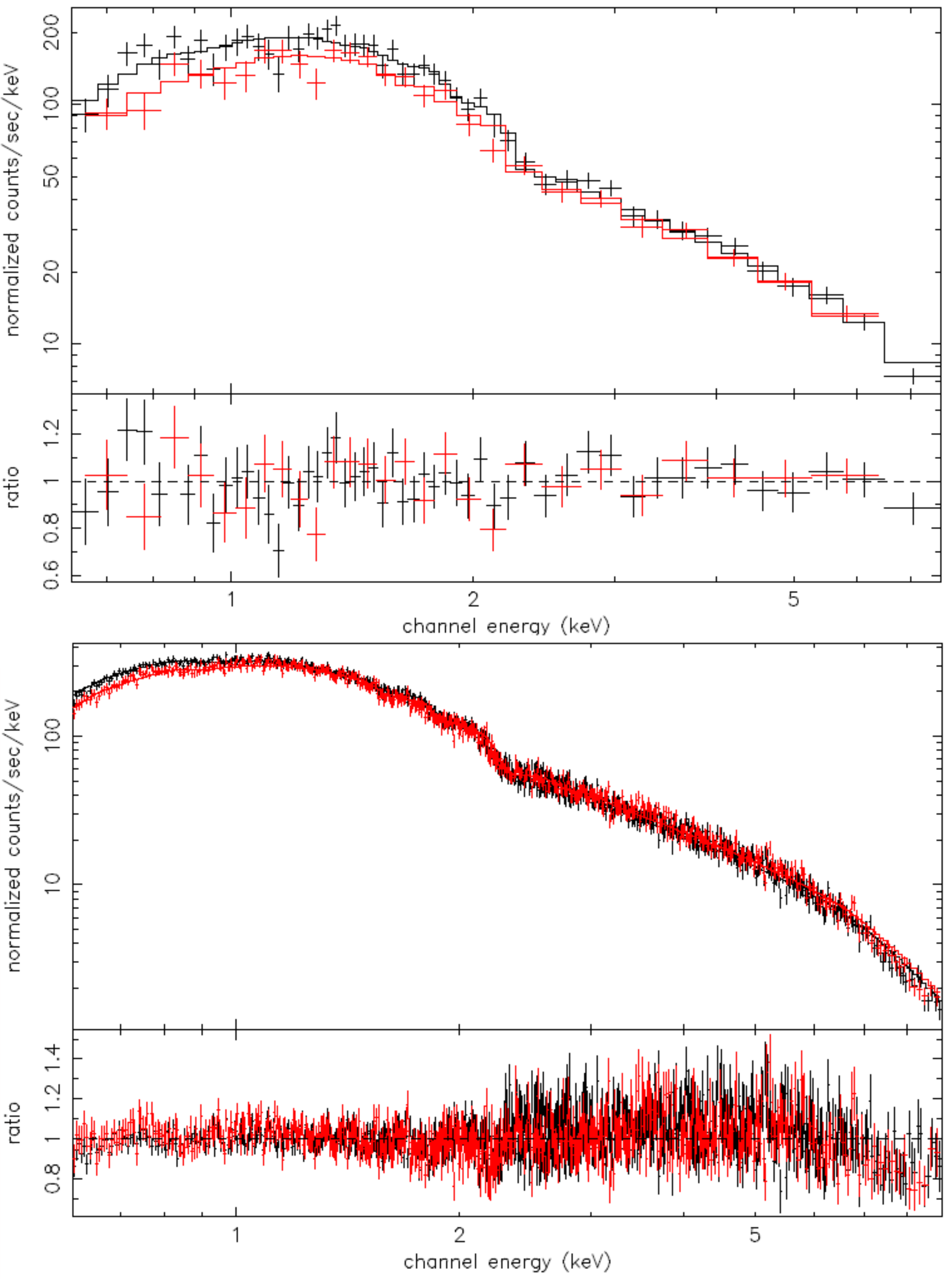}
\caption{PSR J0534+2200 {\it XWW-Newton} Spectrum. {\it Upper panel:} Spectra of the pulsar region covering the 0.65-8.5 keV
energy band. {\it Lower panel:} Spectra of the nebula region covering the 0.6-9.0 keV energy band. See Kirsch et al. 2006 for details
\label{cra-sp}}
\end{figure}

\clearpage

{\bf J0610-2100 - type 0 RL MSP} % Nuova!

J0610-2100 was one of the last millisecond radio
pulsars discovered by the {\it Fermi} collaboration.
It's in a binary system with an orbital period of 0.286d and
its distance, based on radio dispersion measurements, is $\sim$ 3.5 kpc (Burgay et al. 2006).

A {\it SWIFT} observation was performed at the position of the {\it Fermi} source
(obs id. 00031646001, 1.80 ks exposure).
No X-ray source was found inside the {\it Fermi} error box.
For a distance of 3.5 kpc, we computed a
rough absorption column value of 3 $\times$ 10$^{20}$ cm$^{-2}$
and using a simple powerlaw spectrum
for PSR+PWN with $\Gamma$ = 2 and a signal-to-noise of 3,
we obtained an upper limit non-thermal unabsorbed flux of 3.46 $\times$ 10$^{-13}$ erg/cm$^2$s,
that translates in an upper limit luminosity of L$_{3.5kpc}^{nt}$ = 5.09 $\times$ 10$^{32}$ erg/s.

{\bf J0613-0200 - type 2* RL MSP} % ho usato l'apec al posto di gau che era nell'articolo

J0613-0200 was discovered during the
Parkes survey for millisecond pulsars.
This $"$recycled$"$ millisecond pulsars is located at 0.48$_{-0.11}^{+0.19}$ kpc,
on the basis of its dispersion measurement and of the analysis of the
$\sim$ 0.3 M$_s$ stellar companion.
Only radio analysis of the source are available in literature
despite of the {\it XMM-Newton} observation performed in September 2008
(obs. id 0406620301).\\
The XMM field of view is centered 2.5$'$ away from the source, well
inside the central chip of the MOS cameras, operating in full frame mode,
but outside the FOV of the PN camera, operating in timing mode.
For all three instruments, the medium optical filter was used.
After standard data processing (using the epproc and emproc tasks) and
screening of high particle background time intervals,
the good, dead-time corrected exposure time is 64.6 ks.\\
The source X-ray best fit position is 06:13:46.640 -01:58:45.76 (5$"$ error radius),
obtained by using both the XIMAGE and SAS dedicated tools.
We extracted the source spectrum from a circle of 20$"$ radius centered on the source while we extracted the background
from a source-free annulus with radii 40$"$ and 80$"$.
We obtained 749 and 624 source counts in the 0.3-10
keV range for the two MOS cameras, 
(with background contributions of 7.4\% and 7.2\%).
In the MOS1 and MOS2 images, the source count distribution
around the radio position of the pulsar is fully consistent
with that of a point source. Thus, no indication of
any diffuse extended emission was found.
In $\gamma$-rays no nebula emission was detected
down to a flux of 7.46 $\times$ 10$^{-12}$ erg/cm$^2$ s.

The pulsar emission is well described (probability of
obtaining the data if the model is correct - p-value - of 0.04) by a
combination of a power law model and an emission spectrum from 
collisionally-ionized diffuse gas (apec). The apec spectrum
is typical of stellar coronal emission coming from a star
superimposed with the pulsar - maybe its companion star. 
The powerlaw component has a photon index $\Gamma$ = 2.05 $_{-0.49}^{+0.22}$ , 
absorbed by a column N$_H$ = $<$ 3.3 $\times$ 10$^{20}$ cm$^{-2}$.
A simple powerlaw model yields a poor fit (p-value $<$ 10$^{-7}$) as well as a simple
apec model or a blackbody model.
Assuming the best fit model, the 0.3-10 keV unabsorbed pulsar flux is 
9.59 $_{-4.43}^{+6.85}$  $\times$ 10$^{-14}$ erg/cm$^2$ s. For a distance
of 480 pc, its luminosity is L$_{480pc}$ = 2.65 $_{-1.22}^{+1.89}$ $\times$ 10$^{30}$ erg/s.

\begin{figure}
\centering
\includegraphics[angle=0,scale=.50]{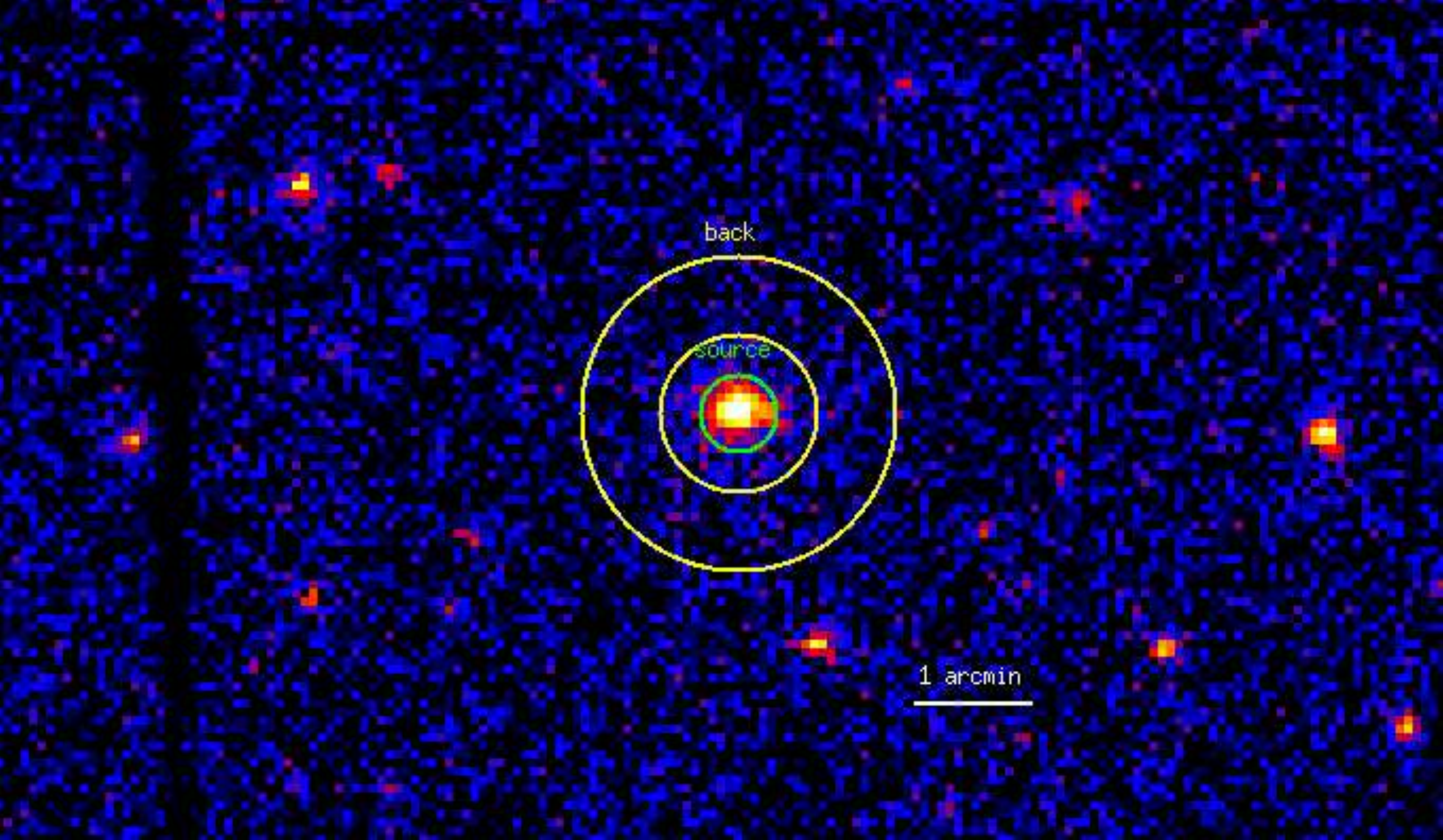}
\caption{PSR J0613-0200 0.3-10 keV MOS Imaging. The two MOS images have been added. 
The green circle marks the pulsar while the yellow annulus the background region used in the analysis.
\label{J0613-im}}
\end{figure}

\begin{figure}
\centering
\includegraphics[angle=0,scale=.50]{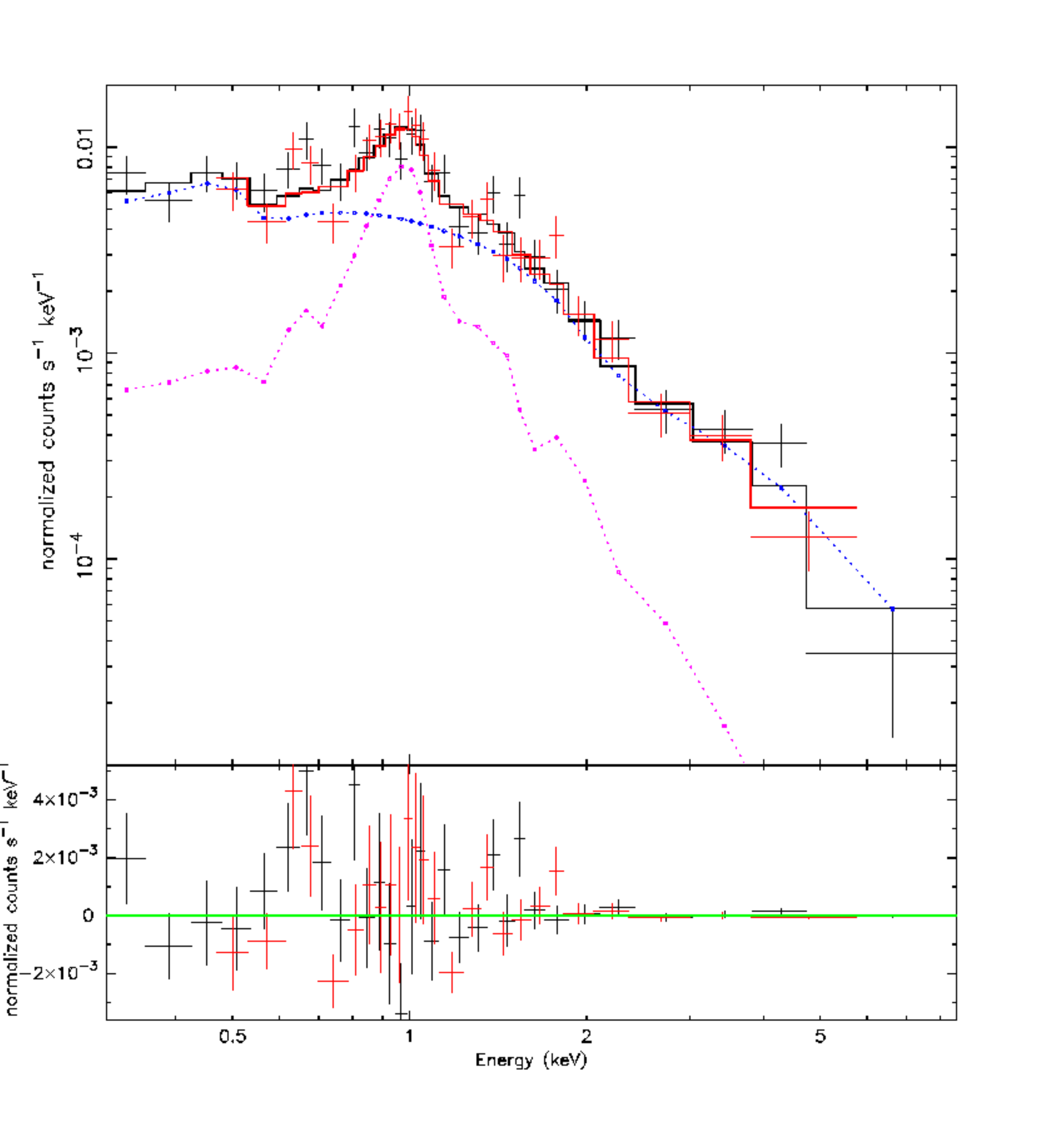}
\caption{PSR J0613-0200 Spectrum. Different colors mark all the different dataset used (see text for details).
Magenta points mark the stellar component while blue points the powerlaw component of the spectrum.
Residuals are shown in the lower panel.
\label{J0613-sp}}
\end{figure}

\clearpage

{\bf J0614-3330 - type 1 RL MSP} % Nuova! osservazione XMM in arrivo

% Ransom et al. 2010
J0614-3330 was detected in Radio by the Pulsar Search Consortium (PSC), a group of approximately
20 LAT-team members and/or pulsar experts associated
with large radio telescopes around the world.
The Green Bank Telescope (GBT) was used to investigate $\gamma$-ray
sources from the {\it Fermi} LAT Bright Source List
(Abdo et al. 2009b, Ransom et al. 2010) with no obvious counterpart.
The distance evaluated on the basis of the radio dispersion measurements is
$\sim$ 1.9 kpc.

Three {\it SWIFT} observations
were performed (obs. id 00031375001,00031375002 and 00031375003) for a total exposure of 15.86 ks.
Using the XIMAGE tools, a source
coincident with the Radio position of the pulsar
at 06:14:10.0 -33:29:53.0 (5$"$ error radius) was found yielding 25 counts.
The spectra obtained in the three observations were added using
mathpha tool and similarly the response
matrix and effective area files using addarf and addrmf.
Due to the low statistic, we used the C-statistic
approach implemented in XSPEC.
No indication on the presence of a PWN can be obtained from the X-ray observation.
However MSPs usually have very faint nebulae - if at all.
Both a simple blackbody and powerlaw spectra are statistically acceptable.
The simple powerlaw model has a photon index $\Gamma$ = 4.91$_{-3.37}^{+4.94}$
absorbed by a column N$_H$ = 4.21$_{-2.37}^{+8.77}$ $\times$ 10$^{21}$ cm$^{-2}$.
The simple blackbody model has a temperature T = 2.89$_{-1.33}^{+0.66}$ $\times$ 10$^6$ K,
with an emitting radius of 218$_{-138}^{+1752}$ m, typical of a thermally emitting hot spot,
absorbed by a column N$_H$ = $<$ 4.24 $\times$ 10$^{21}$ cm$^{-2}$.
Using the powerlaw model best fit the pulsar unabsorbed flux is 1.50 $\pm$ 0.56 $\times$ 10$^{-12}$ erg/cm$^2$ s
Using the thermal model best fit the pulsar unabsorbed flux is 5.18 $\pm$ 2.01 $\times$ 10$^{-14}$ erg/cm$^2$ s
For a distance of 1.9 kpc, this translates into luminosities of 
L$_{1.9kpc}^{nt}$ = 6.50 $\pm$ 2.43 $\times$ 10$^{32}$ or L$_{1.9kpc}^{bol}$ = 2.24 $\pm$ 0.87 $\times$ 10$^{31}$ erg/s.

\begin{figure}
\centering
\includegraphics[angle=0,scale=.30]{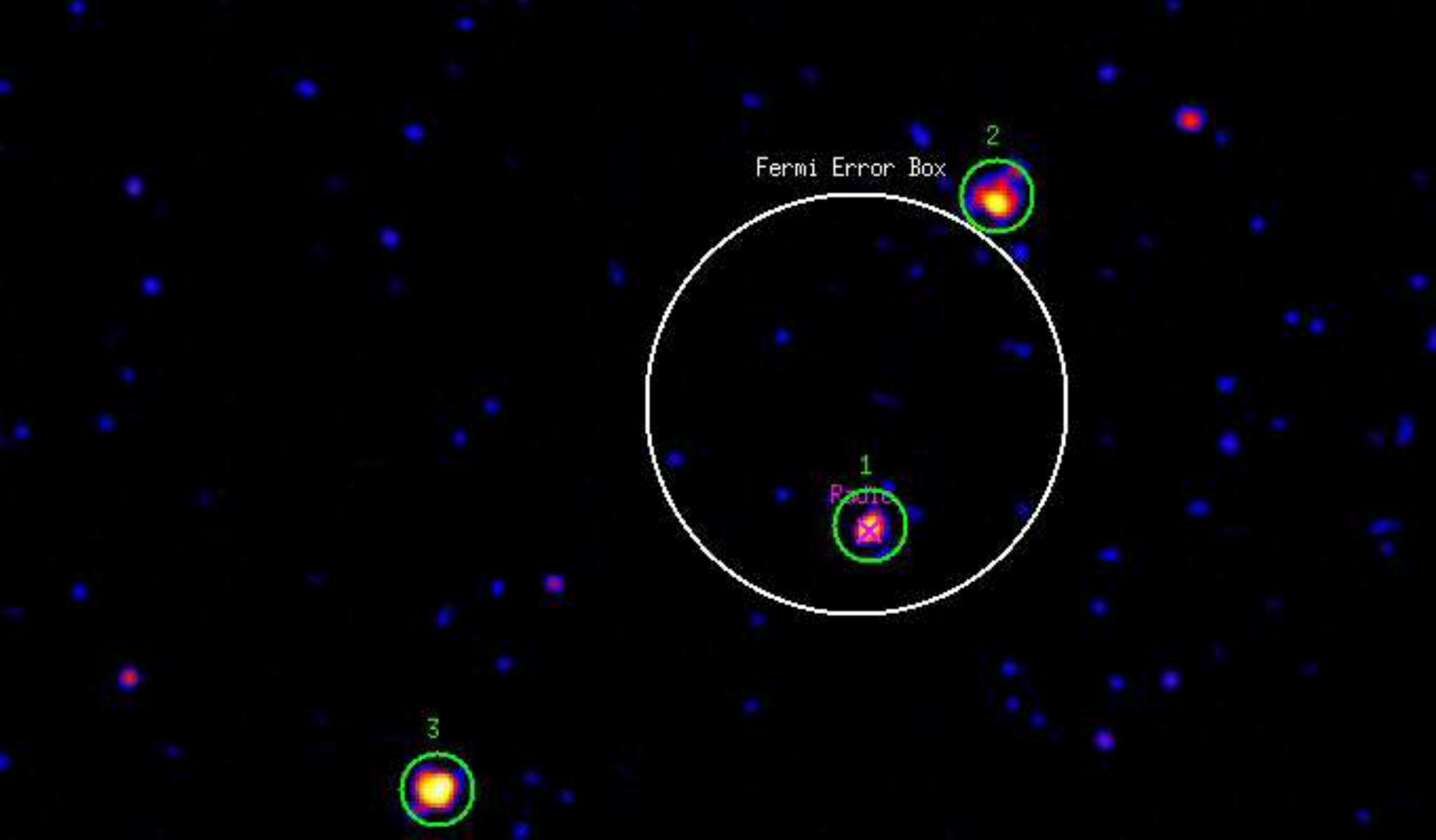}
\caption{PSR J0614-3330 0.3-10 keV XRT Imaging. The XRT images have been added. 
The green circles marks the detected source while the white annulus the Fermi error box from the First Fermi Catalogue (Abdo et al. 2009c).
\label{J0614-im}}
\end{figure}

\begin{figure}
\centering
\includegraphics[angle=0,scale=.40]{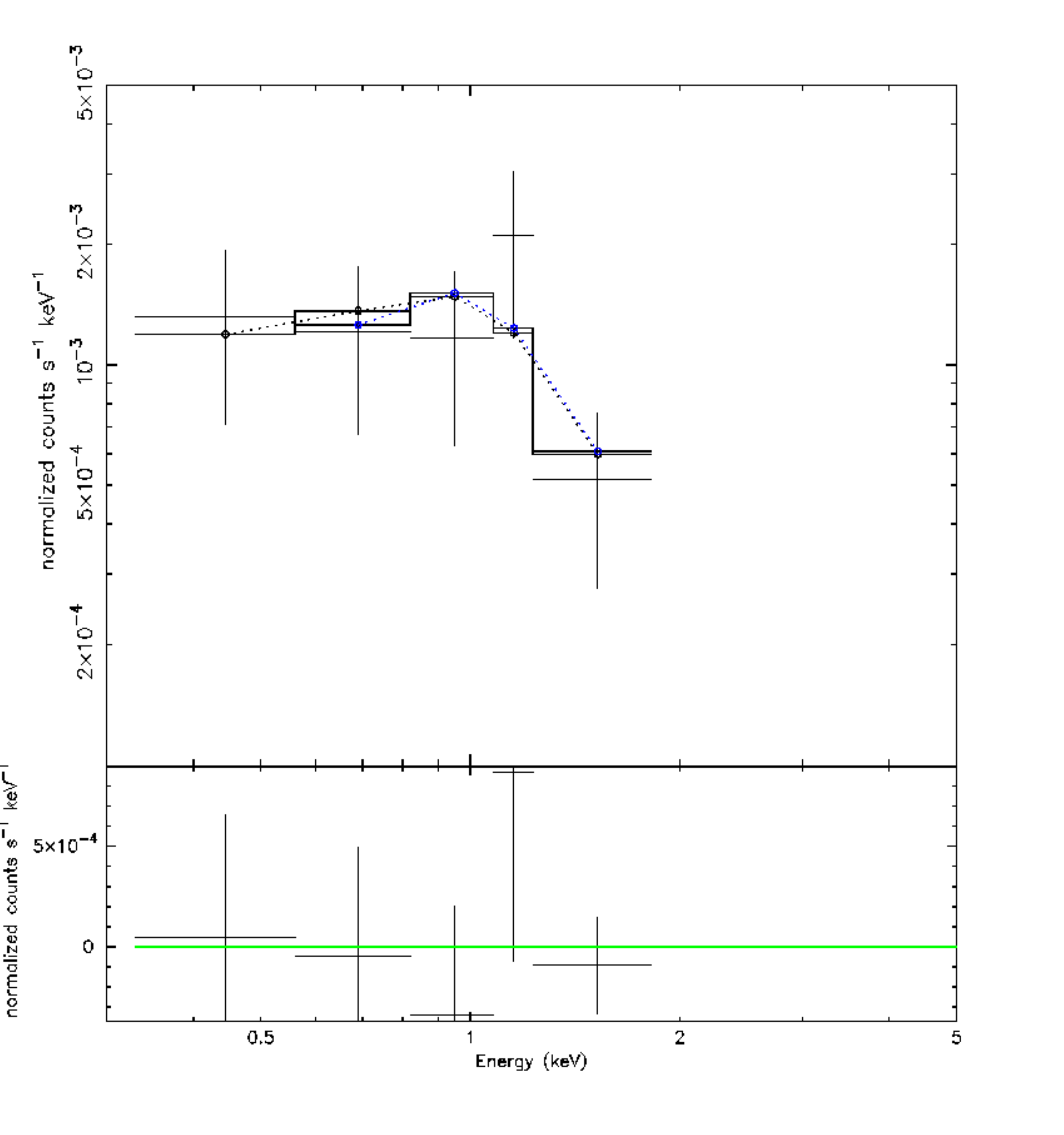}
\caption{PSR J0614-3330 XRT Spectrum (see text for details).
Residuals are shown in the lower panel.
\label{J0614-sp}}
\end{figure}

\clearpage

{\bf J0631+1036 - type 0 RLP}

PSR J0631+1036 is a young ($\tau_c$ = 44 kyr), 288ms pulsar
in the direction of the Galactic anticentre. First reported
by Zepka et al. (1996), it was discovered in a radio search
of the error circle of 2E 0628.7+1037, an Einstein X-ray
source. PSR J0631+1036 has an unusually high dispersion
measure of 125 pc cm$^{-3}$, which corresponds to
a distance of 6.5 kpc (using the model of Taylor \& Cordes
1993). This distance is one of the largest of the Galactic anticentre
pulsars. The association of this source with the dark
cloud LDN 1605 (Lynds 1962), however, suggests that
this high DM is due to ionized gas in or near the cloud.
The actual distance to PSR J0631+1036 is suggested to
be 2.185 $\pm$ 1.440 (Abdo et al. 2009c).
A detection of a 288ms sinusoidal modulation in the ASCA
lightcurve appeared
to confirm the association of the X-ray source and the pulsar (Torii et al. 2001).
Its X-ray spectrum was said to be similar to that of
middle aged $\gamma$-ray pulsars such as Geminga. However, an
{\it XMM-Newton} observation of the PSR J0631+1035 field (obs. id 0103260401,
8 ks exposure after screening of high particle background time intervals),
along with a re-analysis of VLA data (Kennea et al. 2002) confirming the timing
position of the pulsar, shows a 75$"$ displacement between
the X-ray source and the pulsar, and therefore these
cannot be the same object. The X-ray source appears to
be the counterpart of an A0 star, detected by the {\it XMM-Newton}
Optical Monitor. No 288ms period was detected
from the bright X-ray
source. In $\gamma$-rays no nebular emission was detected
down to a flux of 2.07 $\times$ 10$^{-11}$ erg/cm$^2$ s.

For a distance of 2.185 kpc we found a
rough absorption column value of 2 $\times$ 10$^{21}$ cm$^{-2}$
and using a simple powerlaw spectrum
for PSR+PWN with $\Gamma$ = 2 and a signal-to-noise of 3,
we obtained an upper limit non-thermal unabsorbed flux of 2.25 $\times$ 10$^{-14}$ erg/cm$^2$ s,
that translates in an upper limit luminosity of L$_{2.185kpc}^{nt}$ = 1.28 $\times$ 10$^{31}$ erg/s.

{\bf J0633+0632 - type 2 RQP}

J0633+0632 was one of the first pulsars discovered using the
blind search technique (Abdo et al. Science 2009).
Immediately after the detection of pulsations, we made
a request of a {\it SWIFT} observation on behalf of the {\it Fermi} LAT
collaboration. Such an observation revealed one potential
X-ray counterpart inside the 6-months {\it Fermi} error box
at 06:33:43.83 +06:32:30.49 (5$"$ error radius).
A precise timing analysis performed by the collaboration
associated the X-ray source to the $\gamma$-ray pulsar (Ray et al. 2011).
The pseudo-distance of the pulsar based on $\gamma$-ray data (Saz Parkinson et al. (2010))
is $\sim$ 1.1 kpc. No nebular emission was detected
in $\gamma$-rays down to a flux of 3.60 $\times$ 10$^{-11}$ erg/cm$^2$ s.

\begin{figure}
\centering
\includegraphics[angle=0,scale=.40]{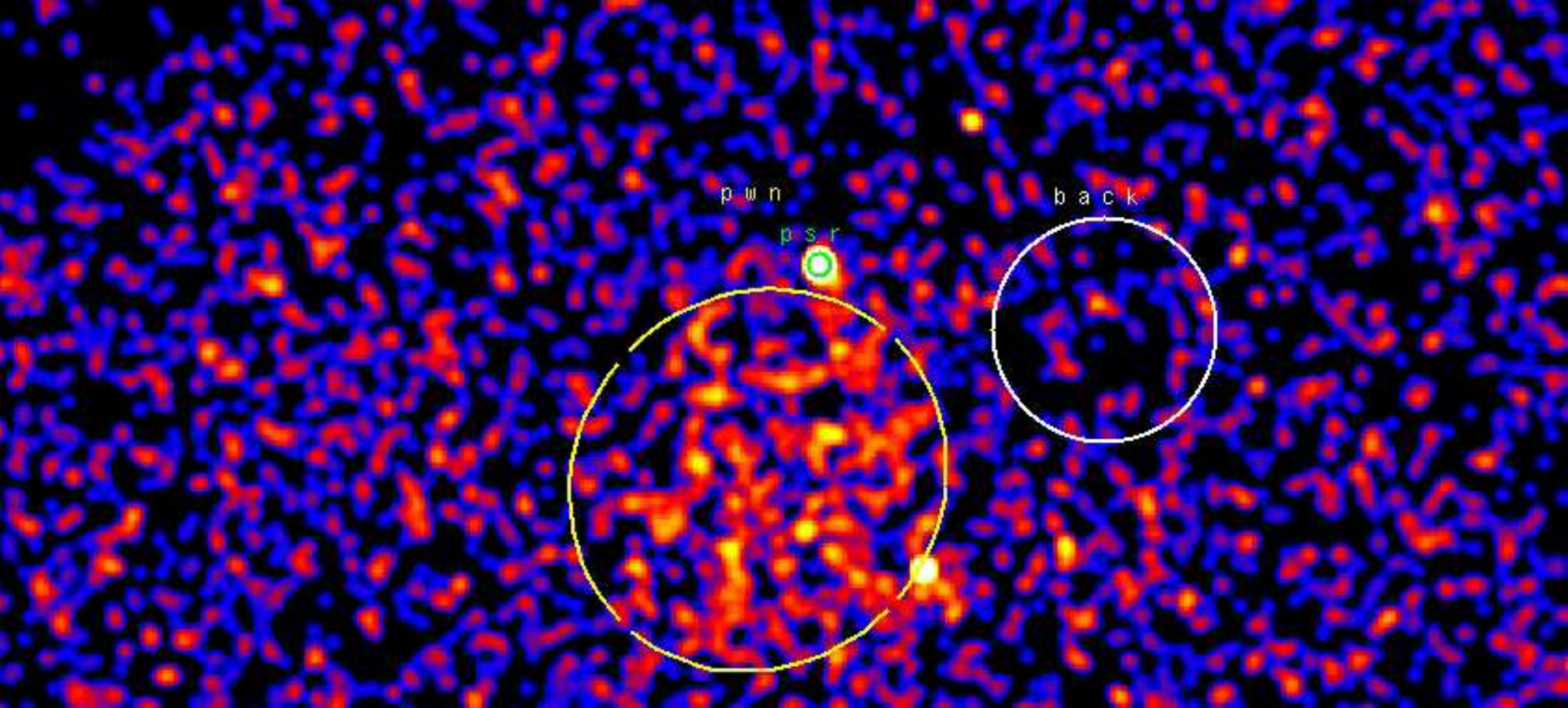}
\caption{PSR J0633+0632 0.3-10 keV {\it Chandra} Imaging. The image has been smoothed with a Gaussian
with Kernel radius of $3"$. The green circle marks the pulsar while the white ellipse the nebular region used in the
analysis. The background region is shown in white.
\label{J0633p0632-im}}
\end{figure}

A {\it Chandra} observation was then requested by the collaboration (obs. id 11123, starting on
2009, December 11 at 19:09:28 UT, 20.2 ks).
By using the celldetect tool inside the CIAO software we found the X-ray counterpart
at 06:33:44.14 +06:32:30.38 (0.95$"$ error radius).
The pulsar spectrum was extracted from a 2$"$ radius circular region around the pulsar
while the background was extracted from a circular region away from the source to avoid
nebular contaminations.
A nebular emission is apparent on the south of the pulsar, with an elliptical shape
of $\sim$ 30$"$ radius (see Figure \ref{J0633p0632-im}).
We obtained 328 pulsar and 527 nebular counts (background contributions
of 0.2\% and 34.3\%).
Due to the low statistic we chose to use the C-statistic only for
the pulsar spectrum.
The best fit model for the pulsar is a combination of a blackbody and a powerlaw
(chisquare value $\chi^2_{red}$ = 1.10, 44 dof)
with a photon index $\Gamma$ = 1.45$_{-0.82}^{+0.76}$ absorbed by a column
N$_H$ = 6.08$_{-6.08}^{+21.91}$ $\times$ 10$^{20}$ cm$^{-2}$.
The thermal component has a temperature of 1.46$_{-0.39}^{+0.28}$ $\times$ 10$^6$ K
and an emitting radius of R$_{1.1kpc}$ = 817$_{-611}^{+8944}$ m.
A simple blackbody model is no statistically allowed while a powerlaw one
is it but the chance probability (10$^{-4}$)
obtained with an f-test seems to disfavor such a model.
The nebula has a photon index $\Gamma$ = 1.19$_{-0.22}^{+0.59}$ absorbed by a column
N$_H$ = 2.11$_{-1.53}^{+2.76}$ $\times$ 10$^{21}$ cm$^{-2}$.
Assuming the best fit model, the 0.3-10 keV unabsorbed thermal pulsar flux is 
1.09 $\pm$ 0.09 $\times$ 10$^{-13}$, the non-thermal pulsar flux is
6.25 $\pm$ 0.50 $\times$ 10$^{-14}$ and the nebular flux is
2.92$_{-0.81}^{+0.79}$ $\times$ 10$^{-13}$ erg/cm$^2$ s. For a distance
of 1.1 kpc, the luminosities are L$_{1.1kpc}^{bol}$ = 1.58 $\pm$ 0.13 $\times$ 10$^{31}$,
L$_{1.1kpc}^{nt}$ = 9.07 $\pm$ 0.73 $\times$ 10$^{30}$ and L$_{1.1kpc}^{pwn}$ = 4.24$_{-1.18}^{+1.15}$ $\times$ 10$^{31}$ erg/s.

\begin{figure}
\centering
\includegraphics[angle=0,scale=.50]{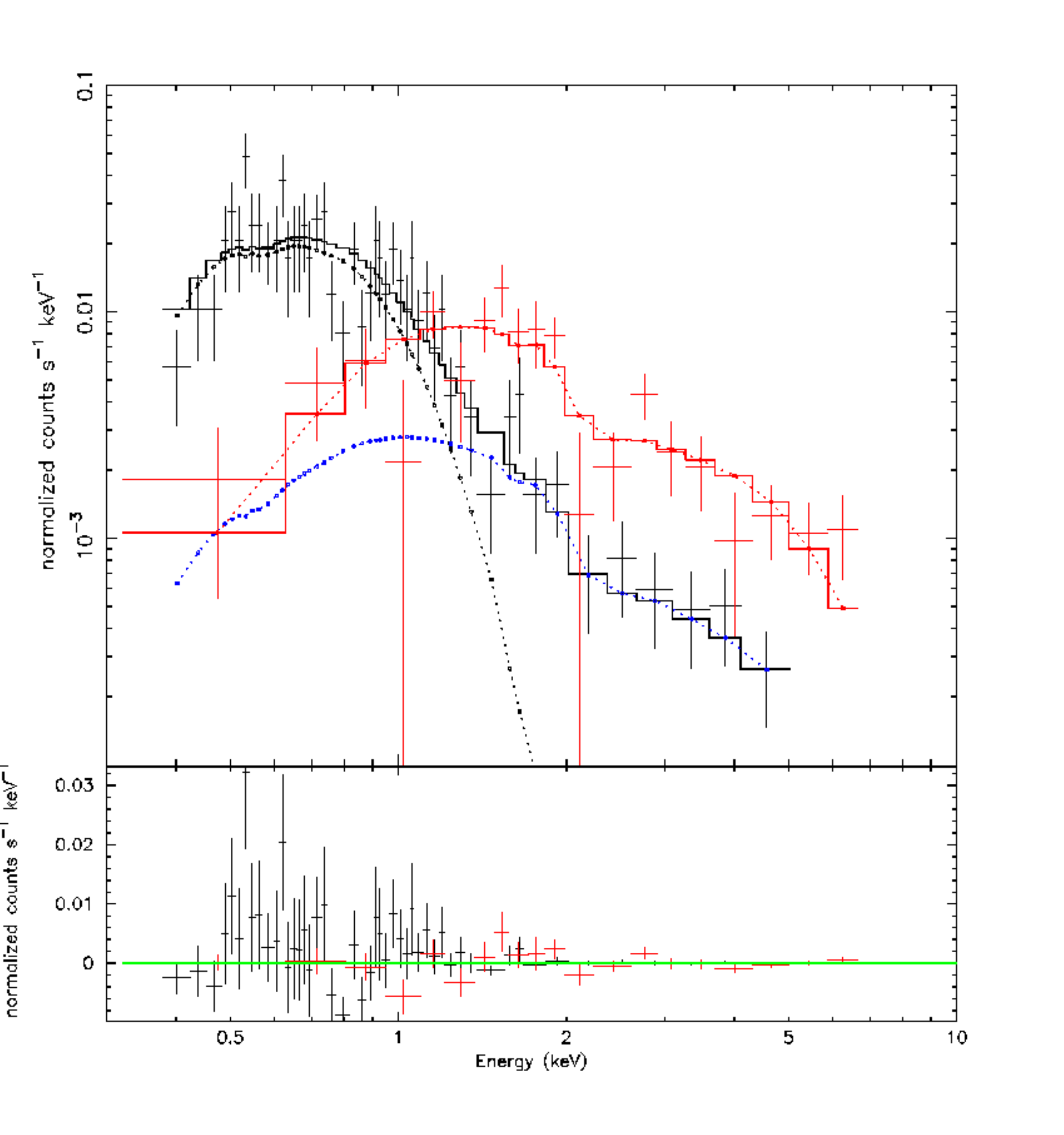}
\caption{PSR J0633+0632 Spectrum. Different colors mark all the different dataset used (see text for details).
Blue points mark the powerlaw component, black points the thermal component of the pulsar spectrum
and red points mark the nebular spectrum.
Residuals are shown in the lower panel.
\label{J0633p0632-sp}}
\end{figure}

\clearpage

{\bf J0633+1746 (Geminga) - type 2* RQP}

Geminga is a nearby, middle-aged isolated neutron star (Bignami \& Caraveo, 1996).
Proximity is a key-parameter for understanding the multiwavelength behaviour
of this source discovered in high-energy gamma-rays and later identified in X-rays
(Bignami et al., 1983) and optical wavelengths (Bignami et al., 1988). The smoking
gun, confirming the previous work based on the interpretation of positional coincidence,
came with the ROSAT discovery of a 237 ms periodicity (Halpern \& Holt, 1992),
immediately seen also at higher energy in contemporary EGRET data (Bertsch et al.
1992) as well as in COS-B archival data (Bignami \& Caraveo, 1992). At the same
time, significant proper motion was discovered at optical wavelengths (Bignami et al.,
1993), definitely linking the proposed counterpart to a fast-moving, pulsar-like object,
the distance to which was later nailed down to 160 pc through its optical parallax
(Caraveo et al., 1996). Thus, Geminga qualifies as a pulsar, and as such is listed in
the radio pulsar catalogue, although it has not been detected at radio wavelengths.

\begin{figure}
\centering
\includegraphics[angle=0,scale=.30]{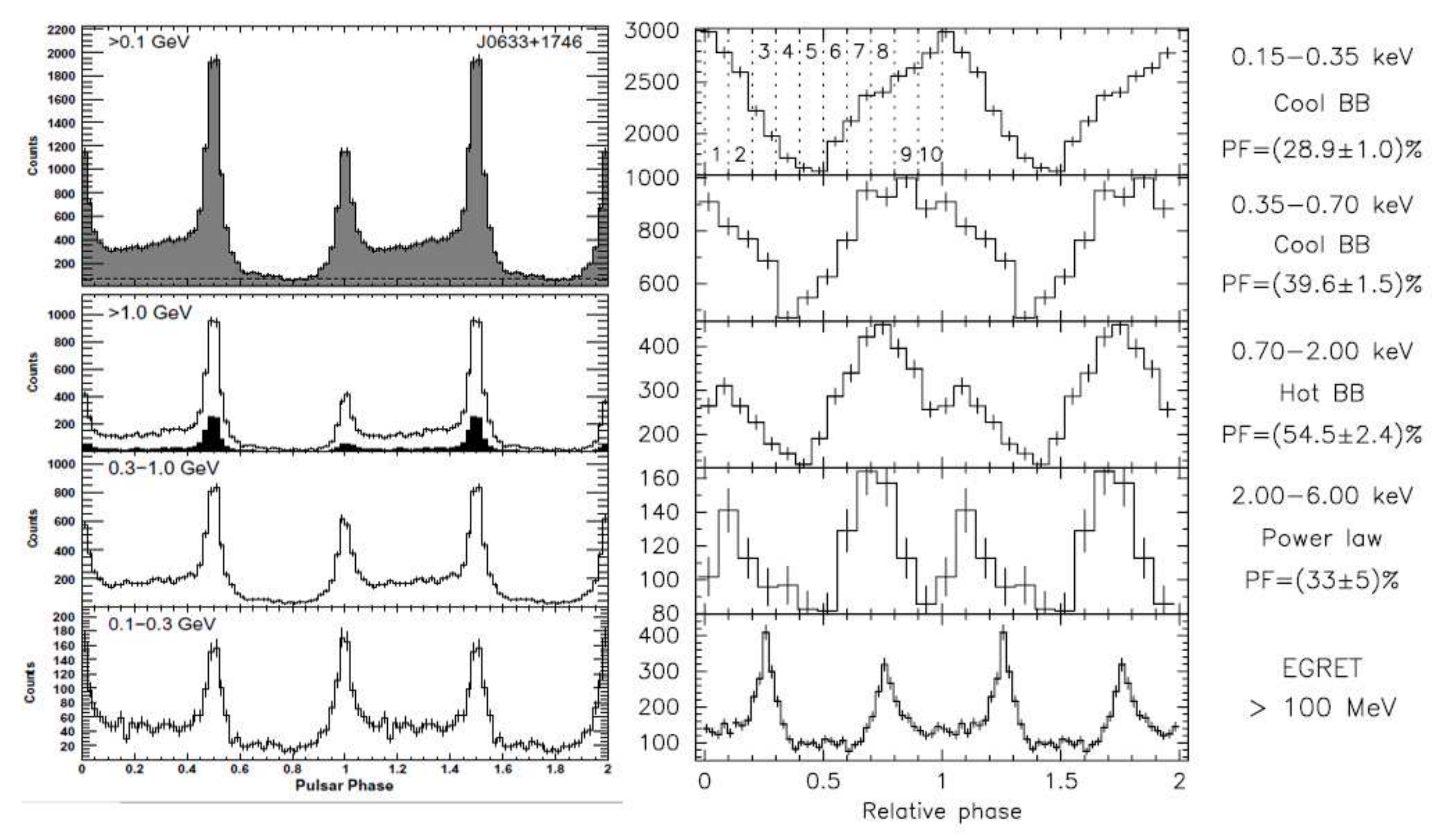}
\caption{PSR J0633+1746 Lightcurve. {\it Left: Fermi} $\gamma$-ray lightcurve folded with Radio
(Abdo et al. catalogue). {\it Right: XMM-Newton} EPIC PN X-ray pulse profiles
for different energy bands, folded with Radio. The choice of phase zero
and the alignment of the X-ray and radio profiles are arbitrary (De Luca et al. 2005).
\label{Geminga-lc}}
\end{figure}

Geminga was observed during a long {\it XMM-Newton} observation
on 2002, April 5. While the MOS cameras were operated
in full-frame mode in order to image on the full field of view of
the telescopes (15' radius), the PN detector
was used to time-tag the photons.
First, an accurate
screening for soft proton flare events was done.
To extract the source photons from the data set taken in imaging
modes (full-frame for MOS and small-window for PN),
a circle of 45$"$ radius has been selected. Background photons
were extracted from suitable regions on the same CCD chip
containing the source.
The best-fitting model is found by combining two
blackbody curves and a power law.
A very faint PWN has been detected, extending  for $\sim$ 2$'$
(see Caraveo et al. 2003); also in $\gamma$-rays
a clear indication of a PWN has been detected,
with a flux of 7.49 $\pm$ 0.22 $\times$ 10$^{-10}$ erg/cm$^2$ s (see PWN catalogue)
and a photon index of 2.24 $\pm$ 0.02.
In 2004, Caraveo et al. studied the phase-resolved spectrum of Geminga,
finding its rotating hot spots.\\
All the X-ray spectral results are taken from De Luca et al. 2005.

\begin{figure}
\centering
\includegraphics[angle=0,scale=.50]{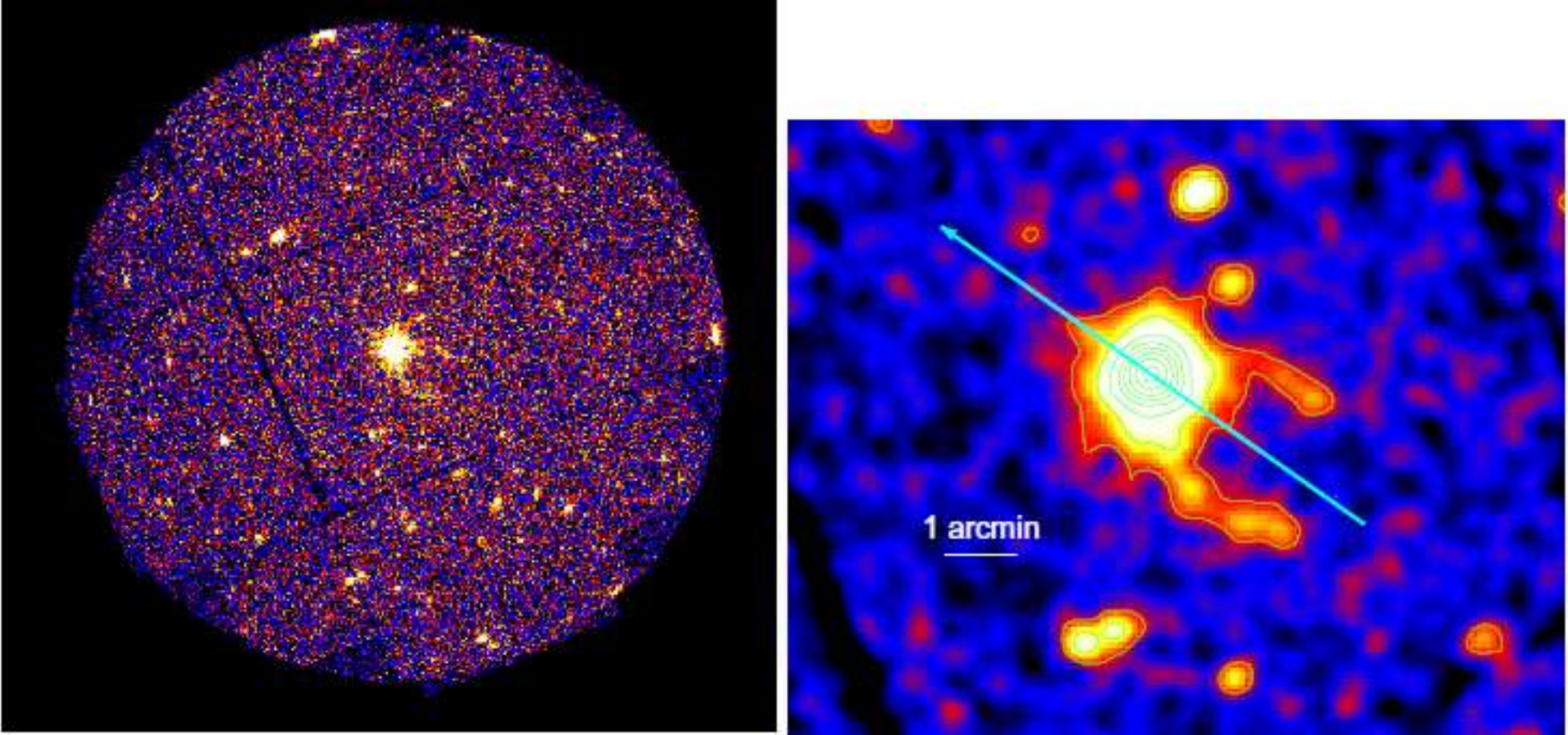}
\caption{PSR J0633+1746 Imaging. {\it Left panel:} The XMM-Newton view of the field of Geminga. Data from the MOS1 and
MOS2 cameras have been merged to produce the image. Events in the 0.3-5 keV range
have been selected. Geminga is the bright
source close to the center of the image; the tails can be seen as two faint diffuse emission
patterns emerging from the source. {\it Right panel:} Inner part of the field, shown after gaussian smoothing. The emission
from Geminga outshines the tails up to $\sim$ 40$"$ from the source. The tails are $\sim$ 2 arcmin
long and cover an area of $\sim$ 2 square arcmin. They show a remarkable symmetry with
respect to the pulsar proper motion direction, marked by the arrow. See Caraveo et al. 2004 for details.
\label{Geminga-im}}
\end{figure}
\begin{figure}
\centering
\includegraphics[angle=0,scale=.50]{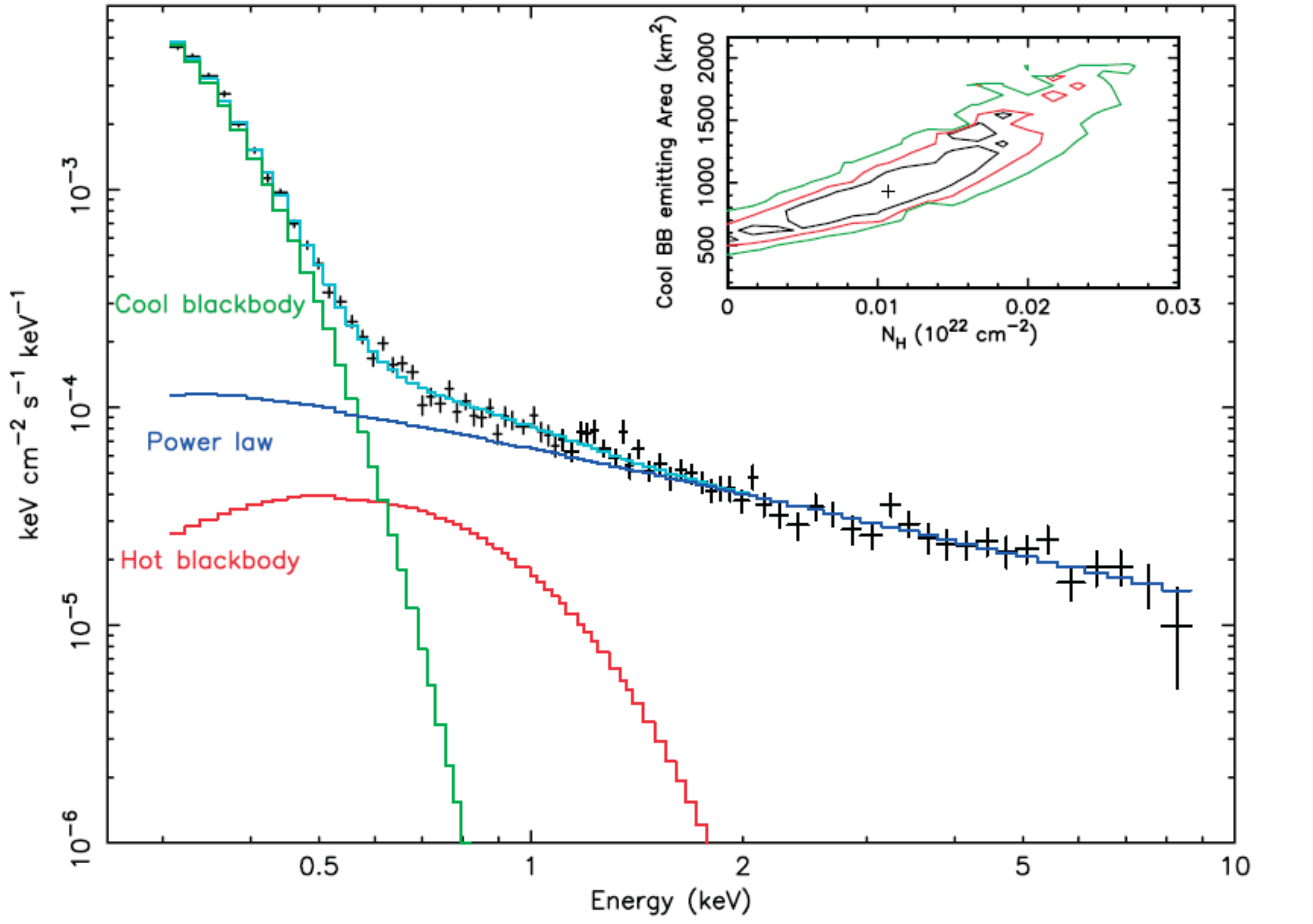}
\caption{PSR J0633+1746 {\it XMM-Newton} Epic PN Spectrum. The best-fitting spectral model is represented by the light blue line. This is obtained by the
sum of a cool blackbody component (green), a hot blackbody component (red), and a power law (blue). The inset shows confidence 
contours for the interstellar column density N$_H$ vs. the emitting surface for the cool blackbody. The 68\%, 90\%, and 99\%
confidence levels for two parameters of interest are plotted. See Caraveo et al. 2004 and De Luca et al. 2005 for details.
\label{Geminga-sp}}
\end{figure}

\clearpage

{\bf J0659+1414 - type 2* RLP}

The definitely bright PSR 0656+14 was first detected by
the Einstein satellite (Cordova et al. 1989). A ROSAT PSPC observation
allowed Finley et al. (1992) to detect the X-ray pulsation
and to measure a pulsed fraction of 14\% $\pm$ 2\%. Further
pointings were then carried out with the ROSAT PSPC detector
in 1992 (Oegelman 1995), for a total exposure time of about
17 ks, collecting $\sim$32000 photons in the 0.1-2.4 keV band. The
overall ROSAT data set was analyzed by Possenti et al. (1996),
who found the bulk of the emission to be of thermal origin , well
described by a blackbody curve (T $\sim$ 9 $\times$ 10$^5$ K) originating
from a large part the star surface (emitting radius $\sim$ 14 km assuming
a distance of 760 pc, corresponding to $\sim$ 5.3 km at the
parallactic distance of 288 pc). A second spectral component
was required to describe the higher energy part of the spectrum,
as well as to explain a significant change of the pulse
profile with energy. It was not possible, however, to discriminate
between a second blackbody component (T $\sim$ 1.9 $\times$ 10$^6$ K)
originating from a region a few hundred times smaller and a
steep ($\Gamma$ $\sim$ 4.5) power law.

\begin{figure}
\centering
\includegraphics[angle=0,scale=.30]{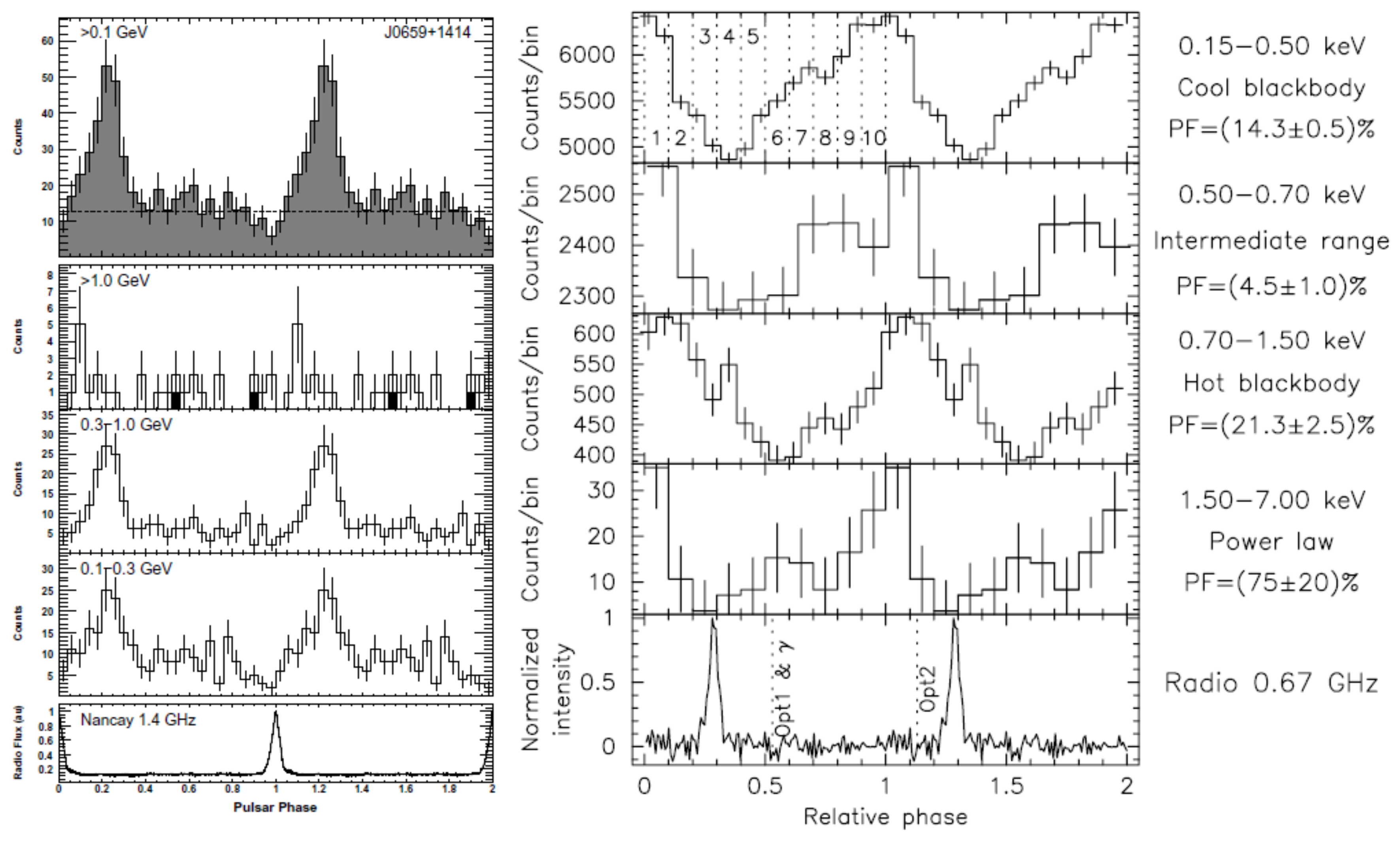}
\caption{PSR J0659+1414 Lightcurve. {\it Left: Fermi} $\gamma$-ray lightcurve folded with Radio
(Abdo et al. catalogue). {\it Right: XMM-Newton} EPIC PN X-ray pulse profiles
for different energy bands, folded with Radio. The choice of phase zero
and the alignment of the X-ray and radio profiles are arbitrary (De Luca et al. 2005).
\label{J0659-lc}}
\end{figure}

J0659+1414 was observed during a long {\it XMM-Newton} observation (obs. id 0112200101)
on 2001, October 23. While the MOS cameras  were operated
in full-frame mode in order to image the full field of view of
the telescopes, the PN detector was used to time-tag the photons in the Small Window mode.
First, an accurate
screening for soft proton flare events was done.
To extract the source photons from the data set taken in imaging
modes (full-frame for MOS and small-window for PN),
a circle of 45$"$ radius has been selected. Background photons
were extracted from suitable regions on the same CCD chip
containing the source.
The best-fitting model is found by combining two
blackbody curves and a power law.
No nebular emission was detected in the {\it XMM-Newton} observation
nor in the {\it Fermi} data (down to a flux of 5.58 $\times$ 10$^{-12}$ erg/cm$^2$ s).
All the X-ray spectral results are taken from De Luca et al. 2005.

\begin{figure}
\centering
\includegraphics[angle=0,scale=.50]{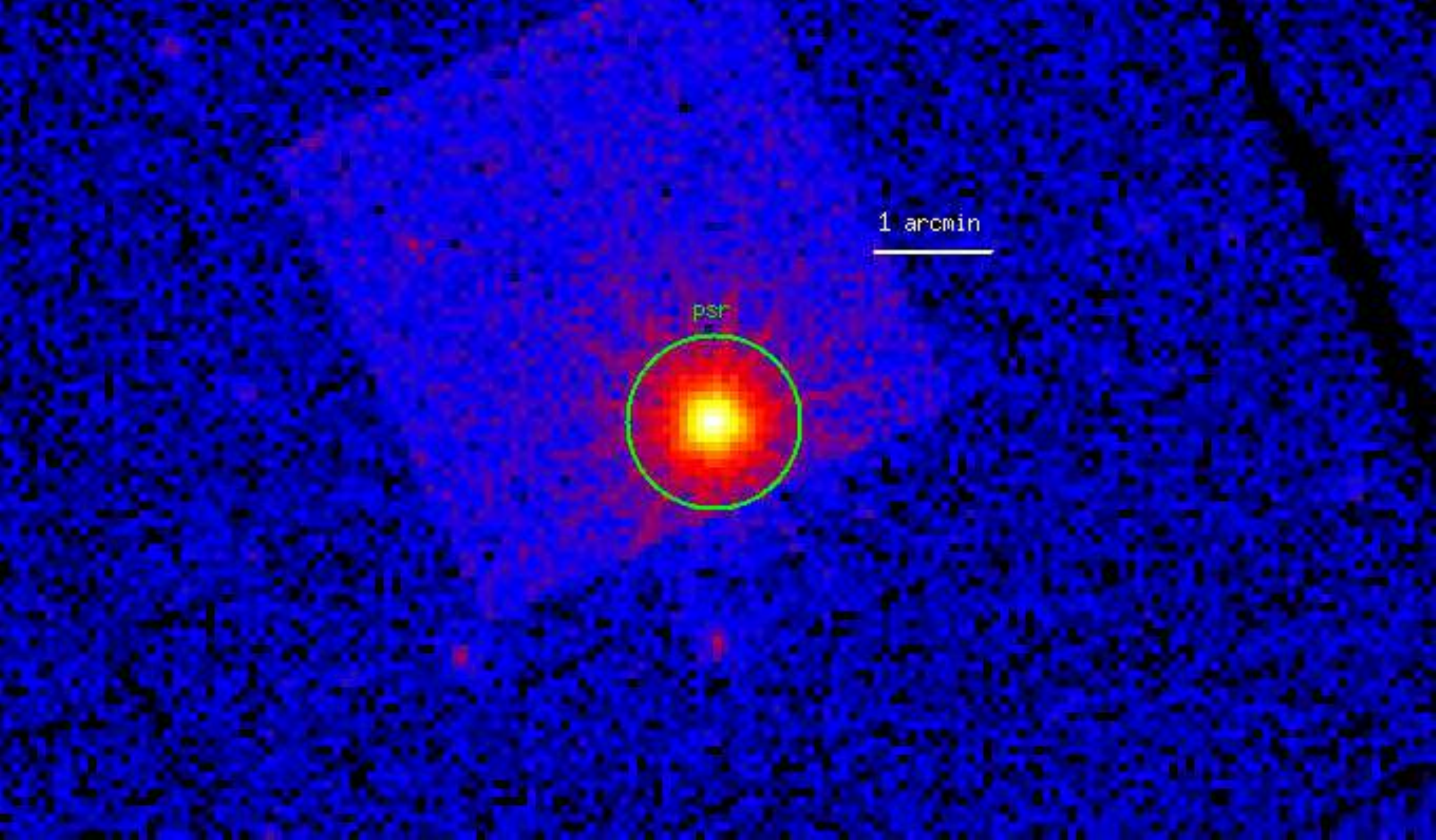}
\caption{PSR J0659+1414 {\it XMM-Newton} Imaging. The PN and the two MOS images have been added. 
The green circle marks the pulsar region used in the analysis.
\label{J0659-im}}
\end{figure}

\begin{figure}
\centering
\includegraphics[angle=0,scale=.50]{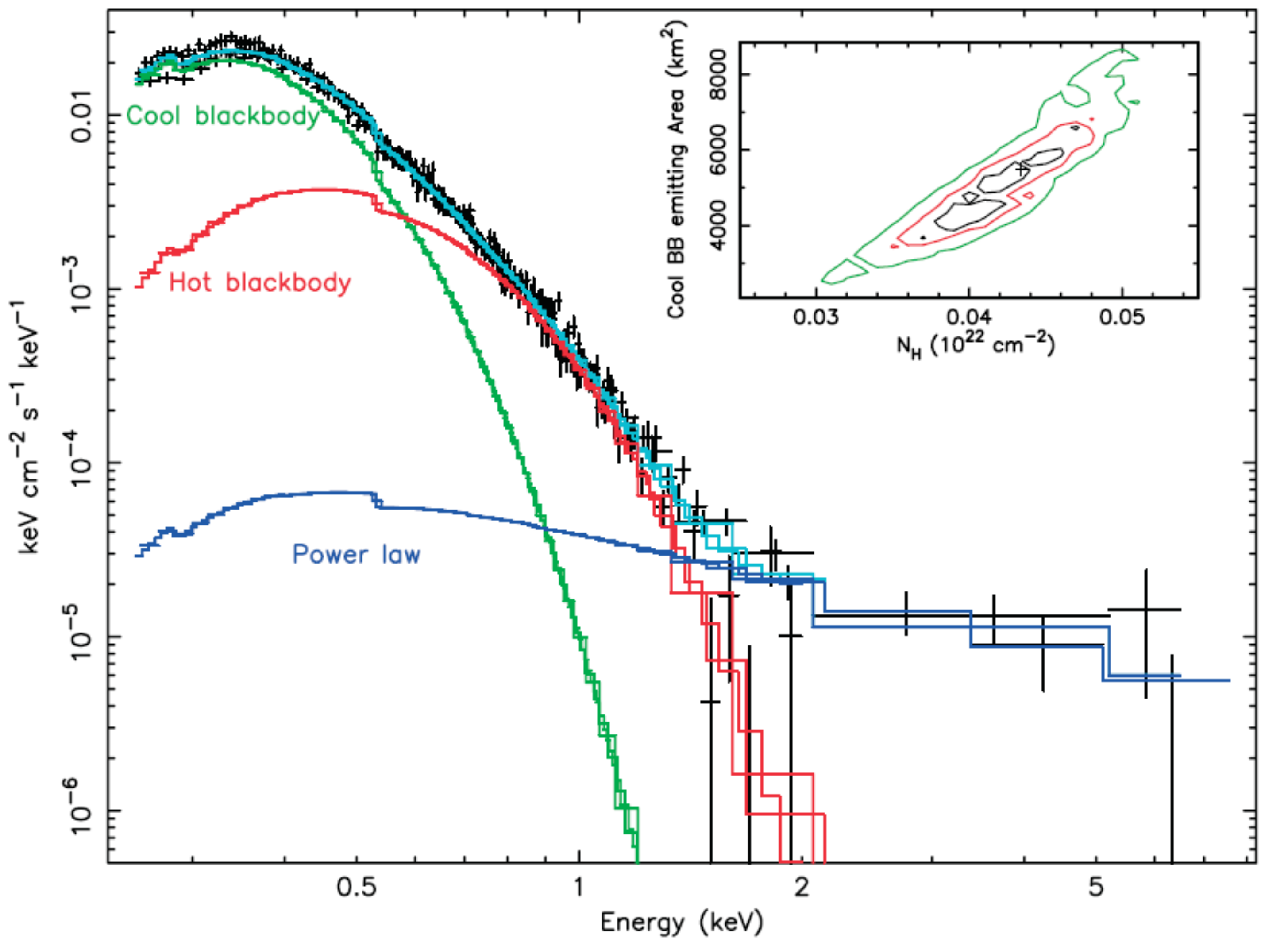}
\caption{PSR J0659+1414 {\it XMM-Newton} Epic PN Spectrum. The best-fitting spectral model is represented by the light blue line. This is obtained by the
sum of a cool blackbody component (green), a hot blackbody component (red), and a power law (blue). The inset shows confidence 
contours for the interstellar column density N$_H$ vs. the emitting surface for the cool blackbody. The 68\%, 90\%, and 99\%
confidence levels for two parameters of interest are plotted. See De Luca et al. 2005 for details.
Residuals are shown in the lower panel.
\label{J0659-sp}}
\end{figure}

\clearpage

{\bf J0734-1559 - type 0 RQP} % Nuova!

Pulsations from J0734-1559 were detected in the last months  using the
blind search technique: it was announced as a pulsar
during the 3rd {\it Fermi} Symposium in Rome.
The pseudo-distance of the object based on $\gamma$-ray data (Saz Parkinson et al. (2010)) is $\sim$ 1.3 kpc.

After the {\it Fermi} detection, we asked for a {\it SWIFT} observation
of the $\gamma$-ray error box (obs id. 00041333001,00041333002, 4.87 ks exposure).
An X-ray source was found at
the edge of the {\it Fermi} error box; however, dedicated timing
analysis performed by the LAT team excluded it
as the counterpart of the $\gamma$-ray pulsar.
For a distance of 1.3 kpc we found a
rough absorption column value of 2 $\times$ 10$^{21}$ cm$^{-2}$
and using a simple powerlaw spectrum
for PSR+PWN with $\Gamma$ = 2 and a signal-to-noise of 3,
we obtained an upper limit non-thermal unabsorbed flux of 2.36 $\times$ 10$^{-13}$ erg/cm$^2$ s
that translates in an upper limit luminosity L$_{1.3kpc}^{nt}$ = 4.78 $\times$ 10$^{31}$ erg/s.

{\bf J0742-2822 - type 0 RLP}

% Weltevrede et al. 2009
PSR J0742-2822 was discovered as a radio
pulsar by Facondi et al. (1973). Koribalski et al.
(1995) determined a kinematic distance between
2.0 $\pm$ 0.6 and 6.9 $\pm$ 0.8 kpc, which is higher than
the distance derived from the DM according to the
Taylor \& Cordes 1993 model (2.1 kpc).\\
Two {\it XMM-Newton} observations were performed
in order to find the X-ray counterpart of such a radio
pulsar:\\
- obs. id 0103260501, start time 2001, April 27 at 21:49:53 UT, exposure 6.0 ks;\\
- obs. id 0103262401, start time 2002, October 11 at 09:42:48.19 UT, exposure 5.0 ks.\\
In both the observations the PN and MOS cameras were
operating in the Full Frame mode.
For all three instruments, the medium optical filter was used.
Both the observations are affected by an high particle
background during the entire exposure, so that no screening was possible.

A source detection was performed in two different energy
bands (0.3-10 keV and 0.3-2 keV) using both the SAS tools
and XIMAGE and no X-ray source was found at the Radio position.\\
For a distance of 2.1 kpc we found a
rough absorption column value of 2 $\times$ 10$^{21}$ cm$^{-2}$ and
using a simple powerlaw spectrum
for PSR+PWN with $\Gamma$ = 2 and a signal-to-noise of 3,
we obtained an upper limit non-thermal unabsorbed flux of 2.25 $\times$ 10$^{-14}$ erg/cm$^2$ s,
that translates in an upper limit luminosity of L$_{2.1kpc}^{nt}$ = 1.19 $\times$ 10$^{31}$ erg/s.

{\bf J0751+1807 - type 2 RL MSP} % cambiato ed aggiunto il bbody

% Webb et al. 2004
PSR J0751+1807 was detected during the first year of {\it Fermi} operations (Abdo et al. 2009c).
It was originally detected in an EGRET source error
box, in September 1993 (Lundgren et al. 1993), using the
radio telescope at Arecibo. Lundgren et al. (1995) combined
the mass function, eccentricity, orbital size and age of the pulsar,
determined from radio data, to predict the expected type
of companion star to the millisecond pulsar. They proposed
the secondary star be a helium white dwarf, with a mass
between 0.12-0.6 M$_s$, in a 6.3 h orbit with the pulsar. However, the
3.49 ms pulse period together with its derivatives cannot provide
the energy to power the $\gamma$-ray source.
PSR J0751+1807 was subsequently detected in the soft X-ray
domain by Becker et al. (1996), using the ROSAT PSPC.
However, there were too few counts to build a spectrum
or detect pulsations. Using the HI survey of Stark et al. (1992),
they deduced an interstellar absorption of 4 $\times$ 10$^{20}$ cm$^{-2}$.
A parallax measurement gave a pulsar distance
of 0.6$_{-0.2}^{+0.6}$ kpc (Nice et al. 2005).\\
In $\gamma$-rays no nebular emission was detected
down to a flux of 1.05 $\times$ 10$^{-11}$ erg/cm$^2$ s.

We used the only {\it XMM-Newton} observation centered on the pulsar (obs. id 0111100301), started on
2000, October 01 at 04:27:17 UT and lasted 36.8 ks.
The PN camera of the EPIC
instrument was operated in Fast Timing mode, while the MOS detectors were set in Full frame mode. For
all three instruments, the thin optical filter was used.
The PN camera dataset wasn't used in our spectral analysis
due to the lack of spatial resolution.
After standard data processing (using the epproc and emproc tasks), no
screening of high particle background time intervals was necessary due to the goodness of the observation.\\
The source X-ray best fit position is 07:51:09.144 +18:07:37.01 (5$"$ error radius)
obtained by using both the XIMAGE and SAS dedicated tools.
we extracted the source spectrum from a circle of 20$"$ radius centered on the source, in order to maximize the signal-to-noise ratio.
we obtained 159 and 186 counts (background contributions of 12.1\% and 11.2\%) inside the source region in the 0.3-10
keV range for the two MOS cameras.
The source count distribution
around the radio position of the pulsar is fully consistent
with that of a point source. Thus, no indication of
any bright diffuse extended emission was found.
Due to the faintness of the source, we decided to use
the C-statistic for the fit of the data.
The pulsar emission is well described ($\chi^2$ value of 71.64
with 63 dof) by a
combination of a power law model and a blackbody.
The powerlaw component has a photon index $\Gamma$ = 1.31$_{-0.60}^{+0.52}$ , 
absorbed by a column N$_H$ = 8.74 $_{-8.74}^{+2.10}$ $\times$ 10$^{20}$ cm$^{-2}$.
The thermal component has a temperature of 1.69$_{-0.54}^{+68}$ $\times$ 10$^6$ K
and an emitting radius of R$_{600pc}$ = 167$_{-145}^{+5689}$ m.
A simple blackbody spectrum yields a very poor fit ($\chi^2_{red}$ $\sim$ 5)
while a simple powerlaw spectrum gives a statistically acceptable
fit ($\chi^2_{red}$ = 1.32) but the chance probability (0.0035)
obtained with an f-test seems to disfavor such a model.
Assuming the best fit model, the 0.3-10 keV unabsorbed thermal flux is 
3.09 $\pm$ 0.48 $\times$ 10$^{-14}$ and the non-thermal flux is
5.92 $\pm$ 0.94 $\times$ 10$^{-14}$ erg/cm$^2$ s. For a distance
of 600 pc, the two luminosities are L$_{600pc}^{bol}$ = 1.33 $\pm$ 0.21 $\times$ 10$^{30}$,
L$_{600pc}^{nt}$ = 2.56 $\pm$ 0.41 $\times$ 10$^{30}$ erg/s.

\begin{figure}
\centering
\includegraphics[angle=0,scale=.50]{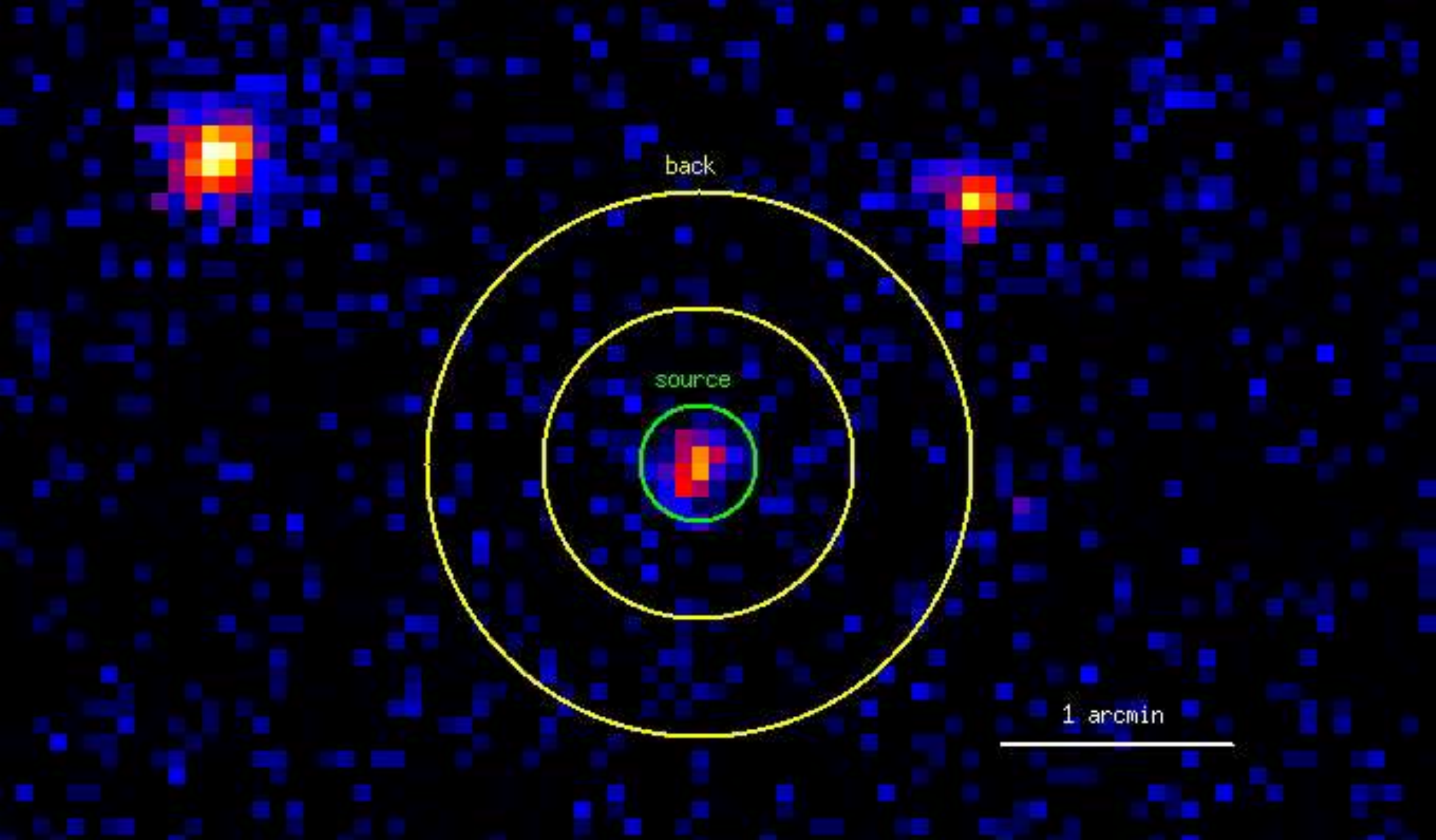}
\caption{PSR J0751+1807 0.3-10 keV MOS Imaging. The two MOS images have been added. 
The green circle marks the pulsar while the yellow annulus the background region used in the analysis.
\label{J0751-im}}
\end{figure}

\begin{figure}
\centering
\includegraphics[angle=0,scale=.50]{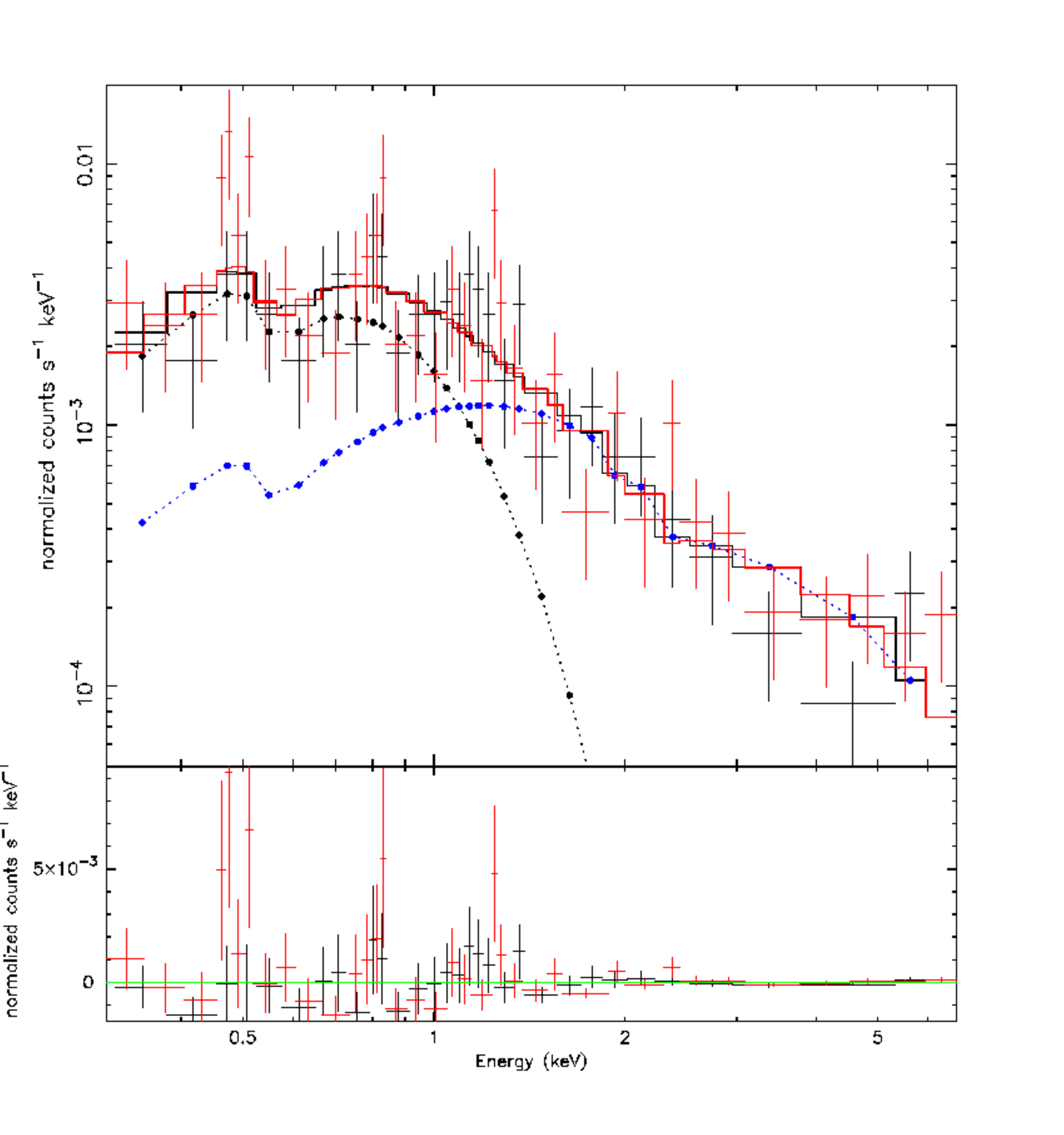}
\caption{PSR J0751+1807 Spectrum. Different colors mark all the different dataset used (see text for details).
Blue points mark the powerlaw component while black points the thermal component of the pulsar spectrum.
Residuals are shown in the lower panel.
\label{J0751-sp}}
\end{figure}

\clearpage

{\bf J0835-4510 (Vela) - type 2* RLP}

% Manzali 2007
The Vela pulsar is one of the best scrutinized neutron stars. It
was the first pulsating radio source observed in the southern
hemisphere (Large et al. 1968), and the swing of the polarization
vector during the radio pulse provided evidence for a rotational
origin of the radio emission (Radhakrishnan et al. 1969). Moreover,
the position of the pulsar, close to the center of the Vela
SNR, confirmed the association of pulsars with rotating neutron
stars born from massive stars' collapse.
About 10 years after the radio discovery, Wallace et al. (1977)
detected a pulsating optical source of V $\sim$ 23.6: at least four
peaks are present in the optical (Gouiffes 1998) and UV (Romani
et al. 2005) light curves, as well as in the hard X-ray energy range
(Harding et al. 2002). HST observations (Caraveo et al. 2001)
of the optical counterpart have allowed the direct measurement
of the pulsar distance (294$^{+76}_{-50}$ pc), later confirmed and refined
through radio observations (287$^{+19}_{-17}$ pc; Dodson et al. 2003).
PSR B0833-45 is also a bright $\gamma$-ray source: pulsations were
detected by SAS 2 (Thompson et al. 1975), COS B (Kanbach et al. 1980; Grenier et al. 1988), the Compton
Gamma Ray Observatory (CGRO) and EGRET (Kanbach et al. 1994)
before the {\it Fermi} detection. {\it Fermi} also detected its nebula
with a flux of 2.10 $\pm$ 0.14 $\times$ 10$^{-10}$ erg/cm$^2$s (Ackermann et al. 2010, Abdo et al. 2011b).

Rontgensatellit (ROSAT) detection of soft X-ray pulsations
(Oegelman et al. 1993) was the last piece of the Vela multiwavelength
puzzle. It turned out to be a difficult observation
since X-rays from the neutron star are embedded in the bright
pulsar wind nebula (PWN), located near the center of the Vela
SNR. A blackbody with temperature $\sim$ (1.5-1.6) $\times$ $10^6$ K and a
radius of 3-4 km could describe the ROSAT spectrum, while the
nebular emission, clearly nonthermal, could be ascribed to synchrotron
emission originating from the interaction of the high energy
particle pulsar wind with the interstellar medium. {\it Chandra}
observations of the Vela pulsar provided high-resolution images
of the X-ray nebula surrounding the neutron star. The nebula turns
out to be quite complex, with spectacular arc and jet-like features
(Helfand et al. 2001; Pavlov et al. 2003) reminiscent of those
observed around the Crab. The thermal nature of the pulsar
spectrum inferred from ROSAT data was confirmed. While high resolution
spectra failed to show absorption features, a nonthermal
harder tail was found in ACIS-S spectra (Pavlov et al. 2001b).
Analyzing {\it XMM-Newton} observations, Mori et al. (2004) found
the same thermal radiation. However, in their preliminary analysis,
they studied only the pulsar phase-averaged thermal emission
below $\sim$ 1 keV. Since thermal emission is a very distinctive
character of middle-aged pulsars (Geminga, PSR B0656+14,
and PSR B1055-52; Becker \& Truemper 1997), Vela's ($\tau$ $\sim$
11400 yr; $\dot{E}$ $\sim$ 7 $\times$ 10$^{36}$ ergs s$^{-1}$) spectral properties make it
more similar to older specimen than to younger ones.

\begin{figure}
\centering
\includegraphics[angle=0,scale=.30]{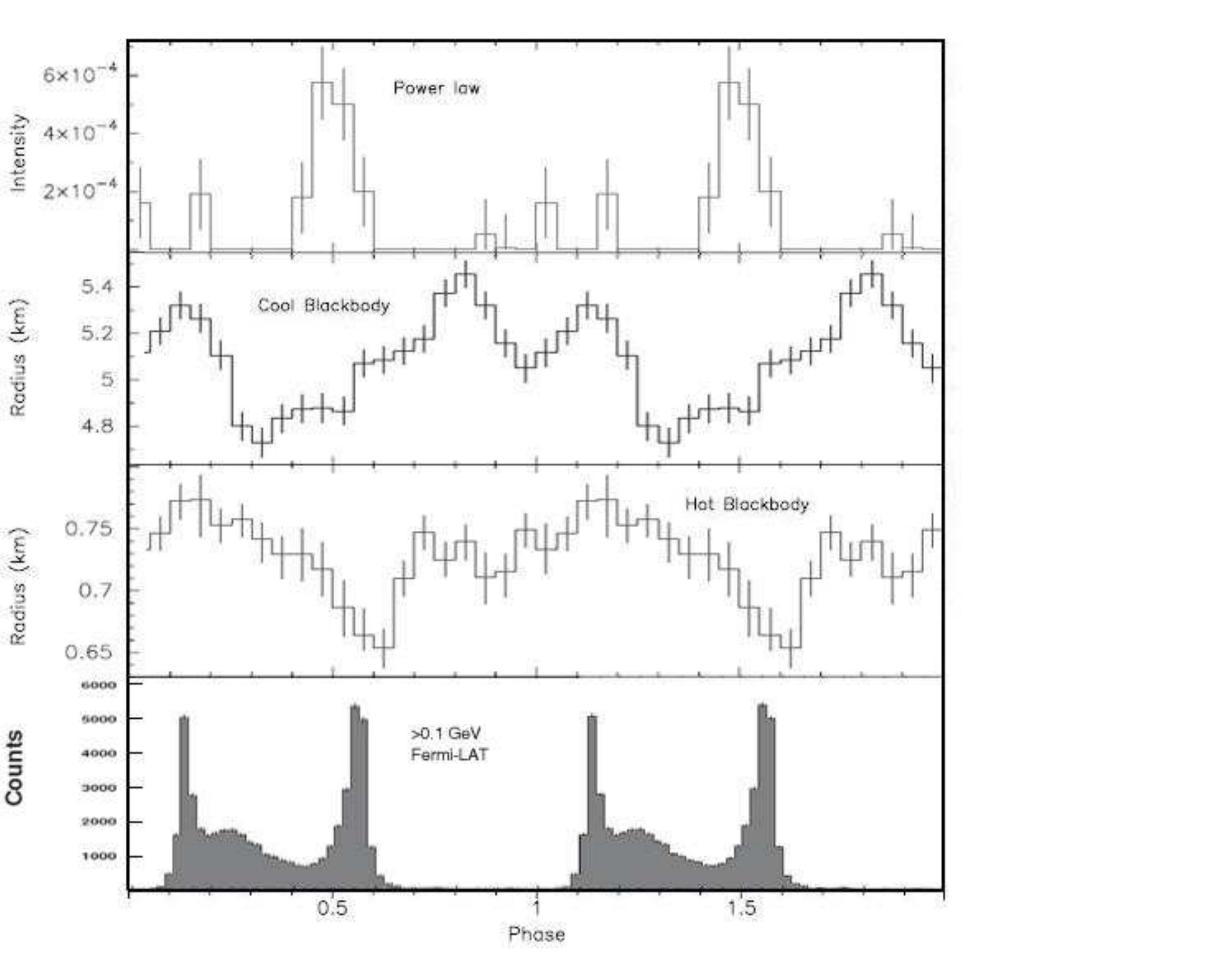}
\caption{PSR J0835-4510 Lightcurve. In the upper panels the
variation of the three spectral components (cool and hot blackbody radius and power-law normalization) as
a function of the rotational phase is shown. The last panel shows the in-phase {\it Fermi} $\gamma$-ray spectrum.
Taken from Manzali et al. 2006, Abdo et al. 2009c.
\label{vela-lc}}
\end{figure}

A complicate method was used in order to avoid the pileup of
the X-ray instruments (both {\it Chandra} and {\it XMM-Newton}) 
and to disentangle the pulsar from its nebular emission.
See Manzali et al. 2007 and Mori et al. 2004.

\begin{figure}
\centering
\includegraphics[angle=0,scale=1.2]{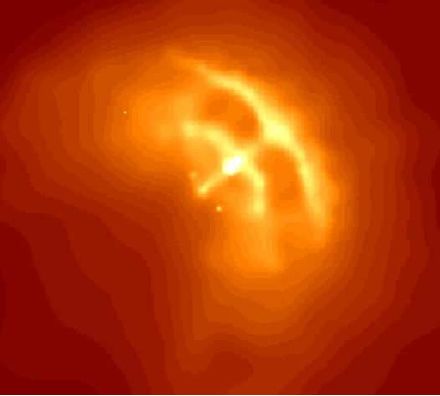}
\caption{PSR J0835-4510 Chandra Imaging. The image has been smoothed with a Gaussian.
\label{vela-im}}
\end{figure}

\clearpage

{\bf J0908-4913 - type 0 RLP} % Nuova! osservazione in arrivo

J0908-4913 was one of the last millisecond $\gamma$-ray pulsars
found by {\it Fermi} using radio ephemerides.
The Radio dispersion measurements set the pulsar distance
at $\sim$ 6.57$^{+1.30}_{-0.90}$ kpc (Taylor \& Cordes 1993).

An {\it XMM-Newton} observation was performed on the position
of the Radio pulsar (obs. id 0103260701, 1.8 ks exposure,
start time on 2001, May 05 at 17:18:21 UT).
Both the PN and MOS cameras were operating in Full Frame mode, using a
medium optical filter.
No screening of high particle background time intervals was
done due to the shortness of the observation.
No X-ray source was found at the Radio position using both the XIMAGE and
SAS tools.
For a distance of a 6.6 kpc, we found a
rough absorption column value of 8 $\times$ 10$^{21}$ cm$^{-2}$
and using a simple powerlaw spectrum
for PSR+PWN with $\Gamma$ = 2 and a signal-to-noise of 3,
we obtained an upper limit 
non-thermal unabsorbed flux of 3.93 $\times$ 10$^{-14}$ erg/cm$^2$ s,
that translates in an upper limit luminosity L$_{6.57kpc}^{nt}$ = 2.04 $\times$ 10$^{32}$ erg/s.

{\bf J0940-5428 - type 0 RLP} % Nuova!

% Crawford & Tiffany 2007
PSRs J0940-5428 is among the
first pulsars discovered in the Parkes Multibeam survey. It has a fast spin
period (88 ms), is
young ($\tau_c$ = 42 ky) and has a large spin-down luminosity
($\dot{E}$ = 1.9 $\times$ 10$^{36}$ erg/s) (Manchester et al. 2001). These
characteristics place it in the category of Vela-like
pulsars, which are generally defined as fast-spinning pulsars
having characteristic ages 10 - 100 kyr and
spin-down luminosities $\dot{E}$ $\sim$ 10$^{36}$ erg/s (e.g., Kramer
et al. 2003).
Radio dispersion measurements found the distance to be $\sim$ 2.95 kpc.

After the {\it Fermi} source detection, a {\it Chandra} observation
of the $\gamma$-ray error box was asked
(obs id. 9077, 10.0 ks exposure).
After the data reduction, no X-ray source were found at
the radio position.
For a distance of 2.95 kpc we found a
rough absorption column value of 5 $\times$ 10$^{21}$ cm$^{-2}$
and using a simple powerlaw spectrum
for PSR+PWN with $\Gamma$ = 2 and a signal-to-noise of 3,
we obtained an upper limit non-thermal unabsorbed flux of 1.29 $\times$ 10$^{-14}$ erg/cm$^2$ s,
that translates in an upper limit luminosity
L$_{2.95kpc}^{nt}$ = 1.35 $\times$ 10$^{31}$ erg/s.

{\bf J1016-5857 - type 2 RLP} % Nuova! osservazione in arrivo

% Camilo et al. 2004
PSR J1016-5857 belongs to the $"$Vela-like$"$ pulsar
class characterized by spin periods P $\sim$ 0.1 s, characteristic
ages 10$^4$ yr $<$ $\tau_c$ $<$ 10$^5$ yr, and spindown
luminosities $\dot{E}$ $>$ 10$^{36}$ erg/s. It was discovered in the Parkes multibeam
pulsar survey of the Galactic plane (e.g. Manchester
et al. 2001) within the error box of an unidentified EGRET source,
3EG J1013-5915 (Camilo et al. 2001). PSR J1016-5857 may
be particularly interesting because it appeared to be coincident
with an Einstein Observatory X-ray source and is positionally
coincident with the very tip of a $"$finger$"$ of radio
emission apparently originating from the SNR G284.3-1.8 (with
an estimated age of $\sim$ 10 ky (Ruinz \& May 1986)).
The pulsar distance obtained from Radio dispersion measurements is $\sim$ 9$_{-2}^{+3}$ kpc (Taylor \& Cordes 1993). % Secondo Camilo et al 2004 la distanza dall'X sarebbe 3kpc

\begin{figure}
\centering
\includegraphics[angle=0,scale=.40]{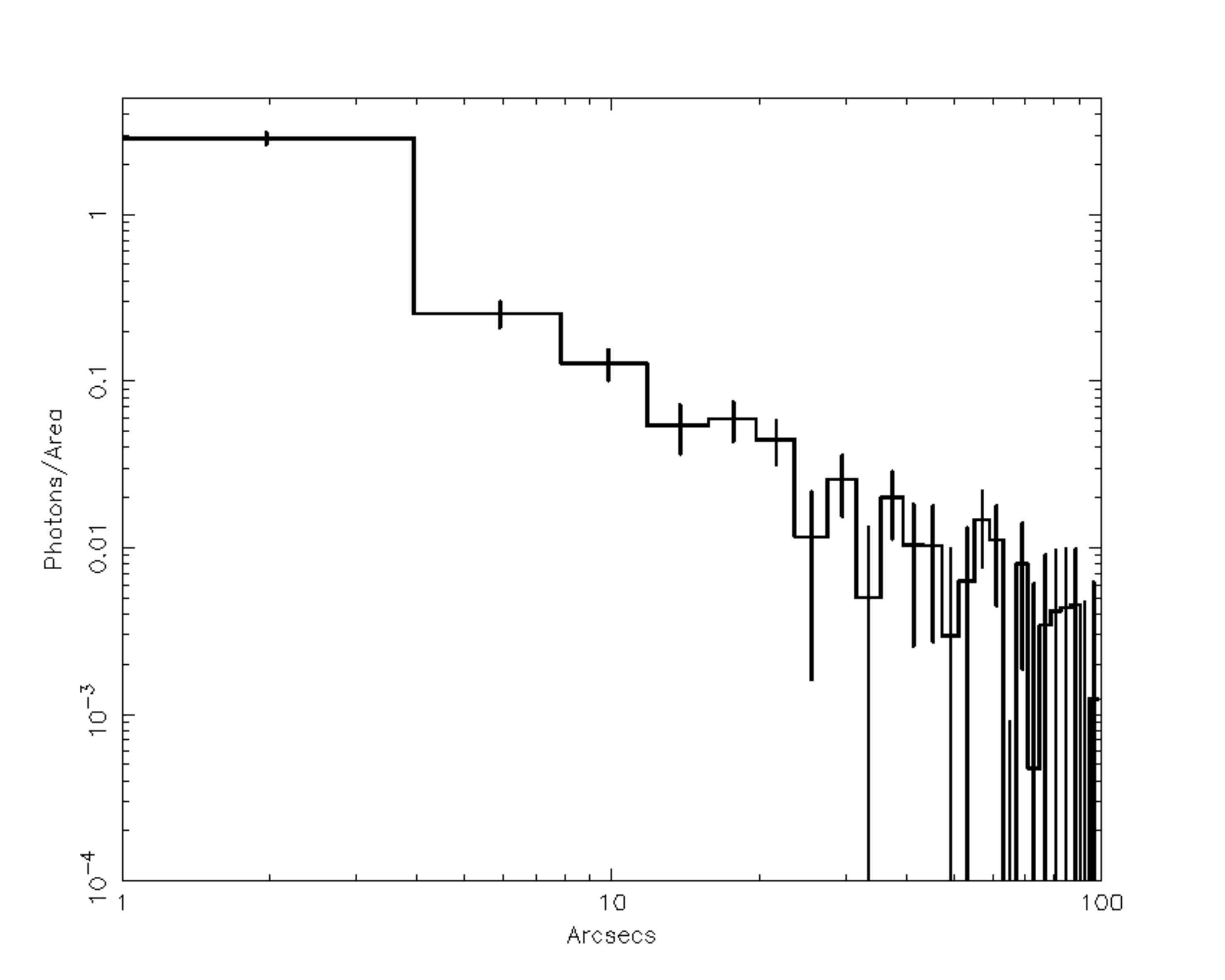}
\caption{PSR J1016-5857 {\it Chandra} radial profile (0.3-10 keV energy range).
{\it Chandra} ACIS On-axis Point Spread Function is almost zero over 2$"$ from the source so that
it's apparent the nebular emission between 2 and 30$"$ from the pulsar.
\label{J1016-psf}}
\end{figure}

Immediately after the detection of pulsations, we re-analyzed the {\it Chandra}
observation pointing J1016-5857 (obs. id 3855, {\it Chandra} ACIS-S observation, very faint mode, starting on 
2006, August 03 at 03:34:43 UT, exposure 19.0 ks).
The source off-axis angle is negligible.
The X-ray source best fit position, found using the celldetect task in the
CIAO distribution, is 10:16:21.27 -58:57:11.22 (1.1$"$ error radius).
Camilo et al. 2004 report the detection of a nebular emission around the pulsar.
we searched for diffuse emission in the immediate
surroundings of the pulsar: results in the 0.3-10 keV energy range are
shown in Figure \ref{J1016-psf}. The presence of an extended nebula is apparent 
up to $\sim$ 30$"$ and possibly extends until $\sim$ 1$'$.

We extracted the pulsar spectrum from a 2$"$ radius circular region, while we extracted the PWN spectrum from an
annulus of radii 2$"$ and 30$"$.
The background was extracted from an circular source-free region away from the source.
We obtained a total of 138 pulsar and 502 nebular counts (the background
contributions are respectively 0.2\% and 13.7\%).
Due to the low statistic, we used the C-statistic
approach implemented in XSPEC.
The pulsar emission is well described by a simple power law model ($\chi^2_{red}$ value
of 1.19, 114 dof fitting both the pulsar and nebular spectra)
with a photon index $\Gamma$ = 1.52$_{-0.20}^{+0.40}$ , absorbed
by a column N$_H$ = 5.75$_{-1.95}^{+2.35}$ $\times$ 10$^{21}$ cm$^{-2}$.
A simple blackbody spectrum yields an unrealistic value of the
temperature ($>$ 10$^7$ K) even if it's statistically
acceptable, while a composite model provides no significative improvement
to the statistic.
The pulsar wind nebula is well described by a simple
powerlaw with a photon index of 1.61$_{-0.36}^{+0.41}$.
Assuming the best fit model, the 0.3-10 keV unabsorbed pulsar flux is 
1.47$_{-1.31}^{+0.40}$ $\times$ 10$^{-13}$ and the nebular flux
is 3.53$_{-2.77}^{+0.26}$ $\times$ 10$^{-13}$ erg/cm$^2$ s.
For a distance
of 9.3 kpc, the two luminosities are L$_{9.3kpc}^{psr}$ = 1.53$_{-1.36}^{+0.42}$ $\times$ 10$^{33}$,
L$_{9.3kpc}^{pwn}$ = 3.66$_{-2.87}^{+0.27}$ $\times$ 10$^{33}$ erg/s.

\begin{figure}
\centering
\includegraphics[angle=0,scale=.50]{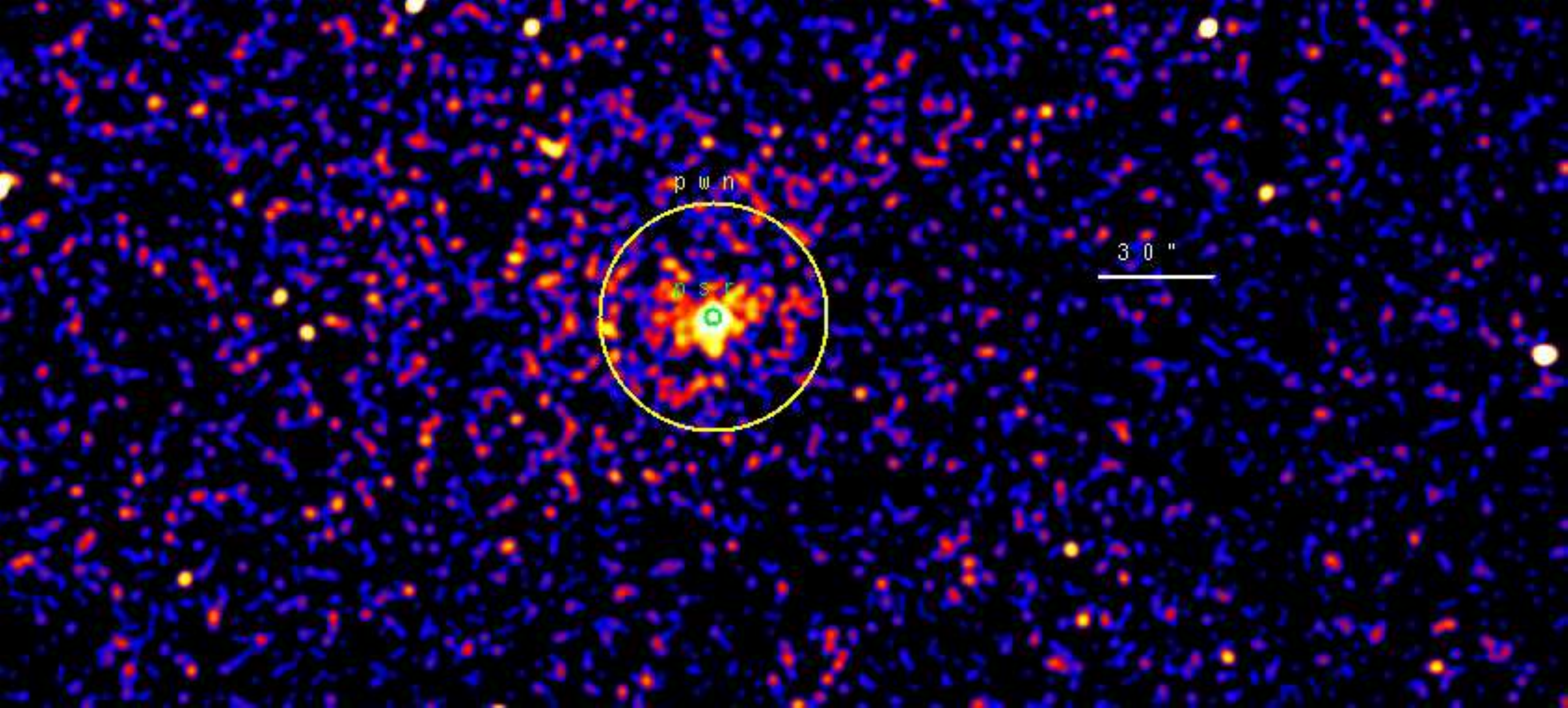}
\caption{PSR J1016-5857 0.3-10 keV {\it Chandra} Imaging. The image has been smoothed with a Gaussian
with Kernel radius of $2"$. The green circle marks the pulsar while the yellow annulus the nebular region used in the analysis.
\label{J1016-im}}
\end{figure}

\begin{figure}
\centering
\includegraphics[angle=0,scale=.50]{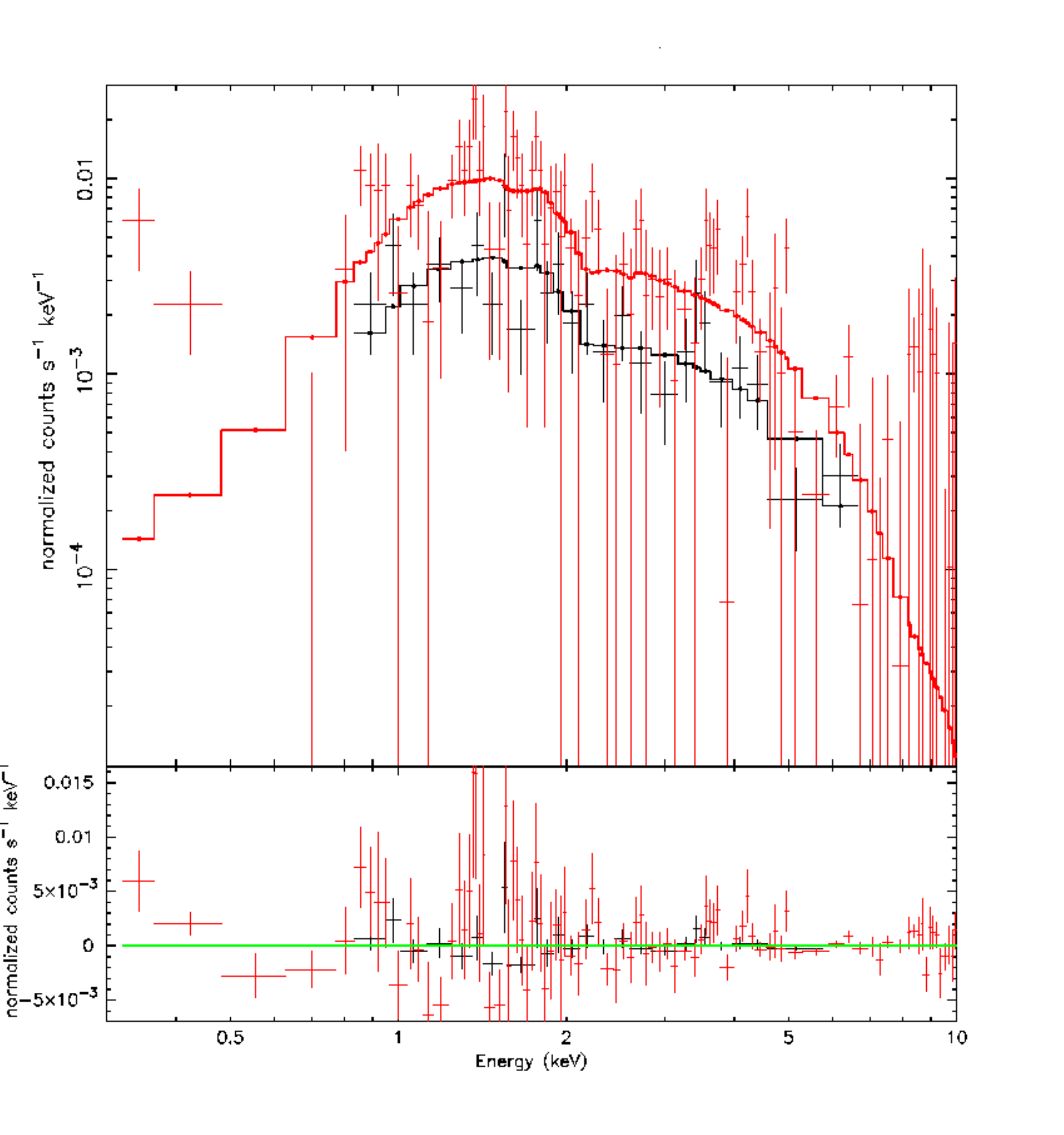}
\caption{PSR J1016-5857 {\it Chandra} Spectrum.
Black points mark the pulsar spectrum while red points the nebular one.
Residuals are shown in the lower panel.
\label{J1016-sp}}
\end{figure}

\clearpage

{\bf J1023-5746 (Westerlund 2) - type 2* RQP} % cambia sull'articolo: c'è la pwn, osservazione in arrivo

J0633+0632 was one of the last {\it Fermi} pulsars discovered using the
blind search technique (Saz Parkinson et al. 2010).
The unofficial name of this pulsar comes from the association
with the Westerlund 2 cluster, subsequently refuted.
{\it Fermi} also detected its nebula with a flux of 
2.76 $\pm$ 1.37 $\times$ 10$^{-11}$ erg/cm$^2$s (Ackermann et al. 2010).
The pseudo-distance of the object based on $\gamma$-ray data (Saz Parkinson et al. (2010))
is $\sim$ 2.4 kpc.

\begin{figure}
\centering
\includegraphics[angle=0,scale=.30]{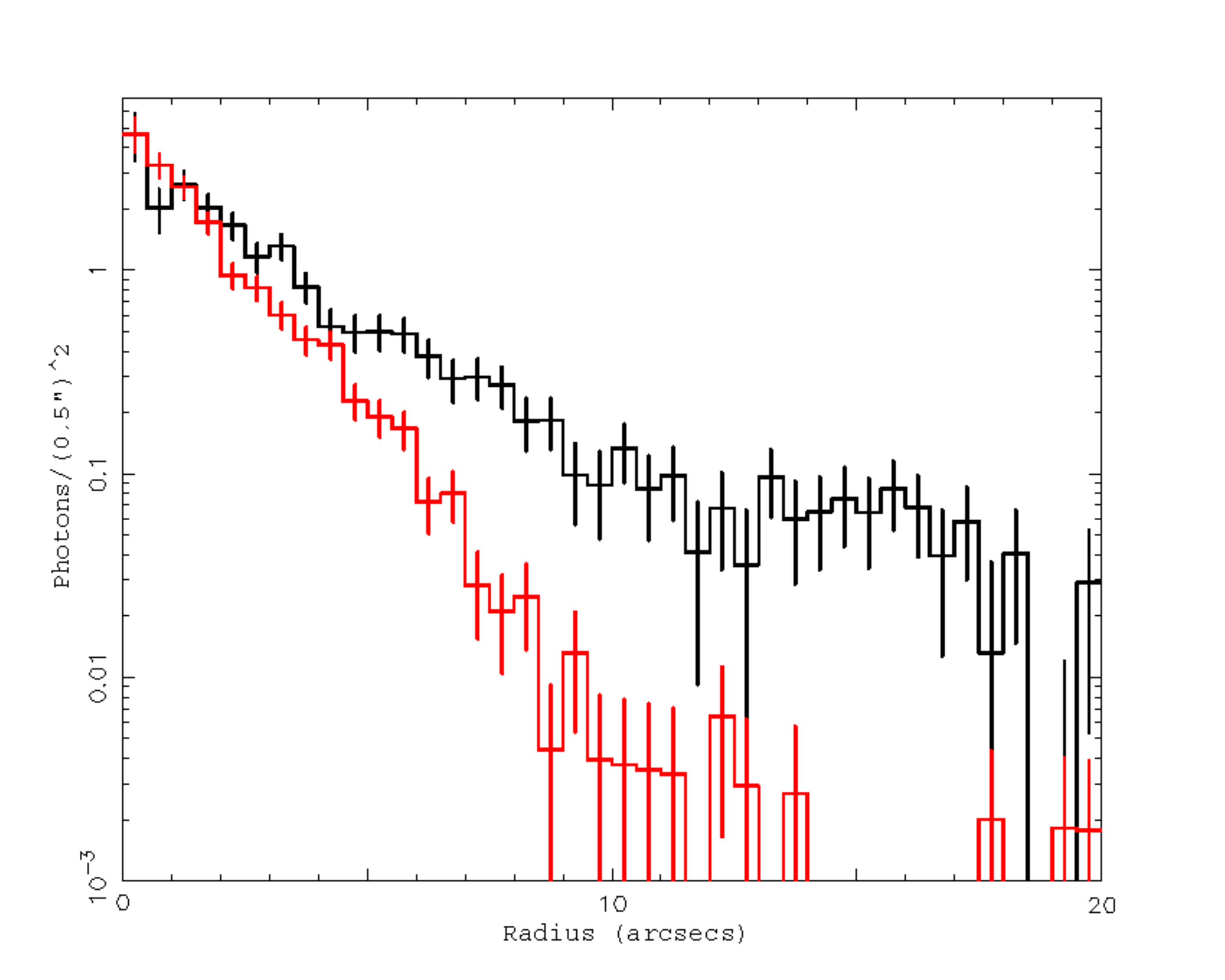}
\caption{PSR J1023-5746 {\it Chandra} Radial Profile (0.3-10 keV energy range) 
(black points) and for a simulated point source 
(red points) having flux, spectrum
and detector coordinates coincident with the ones of the pulsar counterpart
(see text for details). The two profiles don't agree for radii between 2 and 15$"$
showing the presence of a nebula.
\label{J1023-psf}}
\end{figure}

Immediately after the detection of pulsations, we analyzed the
three {\it Chandra} ACIS-I datasets with the {\it Fermi} source inside their
field of view:\\
- obs. id 3501, start time 2006, July 07 at 14:29:26 UT, exposure 36.7 ks;\\
- obs. id 6410, start time 2006, September 07 at 21:40:25 UT, exposure 50.0 ks;\\
- obs. id 6411, start time 2006, September 29 at 18:28:05 UT, exposure 50.0 ks.\\
In all the datasets the off-axis angle is high, $\sim$ 7.5$'$, near the edge
of the instrument FOV. The {\it Chandra} point spread function depends on
the off-axis angle so that it's very difficult to disentangle the pulsar
from its nebula, if present.
The X-ray source best fit position is 10:23:02.95 -57:46:08.00 (4.3$"$ error radius).
We searched for diffuse emission in the immediate
surroundings of the pulsar, by comparing the source intensity profile to the expected ACIS
Point Spread Function (PSF). Assuming the pulsar best fit spectral model, we simulated a
PSF using the ChaRT and MARX packages. Results in the 0.3-10 keV energy range are
shown in Figure \ref{J1023-psf}. The presence of an extended nebula is apparent 
up to $\sim$ 15$"$ from the pulsar while in a $\sim$ 5$"$ circle the pulsar emission
is predominant.

We extracted the pulsar spectrum from a circle of 5$"$ radius, in order to maximize the signal-to-noise ratio
and minimize any nebular contamination, while we extracted the PWN spectrum from an
annulus with radii 5$"$ and 15$"$.
The background was extracted from an annulus with radii 20 and 30$"$.
The spectra obtained in the three observations were added using
mathpha tool and, similarly, the response
matrix and effective area files using addarf and addrmf.
We obtained a total of 368 and 463 source counts for
pulsar and nebula (the background
contributions are respectively 4.3\% and 30.0\%).
Due to the low statistic, we used the C-statistic
approach implemented in XSPEC (requiring no spectral grouping, nor background
subtraction), well suited to study sources with low photon statistics.
We simultaneously fitted the pointlike and nebular spectra
linking the absorption value of the two.
The pulsar emission is well described ($\chi^2_{red}$ value
of 1.049) by an simple power law model
with a photon index of 1.26 $\pm$ 0.30, absorbed
by a column N$_H$ = 1.17 $_{-0.33}^{+0.37}$ $\times$ 10$^{22}$ cm$^{-2}$.
Such an high value of the column density, comparable with the
galactic one in the pulsar direction ($\sim$ 1.3 $\times$ 10$^{22}$ cm$^{-2}$,
using the HEASARC web tools) points to a fairly large distance (possibly greater than 10 kpc).
A simple blackbody spectrum yields an unrealistic value of the
temperature ($>$ 10$^7$ K) even if it's statistically
acceptable. A combined spectrum yields no statistical improvement.
The pulsar wind nebula is well described by a simple
powerlaw with a photon index of 1.54 $\pm$ 0.33.
Assuming the best fit model, the 0.3-10 keV unabsorbed pulsar flux is 
9.42$_{-5.95}^{+1.90}$ $\times$ 10$^{-14}$ and the nebular flux
is 8.53$_{-5.93}^{+1.93}$ $\times$ 10$^{-14}$ erg/cm$^2$ s.
For a distance
of 2.4 kpc, the two luminosities are L$_{2.4kpc}^{psr}$ = 6.51$_{-4.11}^{+1.31}$ $\times$ 10$^{31}$ and
L$_{2.4kpc}^{pwn}$ = 5.89$_{-4.09}^{+1.33}$ $\times$ 10$^{31}$ erg/s.

\begin{figure}
\centering
\includegraphics[angle=0,scale=.40]{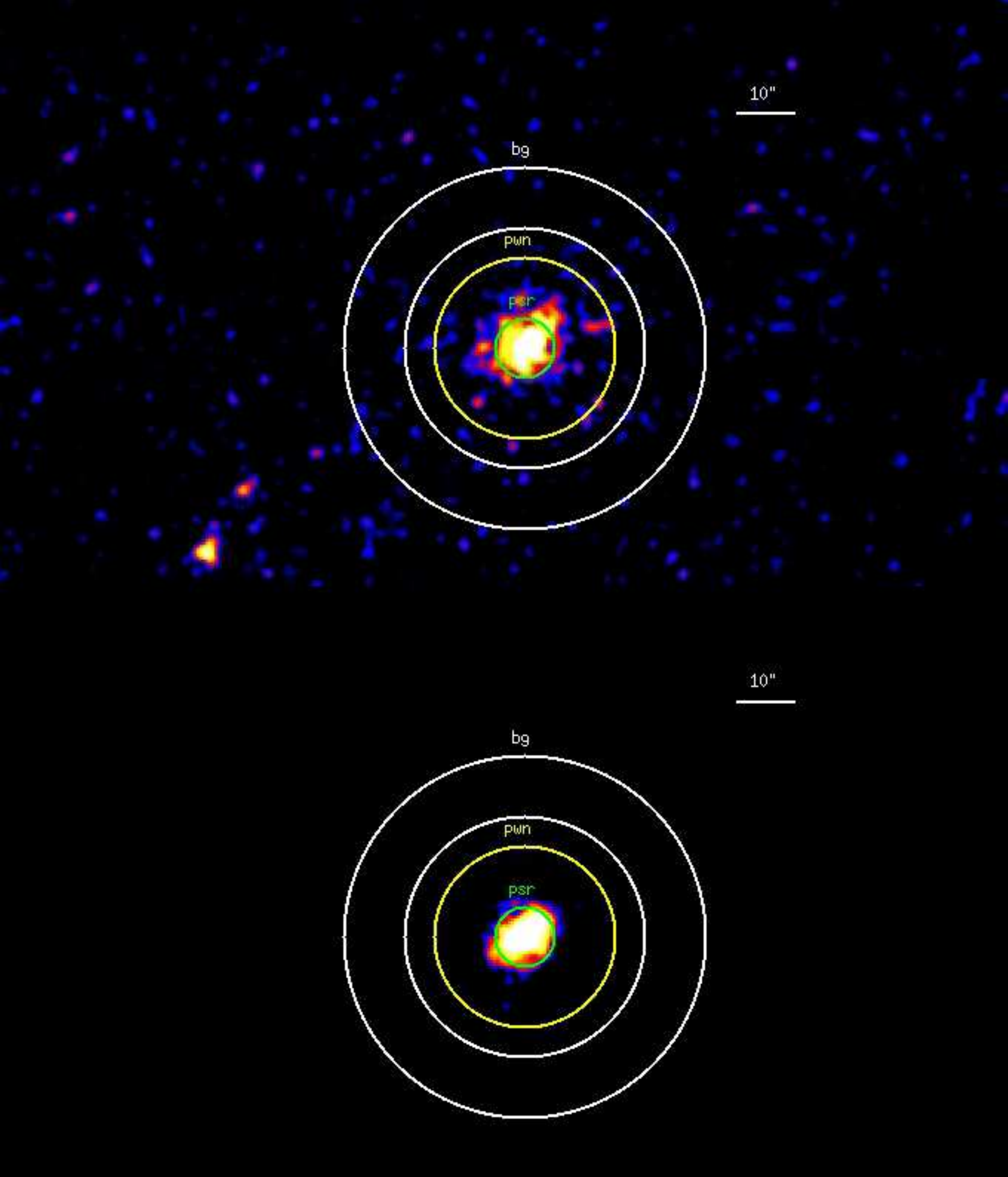}
\caption{{\it Upper Panel:} PSR J1023-5746 0.3-10 keV {\it Chandra} Imaging. The image has been smoothed with a Gaussian
with Kernel radius of $2"$. The green circle marks the pulsar, the yellow annulus the nebular and
the white annulus the background region used in the analysis.
{\it Lower Panel:} Imaging of the simulated point source having flux, spectrum
and detector coordinates coincident with the ones of the pulsar counterpart
(see text for details).
\label{J1023-im}}
\end{figure}

\begin{figure}
\centering
\includegraphics[angle=0,scale=.50]{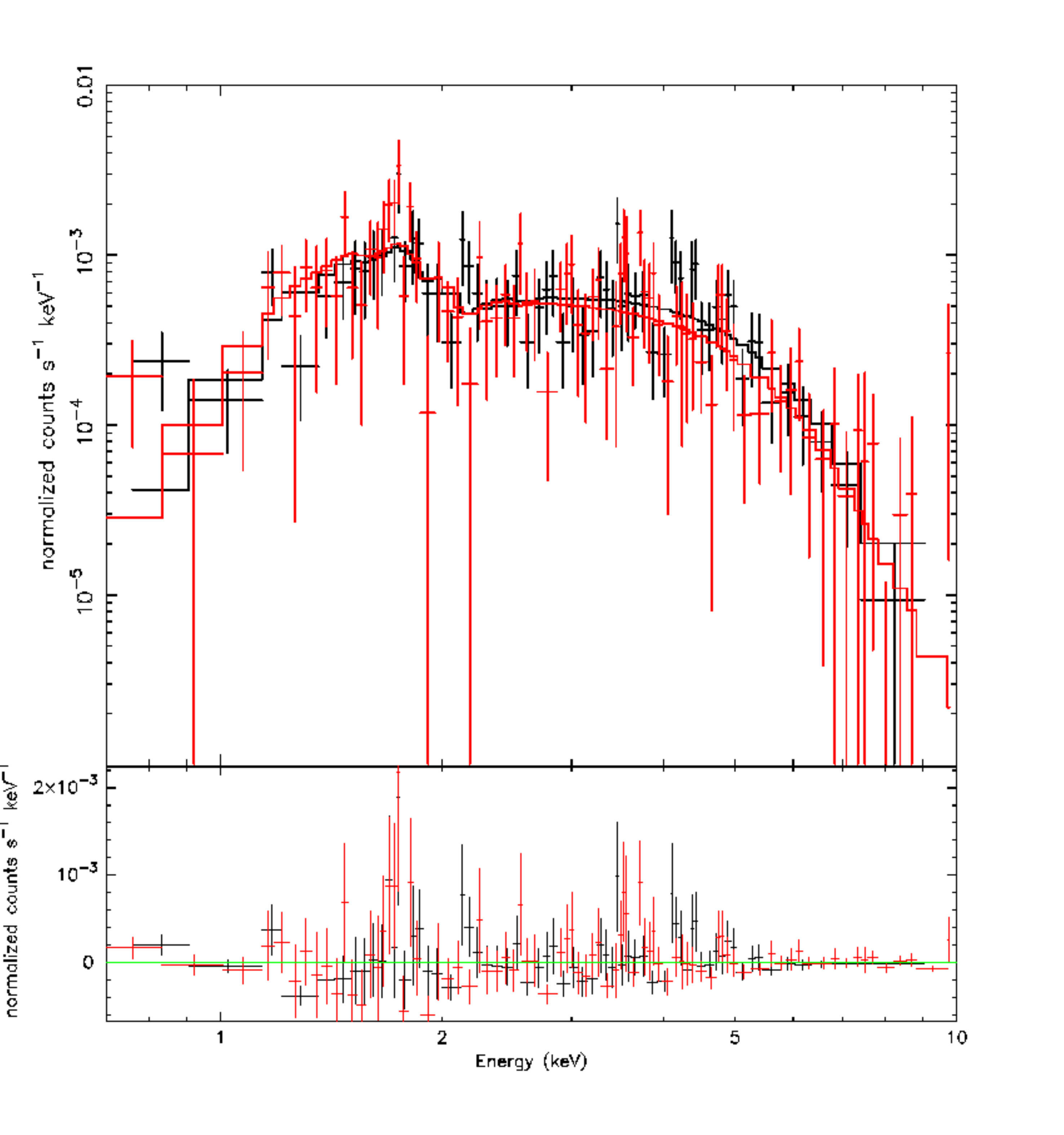}
\caption{PSR J1023-5746 {\it Chandra} Spectrum.
Black points mark the pulsar spectrum while red points the nebular one.
Residuals are shown in the lower panel.
\label{J1023-sp}}
\end{figure}

\clearpage

{\bf J1024-0719 - type 0 RL MSP} % Nuova!

% Becker \& Trumper 1998 e Hui \& Becker 2006
J1024-0719 was discovered
during the Parkes 436 MHz survey of the southern sky (Bailes
et al. 1997). It has a rotation period of 5.16 ms and a period
derivative of $\dot{P}$ = 1.84 $\times$ 10$^{-20}$ s s$^{-1}$, implying a
characteristic age of $\tau_c$ = 4.4 $\times$ 10$^9$ years and a rotational
energy loss of $\dot{E}$ = 5.3 $\times$ 10$^{33}$ erg/s.
X-ray emission from this pulsar was reported by
Becker \& Trumper (1999) in ROSAT HRI data.
The Taylor \& Cordes model
implies d = 530$\pm$120 pc based on the pulsar's dispersion measure
(Bailes et al. 1997).

An XMM observation of this source has been performed, obs. id 0112320301,
start time 2003, December 02 at 02:13:54 UT, with an exposure of 76.6 ks.
The PN camera of the EPIC
instrument was operated in Fast Timing mode, while the MOS detectors were set in Full frame mode. For
all three instruments, the thin optical filter was used.
The PN camera wasn't used in order to perform the spectral analysis
due to the lack of spatial resolution.
No screening for soft proton flare events was done,
due to the goodness of the observation.
The X-ray source best fit position, obtained by using the SAS tools,
is 10:24:38.64 -07:19:19.20 (5$"$ error radius).
No nebular emission was detected in the {\it XMM-Newton} observation.
We extracted the XMM spectrum from a 20$"$ radius circular region
while the background was taken from a circular source-free region
near the pulsar.
Due to the low statistic the two MOS spectra were added using
mathpha tool and, similarly, the response
matrix and effective area files using addarf and addrmf. 
we obtained a total of 309 pulsar counts from the two MOS
cameras (background contribution of 25.2\%).
Both a simple powerlaw and a simple blackbody models are statistically acceptable
(reduced chiquare values of $\chi^2_{red-pow}$ = 0.91 and $\chi^2_{red-bb}$ = 1.03, 8 dof).
The powerlaw model has a photon index $\Gamma$ = 4.12$_{-0.84}^{+1.03}$ , 
absorbed by a column N$_H$ = 2.65$_{-1.29}^{+1.60}$ $\times$ 10$^{21}$ cm$^{-2}$.
Such value of the absorbing column isn't in agreement with the galactic N$_H$
obtained using the Heasarc Web Tools ($\sim$ 4 $\times$ 10$^{20}$ cm$^{-2}$),
so that the simple powerlaw model can be excluded.
The thermal model has a temperature of 2.83$_{-0.37}^{+0.44}$ $\times$ 10$^6$ K
and an emitting radius of R$_{530pc}$ = 29.5$_{-5.9}^{+25.8}$ m and is absorbed
by a column N$_H$ = 0$_{-0}^{+3.58}$ $\times$ 10$^{20}$ cm$^{-2}$.
Such a small emitting radius
is probably due to an hot spot thermal emission, quite typical for such millisecond pulsars.
Assuming the best fit model, the 0.3-10 keV unabsorbed hot spot thermal flux is 
1.11 $\pm$ 0.51 $\times$ 10$^{-14}$ erg/cm$^2$ s.
For a distance
of 350 pc, the pulsar luminosity is L$_{530pc}^{th}$ = 3.74 $\pm$ 1.72 $\times$ 10$^{29}$ erg/s.

\begin{figure}
\centering
\includegraphics[angle=0,scale=.30]{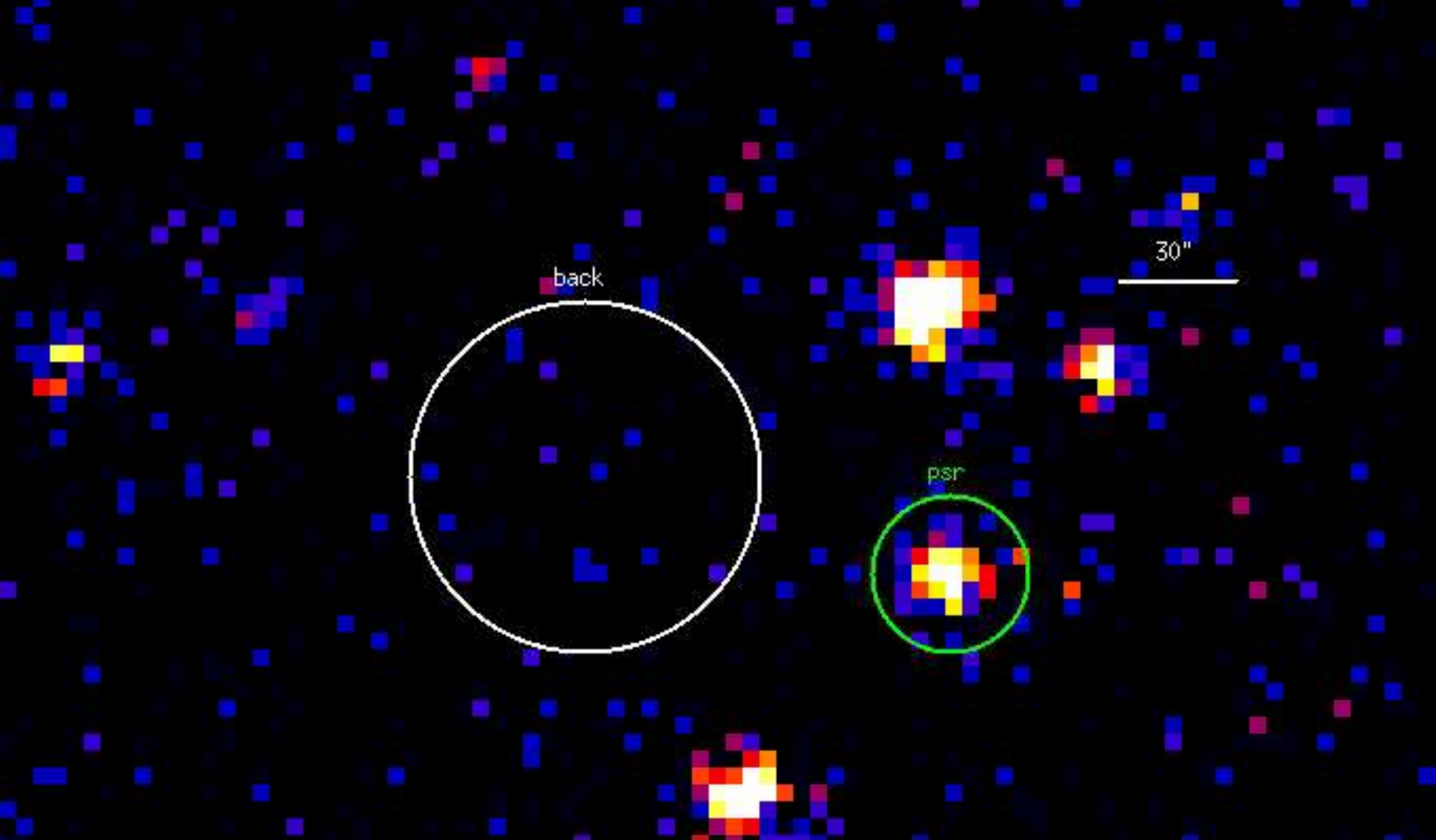}
\caption{PSR J1024-0719 0.3-10 keV MOS Imaging. The two MOS images have been added. 
The green circle marks the pulsar while the yellow annulus the background region used in the analysis.
\label{J1024-im}}
\end{figure}

\begin{figure}
\centering
\includegraphics[angle=0,scale=.30]{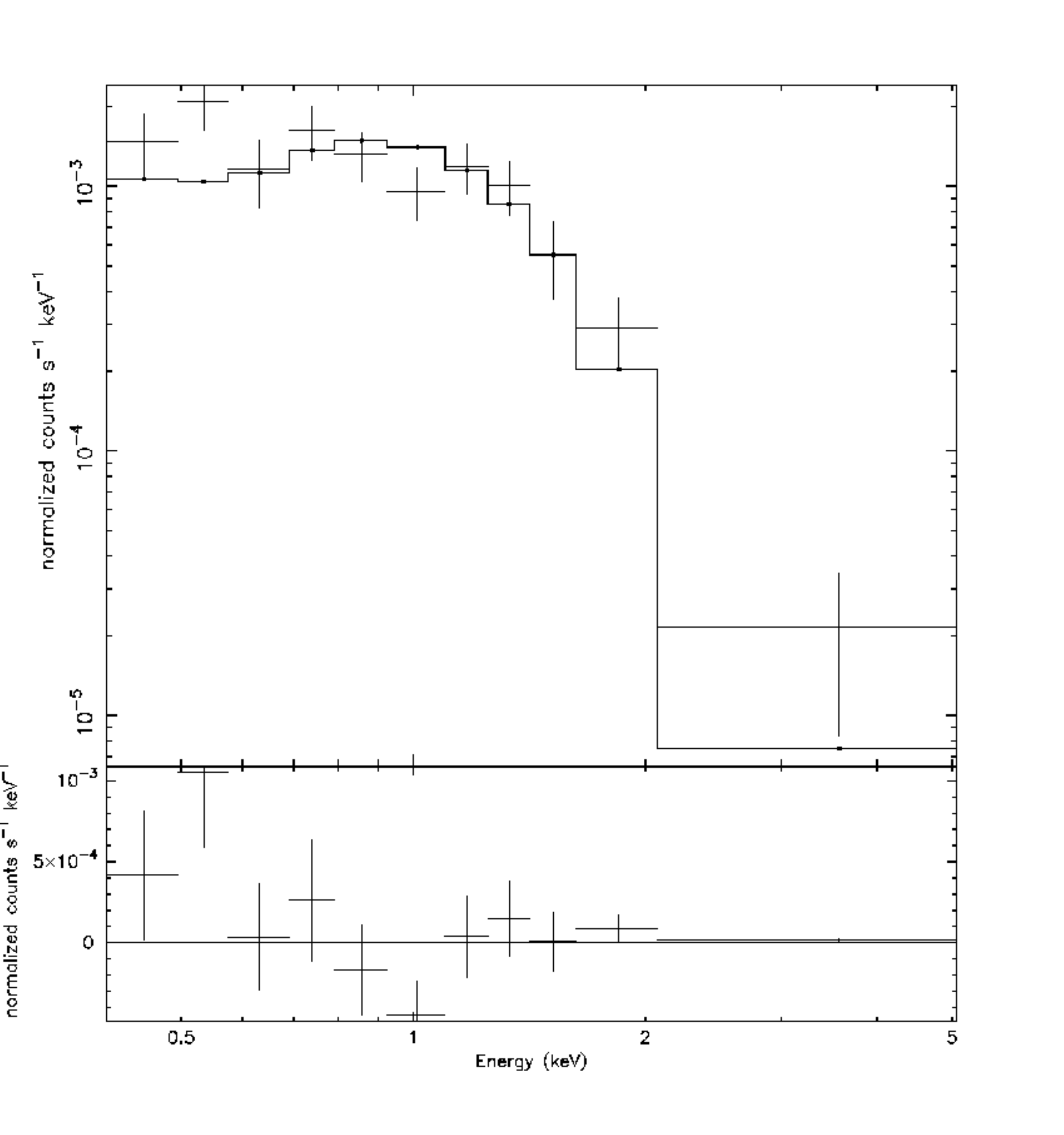}
\caption{PSR J1024-0719 {\it XMM-Newton} MOS Spectrum (see text for details).
Residuals are shown in the lower panel.
\label{J1024-sp}}
\end{figure}

\clearpage

{\bf J1028-5819 - type 1 RLP} % osservazione in arrivo

%Keith et al. 2008
PSR J1028-5819 was detected on 2008 April 6 by Parkes radio telescope,
with a dispersion measure of
96 pc cm$^{-3}$ (Keith et al. 2008).
The detected pulse period of 91.4 ms immediately
indicated that the pulsar was likely to be a young energetic
pulsar and potentially associated with the EGRET
source 3EG J1027-5817.
The estimated distance based on dispersion measurements
is 2.33 $\pm$ 0.70 kpc.
{\it Fermi} detected no nebular emission down to a flux of 9.83 $\times$ 10$^{-11}$ erg/cm$^2$ s.

A {\it SWIFT} TOO observation (obs. id 00031298001)
was requested after the {\it Fermi} discovery of the pulsar.
In the 7.87 ks-long {\it SWIFT} observation we found a source
coincident with the Radio position of the pulsar.
The X-ray statistic was too low to obtain a clear spectrum
of the source nor an indication as the presence of a PWN.
The net countrate of the source is 2.41 $\pm$ 0.69 $\times$ 10$^{-3}$ c/s.
For a distance of 2.3 kpc, we found a
rough absorption column value of 5 $\times$ 10$^{21}$ cm$^{-2}$,
using a simple powerlaw spectrum
for PSR+PWN with $\Gamma$ = 2 and taking in account a 30\% of the resulting flux coming
from the PWN and thermal components,
we obtained an unabsorbed flux of 1.5 $\pm$ 0.5 $\times$ 10$^{-13}$ erg/cm$^2$ s,
leading to a luminosity L$_{2.3kpc}$ = 9.52 $\pm$ 3.17 $\times$ 10$^{31}$ erg/s.

\begin{figure}
\centering
\includegraphics[angle=0,scale=.50]{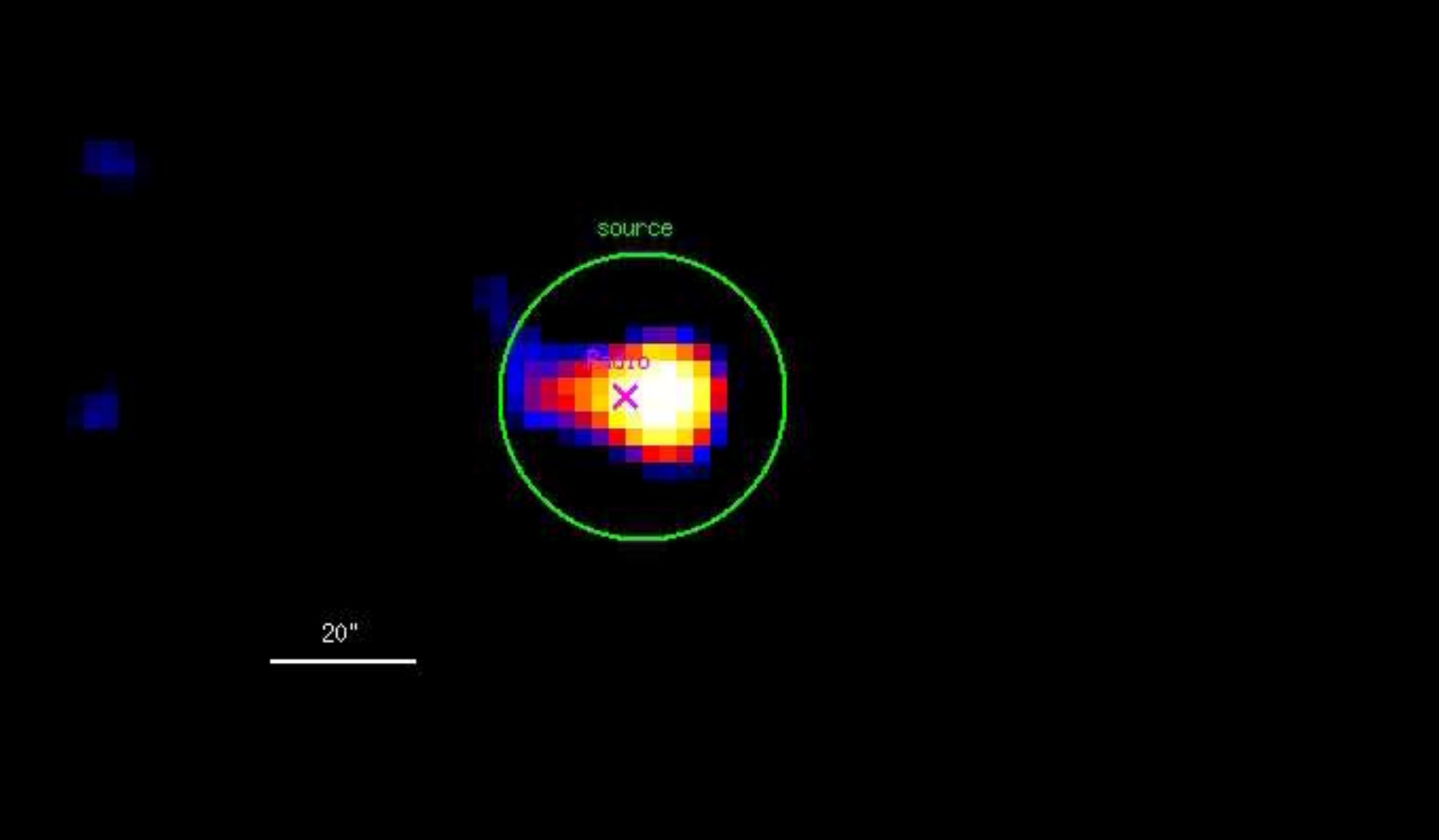}
\caption{PSR J1028-5819 0.3-10 keV {\it SWIFT} XRT Imaging. All the XRT images have been added. 
The green circle marks the source position found by using XIMAGE tools. The magenta cross marks the
Radio position of the source.
\label{J1028-im}}
\end{figure}

\clearpage

{\bf J1044-5737 - type 0 RQP}

Pulsations from J1044-5737 were detected by {\it Fermi} using the
blind search technique (Saz Parkinson et al. 2010).
No $\gamma$-ray nebular emission was detected down to a flux of 
1.32 $\times$ 10$^{-11}$ erg/cm$^2$s (Ackermann et al. 2010).
The pseudo-distance of the object based on $\gamma$-ray data (Saz Parkinson et al. (2010))
is $\sim$ 1.5 kpc.

After the {\it Fermi} detection, we asked for a {\it SWIFT} observation
of the $\gamma$-ray error box (obs id. 00031546001, 4.22 ks exposure)
which yielded an X-ray source at
the edge of the {\it Fermi} error box. However, dedicated timing
analysis performed by the LAT team excluded it
as the counterpart of the $\gamma$-ray pulsar.
For a distance of 1.5 kpc we found a
rough absorption column value of 5 $\times$ 10$^{21}$ cm$^{-2}$
and using a simple powerlaw spectrum
for PSR+PWN with $\Gamma$ = 2 and a signal-to-noise of 3,
we obtained an upper limit non-thermal unabsorbed flux of 3.93 $\times$ 10$^{-13}$ erg/cm$^2$ s,
leading to an upper limit luminosity L$_{1.5kpc}^{nt}$ = 1.06 $\times$ 10$^{32}$ erg/s.

{\bf J1048-5832 - type 2 RLP}

% Gonzalez et al. 2006
The radio pulsar B1046-58 was discovered during a
Parkes survey of the Galactic plane (Johnston et al. 1992). It
has a period of P = 124 ms and period derivative $\dot{P}$ = 9.6 $\times$
10$^{14}$ s s$^{-1}$. These values imply a characteristic age of
20.4 kyr and a spin-down luminosity of 2.0 $\times$ $10^{36}$ erg/s. These properties are
similar to those of other young neutron stars typified by the Vela
pulsar. The dispersion measure of 129 pc cm$^{-3}$ toward the pulsar
implies a distance of 2.71 $\pm$ 0.81 kpc (Taylor \& Cordes 1993).
Deep radio observations did not detect extended emission associated
with PSR B1046-58 (Stappers et al. 1999). X-ray observations
with ASCA and the Rontgensatellit (ROSAT) detected
emission near the pulsar and suggested the presence of large scale
structures, surrounded by faint emission (Pivovaroff et al.
2000). However, the poor angular resolution of these observations
prevented conclusive interpretation of the data.
The EGRET source 2EG J1049-5827 is coincident
with the radio coordinates of PSR B1046-58
and lately the association was confirmed by {\it Fermi}.
No $\gamma$-ray nebular emission was detected by {\it Fermi} down to a flux of 
1.70 $\times$ 10$^{-11}$ erg/cm$^2$s (Ackermann et al. 2010).

\begin{figure}
\centering
\includegraphics[angle=0,scale=.40]{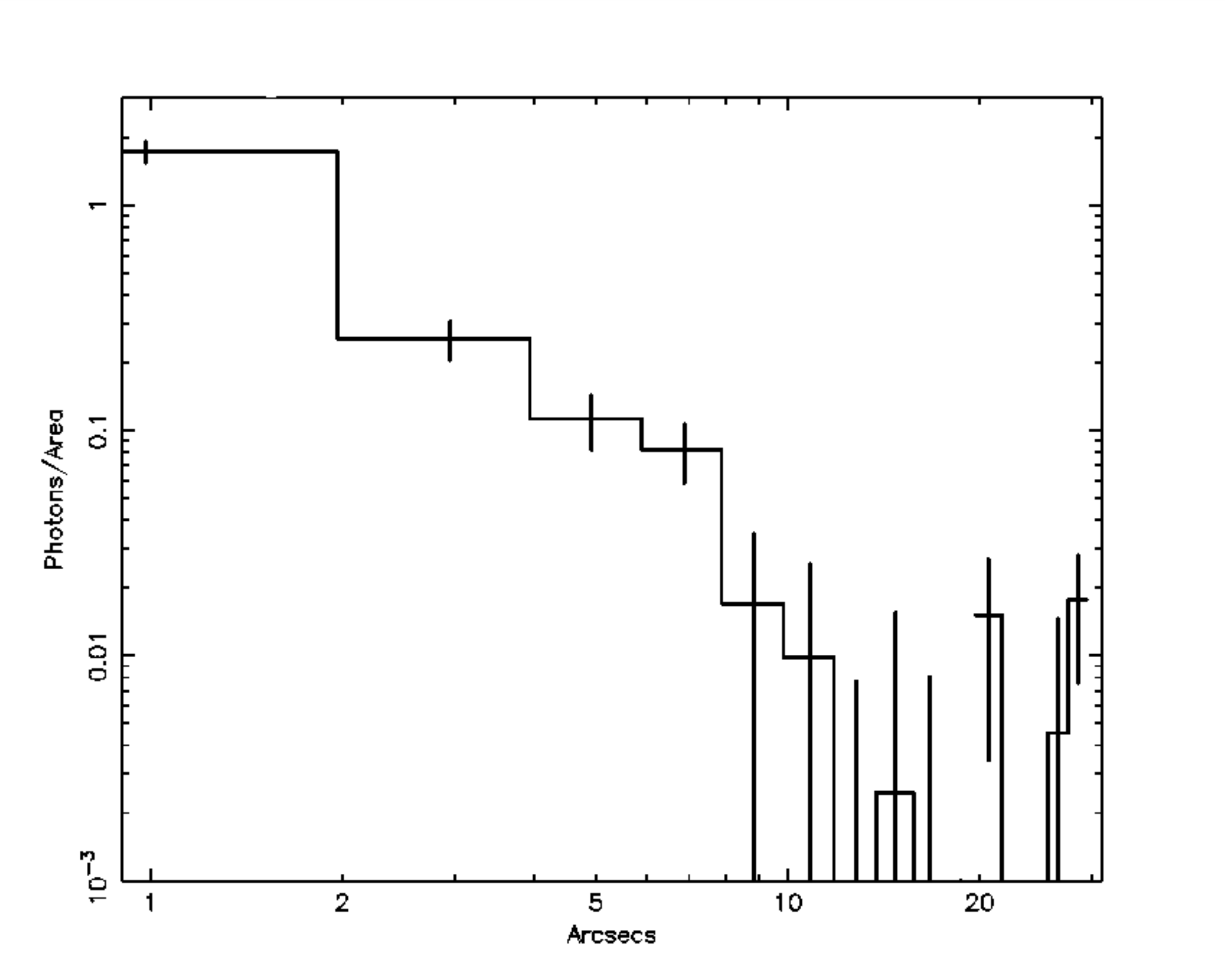}
\caption{PSR J1048-5832 {\it Chandra} radial profile (0.3-10 keV energy range).
{\it Chandra} ACIS On-axis Point Spread Function is almost zero over 2$"$ from the source so that
it's apparent the nebular emission between 2 and 10$"$ from the pulsar.
\label{J1048-psf}}
\end{figure}

We used the only {\it XMM-Newton} observation
centered on the pulsar (obs. id 0054540101), which started on
2002, August 10 at 08:54:50.70 UT and lasted 23.7 ks.
The PN camera of the EPIC
instrument was operated in Small Window mode, while the MOS detectors were set in Full frame mode.
For the PN camera the thin optical filter was used while for the two MOS the medium
filter was used.
The screening of high particle background time intervals was
not necessary, owing to the goodness of the observation.
We also used the {\it Chandra} ACIS-S observation of J1048-5832,
obs. id 3842 started on 2003, October 08 at 22:38:34 UT
for a net exposure of 36.6 ks.
The pulsar position was imaged on the back-illuminated ACIS
S3 chip and the VFAINT exposure mode was adopted. The off-axis angle is negligible.
The X-ray source best fit position, using the celldetect tool inside
the CIAO software, is 10:48:12.63 -58:32:03.50 (0.9$"$ error radius).
we searched for diffuse emission in the immediate
surroundings of the pulsar: results in the 0.3-10 keV energy range are
shown in Figure \ref{J1048-psf}. The presence of an extended nebula is apparent 
between 2 and $\sim$ 10$"$.

In the XMM observation we extracted the source spectrum from a circle of 20$"$ radius centered on the source, containing both the
pulsar and the nebula, while we extracted the background
from a source-free region on the same chip of the source: the presence of a nearby bright source
prevent me to use an annular region.
We extracted the {\it Chandra} pulsar spectrum from a circle of 2$"$ radius centered on the source and the PWN spectrum from an
annulus of radii 2$"$ and 10$"$.
The background was extracted from an annulus with radii 15 and 20$"$.
We obtained 202, 60 and 45 pulsar+pwn counts (the background
contributions are respectively 0.505, 0.830 and 0.809) in the three
{\it XMM-Newton} cameras. 78 pulsar counts and 183 nebular counts
have been extracted from the {\it Chandra} dataset (background
contributions of 1.9\% and 22.9\%).
Owing to the poor statistic, we used the C-statistic
approach implemented in XSPEC and
we simultaneously fitted the five spectra.
The pulsar emission is well described ($\chi^2_{red}$ value
of 1.319, 103 dof using both the pulsar and nebular spectra) by a simple power law model
with a photon index of 1.35 $\pm$ 0.45, absorbed
by a column N$_H$ = 4.60 $\pm$ 0.23 $\times$ 10$^{21}$ cm$^{-2}$.
A simple blackbody spectrum yields an unrealistic value of the
temperature ($>$ 10$^7$ K) even if it's statistically
acceptable; a composite model gives no significative improvement to the fit.
The pulsar wind nebula is well described by a simple
powerlaw with a photon index of 1.22 $\pm$ 0.46.
Assuming the best fit model, the 0.3-10 keV unabsorbed pulsar flux is 
4.90$_{-3.42}^{+1.81}$ $\times$ 10$^{-14}$ and the nebular flux
is 6.08$_{-4.26}^{+2.24}$ $\times$ 10$^{-14}$ erg/cm$^2$ s.
For a distance
of 2.7 kpc, the two luminosities are L$_{kpc}^{psr}$ = 4.29$_{-2.99}^{+1.58}$ $\times$ 10$^{31}$ and
L$_{kpc}^{pwn}$ = 5.32$_{-3.73}^{+1.96}$ $\times$ 10$^{31}$ erg/s.

\begin{figure}
\centering
\includegraphics[angle=0,scale=.40]{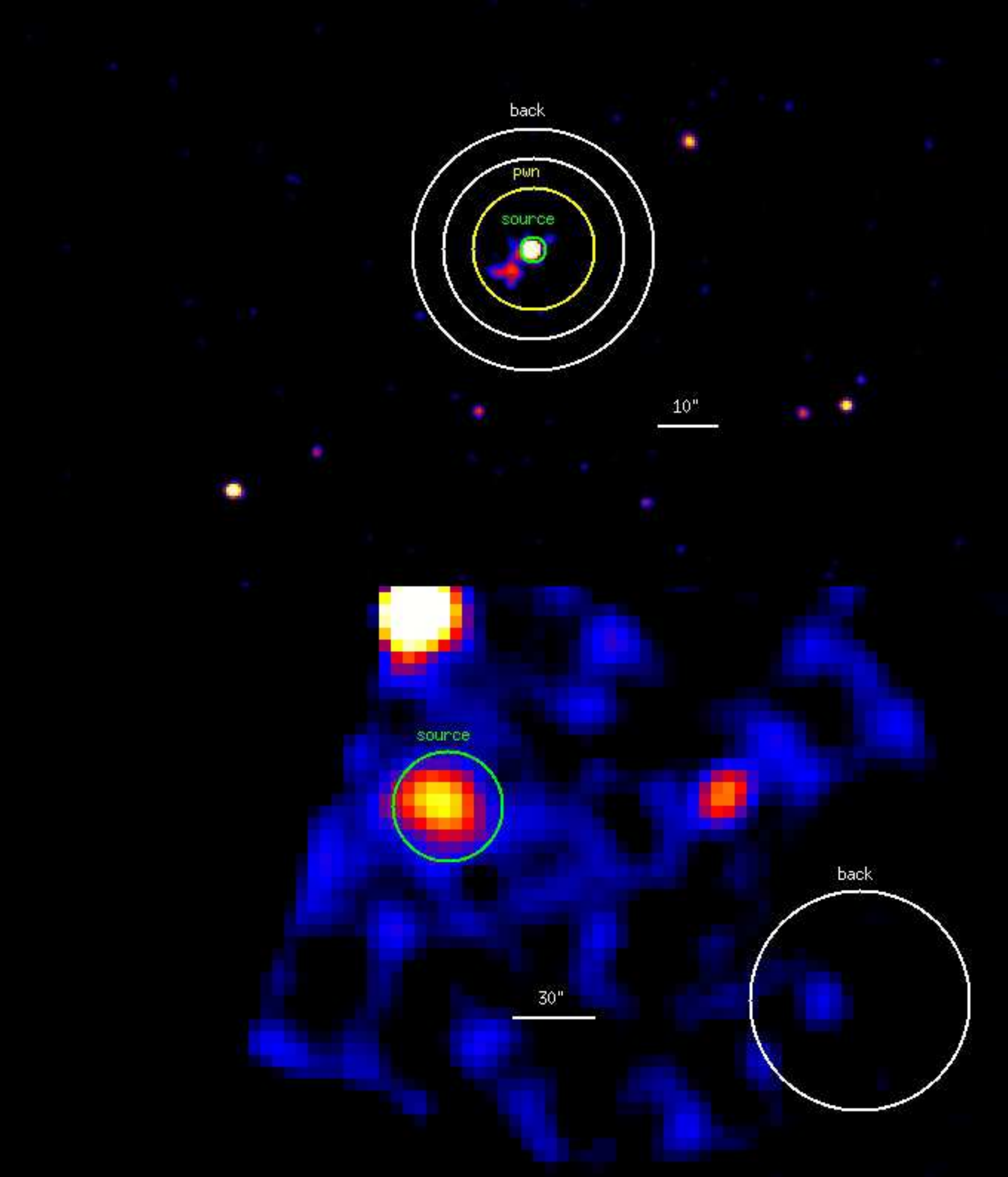}
\caption{{\it Upper Panel:} PSR J1048-5832 0.3-10 keV {\it Chandra} Imaging.
The green circle marks the pulsar, the yellow annulus the nebular region and the white annulus the background used in the analysis.
{\it Lower Panel:} PSR J1048-5832 0.3-10 keV EPIC Imaging. The PN and two MOS images have been added. 
The green circle marks the pulsar while the yellow one the background region used in the analysis.
\label{J1048-im}}
\end{figure}

\begin{figure}
\centering
\includegraphics[angle=0,scale=.50]{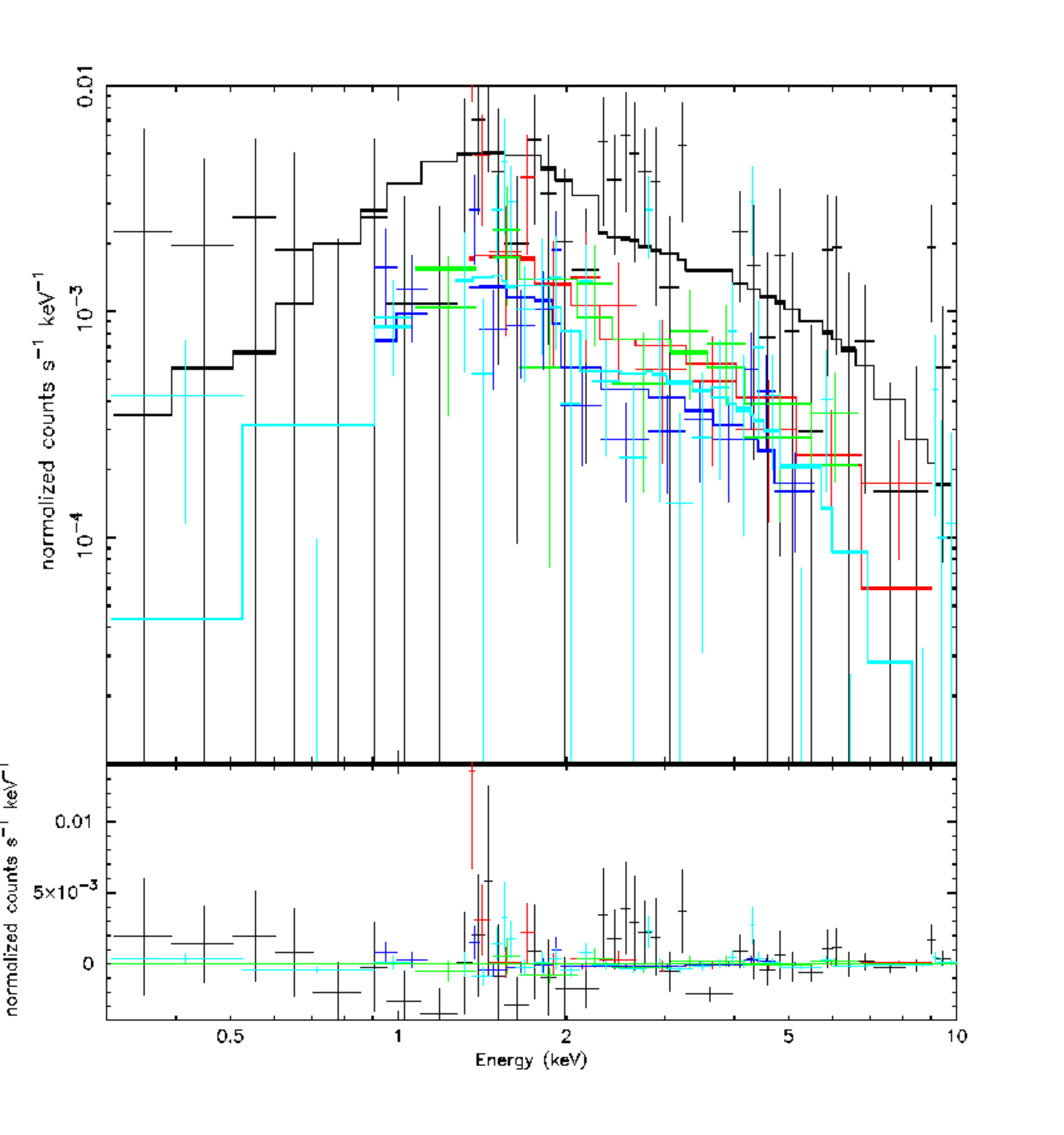}
\caption{J1048-5832 pulsar and nebular spectra. Different colors mark all the different dataset used (see text for details).
Residuals are shown in the lower panel.
\label{J1048-sp}}
\end{figure}

\clearpage

{\bf J1057-5226 - type 2* RLP}

After the Einstein observatory discovery of X-ray emission
from this radio pulsar (Cheng \& Helfand 1983), PSR B1055-52
was observed with ROSAT, both with the HRI (8.6 ks
yielding $\sim $570 source photons) and with the PSPC (15.6 ks, for
a total of $\sim$ 5500 source photons) in 1990-1992 (Oegelman \&
Finley 1993). The timing analysis unveiled the source pulsation,
with a pulsed fraction increasing from $\sim$ 11 \% for energies
$<$0.5 keV to $\sim$ 63 \% above 0.5 keV. The spectrum was best described
by a two-thermal components model. An ASCA observation could add only
$\sim$ 200 photons in the 0.5-10 keV range (Greiveldinger et al.
1996). The radio dispersion measurements suggest a distance
of 720 $\pm$ 200 pc.
No $\gamma$-ray nebular emission was detected by {\it Fermi} down to a flux of 
1.14 $\times$ 10$^{-11}$ erg/cm$^2$s (Ackermann et al. 2010).

\begin{figure}
\centering
\includegraphics[angle=0,scale=.40]{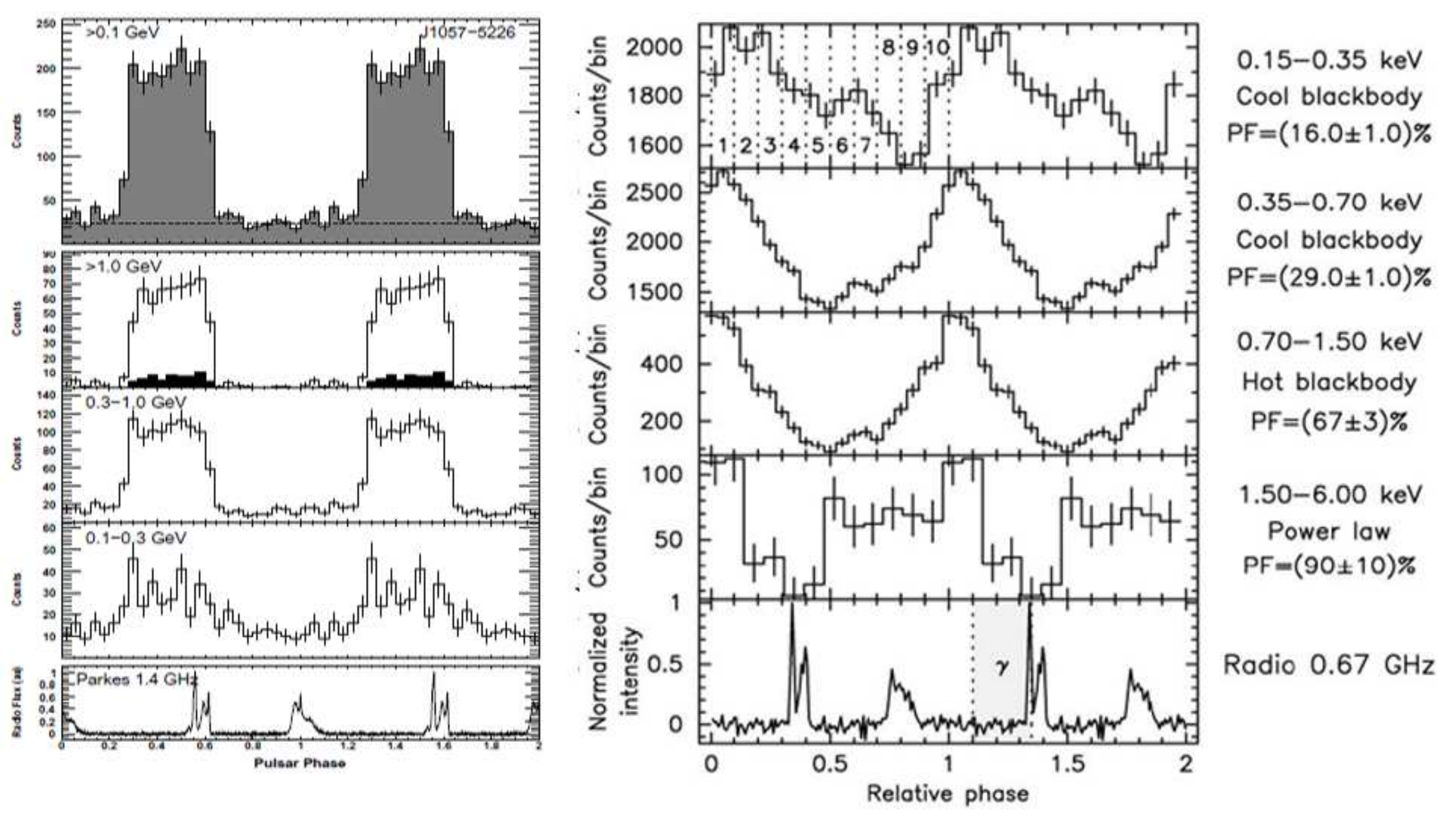}
\caption{PSR J1057-5226 Lightcurve.{\it Left: Fermi} $\gamma$-ray lightcurve folded with Radio
(Abdo et al. catalogue). {\it Right: XMM-Newton} EPIC PN X-ray pulse profiles
for different energy bands, folded with Radio. The choice of phase zero
and the alignment of the X-ray and radio profiles are arbitrary (De Luca et al. 2005).
\label{J1057-lc}}
\end{figure}

J1057-5226 was observed during two {\it XMM-Newton} observation
on 2000, December 14 at 22:18:30 UT and 2000, December 15 at 17:37:26 UT
(exposures of 19.9 ks and 52.7 ks).
The PN camera of the EPIC
instrument was operated in Fast Timing mode, while the MOS detectors were set in Full frame mode. For
all three instruments, the medium optical filter was used.
No screening for soft proton flare events was done,
owing to the goodness of both the observations.
we also used the {\it Chandra} ACIS-S observation
752 started on 2000, January 05 at 10:01:04 UT
for a net exposure of 40.6 ks.
This observation was performed in Continuous Clocking Faint mode;
this is provided to allow 3 msec timing at the expense of one
dimension of spatial resolution. In this mode, one obtains 1 pixel $\times$ 1024 pixel images, each
with an integration time of 2.85 msec. For this reason,
no {\it Chandra} image of the source was taken.
The X-ray source best fit position, using the SAS and XIMAGE dedicated tools,
is 10:57:59.09 -52:26:57.00 (5$"$ error radius).
As reported in De Luca et al. 2004 % controlla che sia vero
no nebular emission was detected in the {\it XMM-Newton} observations.
Since a faint source at an angular distance of 32$"$ was
detected in the northeast direction (see Fig. 17 of Becker\&Aschenbach 2002),
an annular background region is not suitable.
We extracted the {\it Chandra} spectrum from a circle of 2$"$ radius centered on the source
and the {\it XMM-Newton} one from a 30$"$ radius in order to
consider most of the counts of such a bright source
and to maximize the signal-to-noise ratio.
The background was extracted from a source-free
circular region on the same CCD of the pulsar.
We obtained a total of 63092, 11071 and 11722 pulsar counts
respectively from PN, MOS1 and MOS2 cameras (background
contribution of 13.5\%, 0.8\% and 0.8\%) and
7622 pulsar counts from the {\it Chandra} dataset (background
contribution of 1.0\%).
As in the case of Geminga (Caraveo et al. 2004), the best-fitting model is found by combining two
blackbody curves and a power law ($\chi^2_{red}$ = 1.20, 889 dof). No other simple model
yields an acceptable fit.
The powerlaw component has a photon index $\Gamma$ = 1.79$_{-0.18}^{+0.24}$ , 
absorbed by a column N$_H$ = 1.23$_{-0.44}^{+0.54}$ $\times$ 10$^{20}$ cm$^{-2}$.
The hot thermal component has a temperature of 1.76$_{-0.16}^{+0.08}$ $\times$ 10$^6$ K
and an emitting radius of R$_{720pc}$ = 462$_{-57}^{+330}$ m.
Such a small emitting radius suggests an hot spot thermal emission.
The cooler thermal component has a temperature of 8.06 $\pm$ 0.20 $\times$ 10$^5$ K
and an emitting radius of R$_{720pc}$ = 9.88$_{-0.85}^{+2.80}$ km,
pointing to a thermal cooling from the entire neutron star surface.
Assuming the best fit model, the 0.3-10 keV unabsorbed hot spot thermal flux is 
1.93$_{-0.09}^{+0.01}$ $\times$ 10$^{-13}$, the cooling thermal flux is
1.64$_{-0.08}^{+0.02}$ $\times$ 10$^{-12}$ and the non-thermal flux is
1.32$_{-0.08}^{+0.01}$ $\times$ 10$^{-13}$ erg/cm$^2$ s. Using a distance
of 720 pc, the luminosities are L$_{720pc}^{hs}$ = 1.20$_{-0.06}^{+0.01}$ $\times$ 10$^{31}$,
L$_{720pc}^{cool}$ = 1.02$_{-0.05}^{+0.01}$ $\times$ 10$^{32}$ and L$_{720pc}^{nt}$ = 8.21$_{-0.50}^{+0.06}$ $\times$ 10$^{30}$
erg/s.

\begin{figure}
\centering
\includegraphics[angle=0,scale=.50]{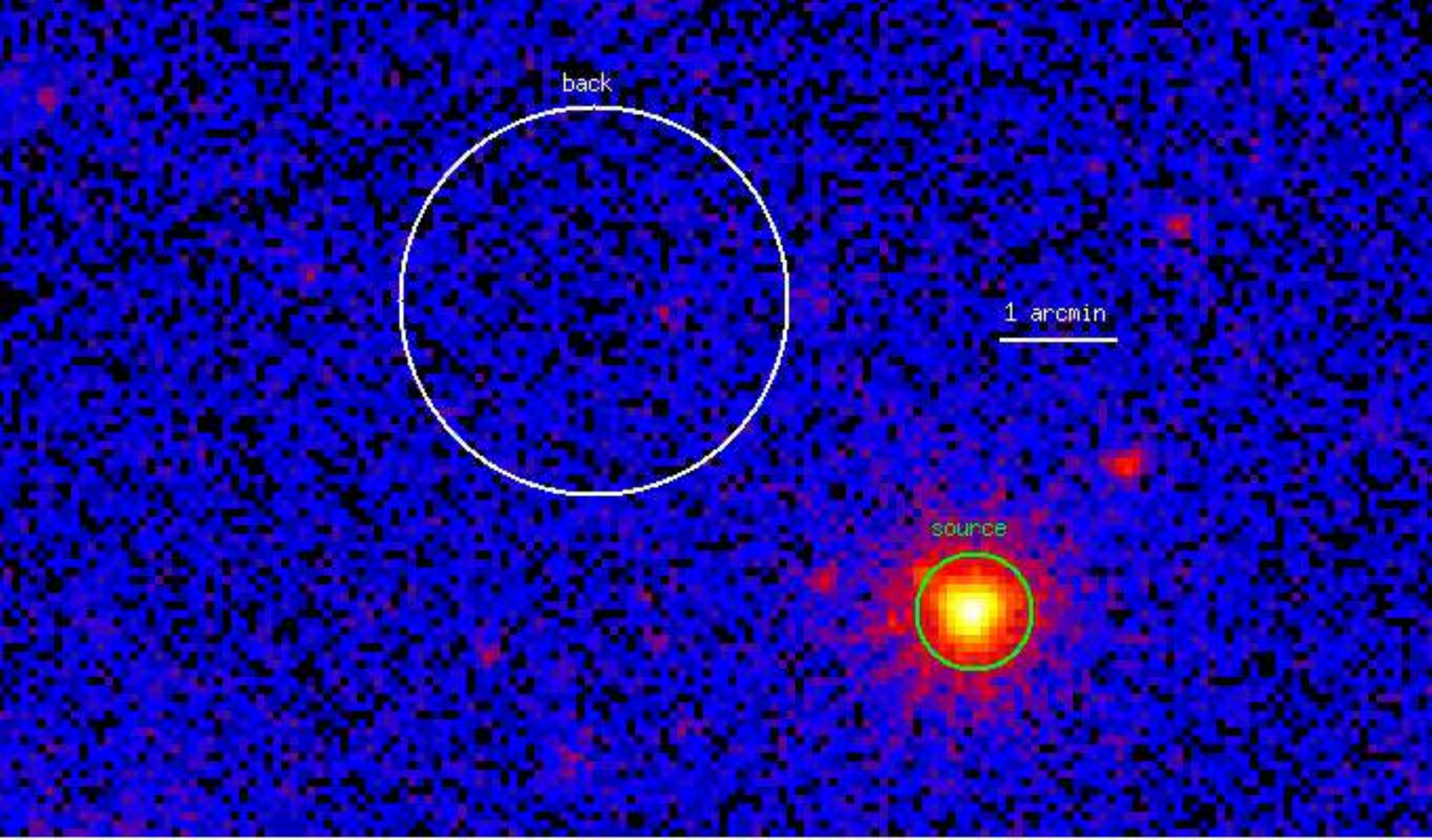}
\caption{PSR J1057-5226 0.3-10 keV MOS Imaging. The two MOS images have been added. 
The green circle marks the pulsar while the yellow one the background region used in the analysis.
\label{J1057-im}}
\end{figure}

\begin{figure}
\centering
\includegraphics[angle=0,scale=.50]{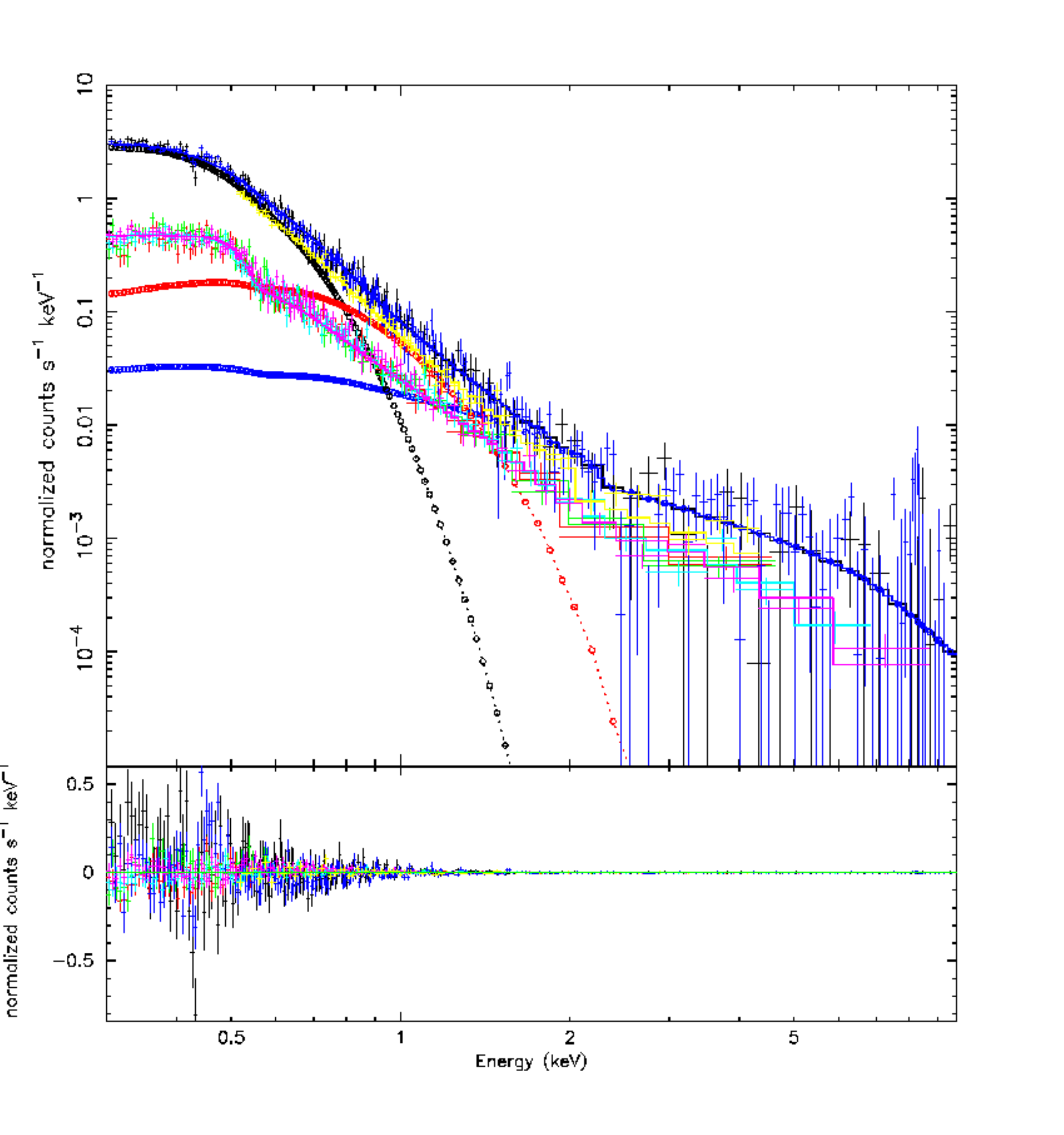}
\caption{PSR J1057-5226 {\it XMM-Newton} Epic PN Spectrum. The best-fitting spectral model is represented by the light blue line. This is obtained by the
sum of a cool blackbody component (green), a hot blackbody component (red), and a power law (blue). The inset shows confidence 
contours for the interstellar column density N$_H$ vs. the emitting surface for the cool blackbody. The 68\%, 90\%, and 99\%
confidence levels for two parameters of interest are plotted. See De Luca et al. 2005 for details.
Residuals are shown in the lower panel.
\label{J1057-sp}}
\end{figure}

\clearpage

{\bf J1119-6127 - type 2 RLP} % Nuova! osservazioni in arrivo

% Safi-Harb \& Kumar 2008
PSR J1119-6127 was
discovered in the Parkes multi-beam pulsar survey (Camilo et al. 2000)
and lies close to the center of
the 15$'$-diameter supernova remnant (SNR) G292.2-0.5 (Crawford et al. 2001; Pivovaroff et al. 2001). HI absorption
measurements imply a kinematic distance for the
remnant of 8.4 $\pm$ 0.4 kpc (Caswell et al. 2004), in agreement
with estimates made from its location in the Carina
spiral arm (Camilo et al. 2000). It has a rotation
period P of 408 ms, a large period derivative $\dot{P}$ of 4 $\times$ 10$^{-12}$, a characteristic age of
1600 yrs and a spin-down luminosity $\dot{E}$ of
2.3 $\times$ 10$^{36}$ ergs s$^{-1}$. No radio PWN was detected around the
pulsar despite its youth (Crawford et al. 2001). The pulsar was found within the supernova
remnant (SNR) G292.2-0.5 of 15' in diameter, discovered with the Australia Telescope
Compact Array (ATCA) (Crawford et al. 2001), and subsequently detected with X-ray
observations acquired with the ROSAT satellite and the Advanced Satellite for Cosmology
and Astrophysics, ASCA (Pivovaroff et al. 2001).
The X-ray counterpart to the radio pulsar was first resolved with {\it Chandra}, which also
revealed for the first time evidence for a compact (3$"$ $\times$ 6$"$) PWN (Gonzalez \& Safi-Harb,
2003). The pulsar was subsequently detected with {\it XMM-Newton} and was
found to exhibit thermal emission with an unusually high pulsed fraction of 74$\pm$14\% in the
0.5-2.0 keV range (Gonzalez et al. 2005). The {\it XMM-Newton}-derived temperature was also
unusually high, in comparison with other pulsars of similar age or even younger, like the
pulsar in the SNR 3C 58 (Slane et al. 2002).

\begin{figure}
\centering
\includegraphics[angle=0,scale=.20]{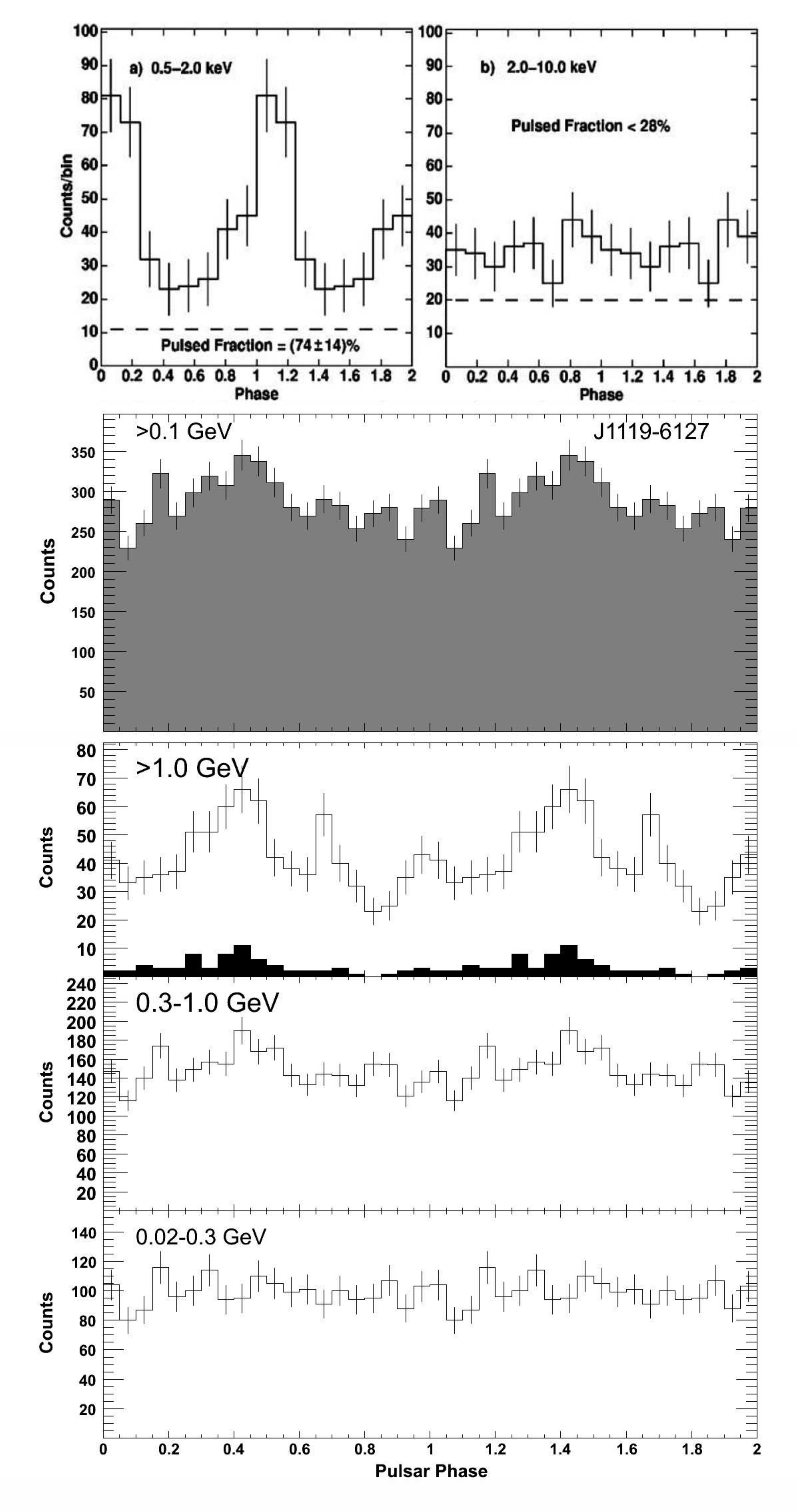}
\caption{PSR J1119-6127 Lightcurve. {\it Upper Panel: XMM-Newton} pulse profiles of PSR J1119-6127 in the 0.5-2 keV (left)
and 2-10 keV (right) ranges. Errors bars are 1$\sigma$ and
two cycles are shown. The peak of the radio pulse is at phase 0. 
The dashed lines represent our estimates for the contribution from the
pulsar's surroundings. See Gonzalez et al. 2008 for details.
{\it Lower Panel: Fermi} $\gamma$-ray lightcurve folded with Radio
(Abdo et al. 2009c).
\label{J1119-lc}}
\end{figure}

Four X-ray observations of J1119-6127 were performed, both by {\it XMM-Newton}
and {\it Chandra}:\\
- obs. id 2833, ACIS-S very faint mode, start time 2000, March 31 at 07:56:13 UT, exposure 57.4 ks;\\
- obs. id 4676, ACIS-S very faint mode, start time 2004, October 31 at 08:37:02 UT, exposure 61.3 ks;\\
- obs. id 6153, ACIS-S very faint mode, start time 2004, November 02 at 19:52:14 UT, exposure 19.1 ks;\\
- obs. id 0150790101, {\it XMM-Newton} observation, start time 2003, June 26 at 05:14:30 UT, exposure 60.7 ks.\\
In the {\it XMM-Newton} observation the PN camera of the EPIC
instrument was operated in Large Window mode (time resolution of $\sim$ 0.9s over a 14'$\times$26' field of view),
while the MOS detectors were set in Full frame mode. For
all three instruments, the medium optical filter was used.
The off-axis angle is negligible in all the observations.
First, an accurate
screening for soft proton flare events was done in the {\it XMM-Newton} observation obtaining a resulting
exposure of 44.8 ks.
The X-ray source best fit position, obtained by using the CIAO celldetect tool,
is 11:19:14.28 -61:27:49.30 (0.7$"$ error radius).
A compact PWN was detected by Gonzalez \& Safi-Harb,
2003. The tail extends up to $\sim$ 30$"$ from the pulsar even if
it's very faint in the last 20$"$.
We chose a 2$"$ radius circular region for the
{\it Chandra} pulsar spectrum and an ad-hoc elliptical region for the nebular emission.
The background spectrum was taken from a nearby source-free region.
The {\it XMM-Newton} source spectrum was extracted from a 20$"$ circular region around the pulsar,
including both the pulsar and most of the nebular counts. The XMM
background spectrum was taken from a source-free region nearby the pulsar.
The spectra obtained in all the {\it Chandra} observations were added using
mathpha tool while were combined the response
matrix and effective area files using addarf and addrmf. 
we obtained a total of 736 pulsar and 569 nebular counts (background
contributions of 0.5\% and 63.5\%) from the {\it Chandra} observations
and 717, 253 and 202 source counts (background
contributions of 24.5\%, 18.1\% and 18.5\%) from the three cameras of the {\it XMM-Newton} satellite.
The best fitting pulsar model is a combination of a powerlaw
and a blackbody (probability of obtaining the data if the model is correct 
- p-value - of 0.72, 86 dof fitting both the pulsar
and the nebula). Both a simple blackbody and a simple powerlaw spectra give
no acceptable fits.
The powerlaw component has a photon index $\Gamma$ = 1.74$_{-0.24}^{+0.45}$, 
absorbed by a column N$_H$ = 1.85$_{-0.38}^{+0.42}$ $\times$ 10$^{22}$ cm$^{-2}$.
The thermal component has a temperature of 2.15 $\pm$ 0.33 $\times$ 10$^6$ K
and an emitting radius of R$_{8.4kpc}$ = 4.96$_{-0.61}^{+5.08}$ km.
The nebular emission has a photon index $\Gamma$ = 1.55 $\pm$ 0.54.
Assuming the best fit model, the 0.3-10 keV unabsorbed thermal flux is 
4.41 $\pm$ 0.63 $\times$ 10$^{-13}$, the non-thermal pulsar flux is
1.48 $\pm$ 0.21 $\times$ 10$^{-13}$ and the non-thermal nebular flux is
6.01 $\pm$ 1.94 $\times$ 10$^{-14}$ erg/cm$^2$ s. 
Using a distance
of 8.4 kpc, the luminosities are L$_{8.4kpc}^{bol}$ = 3.72 $\pm$ 0.54 $\times$ 10$^{33}$,
L$_{8.4kpc}^{nt}$ = 1.25 $\pm$ 0.18 $\times$ 10$^{33}$ and L$_{8.4kpc}^{pwn}$ = 5.09 $\pm$ 1.66 $\times$ 10$^{32}$ erg/s.

\begin{figure}
\centering
\includegraphics[angle=0,scale=.50]{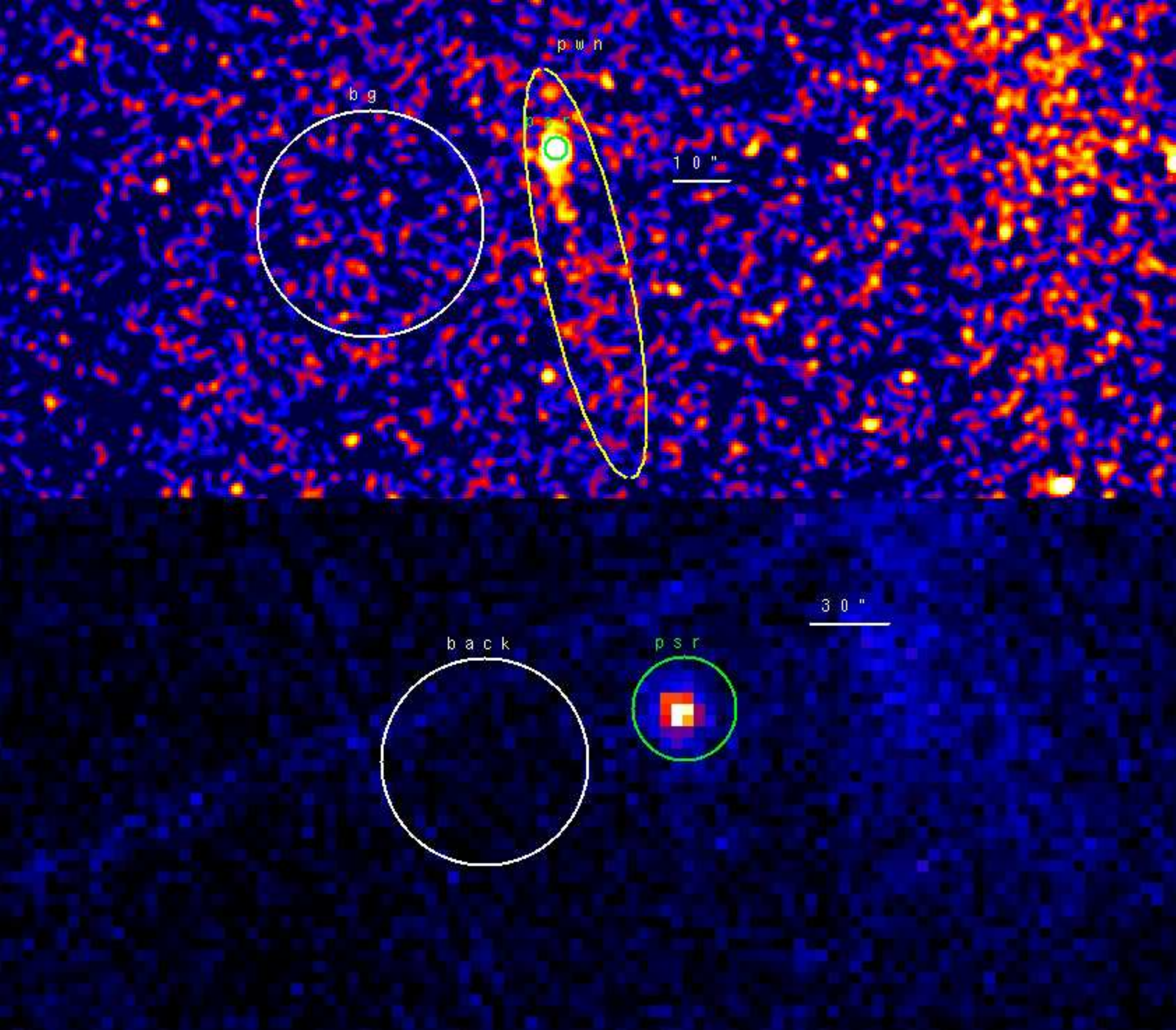}
\caption{{\it Upper Panel:} PSR J1119-6127 0.3-10 keV {\it Chandra} Imaging. The image has been smoothed with a Gaussian
with Kernel radius of $2"$. The green circle marks the pulsar, the yellow ellipse the nebular region and the
white circle the background region used in the analysis.
{\it Lower Panel:} PSR J1119-6127 0.3-10 keV {\it XMM-Newton} EPIC Imaging. The PN and the two MOS images have been added. 
The green circle marks the pulsar while the white annulus the background region used in the analysis.
\label{J1119-im}}
\end{figure}

\begin{figure}
\centering
\includegraphics[angle=0,scale=.50]{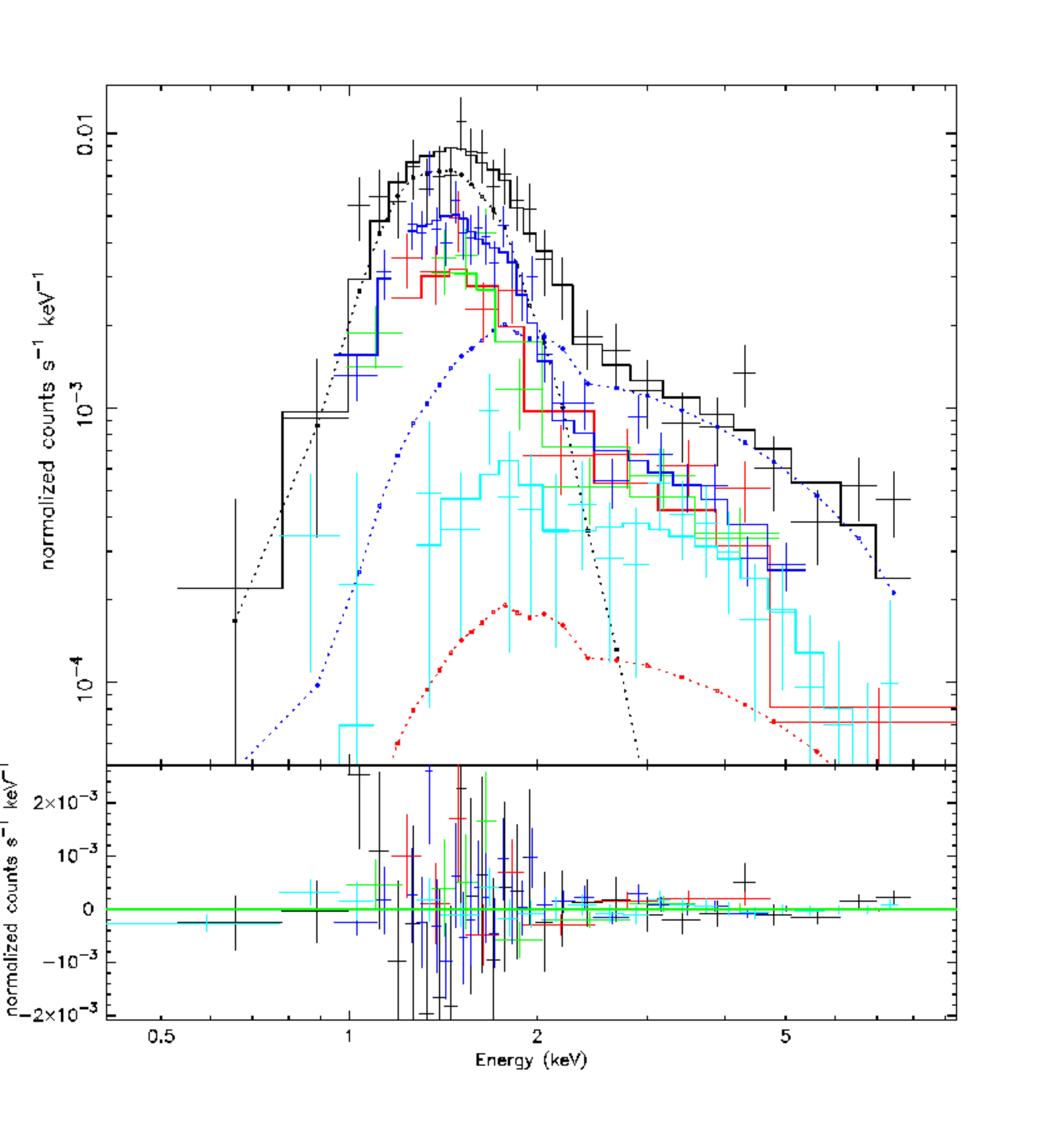}
\caption{PSR J1119-6127 Spectrum. Different colors mark all the different dataset used (see text for details).
Blue points mark the powerlaw component, black points the thermal component of the pulsar spectrum
and red points mark the nebular spectrum.
Residuals are shown in the lower panel.
\label{J1119-sp}}
\end{figure}

\clearpage

{\bf J1124-5916 - type 2 RLP} % ho rifatto lo spettro e trovato il termico

% camilo et al. 2002
J1124-5916 is situated inside the bright supernova remnant
G292.0+1.8. At radio wavelengths,
G292.0+1.8 has the appearance of a composite SNR,
with a central peak and a shell $\sim$ 10' in diameter (Braun et al.
1986). ASCA X-ray observations (Torii, Tsunemi \& Slane
1998) detected a nonthermal nebula coincident with the central
radio component. Recent {\it Chandra} ACIS-S observations have
shown this nebula to be $\sim$ 2' in extent with a resolved
compact source located near its peak. This discovery, together
with the energetics of the nebula, provides nearly incontrovertible
evidence for the existence of a pulsar powering the
nebula (Hughes et al. 2001).
Camilo et al. 2001 found the radio pulsation of the candidate
neutron star using the Parkes telescope and definitively identified
it as a pulsar. Hughes et al. 2001 found the X-ray pulsation
by using the High Resolution Camera on the {\it Chandra} X-Ray Observatory.
Further measurements based on the column density
fitted for the entire SNR in both optical and X-ray bands yield
a distance of 4.8$_{-1.2}^{+0.7}$ kpc (Gonzalez et al. 2004).
No $\gamma$-ray nebular emission was detected by {\it Fermi} down to a flux of 
1.34 $\times$ 10$^{-11}$ erg/cm$^2$s (Ackermann et al. 2010).

\begin{figure}
\centering
\includegraphics[angle=0,scale=.30]{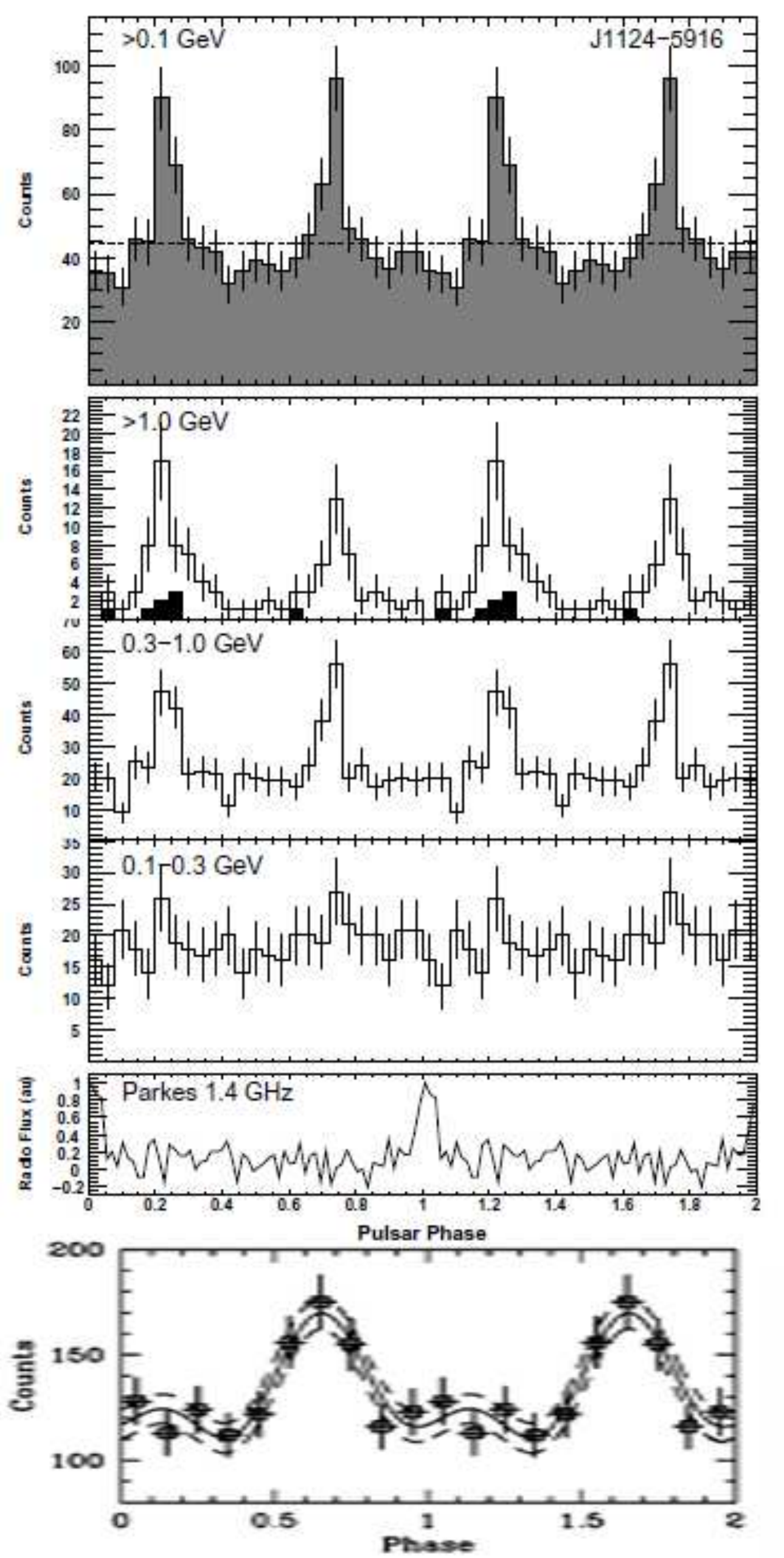}
\caption{PSR J1124-5916 Lightcurve.{\it Upper Panel: Fermi} $\gamma$-ray lightcurve folded with Radio
(Abdo et al. 2009c). {\it Lower Panel: Chandra} pulse-phase light curve for PSR J1124-5916 folded modulo the
best-fit period. Two complete periods are shown. Note the
suppressed zero on the y-axis. Also plotted are the Fourier series estimator
(de Jager et al. 1986) of the light curve (solid curve) and its 1$\sigma$ uncertainty
(dashed curves). See Huges et al. 2003 for Details.
\label{J1124-lc}}
\end{figure}

Many X-ray observations of J1124-5916 were performed, both by {\it XMM-Newton}
and {\it Chandra}. Due to the presence of a PWN and a SNR {\it XMM-Newton} spatial resolution is not enough to disentangle
the three different emissions. This is the reason for which we used only the following
{\it Chandra} ACIS observations:\\
- obs. id 126, ACIS-S faint mode, start time 2000, March 11 at 00:05:46 UT, exposure 44.2 ks;\\
- obs. id 6677, ACIS-I very faint mode, start time 2006, October 16 at 17:52:21, exposure 161.2 ks;\\
- obs. id 6678, ACIS-I very faint mode, start time 2006, October 02 at 05:39:04 UT, exposure 44.3 ks;\\
- obs. id 6679, ACIS-I very faint mode, start time 2006, October 03 at 14:05:44 UT, exposure 156.0 ks;\\
- obs. id 6680, ACIS-I very faint mode, start time 2006, September 13 at 15:42:26 UT, exposure 40.0 ks;\\
- obs. id 8221, ACIS-I very faint mode, start time 2006, October 20 at 12:12:56 UT, exposure 65.8 ks;\\
- obs. id 8447, ACIS-I very faint mode, start time 2006, October 07 at 11:21:49 UT, exposure 48.3 ks.\\
The off-axis angle is negligible in all the observations.
The X-ray source best fit position is 11:24:39.16 -59:16:19.51 (0.5$"$ error radius).
Both the 4' radius thermal emitting SNR and a bright 5$"$ radius PWN
(that can be disentangled from the remnant due to its hard spectrum) are
superimposed to the pulsar. A fainter non-thermal halo extends up to $\sim$2' from the pulsar.
We chose a 2$"$ radius circular region for the
pulsar spectrum and an annulus with radii 2 and 5$"$ to obtain the
inner, brighter nebular spectrum. In the pulsar region the thermal contribution coming from the
remnant results to be negligible; however, the pwn spectrum shows a significative
thermal contribute for energies below $\sim$ 1.5 keV. For this reason, in the
nebular spectrum I've rejected all the events below such an energy.
The background was taken from a region far away the pulsar, outside the remnant.
The spectra obtained in all the ACIS-I observations were added using
mathpha tool and the response
matrix and effective area files using addarf and addrmf. 
We obtained a total of 33017 and 4497 pulsar counts from
ACIS-I and ACIS-S spectra (background
contributions of 1.6\% and 12.1\%) and
9927 plus 1126 nebular counts (background
contributions of 8.1\% and 18.4\%).
The best fitting pulsar model is a combination of a powerlaw
and a blackbody (probability of obtaining the data if the model is correct 
- p-value - of 0.74, 764 dof fitting both the pulsar
and the nebula). A simple blackbody model
yields no acceptable fit while an f-test performed comparing
a simple powerlaw with the composite spectrum gives a
chance probability of 1.7 $\times$ 10$^{-5}$, pointing
to a significative improvement by adding the blackbody component.
A composite powerlaw+blackbody spectrum can't fit the thermal component
seen in the lower energy band of the nebular spectrum and coming from the SNR so that
such a thermal component is not related to the remnant.
The powerlaw component has a photon index $\Gamma$ = 1.54$_{-0.17}^{+0.09}$ , 
absorbed by a column N$_H$ = 3.00$_{-0.48}^{+0.28}$ $\times$ 10$^{21}$ cm$^{-2}$.
The thermal component has a temperature of 4.95$_{-0.21}^{+0.40}$ $\times$ 10$^6$ K
and an emitting radius of R$_{4.8kpc}$ = 274$_{-77}^{+89}$ m.
Such a small emitting radius seems to suggest an hot spot thermal emission.
The nebula has a photon index of $\Gamma$ = 1.78$_{-0.05}^{+0.03}$.
Assuming the best fit model, the 0.3-10 keV unabsorbed hot spot thermal flux is 
1.12$_{-0.23}^{+0.14}$ $\times$ 10$^{-13}$, the non-thermal pulsar flux is
9.78$_{-1.03}^{+1.18}$ $\times$ 10$^{-13}$ and the non-thermal nebular flux is
5.17$_{-0.30}^{+0.24}$ $\times$ 10$^{-13}$ erg/cm$^2$ s. 
Using a distance
of 4.8 kpc, the luminosities are L$_{4.8kpc}^{bol}$ = 3.10$_{-0.64}^{+0.39}$ $\times$ 10$^{32}$,
L$_{4.8kpc}^{nt}$ = 2.70$_{-0.28}^{+0.33}$ $\times$ 10$^{33}$ and L$_{4.8kpc}^{pwn}$ = 1.43$_{-0.08}^{+0.07}$ $\times$ 10$^{32}$
erg/s.

\begin{figure}
\centering
\includegraphics[angle=0,scale=.40]{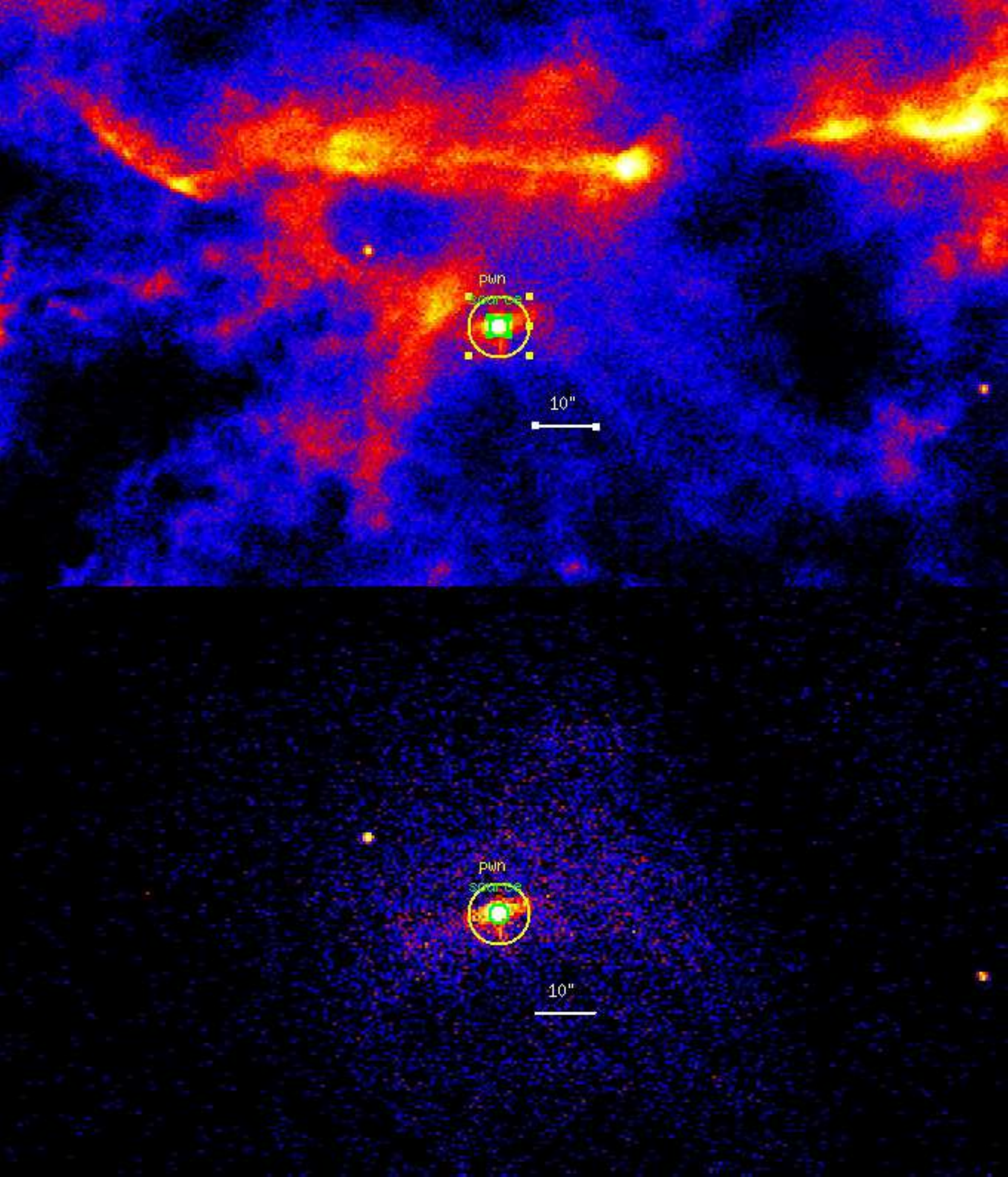}
\caption{PSR J1124-5916 {\it Chandra} Imaging. {\it Upper Panel:} 0.3-10 keV energy range
{\it Lower Panel:} 3-10 keV energy range.
The green circle marks the pulsar while the yellow annulus the nebular region used in the analysis.
\label{J1124-im}}
\end{figure}

\begin{figure}
\centering
\includegraphics[angle=0,scale=.50]{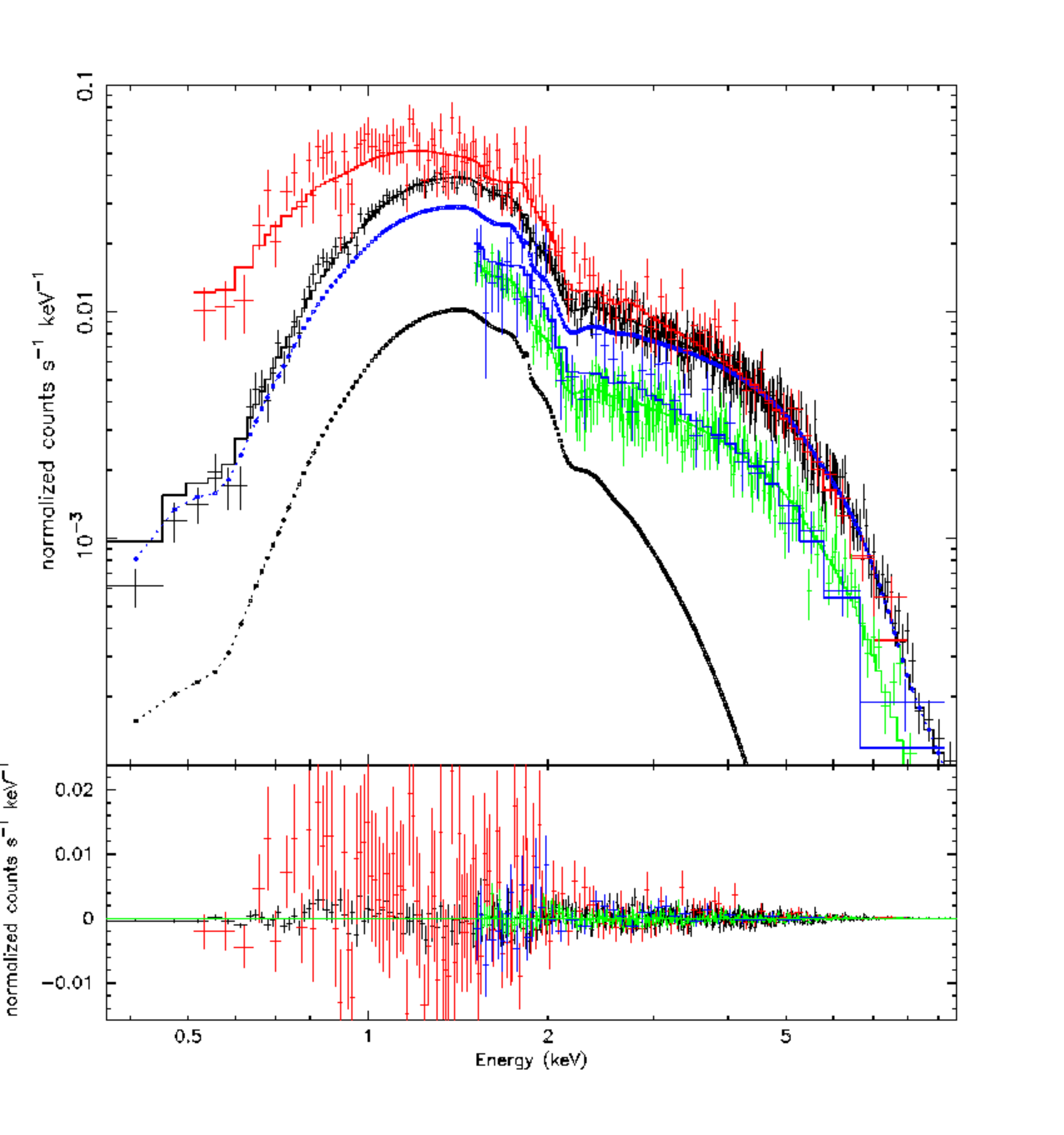}
\caption{PSR J1124-5916 Spectrum. Different colors mark all the different dataset used (see text for details).
Blue points mark the powerlaw component, black points the thermal component of the pulsar spectrum. Red points
mark the nebular spectrum.
Residuals are shown in the lower panel.
\label{J1124-sp}}
\end{figure}

\clearpage

{\bf J1135-6055 - type 2 RQP} % Nuova!

Pulsations from J1135-6055 were detected by {\it Fermi} in the last months using the
blind search technique. Such a pulsar was announced during the 3rd {\it Fermi} Symposium in Rome.
The pseudo-distance of the object based on $\gamma$-ray data (Saz Parkinson et al. (2010))
is $\sim$ 2.88 kpc.

We analyzed the only {\it Chandra} observation covering on such a pulsar - obs. id 3924,
{\it Chandra} ACIS-S observation, very faint mode, start time 2003, August 24 at 04:53:56 UT, exposure 36.6 ks.
The pulsar position was imaged on the back-illuminated ACIS
S3 chip and the VFAINT exposure mode was adopted. The off-axis angle is negligible.
Many pointlike sources are present inside the {\it Fermi} Error Box but only one shows an apparent jets-like nebular emission.
Such source was indicated as the X-ray counterpart after a dedicated timing analysis by the {\it Fermi} LAT group.
By using the CIAO dedicated tools, the pulsar best fit position is 11:35:08.45 -60:55:36.99 (1.7$"$ error radius).
A nebular emission is apparent in the {\it Chandra} obtained image: two jets extending in the north and north-west direction
with a length of about 30 and 40$"$. A fainter emission could be present at the edge of the jet-like nebulae.
We chose a 2$"$ radius circular region for the
pulsar spectrum and an ad hoc double-ellipse as extraction region for the nebular emission.
The background spectrum was taken from a source-free circular region near the pulsar.
We obtained a total of 60 pulsar and 684 nebular counts from
ACIS-I and ACIS-S spectra (background
contributions of 3.1\% and 23.8\%).
Due to the low statistic, we used the C-statistic
approach implemented in XSPEC.
The best fitting pulsar model is a simple powerlaw 
(reduced chisquare value of $\chi^2_{red}$ = 0.94, 33 dof)
with a photon index $\Gamma$ = 1.15$_{-0.50}^{+0.52}$, 
absorbed by a column N$_H$ = 4.19$_{-1.52}^{+1.89}$ $\times$ 10$^{21}$ cm$^{-2}$.
A simple blackbody model yields no acceptable fit and a composite thermal plus nonthermal model
is not statistically compelling.
The nebular emission has a photon index $\Gamma$ = 1.84$_{-0.36}^{+0.41}$.
An analysis of the two jets led to similar photon indexes $\Gamma^{jet1}$ = 1.85$_{-0.25}^{+0.52}$
and $\Gamma^{jet2}$ = 2.69$_{-0.62}^{+0.77}$. A longer observation by {\it Chandra} would be able to
find spectral differences (if any) between the two jets.
Assuming the best fit model, the 0.3-10 keV unabsorbed pulsar flux is 
3.70$_{-3.21}^{+1.45}$ $\times$ 10$^{-14}$ and the non-thermal nebular flux is
1.87$_{-1.05}^{+0.39}$ $\times$ 10$^{-13}$ erg/cm$^2$ s.
Using a distance of 2.9 kpc, the luminosities are 
L$_{2.9kpc}^{nt}$ = 3.72$_{-3.25}^{+1.46}$ $\times$ 10$^{31}$ and L$_{2.9kpc}^{pwn}$ = 1.89$_{-1.06}^{+0.39}$ $\times$ 10$^{32}$ erg/s.

\begin{figure}
\centering
\includegraphics[angle=0,scale=.40]{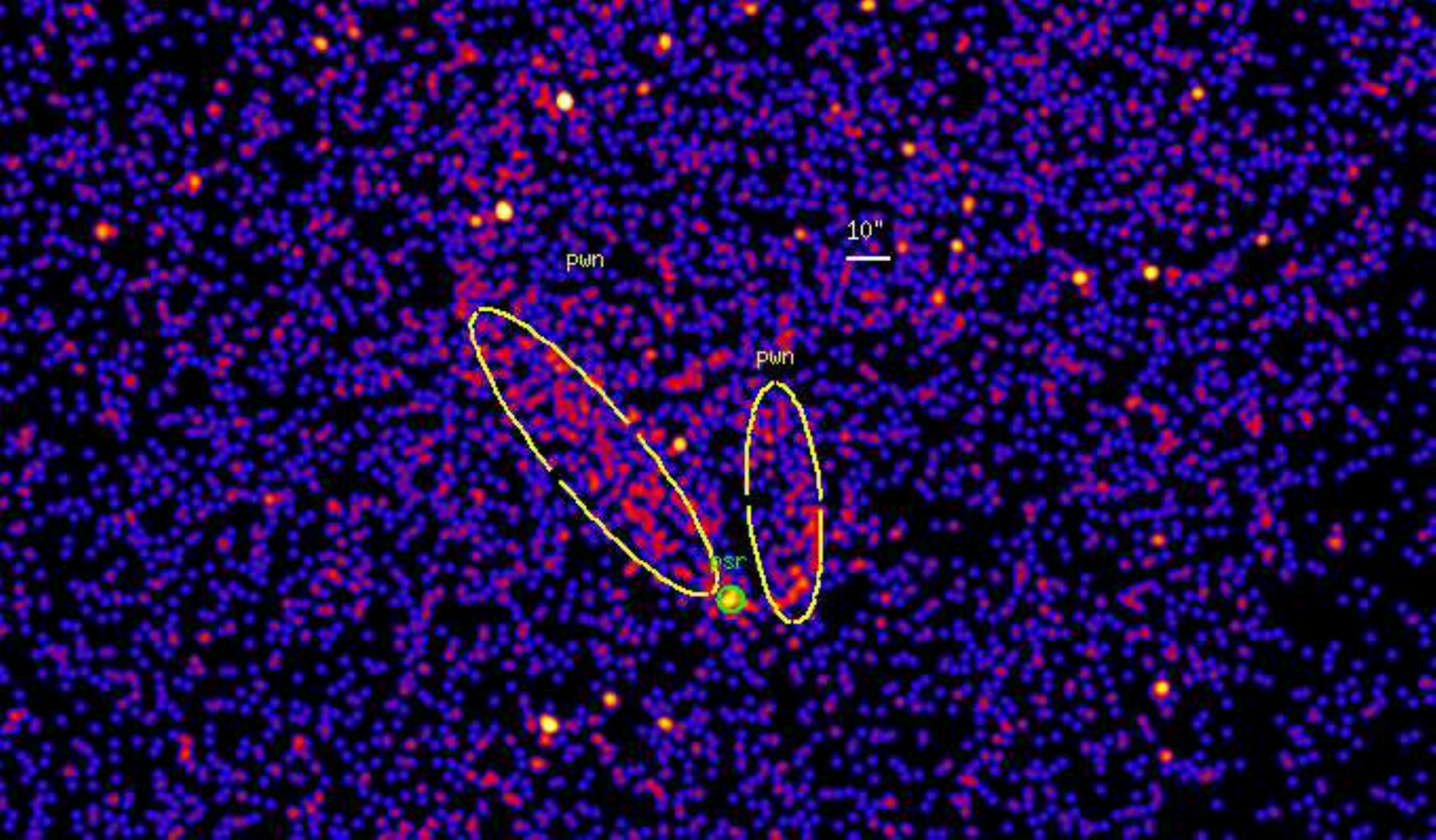}
\caption{PSR J1135-6055 0.3-10 keV {\it Chandra} Imaging. The green circle marks the pulsar
while the yellow ellipses the nebular region used in the analysis.
\label{J1135-im}}
\end{figure}

\begin{figure}
\centering
\includegraphics[angle=0,scale=.40]{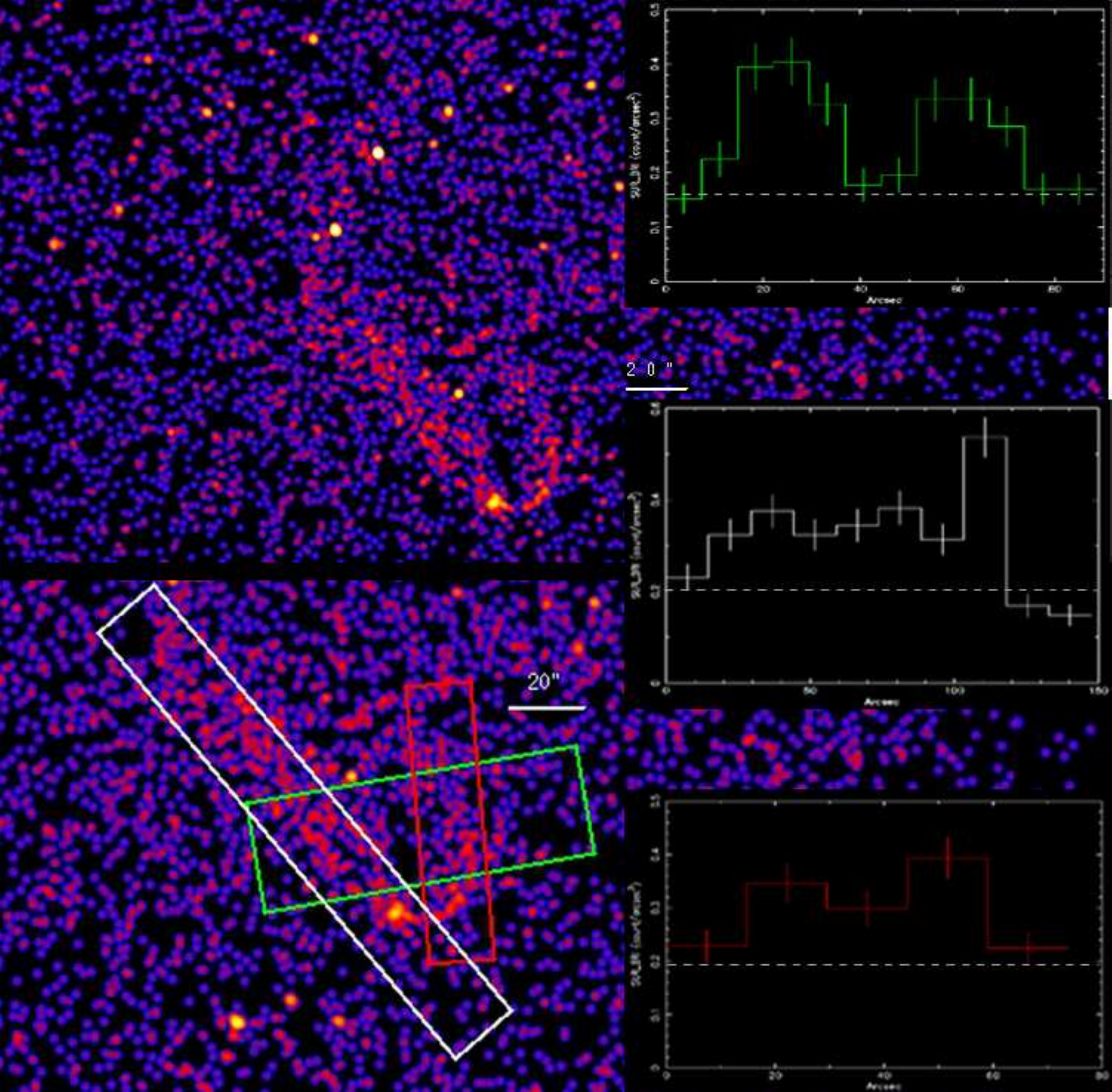}
\caption{Here are showed J1135-6055 and its jets in the 0.3-10 keV band as seen by Chandra. In the three
panels on the right are showed the linear brilliance profiles of the two jets (jet1 in white and jet2 in red)
and the regions from where they are taken. In green is reported the brilliance profile perpendicular to the
nebula: the two-jets structure of the extended emission is apparent.
\label{J1135-psf}}
\end{figure}

\begin{figure}
\centering
\includegraphics[angle=0,scale=.40]{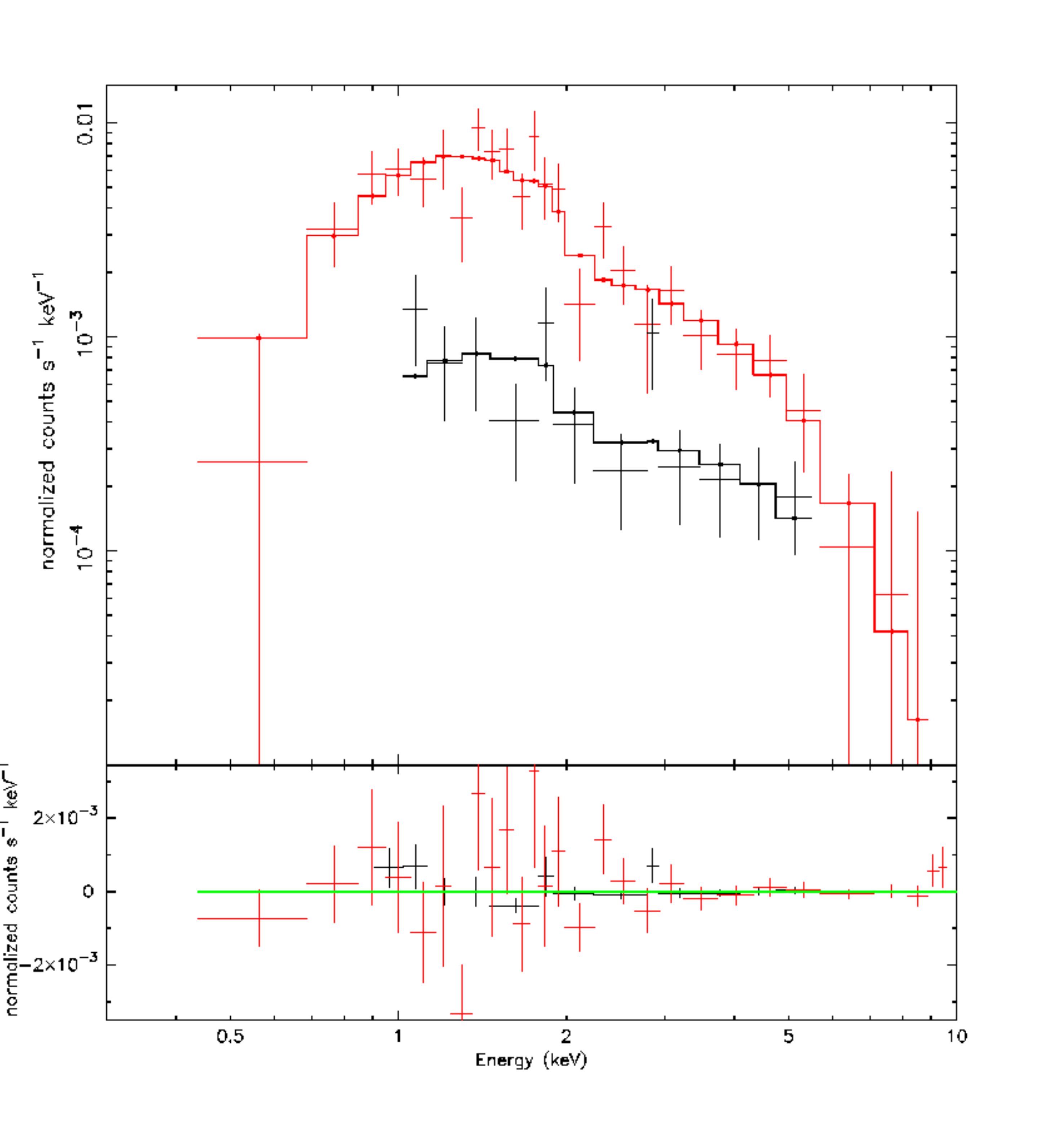}
\caption{PSR J1135-6055 {\it Chandra} Spectrum is marked in black while the nebular spectrum in red.
Residuals are shown in the lower panel.
\label{J1135-sp}}
\end{figure}

\clearpage

{\bf J1231-1411 - type 2 RLP} % Nuova!

% Ransom et al. 2010
J1231-1411 appears to be a nearby ($<$ 2kpc) pulsar, detected in Radio 
by Pulsar Search Consortium.
The Green Bank Telescope (GBT) was used to cover unassociated
sources from the {\it Fermi} LAT Bright Source List
(Abdo et al. 2009b, Ransom et al. 2010).
J1231-1411 has an orbital period of 1.9 days,
with companions of mass $\sim$ 0.2-0.3M$_s$.
Dispersion measurements place the source at $\sim$ 0.4 kpc.

J1231-1411 was observed during one {\it XMM-Newton} observation
on 2009, July 15 at 09:07:20 UT (exposure of 25.6 ks).
In the observation the PN camera of the EPIC
instrument was operated in Small Window mode, while the MOS detectors were set in Full frame mode. For
the PN camera, a thin optical filter was used while for the two MOS cameras the medium optical filter was used.
First, an accurate
screening for soft proton flare events was done in the {\it XMM-Newton} observation obtaining a resulting
exposure of 19.1 ks.
The X-ray source best fit position, obtained by using the XIMAGE and SAS dedicated tools,
is 12:31:11.56 -14:11:44.36 (5$"$ error radius).
We searched for diffuse emission in the immediate
surroundings of the pulsar in the {\it XMM-Newton} MOS cameras data. The resulting graph was fitted
using the prescriptions of the XMM calibration document CAL-TN-0052 in order to find the
Point Spread Function and to find any possible excess of counts. No X-ray nebular emission was detected.

\begin{figure}
\centering
\includegraphics[angle=0,scale=.40]{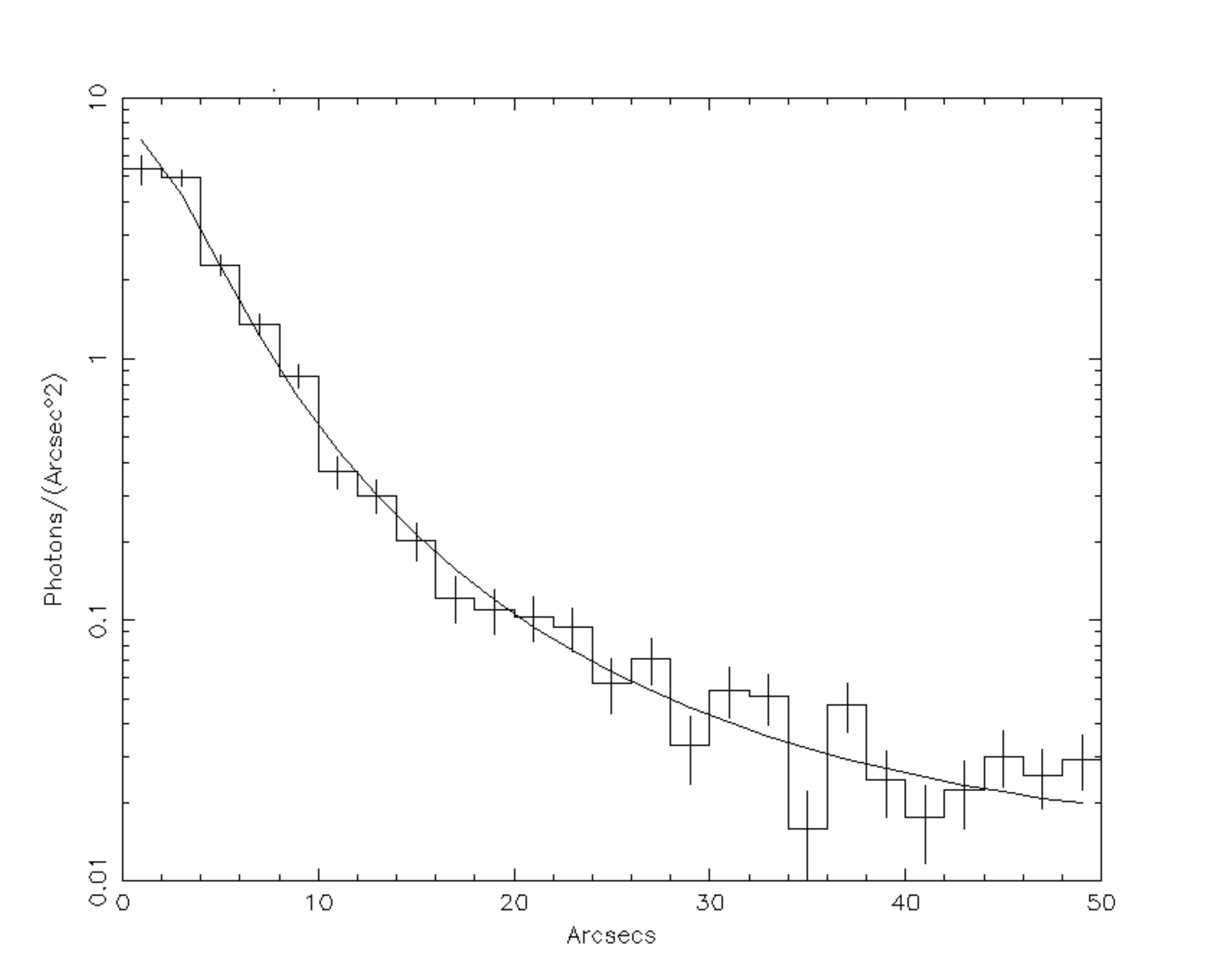}
\caption{PSR J1231-1411 {\it XMM-Newton} PN radial profile (0.3-10 keV energy range).
This was fitted by using the XMM calibration document CAL-TN-0052. No excess has been found.
\label{J1231-psf}}
\end{figure}

We chose a 20$"$ radius circular region for the
pulsar spectrum while the background spectrum 
was taken from a source-free region nearby the pulsar.
We obtained a total of 1691, 489 and 442 pulsar counts from the three
cameras of {\it XMM-Newton} (background
contributions of 5.0\%, 4.5\% and 4.3\%).
Even if a simple powerlaw model is statistically acceptable, Maeda et al. 2011
found a clear indication of a thermal component by using the SUZAKU telescope.
Our best fit powerlaw component has a photon index $\Gamma$ = 3.84$_{-0.50}^{+0.40}$, 
absorbed by a column N$_H$ = 1.13$\pm$ 0.51 $\times$ 10$^{21}$ cm$^{-2}$.
The thermal component has a temperature of 2.05$_{-0.56}^{+0.68}$ $\times$ 10$^6$ K
and an emitting radius of R$_{0.40kpc}$ = 79.7$_{-9.4}^{+7.8}$ m, typical of an hot
spot emission.
Assuming the best fit model, the 0.3-10 keV unabsorbed thermal flux is 
3.70$_{-0.80}^{1.60}$ $\times$ 10$^{-14}$ and the non-thermal pulsar flux is
4.12$_{-1.74}^{+0.88}$ $\times$ 10$^{-13}$ erg/cm$^2$ s. 
Using a distance
of 0.40 kpc, the luminosities are L$_{0.55kpc}^{bol}$ = 9.75$_{-2.11}^{+4.22}$ $\times$ 10$^{29}$ and
L$_{0.55kpc}^{nt}$ = 1.09$_{-0.46}^{0.23}$ $\times$ 10$^{31}$ erg/s.

\begin{figure}
\centering
\includegraphics[angle=0,scale=.50]{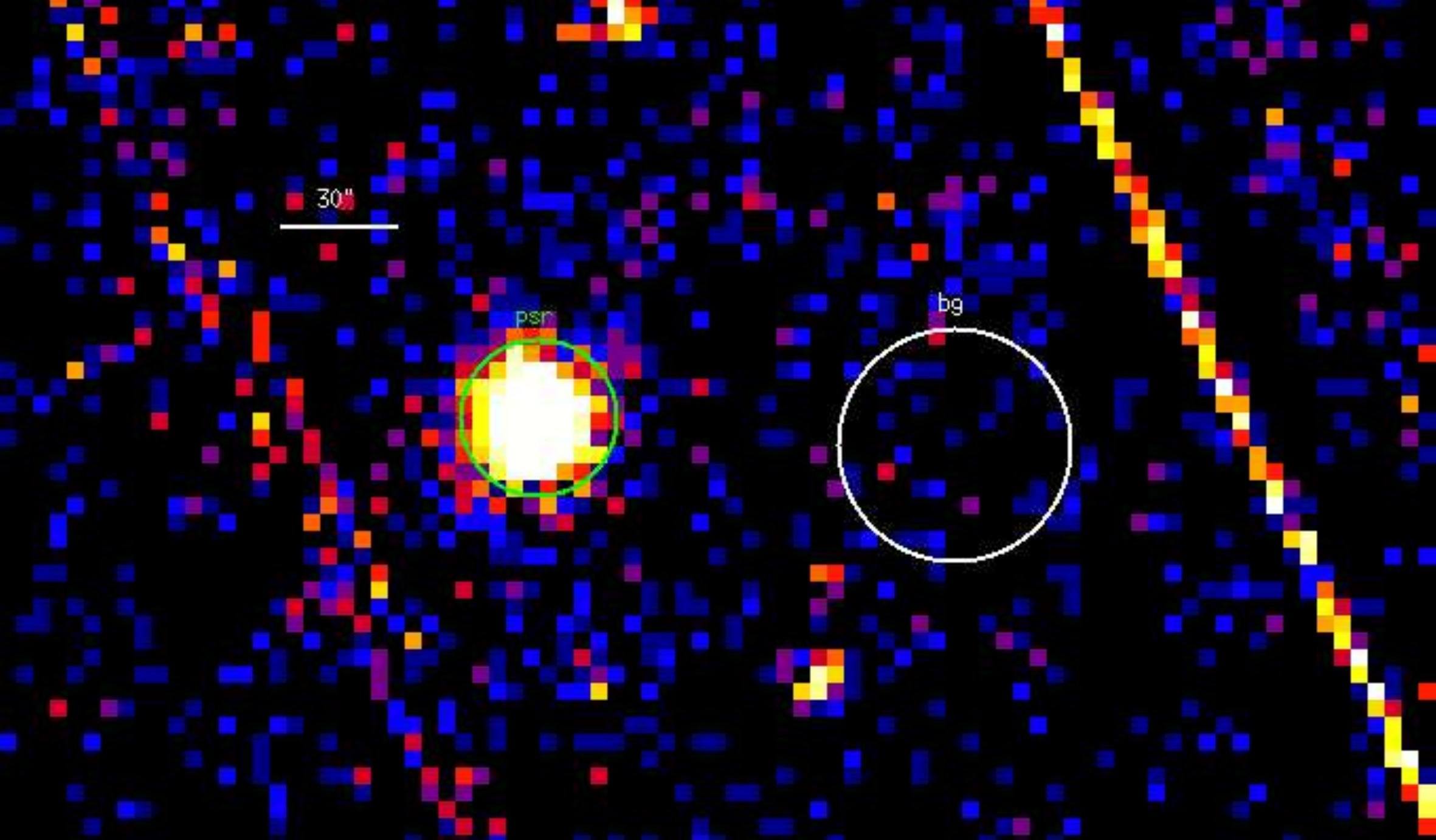}
\caption{PSR J1231-1411 0.3-10 keV {\it XMM-Newton} EPIC Imaging. The PN and two MOS images have been added. 
The green circle marks the pulsar while the white one the background region used in the analysis.
\label{J1231-im}}
\end{figure}

\begin{figure}
\centering
\includegraphics[angle=0,scale=.50]{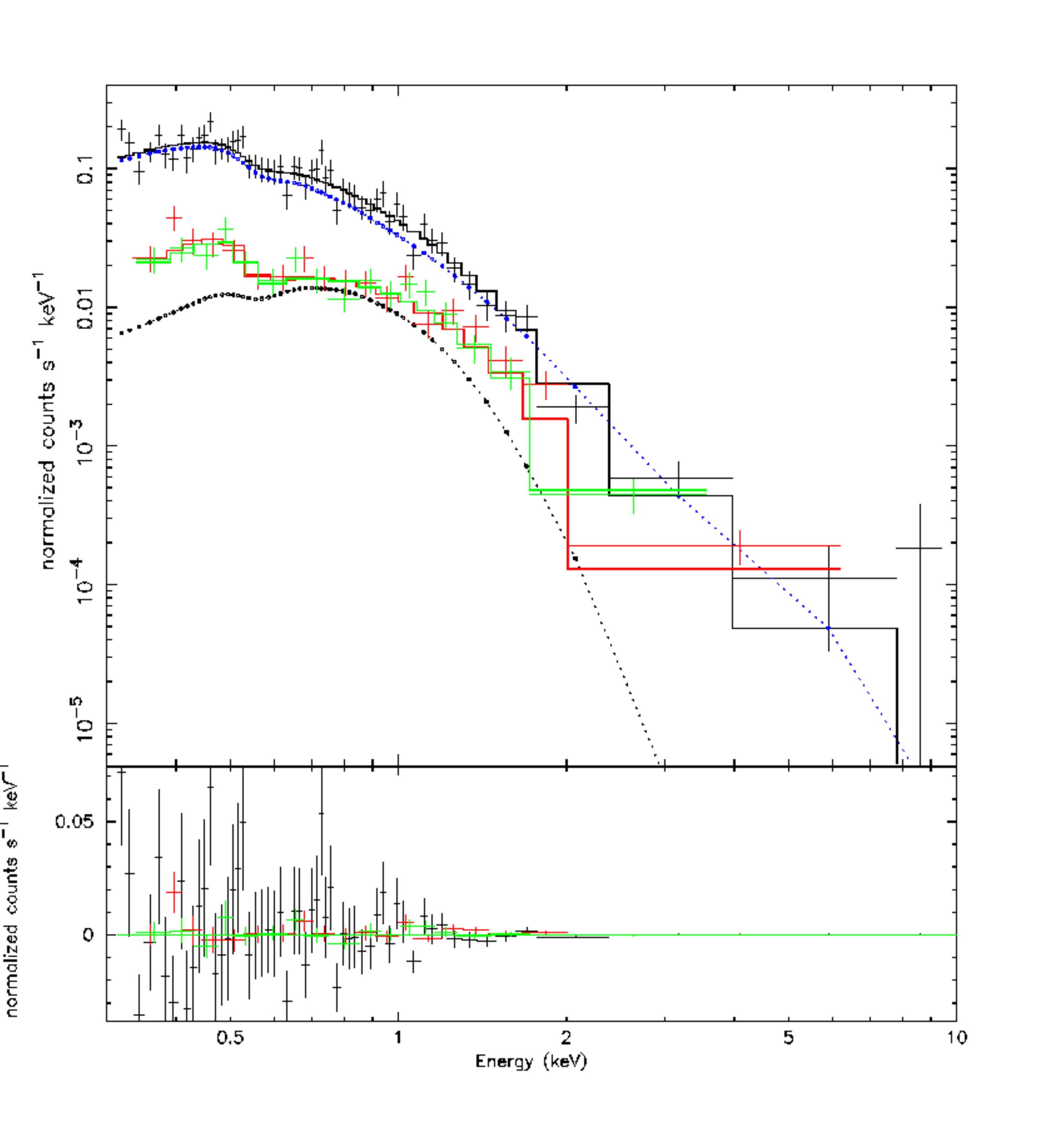}
\caption{PSR J1231-1411 Spectrum. Different colors mark all the different dataset used (see text for details).
Blue points mark the powerlaw component while black points the thermal component of the pulsar spectrum.
Residuals are shown in the lower panel.
\label{J1231-sp}}
\end{figure}

\clearpage

{\bf J1357-6429 - type 2 RLP} % Nuova!, osservazione in arrivo

% Abdo et al. 2011
PSR J1357-6429 is a very young and energetic pulsar discovered
during the Parkes multibeam survey of the Galactic
plane as reported in Camilo et al. (2004). From its dispersion
measure of (127.2 $\pm$ 0.5) cm$^{-3}$ pc, the NE2001 model of the
Galactic electron distribution (Taylor \& Cordes, 1993) assigns
PSR J1357-6429 a distance of 2.4 $\pm$ 0.6 kpc. With a spin
period of 166 ms, a period derivative of 3.6 $\times$ 10$^{-13}$ and
a characteristic age of 7300 yr, this pulsar may be, after the
Crab, the nearest very young pulsar known. No EGRET source
is coincident with PSR J1357-6429. Recently, Very Large
Telescope data was used to search for optical emission from
PSR J1357-6429 (Mignani et al., 2011). No counterpart was
found, and an upper limit of V $>$ 27.3 was determined.

\begin{figure}
\centering
\includegraphics[angle=0,scale=.30]{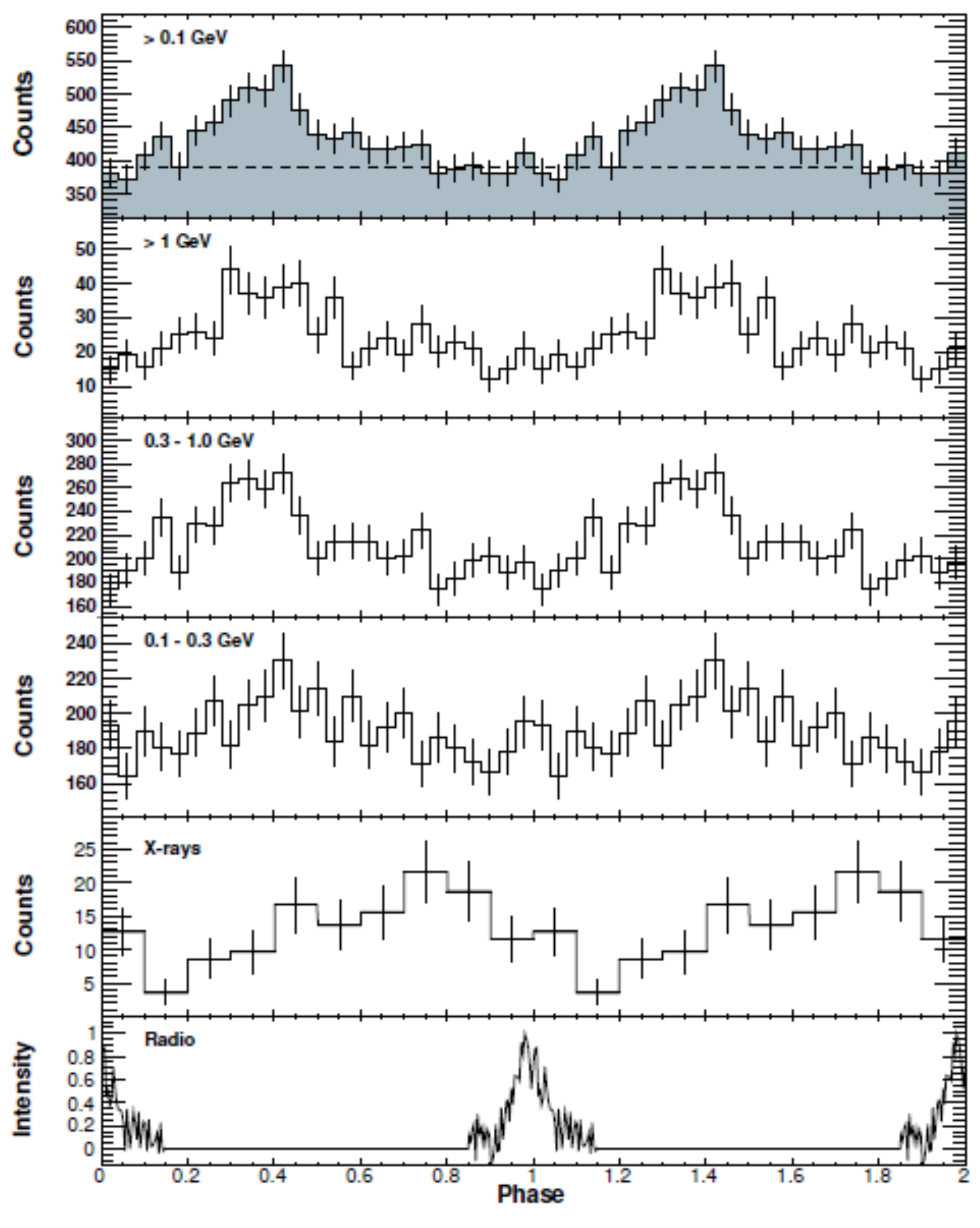}
\caption{{\it Top panel:} Phase-aligned histogram of PSR J1357-6429 above
0.1 GeV and within an energy-dependent circular region as defined
in section Abdo et al. in preparation. Two rotations are plotted with 25 bins per period. The
dashed line shows the background level, as estimated from a ring surrounding
the pulsar during the off-pulse phase interval. {\it Three following
panels:} Phase histograms for PSR J1357-6429 in the three indicated
energy ranges, each with 25 bins per pulse period. {\it Second
panel from bottom:} X-ray pulse profile extracted from {\it Chandra} HRCS
data (Zavlin, 2007). {\it Bottom panel:} Radio pulse profile based on
Parkes observations at a center frequency of 1.4 GHz with 256 phase
bins (Johnston \& Weisberg, 2006).
Figure taken from Abdo et al. in preparation.
\label{J1357-lc}}
\end{figure}

\begin{figure}
\centering
\includegraphics[angle=0,scale=.50]{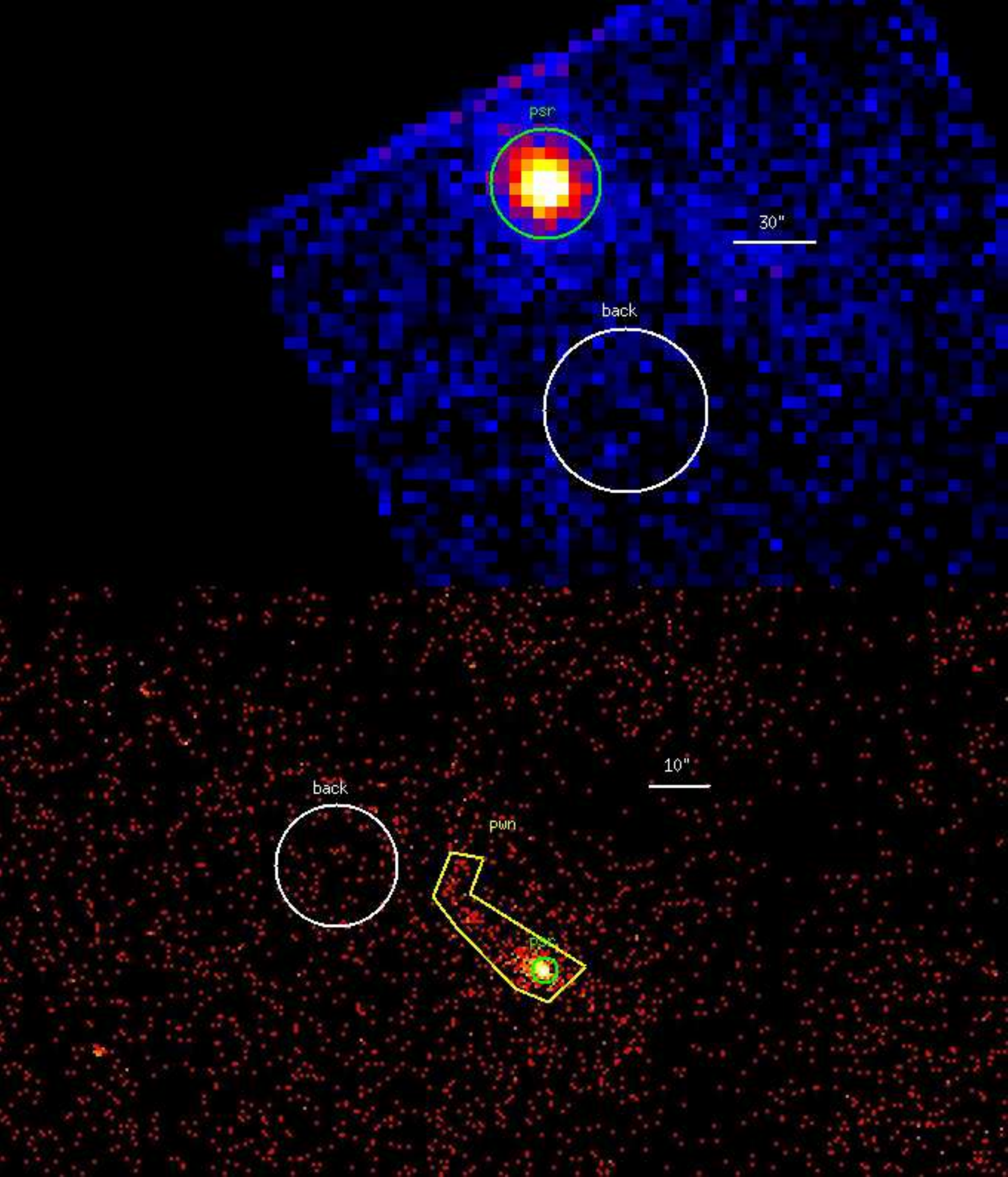}
\caption{{\it Upper Panel:} PSR J1357-6429 0.3-10 keV {\it XMM-Newton} EPIC Imaging. The PN and the two MOS images have been added. 
The green circle marks the source while the yellow annulus the background region used in the analysis.
{\it Lower Panel:} PSR J1357-6429 0.3-10 keV {\it Chandra} Imaging.
The green circle marks the pulsar while the yellow polygon the nebular region used in the analysis. The white circle marks the background region.
\label{J1357-im}}
\end{figure}

J1357-6429 was observed during two {\it XMM-Newton} observations and one {\it Chandra}:\\
- obs. id 0306910101, starting on 2005, August 05 at 07:38:55 UT, exposure 13.0 ks;\\
- obs. id 0603280101, starting on 2009, August 14 at 15:15:55 UT, exposure 79.3 ks;\\
- obs. id 0603280101, {\it Chandra} ACIS-I observation, vfaint mode, starting on 2009, October 08 at 06:56:33 UT, exposure 60.0 ks.\\
In the first observation the PN and MOS cameras were set in Full frame mode and a medium optical filter was used.
In the second observation the PN camera of the EPIC
instrument was operated in Small Window mode, while the MOS detectors were set in Full frame mode. For
both the instruments a medium optical filter was used.
No screening of soft proton flares was done owing to the goodness of the observations.
The X-ray source best fit position, obtained by using the celldetect tool inside the CIAO software,
is 13:57:02.52 -64:29:29.98 (1$"$ error radius).
A nebular emission is apparent in the {\it Chandra} observation around the pulsar plus
a faint trail-like emission extending until $\sim$ 20$"$ from it.
For the {\it Chandra} observation we chose a circular 2$"$ radius region for the pulsar
spectrum. The nebular emission in such a region is not negligible so that
this spectrum will be fitted with a two-powerlaw model in order to take in
account both the nebular and pulsar flux. The nebular spectrum has been extracted
from an ad-hoc region (see Figure \ref{J1357-im}). The background spectrum
was extracted from a circular region on the same CCD.
We chose a 20$"$ radius circular region for the
{\it XMM-Newton} pulsar plus nebula spectrum while the background spectrum 
was taken from a source-free region nearby the pulsar.
We obtained a total of 1050, 225 and 227 pulsar counts from the three
cameras of {\it XMM-Newton} in the first observation (background
contributions of 52.9\%, 31.6\% and 27.8\%).
We obtained a total of 3620, 1254 and 1173 pulsar counts from the three
cameras of {\it XMM-Newton} in the second observation (background
contributions of 34.7\%, 23.2\% and 25.1\%).
We also obtained 448 and 255 counts (background contributions of 0.4\% and 13.9\%) from the considered regions.
The best fit spectrum is a combination of a blackbody and a powerlaw
(probability of obtaining the data if the model is correct 
- p-value - of 0.69, 303 dof using both the pulsar and nebula)
with a photon index $\Gamma$ = 1.48 $\pm$ 0.30, 
absorbed by a column N$_H$ = 1.89$_{-0.45}^{+0.48}$ $\times$ 10$^{21}$ cm$^{-2}$.
The thermal component has a temperature of 2.27$_{-0.30}^{+0.32}$ $\times$ 10$^6$ K
and an emitting radius of R$_{2.4kpc}$ = 478$_{-45}^{+914}$ m, typical of an hot
spot emission.
The nebular photon index is $\Gamma$ = 1.30$_{-0.11}^{+0.21}$.
A simple powerlaw model is statistically acceptable but
an f-test performed comparing
a simple powerlaw with the composite spectrum gives a
chance probability of less than 10$^{-10}$, pointing
to a significative improvement by adding the blackbody component.
Assuming the best fit model, the 0.3-10 keV unabsorbed thermal flux is 
5.66 $\pm$ 2.09 $\times$ 10$^{-14}$, the non-thermal pulsar flux is
4.19 $\pm$ 1.56 $\times$ 10$^{-14}$ and the nebular flux 3.81$_{-0.52}^{+0.38}$ $\times$ 10$^{-13}$ erg/cm$^2$ s. 
Using a distance
of 2.4 kpc, the luminosities are L$_{2.4kpc}^{bol}$ = 3.91 $\pm$ 1.44 $\times$ 10$^{31}$,
L$_{2.4kpc}^{nt}$ = 2.90 $\pm$ 1.08 $\times$ 10$^{31}$ and L$_{2.4kpc}^{pwn}$ = 2.63$_{-0.36}^{+0.26}$ $\times$ 10$^{32}$ erg/s.

\begin{figure}
\centering
\includegraphics[angle=0,scale=.50]{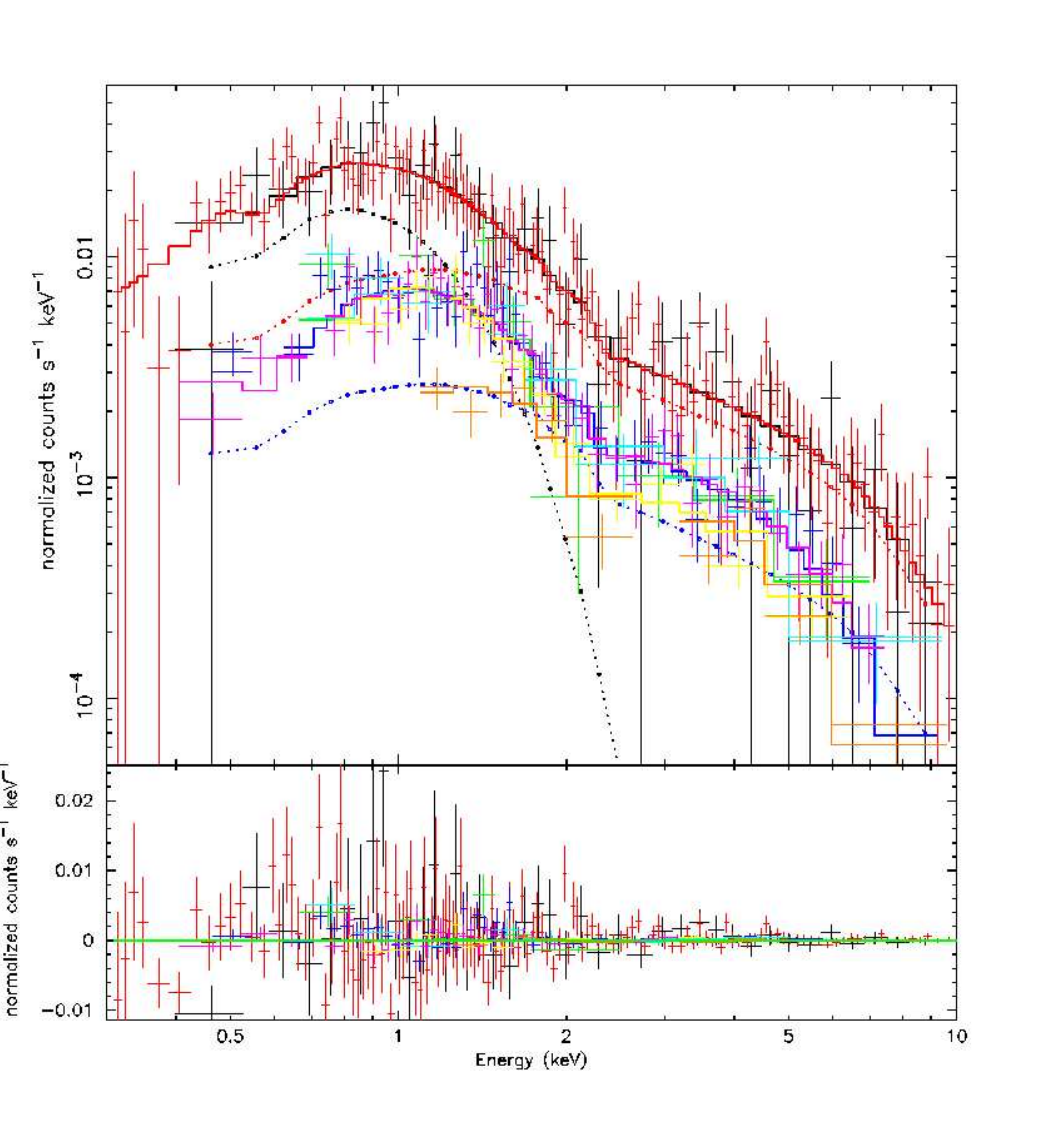}
\caption{PSR J1357-6429 Spectrum. Different colors mark all the different dataset used (see text for details).
Blue points mark the powerlaw component while black points the thermal component of the pulsar spectrum.
Red points mark the nebular spectrum.
Residuals are shown in the lower panel.
\label{J1357-sp}}
\end{figure}

\clearpage

{\bf J1410-6132 - type 0 RLP} % Nuova!

% O'Brien et al. 2008
J1410-6132 is a 50-ms pulsar found by O'Brien et al. 2008 during an high frequency
survey of the Galactic plane, using a 7-beam 6.3-GHz receiver on the 64-m
Parkes radio telescope.
The pulsar lies within the error box of the unidentified EGRET
source 3EG J1410-6147, has a characteristic age of 26 kyr and a spin-down energy
of 10$^{37}$ erg s$^{-1}$. It has a very high dispersion measure of 960 $\pm$ 10 cm$^{-3}$ pc and one of the
largest rotation measure of any pulsar, RM=2400 $\pm$ 30 rad m$^{-2}$, pointing to a distance
of 15.6 kpc. The pulsar is very
scatter-broadened at frequencies of 1.4 GHz and below, making pulsed emission almost
impossible to detect.

No X-ray observations were carried on the pulsar position.

{\bf J1413-6205 - type 0 RQP} % osservazione in arrivo a maggio

Pulsations from J1413-6205 were detected by {\it Fermi} using the
blind search technique (Saz Parkinson et al. 2010).
No $\gamma$-ray nebular emission was detected down to a flux of 
5.34 $\times$ 10$^{-12}$ erg/cm$^2$s (Ackermann et al. 2010).
The pseudo-distance of the object based on $\gamma$-ray data (Saz Parkinson et al. (2010))
is $\sim$ 1.4 kpc.

After the {\it Fermi} detection, we asked for a {\it SWIFT} observation
of the $\gamma$-ray error box (obs id. 00031410001-00031410002, 4.30 ks exposure).
After the data reduction, no X-ray source was found inside
the {\it Fermi} error box.
For a distance of 1.4 kpc we found a
rough absorption column value of 4 $\times$ 10$^{21}$ cm$^{-2}$
and using a simple powerlaw spectrum
for PSR+PWN with $\Gamma$ = 2 and a signal-to-noise of 3,
we obtained an upper limit non-thermal unabsorbed flux of 4.9 $\times$ 10$^{-13}$ erg/cm$^2$ s,
that translates in an upper limit luminosity of L$_{1.4kpc}^{nt}$ = 1.15 $\times$ 10$^{32}$ erg/s.

{\bf J1418-6058 (Rabbit) - type 2* RQP} % nonostante Ng sono ancora incerto su quale sia la controparte, osservazione in arrivo

The radio/X-ray morphology of this region is complex.
Some X-ray observations were done in order to search for
the counterpart of the EGRET source 3EG J1420-6038.
Two interesting sources were found in the X-rays: the
radio pulsar J1420-6048 and an extended non-thermal emission
situated $\sim$ 10' away from it and named $"$Rabbit$"$.
Within the Rabbit Nebula, there were only two detected point sources
but it was not possible to definitively identify one of them as
the pulsar responsible for the nebula nor identify it with the EGRET
source. Yadigaroglu \& Romani 1997 and Ng et al. 2005 place the
pulsar at 3.5 $\pm$ 1.5 kpc.
No $\gamma$-ray nebular emission was detected down to a flux of 
8.60 $\times$ 10$^{-11}$ erg/cm$^2$s (Ackermann et al. 2010).

\begin{figure}
\centering
\includegraphics[angle=0,scale=.40]{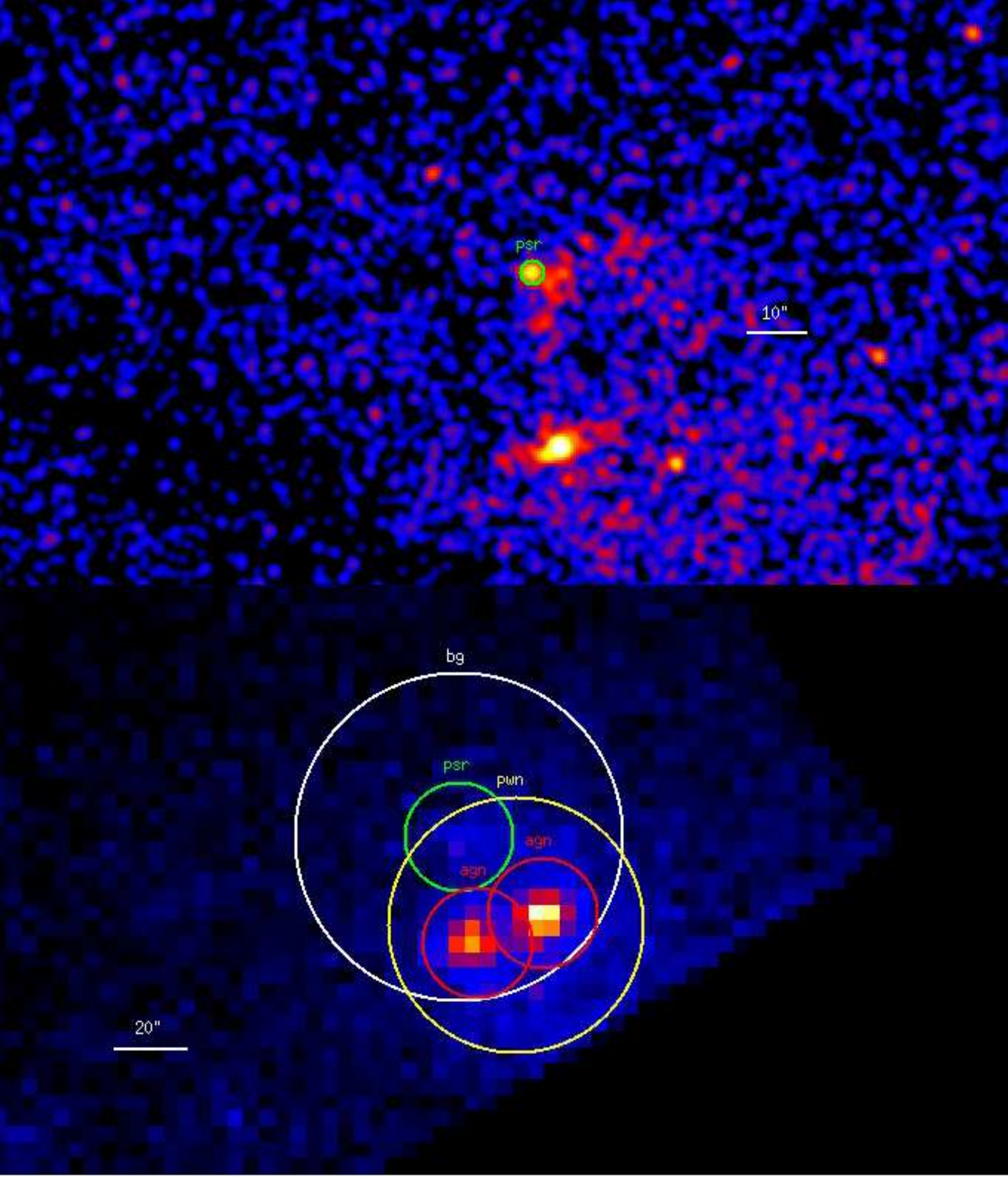}
\caption{{\it Lower Panel:} PSR J1418-6058 0.3-10 keV {\it XMM-Newton} EPIC Imaging. All the PN and the two MOS images have been added. 
The green circle marks the source while the white annulus the background region used in the analysis. The yellow region has been used
in order to analyze the nebular spectrum. The two red sources have been subtracted for all the analyses.
{\it Upper Panel:} PSR J1418-6058 0.3-10 keV {\it Chandra} Imaging.
The green circle marks the pulsar region used in the analysis. The lack of the {\it XMM}-detected bottom-right bright source is apparent and
marks it as a background AGN.
\label{rabbit-im}}
\end{figure}

J1418-6058 was observed during two {\it XMM-Newton} observation
on 2003, March 10 at 12:01:05 UT and 2009, February 21 at 16:59:25.59 UT
(exposures of 26.8 and 124.7 ks).
In both the observations the PN camera of the EPIC
instrument was operated in Small Window mode, while the MOS detectors were set in Full frame mode. For
all three instruments, the medium optical filter was used.
No screening for soft proton flare events was done,
owing to the goodness of both the observations.
we also used the {\it Chandra} ACIS-I observation of J1418-6058,
obs. id 7540 started on 2007, June 14 at 21:30:14 UT
for a net exposure of 71.1 ks.
Three sources have been detected inside the nebula
using both the ciao celldetect tool on the {\it Chandra} image
and the SAS tools on the XMM images:\\
- source 1 : 14:18:42.66 , -60:58:02.93 (error 1.5$"$);\\
- source 2 : 14:18:43.34 , -60:57:33.90 (error 2$"$);\\
- source 3 : 14:18:40.16 , -60:57:55.06 (error 5$"$).\\
Source 3 was excluded due to its non-detection in the {\it Chandra} image:
such a variation is typical of AGNs but it's not compatible
with an isolated neutron star. Both the remaining two sources
are acceptable after a dedicated timing analysis with {\it Fermi} (Ray et al. 2010)
nor show any hint of variation in the three available X-ray observations.
No clear optical counterpart has been found in the NOMAD optical catalogue (Zacharias et al. 2004).
After a spectral and timing analysis on the {\it Chandra} observation,
Ng et al. 2004 excluded the source 1 to be a pulsar, pointing
towards an AGN interpretation. Moreover, a 2MASS infrared source is
positionally consistent with such an X-ray source (2MASS 14184282-6058030
or GLIMPSE G313.3235+00.1312, Jmag=16.6, Hmag=15.1, Kmag=15.0). No associated
optical source was found: an obscured AGN is expected to reach $\sim$ 24 Vmag
so that it would not be detected in the optical catalogues. Anyway, such an AGN is
brighter in the red optical band (or infrared).\\
We extracted the XMM pulsar spectrum from a 15$"$ radius circle centered on it, in order to exclude the near sources.
We extracted the XMM nebular spectrum from a 35$"$ radius circle (see Figure \ref{rabbit-im}) and we excluded the pointlike sources;
such a region doesn't comprehend all the nebula so that its flux cannot be used in future analyses.
We extracted the XMM background spectrum from an annulus with radii 15$"$ and 45$"$ from which the nebula was excluded.
The {\it Chandra} pulsar spectrum was extracted from a 2$"$ radius circle centered on the pulsar and the background from a
source-free circular region on the same CCD. Due to the position near the edge of the CCD we didn't analyzed the
nebular emission on {\it Chandra}.
The spectra obtained in the two {\it XMM-Newton} observations were added using
mathpha tool and, similarly, the response
matrix and effective area files using addarf and addrmf.
We obtained 50 pulsar counts from the {\it Chandra} observation (background contribution of 4.4\%),
5140, 802 and 702 source counts from the PN and two MOS cameras of the first {\it XMM-Newton} observation (background contributions of 91.7\%, 73.3\% and 61.5\%)
and 14367, 2616 and 2214 nebular counts from the second {\it XMM-Newton} observation (background contributions of 82.6\%, 72.1\% and 67.4\%).
The best fit model is a simple powerlaw (probability of obtaining the data if the model is correct 
- p-value - of 0.64, 106 dof using both the pulsar and nebula) with a photon index
$\Gamma$ = 1.80$_{-0.30}^{+0.47}$ , absorbed by a column N$_H$ = 2.25$_{-0.45}^{+0.52}$ $\times$ 10$^{22}$ cm$^{-2}$.
The PWN has a photon index $\Gamma$ = 1.70$_{-0.19}^{+0.14}$.
A simple blackbody model is not statistically acceptable while a composite model gives no statistically
significative improvement.
Assuming the best fit model, the pulsar flux is
3.59$\pm$1.44 $\times$ 10$^{-14}$ erg/cm$^2$ s. 
Using a distance
of 3.5 kpc, the pulsar luminosity is L$_{3.5kpc}^{nt}$ = 5.28 $\pm$ 2.12 $\times$ 10$^{31}$ erg/s.

\begin{figure}
\centering
\includegraphics[angle=0,scale=.50]{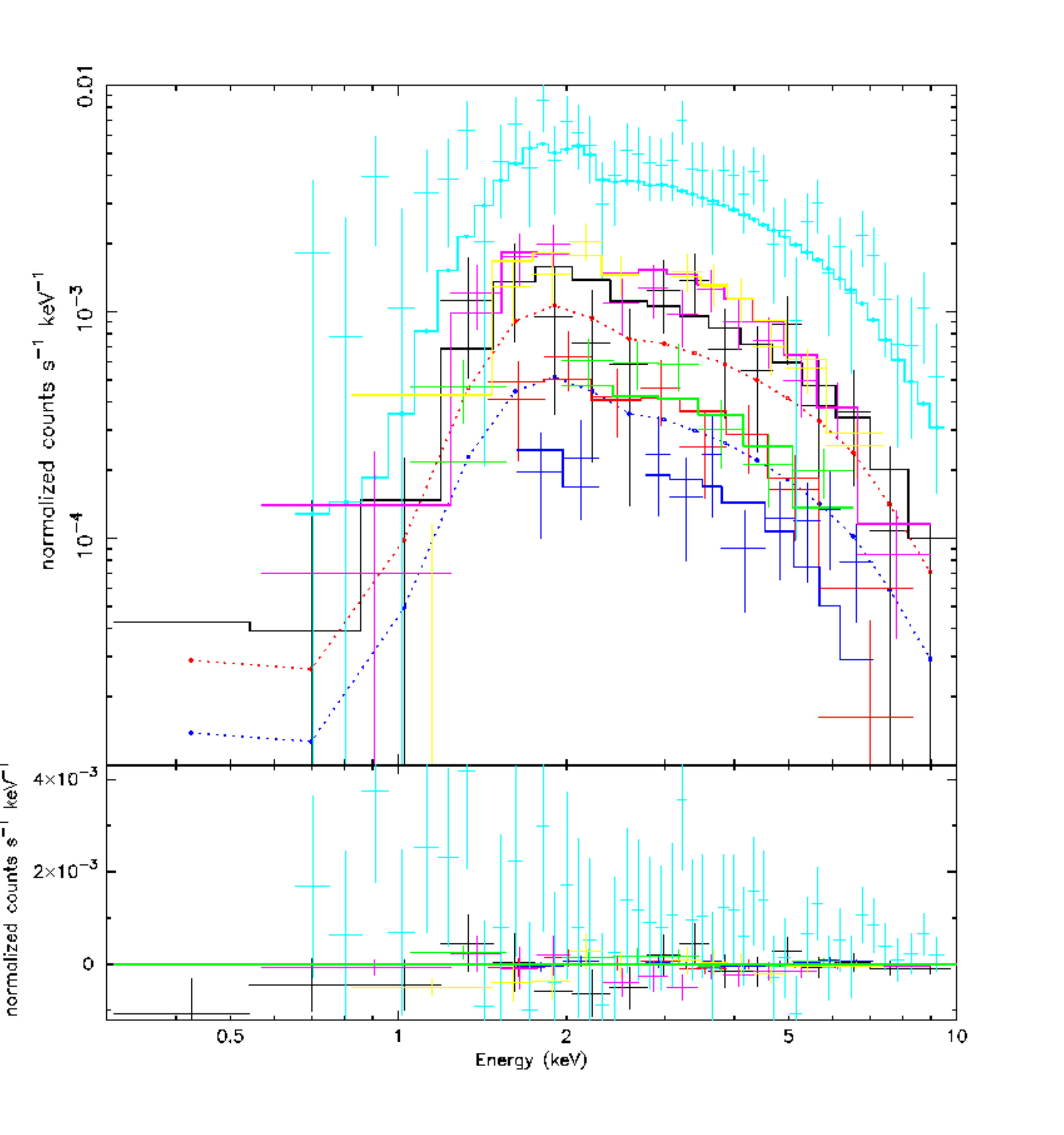}
\caption{PSR J1418-6058 Spectrum. Different colors mark all the different dataset used (see text for details).
Blue points mark the pulsar spectrum while red points the nebular one.
Residuals are shown in the lower panel.
\label{rabbit-so}}
\end{figure}

\clearpage

{\bf J1420-6048 - type 2 RLP} %ha la pwn, vedi Van Etten & Romani, ma è talmente tenue da essere inutile, osservazione in arrivo

%Van Etten & Romani 2010
The campaign to identify 3EG J1420-6038 has revealed sources across the electromagnetic
spectrum from radio to VHE $\gamma$-rays. The complex of compact and extended radio
sources in this region is referred to as the Kookaburra (Roberts et al. 1999), and covers nearly
a square degree along the Galactic plane. Within a Northeasterly excess in this complex
D'Amico et al. (2001) discovered PSR J1420-6048, a young
energetic pulsar with period 68 ms, characteristic age $\tau_c$ = 13 kyr, and spin down energy
$\dot{E}$ = 1.0 $\times$ 10$^{37}$ erg/s. Aharonian et al.
(2006) report on the discovery of two bright VHE $\gamma$-ray sources coincident with the Kookaburra
complex. HESS J1420-607 is centered just north of J1420-6048, with best fit extension
overlapping the pulsar position. The other H.E.S.S. source appears to correspond to the
Rabbit nebula half a degree southwest, which is also observed in the radio (Roberts et al.
1999) and X-ray (Roberts et al. 2001a).
The radio dispersion measurement gives a pulsar distance of 5.6$\pm$1.7 kpc (D'Amico et al. 2001).
No $\gamma$-ray nebular emission was detected by {\it Fermi} down to a flux of 
1.39 $\times$ 10$^{-10}$ erg/cm$^2$s (Ackermann et al. 2010).

An {\it XMM-Newton} observation of J1420-6048 was performed: obs. id. 0505840101,
start time 2008, February 15 at 17:57:41 UT, exposure 28.0 ks.
Both the PN and MOS cameras were set in the Full Frame mode and a medium optical
filter was used. First, an accurate
screening for soft proton flare events was done obtaining a resulting
exposure of 18.9 ks. Due to the badness of the observation,
we cannot remove all the proton flares in the dataset.
The X-ray source best fit position is 14:20:08.232 -60:48:16.56 (5$"$ error radius).
Van Etten \& Romani 2010 found a faint nebular emission in a long
Suzaku observation of the pulsar. Such an emission is therefore so faint 
(very similar to the pulsar's one on a 3' radius halo) that it can be
ignored in order to make a spectrum of the source.
we extracted pulsar spectrum from a 20$"$ radius circle and
the background was taken from a source-free region far away the pulsar, outside the faint halo.
we obtained a total of 2319, 526 and 453 pulsar counts from
the three {\it XMM-Newton} cameras 
in the 0.3-10 keV energy range(background
contributions of 82.2\%, 62.7\% and 52.9\%).
Such an high background contribution is
due to the incomplete proton flares screening
for the badness of the observations; it also requires
an event grouping in the PN spectrum of 100 instead of 25.
The best fitting pulsar model is a simple powerlaw
(probability of obtaining the data if the model is correct 
- p-value - of 0.16, 59 dof).
A simple blackbody model gives a too high value of the temperature
(T $>$ 2 $\times$ 10$^6$ K) while the add of
a blackbody component gives not a significative improvement in the fit. 
The pulsar spectrum has a photon index $\Gamma$ = 0.84$_{-0.37}^{+0.55}$ , 
absorbed by a column N$_H$ = 2.02$_{-1.06}^{+1.61}$ $\times$ 10$^{22}$ cm$^{-2}$.
Assuming the best fit model, the pulsar flux is
(1.6$\pm$0.7) $\times$ 10$^{-13}$ erg/cm$^2$ s. 
Using a distance
of 5.6 kpc, the pulsar luminosity is L$_{5.6kpc}^{nt}$ = 6.02 $\pm$ 2.63 $\times$ 10$^{32}$ erg/s.

\begin{figure}
\centering
\includegraphics[angle=0,scale=.50]{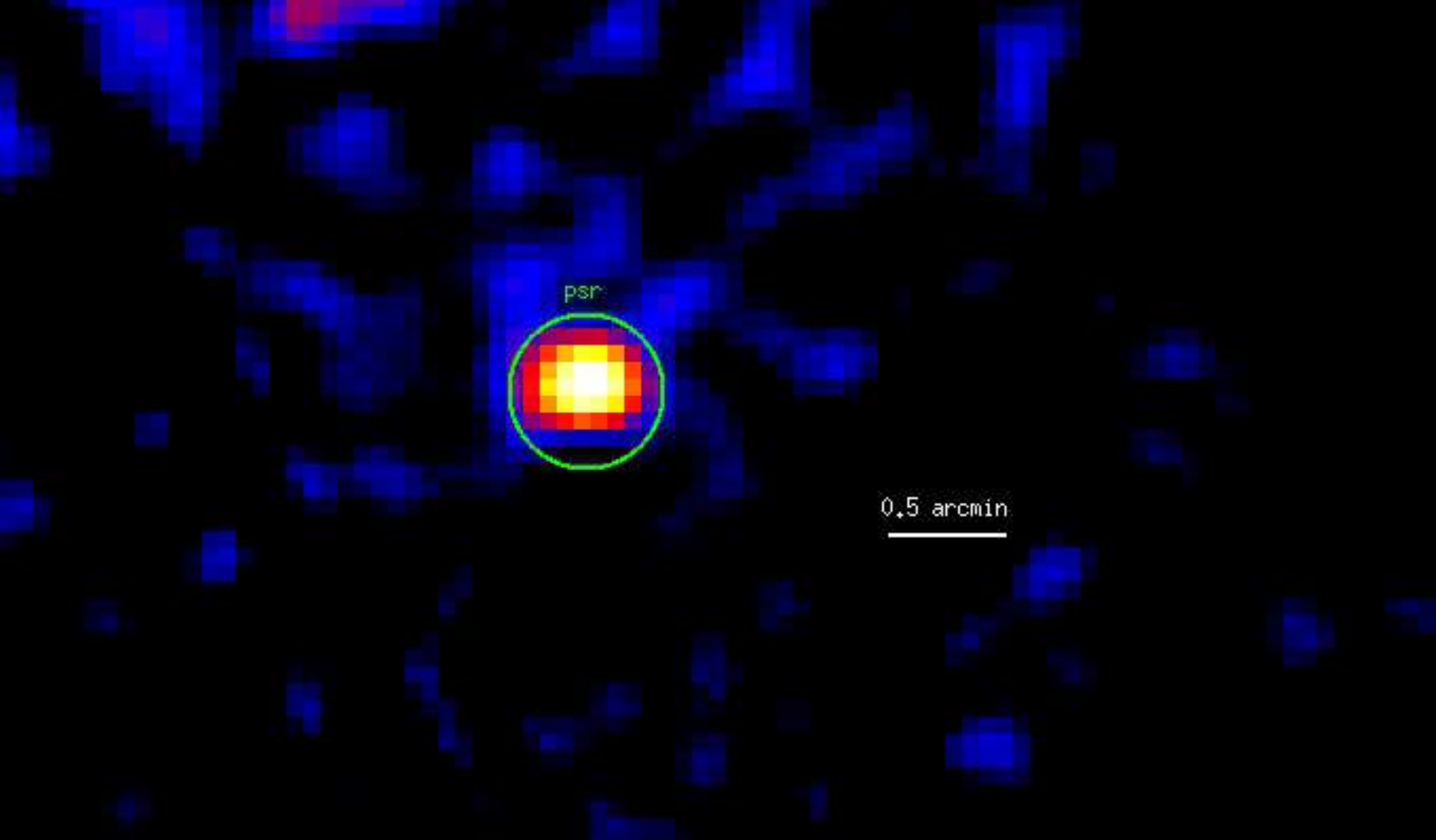}
\caption{PSR J1420-6048 0.3-10 keV {\it XMM-Newton} EPIC Imaging. The PN and the two MOS images have been added. 
The green circle marks the source region used for the analysis.
\label{J1420-im}}
\end{figure}

\begin{figure}
\centering
\includegraphics[angle=0,scale=.50]{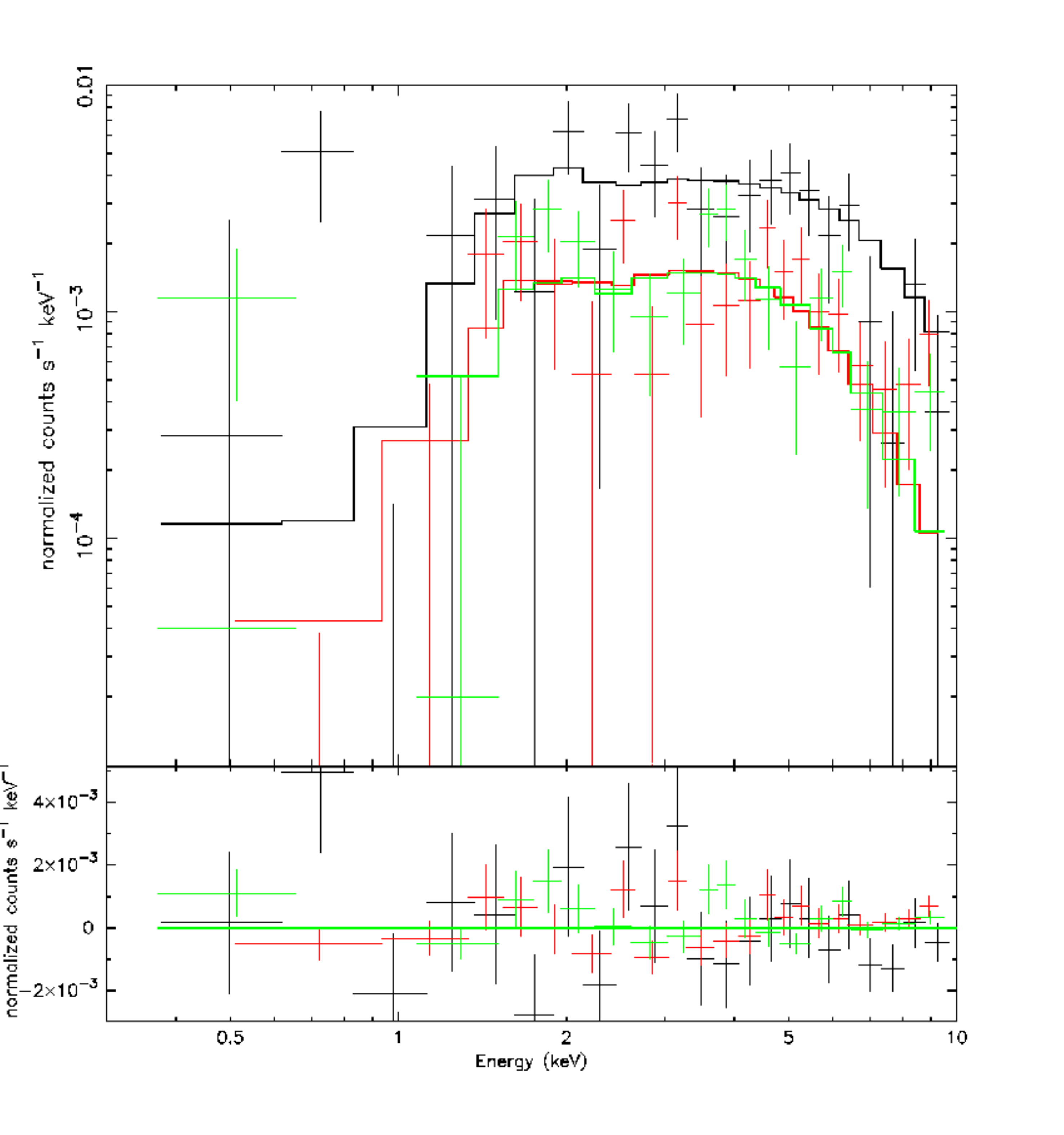}
\caption{PSR J1420-6048 Spectrum. Different colors mark all the different dataset used (see text for details).
Residuals are shown in the lower panel.
\label{J1420-sp}}
\end{figure}

\clearpage

{\bf J1429-5911 - type 0 RQP} % ho trovato un nuovo upper limit, corretto

Pulsations from J1429-5911 were detected by {\it Fermi} using the
blind search technique (Saz Parkinson et al. 2010).
No $\gamma$-ray nebular emission was detected down to a flux of 
2.35 $\times$ 10$^{-11}$ erg/cm$^2$s (Ackermann et al. 2010).
The pseudo-distance of the object based on $\gamma$-ray data (Saz Parkinson et al. (2010))
is $\sim$ 1.6 kpc.

After the {\it Fermi} detection, we asked for a {\it SWIFT} observation
of the $\gamma$-ray error box (obs id. 00031529001-00031660001, 6.99 ks exposure).
After the data reduction, no X-ray source was found.
For a distance of 1.6 kpc we found a
rough absorption column value of 8 $\times$ 10$^{21}$ cm$^{-2}$
and using a simple powerlaw spectrum
for PSR+PWN with $\Gamma$ = 2 and a signal-to-noise of 3,
we obtained an upper limit non-thermal unabsorbed flux of 5.76 $\times$ 10$^{-13}$ erg/cm$^2$ s,
that translates in an upper limit luminosity L$_{1.6kpc}^{nt}$ = 1.77 $\times$ 10$^{32}$ erg/s.

{\bf J1459-60 - type 0 RQP} % osservazione in arrivo

J1459-5911 was one of the first pulsars discovered using the
blind search technique (Abdo et al. 2009).
No $\gamma$-ray nebular emission was detected down to a flux of 
2.56 $\times$ 10$^{-11}$ erg/cm$^2$s (Ackermann et al. 2010).
The pseudo-distance of the object based on $\gamma$-ray data (Saz Parkinson et al. (2010))
is $\sim$ 1.5 kpc.

After the {\it Fermi} detection, we asked for a {\it SWIFT} observation
of the $\gamma$-ray error box (obs id. 00031359001-00031359002, 10.13 ks exposure).
After the data reduction, two X-ray source were found inside
the {\it Fermi} error box. A dedicated timing analysis
on the {\it Fermi} data excluded both the sources as potential
counterparts (Ray et al. 2011).
For a distance of 1.5 kpc we found a
rough absorption column value of 1 $\times$ 10$^{22}$ cm$^{-2}$
and using a simple powerlaw spectrum
for PSR+PWN with $\Gamma$ = 2 and a signal-to-noise of 3,
we obtained an upper limit non-thermal unabsorbed flux of 3.93 $\times$ 10$^{-13}$ erg/cm$^2$ s,
leading to an upper limit luminosity L$_{1.5kpc}^{nt}$ = 1.06 $\times$ 10$^{32}$ erg/s.

{\bf J1509-5850 - type 2 RLP} % rifatto unendo i files e usando la cstatistic

%Hui & Becker 2007
PSR J1509-5850 was discovered by Manchester
et al. (2001) in the Parkes Multibeam Pulsar Survey. The
pulsar has a rotation period of P = 88.9 ms and a period derivative
of $\dot{P}$ = 9.17 $\times$ 10$^{-15}$. These spin parameters imply a
characteristic age of 1.54 $\times$ 10$^5$ yrs, a dipole surface magnetic
field of B = 9.14 $\times$ 10$^{11}$ G and a spin-down luminosity of
5.1 $\times$ 10$^{35}$ erg s$^{-1}$. The radio dispersion measure
yields a distance of approximately 3.81 kpc based on the
galactic free electron model of Taylor \& Cordes (1993). Using
the model of Taylor \& Cordes (1993) the dispersion measure
based distance is estimated to be 2.6 $\pm$ 0.8 kpc. The proper motion
of this pulsar is not yet known. Kargaltsev
et al. (2006) reported that an X-ray trail-like pulsar wind
nebula associated with PSR J1509-5850 was observed in
by {\it Chandra}. The X-ray nebula was found to be
extended in the south-west direction.
No $\gamma$-ray nebular emission was detected down to a flux of 
2.82 $\times$ 10$^{-11}$ erg/cm$^2$s (Ackermann et al. 2010).

Three different X-ray observations of J1509-5850 were performed, two
by {\it XMM-Newton} and one by {\it Chandra}:\\
- obs. id 3513, {\it Chandra} ACIS-S very faint mode, start time 2003, February 09 at 12:33:29 UT, exposure 40.1 ks;\\
- obs. id 0500630101, {\it XMM-Newton} observation, start time 2008, January 28 at 16:22:28 UT, exposure 49.4 ks;\\
- obs. id 0500630301, {\it XMM-Newton} observation, start time 2008, March 01 at 21:39:25 UT, exposure 47.4 ks;\\
In both the XMM observations the PN and MOS cameras were
operating in the Full Frame mode. For
all three instruments, the medium optical filter was used.
The off-axis angle is negligible in all the observations.
The X-ray source best fit position, obtained by using the celldetect tool
inside the CIAO distribution, is 15:09:27.159 -58:50:56.09 (1$"$ error radius).
First, an accurate
screening for soft proton flare events was done in the second {\it XMM-Newton} observation, obtaining a resulting
exposure of 18.9 ks. The first XMM observation requests no screening
owing to the goodness of the dataset.
For the {\it Chandra} observation, we chose a 2$"$ radius circular region for the
pulsar spectrum and the background is extracted from a circular source-free region
away from the source, in order to exclude the nebular emission.
For the {\it XMM-Newton} observation, we chose a 10$"$ radius circular region around
the pulsar in order to minimize the nebular contribution; 
the background is extracted from a circular source-free region on the same CCD,
away from the source, in order to exclude the nebular emission.
The spectra obtained in the two {\it XMM-Newton} observations were added using
mathpha tool and, similarly, the response
matrix and effective area files using addarf and addrmf. 
we obtained a total of 82, 389, 146 and 148 pulsar counts
from respectively {\it Chandra}, PN camera and the two MOS cameras
(background contributions of 0.4\%, 14.0\%, 3.6\% and 2.9\%).
Due to the low statistic, we used the C-statistic
approach implemented in XSPEC.
The best fitting pulsar model is a simple powerlaw ($\chi^2_{red}$ = 1.105,
145 dof) with a
photon index $\Gamma$ = 1.36 $\pm$ 0.20, absorbed by a column
N$_H$ = 7.95$_{-1.65}^{+2.21}$ $\times$ 10$^{21}$ cm$^{-2}$.
A simple blackbody model gives an unrealistic value of the
temperature (T $>$ 2 $\times$ 10$^7$ K) while the add of a thermal
model to the powerlaw gives no significant improvement
in the statistic of the fit.
We fitted separately the nebular emission and we found a photon
index of $\Gamma$ = 1.17 $\pm$ 0.13.
Assuming the best fit model, the 0.3-10 keV pulsar flux is
5.34$_{-1.80}^{+1.97}$ $\times$ 10$^{-14}$ and the nebular flux
is 2.47$_{-0.54}^{+0.32}$ $\times$ 10$^{-13}$ erg/cm$^2$ s. 
Using a distance
of 2.6 kpc, the pulsar luminosity is L$_{2.6kpc}^{nt}$ = 4.33$_{-1.46}^{+1.60}$ $\times$ 10$^{31}$
and the nebular luminosity is L$_{2.6kpc}^{pwn}$ = 2.00$_{-0.44}^{+0.26}$ $\times$ 10$^{32}$ erg/s.

\begin{figure}
\centering
\includegraphics[angle=0,scale=.40]{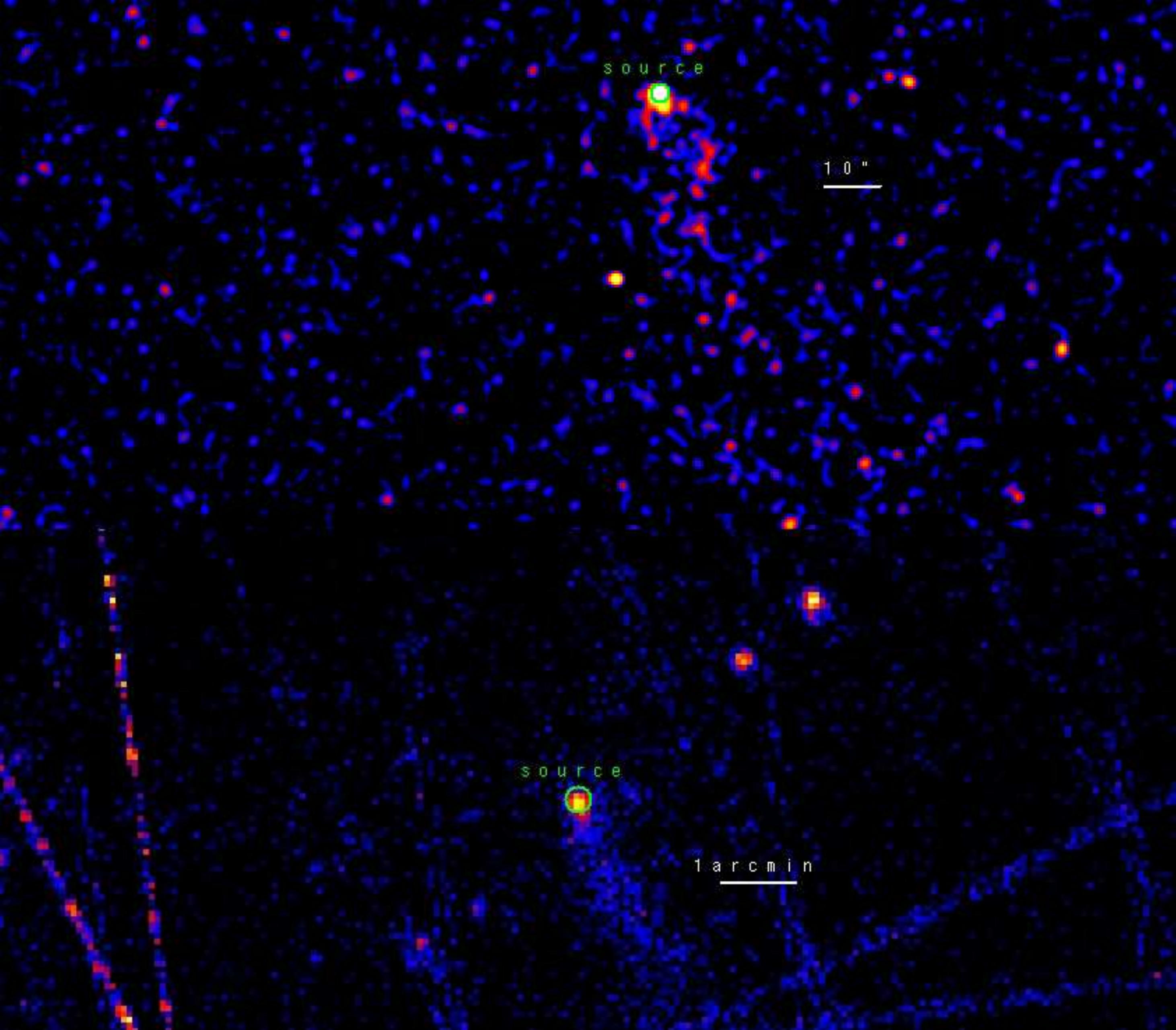}
\caption{{\it Upper Panel:} PSR J1509-5850 0.3-10 keV {\it Chandra} Imaging. The image has been smoothed with a Gaussian
with Kernel radius of $2"$. The green circle marks the pulsar region used in the analysis.
{\it Lower Panel:} PSR J1509-5850 0.3-10 keV {\it XMM-Newton} EPIC Imaging. The PN and the two MOS images have been added. 
\label{J1509-im}}
\end{figure}

\begin{figure}
\centering
\includegraphics[angle=0,scale=.50]{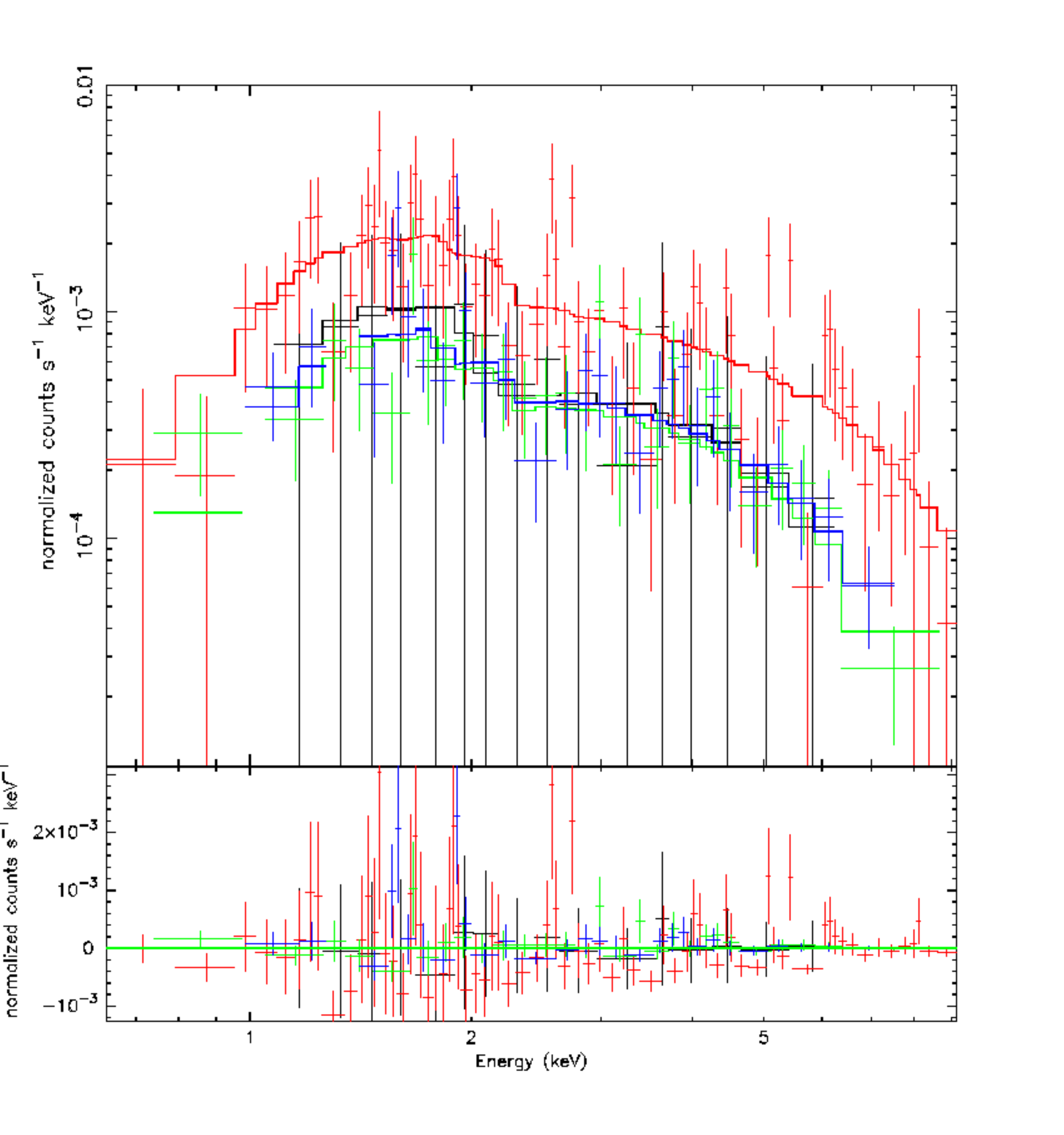}
\caption{PSR J1509-5850 Spectrum. Different colors mark all the different dataset used (see text for details).
Residuals are shown in the lower panel.
\label{J1509-sp}}
\end{figure}

\clearpage

{\bf J1513-5908 - (type 2*) RL MSP} % Nuova! sul file fits la segno come una ibis, quindi con -1

% Abdo et al. 2010
The 150 ms rotation period of J1513-5908 
was discovered by the Einstein satellite
(Seward \& Harden 1982) and soon thereafter confirmed
in the radio domain (Manchester et al. 1982). With a
large period derivative (1.5 $\times$ 10$^{-12}$ s s$^{-1}$), this pulsar is
one of the youngest and most energetic pulsars known in
the Galaxy with a characteristic age of 1700 years and a
spin-down power $\dot{E}$ of 1.8 $\times$ 10$^{37}$ erg/s. The inferred
surface magnetic field is 1.5 $\times$ 10$^{13}$ G derived under the 
assumption of a dipolar magnetic field. The measurement
of the pulsar braking index n = 2.839 shows that this
assumption is reasonable (e.g. Livingstone et al. 2005).
Therefore, the high magnetic field is not much below the
quantum-critical magnetic field of 4.413 $\times$ 10$^{13}$ G, the
domain of the so-called high-B-field pulsars and magnetars.
The distance is estimated at 5.2 $\pm$ 1.4 kpc using 
HI absorption measurements (Gaensler et al. 1999).
This is consistent with the value of 4.2 $\pm$ 0.6 kpc derived
from the dispersion measure (Taylor \& Cordes 1993).
PSR B1509-58 has been studied by all major X-ray
and $\gamma$-ray observatories yielding a broad-band spectral
energy distribution and pulse profiles as a function of
energy. Its detection by COMPTEL (0.75 - 30 MeV,
Kuiper et al. 1999) and non-detection with the Energetic
Gamma-Ray Experiment Telescope (EGRET) in
the 30 MeV - 30 GeV energy range, both aboard the
Compton Gamma Ray Observatory (CGRO), indicate
an abrupt spectral break between 10 and 30 MeV. This
break is well below the break energies of most $\gamma$-ray pulsars
detected by {\it Fermi} which are typically around a few
GeV (Abdo et al. 2010a). More recently, the detection
of pulsed $\gamma$-rays from PSR B1509-58 at 4$\sigma$ level above
100MeV was reported by AGILE (Pellizzoni et al. 2009).
Einstein X-ray observations revealed
an elongated non-thermal source centered on the pulsar
(Seward \& Harden 1982), later confirmed by ROSAT
and interpreted as a pulsar wind nebula powered by
PSR B1509-58 (Trussoni et al. 1996). This PWN, composed
of arcs and bipolar jets, is especially bright and
extended in X-rays, and at very high energies. The dimensions of the PWN as
observed by ROSAT (Trussoni et al. 1996) and HESS
(Aharonian et al. 2005) are 10' $\times$ 6' and 6.4' $\times$ 2.3' 
respectively. The pulsar is associated with the 13 ky old
SNR G320.4-1.2.
The detection of pulsed $\gamma$-rays
from PSR B1509-58 at 4$\sigma$ level above 100 MeV was
reported by AGILE (Pellizzoni et al. 2009). On the
other hand, {\it Fermi} reported only an upper limit on the
pointlike source detection.
More recently, in one of the first versions of the 2nd {\it Fermi} LAT catalogue
the pointlike source was detected. Due to the doubts about its detection
such a pulsar will not be considered in the general analyses.

\begin{figure}
\centering
\includegraphics[angle=0,scale=.40]{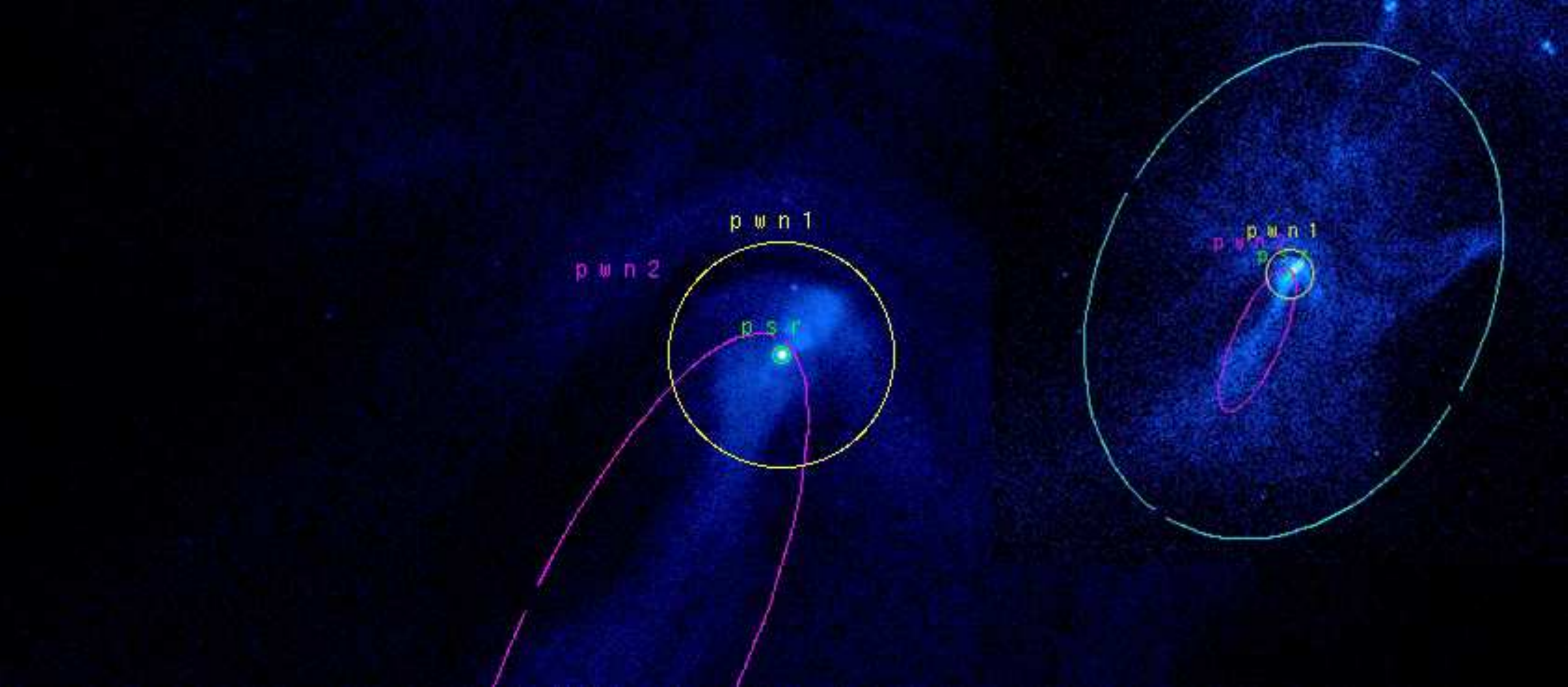}
\caption{PSR J1513-5908 0.3-10 keV {\it Chandra} Imaging. The green circle marks the pulsar while the yellow ellipses
the nebular regions used in the analysis.
\label{J1513-im}}
\end{figure}

Many X-ray observations of this source were performed in the last years.
Due to the complexity of the nebular emission, we decided to use only {\it Chandra}
data due to their high space resolution. we analyzed the following observations:\\
- obs. id 5562, {\it Chandra} ACIS-S very faint mode, start time 2004, December 31 at 09:13:51 UT, exposure 30.0 ks;\\
- obs. id 9138, {\it Chandra} ACIS-S very faint mode, start time 2008, June 19 at 18:48:12 UT, exposure 59.7 ks;\\
- obs. id 3833, {\it Chandra} ACIS-I very faint mode, start time 2003, October 18 at 00:37:56 UT, exposure 19.4 ks;\\
- obs. id 3834, {\it Chandra} ACIS-I very faint mode, start time 2003, April 21 at 17:01:28 UT, exposure 9.6 ks;\\
- obs. id 4384, {\it Chandra} ACIS-I very faint mode, start time 2003, April 28 at 06:06:33 UT, exposure 10.0 ks;\\
- obs. id 5534, {\it Chandra} ACIS-I very faint mode, start time 2004, December 28 at 10:27:39 UT, exposure 50.1 ks;\\
- obs. id 5535, {\it Chandra} ACIS-I very faint mode, start time 2005, February 07 at 15:16:11 UT, exposure 43.1 ks;\\
- obs. id 6116, {\it Chandra} ACIS-I very faint mode, start time 2005, April 29 at 03:41:58 UT, exposure 47.6 ks;\\
- obs. id 6117, {\it Chandra} ACIS-I very faint mode, start time 2005, October 18 at 00:12:41 UT, exposure 46.1 ks;\\
- obs. id 754, {\it Chandra} ACIS-I very faint mode, start time 2000, August 14 at 13:32:51 UT, exposure 19.3 ks.\\
The off-axis angle is negligible in all the ACIS-I observations. The ACIS-S observations'
off-axis angles are too high to be neglected: due to the complexity of the source and the poor spatial resolution
due to the high off-axis angle, we decided not to analyze the spectrum of such two observations.
The X-ray source best fit position is 15:13:55.63 -59:08:09.29 (0.5$"$ error radius), obtained by using the celldetect
tool inside the CIAO distribution.
An extended and composite nebula is apparent. A single powerlaw spectrum is unable to describe all the nebula
so that we divided it into three different regions: the tail, the inner and the outer regions (see Figure \ref{J1513-im}).
Due to the extreme pileup of the pulsar, all the spectral information about it are taken from Gaensler et al. 2001 and
DeLaney et al. 2005. Anyway, we analyzed the nebula.
we obtained a total of 79900, 153950 and 1.12589e+06 counts from the
three nebular regions (background contributions of 1.5\%, 3.7\% and 13.0\%).
The best fitting column density is
N$_H$ = 9.18 $\pm$ 0.02 $\times$ 10$^{21}$ cm$^{-2}$.
The tail has a photon index $\Gamma$ = 1.49 $\pm$ 0.02.
The inner part of the nebula has a photon index $\Gamma$ = 1.60 $\pm$ 0.02.
The outer part of the nebula has a photon index $\Gamma$ = 1.96 $\pm$ 0.01.
Assuming the best fit model, the 0.3-10 keV tail flux is
8.65$_{-0.17}^{+0.19}$ $\times$ 10$^{-12}$, the inner nebula flux is
1.63 $\pm$ 0.01 $\times$ 10$^{-11}$ and the outer nebula flux is
1.21 $\pm$ 0.01 $\times$ 10$^{-10}$ erg/cm$^2$s.
Using a distance
of 5.2 kpc, the tail luminosity is L$_{5.2kpc}^{tail}$ = 2.80$_{-0.06}^{+0.07}$ $\times$ 10$^{34}$,
and the inner nebula luminosity is L$_{5.2kpc}^{in}$ = 5.29 $\pm$ 0.04 $\times$ 10$^{34}$
and the outer nebula luminosity is L$_{5.2kpc}^{out}$ = 3.93 $\pm$ 0.03 $\times$ 10$^{35}$ erg/s.
From the previously cited papers, the pulsar best fit model is a powerlaw with $\Gamma$ = 2.05 $\pm$ 0.04,
with an unabsorbed flux of 5.2 $\pm$ 1.8 $\times$ 10$^{-11}$ erg/cm$^2$ s.
The pulsar luminosity is L$_{5.2kpc}^{psr}$ = 1.69 $\pm$ 0.58 $\times$ 10$^{35}$ erg/s.

\clearpage

{\bf J1531-5610 - type 1 RLP} % Nuova! osservazione in arrivo

J1531-5610 was found during the Parkes Multibeam Pulsar Survey (Kramer et al. 2003)
as a radio pulsar with a period of $\sim$ 84ms and a period derivative of 1.37 $\times$ 10$^{-14}$.
The pulsar distance, based on radio dispersion measurements, is $\sim$ 2.1 kpc.

There is only one {\it Chandra} ACIS-I observation of J1531-5610, obs. id 9078, starting on
2008, January 04 at 12:07:09 UT, for a total exposure of 10.0 ks.
The pulsar is located 1.5$'$ from the center of the FOV, at
15:31:27.91 -56:10:56.00 (2$"$ error radius). No search for diffuse emission
is possible due to the low statistic. The pulsar spectrum was extracted
from a 2$"$ radius circular region around the source while the background from
an annulus with radii 5$"$ and 10$"$ centered on the pulsar.
The net countrate of the source is 9.77 $\pm$ 3.19 $\times$ 10$^{-4}$ c/s.
Using a distance of 2.1 kpc, we infer a
rough absorption column value of 4 $\times$ 10$^{21}$ cm$^{-2}$.
Moreover, using powerlaw spectrum
for PSR+PWN with $\Gamma$ = 2 and assigning a 30\% of the resulting flux to
the PWN and thermal components, we obtained an
unabsorbed flux of 2.31 $\pm$ 0.75 $\times$ 10$^{-14}$ erg/cm$^2$ s
and a luminosity of L$_{2.1kpc}$ = 1.48 $\pm$ 0.48 $\times$ 10$^{31}$ erg/s.

\begin{figure}
\centering
\includegraphics[angle=0,scale=.50]{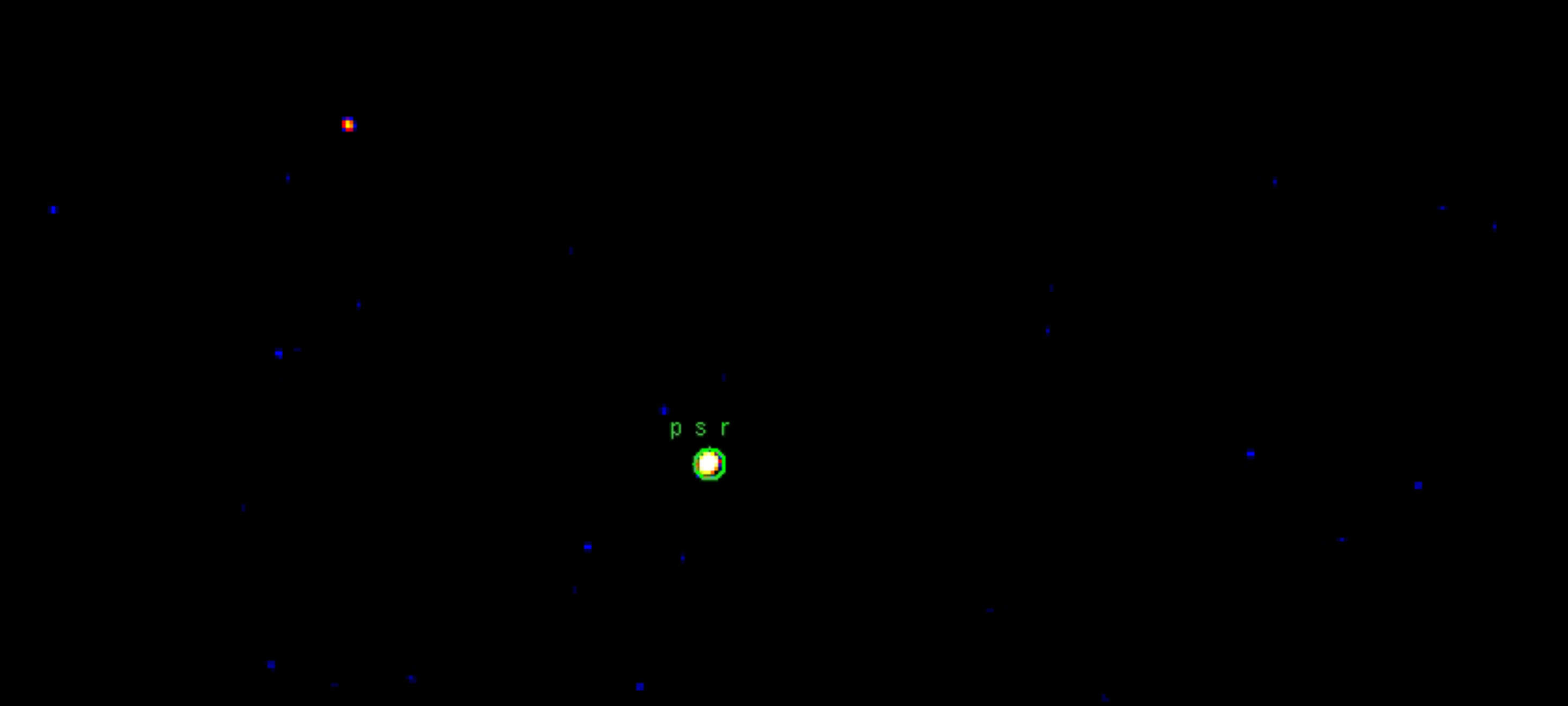}
\caption{PSR J1531-5610 0.3-10 keV {\it Chandra} Imaging. The image has been smoothed with a Gaussian
with Kernel radius of $2"$. The green circle marks the pulsar.
J1531 image. \label{J1531-im}}
\end{figure}

\clearpage

{\bf J1600-3053 - type 0 RL MSP} % Nuova!

% Jacoby et al. 2007
J1600-3053 was found in 2007 during
a 21-cm survey using the Parkes radio telescope (Jacoby et al. 2007).
It's a 3.6 millisecond pulsar with a known proper motion ($\sim$ 5 mas/yr)
in a binary system - measurement of the Shapiro delay constrains
the orbital inclination to be between 59 and 70$^{\circ}$ (95\% confidence).
The radio dispersion measurements place this pulsar at a distance
of $\sim$ 1.53 kpc (Jacoby et al. 2006).

J1600-3053 has a dedicated {\it XMM-Newton} observation performed on 2008,
February 17 at 10:24:56 UT, for a total exposure of 30.5 ks.
The PN camera of the EPIC
instrument was operated in Fast Timing mode, while the MOS detectors were set in Full frame mode.
Both for the PN and MOS cameras a medium optical filter was used.
First, an accurate
screening for soft proton flare events was done in the {\it XMM-Newton} observations obtaining a total
exposure of 26.5 ks.
We chose a 20$"$ radius circular region around
the pulsar for its spectrum while for the background an annulus of radii 30 and 50$"$ was chosen.
The PN dataset wasn't used due to the faintness of the source united with the
lack of spatial information coming from the Timing Mode.
The spectra obtained in the two MOS datasets were added using
mathpha tool and, similarly, the response
matrix and effective area files using addarf and addrmf. 
We obtained a total of 100 source counts (background contribution of 64.7\%).
The data can be fitted only by using a simple blackbody model,
where the N$_H$ value was frozen to 1 $\times$ 10$^{21}$ cm$^{-2}$,
the galactic value in that direction.
The temperature of the thermal emission results to be 
T = 4.10$_{-1.31}^{+4.10}$ $\times$ 10$^6$ K.
The blackbody radius R$_{1.53kpc}$ = 99$_{-89}^{+394}$ m determined from the
model parameters suggests that the emission is from a hot spot.
A simple powerlaw model is not statistically acceptable.
Assuming the best fit model, the 0.3-10 keV pulsar flux is
6.85 $\pm$ 2.14 $\times$ 10$^{-15}$. 
Using a distance
of 1.53 kpc, the pulsar luminosity is L$_{1.53kpc}^{bol}$ = 1.92 $\pm$ 0.60 $\times$ 10$^{30}$ erg/s.

\begin{figure}
\centering
\includegraphics[angle=0,scale=.40]{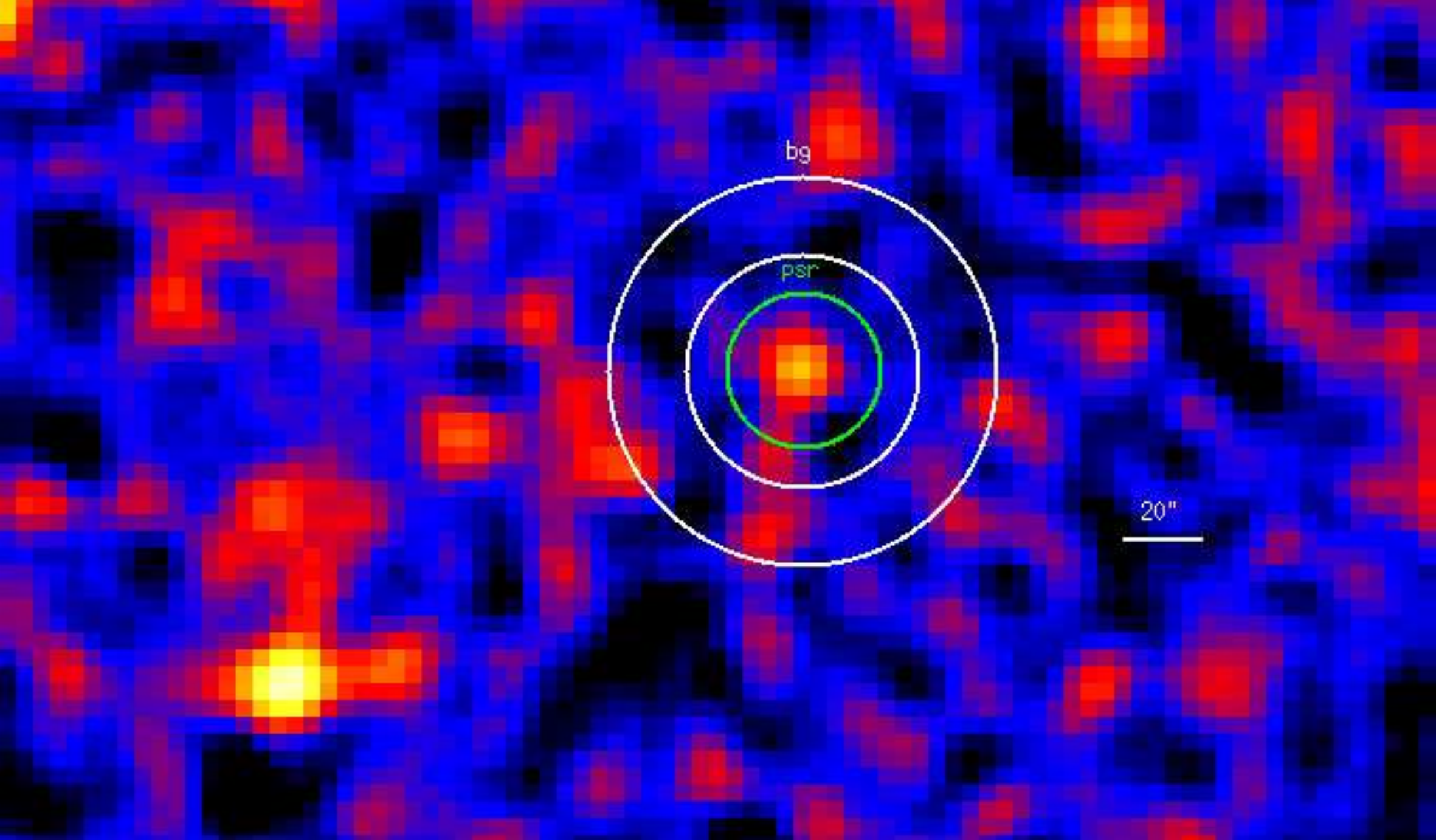}
\caption{PSR J1600-3053 0.3-10 keV MOS Imaging. The two MOS images have been added.
The image has been smoothed with a Gaussian with Kernel radius of $5"$.
The green circle marks the pulsar while the yellow annulus the background region used in the analysis.
\label{J1600-im}}
\end{figure}

\begin{figure}
\centering
\includegraphics[angle=0,scale=.30]{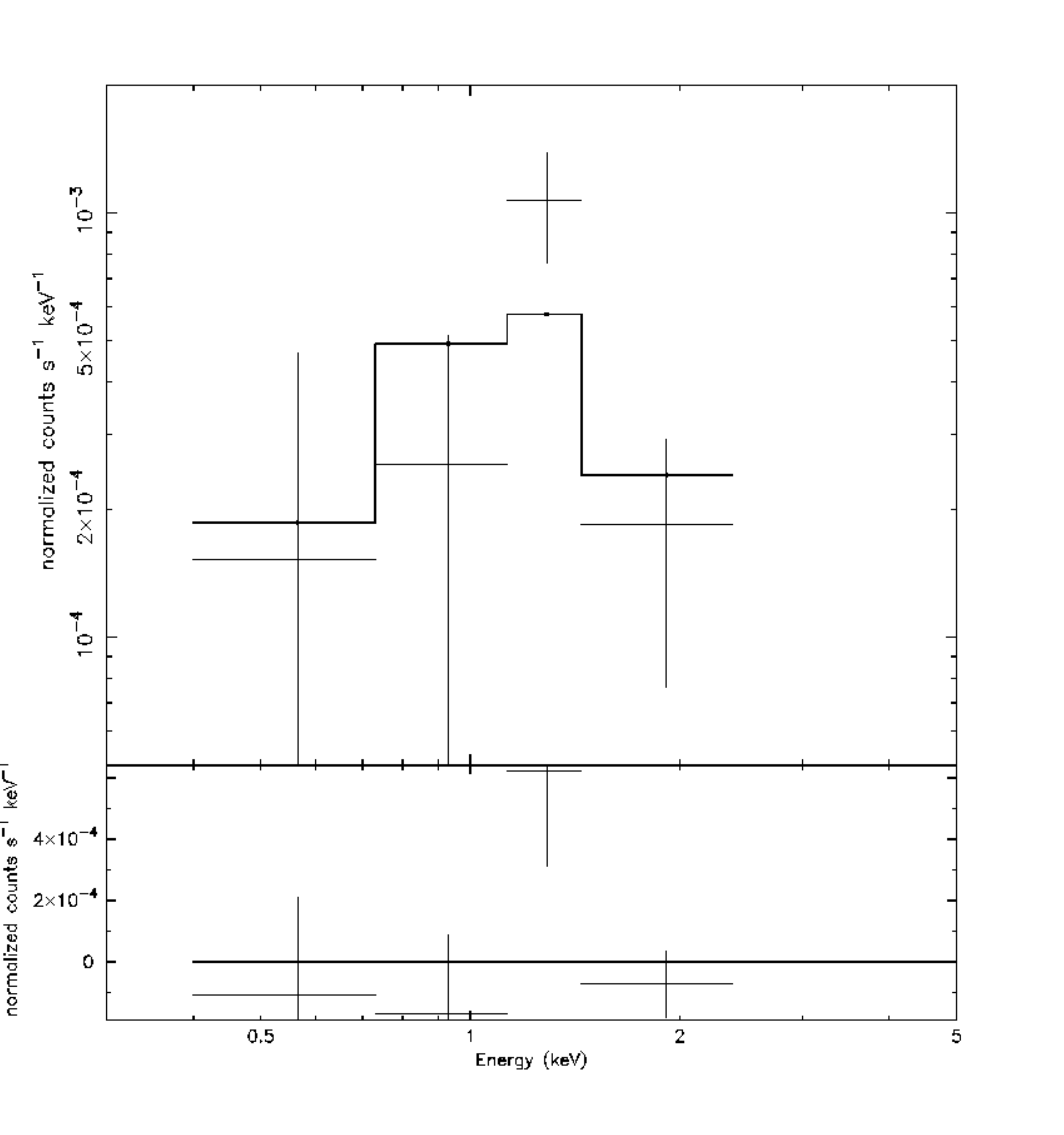}
\caption{PSR J1600-3053 {\it XMM-Newton} MOSs' Spectrum (see text for details). \label{J1600-sp}}
Residuals are shown in the lower panel.
\end{figure}

\clearpage

{\bf J1614-2230 - type 0 RL MSP}

% Hessels et al. 2004
PSR J1614-2230 is a recycled millisecond pulsar with a very low inferred magnetic
field (B = 1.8 $\times$ 10$^8$ G). It has one of the highest minimum companion mass of
the binary pulsars with spin periods lower than 8ms, suggesting a possible
non-standard evolution. Compared with the orbital period versus companion
mass relationship of Rappaport et al. 1995 (MNRAS, 273, 731) the orbital period
of PSR J1614-2230 is short by a factor $>$10 for its companion mass. The
companion may be a CNO white dwarf or a low-mass degenerate dwarf. The
spin-down rate of PSR J1614-2230 gives an $\dot{E}$ $\sim$ 1.3 $\times$ 10$^{34}$ erg s$^{-1}$.
A dispersion measurement based on Radio data yields a pulsar
distance of 1.27 $\pm$ 0.39 kpc.
No $\gamma$-ray nebular emission was detected down to a flux of 
2.44 $\times$ 10$^{-11}$ erg/cm$^2$s (Ackermann et al. 2010).

Three different X-ray observations of J1509-5850 were performed, two
by {\it XMM-Newton} and one by {\it Chandra}:\\
- obs. id 7509, {\it Chandra} ACIS-S very faint mode, start time 2007, April 26 at 00:54:49 UT, exposure 20.0 ks;\\
- obs. id 0304960101, {\it XMM-Newton} observation, start time 2005, August 17 at 06:30:12 UT, exposure 7.8 ks;\\
- obs. id 0404790101, {\it XMM-Newton} observation, start time 2007, February 08 at 17:43:22 UT, exposure 56.2 ks.\\
In both the XMM observations the PN and MOS cameras were
operating in the Full Frame mode. For
all three instruments, the medium optical filter was used.
The off-axis angle is negligible in all the observations.
The X-ray source best fit position is 16:14:36.50 -22:30:31.13 (1$"$ error radius).
First, an accurate
screening for soft proton flare events was done in the {\it XMM-Newton} observations obtaining a resulting total
exposure of 44.0 ks in the two observations.
No indications of the presence of an X-ray extended nebula were found
both in the {\it XMM-Newton} and {\it Chandra} observations.
For the {\it Chandra} dataset, we chose a 2$"$ radius circular region for the
pulsar spectrum and the background is extracted from a circular source-free region
away from the source, in order to exclude a faint source near the pulsar.
For the {\it XMM-Newton} observation, we chose a 15$"$ radius circular region around
the pulsar in order to minimize the contribution of a nearby source; 
the background is extracted from a circular source-free region on the same CCD,
away from the source.
The spectra obtained in the two {\it XMM-Newton} observations were added using
mathpha tool and, similarly, the response
matrix and effective area files using addarf and addrmf. 
We obtained a total of 989, 151, 151 and 52 pulsar counts
from respectively PN camera, the two MOS cameras and {\it Chandra}
(background contributions of 35.4\%, 20.8\%, 18.8\% and 1.0\%).
The best fitting pulsar model is a simple blackbody 
(probability of obtaining the data if the model is correct 
- p-value - of 0.61, 48 dof) with a temperature of
T = 2.84$_{-0.22}^{+0.12}$ $\times$ 10$^6$ K, absorbed by a column
N$_H$ = $<$ 2.45 $\times$ 10$^{20}$ cm$^{-2}$.
A simple powerlaw model is not statistically acceptable. A combination
of a powerlaw and a blackbody model is statistically acceptable but
gives no significative improvement with the respect to the simple thermal model.
Assuming the best fit model, the 0.3-10 keV pulsar flux is
2.86$_{-0.86}^{+0.15}$ $\times$ 10$^{-14}$ erg/cm$^2$ s. 
Using a distance
of 1.27 kpc, the pulsar luminosity is L$_{1.27kpc}^{bol}$ = 5.53$_{-1.66}^{+0.29}$ $\times$ 10$^{30}$ erg/s.

\begin{figure}
\centering
\includegraphics[angle=0,scale=.40]{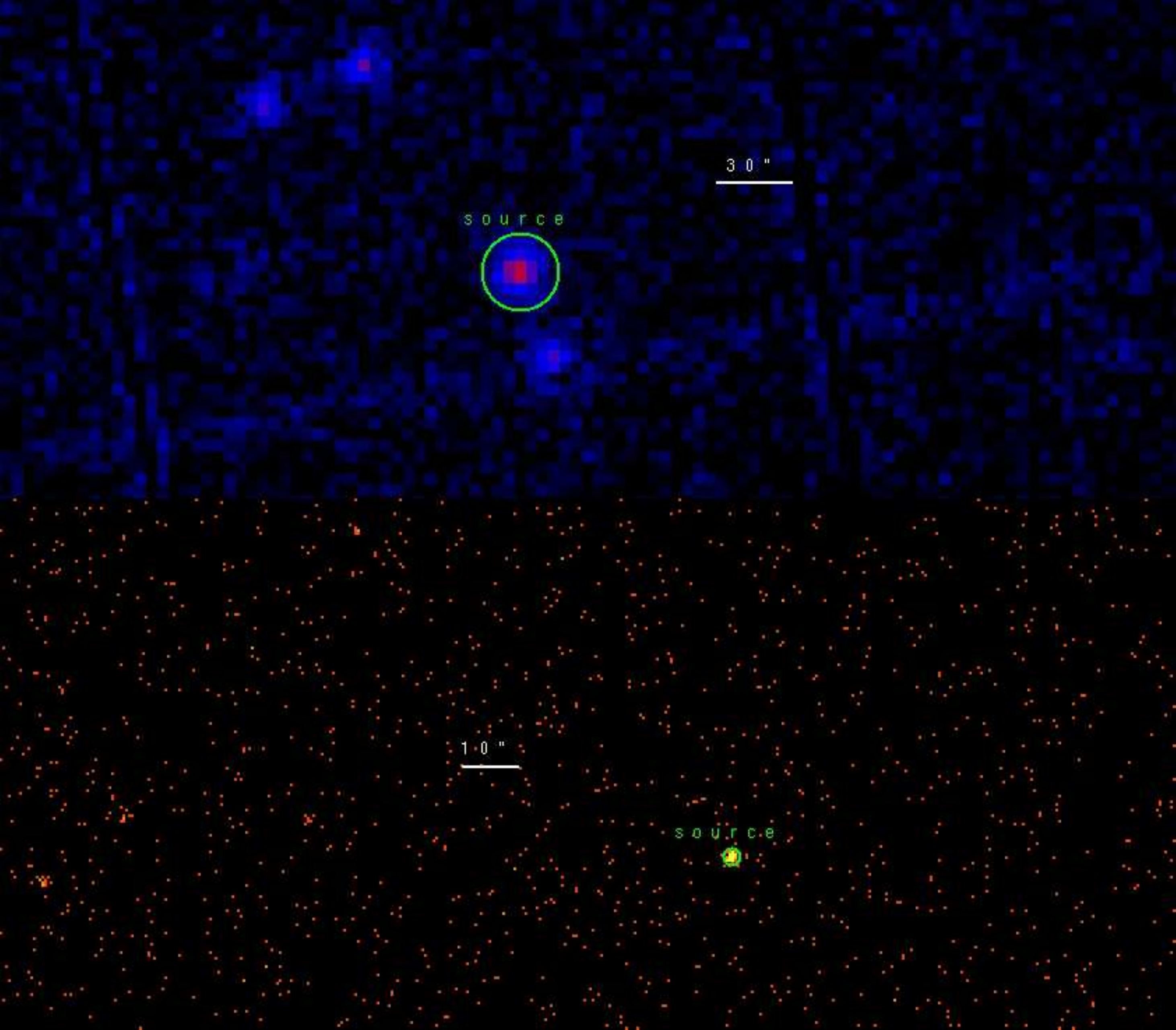}
\caption{{\it Upper Panel:} PSR J1614-2230 0.3-10 keV {\it XMM-Newton} Imaging. The PN and the two MOS images have been added. 
The green circle marks the pulsar region used in the analysis.
{\it Lower Panel:} PSR J1614-2230 0.3-10 keV {\it Chandra} Imaging. The green circle marks the pulsar region used in the analysis.
\label{J1614-im}}
\end{figure}

\begin{figure}
\centering
\includegraphics[angle=0,scale=.50]{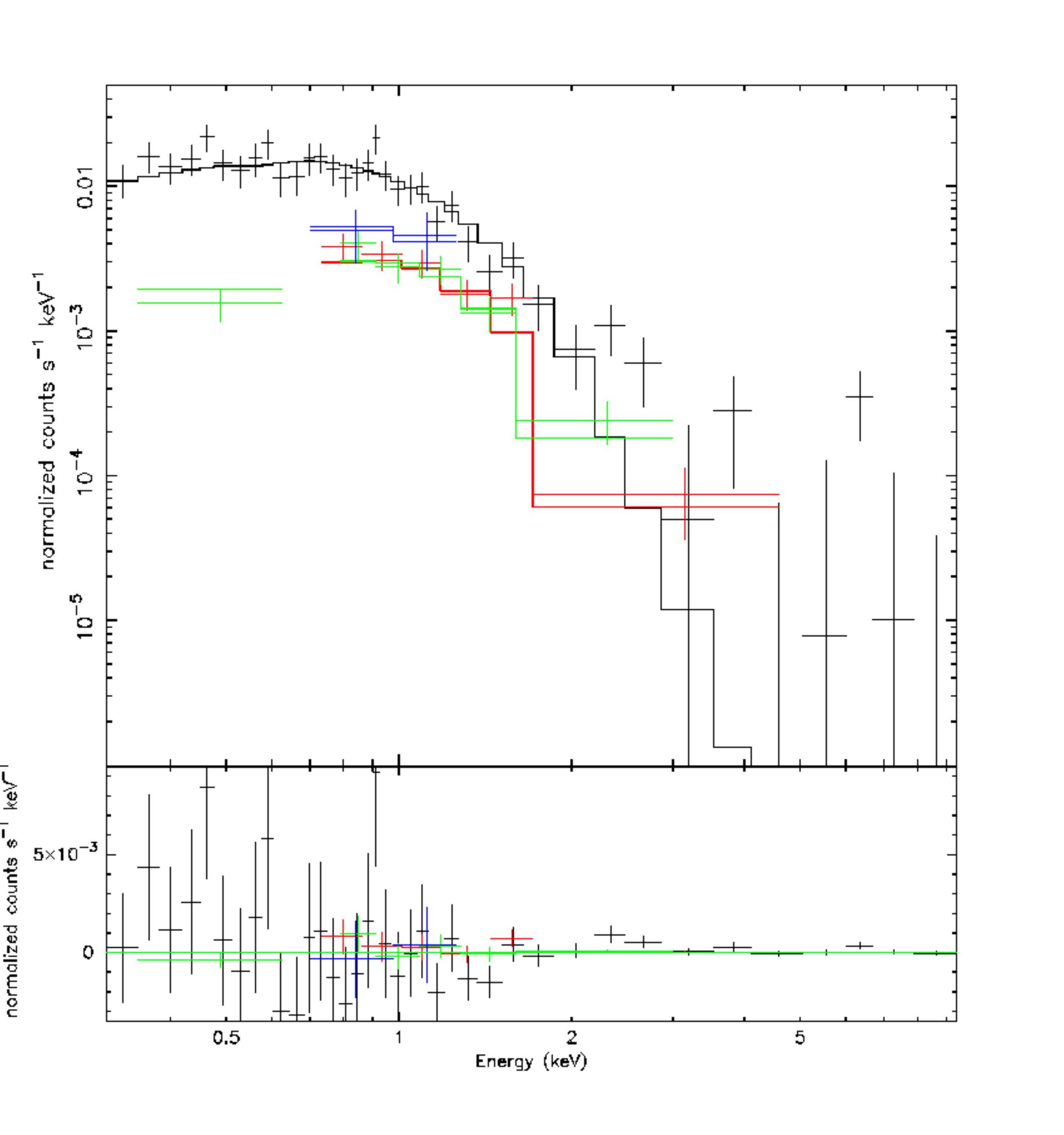}
\caption{PSR J1614-2230 Spectrum. Different colors mark the different instruments used (see text for details). \label{J1614-sp}}
Residuals are shown in the lower panel.
\end{figure}

\clearpage

{\bf J1648-4611 - type 0 RLP} % Nuova!

J1648-4611 is a radio pulsar found during the Parkes Multibeam Pulsar Survey (Kramer M. et al. 2003).
Radio dispersion measurements place the pulsar at a distance of $\sim$ 5.71 kpc.

A {\it Chandra} observation was pointed on the 3-months {\it Fermi} source position (obs. id 11836)
on 2010, January 24 at 03:00:42 UT for a total exposure of 10.1 ks.
No X-ray source was found at the radio position of the pulsar.
For a distance of 5.71 kpc we found a
rough absorption column value of 1 $\times$ 10$^{22}$ cm$^{-2}$
and using a simple powerlaw spectrum
for PSR+PWN with $\Gamma$ = 2 and a signal-to-noise of 3,
we obtained an upper limit
non-thermal unabsorbed flux of 2.16 $\times$ 10$^{-14}$ erg/cm$^2$ s,
that translates into a luminosity of L$_{5.71kpc}$ = 8.47 $\times$ 10$^{31}$ erg/s.

{\bf J1658-5324 - type 0 RL MSP} % Nuova! osservazione in arrivo

J1658-5324 is an isolated millisecond pulsar found by the Pulsar Search
Consortium. Its distance obtained from radio dispersion measurements is
$\sim$ 0.9 kpc.

Only a {\it SWIFT} observation is available for PSR J1658-5324
(obs. id 00031703001, 2.9 ks exposure).
No X-ray source was found at the radio position of the pulsar.
For a distance of 0.9 kpc, we found a rough absorption column value of 2 $\times$ 10$^{21}$ cm$^{-2}$
and using a simple powerlaw spectrum
for PSR+PWN with $\Gamma$ = 2 and a signal-to-noise of 3,
we obtained an upper limit
non-thermal unabsorbed flux of 2.81 $\times$ 10$^{-13}$ erg/cm$^2$ s,
that translates in an upper limit luminosity L$_{0.9kpc}^{nt}$ = 2.73 $\times$ 10$^{31}$ erg/s.

{\bf J1709-4429 - type 2 RLP}

% McGowan et al. 2004
PSR B1706-44 is a young ($\tau_c$ = 1.7 $\times$ 10$^4$ yr), energetic
102 ms pulsar originally discovered by Johnston et al. (1992).
It is one of several sources with spin-down ages of 10$^4$-10$^5$ yr
that are referred to as Vela-like pulsars because of their similar
emission properties. The source is
known to display glitches (Johnston et al. 1995) and is
plausibly associated with the supernova remnant G343.1-2.3
(McAdam, Osborne \& Parkinson 1993).
According to the Taylor \& Cordes (1993) model, the
pulsar is $\sim$ 2.3 kpc away, which agrees with the kinematic
distance in the range 2.4-3.2 kpc inferred from H I absorption
(Koribalski et al. 1995). Very Large Array images (Frail et al. 1994; Giacani et al. 2001) indicate that the pulsar
is located inside a synchrotron plerionic nebula about 3.5$"$ $\times$ 2.5$"$
in size. Evidence for a more extended X-ray compact nebula
(with radius $\sim$ 20$"$-30$"$) was also found in the ROSAT HRI and
{\it Chandra} images (Finley et al. 1998; Dodson \& Golap 2002).
An unpulsed X-ray source at the radio pulsar position had
been detected with the ROSAT PSPC (Becker, Brazier \&
Trumper 1995), ASCA (Finley et al. 1998), and {\it BeppoSAX}
(Mineo et al. 2002). More recently, deeper {\it Chandra} observations
have been presented by Gotthelf, Halpern \& Dodson
(2002). These authors discovered a broad sinusoidal X-ray
pulsation at the radio period, with a pulsed fraction of
23\% $\pm$ 6\%. The phasing of the radio pulse was consistent with
that of the center of the broad X-ray peak. The high spectral and
spatial resolution of {\it Chandra} allowed a multicomponent fit of
the X-ray spectrum, revealing the presence of a thermal component;
the X-ray spectrum was found to be well fitted with a
two-component model (blackbody plus powerlaw).
The blackbody radius determined from the
model parameters suggests that the emission is from a hot spot.
No $\gamma$-ray nebular emission was detected by {\it Fermi} down to a flux of 
3.94 $\times$ 10$^{-11}$ erg/cm$^2$s (Ackermann et al. 2010).

\begin{figure}
\centering
\includegraphics[angle=0,scale=.40]{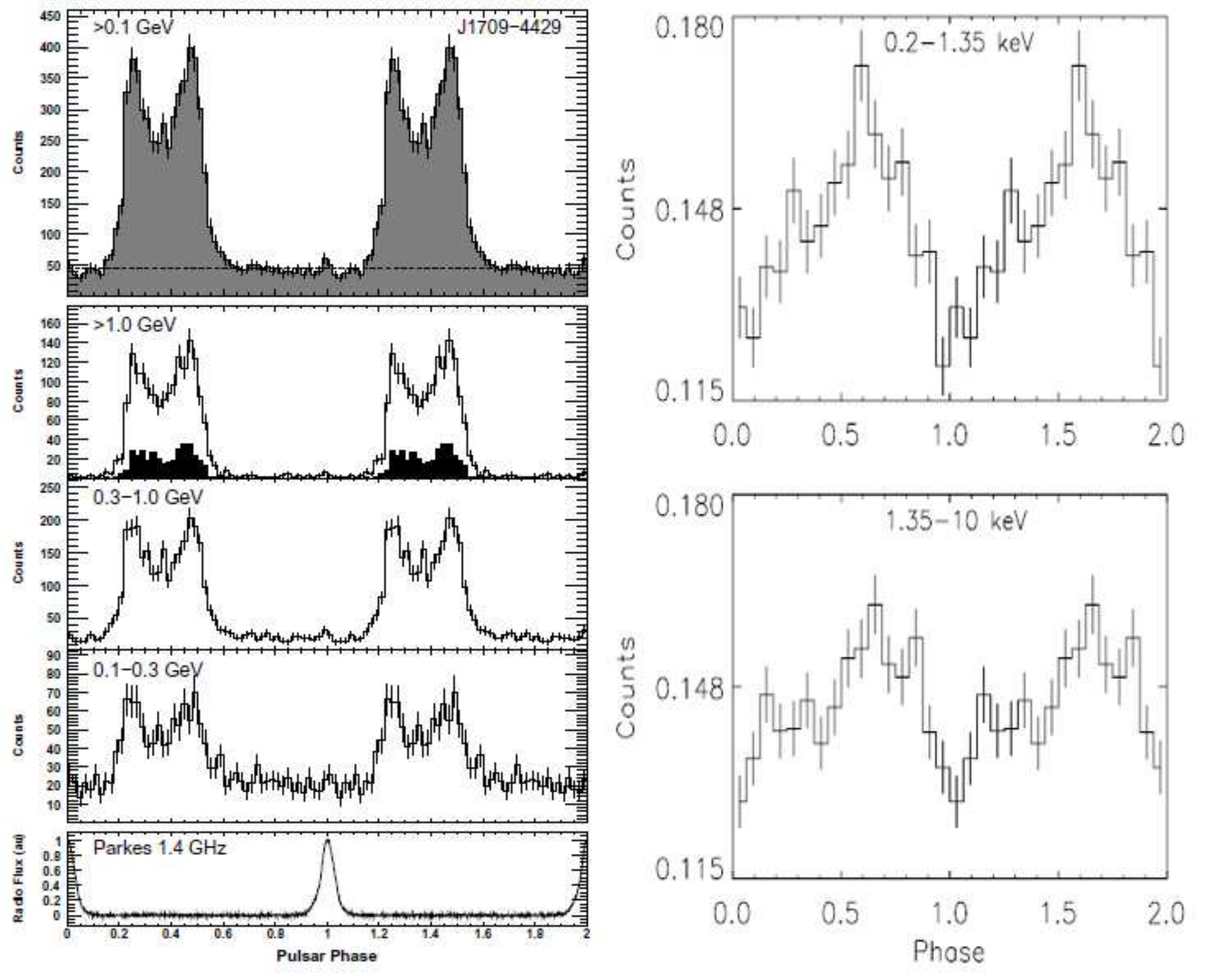}
\caption{PSR J1709-4429 Lightcurve. {\it Left: Fermi} $\gamma$-ray lightcurve folded with Radio
(Abdo et al. 2009c).  {\it Right: XMM-Newton} pulse profiles for combined PN and MOS2 data
in different energy ranges. Top: 0.2-1.35 keV. Bottom: 1.35-10.0 keV. See McGowan et al. 2004
for details.
\label{J1709-lc}}
\end{figure}

Three different X-ray observations of J1709-4429 were performed, two
by {\it XMM-Newton} and one by {\it Chandra}:\\
- obs. id 4608, {\it Chandra} ACIS-I very faint mode, start time 2004, February 01 at 02:05:30 UT, exposure 100.1 ks;\\
- obs. id 0112200601, {\it XMM-Newton} observation, start time 2002, March 12 at 09:31:17.32 UT, exposure 37.3 ks;\\
- obs. id 0112200701, {\it XMM-Newton} observation, start time 2002, March 13 at 21:18:43.31 UT, exposure 44.2 ks.\\
In both the XMM observations the PN camera was operating in the Small Window mode and MOS1 camera was
operating in the Full Frame mode. The MOS2 camera was operating in Fast Uncompressed mode so that
it can't be used to extract an image (and a spectrum) of the source
For the PN camera a thin optical filter was used while for the MOS cameras the medium optical filter was used.
The X-ray source best fit position, obtained by using the celldetect tool inside the CIAO
software, is 17:09:42.72 -44:29:08.69 (1$"$ error radius).
First, an accurate
screening for soft proton flare events was done in the {\it XMM-Newton} observations obtaining a resulting total
exposure of 52.8 ks in the two observations.
A nebular emission of radius $\sim$ 25$"$ is apparent in the {\it Chandra} observation.
For the {\it Chandra} observation, we chose a 1.5$"$ radius circular region for the
pulsar spectrum in order to minimize the contribution of the pulsar wind nebula
while the nebular spectrum was extracted by an annular region 
with radii of 1.5 and 25$"$, respectively. The background is extracted from a circular source-free region
away from the source, in order to exclude both the nebular emission and the edge of the CCD.
For the {\it XMM-Newton} observation, we chose a 30$"$ radius circular region around
the pulsar in order to take in account both the pulsar and the nebular emission;
the background is extracted from a circular source-free region on the same CCD,
away from the source.
We obtained a total of 5263 and 2091 counts from respectively observation 1 PN and MOS cameras
(background contributions of 16.7\% and 8.5\%), 7791 and 3206 counts from {\it XMM-Newton} observation 2
(background contributions of 17.9\% and 8.1\%) and 2829 and 3584 counts from {\it Chandra} pulsar
and nebular emission (background contributions of 1.0\% and 5.5\%).
The best fitting pulsar model is a combination of a blackbody and a powerlaw
(probability of obtaining the data if the model is correct 
- p-value - of 0.35, 828 dof using both the pulsar and nebular spectra).
The powerlaw component has a photon index $\Gamma$ = 1.88 $\pm$ 0.21
absorbed by a column N$_H$ = 4.56$_{-0.29}^{+0.44}$ $\times$ 10$^{21}$ cm$^{-2}$.
The thermal component has a temperature of T = 1.93 $\pm$ 0.14 $\times$ 10$^6$ K.
The blackbody radius is R$_{2.5kpc}$ = 4.30$_{-0.86}^{+1.72}$ km.
The nebular emission has a photon index $\Gamma$ = 1.44 $\pm$ 0.05.
A simple powerlaw or blackbody model are not statistically acceptable.
Assuming the best fit model, the 0.3-10 keV unabsorbed thermal pulsar flux is
5.26$_{-1.31}^{+0.50}$ $\times$ 10$^{-13}$, the non-thermal flux is 
3.78$_{-0.94}^{+0.37}$ $\times$ 10$^{-13}$, and the non-thermal nebular flux is
8.36$_{-0.67}^{+0.52}$ $\times$ 10$^{-13}$ erg/cm$^2$ s. 
Using a distance
of 2.5 kpc, the luminosities are L$_{2.5kpc}^{bol}$ = 3.94$_{-0.98}^{+0.37}$ $\times$ 10$^{32}$,
L$_{2.5kpc}^{nt}$ = 2.83$_{-0.70}^{+0.28}$ $\times$ 10$^{32}$ and L$_{2.5kpc}^{pwn}$ = 6.27$_{-0.50}^{+0.39}$ $\times$ 10$^{32}$ erg/s.

\begin{figure}
\centering
\includegraphics[angle=0,scale=.50]{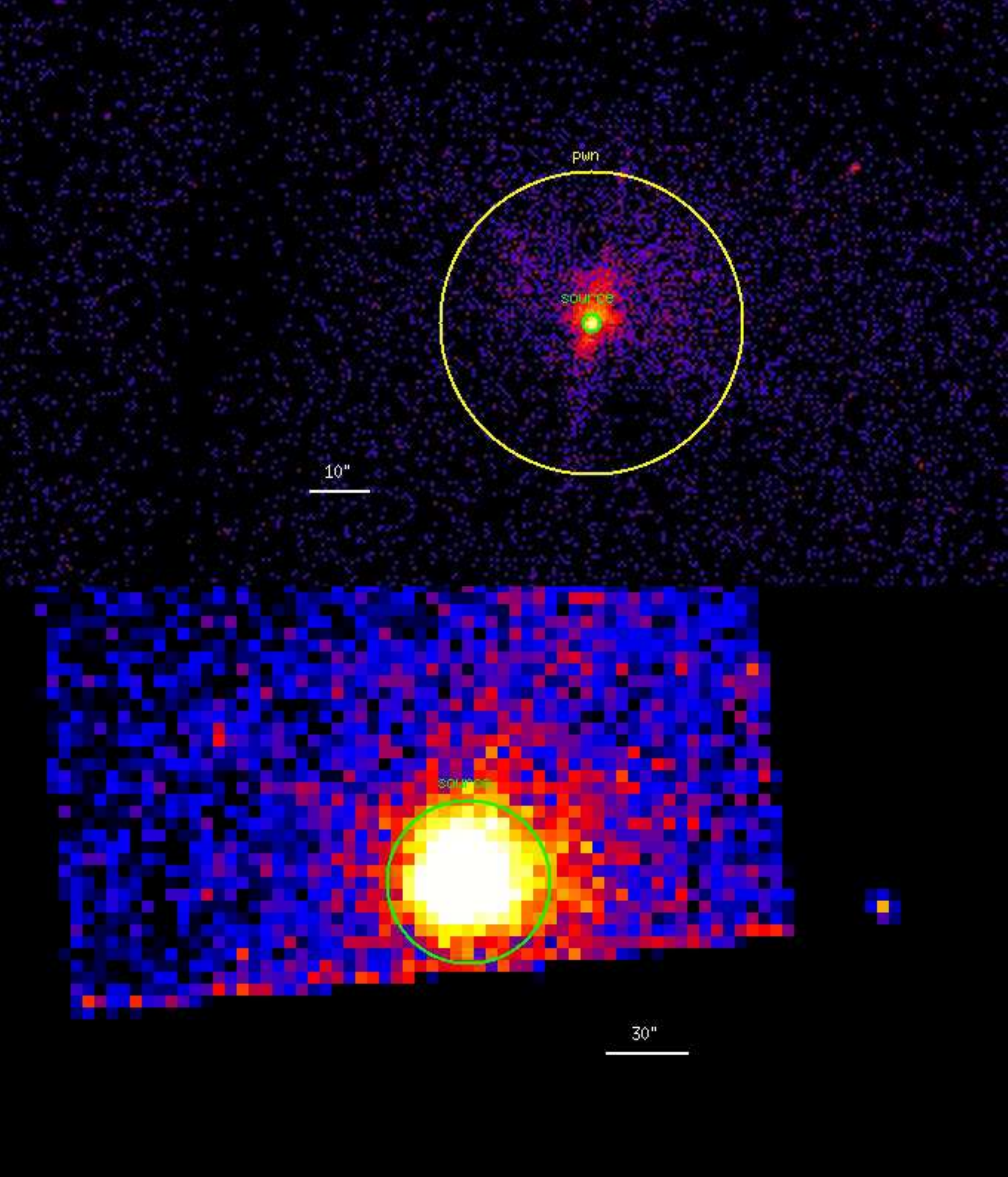}
\caption{{\it Upper Panel:} PSR J1709-4429 0.3-10 keV {\it Chandra} Imaging.
The green circle marks the pulsar while the yellow annulus the nebular region used in the analysis.
{\it Lower Panel:} PSR J1709-4429 0.3-10 keV {\it XMM-Newton} EPIC Imaging. The PN and the two MOS images have been added. 
The green circle marks the source region used in the analysis.
\label{J1709-im}}
\end{figure}

\begin{figure}
\centering
\includegraphics[angle=0,scale=.50]{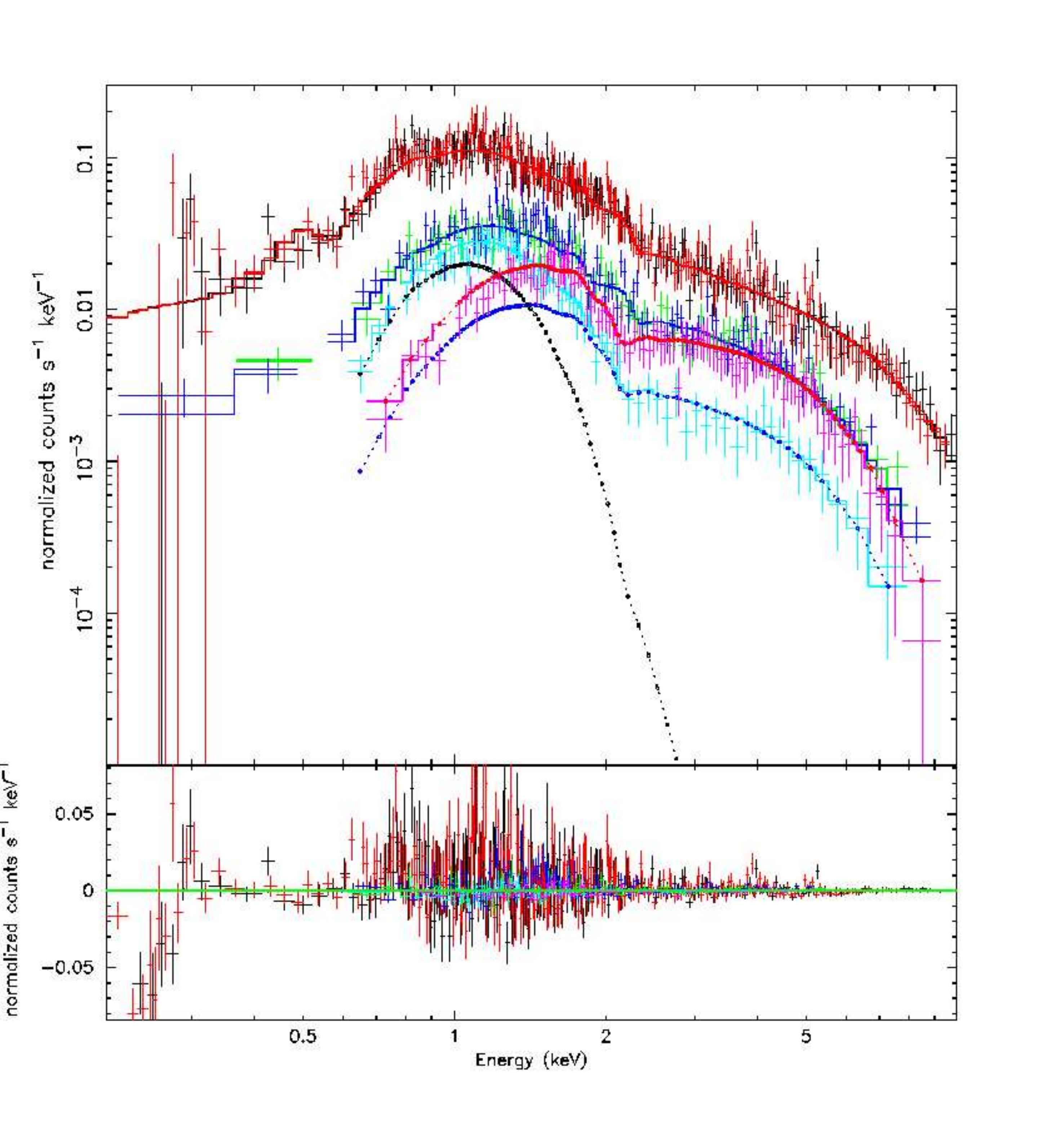}
\caption{PSR J1709-4429 Spectrum. Different colors mark all the different dataset used (see text for details).
Blue points mark the powerlaw component while black points the thermal component of the pulsar spectrum.
Red points mark the nebular spectrum.
Residuals are shown in the lower panel.
\label{J1709-sp}}
\end{figure}

\clearpage

{\bf J1713+0747 - type 0 RL MSP} % Nuova!

J1713+0747 is a 4.6ms millisecond pulsar found by Foster et al. 1993 by using the Arecibo Telescope.
Radio dispersion measurements placed the pulsar at $\sim$ 1.05 kpc.

No X-ray observations were performed at the radio position of the pulsar.

{\bf J1718-3825 - type 2 RLP} % c'è una nuova osservazione chandra e anche una nuova distanza presa da Weltevrede, osservazione in arrivo

% Weltevrede et al. 2010
PSR J1718-3825 is a 75 ms pulsar discovered
by Manchester et al. (2001) at radio wavelengths.
It has an associated X-ray nebula
(Hinton et al. 2007) and an associated HESS
source (Aharonian et al. 2007). Its distance derived
from the DM is 3.82 $\pm$ 1.15 kpc according to
Manchester et al. 2001.
No $\gamma$-ray nebular emission was detected by {\it Fermi} down to a flux of 
9.44 $\times$ 10$^{-12}$ erg/cm$^2$s (Ackermann et al. 2010).

A total of two different observations of J1718-3825 have been performed:
- obs. id 0401960101, {\it XMM-Newton} observation, start time 2006, September 04 at 15:13:23 UT, exposure 22.3 ks;\\
- obs. id 9079, {\it Chandra} ACIS-I very faint mode, start time 2009, January 30 at 03:09:03 UT, exposure 100.0 ks.\\
In the XMM observation both the PN and MOS cameras were operating in Full Frame mode and
a medium optical filter was used.
The X-ray source best fit position is 17:18:13.55 -38:25:17.68 (1$"$ error radius),
obtained by using the celldetect tool inside the CIAO software.
First, an accurate
screening for soft proton flare events was done in the {\it XMM-Newton} observations with a resulting
exposure of 13.1 ks.
A nebular emission of radius $\sim$ 10$"$ is apparent in the {\it Chandra} observation,
probably extending up to 15$"$.

\begin{figure}
\centering
\includegraphics[angle=0,scale=.30]{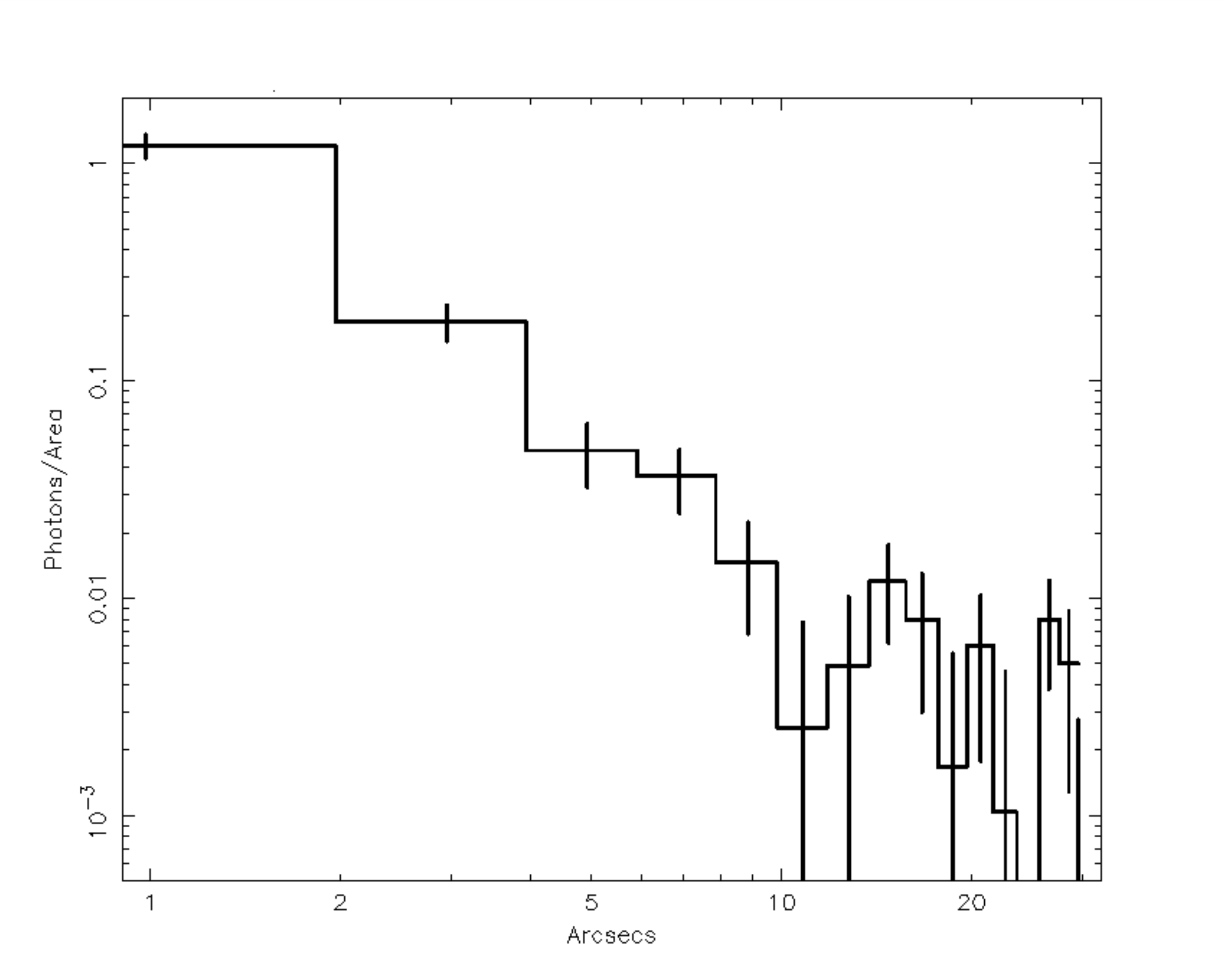}
\caption{PSR J1718-3825 {\it Chandra} radial profile (0.3-10 keV energy range).
{\it Chandra} ACIS On-axis Point Spread Function is almost zero over 2$"$ from the source so that
it's apparent the nebular emission between 2 and 10-15$"$ from the pulsar.
\label{J1718-psf}}
\end{figure}

For the {\it Chandra} observation, we chose a 2$"$ radius circular region for the
pulsar spectrum while the nebular spectrum was extracted from an annular region 
with radii of 2 and 15$"$, respectively. The background was extracted from
an annular region with radii of 15 and 30$"$, respectively.
For the {\it XMM-Newton} observation, we chose a 20$"$ radius circular region around
the pulsar in order to take in account both the pulsar and the nebular emission;
the background is extracted from a circular source-free region on the same CCD,
away from the source, in order to avoid the contamination from a faint source near the pulsar.
we obtained a total of 388, 77, 85, 45 and 83 counts from XMM spectra (PN, MOS1 and 2),
{\it Chandra} pulsar and nebular spectrum (background contributions of 18.5\%, 8.1\%, 18.1\%, 0.9\% and 38.3\%).
Due to the low statistic, we used the C-statistic
approach implemented in XSPEC, well suited to study sources with low photon statistics.
The best fitting pulsar model is a simple powerlaw (reduced chisquare $\chi^2_{red}$ =
0.93, 126 dof using both the pulsar and nebular spectra) with a photon index 
$\Gamma$ = 1.55 $\pm$ 0.48 absorbed by a column N$_H$ = 4.07$_{-1.55}^{+1.46}$ $\times$ 10$^{21}$ cm$^{-2}$.
The nebular emission has a photon index $\Gamma$ = 1.18 $\pm$ 0.40.
A simple blackbody model is not statistically acceptable while a composite
blackbody plus powerlaw spectrum gives no relevant statistical improvement.
Assuming the best fit model, the 0.3-10 keV unabsorbed pulsar flux is 
1.18$_{-0.97}^{+0.58}$ $\times$ 10$^{-13}$, and the nebular flux is
1.33$_{-0.95}^{+0.55}$ $\times$ 10$^{-13}$ erg/cm$^2$ s. 
Using a distance
of 3.8 kpc, the luminosities are 
L$_{3.8kpc}^{nt}$ = 2.04$_{-1.68}^{+1.00}$ $\times$ 10$^{32}$ and L$_{3.8kpc}^{pwn}$ = 2.30$_{-1.64}^{+0.95}$ $\times$ 10$^{32}$ erg/s.

\begin{figure}
\centering
\includegraphics[angle=0,scale=.50]{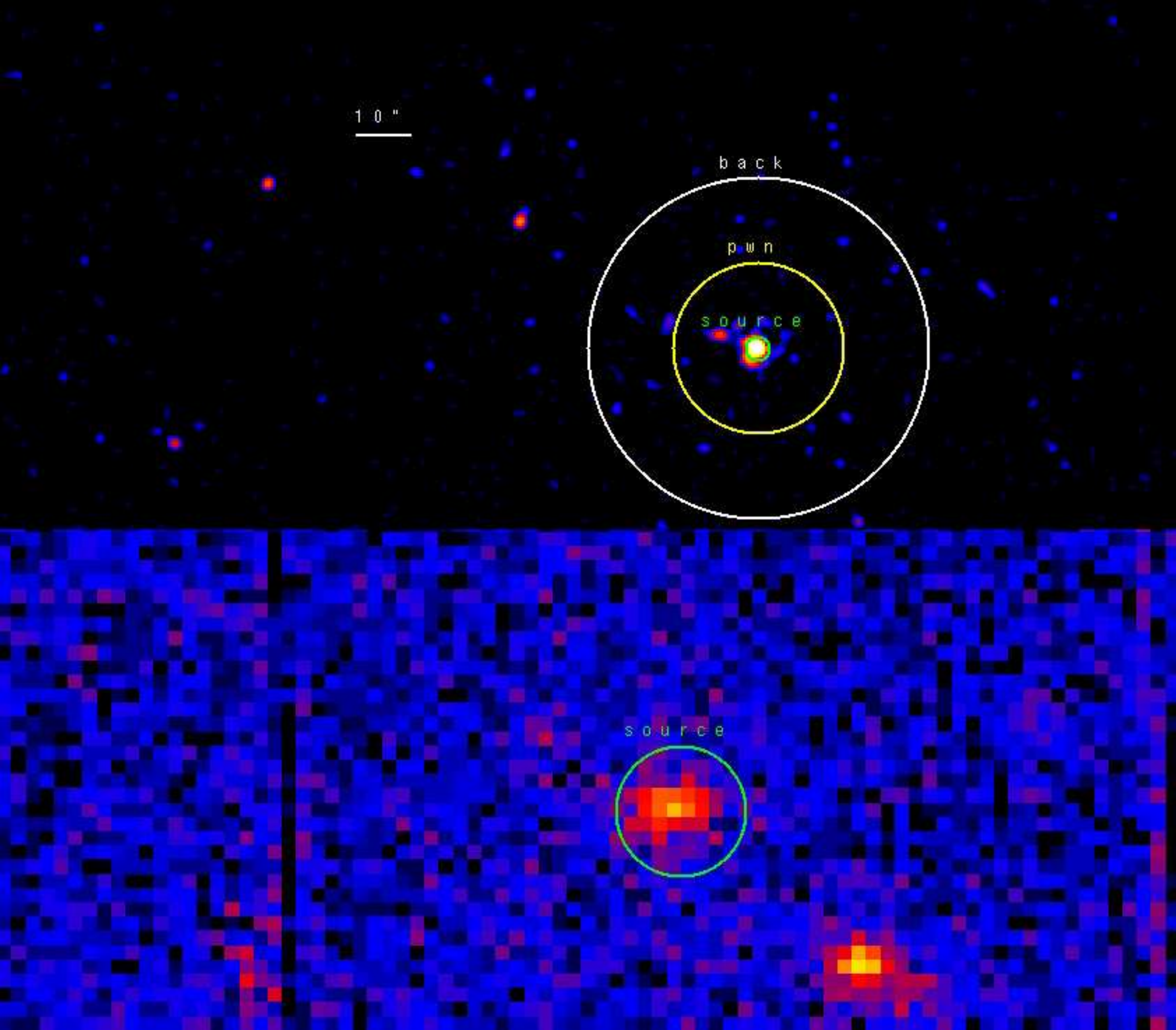}
\caption{{\it Upper Panel:} PSR J1718-3825 0.3-10 keV {\it Chandra} Imaging.
The green circle marks the pulsar, the yellow annulus the nebular region,
while the white annulus the background region used in the analysis.
{\it Lower Panel:} PSR J1718-3825 0.3-10 keV {\it XMM-Newton} EPIC Imaging. The PN and the two MOS images have been added. 
The green circle marks the source region used in the analysis.
\label{J1718-im}}
\end{figure}

\begin{figure}
\centering
\includegraphics[angle=0,scale=.50]{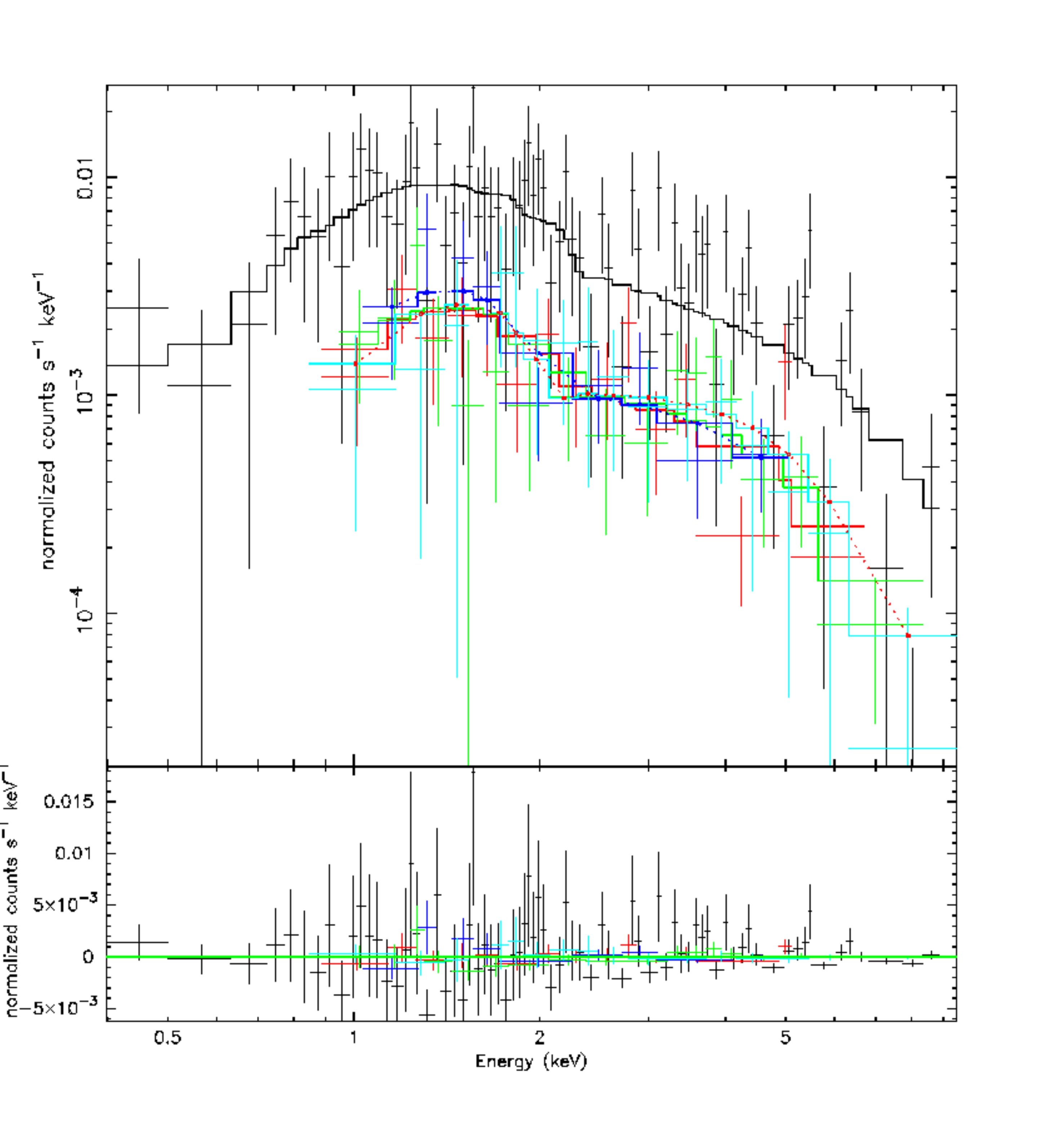}
\caption{PSR J1718-3825 Spectrum. Different colors mark all the different dataset used (see text for details).
Black points mark the pulsar spectrum while red points mark the nebular spectrum.
Residuals are shown in the lower panel.
\label{J1718-sp}}
\end{figure}

\clearpage

{\bf J1730-3350 - type 0 RLP} % Nuova!

PSR J1730-3350 (B1727-33) has a period of 139 ms and is young, with a characteristic
age of 26 kyr (Johnston et al. 1995). It's the central source of the SNR G354.1+00.1,
at a distance of $\sim$ 5 kpc. The pulsar distance derived from
radio dispersion measurements is $\sim$ 4.24 kpc (Becker \& Trumper 1997).

We used three different observations performed both by {\it XMM-Newton} and {\it Chandra}:\\
- obs. id 0112200801, {\it XMM-Newton} observation, start time 2002, February 26 at 19:05:10 UT, exposure 7.2 ks;\\
- obs. id 0112201401, {\it XMM-Newton} observation, start time 2002, September 09 at 23:08:38 UT, exposure 5.0 ks;\\
- obs. id 9080, {\it Chandra} ACIS-I very faint mode, start time 2008, October 27 at 16:22:56 UT, exposure 10.1 ks.\\
In the XMM observations both the MOS cameras were operating in Full Frame mode. Only in the second observation
also the PN camera was operative (in the Full Frame mode).
A medium optical filter was used for both PN and MOS cameras.
No source was found at the radio position of the pulsar.
Due to the presence of an extended thermal emission all the upper limits
were obtained by using the 2-10 keV energy band taking advantage of XSPEC to convert
the fluxes to 0.3-10 keV energy beam.
We used both the {\it Chandra} and {\it XMM-Newton} observations and adopted
the lowest limit obtained.

For a distance of 4.24 kpc we found a
rough absorption column value of 1 $\times$ 10$^{22}$ cm$^{-2}$
and using a simple powerlaw spectrum
for PSR+PWN with $\Gamma$ = 2 and a signal-to-noise of 3,
we obtained an upper limit non-thermal unabsorbed flux of 5.62 $\times$ 10$^{-15}$ erg/cm$^2$ s,
that translates in an upper limit luminosity L$_{4.24kpc}^{nt}$ = 1.21 $\times$ 10$^{31}$ erg/s.

{\bf J1732-31 - type 2 RQP} % osservazione in arrivo

J1732-31 was one of the first pulsars discovered using the
blind search technique (Abdo et al. 2009).
No $\gamma$-ray nebular emission was detected down to a flux of 
8.19 $\times$ 10$^{-12}$ erg/cm$^2$s (Ackermann et al. 2010).
The pseudo-distance of the object based on $\gamma$-ray data (Saz Parkinson et al. (2010))
is $\sim$ 0.6 kpc.

After the {\it Fermi} detection, we asked for a {\it SWIFT} observation
of the $\gamma$-ray error box (obs id. 00031369001-00090194001, 13.09 ks exposure).
A {\it Chandra} observation was then requested by the collaboration (obs. id 11125, starting on
2010, July 25 at 18:40:02 UT, exposure of 20.0 ks).
By using the celldetect tool inside the CIAO software we found an X-ray source
at 17:32:33.56 -31:31:23.95 (0.9$"$ error radius), that was confirmed by timing analysis
to be the counterpart of the $\gamma$-ray pulsar (Ray et al. 2011).
No nebular emission is present around the pulsar.
The pulsar spectrum was extracted from a 2$"$ radius circular region around the pulsar
while the background was extracted from an annular region around the pulsar.
We obtained 68 pulsar counts (background contribution of 1.2\%).
Due to the low statistic we chose to use the C-statistic.
The best fit model for the pulsar is a simple powerlaw
(chisquare value $\chi^2_{red}$ = 1.38, 10 dof)
with a photon index $\Gamma$ = 2.01$_{-0.62}^{+0.90}$ absorbed by a column
N$_H$ = 9.39$_{-9.39}^{+28.58}$ $\times$ 10$^{20}$ cm$^{-2}$.
A simple blackbody model is not statistically allowed while a composite model
cannot be studied due to the low statistic.
Assuming the best fit model, the 0.3-10 keV unabsorbed thermal pulsar flux is 
3.69 $\pm$ 1.30 $\times$ 10$^{-14}$ erg/cm$^2$ s. Using a distance
of 600 pc, the pulsar luminosity is L$_{0.6kpc}^{ny}$ = 1.59 $\pm$ 0.56 $\times$ 10$^{30}$ erg/s.

\begin{figure}
\centering
\includegraphics[angle=0,scale=.50]{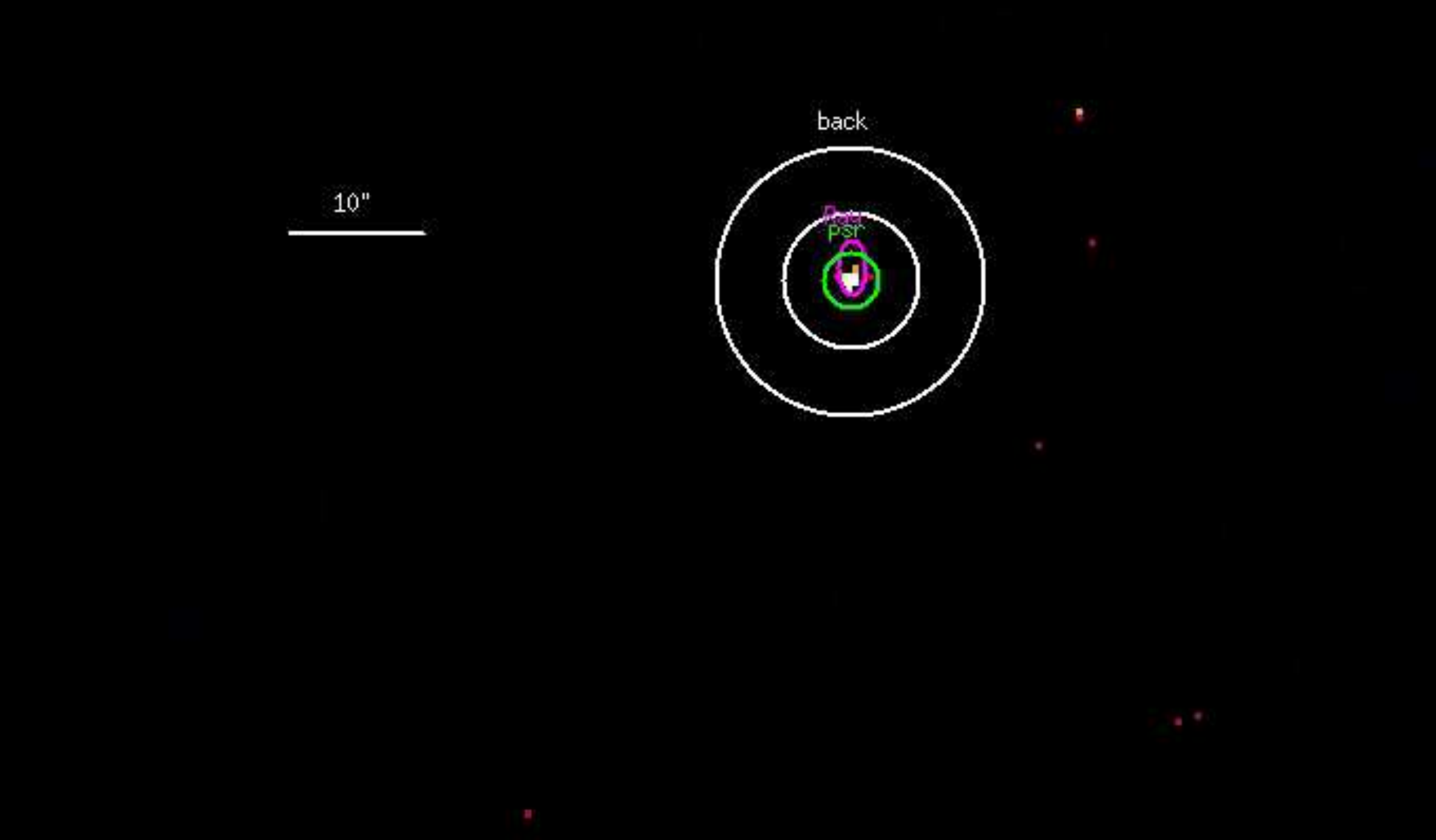}
\caption{PSR J1732-31 {\it Chandra} Imaging.
The green circle marks the pulsar while the white annulus the background region used in the analysis.
\label{J1732-im}}
\end{figure}

\begin{figure}
\centering
\includegraphics[angle=0,scale=.50]{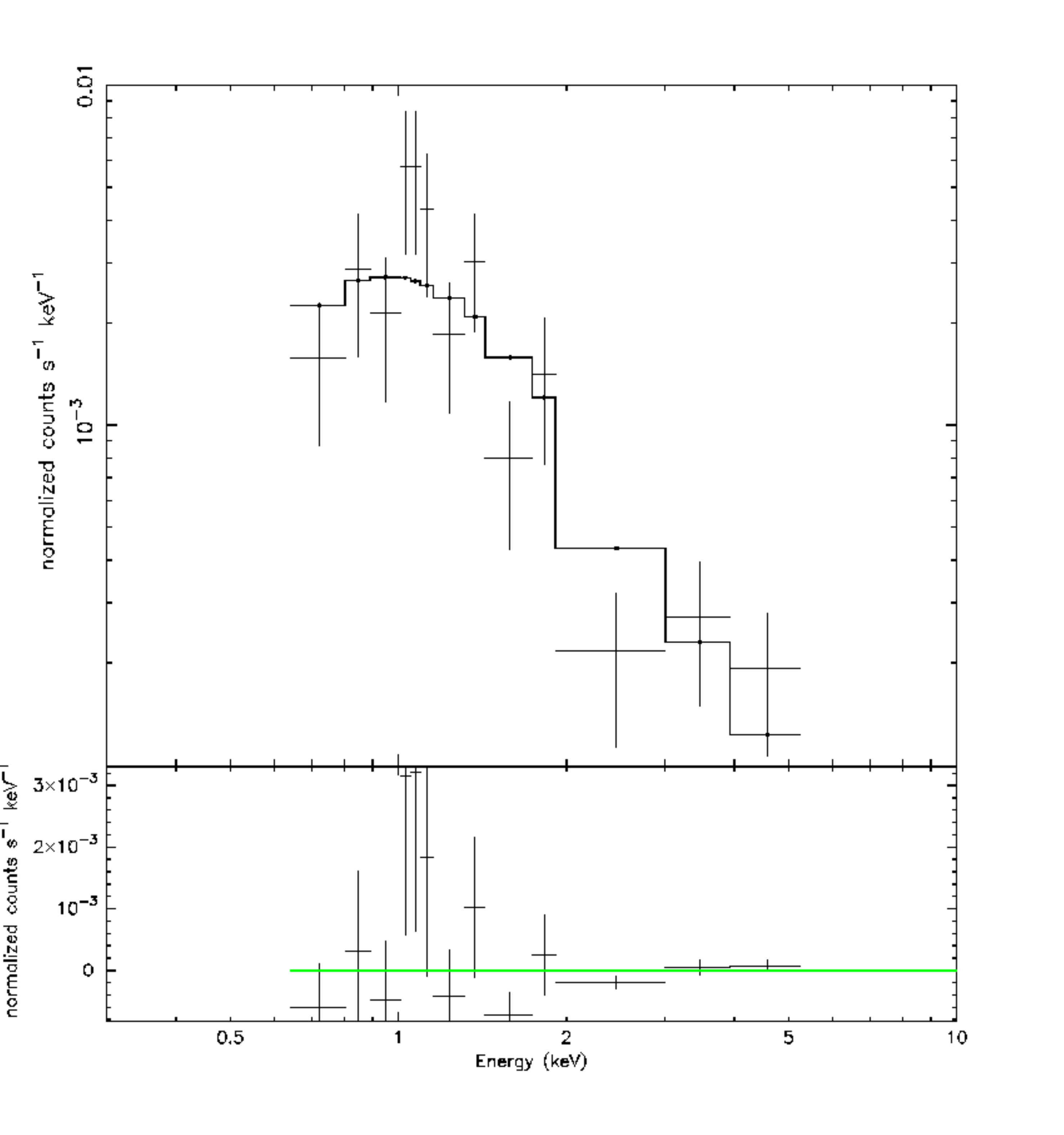}
\caption{PSR J1732-31 {\it Chandra} Spectrum (see text for details).
Residuals are shown in the lower panel.
\label{J1732-sp}}
\end{figure}

\clearpage

{\bf J1741-2054 - type 2 RLP}

% Romani et al. 2010
PSR J1741-2054, discovered in a blind pulsation search of
a {\it Fermi} point source (Abdo et al. 2009), is a P = 413ms,
characteristic age $\tau_c$ = 391 kyr $\gamma$-ray pulsar. The pulsar
was then detected in archival Parkes radio observations and
subsequently studied at GBT (Camilo et al. 2009), showing
it to be very faint and heavily affected by scintillation. The radio observations
also gave a remarkably low dispersion measure
DM = 4.7 pc cm$^{-3}$, which, for a standard Galactic electron
density model (Taylor \& Cordes 1993), implies a distance of
$\sim$ 0.38 kpc. This makes PSR J1741-2054 one of the closest energetic pulsars
known. At this distance, the low flux density observed
at pulsar discovery, S$_{1.4}$ $\sim$ 160 μJy, also makes it the least
luminous radio pulsar known. Indeed, since the source scintillates
to undetectability at many epochs, the time average radio
luminosity is L$_{1.4}$ $<$ 0.025 d$^2_{0.4}$ mJy kpc$^2$. The low distance
was already suspected from the relatively high $\gamma$-ray flux (for
the modest $\dot{E}$ = 9.4 $\times$ 10$^{33}$ erg s$^{-1}$ spin-down luminosity). At
the DM-estimated distance, the pulsar would emit 28\% of its
spin-down power in $\gamma$-rays, if the radiation were isotropic.
No $\gamma$-ray nebular emission was detected by {\it Fermi} down to a flux of 
1.16 $\times$ 10$^{-11}$ erg/cm$^2$s (Ackermann et al. 2010).

\begin{figure}
\centering
\includegraphics[angle=0,scale=.50]{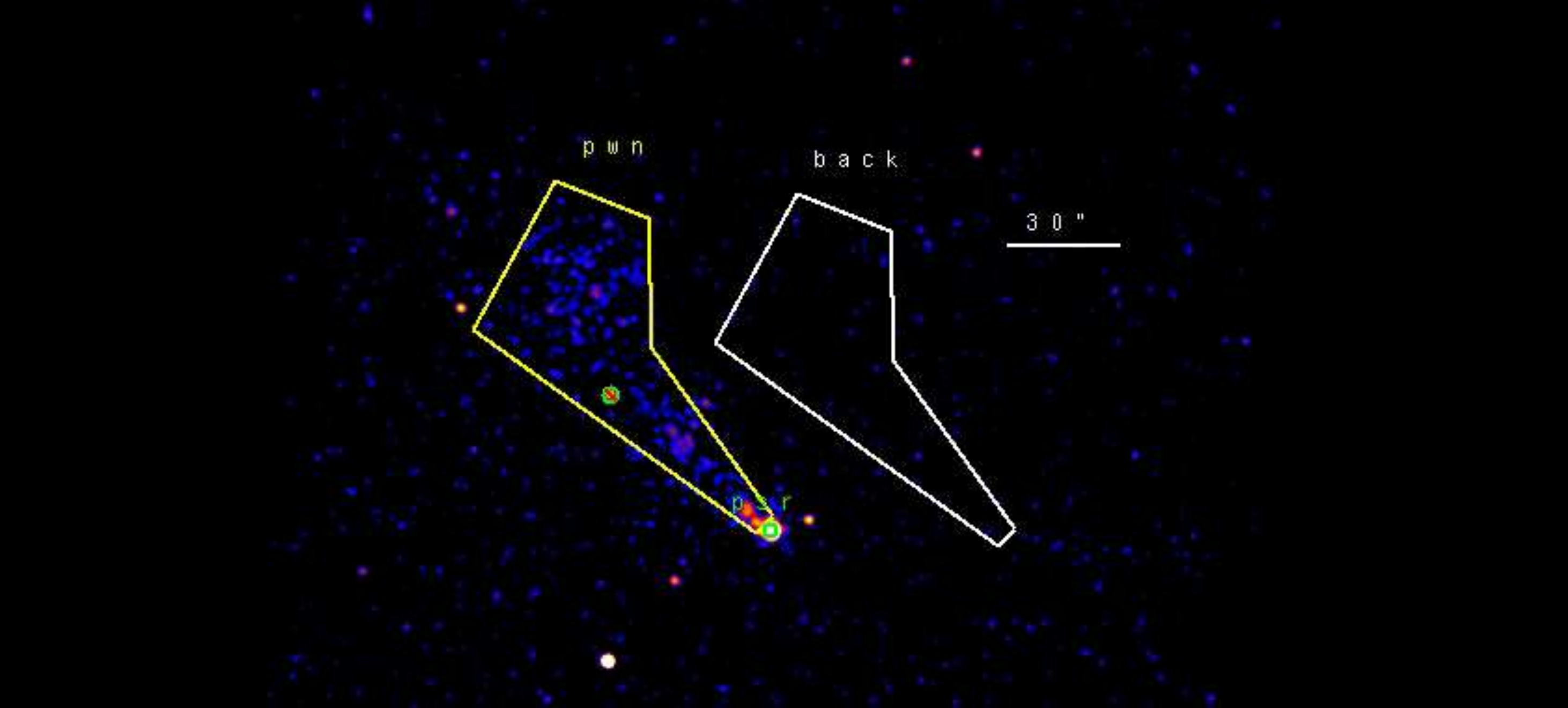}
\caption{PSR J1741-2054 0.3-10 keV {\it Chandra} Imaging. The image has been smoothed with a Gaussian
with Kernel radius of $3"$. The green circle marks the pulsar while the yellow polygon the nebular region used in the analysis.
The white polygon marks the background region used in the analysis.
\label{J1741-im}}
\end{figure}

We detected the J1741-2054 X-ray counterpart using {\it SWIFT} data (Abdo et al.(2009b)).
Later, Sivakoff G. obtained a long {\it Chandra} observation of this pulsar (Sivakoff et al. in preparation).
The {\it Chandra} observation started on 2010, May 21 at 02:25:03 UT for a total exposure
of 50.0 ks.
In this observation (obs. id 11251) the pulsar position was imaged on the back-illuminated ACIS
S3 chip and the VFAINT exposure mode was adopted. The off-axis angle is negligible.
The X-ray source best fit position (obtained by using the celldetect
tool inside the CIAO distribution) is 17:41:57.30 -20:54:12.51 (1$"$ error radius).
A nebular trail emission is apparent in the {\it Chandra} observation.
We chose a 2$"$ radius circular region for the
pulsar spectrum
while the nebular spectrum was extracted by an ad-hoc region (see figure \ref{J1741-im})
in order to maximize the signal-to-noise ratio (and a pointlike serendipitous source was removed).
The background is extracted from a region
away from the source, with the same shape of the nebular region in order to minimize
every possible systematic error.
We obtained a total of 3396 and 1349 counts from the pulsar and the trail
(background contributions of less than 0.1\% and 46.1\%).
The best fitting pulsar model is a combination of a blackbody and a powerlaw
(probability of obtaining the data if the model is correct 
- p-value - of 0.86, 138 dof using both the pulsar and nebular spectra).
The powerlaw component has a photon index $\Gamma$ = 2.71$_{-0.11}^{+0.14}$ 
absorbed by a column N$_H$ = 1.53$_{-0.36}^{+0.51}$ $\times$ 10$^{21}$ cm$^{-2}$.
The thermal component has a temperature of T = 8.29$_{-0.84}^{+0.80}$ $\times$ 10$^5$ K.
The blackbody radius R = 2.60$_{-1.64}^{+4.27}$ km determined from the
model parameters suggests that the emission is from a hot spot.
The nebular emission has a photon index $\Gamma$ = 1.67$_{-0.14}^{+0.22}$ .
A simple powerlaw or blackbody models are not statistically acceptable.
Assuming the best fit model, the 0.3-10 keV unabsorbed thermal pulsar flux is
4.83$_{-0.88}^{+0.26}$ $\times$ 10$^{-13}$, the non-thermal flux is 
6.24$_{-1.14}^{+0.34}$ $\times$ 10$^{-13}$, and the non-thermal nebular flux is
1.68$_{-0.34}^{+0.28}$ $\times$ 10$^{-13}$ erg/cm$^2$ s. 
Using a distance
of 0.38 kpc, the luminosities are L$_{0.38kpc}^{bol}$ = 8.37$_{-1.52}^{+0.45}$ $\times$ 10$^{30}$,
L$_{0.38kpc}^{nt}$ = 1.08$_{-0.20}^{+0.06}$ $\times$ 10$^{31}$ and L$_{0.38kpc}^{pwn}$ = 2.91$_{-0.59}^{+0.48}$ $\times$ 10$^{30}$ erg/s.

\begin{figure}
\centering
\includegraphics[angle=0,scale=.50]{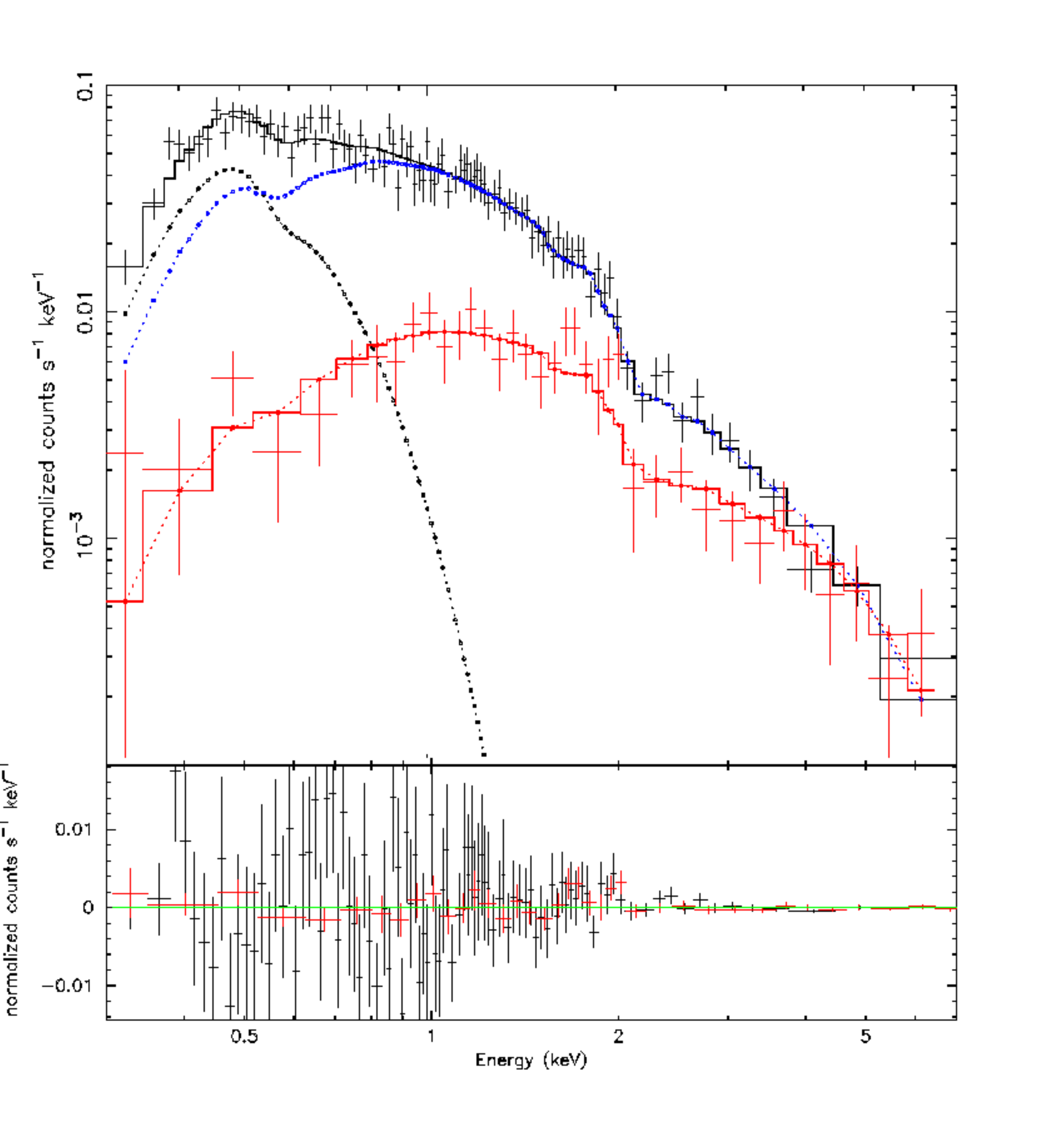}
\caption{PSR J1741-2054 Spectrum. Different colors mark all the different dataset used (see text for details).
Blue points mark the powerlaw component while black points the thermal component of the pulsar spectrum.
Red points mark the nebular spectrum.
Residuals are shown in the lower panel.
\label{J1741-sp}}
\end{figure}

\clearpage

{\bf J1744-1134 - type 0 RL MSP}  % rifatta con la cstatistic

% Becker & Trumper 1998
PSR J1744-1134 is a millisecond isolated pulsar. It was discovered
by Bailes et al. (1997) during the Parkes 436MHz survey
of the southern sky. J1744-1134 has a rotation period of
4.1 ms and a period derivative $\dot{P}$ = 7 $\times$ 10$^{-6}$,
implying a characteristic age of 9 $\times$ 10$^6$ yr and a rotational
energy loss of 4 $\times$ 10$^{33}$ erg/s. This millisecond pulsar
is relatively close to us: 
the radio parallax yielded a distance of
357$_{-35}^{+43}$ pc (Toscano et al., 1999).
No $\gamma$-ray nebular emission was detected by {\it Fermi} down to a flux of 
3.06 $\times$ 10$^{-11}$ erg/cm$^2$s (Ackermann et al. 2010).

Only one X-ray observation has J1744-1134 in its field of view.
This {\it Chandra} ACIS-S observation (obs. id 7646) started on 2007, July 22 at
00:50:09 UT for a total exposure of 64.1 ks.
In this observation (obs. id 11251) the pulsar position was imaged on the back-illuminated ACIS
S3 chip and the VFAINT exposure mode was adopted. The off-axis angle is negligible.
The X-ray source best fit position (obtained by using the celldetect
tool inside the Ciao distribution) is 17:44:29.40 -11:34:55.00 (0.95$"$ error radius).
No nebular emission was detected.
We used a 2$"$ radius circular region for the
pulsar spectrum and we extracted the background from an annular region with radii of 5 and 10$"$.
Due to the low statistic, we used the C-statistic
approach implemented in XSPEC.
we obtained a total of 299 pulsar counts with a background contribution of 0.4\%.
The best fitting pulsar model is a simple blackbody (reduced chisquare of
$\chi^2_{red}$ = 1.32) with a temperature of 
T = 3.31$_{-0.46}^{+0.50}$ $\times$ 10$^6$ K absorbed by a column
N$_H$ = 9.40$_{-9.40}^{+11.5}$ $\times$ 10$^{20}$ cm$^{-2}$.
The blackbody radius R$_{357pc}$ = 22.0$_{-11.6}^{+34.2}$ m,
pointing to an hot spot emission mechanism.
A simple powerlaw model is not statistically acceptable and a combination
of blackbody and powerlaw doesn't improve significatively the statistic.
Assuming the best fit model, the 0.3-10 keV unabsorbed thermal pulsar flux is
2.56$_{-1.14}^{+0.20}$ $\times$ 10$^{-14}$ erg/cm$^2$ s. 
Using a distance
of 357 pc, the pulsar luminosity is L$_{357pc}^{bol}$ = 3.91$_{-1.74}^{+0.31}$ $\times$ 10$^{29}$ erg/s.

\begin{figure}
\centering
\includegraphics[angle=0,scale=.40]{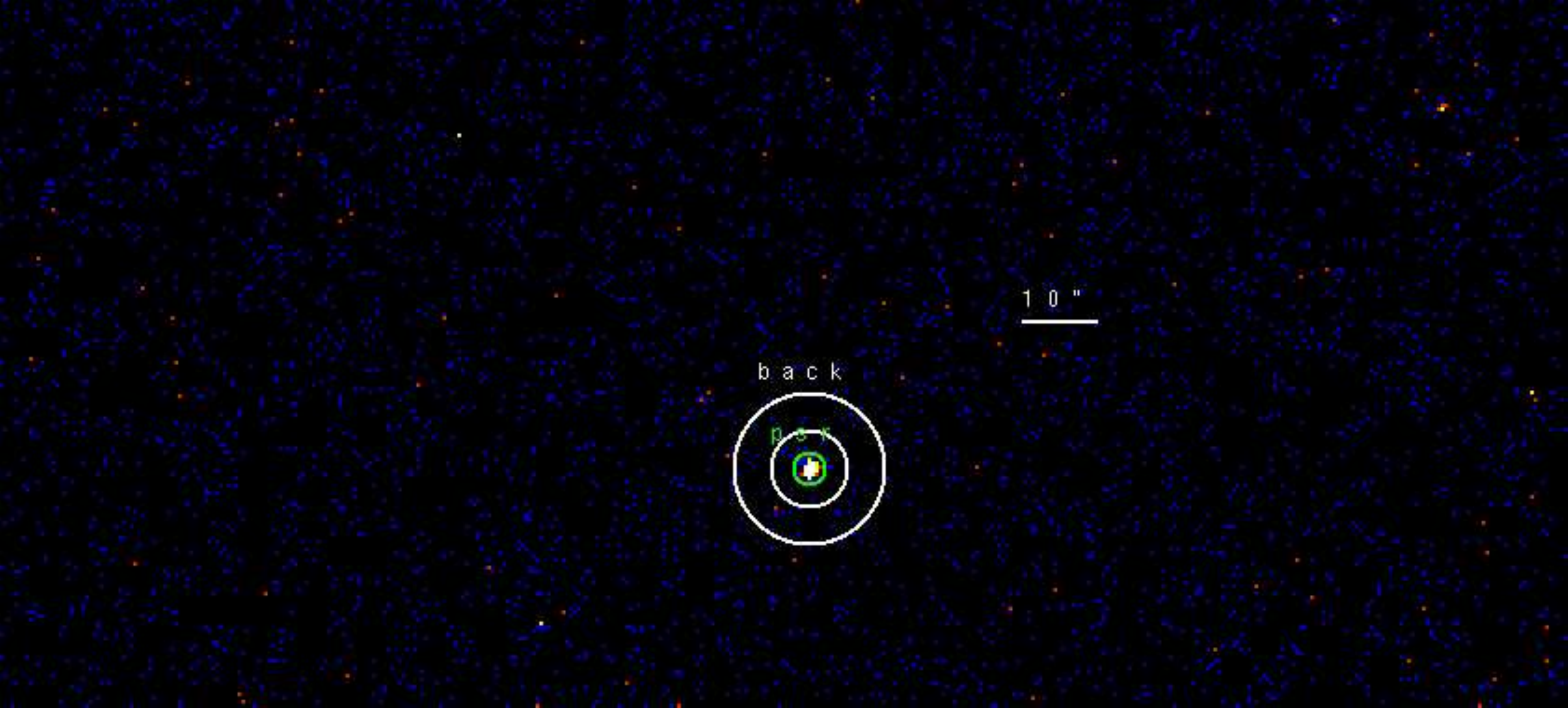}
\caption{PSR J1744-1134 0.3-10 keV {\it Chandra} Imaging.
The green circle marks the pulsar while the white annulus the background region used in the analysis.
\label{J1744-im}}
\end{figure}

\begin{figure}
\centering
\includegraphics[angle=0,scale=.30]{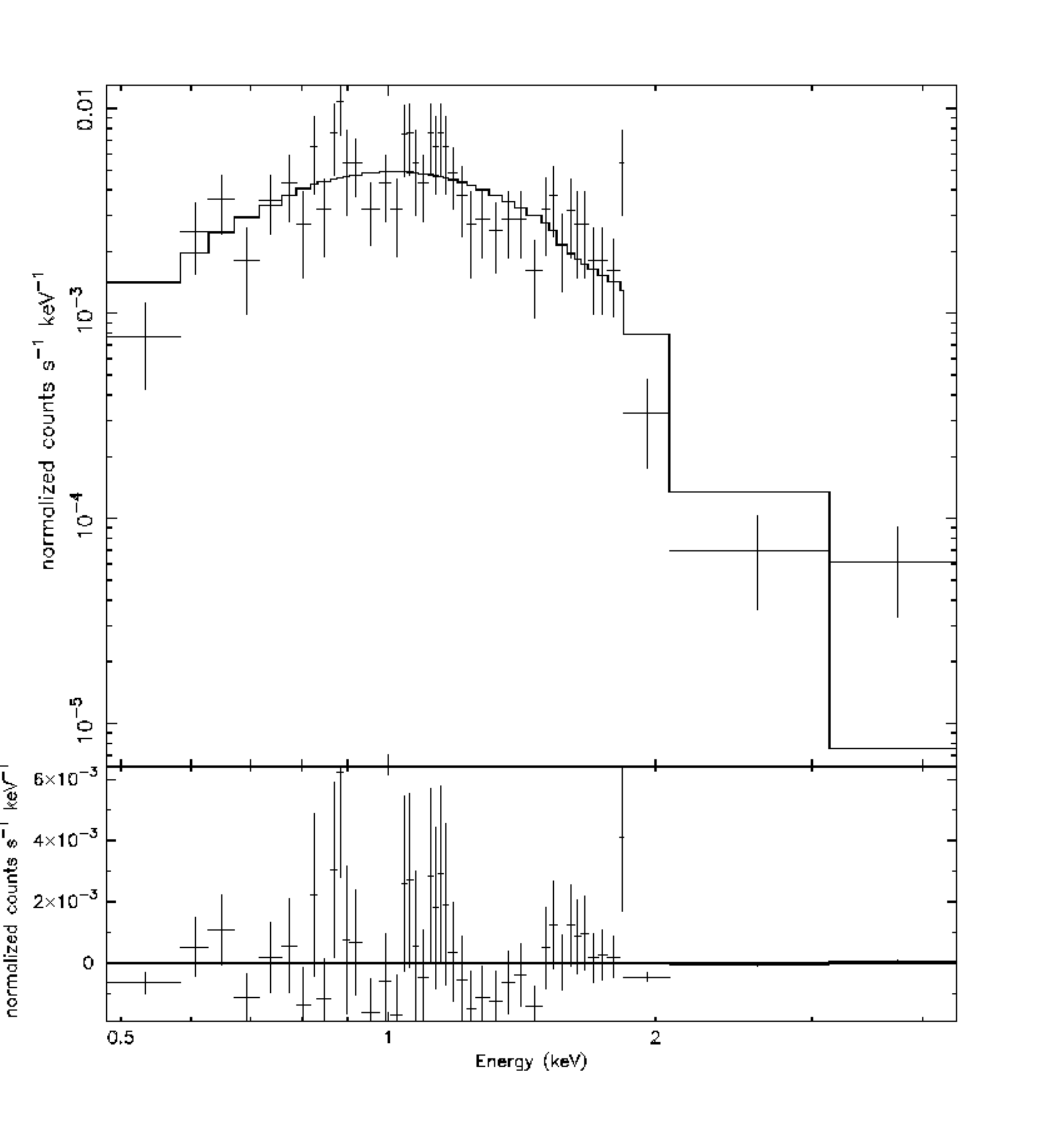}
\caption{PSR J1744-1134 {\it Chandra} Spectrum (see text for details).
Residuals are shown in the lower panel.
\label{J1744-sp}}
\end{figure}

\clearpage

{\bf J1747-2958 (mouse) - type 2 RLP}

Camilo et al. 2002 found J1747-2958 using the
64 m Parkes radio telescope. They successfully identified a
99 ms pulsar, PSR J1747-2958, with a characteristic age $\tau_c$ =
26 kyr and a spin-down luminosity $\dot{E}$ = 2.51 $\times$ 10$^{36}$ erg/s.
Such a detection was the result of an intense campaign of observations
done in order to identify the source powering the long-time-known radio nebula
named $"$mouse$"$ (Yusef-Zadeh \& Bally 1987). A dispersion measurement
based on radio data gives a distance of 2.0 $\pm$ 0.6 kpc (Camilo et al. 2002).

\begin{figure}
\centering
\includegraphics[angle=0,scale=.40]{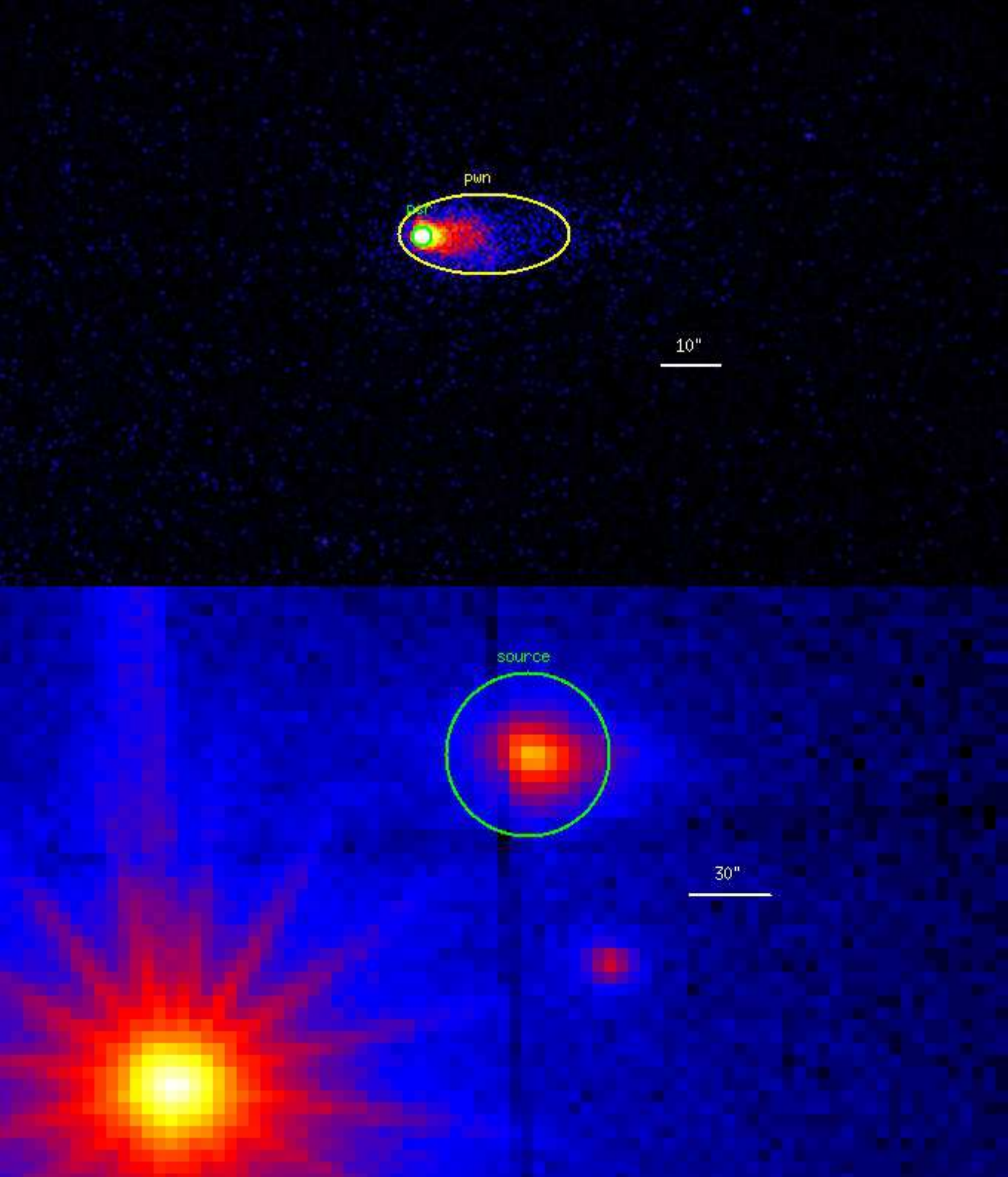}
\caption{{\it Upper Panel:} PSR J1747-2958 0.3-10 keV {\it Chandra} Imaging.
The green circle marks the pulsar while the yellow annulus the nebular region used in the analysis.
{\it Lower Panel:} PSR J1747-2958 0.3-10 keV {\it XMM-Newton} EPIC Imaging. The PN and the two MOS images have been added. 
The green circle marks the source region used in the analysis.
\label{J1747-im}}
\end{figure}

Two different observations of J1747-2958 have been performed:\\
- obs. id 0152920101, {\it XMM-Newton} observation, start time 2003, April 02 at 20:05:06 UT, exposure 50.2 ks;\\
- obs. id 2834, {\it Chandra} ACIS-S faint mode, start time 2002, October 23 at 20:43:10 UT, exposure 36.8 ks.\\
In the XMM observation both the PN and MOS cameras were operating in Full Frame mode and
a thick optical filter was used.
The X-ray source best fit position, obtained by using the celldetect tool inside the ciao distribution, is 17:47:15.86 -29:58:01.55 (1$"$ error radius).
No screening for soft proton flares was done owing to the goodness of the XMM observation.
An elliptical nebular emission is apparent in the {\it Chandra} observation. The major axis of the ellipse
is about 40$"$ (Gaensler et al. 2004).
For the {\it Chandra} observation, we chose a 1.5$"$ radius circular region for the
pulsar spectrum in order to minimize the contribution of the pulsar wind nebula
while the nebular spectrum was extracted by an elliptical region 
with a semimajor axis of $\sim$ 15$"$ (see Figure \ref{J1747-im}), that comprehend the 
brighter part of the nebula. The background was extracted from a circular source-free
region away from the source and the nebula.
For the {\it XMM-Newton} observation, we chose a 30$"$ radius circular region around
the pulsar in order to take into account both the pulsar and the nebular emission;
the background was extracted from a circular source-free region on the same CCD,
away from the source, in order to avoid the contamination of a faint source near the pulsar.
We obtained a total of 21858, 9861 and 11088 counts from both the pulsar and the nebula
in the XMM observation (background contributions of 0.02, 0.018 and 0.018). we also
obtained 5220 counts from the pulsar region and 6327 photons from the nebula in the
{\it Chandra} observation (background contributions of less than 0.1\% and 1.0\%).
In this case, the brightness of the nebula makes it necessary to consider the PWN contribution
also in the spectrum taken inside the 1.5$"$ radius circular region around the pulsar.
The best fitting pulsar model is a simple powerlaw (probability of obtaining the data if the model is correct 
- p-value - of 0.062, 1684 dof using both the pulsar and nebular spectra) with a photon index
$\Gamma$ = 1.51$_{-0.44}^{+0.12}$ absorbed by a column N$_H$ = 2.56$_{-0.06}^{+0.09}$ $\times$ 10$^{22}$ cm$^{-2}$.
The nebular emission has a photon index $\Gamma$ = 2.04 $\pm$ 0.06.
A simple blackbody model is not statistically acceptable while a composite
blackbody plus powerlaw spectrum yields no relevant statistical improvement.
Assuming the best fit model, the 0.3-10 keV unabsorbed pulsar flux is 
4.87$_{-0.60}^{+2.13}$ $\times$ 10$^{-12}$, and the nebular flux is
8.45$_{-0.40}^{+1.02}$ $\times$ 10$^{-12}$ erg/cm$^2$ s. 
Using a distance
of 2.0 kpc, the luminosities are 
L$_{2kpc}^{nt}$ = 2.34$_{-0.29}^{+1.02}$ $\times$ 10$^{33}$ and L$_{2kpc}^{pwn}$ = 4.06$_{-0.19}^{+0.49}$ $\times$ 10$^{33}$ erg/s.

\begin{figure}
\centering
\includegraphics[angle=0,scale=.50]{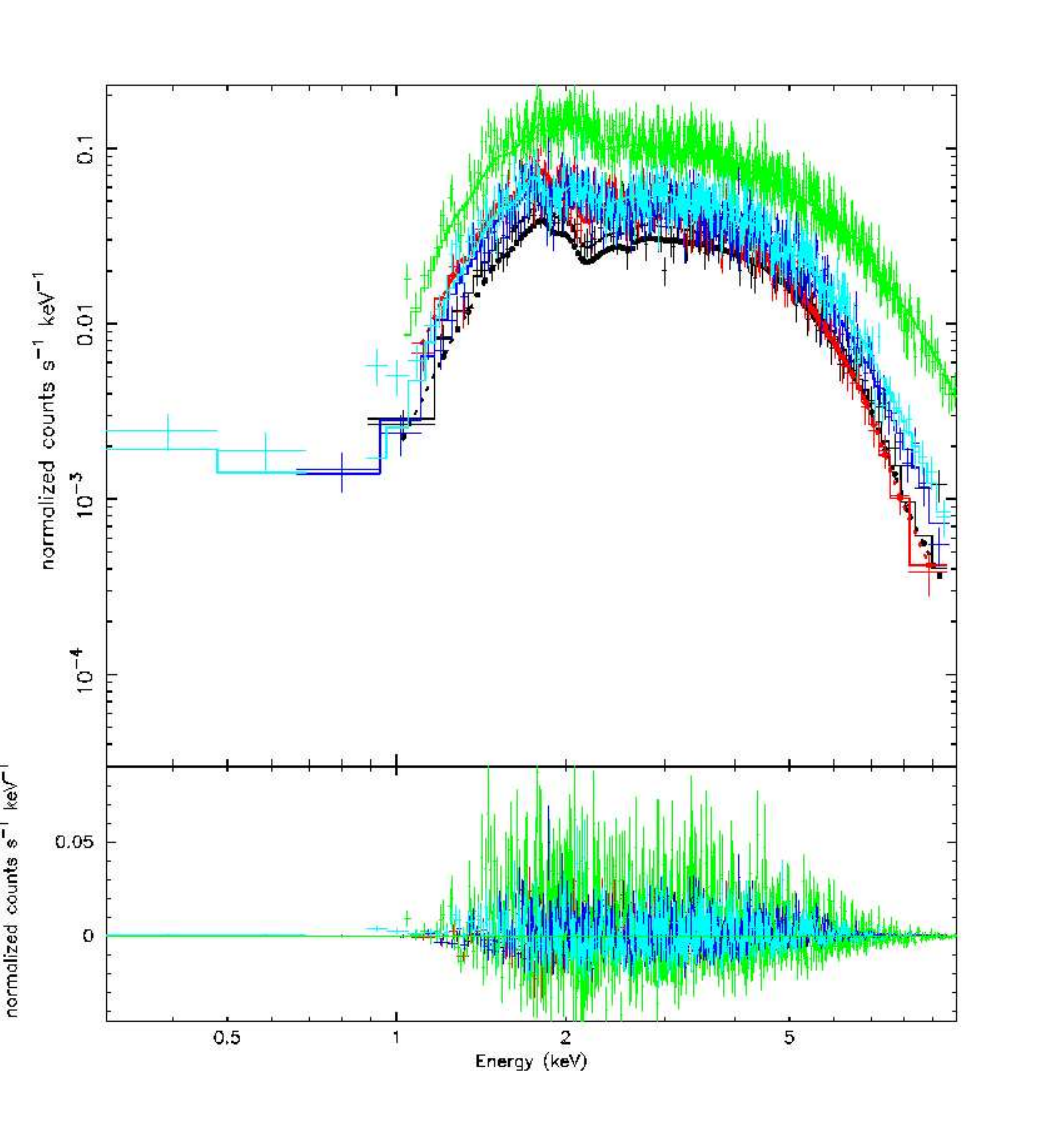}
\caption{PSR J1747-2958 Spectrum. Different colors mark all the different dataset used (see text for details).
Black points mark the pulsar spectrum while red points the nebular one.
Residuals are shown in the lower panel.
\label{J1747-sp}}
\end{figure}

\clearpage

{\bf J1801-2451 (Duck) - type 2 RLP} % Nuova!

% Zeiger et al. 2006
PSR B1757-24 is an energetic young pulsar, with
a period P = 125 ms, a spindown energy loss rate
$\dot{E}$ = 2.5 $\times$ 10$^{36}$ erg/s, and a characteristic age $\tau_c$
= 15.5 kyr. Its location
and the morphology of its PWN suggest that it is escaping
the circular SNR G5.4-1.2 (Frail \& Kulkarni 1991;
Manchester et al. 1991), whose asymmetric brightness, in
conjunction with the PWN produced by the pulsar, produces
the structure known as $"$the Duck$"$.
Due to the lack of detection of proper motion, the association of PSR B1757-24
with SNR G5.4-1.2 is unlikely for the pulsar characteristic age, although an association
cannot be excluded for a significantly larger age (Zeiger et al. 2008).
Based on the dispersion measure, PSR B1757-24
is estimated to lie at a distance of 5.2 $\pm$ 0.5 kpc
(Taylor \& Cordes 1993), while the distance to G5.27-1.2
is greater than 4.3 kpc based on H I absorption
(Frail et al. 1994; Thorsett et al. 2002). Here we adopt
a distance of 5 kpc.

\begin{figure}
\centering
\includegraphics[angle=0,scale=.40]{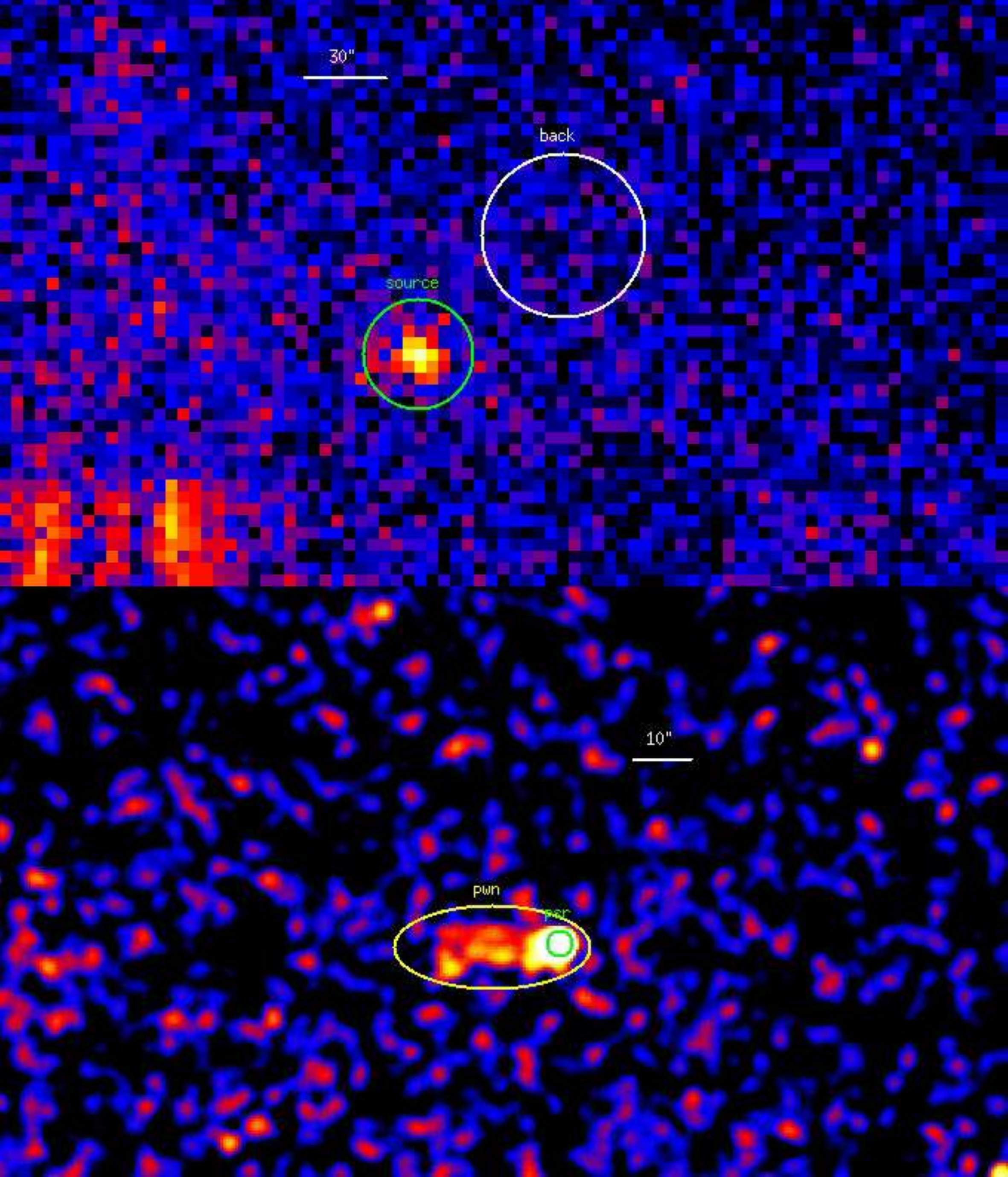}
\caption{{\it Upper Panel:} PSR J1801-2451 0.3-10 keV {\it XMM-Newton} EPIC Imaging. The PN and the two MOS images have been added. 
The green circle marks the source while the white circle the background region used in the analysis.
{\it Lower Panel:} PSR J1801-2451 0.3-10 keV {\it Chandra} Imaging.
The green circle marks the pulsar while the yellow annulus the nebular region used in the analysis.
\label{J1801-im}}
\end{figure}

Two X-ray observations of the pulsar were found:\\
- obs. id 0103261901, {\it XMM-Newton} observation, start time 2002, March 16 at 21:26:28 UT, exposure 5.0 ks;\\
- obs. id 753, {\it Chandra} ACIS-S faint mode, start time 2000, April 12 at 12:41:23 UT, exposure 20.0 ks.\\
In the XMM observation both the PN and MOS cameras were operating in Full Frame mode and
a medium optical filter was used.
The X-ray source best fit position, obtained by using the celldetect tool inside the ciao distribution, is 
18:01:00.015 -24:51:27.73 (1$"$ error radius).
No screening for soft proton flares was needed.
A trail-like nebular emission is apparent in the {\it Chandra} observation with the major axis of the ellipse
of about 15$"$.
For the {\it Chandra} observation, we chose a 2$"$ radius circular region for the
pulsar spectrum while the nebular spectrum was extracted by an elliptical region 
with a semimajor axis of 15$"$ (see Figure \ref{J1801-im}). The background was extracted from a circular source-free
region away from the source and the nebula.
For the {\it XMM-Newton} observation, we chose a 20$"$ radius circular region around
the pulsar while the background is extracted from a circular source-free region on the same CCD.
Due to the low statistic, the spectra obtained from the two MOS datasets were added using
mathpha tool and, similarly, the response
matrix and effective area files using addarf and addrmf. 
We obtained a total of 357 and 101 pulsar and nebular counts from the {\it Chandra} observation
(background contributions of 0.4\% and 29.3\%); we also obtained 300 and 351 source
counts from the PN and MOS cameras, respectively (background contributions of 44.9\% and 27.0\%).
The best fitting pulsar model is a simple powerlaw (probability of obtaining the data if the model is correct 
- p-value - of 0.40, 38 dof using both the pulsar and nebular spectra) with a photon index
$\Gamma$ = 1.54$_{-0.44}^{+0.28}$ absorbed by a column N$_H$ = 3.74$_{-1.08}^{+1.20}$ $\times$ 10$^{22}$ cm$^{-2}$.
The nebular emission has a photon index $\Gamma$ = 1.94$_{-0.81}^{+0.87}$.
A simple blackbody model is not statistically acceptable while a composite
blackbody plus powerlaw spectrum provides no relevant statistical improvement.
Assuming the best fit model, the 0.3-10 keV unabsorbed pulsar flux is 
9.97 $\pm$ 2.02 $\times$ 10$^{-13}$ and the nebular flux is
3.27 $\pm$ 1.24 $\times$ 10$^{-13}$ erg/cm$^2$ s. 
Using a distance
of 5.0 kpc, the luminosities are 
L$_{5kpc}^{nt}$ = 2.99 $\pm$ 0.61 $\times$ 10$^{33}$ and L$_{5kpc}^{pwn}$ = 9.81 $\pm$ 3.72 $\times$ 10$^{32}$ erg/s.

\begin{figure}
\centering
\includegraphics[angle=0,scale=.50]{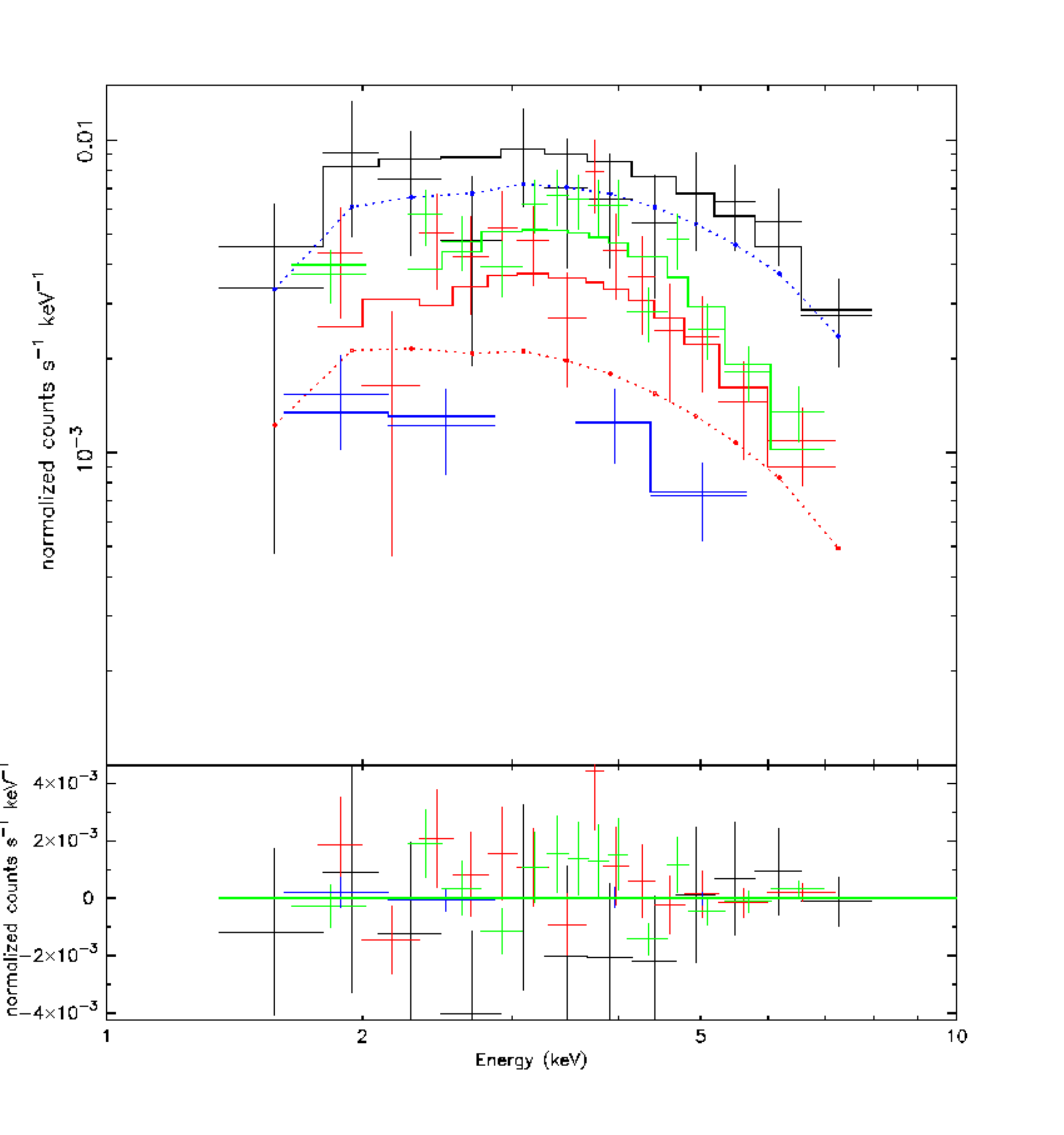}
\caption{PSR J1801-2451 Spectrum. Different colors mark all the different dataset used (see text for details).
Black points mark the pulsar spectrum of the pulsar spectrum while red points mark the nebular emission
Residuals are shown in the lower panel.
\label{J1801-sp}}
\end{figure}

\clearpage

{\bf J1809-2332 (Tazmanian Devil) - type 2 RQP} % osservazione in arrivo

% Roberts & Brogan 2008
In the EGRET era 
3EG J1809-2328 was one of the brightest sources of GeV
emission in the Galaxy (Hartman et al. 1999). ASCA imaging of the EGRET error box revealed an
extended hard X-ray source, which (Braje et al. 2002)
subsequently resolved into an apparent rapidly moving PWN
and a young stellar cluster using a short {\it Chandra} observation
(see Gaensler \& Slane 2006 for general
reviews of PWN). Radio imaging with the VLA and observations
with {\it XMM-Newton} confirm the PWN nature of the X-ray source
(Braje et al. 2002). Due to the radio nebula's distinctive funnel shape
(presumably imposed by its motion), its unusually powerful $\gamma$-ray
emission, and the growing tradition of naming PWNe after animals,
this nebula is sometimes referred to as $"$Taz$"$ (short for Tasmanian
devil). Johnston et al. 1996 found a distance of 1.7 $\pm$ 1.0 kpc,
evaluated using kinematic models.
No $\gamma$-ray nebular emission was detected by {\it Fermi} down to a flux of 
2.12 $\times$ 10$^{-11}$ erg/cm$^2$s (Ackermann et al. 2010).

\begin{figure}
\centering
\includegraphics[angle=0,scale=.30]{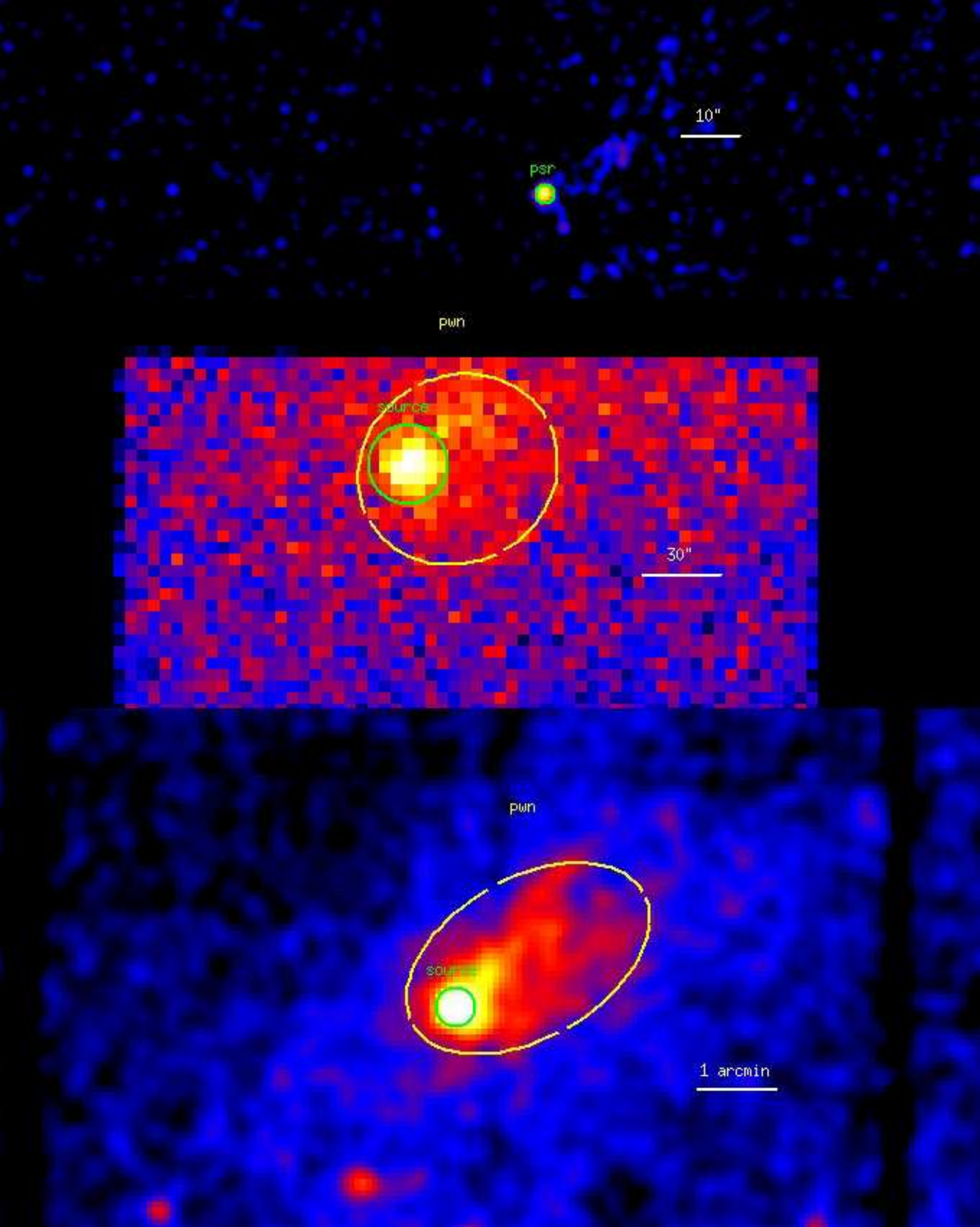}
\caption{{\it Upper Panel:} PSR J1809-2332 0.3-10 keV {\it Chandra} Imaging.
The green circle marks the pulsar region used in the analysis.
{\it Medium Panel:} PSR J1809-2332 0.3-10 keV {\it XMM-Newton} PN Imaging.
The green circle marks the source while the yellow ellipse the nebular region used in the analysis.
{\it Lower Panel:} PSR J1809-2332 0.3-10 keV {\it XMM-Newton} MOS Imaging. The two MOS images have been added.
The green circle marks the source while the yellow ellipse the nebular region used in the analysis.
\label{J1809-im}}
\end{figure}

Three different X-ray observations of J1709-4429 were performed, two
by {\it XMM-Newton} and one by {\it Chandra}:\\
- obs. id 4608, {\it Chandra} ACIS-I faint mode, start time 2000, August 16 at 03:16:46 UT, exposure 9.8 ks;\\
- obs. id 0141610601, {\it XMM-Newton} observation, start time 2002, September 21 at 23:49:04.60 UT, exposure 15.6 ks;\\
- obs. id 0201270101, {\it XMM-Newton} observation, start time 2004, October 02 at 07:44:01.39 UT, exposure 69.1 ks.\\
In both the XMM observations the PN camera was operating in the Small Window mode and MOS cameras were
operating in the Full Frame mode.
For the PN camera a thin optical filter was used while for the MOS cameras the medium optical filter was used.
No screening for soft proton flare events was needed.
The X-ray source best fit position (obtained by using the celldetect
tool inside the Ciao distribution) is 18:09:50.22 -23:32:22.37 (1$"$ error radius).
The {\it Chandra} observation is too short to search for faint short-scale nebulae.
In the XMM observations a nebular
emission in apparent as a trail with a length of more than 2$'$. A fainter halo is present in most of the central
CCD of the MOS cameras.
For the {\it Chandra} observation, we chose a 2$"$ radius circular region for the
pulsar spectrum. The background is extracted from a circular source-free region
away from the source, in order to exclude any nebular emission.
For the {\it XMM-Newton} observation, we chose a 15$"$ radius circular region around
the pulsar. The nebular emission was extracted from an ellipse with a semimajor
axis of 38.5$"$ in the PN dataset while we chose a larger region in the MOS datasets
in order to comprehend all the bright trail (ellipse with a semimajor axis of 100$"$,
see Figure \ref{J1809-im}). The background was extracted from a circular source-free region on the same CCD,
away from the source.
We obtained a total of 3532, 1082, 954 and 51 pulsar counts respectively from PN, the two
MOS cameras and the {\it Chandra} observation (background contributions of 45.7\%, 19.6\%, 21.4\% and 0.2\%).
We also obtained 11378, 8723 and 8223 nebular counts respectively from PN, the two
MOS cameras and the {\it Chandra} observation (background contributions of 71.7\%, 65.7\% and 63.5\%).
The best fitting pulsar model is a combination of a blackbody and a powerlaw
(probability of obtaining the data if the model is correct 
- p-value - of 0.11, 475 dof using both the pulsar and nebular spectra).
The powerlaw component has a photon index $\Gamma$ = 1.85$_{-0.36}^{+1.89}$ 
absorbed by a column N$_H$ = 6.1$_{-0.8}^{+0.9}$ $\times$ 10$^{21}$ cm$^{-2}$.
The thermal component has a temperature of T = (2.20 $\pm$ 0.29) $\times$ 10$^6$ K.
The blackbody radius R = 1.54$_{-0.44}^{+1.26}$ km determined from the
model parameters, suggests that the emission is from a hot spot.
The nebular emission has a photon index $\Gamma$ = 1.43 $\pm$ 0.10.
A simple powerlaw or blackbody model are not statistically acceptable.
Assuming the best fit model, the 0.3-10 keV unabsorbed non-thermal pulsar flux is
1.40$_{-0.23}^{+0.25}$ $\times$ 10$^{-13}$, the thermal flux is 
1.74$_{-0.30}^{+0.32}$ $\times$ 10$^{-13}$, and the total nebular flux (coming from the MOS region) is
1.11 $\pm$ 0.12 $\times$ 10$^{-12}$ erg/cm$^2$ s.
Using a distance
of 1.7 kpc, the luminosities are L$_{1.7kpc}^{nt}$ = 4.85$_{-0.80}^{+0.87}$ $\times$ 10$^{31}$,
L$_{1.7kpc}^{bol}$ = 6.03$_{-1.04}^{+1.11}$ $\times$ 10$^{31}$ and L$_{1.7kpc}^{pwn}$ = 3.84 $\pm$ 0.42 $\times$ 10$^{32}$ erg/s.

\begin{figure}
\centering
\includegraphics[angle=0,scale=.50]{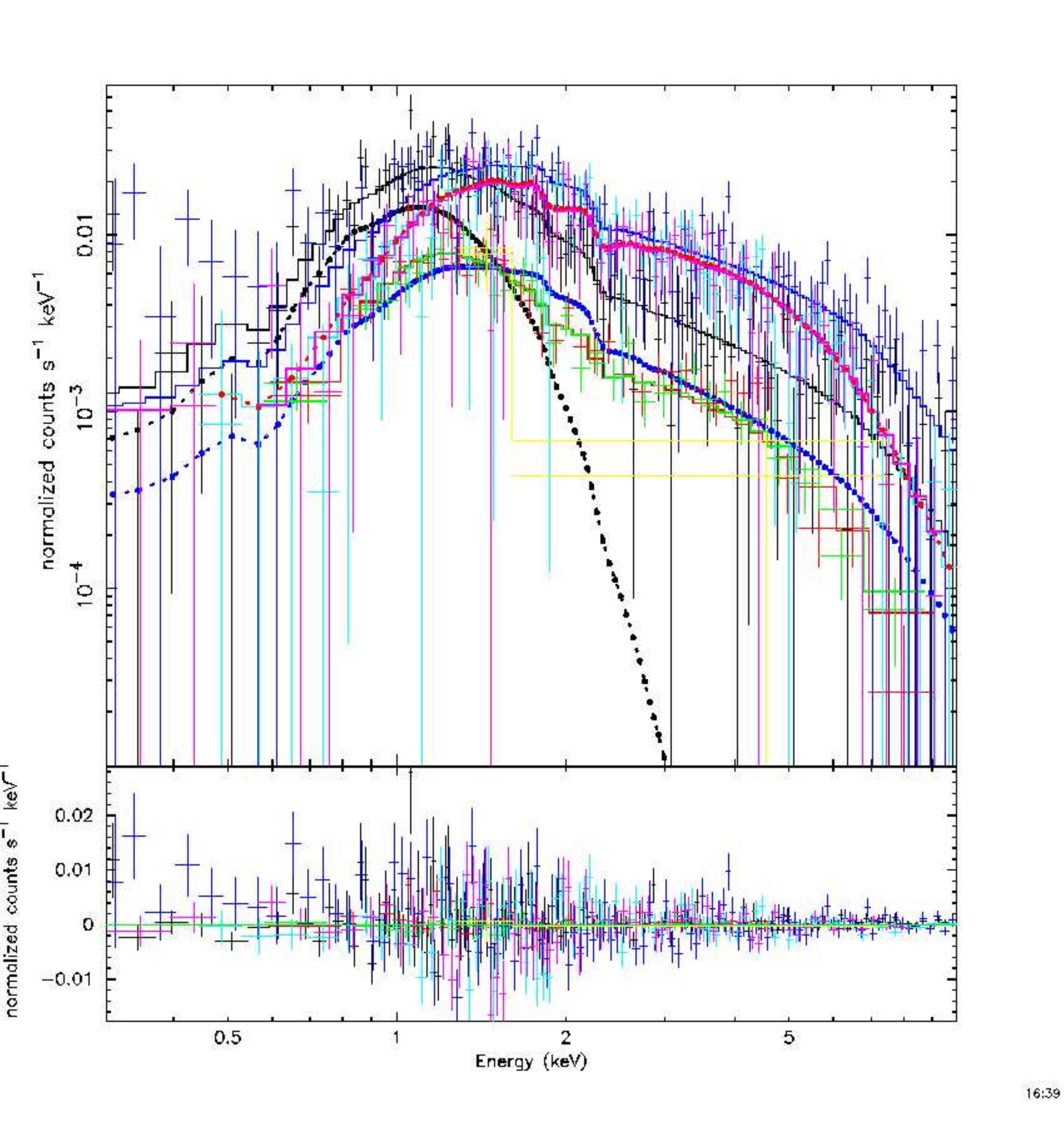}
\caption{PSR J1809-2332 Spectrum. Different colors mark all the different dataset used (see text for details).
Blue points mark the powerlaw component while black points the thermal component of the pulsar spectrum.
Red points mark the nebular spectrum.
Residuals are shown in the lower panel.
\label{J1809-sp}}
\end{figure}

\clearpage

{\bf J1813-1246 - type 1 RQP} % migliorato con la cstatistic

J1813-1246 was one of the first pulsars discovered using the
blind search technique (Abdo et al. Science 2009).
The pseudo-distance of the pulsar based on $\gamma$-ray data (Saz Parkinson et al. (2010))
is $\sim$ 2.0 kpc. A nebular emission was detected in the $\gamma$-ray band
at a flux of 1.19 $\pm$ 0.09 $\times$ 10$^{-10}$ erg/cm$^2$ s.

Immediately after the detection of pulsations, we made
a request of a {\it SWIFT} observation on behalf of the {\it Fermi} LAT
collaboration (obs id. 00031381001, 00031381002, 00031381003 and 00090197003
for a total exposure of 13.25 ks). Such observations revealed one potential
X-ray counterpart inside the 6-months {\it Fermi} error box
at 18:13:23.40 -12:45:58.82 (10$"$ error radius).
A precise timing analysis performed by the collaboration
(Ray et al. 2011) confirmed the X-ray source to be the counterpart of the $\gamma$-ray pulsar.
We used a 20$"$ radius circle in order to extract the pulsar
spectrum and a circular region away from the source as background.
No study for an extended emission is possible due to the low statistic;
anyway, the high $\dot{E}$ of PSR J1813-1246 could certainly power a bright pulsar wind nebula.
We obtained 81 counts from the {\it SWIFT} observations (background contribution of 1.8\%).
Due to the low statistic, we used the C-statistic
approach implemented in XSPEC.
The best fitting pulsar model is a simple powerlaw
(reduced chisquare of $\chi^2_{red}$ = 0.94, 13 dof)
with a photon index $\Gamma$ = 1.72$_{-0.87}^{+1.56}$ 
absorbed by a column N$_H$ = 3.96$_{-1.95}^{+3.79}$ $\times$ 10$^{22}$ cm$^{-2}$.
A simple blackbody model is statistically acceptable
but yields an unrealistic value of the temperature ($>$ 10$^7$ K).
The statistic is too low to study composite models.
Attributing the 30\% of the flux to thermal and/or
nebular emission and assuming the best fit model, the 0.3-10 keV unabsorbed source flux is
1.51 $\pm$ 0.38 $\times$ 10$^{-12}$ erg/(cm$^2$ s).
Using a distance
of 2.0 kpc, the source luminosity is L$_{2kpc}^{nt}$ = 7.25 $\pm$ 1.82 $\times$ 10$^{32}$ erg/s.

\begin{figure}
\centering
\includegraphics[angle=0,scale=.30]{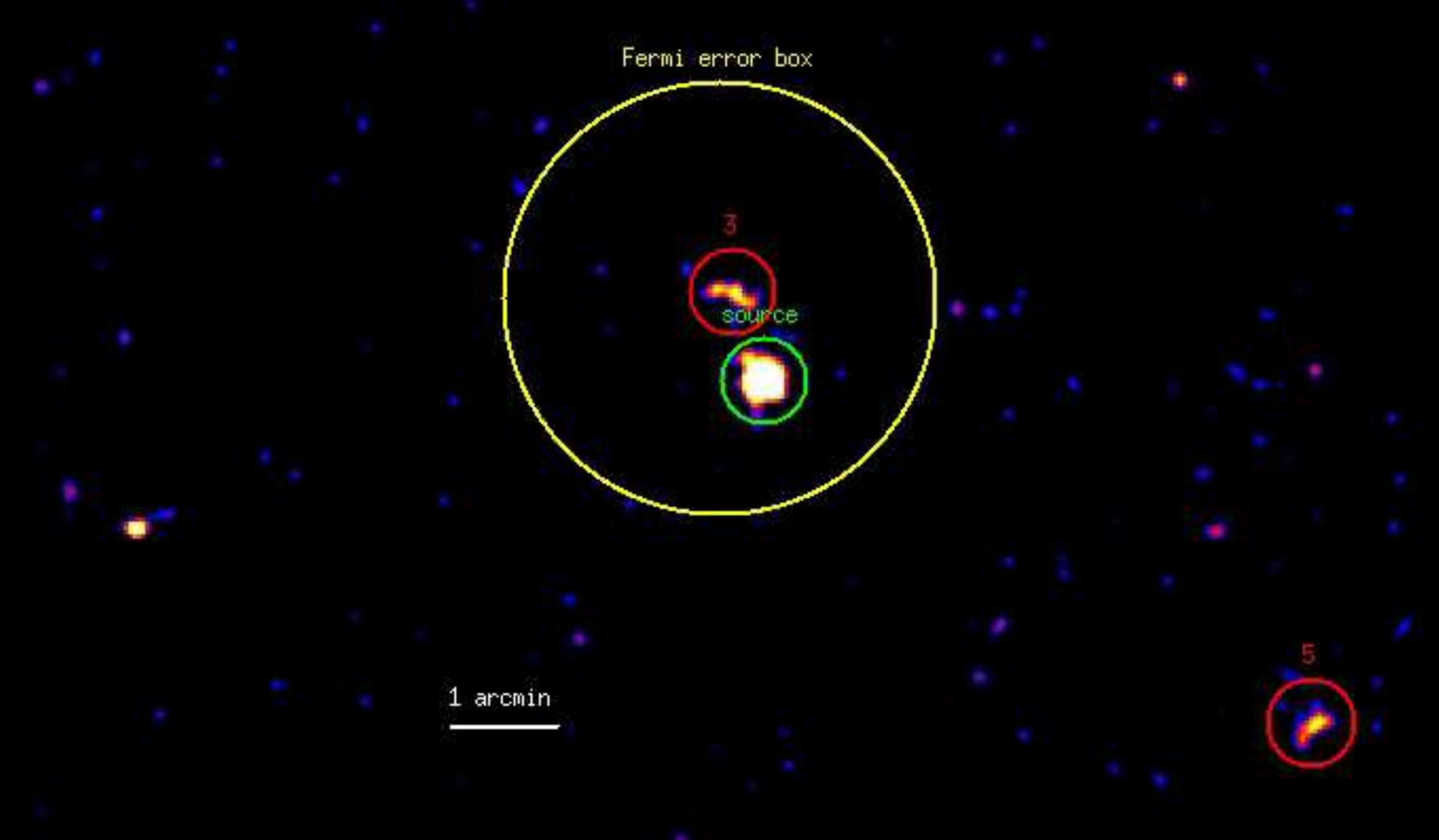}
\caption{PSR J1813-1246 0.3-10 keV {\it SWIFT} XRT Imaging. The image has been smoothed with a Gaussian
with Kernel radius of $5"$. The green circle marks the pulsar, the red one a background source
while the white circle the {\it Fermi} error box.
\label{J1813-im}}
\end{figure}

\begin{figure}
\centering
\includegraphics[angle=0,scale=.40]{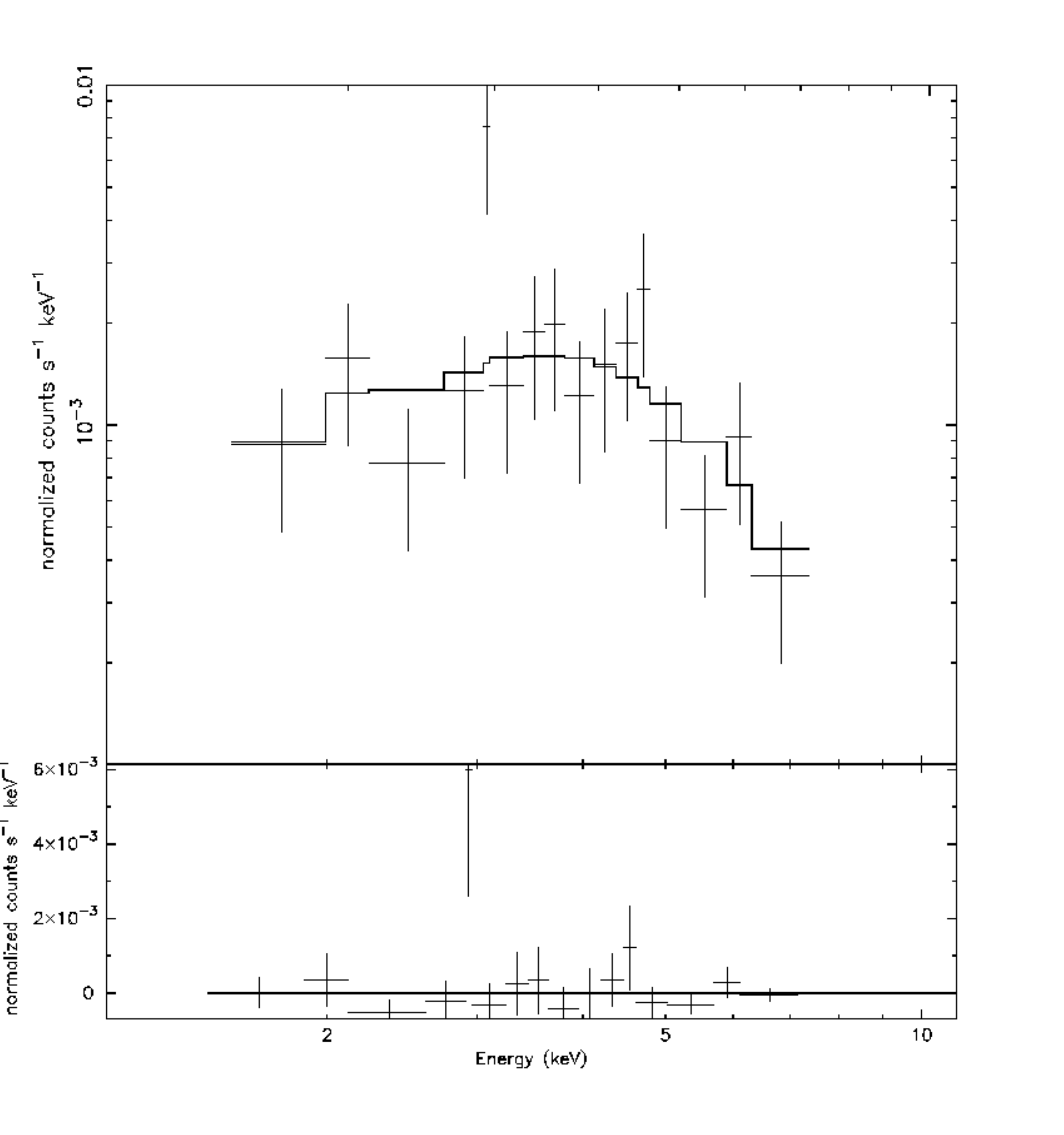}
\caption{PSR J1813-1246 {\it SWIFT} XRT Spectrum (see text for details).
Residuals are shown in the lower panel.
\label{J1813-sp}}
\end{figure}

\clearpage

{\bf J1823-3021A - type 0 RL MSP} % Nuova!

PSR J1823-3021A is an accreting NS low mass X-ray binary in the globular
cluster NGC 6624, which has a well-determined distance (7.6 $\pm$ 0.4 kpc
Kuulkers et al. 2003).
It was discovered by Biggs et al. 1990 in a survey of
globular clusters using the 76-m Lovell telescope at Jodrell Bank
at 610 MHz (Biggs, Lyne \& Brinklow 1989, Biggs, Lyne \& Johnston
1989). Radio dispersion measurements united with the known
distance of the cluster place this pulsar at $\sim$ 7.9 kpc.

The X-ray spectrum undetectable due to the pulsar closeness to the very bright
low-mass X-ray binary 4U 1820-30.

{\bf J1826-1256 (Eel) - type 2 RQP} % rifatta con la cstat, oramai di tipo 3

J1826-1256 was one of the first pulsars discovered using the
blind search technique (Abdo et al. Science 2009).
Analyzing two archival {\it Chandra} observations
with the {\it Fermi} error box inside their FOV, an extended source
surrounding a pointlike one has been detected.
A precise timing analysis performed by the collaboration
(Ray et al. 2011) confirmed the X-ray source to be the
counterpart of the $\gamma$-ray pulsar.
The pseudo-distance based on $\gamma$-ray data (Saz Parkinson et al. (2010))
is $\sim$ 1.2 kpc.
No $\gamma$-ray nebular emission was detected by {\it Fermi} down to a flux of 
1.61 $\times$ 10$^{-10}$ erg/cm$^2$s (Ackermann et al. 2010).

Two different {\it Chandra} ACIS-I observations were performed with the X-ray counterpart
inside their FOV:\\
- obs. id 3851, start time 2003, February 17 at 20:04:48 UT, exposure 15.1 ks;\\
- obs. id 7641, start time 2007, July 26 at 16:05:20 UT, exposure 74.8 ks.\\
The X-ray source best fit position (obtained by using the celldetect
tool inside the Ciao distribution) is 18:26:08.545 -12:56:34.46 (1$"$ error radius).
I've obtained a radial profile around the pointlike source. A nebular emission
is present until $\sim$ 20$"$ from the pointlike source. A possible fainter emission
extends until $\sim$ 1'. xyz

\begin{figure}
\centering
\includegraphics[angle=0,scale=.25]{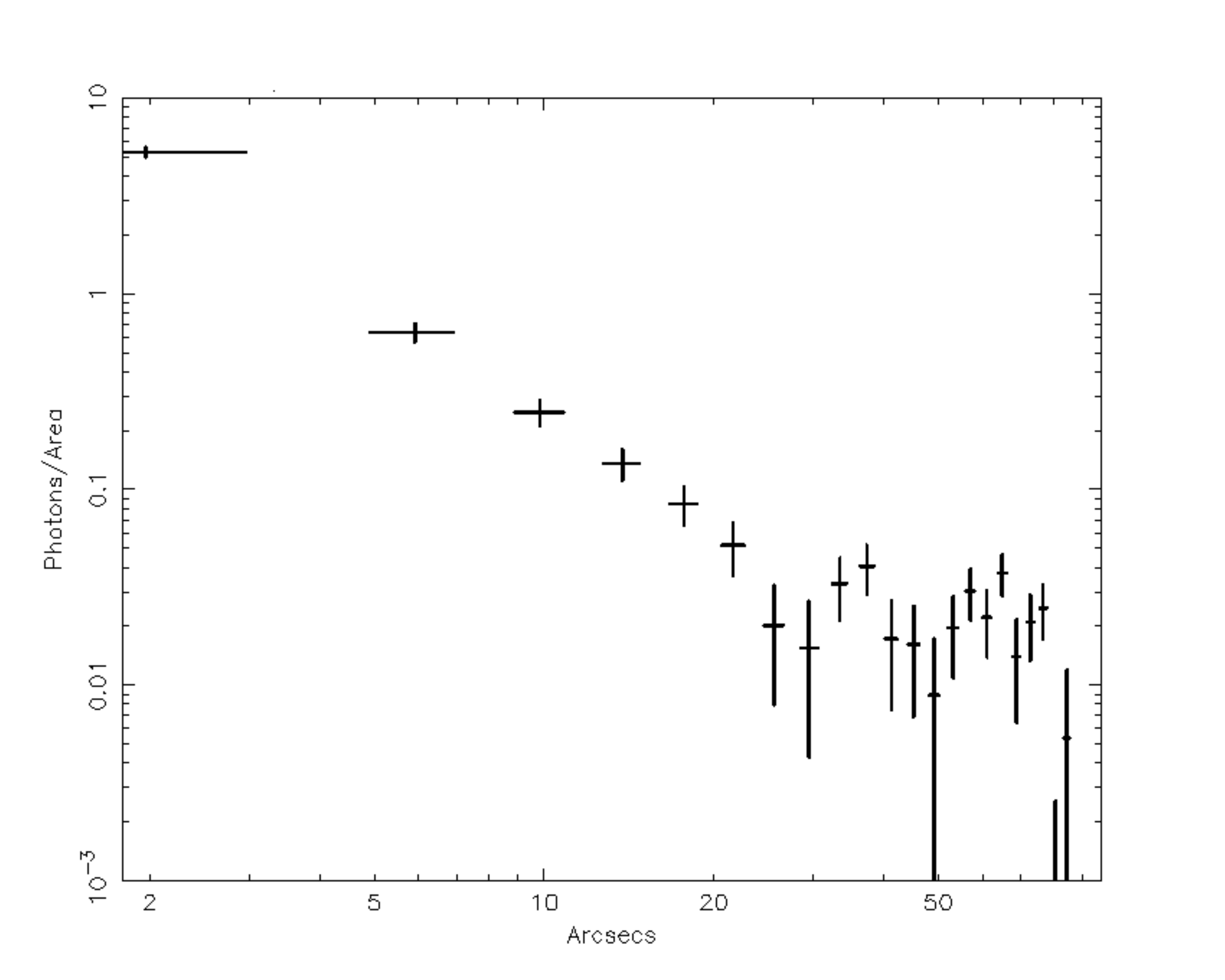}
\caption{PSR J1826-1256 {\it Chandra} radial profile (0.3-10 keV energy range).
{\it Chandra} ACIS On-axis Point Spread Function is almost zero over 2$"$ from the source so that
it's apparent the nebular emission between 2 and 20$"$ from the pulsar.
\label{J1826-psf}}
\end{figure}

For the {\it Chandra} observation, we chose a 2$"$ radius circular region for the
pulsar spectrum. The background is extracted from a circular source-free region
away from the source, in order to exclude any nebular contamination.
The nebular spectrum was extracted from an annulus with radii 2 and 20$"$
around the pulsar.
The spectra obtained in the two {\it Chandra} observations were added using the
mathpha tool and, similarly, the response
matrix and effective area files using addarf and addrmf. 
We obtained a total of 264 pulsar and 640 nebular counts
(background contributions of 0.8\% and 44.5\%).
The best fitting pulsar model is a flat powerlaw
(reduced chisquare of $\chi^2_{red}$ = 1.01, 158 dof
using both the pulsar and nebular spectra) with a
photon index $\Gamma$ = 0.79 $\pm$ 0.39
absorbed by a column N$_H$ = 1.26$_{-0.46}^{+0.53}$ $\times$ 10$^{22}$ cm$^{-2}$.
A simple blackbody model is not statistically acceptable while
a composite blackbody plus powerlaw spectrum is not statistically needed.
The nebular emission has a photon index $\Gamma$ = 0.86 $\pm$ 0.39.
Assuming the best fit model, the 0.3-10 keV unabsorbed pulsar flux is
(1.12 $\pm$ 0.25) $\times$ 10$^{-13}$ and the nebular flux is
(1.52 $\pm$ 0.33) $\times$ 10$^{-13}$ erg/cm$^2$ s.
Using a distance
of 1.2 kpc, the luminosities are L$_{1.2kpc}^{nt}$ = (1.94 $\pm$ 0.43) $\times$ 10$^{31}$
and L$_{1.2kpc}^{pwn}$ = (2.63 $\pm$ 0.57) $\times$ 10$^{31}$ erg/s.

\begin{figure}
\centering
\includegraphics[angle=0,scale=.30]{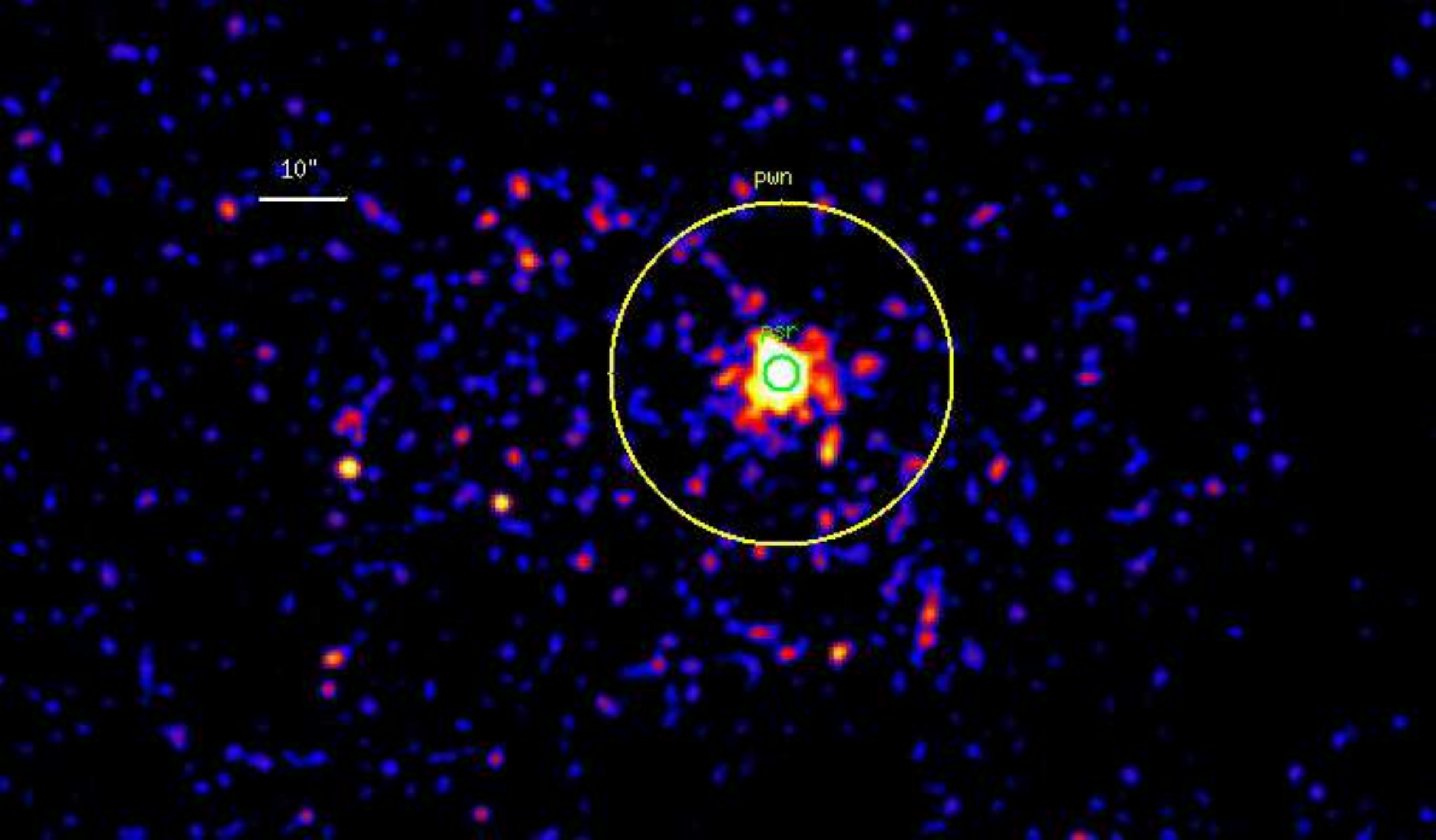}
\caption{PSR J1826-1256 0.3-10 keV {\it Chandra} Imaging. The image has been smoothed with a Gaussian
with Kernel radius of $2"$. The green circle marks the pulsar while the yellow annulus the nebular region used in the analysis.
\label{J1826-im}}
\end{figure}

\begin{figure}
\centering
\includegraphics[angle=0,scale=.50]{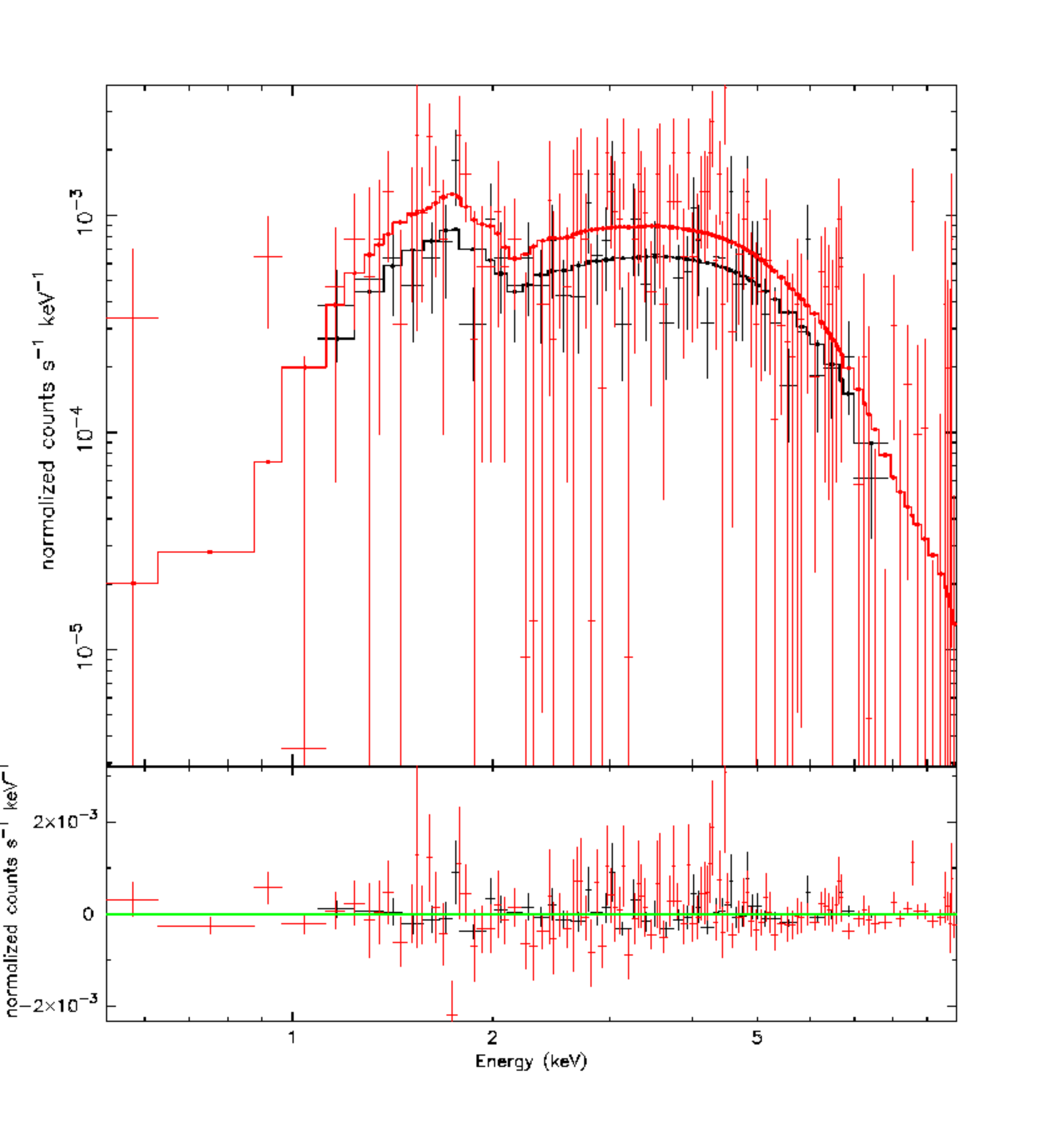}
\caption{PSR J1826-1256 Spectrum. The black points mark the pulsar spectrum while the red ones the nebular one.
Residuals are shown in the lower panel.
\label{J1826-sp}}
\end{figure}

\clearpage

{\bf J1833-1034 - type 2 RLP} % risistemato lo spettro

% Matheson & Safi-Harb 2010
In radio, Gupta et al. (2005) and Camilo et al. (2006) independently discovered the
long sought pulsar PSR J1833-1034 associated with the SNR G21.5-0.9. They found P = 61.86 ms,
$\dot{P}$ = 2.0 $\times$ 10$^{-13}$, a characteristic age of 4.7 kyr,
and a spin-down luminosity of $\dot{E}$ = 3.4 $\times$ 10$^{37}$ erg/s. Bietenholz \& Bartel (2008), using VLA observations from 1991
and 2006, derived an expansion speed of 910 $\pm$ 160 km s$^{-1}$ with respect to the centre of
the nebula and estimated the age of G21.5-0.9 as 870$^{+200}_{-150}$ yr, making it one of the youngest
known Galactic PWNe.
Camilo et al. (2006) used a compilation of studies to conclude that the best estimate of
the distance to G21.5-0.9 was 4.7 $\pm$ 0.4 kpc. By comparing HI spectra with $^{13}$CO emission
spectra, Tian \& Leahy (2008) also find a kinematic distance for G21.5-0.9 of $\sim$ 4.8 kpc.
No $\gamma$-ray nebular emission was detected by {\it Fermi} down to a flux of 
1.04 $\times$ 10$^{-11}$ erg/cm$^2$s (Ackermann et al. 2010).

Many X-ray observations of J1833-1034 were performed:\\
- obs. id 0122700101, {\it XMM-Newton} observation, start time 2000, April 07 at 13:50:12 UT, exposure 30.5 ks;\\
- obs. id 0122700201, {\it XMM-Newton} observation, start time 2000, April 09 at 13:12:00 UT, exposure 28.4 ks;\\
- obs. id 0122700301, {\it XMM-Newton} observation, start time 2000, April 11 at 13:15:22 UT, exposure 28.5 ks;\\
- obs. id 1233, {\it Chandra} ACIS-I faint mode, start time 1999, November 05 at 20:17:53 UT, exposure 14.2 ks;\\
- obs. id 1441, {\it Chandra} ACIS-I faint mode, start time 1999, November 15 at 13:25:35 UT, exposure 9.1 ks;\\
- obs. id 1442, {\it Chandra} ACIS-I faint mode, start time 1999, November 15 at 16:32:16 UT, exposure 9.7 ks;\\
- obs. id 1443, {\it Chandra} ACIS-I faint mode, start time 1999, November 15 at 19:32:16 UT, exposure 9.7 ks;\\
- obs. id 1719, {\it Chandra} ACIS-I faint mode, start time 2000, May 23 at 14:09:39 UT, exposure 7.7 ks;\\
- obs. id 1720, {\it Chandra} ACIS-I faint mode, start time 2000, May 23 at 16:50:59 UT, exposure 7.5 ks;\\
- obs. id 1721, {\it Chandra} ACIS-I faint mode, start time 2000, May 23 at 19:13:09 UT, exposure 7.6 ks;\\
- obs. id 1722, {\it Chandra} ACIS-I faint mode, start time 2000, May 23 at 21:35:19 UT, exposure 7.6 ks;\\
- obs. id 1723, {\it Chandra} ACIS-I faint mode, start time 2000, May 23 at 23:57:29 UT, exposure 7.6 ks;\\
- obs. id 1724, {\it Chandra} ACIS-I faint mode, start time 2000, May 24 at 02:19:46 UT, exposure 7.6 ks;\\
- obs. id 1725, {\it Chandra} ACIS-I faint mode, start time 2000, May 24 at 04:41:49 UT, exposure 7.6 ks;\\
- obs. id 1726, {\it Chandra} ACIS-I faint mode, start time 2000, May 24 at 07:03:59 UT, exposure 7.6 ks;\\
- obs. id 1433, {\it Chandra} ACIS-S faint mode, start time 1999, November 15 at 22:32:21 UT, exposure 15.0 ks;\\
- obs. id 1716, {\it Chandra} ACIS-S faint mode, start time 2000, May 23 at 06:51:09 UT, exposure 7.7 ks;\\
- obs. id 1717, {\it Chandra} ACIS-S faint mode, start time 2000, May 23 at 09:25:19 UT, exposure 7.5 ks;\\
- obs. id 1718, {\it Chandra} ACIS-S faint mode, start time 2000, May 23 at 11:47:29 UT, exposure 7.6 ks.\\
In all the XMM observations the PN and MOS cameras were operating in the Full Frame mode and
a medium optical filter was used.
First, an accurate
screening for soft proton flare events was done in the {\it XMM-Newton} observations obtaining a resulting total
exposure of 64.9 ks.
The X-ray source best fit position (obtained by using the celldetect
tool inside the Ciao distribution) is 18:33:33.57 -10:34:07.02 (0.5$"$ error radius).
A nebular emission of radius $\sim$ 40$"$ is apparent in the {\it Chandra} observations.
In the {\it XMM-Newton} observation the nebular emission in present until $\sim$ 2.7$'$ from pulsar.
For the {\it Chandra} observation, we chose a 2$"$ radius circular region while the nebular spectrum was extracted by an annular region 
with radii of 2 and 50$"$. The background was extracted from a circular source-free region
away from the source, in order to exclude the nebular emission.
For the {\it XMM-Newton} observation, we chose a 50$"$ radius circular region around
the pulsar in order to take in account both the pulsar and the nebular emission;
the background was extracted from a circular source-free region on the same CCD,
away from the source.
The spectra obtained in the {\it Chandra} ACIS-I observations were added using
mathpha tool and, similarly, the response
matrix and effective area files using addarf and addrmf. We made the same thing
for the {\it Chandra} ACIS-S and {\it XMM-Newton} observations.
we obtained a total of 14782 and 8819 pulsar counts from respectively {\it Chandra} ACIS-I
and ACIS-S observations (background contributions of less than 0.1\%).
we obtained a total of 225935 and 105781 nebular counts from respectively {\it Chandra} ACIS-I
and ACIS-S observations (background contributions of 1.5\% and 2.2\%).
we obtained a total of 270934, 111727 and 111576 counts from the PN and the two MOS observations
(background contributions of 1.1\%, 0.9\% and 1.0\%).
The best fitting pulsar model is a simple powerlaw
(reduced chisquare of $\chi^2_{red}$ = 1.52, 3838 dof using both the pulsar and nebular spectra).
The powerlaw has a photon index $\Gamma$ = 1.52 $\pm$ 0.02 
absorbed by a column N$_H$ = (2.10 $\pm$ 0.01) $\times$ 10$^{22}$ cm$^{-2}$.
The nebula has a photon index $\Gamma$ = 1.85 $\pm$ 0.01.
A simple blackbody model is not statistically acceptable
while a combination of powerlaw and blackbody gives not any statistically
significative improvement.
Assuming the best fit model, the 0.3-10 keV unabsorbed pulsar flux is
(6.63 $\pm$ 0.15) $\times$ 10$^{-12}$ and the total nebular flux is
(7.21 $\pm$ 0.05) $\times$ 10$^{-11}$ erg/cm$^2$ s.
Using a distance
of 4.7 kpc, the luminosities are L$_{4.7kpc}^{nt}$ = (1.76 $\pm$ 0.04) $\times$ 10$^{34}$
and L$_{4.7kpc}^{pwn}$ = (1.91 $\pm$ 0.01) $\times$ 10$^{35}$ erg/s.

\begin{figure}
\centering
\includegraphics[angle=0,scale=.40]{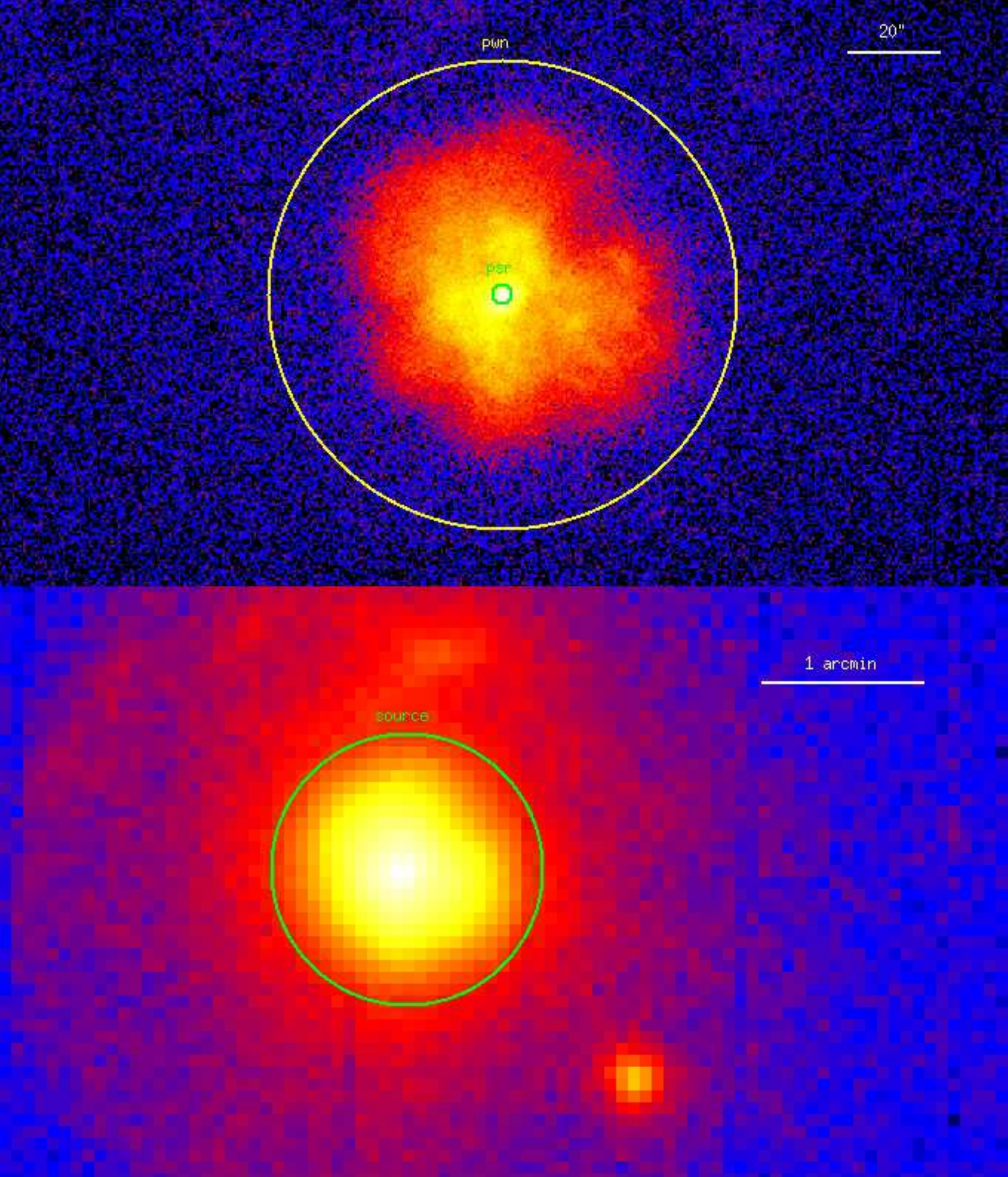}
\caption{{\it Upper Panel:} PSR J1833-1034 0.3-10 keV {\it Chandra} Imaging.
The green circle marks the pulsar region while the yellow annulus the nebular used in the analysis.
{\it Lower Panel:} PSR J1833-1034 0.3-10 keV {\it XMM-Newton} Imaging. The PN and the two MOS images have been added.
The green circle marks the source region used in the analysis.
\label{J1833-im}}
\end{figure}

\begin{figure}
\centering
\includegraphics[angle=0,scale=.50]{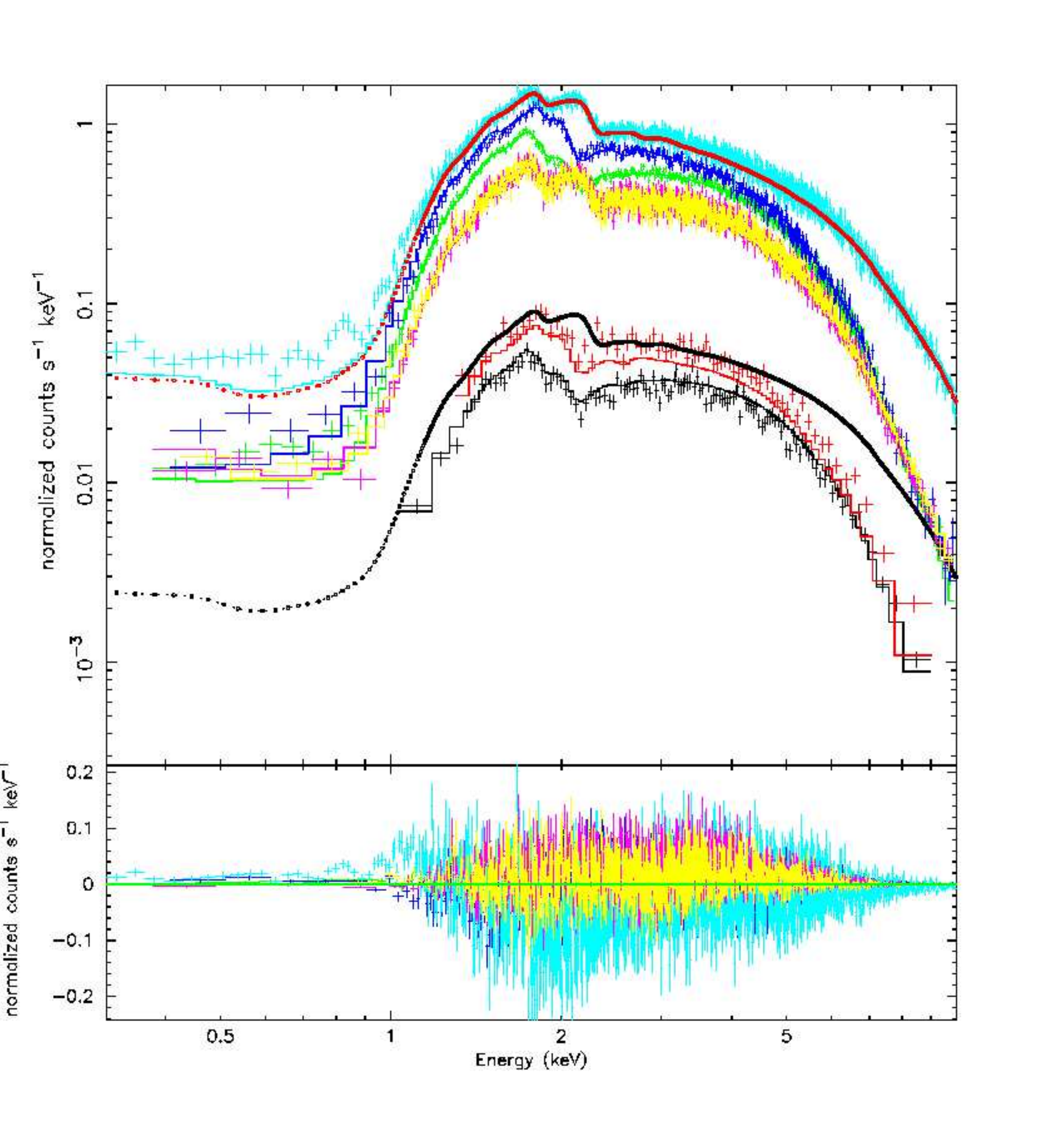}
\caption{PSR J1833-1034 Spectrum. Different colors mark all the different dataset used (see text for details).
Black points mark the pulsar spectrum while red points the nebular one.
Residuals are shown in the lower panel.
\label{J1833-sp}}
\end{figure}

\clearpage

{\bf J1836+5925 (Next Geminga) - type 2 RQP} % ho rifatto l'analisi

% Abdo et al. 2010
Since its discovery by EGRET (Lin et al. 1992), the bright $\gamma$-ray source GRO J1837+59
defied straightforward identification. It was reported as a persistent source with a varying flux of
3-8 $\times$ 10$^{-7}$ ph cm$^{-2}$ s$^{-1}$ and a relatively hard spectrum of photon index 1.7 in non-consecutive,
typically 2-3 week-long, observing periods. Its location at high Galactic latitude in conjunction
with early reports of $\gamma$-ray variability (later questioned by Nolan et al. 1996; Reimer et al.
2001) suggested it might be a blazar. However, the lack of a radio-bright counterpart, common to
EGRET-detected blazars, cast doubts on such an interpretation.
With the detection of faint X-ray counterpart candidates in the error contour of 3EG J1835+5918 (Reimer et al.
2000), the interpretation focused increasingly on a nearby radio-quiet neutron star. The complete
characterization of all but one of the ROSAT HRI X-ray sources was presented by Mirabal et al.
(2000) and Reimer et al. (2001), who singled out RX J1836.2+5925 as the most probable counterpart
of 3EG J1835+5918. Subaru/FOCAS observations in the B- and U-bands proposed possible
optical counterparts (Totani et al. 2002), while Hubble Space Telescope observations set an optical
upper limit of V $>$ 28.5 (Halpern et al. 2002). A scenario of a thermally emitting neutron star which
was either older or more distant than the archetypal radio-quiet $\gamma$-ray pulsar Geminga emerged
as the most plausible explanation for the source (Halpern \& Ruderman 1993; Bignami \& Caraveo
1996; Mirabal \& Halpern 2001), with an upper limit on the distance of 800 pc, determined from
X-ray observations (Halpern et al. 2002). Using {\it Chandra} observations separated by three years,
Halpern et al. (2007) were also able to determine an upper limit on the proper motion of 0.14$"$ per
year, or v$_t$ $<$ 530 km s$^{-1}$ at 800 pc. However, a timing signature, which would settle the nature
of this source, was never found in the EGRET data (Chandler et al. 2001; Ziegler et al. 2008), nor
in repeated observations by {\it Chandra} (Halpern et al. 2002, 2007), nor in a 24-hr observation with
NRAO's Green Bank Telescope (GBT) (Halpern et al. 2007). 
A $\gamma$-ray nebular emission was detected with a flux of 
(542.16 $\pm$ 34.03) $\times$ 10$^{-12}$ erg/cm$^2$s (Ackermann et al. 2010).

Three different X-ray observations of J1836+5925 were performed, two
by {\it XMM-Newton} and one by {\it Chandra}:\\
- obs. id 2764, {\it Chandra} ACIS-S very faint mode, start time 2002, March 06 at 02:22:00 UT, exposure 28.1 ks;\\
- obs. id 0511581701, {\it XMM-Newton} observation, start time 2008, May 18 at 09:55:31 UT, exposure 15.0 ks;\\
- obs. id 0511581801,  {\it XMM-Newton} observation, start time 2008, June 25 at 06:27:07 UT, exposure 13.0 ks.\\
In both the XMM observations the PN and MOS cameras were operating in the Full Frame mode
and a thin optical filter was used.
First, an accurate
screening for soft proton flare events was done in the {\it XMM-Newton} observations obtaining a resulting total
exposure of 24.4 ks.
The X-ray source best fit position (obtained by using the celldetect
tool inside the Ciao distribution) is 18:36:13.70 +59:25:30.12 (0.9$"$ error radius).
No nebular emission was detected in both the {\it Chandra} and XMM observations.
For the {\it Chandra} observation, we chose a 2$"$ radius circular region for the
pulsar spectrum; the background was extracted from an annular region
with radii 10 and 15$"$.
For the {\it XMM-Newton} observation, we chose a 15$"$ radius circular region around
the pulsar. The background was extracted from a circular source-free region on the same CCD
due to a faint source near the pulsar.
we obtained a total of 747, 177, 125 and 131 pulsar counts respectively from PN, the two
MOS cameras and the {\it Chandra} observation (background contributions of 57.5\%, 36.2\%, 47.2\% and 0.2\%).
The best fitting pulsar model is a combination of a blackbody and a powerlaw
(probability of obtaining the data if the model is correct 
- p-value - of 0.50, 40 dof).
The powerlaw component has a photon index $\Gamma$ = 2.05$_{-0.32}^{+0.54}$ 
absorbed by a column N$_H$ = 0.07$_{-0.07}^{+1.06}$ $\times$ 10$^{21}$ cm$^{-2}$.
The thermal component has a temperature of T = 6.31$_{-2.60}^{+2.35}$ $\times$ 10$^5$ K.
The blackbody radius R$_{0.4kpc}$ = 982 $\pm$ 868 m determined from the
model parameters suggests that the emission is from a hot spot.
A simple blackbody model is not statistically acceptable;
a simple powerlaw model is acceptable but
an f-test performed comparing
a simple powerlaw with the composite spectrum gives a
chance probability of 1.30 $\times$ 10$^{-3}$, pointing
to a significative improvement by adding the blackbody component.
Assuming the best fit model, the 0.3-10 keV unabsorbed non-thermal pulsar flux is
3.11$_{-2.14}^{+0.36}$ $\times$ 10$^{-14}$ and the thermal flux is 
1.06$_{-0.73}^{+0.27}$ $\times$ 10$^{-14}$ erg/cm$^2$ s.
Using a distance
of 0.4 kpc, the luminosities are L$_{0.4kpc}^{nt}$ = 5.97$_{-4.11}^{+0.69}$ $\times$ 10$^{29}$ and
L$_{0.4kpc}^{bol}$ = 2.03$_{-1.40}^{+0.52}$ $\times$ 10$^{29}$ erg/s.

\begin{figure}
\centering
\includegraphics[angle=0,scale=.50]{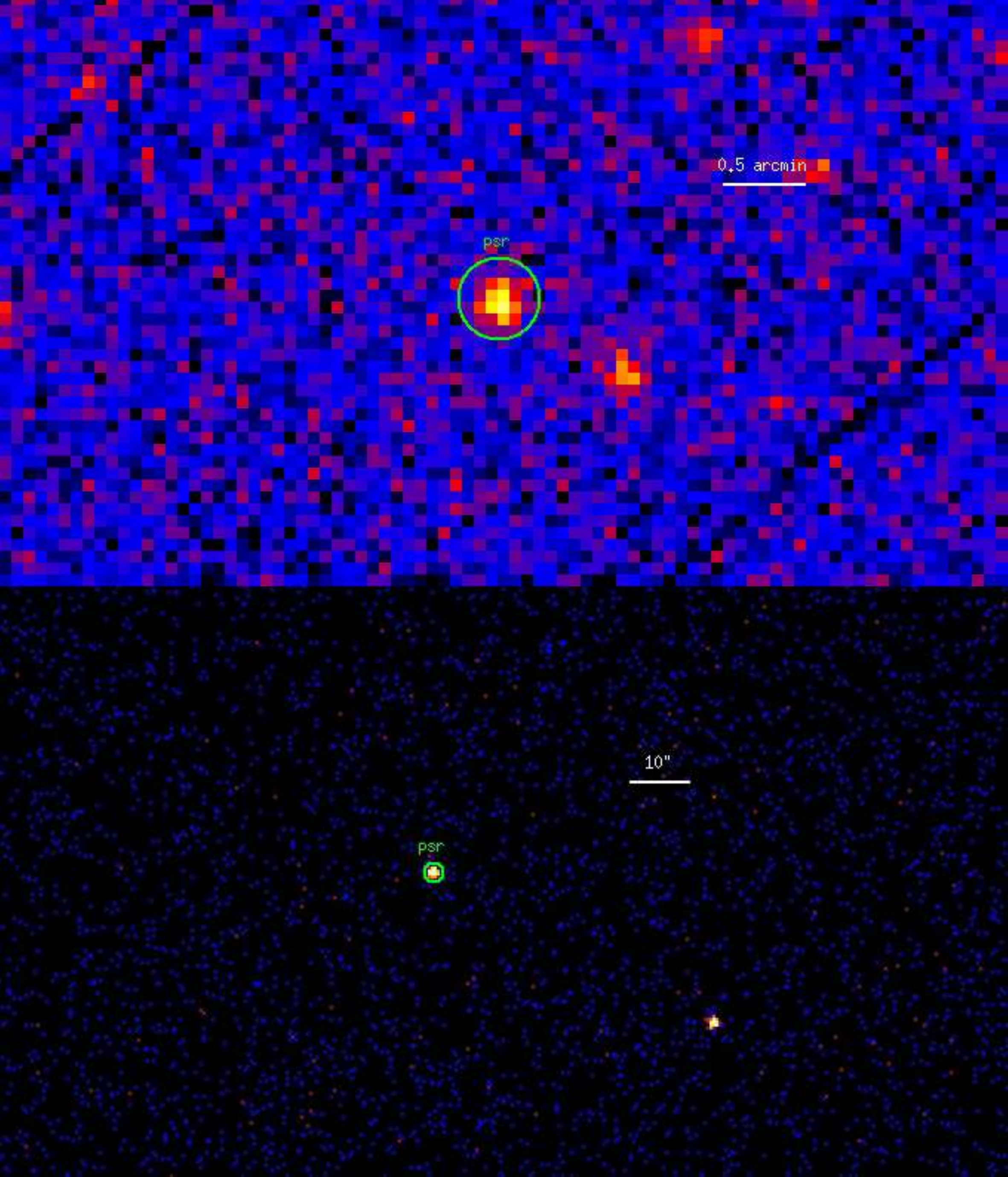}
\caption{{\it Upper Panel:} PSR J1836+5925 0.3-10 keV {\it XMM-Newton} Imaging. The PN and the two MOS images have been added.
The green circle marks the source region used in the analysis.
{\it Lower Panel:} PSR J1836+5925 0.3-10 keV {\it Chandra} Imaging.
The green circle marks the pulsar region used in the analysis.
\label{J1836-im}}
\end{figure}

\begin{figure}
\centering
\includegraphics[angle=0,scale=.50]{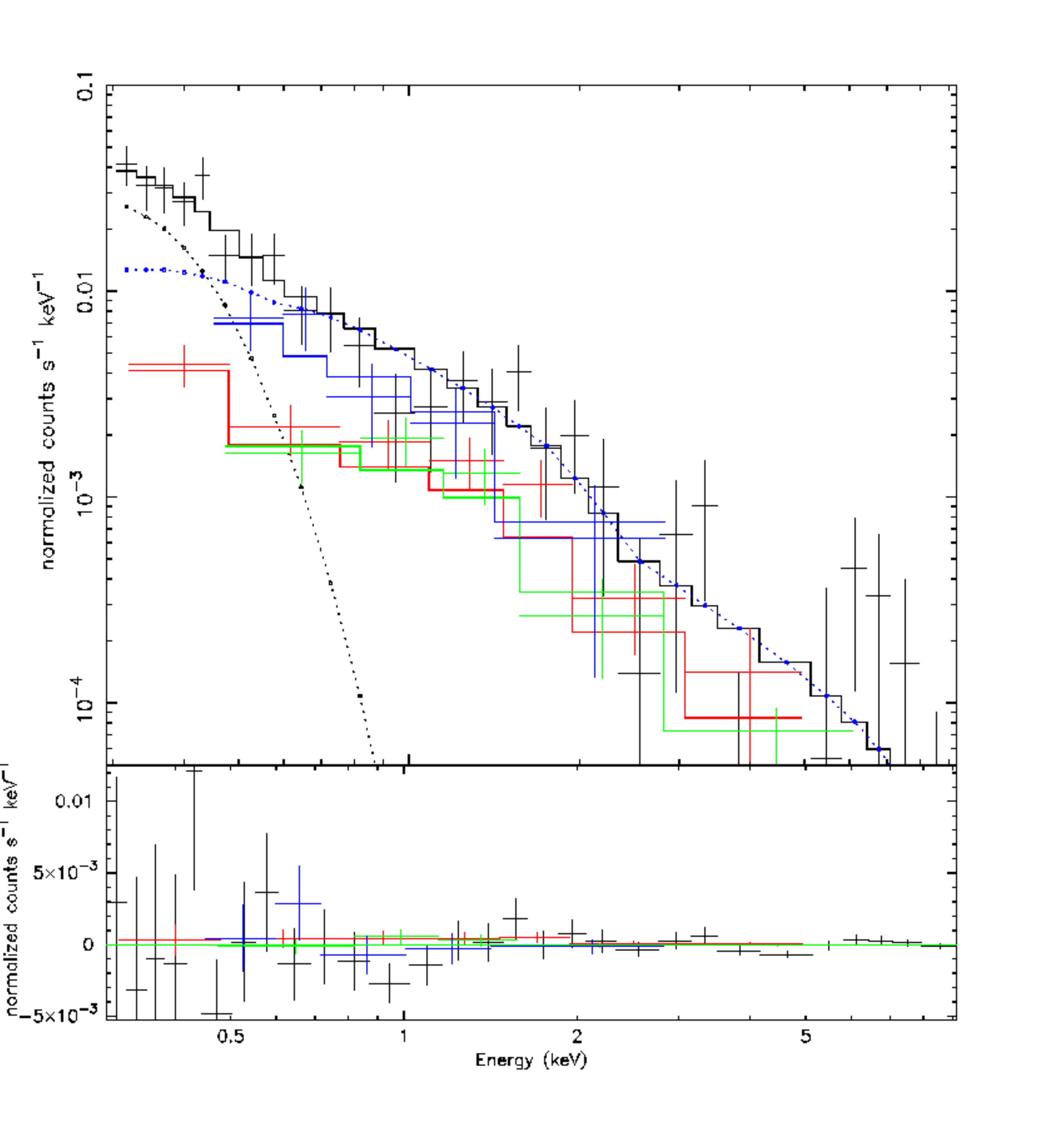}
\caption{PSR J1836+5925 Spectrum. Different colors mark all the different dataset used (see text for details).
Blue points mark the powerlaw component while black points the thermal component of the pulsar spectrum.
Residuals are shown in the lower panel.
\label{J1836-sp}}
\end{figure}

\clearpage

{\bf J1846+0919 - type 0 RQP}

J1732-31 was one of the first pulsars discovered using the
blind search technique (Saz Parkinson et al. 2010).
No $\gamma$-ray nebular emission was detected down to a flux of 
5.30 $\times$ 10$^{-12}$ erg/cm$^2$s (Ackermann et al. 2010).
The pseudo-distance of the object based on $\gamma$-ray data (Saz Parkinson et al. (2010))
is $\sim$ 1.2 kpc.

After the {\it Fermi} detection, we asked for a {\it SWIFT} observation
of the $\gamma$-ray error box (obs id. 00031460001, 3.71 ks exposure).
After the data reduction, no X-ray source were found inside
the {\it Fermi} error box.
For a distance of 1.2 kpc we found a
rough absorption column value of 2 $\times$ 10$^{21}$ cm$^{-2}$
and using a simple powerlaw spectrum
for PSR+PWN with $\Gamma$ = 2 and a signal-to-noise of 3,
we obtained an upper limit non-thermal unabsorbed flux of 2.92 $\times$ 10$^{-13}$ erg/cm$^2$ s,
that translates in an upper limit luminosity of L$_{1.2kpc}^{nt}$ = 5.05 $\times$ 10$^{31}$ erg/s.

{\bf J1902-5105 - type 0 RL MSP} % Nuova!

J1902-5105 is a millisecond pulsar in a binary system
found by the Pulsar Search Consortium.
Radio dispersion measurements found the distance to be $\sim$ 1.2 kpc.

After the {\it Fermi} source detection, a {\it SWIFT} observation
of the $\gamma$-ray error box was asked
(obs id. 00031727001, 4.22 ks exposure).
After the data reduction, no X-ray source were found at
the radio position.
For a distance of 1.2 kpc we found a
rough absorption column value of 3 $\times$ 10$^{20}$ cm$^{-2}$
and using a simple powerlaw spectrum
for PSR+PWN with $\Gamma$ = 2 and a signal-to-noise of 3,
we obtained an upper limit non-thermal unabsorbed flux of 1.54 $\times$ 10$^{-13}$ erg/cm$^2$ s,
that translates in an upper limit luminosity
L$_{1.2kpc}^{nt}$ = 2.66 $\times$ 10$^{31}$ erg/s.

{\bf J1907+0601 - type 1 RLP} % osservazione in arrivo

% Abdo et al. 2010
The TeV source MGRO J1908+06 was discovered by the {\it MILAGRO} collaboration at a median
energy of 20 TeV in their survey of the northern Galactic Plane (Abdo et al. 2007) with
a flux $\sim$ 80\% of the Crab at these energies. It was subsequently detected in the 300 GeV -
20 TeV range by the HESS (Aharonian et al. 2009) and VERITAS (Ward 2008) experiments.
The HESS observations show the source HESS J1908+063 to be clearly extended, spanning
$\sim$ 0.3$^{\circ}$ of a degree on the sky with hints of energy-dependent substructure. A decade earlier
Lamb \& Macomb (1997) catalogued a bright source of GeV emission from the EGRET data,
GeV J1907+0557, which is positionally consistent with MGRO J1908+06. It is near, but
inconsistent with, the third EGRET catalogue (Hartman et al. 1999) source 3EG J1903+0550
(Roberts et al. 2001).The {\it Fermi}
Bright Source List (Abdo et al. 2009b), based on 3 months of survey data, contains 0FGL
J1907.5+0602 which is coincident with GeV J1907+0557.
0FGL J1907.5+0602 was found to pulse with a period of 106.6 ms,
have a spin-down energy of $\sim$ 2.8 $\times$ 10$^{36}$ erg s$^{-1}$, and was given a preliminary designation
of PSR J1907+06.
No $\gamma$-ray nebular emission was detected by {\it Fermi} down to a flux of 
1.88 $\times$ 10$^{-11}$ erg/cm$^2$s (Ackermann et al. 2010).
The pseudo-distance of the object based on $\gamma$-ray data (Saz Parkinson et al. (2010))
is $\sim$ 1.3 kpc.

Only one {\it Chandra} ACIS-S observation was performed with the X-ray counterpart
inside their FOV, obs. id 11124 starting on 2009, August 19 at 02:34:14 UT
with an exposure of 19.1 ks.
The X-ray source best fit position (obtained by using the celldetect
tool inside the Ciao distribution) is 19:07:54.76 +06:02:14.71 (1$"$ error radius).
No study for an extended emission is possible due to the low statistic.
we chose a 2$"$ radius circular region for the
pulsar spectrum while the background is extracted from an annular region with radii
7 and 15$"$ around the pulsar.
we obtained a total of 30 pulsar counts (background contribution 3.4\%).
We used the C-statistic
approach implemented in XSPEC.
The best fitting pulsar model is a powerlaw
(reduced chisquare of $\chi^2_{red}$ = 0.69, 3 dof) with a
photon index $\Gamma$ = 3.16$_{-2.28}^{+2.76}$ 
absorbed by a column N$_H$ = 3.98$_{-3.75}^{+4.68}$ $\times$ 10$^{22}$ cm$^{-2}$.
A simple blackbody model is statistically acceptable but gives an unrealistic
high temperature (T $>$ 10$^7$ K).
By taking in account 30\% of the flux coming from thermal and/or nebular contributions and
assuming the best fit model, the 0.3-10 keV unabsorbed pulsar flux is
2.75 $\pm$ 1.01 $\times$ 10$^{-13}$ erg/cm$^2$s and 
for a distance
of 1.3 kpc, its luminosity is L$_{1.3kpc}^{nt}$ = 5.58 $\pm$ 2.06 $\times$ 10$^{31}$ erg/s.

\begin{figure}
\centering
\includegraphics[angle=0,scale=.50]{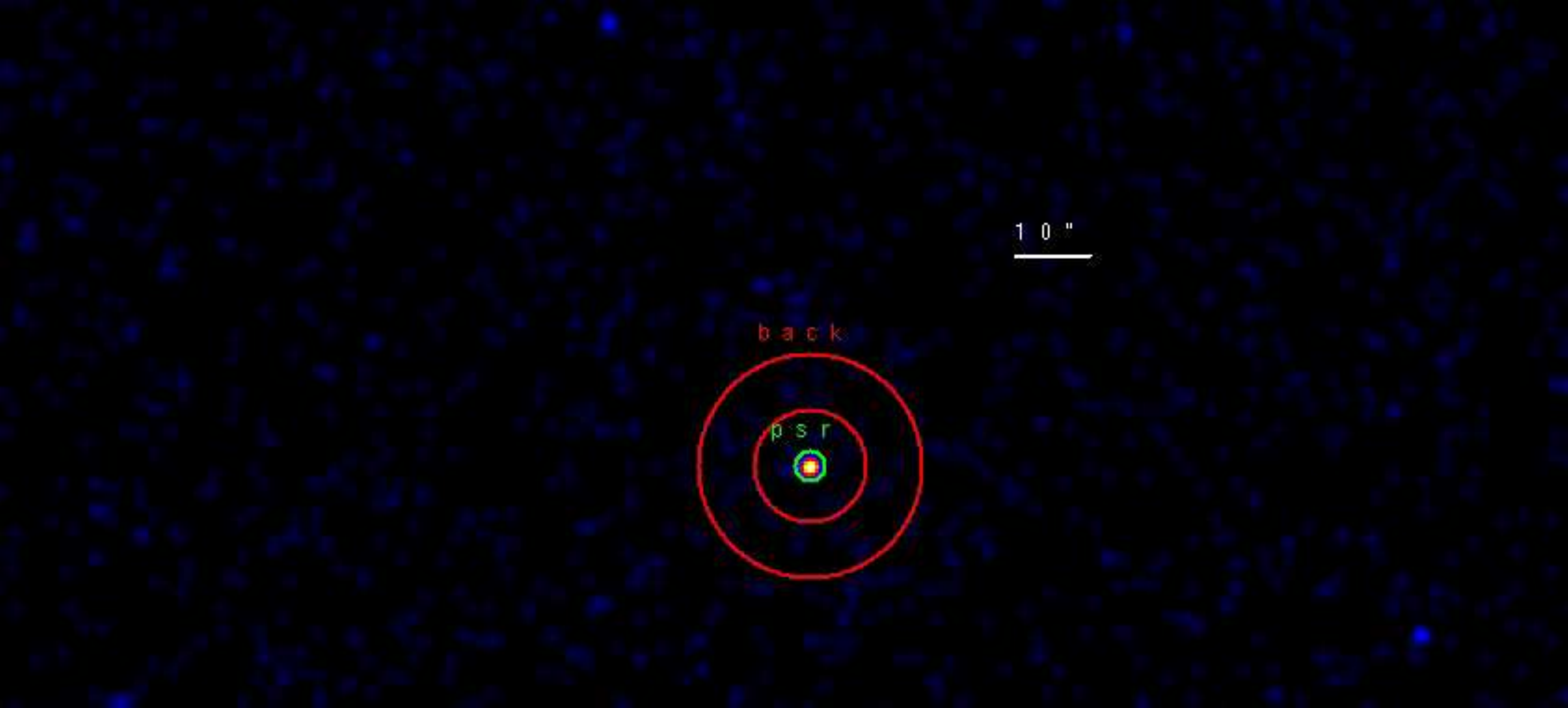}
\caption{PSR J1907+0601 0.3-10 keV {\it Chandra} Imaging.
The green circle marks the pulsar while the red circle the background region used in the analysis.
\label{J1907-im}}
\end{figure}

\begin{figure}
\centering
\includegraphics[angle=0,scale=.50]{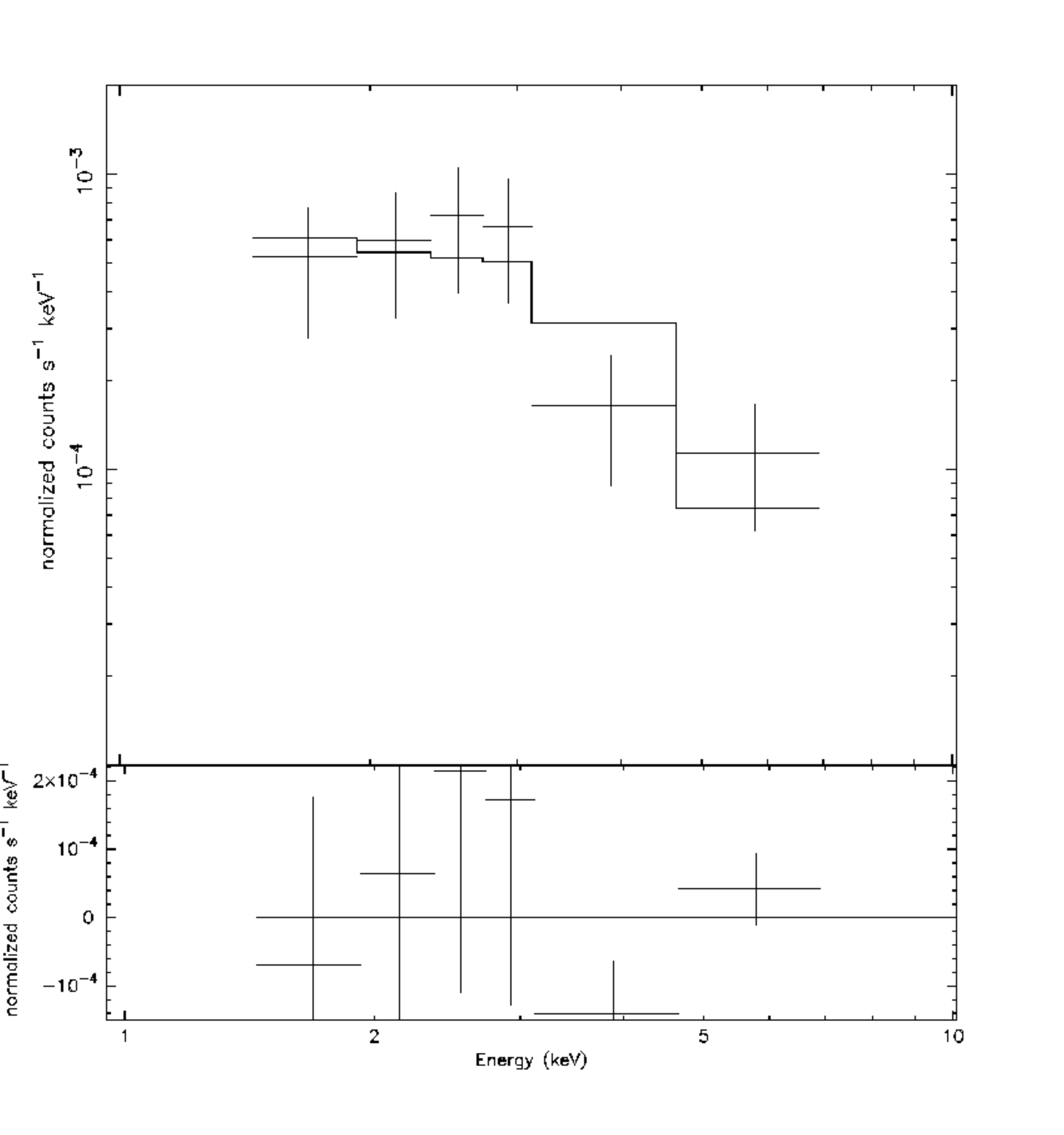}
\caption{PSR J1907+0601 {\it Chandra} Spectrum (see text for details).
Residuals are shown in the lower panel.
\label{J1907-sp}}
\end{figure}

\clearpage

{\bf J1939+2134 - type 2 RL MSP} % Nuova! Con dati che diverranno pubblici in aprile osservazione in arrivo

% Cusumano et al. 2003
PSR B1937+21 was the first MSP discovered (Backer
\& Sallmen 1982) and, with the period of 1.56 ms, it remains the
most rapidly rotating neutron star presently known. The
distance estimated from the observed dispersion measure
(DM) and from a model for the Galactic free electron distribution
(Taylor \& Cordes 1993, Cordes \& Lazio 2002)
is 7.7 $\pm$ 3.8 kpc (see also discussion in Nicastro et al. 2003).
Its spin down luminosity is $\dot{E}$ $\sim$ 1.1 $\times$ 10$^{36}$ erg/s and
the dipolar magnetic field component at the star surface
is $\sim$ 4.1 $\times$ 10$^8$ G. Like the Crab pulsar (Lundgren et al.
1995), PSR B0540-69 (Johnston \& Romani 2003) and the
other MSP PSR B1821-24 (Romani \& Johnston 2001), PSR
B1937+21 exhibits sporadic emission of giant pulses in the
radio band (Sallmen \& Backer 1995; Cognard et al. 1996,
Kinkhabwala \& Thorsett 2000). Such pulses are extremely short
events ($<$ 0.3 $\mu$s at 2.38 GHz) confined to small phase
windows trailing the main and interpulse.
X-ray emission from this pulsar was detected by ASCA
(Takahashi et al. 2001) above 2 keV, with a pulse profile
characterized by a single sharp peak and a pulsed fraction
of 44\%. Comparing the X-ray and radio phase at
rival times, these authors claimed that the X-ray pulse is
aligned with the radio interpulse. Later, BeppoSAX detected
pulsed emission from PSR B1937+21 (Nicastro et
al. 2002, 2003) and the pulse profile was found to show
a double peak pattern with a phase separation from P1
to P2 of 0.52 $\pm$ 0.02 and a significance of the second peak
of $\sim$ 5$\sigma$.

\begin{figure}
\centering
\includegraphics[angle=0,scale=.30]{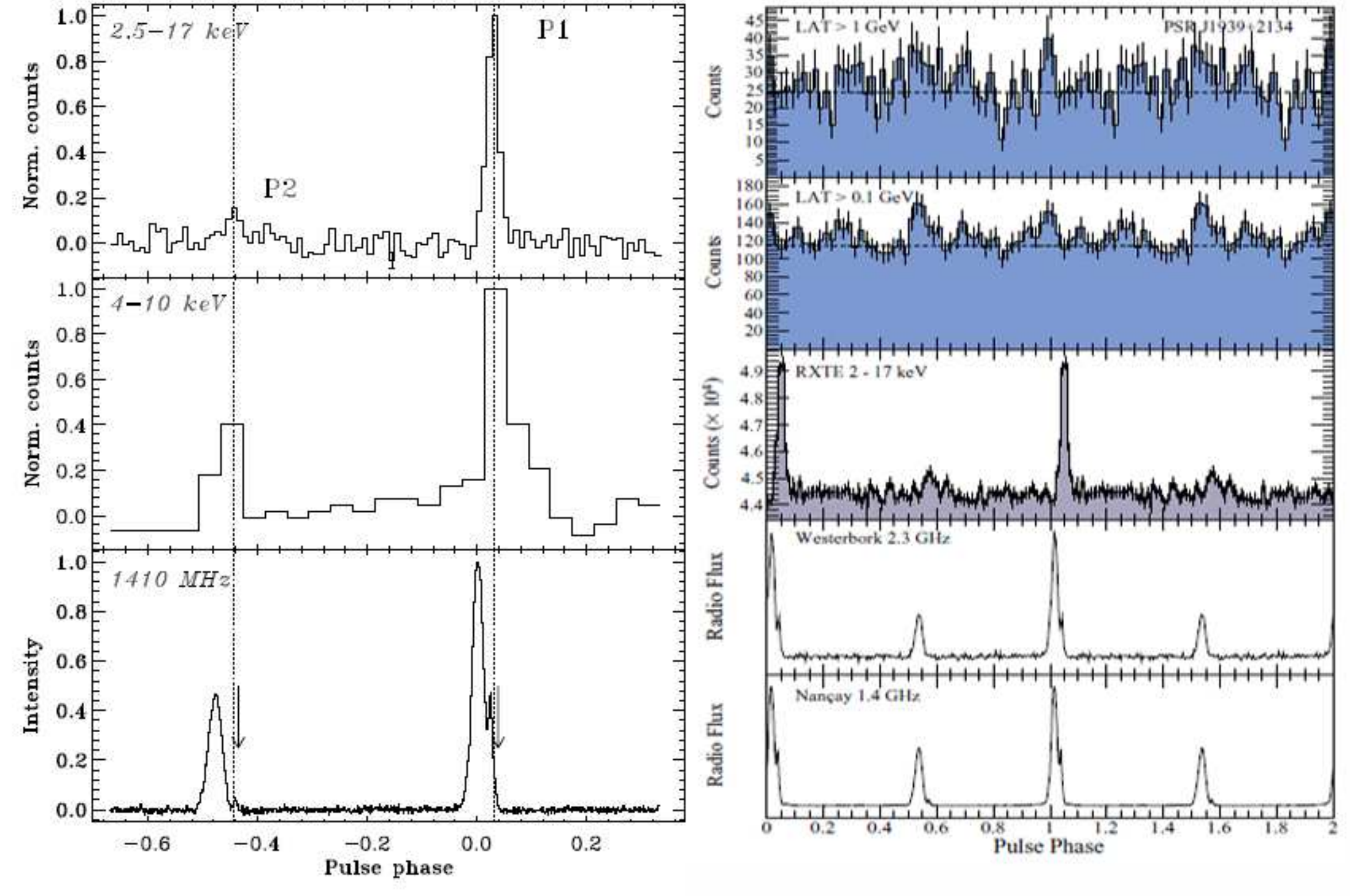}
\caption{PSR J1939+2134 Lightcurve. {\it Left:} The {\it RXTE} pulse profile
in the 2-17 keV energy band; the 4-10 keV {\it BeppoSAX} profile
(Nicastro et al. 2002) with the P1 aligned with the P1
phase in the top panel; radio pulse profile at 1.6 GHz.
See Cusumano et al. 2008 for details.
{\it Right: Fermi} $\gamma$-ray lightcurve folded with Radio
(Abdo et al. 2 millisecond pulsars in preparation). 
\label{J1939-lc}}
\end{figure}

Only one X-ray observation of the source is public, performed by {\it Chandra}
ACIS-S in the faint mode (obs. id 5516), starting on 2005, June 28 at 22:40:54 UT,
for a total exposure of 50.1 ks.
The X-ray source best fit position (obtained by using the celldetect
tool inside the Ciao distribution) is 19:39:38.57 +21:34:59.30 (1$"$ error radius).
we searched for diffuse emission in the immediate
surroundings of the pulsar, by comparing the source intensity profile to the expected ACIS
Point Spread Function (PSF). Assuming the pulsar best fit spectral model, we simulated a
PSF using the ChaRT and MARX packages. No nebular emission was detected.
we chose a 2$"$ radius circular region for the
pulsar spectrum; the background was extracted from an annular region
with radii 20 and 30$"$.
we obtained a total of 550 pulsar counts (background contribution of 0.1\%).
we used the C-statistic approach implemented in XSPEC.
The best fitting pulsar model is a simple powerlaw (reduced chisquare value $\chi^2_{red}$ = 1.4, 18 dof)
with a photon index $\Gamma$ = 1.15$_{-0.32}^{+0.46}$ 
absorbed by a column N$_H$ = 1.09$_{-0.44}^{+0.63}$ $\times$ 10$^{22}$ cm$^{-2}$.
A simple blackbody model isn't statistically acceptable while a composite model
gives no significative improvement in the fit.
Assuming the best fit model, the 0.3-10 keV unabsorbed non-thermal pulsar flux is
3.95 $\pm$ 0.71 $\times$ 10$^{-13}$ erg/cm$^2$ s.
Using a distance
of 7.7 kpc, the pulsar luminosity is L$_{7.7kpc}^{nt}$ = 2.81 $\pm$ 0.51 $\times$ 10$^{33}$ erg/s.

\begin{figure}
\centering
\includegraphics[angle=0,scale=.50]{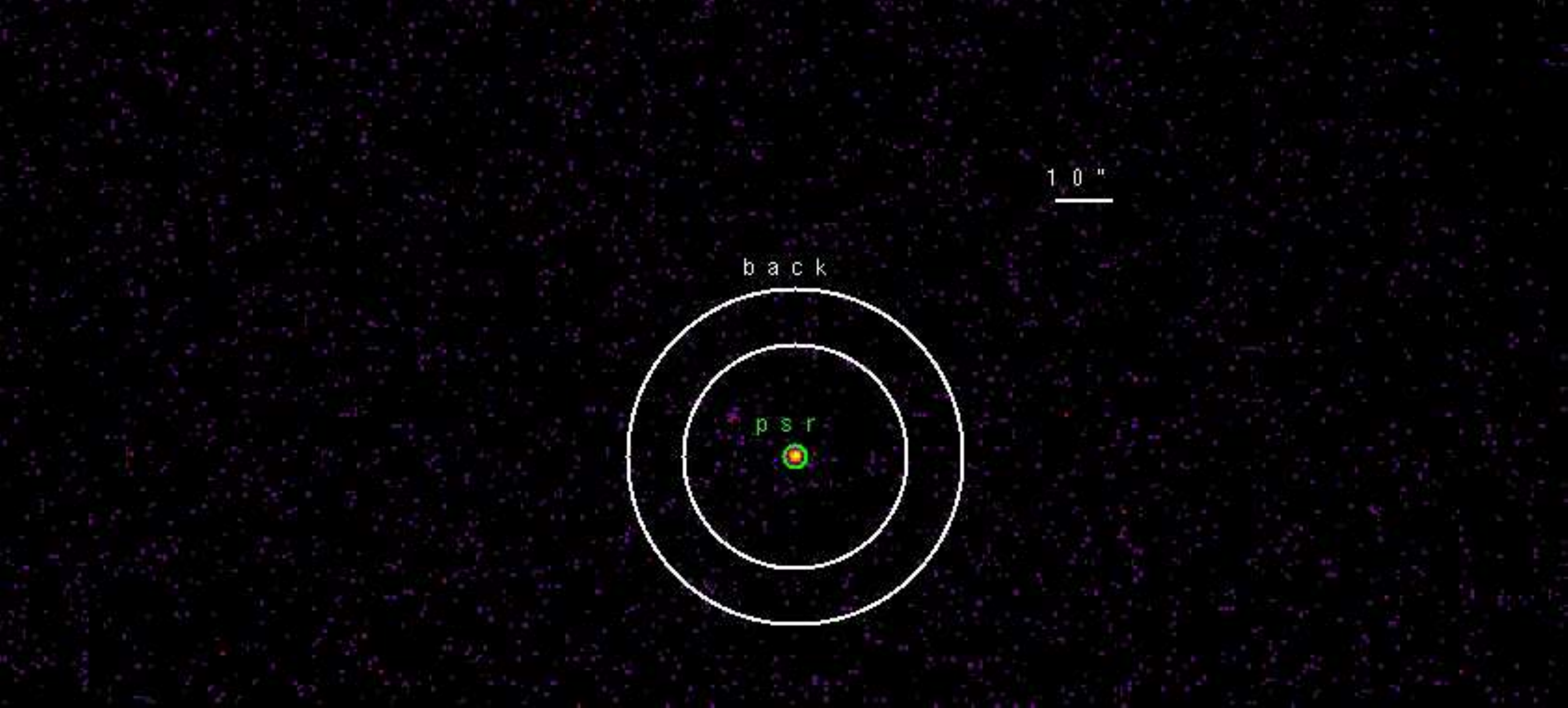}
\caption{PSR J1939+2134 0.3-10 keV {\it Chandra} Imaging.
The green circle marks the pulsar while the white annulus the background region used in the analysis.
\label{J1939-im}}
\end{figure}

\begin{figure}
\centering
\includegraphics[angle=0,scale=.50]{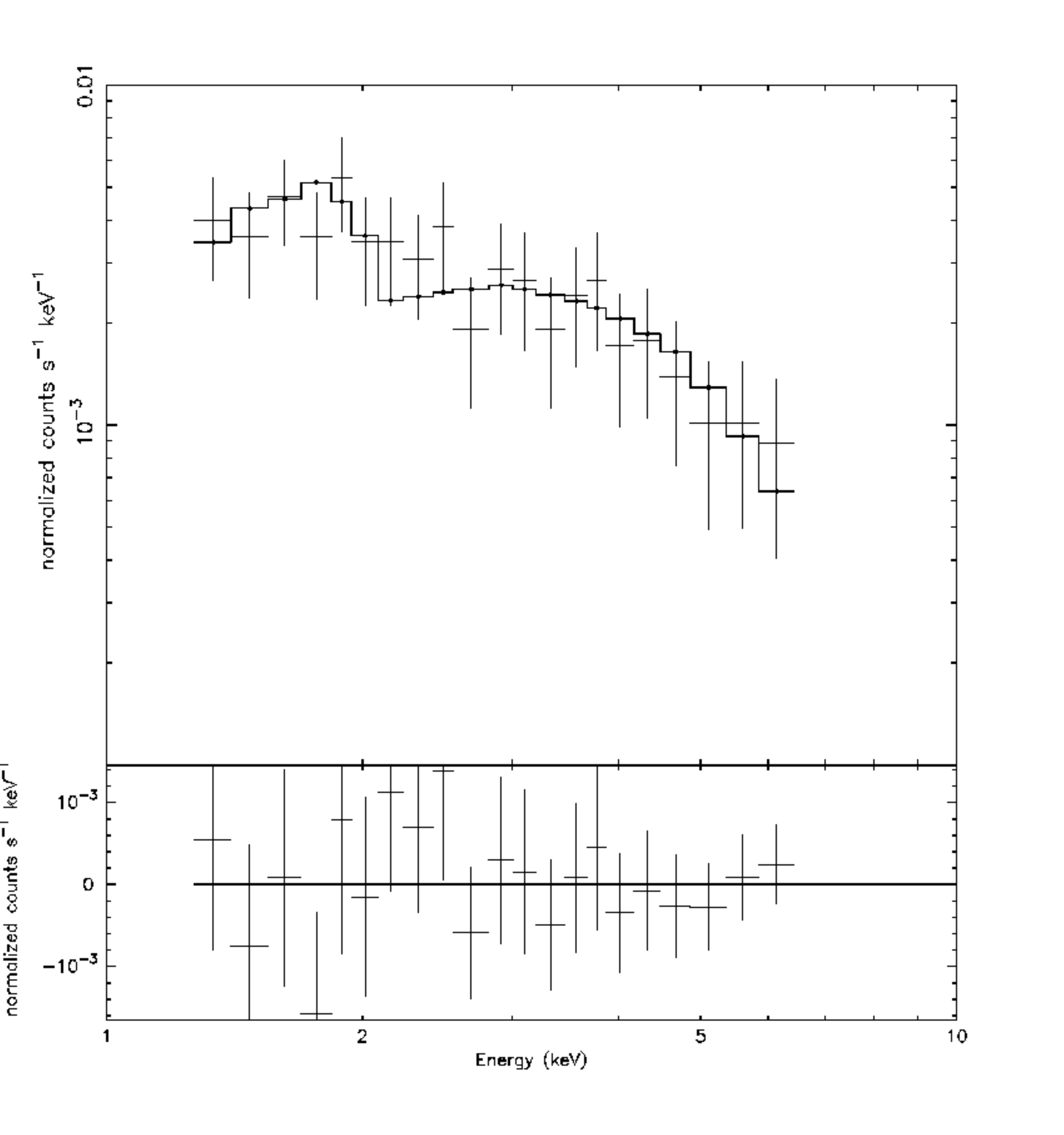}
\caption{PSR J1939+2134 {\it Chandra} Spectrum (see text for details).
Residuals are shown in the lower panel.
\label{J1939-sp}}
\end{figure}

\clearpage

{\bf J1952+3252 (CTB80) - type 2 RLP} % rifatta l'analisi osservazione in arrivo

% Zeiger et al. 2008
PSR B1951+32, a 39.5 ms pulsar with $\tau_c$ = 107 kyr and a
spin-down energy-loss rate $\dot{E}$ = 4 $\times$ 10$^{36}$ erg s$^{-1}$, lies near the southwestern
edge of the approximately circular infrared shell of CTB
80 (Fesen et al. 1988). The pulsar is surrounded by a $\sim$ 30$"$ diameter
asymmetric PWN at the western edge of an 8' $\times$ 4' east-
west plateau of emission (Castelletti et al. 2003). CTB 80, observed
at 1.4 GHz to have three arms that cover 1.8 deg$^2$ and converge
near B1951+32 (Castelletti et al. 2003), has an expanding H I
shell that yields a dynamical age estimate of 77 kyr for the
pulsar-SNR system (Koo et al. 1990). From its dispersion measure,
the distance to B1951+32 is estimated as 3.1 $\pm$ 0.2 kpc
(Taylor \& Cordes 1993), while the distance to CTB 80 is estimated
as 2.0 $\pm$ 0.5 kpc from H I absorption (Strom \& Stappers 2000). Here
we adopt a distance of 2 kpc. Kulkarni et al. (1988) estimated
a speed of 300 km s$^{-1}$ for the pulsar from scintillation measurements,
while Migliazzo et al. (2002) directly measure a
proper motion $\mu$ = 25 $\pm$ 4 mas yr$^{-1}$ at a position angle 252$^{\circ}$ $\pm$
7$^{\circ}$ after correcting for the effects of differential Galactic rotation.
Their implied transverse velocity of the pulsar is V$_{2kpc}$ =
(240 $\pm$ 40) km s$^{-1}$.
In the direction of the proper motion vector measured by
Migliazzo et al. (2002), Moon et al. (2004) observe a cometary
X-ray synchrotron nebula and an H$\alpha$ bow shock. The X-ray
emission peaks at the pulsar location and is confined within the
H$\alpha$ structure (Hester 2000) that is clearly defined at an angular
separation of $\sim$ 7$"$ from the pulsar. The X-rays are produced by
synchrotron emission from the pulsar wind, confined within a bow
shock produced by the ram pressure of the wind interacting with
the SNR wall.
In radio images, B1951+32 appears to be situated just inside
a limb-brightened bubble $\sim$ 30$"$ in diameter.
A 5$"$ portion of the bubble nearest to the pulsar is substantially
brighter than any other portion of the shell. In MERLIN observations
at 1.6 GHz with 0.15$"$ resolution, Golden et al. (2005)
find that the radio-bright arc shows compact structure, resembling
a radio bow shock. The structure is $\sim$ 2.5$"$ from the pulsar,
which puts it at the head of the cometary X-ray nebula, but
enclosed within the H$\alpha$ bow shock which is $\sim$ 7$"$ away from the
pulsar (Moon et al. 2004).
No $\gamma$-ray nebular emission was detected by {\it Fermi} down to a flux of 
1.87 $\times$ 10$^{-11}$ erg/cm$^2$s (Ackermann et al. 2010).

Two different observations of J1952+3252 were performed, one by
{\it XMM-Newton} and one by {\it Chandra}:\\
- obs. id 1984, {\it Chandra} ACIS-S very faint mode, start time 2001, July 12 at 06:24:50 UT, exposure 78.2 ks;\\
- obs. id 0204070101, {\it XMM-Newton} observation, start time 2004, May 11 at 12:49:28 UT, exposure 10.0 ks.\\
In the XMM observation the PN camera was operating in Small Window mode while the  MOS cameras in the Full Frame mode.
For PN and MOS cameras a thin optical filter was used.
No screening for soft proton flare events was done in the {\it XMM-Newton} observation due to the 
goodness of the observation.
The X-ray source best fit position (obtained by using the celldetect
tool inside the Ciao distribution) is 19:52:58.20 +32:52:40.57 (1$"$ error radius).
A bright $\sim$ 35$"$ nebular emission was detected in both the {\it Chandra} and XMM observations.
A fainter halo extends until $\sim$ 80$"$, visible in the {\it XMM-Newton} observation.
For the {\it Chandra} observation, we chose a 2$"$ radius circular region for the
pulsar spectrum; the nebular spectrum is extracted from an annular region
with radii 2 and 40$"$. The background is extracted from a source-free region away from the pulsar.
For the {\it XMM-Newton} observation, we chose a 40$"$ radius circular region around
the pulsar. The background was extracted from a circular source-free region on the same CCD.
We obtained a total of 21463 pulsar counts and 43510 nebular counts in the {\it Chandra} observation
(background contributions of 0.5\% and 4.6\%). we also obtained 14190, 5883 and 6091 counts
from the PN and MOS cameras in the {\it XMM-Newton} observation (background contributions of 14.9\%, 7.0\% and 6.7\%).
The best fitting pulsar model is a combination of a blackbody and a powerlaw
(reduced chisquare value $\chi^2_{red}$ = 1.14, 1552 dof).
The powerlaw component has a photon index $\Gamma$ = 1.71 $\pm$ 0.03 
absorbed by a column N$_H$ = (3.33 $\pm$ 0.09) $\times$ 10$^{21}$ cm$^{-2}$.
The thermal component has a temperature of T = 1.61$_{-0.15}^{+0.16}$ $\times$ 10$^6$ K and a
radius R$_{2kpc}$ = 4.18 $_{-1.85}^{+2.95}$ km.
The nebular powerlaw has a photon index $\Gamma$ = 1.81 $\pm$ 0.02.
Both a simple powerlaw and a simple blackbody models aren't statistically acceptable.
Assuming the best fit model, the 0.3-10 keV unabsorbed non-thermal pulsar flux is
(4.07 $\pm$ 0.15) $\times$ 10$^{-12}$, the thermal flux is 
(3.30 $\pm$ 0.08) $\times$ 10$^{-13}$ and the nebular flux is
(7.77 $\pm$ 0.15) $\times$ 10$^{-12}$ erg/cm$^2$ s.
Using a distance
of 2.0 kpc, the luminosities are L$_{2kpc}^{nt}$ = (1.95 $\pm$ 0.07) $\times$ 10$^{34}$ and
L$_{2kpc}^{bol}$ = (1.58 $\pm$ 0.04) $\times$ 10$^{33}$ and L$_{2kpc}^{pwn}$ = (3.73 $\pm$ 0.07) $\times$ 10$^{34}$ erg/s.

\begin{figure}
\centering
\includegraphics[angle=0,scale=.50]{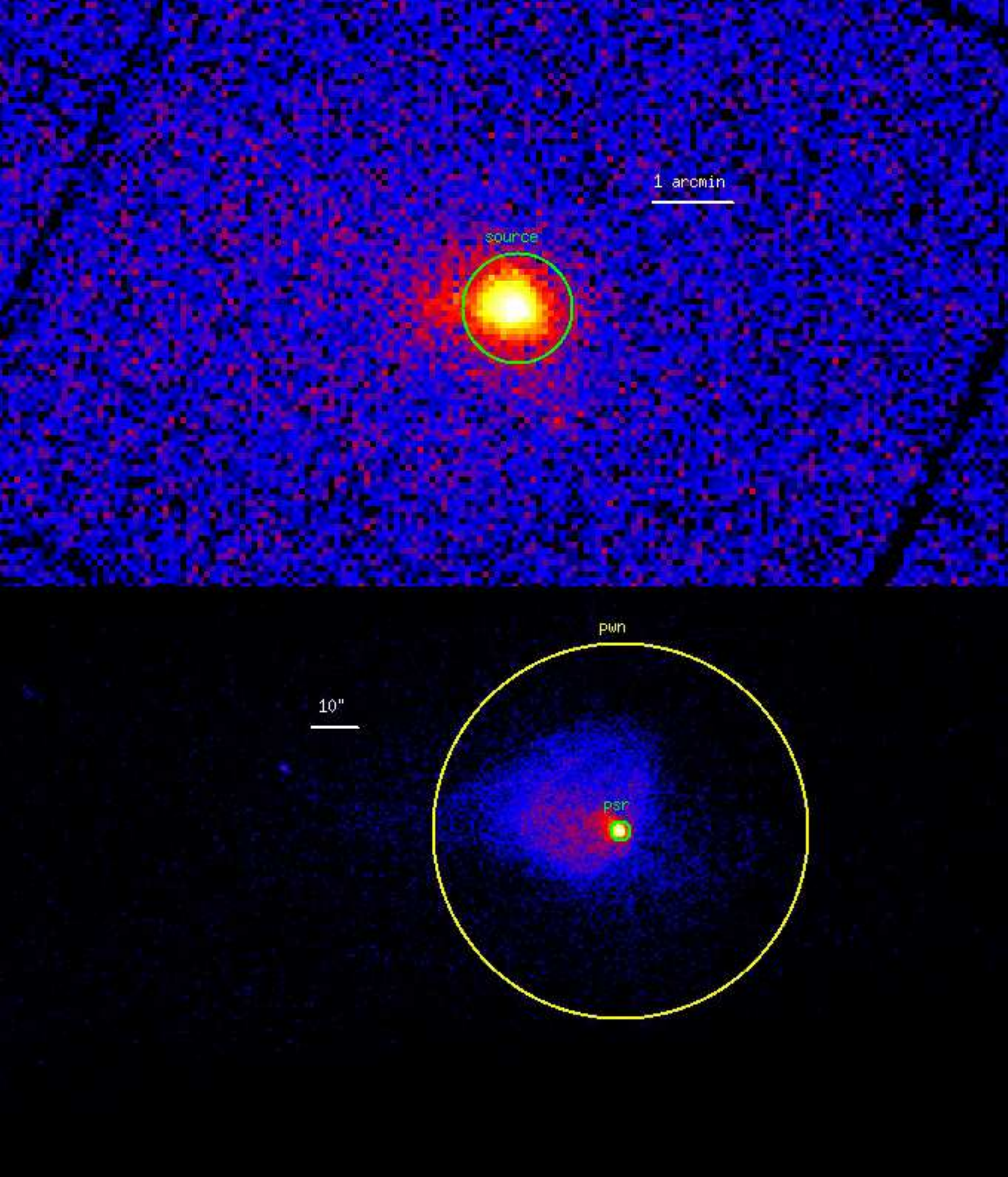}
\caption{{\it Upper Panel:} PSR J1952+3252 0.3-10 keV {\it XMM-Newton} Imaging. The PN and the two MOS images have been added.
The green circle marks the source region used in the analysis.
{\it Lower Panel:} PSR J1952+3252 0.3-10 keV {\it Chandra} Imaging.
The green circle marks the pulsar region while the yellow annulus the background used in the analysis.
\label{J1952-im}}
\end{figure}

\begin{figure}
\centering
\includegraphics[angle=0,scale=.50]{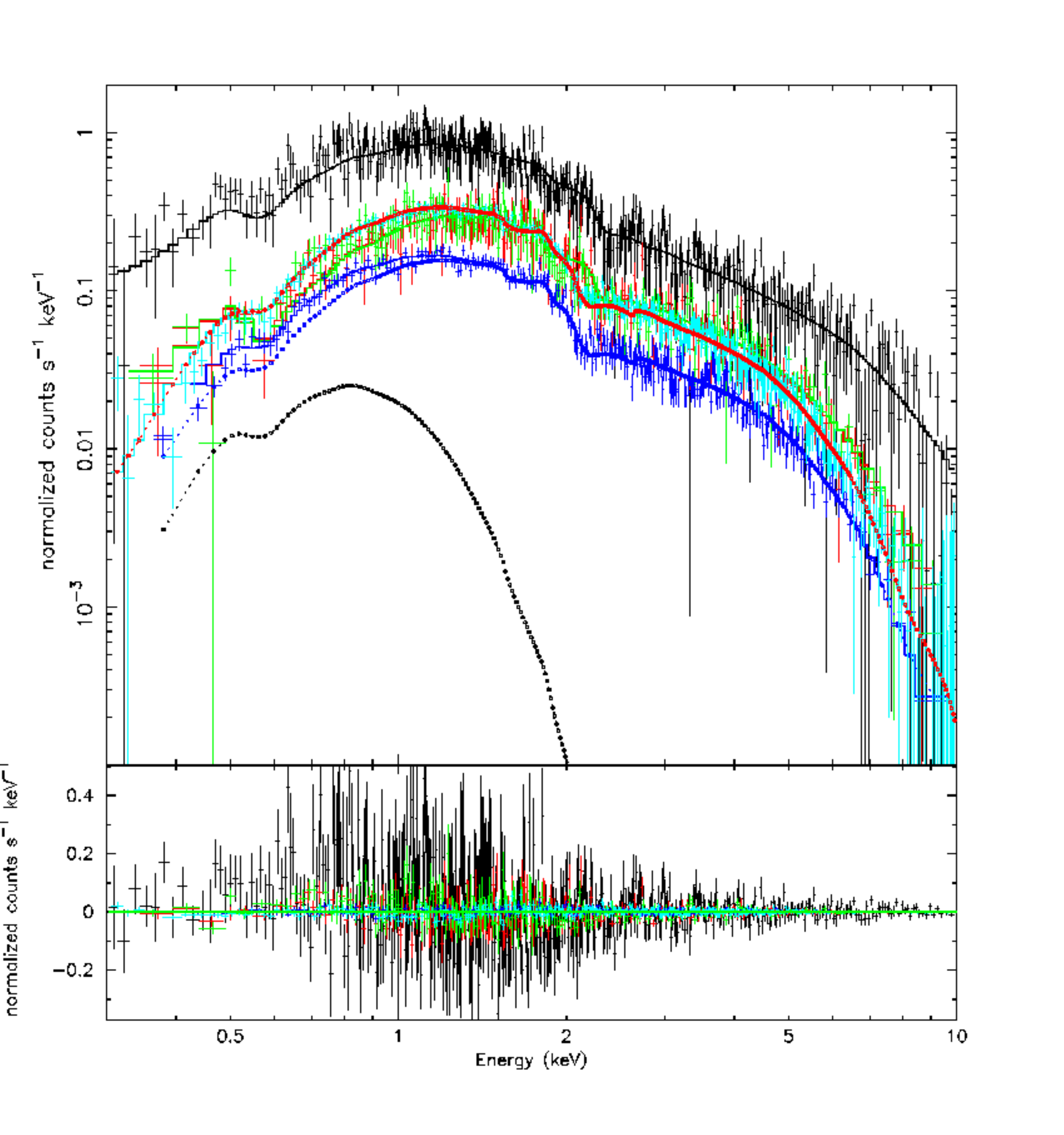}
\caption{PSR J1952+3252 Spectrum. Different colors mark all the different dataset used (see text for details).
Blue points mark the powerlaw component while black points the thermal component of the pulsar spectrum.
Red points mark the nebular spectrum.
Residuals are shown in the lower panel.
\label{J1952-sp}}
\end{figure}

\clearpage

{\bf J1954+2836 - type 0 RQP}

Pulsations from J1954+2836 were discovered by {\it Fermi} using the
blind search technique (Saz Parkinson et al. 2010).
No $\gamma$-ray nebular emission was detected down to a flux of 
2.38 $\times$ 10$^{-11}$ erg/cm$^2$s (Ackermann et al. 2010).
The pseudo-distance of the object based on $\gamma$-ray data (Saz Parkinson et al. (2010))
is $\sim$ 1.7 kpc.

After the {\it Fermi} detection, we asked for a {\it SWIFT} observation
of the $\gamma$-ray error box (obs id. 00031555001, 6.28 ks exposure).
After the data reduction, no X-ray source were found inside
the {\it Fermi} error box.
For a distance of 1.7 kpc we found a
rough absorption column value of 5 $\times$ 10$^{21}$ cm$^{-2}$
and using a simple powerlaw spectrum
for PSR+PWN with $\Gamma$ = 2 and a signal-to-noise of 3,
we obtained an upper limit non-thermal unabsorbed flux of 3.65 $\times$ 10$^{-13}$ erg/cm$^2$ s,
that translates in an upper limit luminosity L$_{1.7kpc}^{nt}$ = 1.27 $\times$ 10$^{32}$ erg/s.

{\bf J1957+5036 - type 0 RQP}

Pulsations from J1957+5036 were detected by {\it Fermi} using the
blind search technique (Saz Parkinson et al. 2010).
No $\gamma$-ray nebular emission was detected down to a flux of 
6.04 $\times$ 10$^{-12}$ erg/cm$^2$s (Ackermann et al. 2010).
The pseudo-distance of the object based on $\gamma$-ray data (Saz Parkinson et al. (2010))
is $\sim$ 0.9 kpc.

After the {\it Fermi} detection, we asked for a {\it SWIFT} observation
of the $\gamma$-ray error box (obs id. 00031484001, 3.56 ks exposure).
After the data reduction, one X-ray source were found inside
the {\it Fermi} error box. After a dedicated timing analysis
on the {\it Fermi} data the source was excluded as potential
counterpart.
For a distance of 0.9 kpc we found a
rough absorption column value of 1 $\times$ 10$^{21}$ cm$^{-2}$
and using a simple powerlaw spectrum
for PSR+PWN with $\Gamma$ = 2 and a signal-to-noise of 3,
we obtained an upper limit non-thermal unabsorbed flux of 2.98 $\times$ 10$^{-13}$ erg/cm$^2$ s,
that translates in an upper limit luminosity L$_{0.9kpc}^{nt}$ = 2.90 $\times$ 10$^{31}$ erg/s.

{\bf J1958+2841 - type 1 RQP} %usata la cstat osservazione in arrivo

J1957+5036 was one of the first pulsars discovered using the
blind search technique (Abdo et al. 2009 Science).
No $\gamma$-ray nebular emission was detected down to a flux of 
1.71 $\times$ 10$^{-11}$ erg/cm$^2$s (Ackermann et al. 2010).
The pseudo-distance of the object based on $\gamma$-ray data (Saz Parkinson et al. (2010))
is $\sim$ 1.4 kpc.

After the {\it Fermi} detection, we asked for a {\it SWIFT} observation
of the $\gamma$-ray error box (obs id. 00031374001, 00031374002, 00090265001 and 00090265003, total 12.15ks exposure).
Such observations revealed one potential
X-ray counterpart inside the 1-year {\it Fermi} error box
at 19:58:46.18 +28:46:02.71 (10$"$ error radius).
A precise timing analysis performed by the collaboration
(Ray et al. 2011) confirmed the X-ray source to be the counterpart of the $\gamma$-ray pulsar.
No indication can be obtained from the X-ray observation
regarding the presence of a PWN.
We used a 20$"$ radius circle in order to extract the pulsar
spectrum and an annular region with radii 40 and 80$"$ around the source.
No study for an extended emission is possible due to the low statistic.
We obtained a total of 20 source counts (background contribution of 9.3\%).
We used the C-statistic approach implemented in XSPEC.
Using a distance of 1.4 kpc we found a
rough absorption column value of 4 $\times$ 10$^{21}$ cm$^{-2}$ and
We freezed the N$_H$ to such value.
The best fitting pulsar model is a simple powerlaw
(reduced chisquare of $\chi^2_{red}$ = 0.60, 2 dof)
with a photon index $\Gamma$ = 1.14$_{-0.84}^{+0.79}$.
A simple blackbody model is statistically acceptable
but gives an unrealistic value of the temperature ($>$ 10$^7$ K).
The statistic is too low to study composite models.
By taking in account the 30\% of the flux coming from thermal and/or nebular emissions and
assuming the best fit model, the 0.3-10 keV unabsorbed source flux is
1.11 $\pm$ 0.71 $\times$ 10$^{-13}$ erg/(cm$^2$ s).
Using a distance
of 1.4 kpc, the source luminosity is L$_{1.4kpc}^{nt}$ = 2.62 $\pm$ 1.68 $\times$ 10$^{31}$ erg/s.

\begin{figure}
\centering
\includegraphics[angle=0,scale=.50]{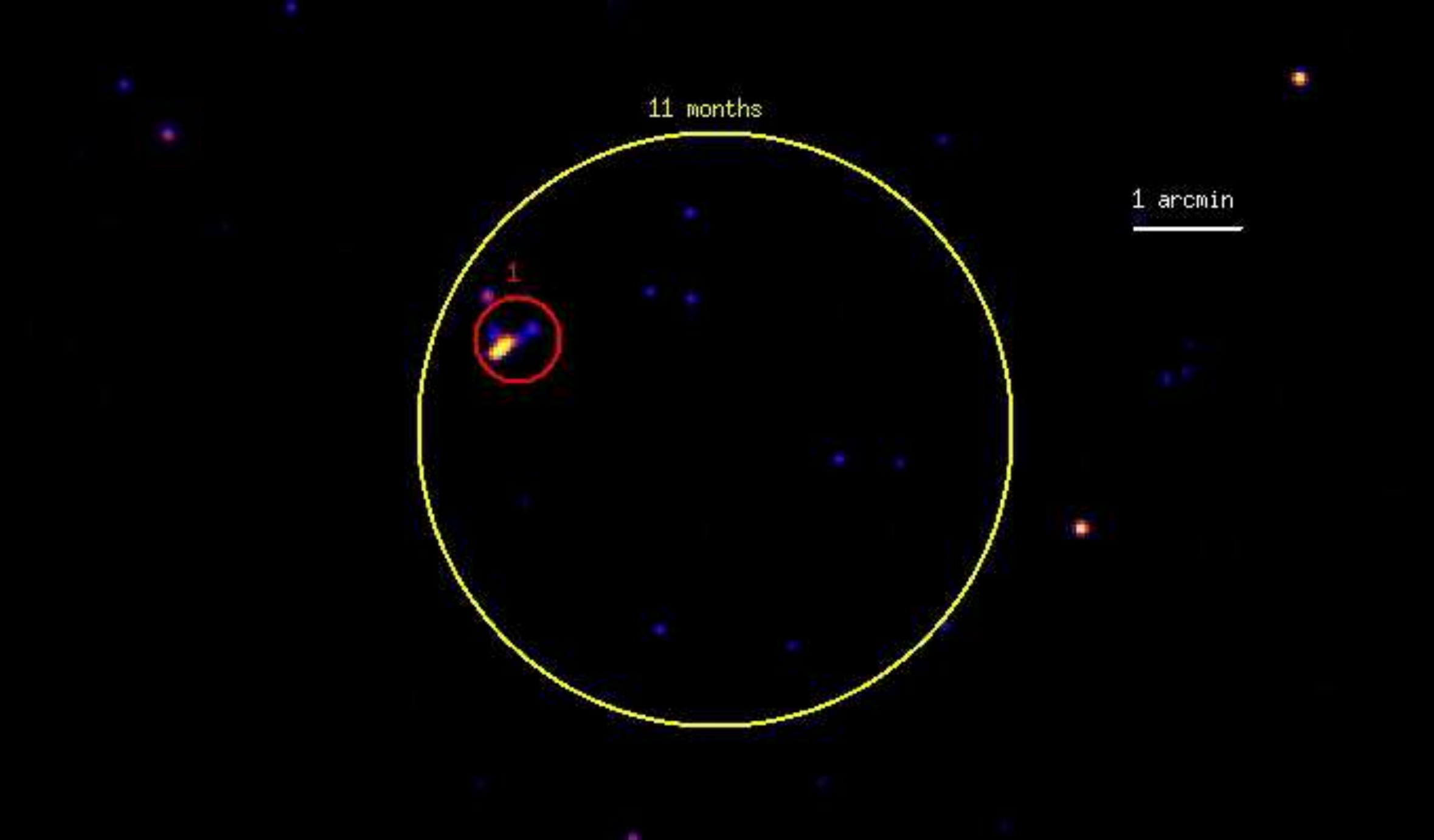}
\caption{PSR J1958+2841 {\it SWIFT} XRT Imaging. The red circle mark the pulsar X-ray counterpart
while the white one the 1FGL {\it Fermi} error box.
\label{J1958-im}}
\end{figure}

\begin{figure}
\centering
\includegraphics[angle=0,scale=.50]{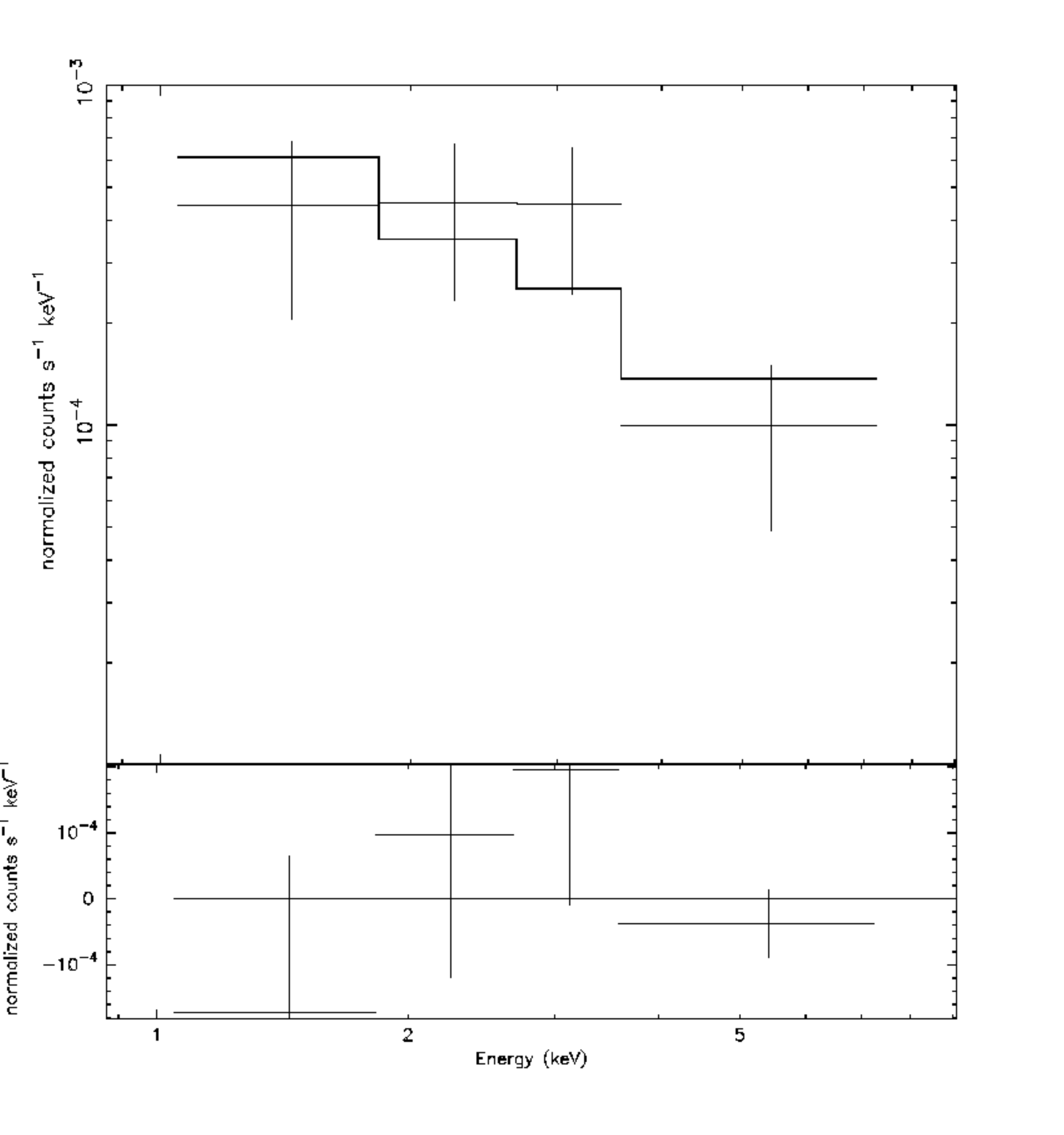}
\caption{PSR J1958+2841 {\it SWIFT} XRT Spectrum (see text for details).
Residuals are shown in the lower panel.
\label{J1958-sp}}
\end{figure}

\clearpage

{\bf J1959+2048 (Black Widow) - type 2 RL MSP} % Nuova!

% Huang & Becker 2006
The millisecond
pulsar PSR B1957+20 has a 0.025 M$_s$ low mass
white dwarf companion (Fruchter et al. 1988). The pulsar has a spin period of 1.6 ms which is one
of the shortest among all known MSPs. The pulsar is orbiting its companion
with an orbit period of 9.16 h. Optical observations
by Fruchter et al. (1988) and Van Paradijs et al. (1988)
revealed that the pulsar wind consisting of electromagnetic
radiation and high-energy particles is ablating and
evaporating its white dwarf companion star. This rarely
observed property gave the pulsar the name black widow
pulsar. Interestingly, the radio emission from the pulsar
is eclipsed for approximately 10\% of each orbit by material
expelled from the white dwarf companion. For a radio
dispersion measure inferred distance of 2.5 $\pm$ 1.0 kpc (Taylor
\& Cordes 1993) the pulsar moves through the sky with
a supersonic velocity of 220 km/sec. The interaction of a
relativistic wind flowing away from the pulsar with the interstellar
medium (ISM) produces an H$\alpha$ bow shock which
was the first one seen around a $"$recycled$"$ pulsar (Kulkarni
\& Hester 1988).

\begin{figure}
\centering
\includegraphics[angle=0,scale=.40]{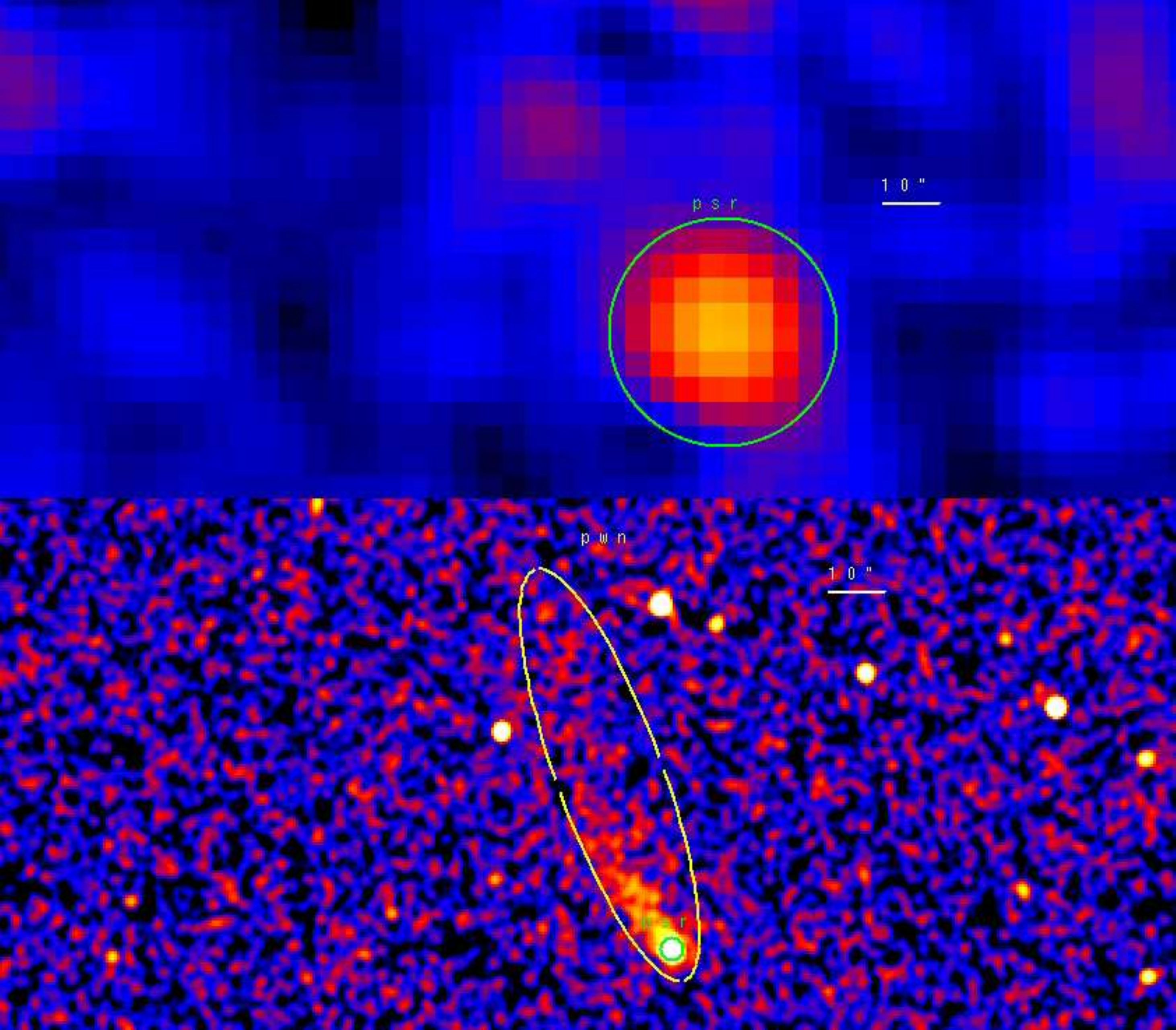}
\caption{{\it Upper Panel:} PSR J1959+2048 0.3-10 keV {\it XMM-Newton} MOS Imaging. The two MOS images have been added.
The green circle marks the source region used in the analysis.
{\it Lower Panel:} PSR J1959+2048 0.3-10 keV {\it Chandra} Imaging.
The green circle marks the pulsar region while the yellow ellipse the nebular region used in the analysis.
\label{J1959-im}}
\end{figure}

Three X-ray observations, both by {\it XMM-Newton} and {\it Chandra}, were performed
in the last years to monitor such an interesting object:\\
- obs. id 1911, {\it Chandra} ACIS-S very faint mode, start time 2001, June 21 at 20:00:51 UT, exposure 43.2 ks;\\
- obs. id 9088, {\it Chandra} ACIS-S very faint mode, start time 2008, August 15 at 17:04:03 UT, exposure 169.1 ks;\\
- obs. id 0204910201, {\it XMM-Newton} observation, start time 2004, October 31 at 23:28:45 UT, exposure 30.3 ks.\\
The PN camera of the EPIC
instrument was operated in Fast Timing mode, while the MOS detectors  were set in Full frame mode.
Due to the faintness of the source and the lack
of spatial resolution in the Fast Timing mode, the PN dataset wasn't used.
For both the PN and MOS cameras a thin optical filter was used.
First, an accurate
screening for soft proton flare events was done in the {\it XMM-Newton} observations obtaining a resulting total
exposure of 25.5 ks.
The X-ray source best fit position (obtained by using the celldetect
tool inside the Ciao distribution) is 19:59:36.75 +20:48:14.81 (0.9$"$ error radius).
A $\sim$ 40$"$ faint trail-like nebular emission is apparent in the {\it Chandra} observation.
For the {\it Chandra} observation, we chose a 2$"$ radius circular region for the
pulsar spectrum; the nebular spectrum is extracted from an elliptical region
with the major axis of 40$"$ (see Figure \ref{J1959-im}).
The background is extracted from a source-free region away from the pulsar.
For the {\it XMM-Newton} observation, we chose a 20$"$ radius circular region around
the pulsar for its spectrum. The background was extracted from a circular source-free region on the same CCD.
we obtained a total of 306 and 1151 pulsar counts (background contributions of 0.2\% and 0.5\%),
177 and 680 nebular counts (background contributions of 78.3\% and 52.3\%) from the two {\it Chandra} observations.
we also obtained 100 and 125 source counts (background contributions of 10.1\% and 8.8\%) from the two MOS cameras.
The best fitting pulsar model is a combination of a blackbody and a powerlaw
(probability of obtaining the data if the model is correct 
- p-value - of 0.02, 85 dof using both the pulsar and nebular spectra).
The powerlaw component has a photon index $\Gamma$ = 1.37$_{-0.48}^{+0.43}$  
absorbed by a column N$_H$ = 3.72$_{-3.72}^{+3.79}$ $\times$ 10$^{20}$ cm$^{-2}$.
The thermal component has a temperature of T = 3.11$_{-0.78}^{+0.70}$ $\times$ 10$^6$ K and
an emitting radius R = 1.56$_{-1.01}^{+8.50}$ km.
The nebular photon index is $\Gamma$ = 1.79$_{-0.36}^{+0.49}$.
A simple blackbody models isn't statistically acceptable
while a simple powerlaw model is acceptable but an f-test performed
comparing it with the composite spectrum gives a
chance probability of 1 $\times$ 10$^{-3}$, pointing
to a significative improvement by adding the thermal component.
Assuming the best fit model, the 0.3-10 keV unabsorbed non-thermal pulsar flux is
5.49$_{-4.40}^{+1.00}$ $\times$ 10$^{-14}$, the thermal flux is 
1.05$_{-0.84}^{+0.20}$ $\times$ 10$^{-14}$ and the nebular flux is
1.68$_{-0.71}^{+0.61}$ $\times$ 10$^{-14}$ erg/cm$^2$ s.
Using a distance
of 2.5 kpc, the luminosities are L$_{2.5kpc}^{nt}$ = 4.12$_{-3.30}^{+0.75}$ $\times$ 10$^{31}$ and
L$_{2.5kpc}^{bol}$ = 7.87$_{-6.30}^{+1.50}$ $\times$ 10$^{30}$ and L$_{2.5kpc}^{pwn}$ = 1.26$_{-0.53}^{+0.46}$ $\times$ 10$^{31}$ erg/s.

\begin{figure}
\centering
\includegraphics[angle=0,scale=.50]{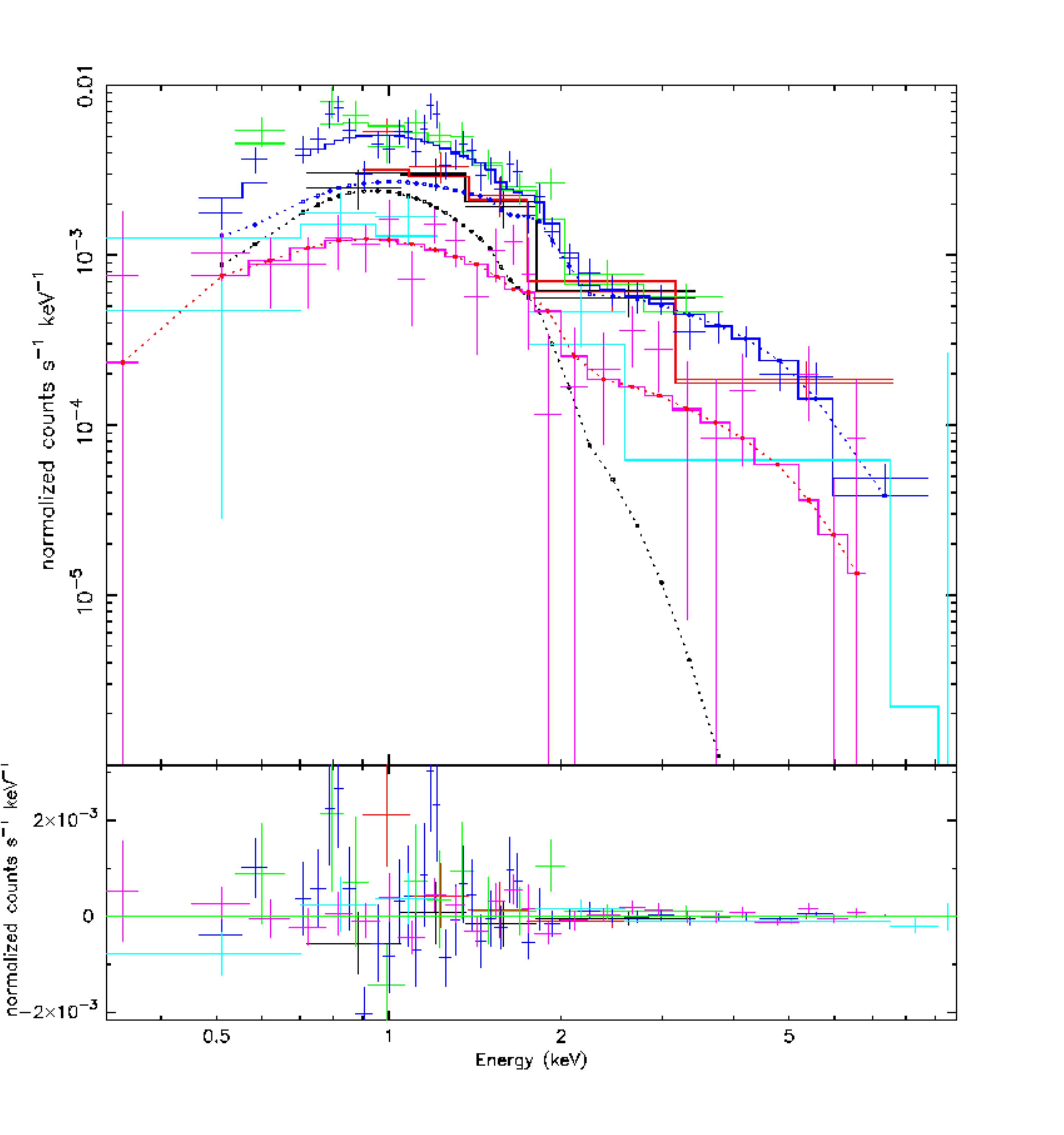}
\caption{PSR J1959+2048 Spectrum. Different colors mark all the different dataset used (see text for details).
Blue points mark the powerlaw component while black points the thermal component of the pulsar spectrum.
Red points mark the nebular spectrum.
Residuals are shown in the lower panel.
\label{J1959-sp}}
\end{figure}

\clearpage

{\bf J2017+0603 - type 0 RL MSP} % Nuova! osservazione in arrivo

J2017+0603 is a millisecond pulsar in a binary system
found by the Pulsar Search Consortium.
Radio dispersion measurements found the distance to be $\sim$ 1.6 kpc.

After the {\it Fermi} detection, three {\it SWIFT} observation
of the $\gamma$-ray error box were asked (obs id. 00038480001-2-3, 16.45 ks exposure).
After the data reduction, no X-ray source were found at the radio position.
For a distance of 1.6 kpc we found a
rough absorption column value of 1 $\times$ 10$^{21}$ cm$^{-2}$
and using a simple powerlaw spectrum
for PSR+PWN with $\Gamma$ = 2 and a signal-to-noise of 3,
we obtained an upper limit non-thermal unabsorbed flux of 8.26 $\times$ 10$^{-14}$ erg/cm$^2$ s,
that translates in an upper limit luminosity L$_{1.6kpc}^{nt}$ = 2.54 $\times$ 10$^{31}$ erg/s.

{\bf J2021+3651 (Dragonfly) - type 2 RLP} % rifatta sommando gli spettri chandra

% Hessels et al. 2004
PSR J2021+3651 (spin period P $\sim$ 103.7 ms, dispersion
measure DM $\sim$ 369 pc cm$^{-3}$, and flux density at 1400 MHz
S$_{1400}$ $\sim$ 100 $\mu$Jy) was discovered by Roberts et al. (2002) with
the 305 m Arecibo radio telescope during a targeted search for
radio pulsations from X-ray sources proposed as counterparts
to unidentified EGRET $\gamma$-ray sources.
Timing observations of PSR J2021+3651 revealed that it is young and energetic (characteristic age
$\tau_c$ = 17 kyr and spin-down energy $\dot{E}$ =
3.4 $\times$ 10$^{36}$ erg/s) and a likely counterpart to the ASCA X-ray source
AX J2021.1+3651 and the EGRET $\gamma$-ray source GeV J2020+3658.
Van Etten et al. 2008 found the pulsar distance to be 2.1$_{-1.0}^{+2.1}$ kpc
by evaluating the column density.
No $\gamma$-ray nebular emission was detected down to a flux of 
1.01 $\times$ 10$^{-10}$ erg/cm$^2$s (Ackermann et al. 2010).

Four different X-ray observations of J2021+3651 were performed, both by
{\it XMM-Newton} and {\it Chandra}:\\
- obs. id 3901, {\it Chandra} ACIS-S very faint mode, start time 2003, February 12 at 05:50:07 UT, exposure 19.9 ks;\\
- obs. id 7603, {\it Chandra} ACIS-S very faint mode, start time 2006, December 29 at 19:30:54 UT, exposure 60.2 ks;\\
- obs. id 8502, {\it Chandra} ACIS-S very faint mode, start time 2006, December 25 at 23:07:39 UT, exposure 34.2 ks;\\
- obs. id 0404540201, {\it XMM-Newton} observation, start time 2006, May 21 at 12:08:43 UT, exposure 28.5 ks.\\
In the XMM observation both the PN and MOS cameras were operating in the Full Frame mode
and a medium optical filter was used.
First, an accurate
screening for soft proton flare events was done in the {\it XMM-Newton} observations obtaining a resulting
exposure of 12.8 ks.
The X-ray source best fit position (obtained by using the celldetect
tool inside the CIAO distribution) is 20:21:05.47 +36:51:04.55 (1$"$ error radius).
A $\sim$ 10$"$ nebular emission is apparent in the {\it Chandra} observation.
For the {\it Chandra} observation, we chose a 2$"$ radius circular region for the
pulsar spectrum while the nebular spectrum was extracted from an annular region
with radii 2 and 15$"$. The background was extracted from a source-free region away from the pulsar.
For the {\it XMM-Newton} observation, we chose a 30$"$ radius circular region around
the pulsar in order to take in account both the pulsar and nebular emissions.
The background was extracted from a circular source-free region on the same CCD.
The spectra obtained in the {\it Chandra} observations were added using
the mathpha tool and, similarly, the response
matrix and effective area files using addarf and addrmf.
we obtained a total of 1759 pulsar counts and 6128 nebular counts in the {\it Chandra} observations
(background contributions of 0.2\% and 5.3\%). we also obtained 2344, 774 and 823 counts
from the PN and MOS cameras in the {\it XMM-Newton} observation (background contributions of 11.2\%, 8.5\% and 8.3\%).
The best fitting pulsar model is a combination of a blackbody and a powerlaw
(probability of obtaining the data if the model is correct 
- p-value - of 0.30, 388 dof using both the pulsar and nebular spectra).
The powerlaw component has a photon index $\Gamma$ = 1.68$_{-0.13}^{+0.18}$  
absorbed by a column N$_H$ = 6.38$_{-0.39}^{+0.50}$ $\times$ 10$^{21}$ cm$^{-2}$.
The thermal component has a temperature of T = 1.66 $\pm$ 0.13 $\times$ 10$^6$ K
and an emitting radius R$_{2.1kpc}$ = 4.60$_{-1.83}^{+5.95}$ km.
The nebular powerlaw has a photon index $\Gamma$ = 1.46$_{-0.04}^{+0.06}$.
Both a simple powerlaw and a simple blackbody models aren't statistically acceptable.
Assuming the best fit model, the 0.3-10 keV unabsorbed non-thermal pulsar flux is
2.15$_{-0.49}^{+0.24}$ $\times$ 10$^{-13}$, the thermal flux is 
3.83$_{-0.86}^{+0.42}$ $\times$ 10$^{-13}$ and the nebular flux is
1.04 $\pm$ 0.06 $\times$ 10$^{-12}$ erg/cm$^2$ s.
Using a distance
of 2.1 kpc, the luminosities are L$_{2.1kpc}^{nt}$ = 1.14$_{-0.26}^{+0.13}$ $\times$ 10$^{32}$ and
L$_{2.1kpc}^{bol}$ = 2.03$_{-0.46}^{+0.22}$ $\times$ 10$^{32}$ and L$_{2.1kpc}^{pwn}$ = 5.50 $\pm$ 0.32 $\times$ 10$^{32}$ erg/s.

\begin{figure}
\centering
\includegraphics[angle=0,scale=.40]{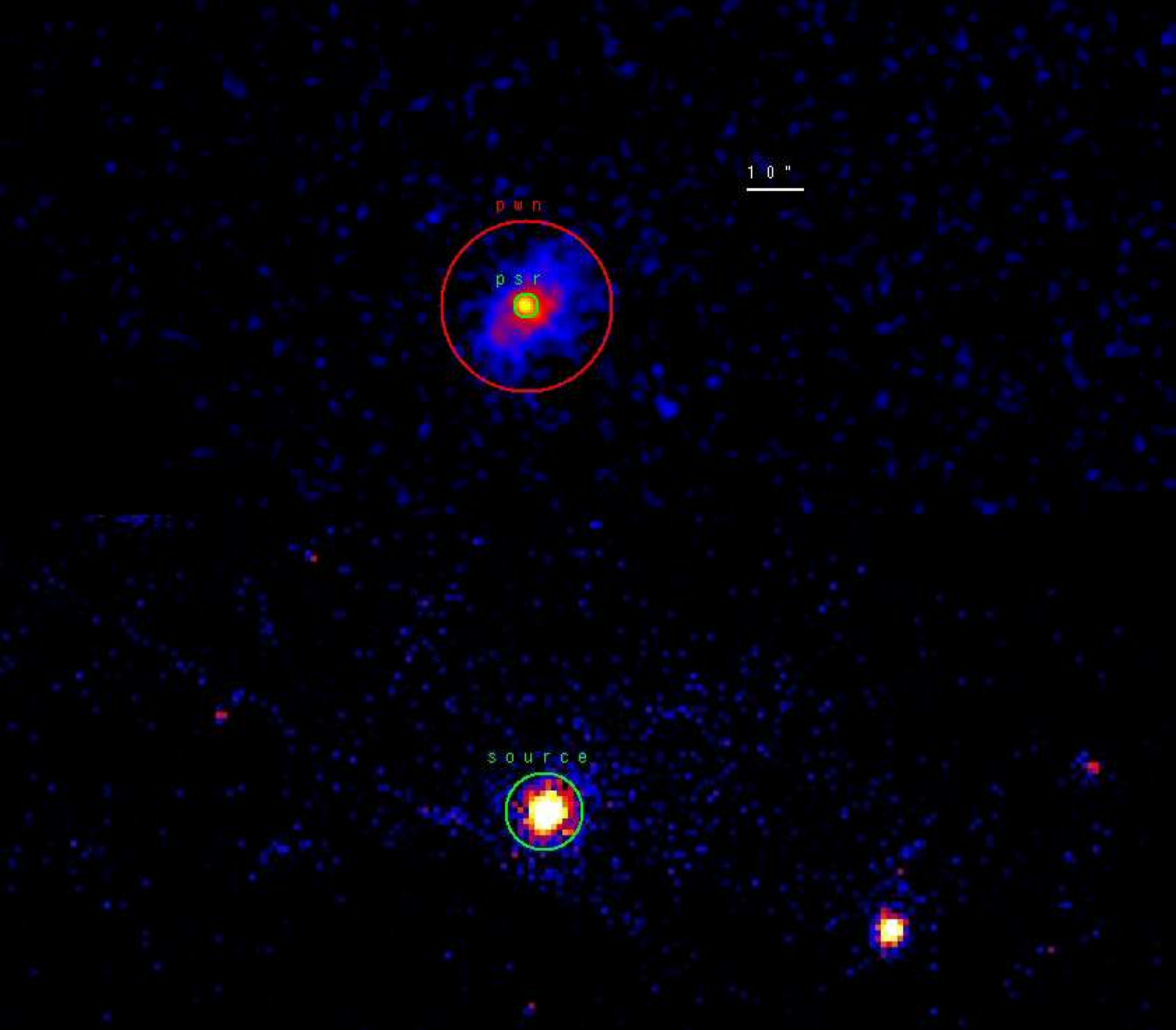}
\caption{{\it Upper Panel:} PSR J2021+3651 0.3-10 keV {\it Chandra} Imaging.
The green circle marks the pulsar region while the red annulus the nebular region used in the analysis.
{\it Lower Panel:} PSR J2021+3651 0.3-10 keV {\it XMM-Newton} EPIC Imaging. The PN and the two MOS images have been added.
The green circle marks the source region used in the analysis.
\label{J2021p3651-im}}
\end{figure}

\begin{figure}
\centering
\includegraphics[angle=0,scale=.50]{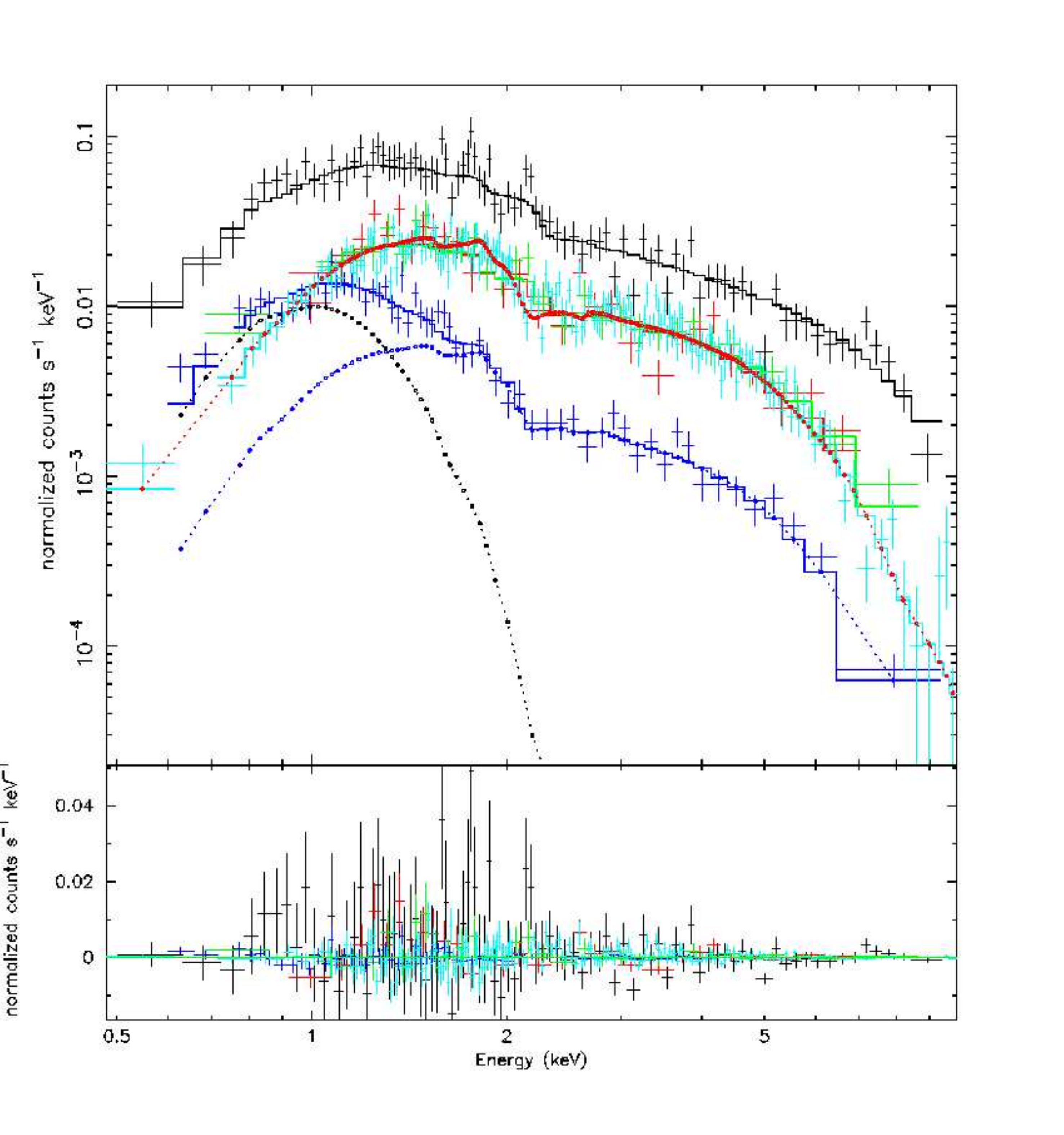}
\caption{PSR J2021+3651 Spectrum. Different colors mark all the different dataset used (see text for details).
Blue points mark the powerlaw component while black points the thermal component of the pulsar spectrum.
Red points mark the nebular spectrum.
Residuals are shown in the lower panel.
\label{J2021p3651-sp}}
\end{figure}

\clearpage

{\bf J2021+4026 (Gamma Cygni) - type 2 RQP} % rifatta con la cstat ed usando anche l'osservazione xmm, risulta di tipo 2 visto che il fit spettrale non mi piace

% Trepl et al. 2010
The $\gamma$-ray detection of PSR J2021+4026 was firstly
reported in the FERMI bright source list with a signal-to-
noise ratio $>$ 10$\sigma$ (Abdo et al. 2009d). The nominal $\gamma$-ray
position of PSR J2021+4026 is located at the edge of the
supernova remnant G78.2+2.1 (Abdo et al. 2009d). Using the first 6 months of the LAT data, the timing
ephemerides of the pulsar were recently reported by Abdo
et al. (2009c). It has a spin period of P = 265 ms and
a spin-down rate of $\dot{P}$ = 5.48 $\times$ 10$^{-14}$. These spin
parameters imply a characteristic age of $\tau_c$ $\sim$ 77 kyr, a
surface magnetic field of $\sim$ 4 $\times$ 10$^{12}$ G and a spin-down
luminosity of $\dot{E}$ = 1.2 $\times$ 10$^{35}$ erg/s.
By using Landecker et al. 1980, Abdo et al. Science 2009 found the pulsar distance
to be 1.50 $\pm$ 0.45 kpc.
A $\gamma$-ray nebular emission was detected with an high confidence and a flux of 
8.88 $\pm$ 0.09 $\times$ 10$^{-10}$ erg/cm$^2$s (Ackermann et al. 2010).

\begin{figure}
\centering
\includegraphics[angle=0,scale=.30]{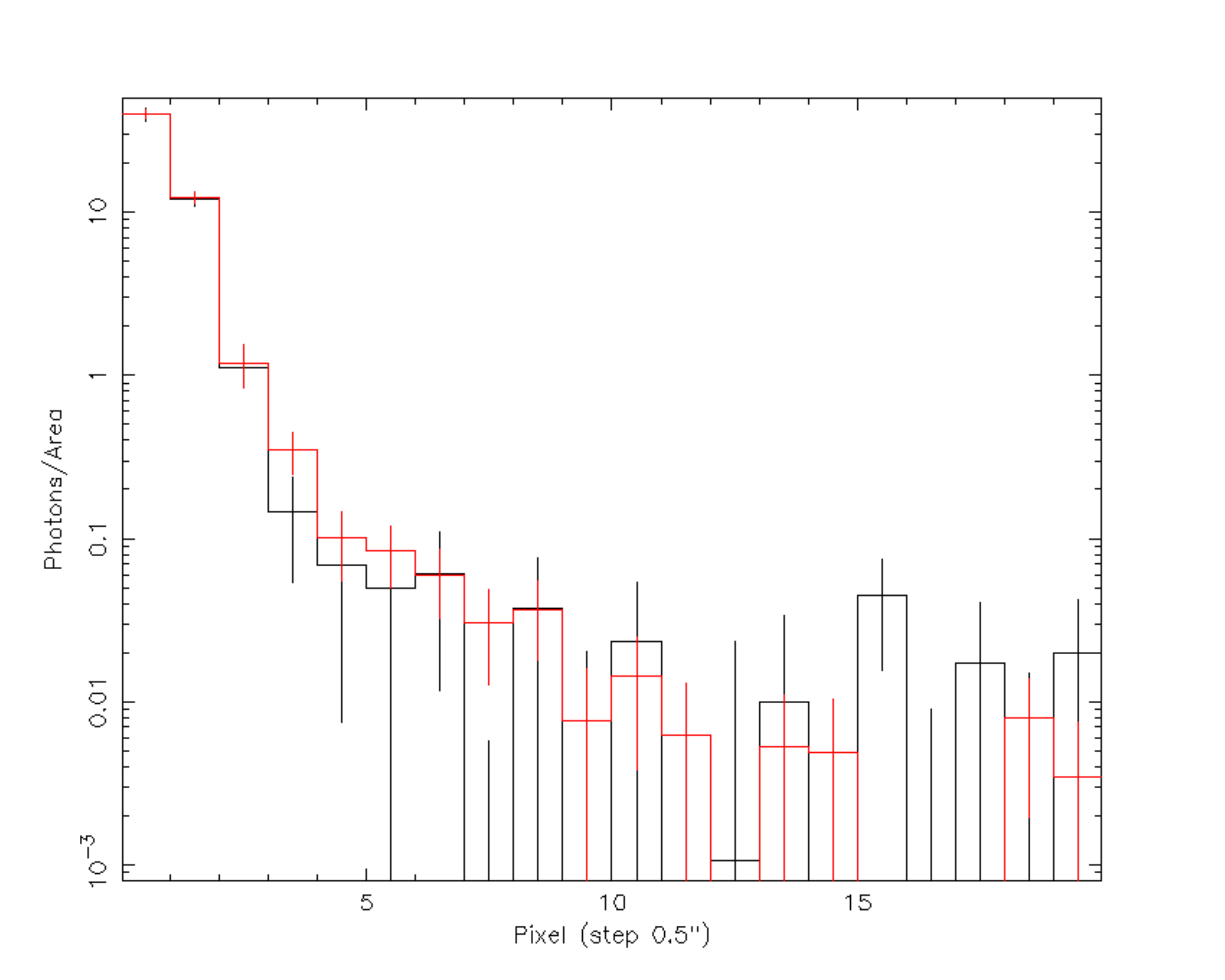}
\caption{PSR J2021+4026 {\it Chandra} Radial Profile (0.3-10 keV energy range) 
(black points) and for a simulated point source 
(red points) having flux, spectrum
and detector coordinates coincident with the ones of the pulsar counterpart
(see text for details). The two profiles perfectly agree excluding any extended emission
around the pulsar.
\label{J2021p4026-psf}}
\end{figure}

Four different X-ray observations of J2021+3651 were performed, both by
{\it XMM-Newton} and {\it Chandra}. Two of them cannot be used in order to obtain the
spectrum of the pulsar: {\it XMM-Newton} 0150960201 is a short observation hardly
dominated by proton flare events while {\it Chandra} 5533 is a short observation that
has the pulsar $\sim$ 10' off-axis, near the edge of the FOV.
we used the remaining two observations:\\
- obs.id 11235, {\it Chandra} ACIS-S very faint mode, start time 2010, August 27 at 12:28:03 UT, exposure 55.8 ks;\\
- obs. id 0150960801, {\it XMM-Newton} observation, start time 2003, December 01 at 21:26:18 UT, exposure 7.54 ks.\\
In the XMM observation both the PN and MOS cameras were operating in the Full Frame mode
and a thin optical filter was used.
First, an accurate
screening for soft proton flare events was done in the {\it XMM-Newton} observations obtaining a resulting
exposure of 5.60 ks.
The X-ray source best fit position (obtained by using the celldetect
tool inside the CIAO distribution) is 20:21:30.74 +40:26:46.04  (1.4$"$ error radius).
we searched for diffuse emission in the immediate
surroundings of the pulsar, by comparing the source intensity profile to the expected ACIS
Point Spread Function (PSF). Assuming the pulsar best fit spectral model, we simulated a
PSF using the ChaRT and MARX packages. Results in the 0.3-10 keV energy range are
shown in Figure \ref{J2021p4026-psf}. No X-ray nebular emission was detected.

For the {\it Chandra} observation, we chose a 2$"$ radius circular region for the
pulsar spectrum and an annular region with radii 10 and 15$"$ for the background one.
For the {\it XMM-Newton} observation, we chose a 20$"$ radius circular region around
the pulsar. The background was extracted from an annular region around the source
with radii 25 and 40$"$.
Due to the very low statistic in the XMM observation, we added the two MOS pulsar spectra using
mathpha tool and, similarly, the response
matrix and effective area files using addarf and addrmf.
We obtained a total of 265, 73 and 30 pulsar counts respectively from the {\it Chandra}, XMM PN camera
and XMM MOS cameras spectra (background contributions of 0.4\%, 62.3\% and 72.2\%).
We used the C-statistic
approach implemented in XSPEC by fitting only the {\it Chandra} data.
The best fitting pulsar model is a combination of a blackbody and a powerlaw
(reduced chisquare value $\chi^2_{red}$ = 0.94, 38 dof).
The powerlaw component has a photon index $\Gamma$ = 0.86$_{-0.86}^{+1.87}$  
absorbed by a column N$_H$ = 6.52$_{-3.73}^{+3.05}$ $\times$ 10$^{21}$ cm$^{-2}$.
The thermal component has a temperature of T = 2.82$_{-0.62}^{+1.00}$ $\times$ 10$^6$ K.
The blackbody radius R$_{1.5kpc}$ = 230$_{-114}^{+516}$ m determined from the
model parameters suggests that the emission is from a hot spot.
Both a simple powerlaw and a simple blackbody models are statistically acceptable but
an f-test performed comparing
a simple blackbody with the composite spectrum gives a
chance probability of 1.8 $\times$ 10$^{-6}$, pointing
to a significative improvement by adding the powerlaw component.
A simple powerlaw model gives an unrealistic high
photon index value ($>$ 4).
Assuming the best fit model, the 0.3-10 keV unabsorbed non-thermal pulsar flux is
1.48 $\pm$ 0.09 $\times$ 10$^{-14}$ and the thermal flux is 
8.17 $\pm$ 0.48 $\times$ 10$^{-14}$ erg/cm$^2$ s.
Using a distance
of 1.5 kpc, the luminosities are L$_{1.5kpc}^{nt}$ = 4.00 $\pm$ 0.24 $\times$ 10$^{30}$ and
L$_{1.5kpc}^{bol}$ = 2.21 $\pm$ 0.13 $\times$ 10$^{31}$ erg/s.

\begin{figure}
\centering
\includegraphics[angle=0,scale=.40]{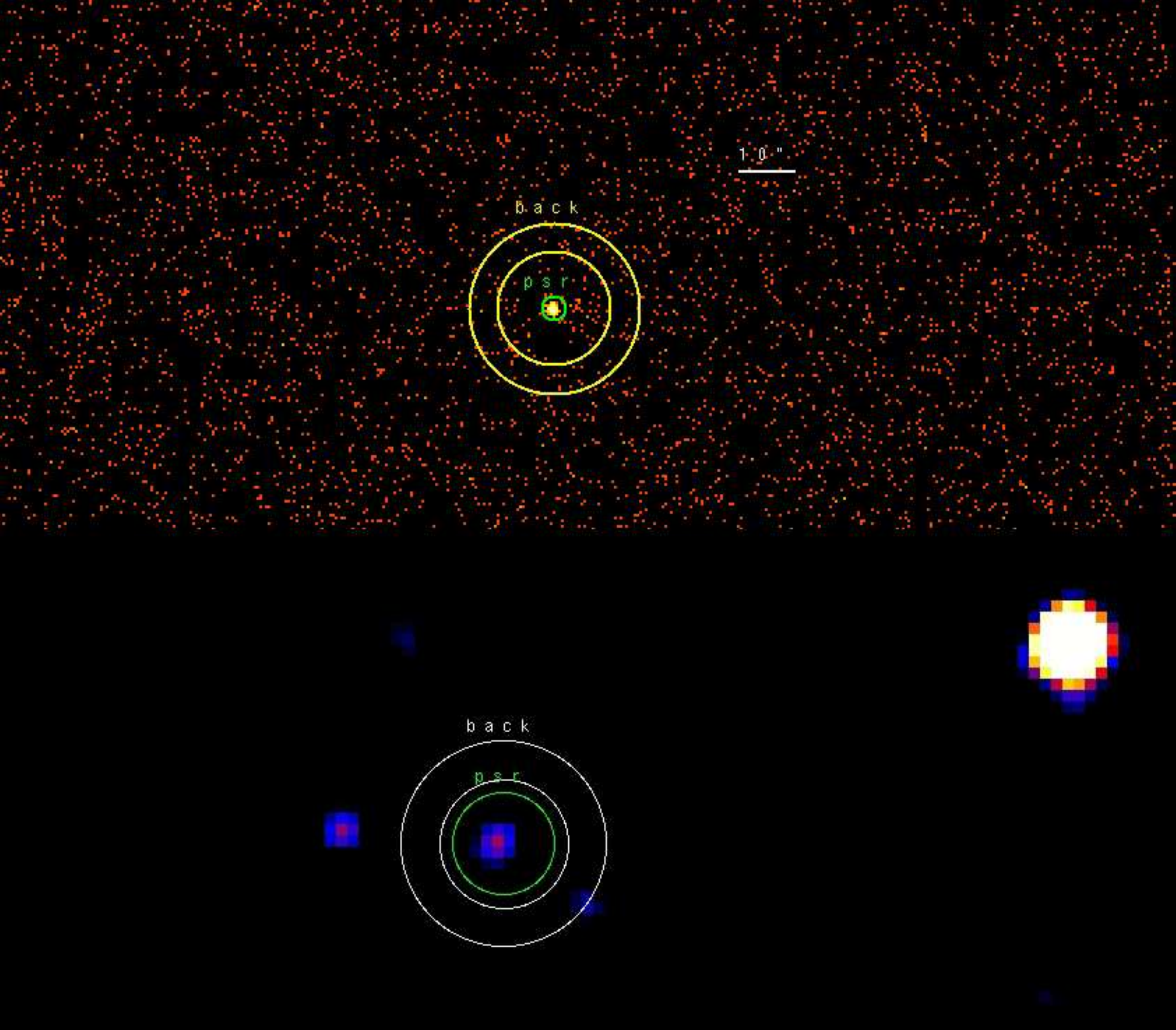}
\caption{{\it Upper Panel:} PSR J2021+4026 0.3-10 keV {\it Chandra} Imaging.
The green circle marks the pulsar region while the white annulus the background region used in the analysis.
{\it Lower Panel:} PSR J2021+4026 0.3-10 keV {\it XMM-Newton} EPIC Imaging. The PN and the two MOS images have been added.
The green circle marks the pulsar region while the white annulus the background region used in the analysis.
The image has been smoothed with a Gaussian with Kernel radius of $5"$.
\label{J2021p4026-im}}
\end{figure}

\begin{figure}
\centering
\includegraphics[angle=0,scale=.50]{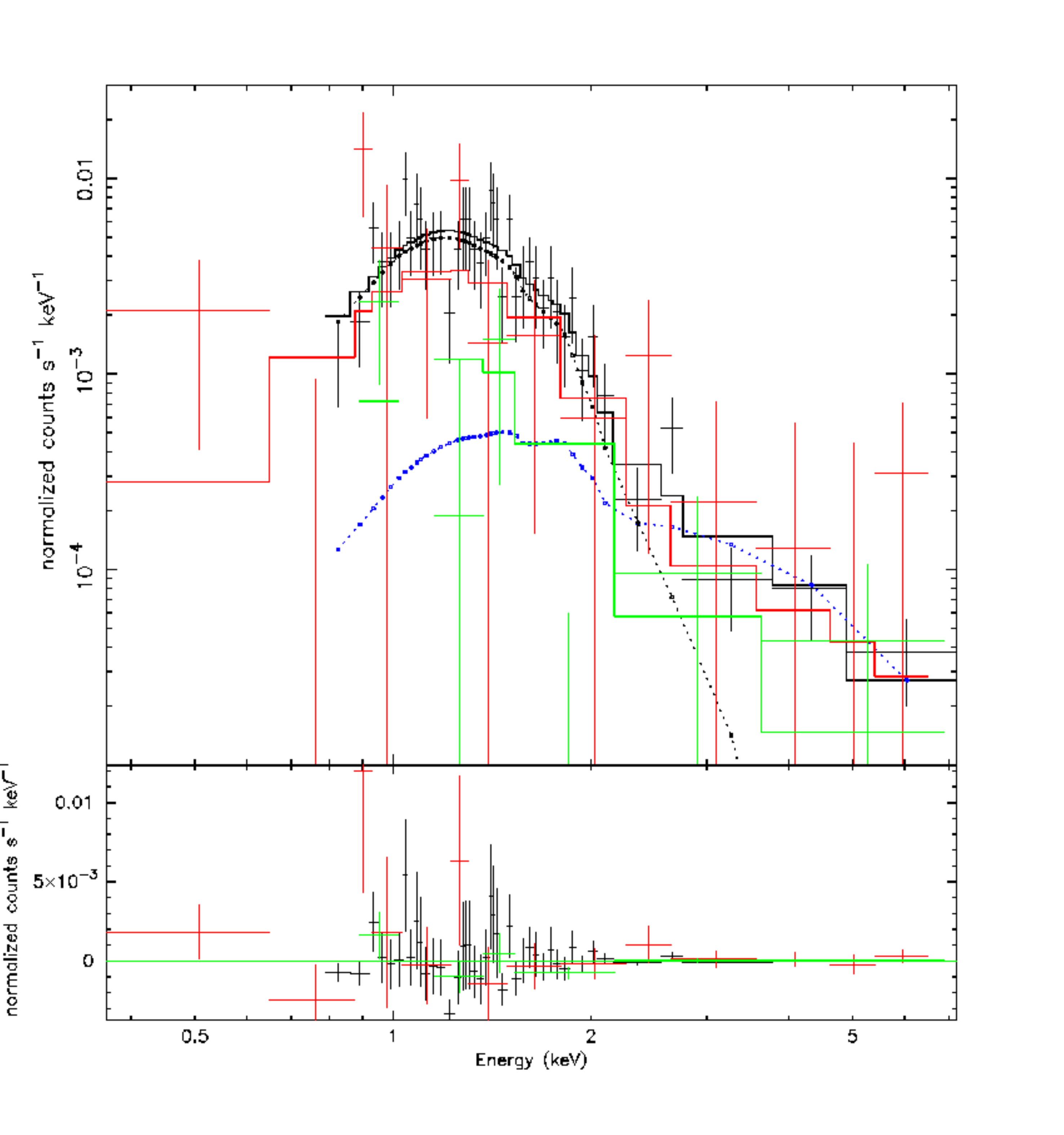}
\caption{PSR J2021+4026 Spectrum. Different colors mark all the different dataset used (see text for details).
Blue points mark the powerlaw component while black points the thermal component of the pulsar spectrum.
Residuals are shown in the lower panel.
\label{J2021p4026-sp}}
\end{figure}

\clearpage

{\bf J2030+3641 - type 0 RLP} % Nuova!

J2030+3641 is an isolated radio pulsar found by the Pulsar Search Consortium.
Radio dispersion measurements found the distance to be $\sim$ 8.1 kpc.

After the {\it Fermi} detection, one {\it SWIFT} observation
of the $\gamma$-ray error box was asked (obs id. 00031711001, 3.78 ks exposure).
After the data reduction, no X-ray source were found at the radio position.
For a distance of 8.1 kpc we found a
rough absorption column value of 8 $\times$ 10$^{21}$ cm$^{-2}$
and using a simple powerlaw spectrum
for PSR+PWN with $\Gamma$ = 2 and a signal-to-noise of 3,
we obtained an upper limit non-thermal unabsorbed flux of 4.52 $\times$ 10$^{-13}$ erg/cm$^2$ s,
that translates in an upper limit luminosity L$_{8.1kpc}^{nt}$ = 3.56 $\times$ 10$^{33}$ erg/s.

{\bf J2032+4127 - type 2 RLP} % rifatta l'analisi con la cstat e passa al tipo 3

J2032+4127 was one of the first pulsars discovered using the
blind search technique (Abdo et al. 2009 Science).
Camilo et al. 2010 found the Radio counterpart of J2032+4127
with a flux density at 2GHz of S$_2$ = 0.12 mJy and at 0.8GHz of S$_{0.8}$ = 0.65 mJy.
No $\gamma$-ray nebular emission was detected down to a flux of 
1.71 $\times$ 10$^{-10}$ erg/cm$^2$s (Ackermann et al. 2010).
The distance of the pulsar based on the Radio dispersion measure is (3.60 $\pm$ 1.08) kpc (Camilo et al. 2009).

Three different X-ray observations of J2032+4127 were performed, both by
{\it XMM-Newton} and {\it Chandra}:\\
- obs.id 4501, {\it Chandra} ACIS-I very faint mode, start time 2004, July 19 at 02:04:33 UT, exposure 49.4 ks;\\
- obs. id 0305560101, {\it XMM-Newton} observation, start time 2005, October 22 at 00:20:26 UT, exposure 26.0 ks;\\
- obs. id 0305560201, {\it XMM-Newton} observation, start time 2005, October 26 at 00:05:20 UT, exposure 26.0 ks.\\
In the XMM observations both the PN and MOS cameras were operating in the Full Frame mode
and a medium optical filter was used. The source is outside the FOV of the MOS1 camera in the second
observation so that it wasn't used.
First, an accurate
screening for soft proton flare events was done in the {\it XMM-Newton} observations obtaining a resulting total
exposure of 43.4 ks.
The X-ray source best fit position (obtained by using the celldetect
tool inside the Ciao distribution) is 20:32:14.72 +41:27:39.45 (1.6$"$ error radius).
No search for diffuse emission is possible in the XMM observations due to the low statistic and the presence of
a nearby source. No hint of diffuse emission is present in the {\it Chandra} observation.
For the {\it Chandra} observation, we chose a 2$"$ radius circular region for the
pulsar spectrum and a circular source-free region away from the source for the background.
For the {\it XMM-Newton} observation, we chose a 15$"$ radius circular region around
the pulsar in order the exclude a nearby source.
The background was extracted from a circular source-free region away from the source.
Due to the low statistic in the XMM observation, we added the two XMM observations' spectra using
mathpha tool and, similarly, the response
matrix and effective area files using addarf and addrmf.
We obtained a total of 64, 323, 65 and 103 pulsar counts respectively from the {\it Chandra}, XMM PN, MOS1 and 2 cameras
(background contributions of 12.5\%, 17.0\%, 20.6\% and 21.9\%).
We used the C-statistic approach implemented in XSPEC.
The best fitting model is a simple powerlaw
(reduced chisquare value $\chi^2_{red}$ = 1.28, 102 dof)
with  a photon index $\Gamma$ = 2.00 $\pm$ 0.33,
absorbed by a column N$_H$ = 4.78$_{-1.49}^{+1.31}$ $\times$ 10$^{21}$ cm$^{-2}$.
A simple blackbody model is not statistically acceptable and
the add of a thermal model to the simple powerlaw gives no statistically
significative improvement.
Assuming the best fit model, the 0.3-10 keV unabsorbed pulsar flux is
2.73$_{-1.56}^{+1.39}$ $\times$ 10$^{-14}$ erg/cm$^2$ s.
Using a distance
of 3.6 kpc, the pulsar luminosity is L$_{3.6kpc}^{nt}$ = 4.25$_{-2.43}^{+2.16}$ $\times$ 10$^{31}$ erg/s.

\begin{figure}
\centering
\includegraphics[angle=0,scale=.40]{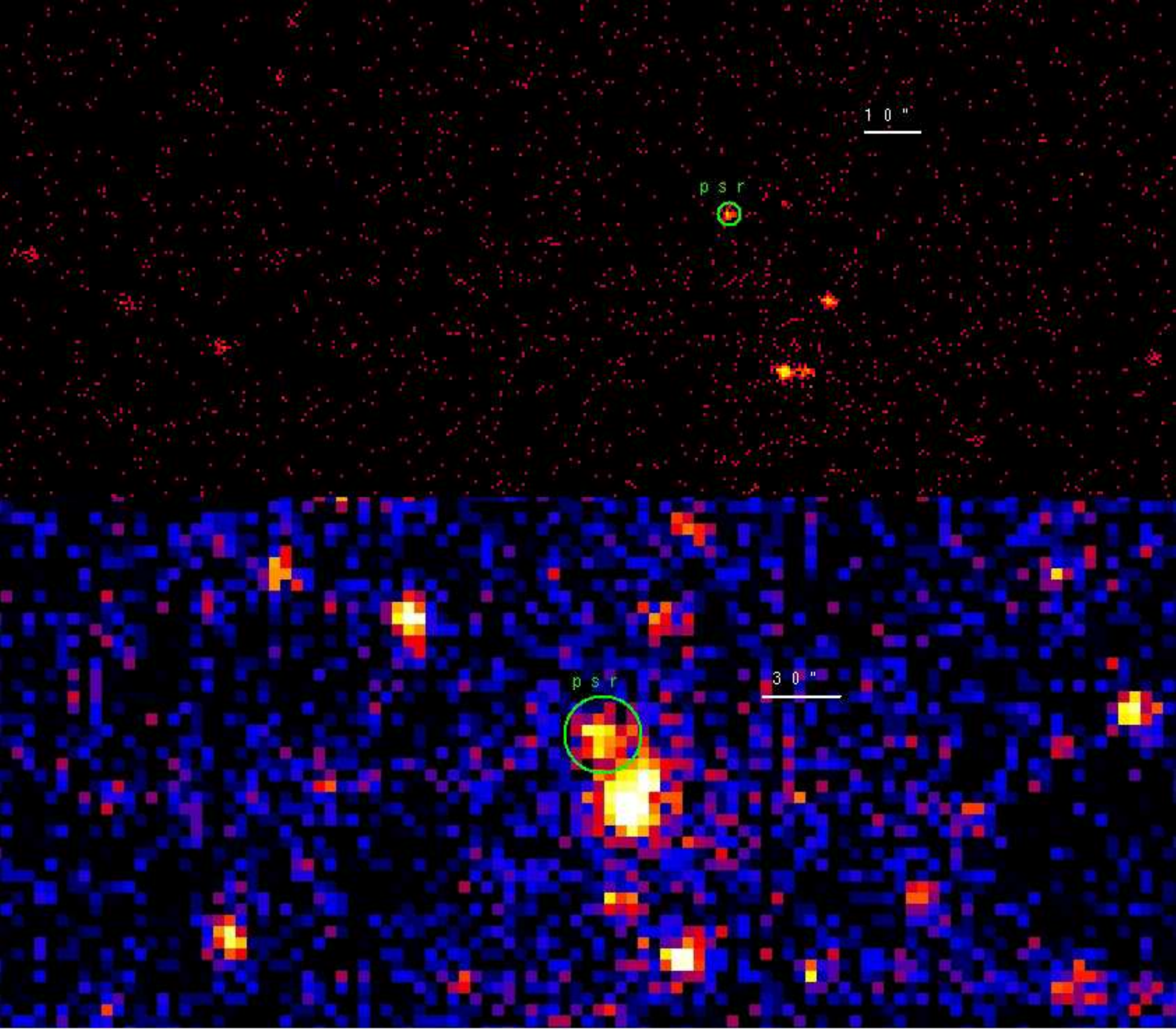}
\caption{{\it Upper Panel:} PSR J2032+4127 0.3-10 keV {\it Chandra} Imaging.
The green circle marks the pulsar region used in the analysis.
{\it Lower Panel:} PSR J2032+4127 0.3-10 keV {\it XMM-Newton} EPIC Imaging. The PN and the two MOS images have been added.
The green circle marks the pulsar region used in the analysis.
\label{J2032-im}}
\end{figure}

\begin{figure}
\centering
\includegraphics[angle=0,scale=.50]{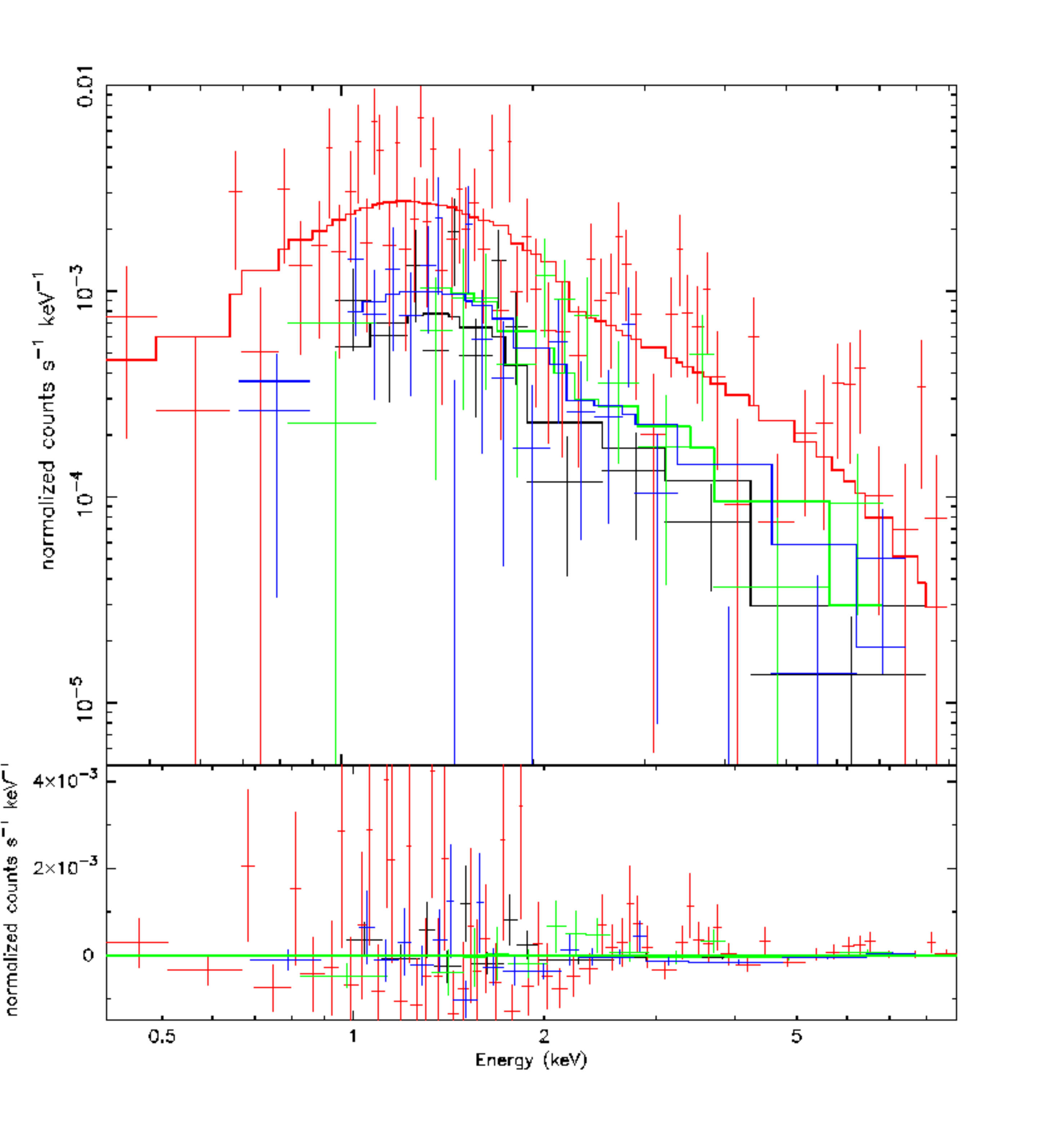}
\caption{PSR J2032+4127 Spectrum. Different colors mark all the different dataset used (see text for details).
Residuals are shown in the lower panel.
\label{J2032-sp}}
\end{figure}

\clearpage

{\bf J2043+1710 - type 0 RL MSP} % Nuova!

J2043+1710 is a millisecond pulsar in a binary system
found by the Pulsar Search Consortium.
Radio dispersion measurements found the distance to be $\sim$ 1.8 kpc.

After the {\it Fermi} detection, two {\it SWIFT} observation
of the $\gamma$-ray error box were asked (obs id. 00031768001-2, 10.58 ks exposure).
After the data reduction, no X-ray source were found at the radio position.
For a distance of 1.8 kpc we found a
rough absorption column value of 6 $\times$ 10$^{20}$ cm$^{-2}$
and using a simple powerlaw spectrum
for PSR+PWN with $\Gamma$ = 2 and a signal-to-noise of 3,
we obtained an upper limit non-thermal unabsorbed flux of 9.80 $\times$ 10$^{-14}$ erg/cm$^2$ s,
that translates in an upper limit luminosity L$_{1.8kpc}^{nt}$ = 3.81 $\times$ 10$^{31}$ erg/s.

{\bf J2043+2740 - type 2 RL MSP} % rifatta usando la cstat

% Noutsos et al. 2010
PSR J2043+2740 was discovered in radio, in the Arecibo
millisecond-pulsar survey at 430 MHz (Thorsett et al. 1994).
Based on its dispersion measure, DM= 21.0 $\pm$ 0.1 pc cm$^{-3}$ (Ray
et al. 1996), and the free-electron density model of
Taylor \& Cordes (1993), the distance estimate for this pulsar is
1.80 $\pm$ 0.54 kpc (Ray et al. 1996, Manchester et al. 2005). PSR J2043+2740 lies
near the south-western shell of the Cygnus Loop ($\sim$ 15 pc outside
the observable edge), perhaps suggesting an association
with the remnant. However, the evidence so far suggests that
such an association is unlikely: the distance to the Cygnus Loop
has been estimated to 540$^{+100}_{-80}$ pc (Blair et al. 2005; Blair et
al. 2009); in addition, assuming that the pulsar was born within
the observable limits of the remnant, the age of the latter ($<$ 12
kyr; Sankrit \& Blair 2002) suggests a transverse velocity of
$>$ 980 km s$^{-1}$ for the pulsar, which is significantly higher than
the average birth velocity of the known pulsar sample (400 $\pm$ 40
km s$^{-1}$; Hobbs et al. 2005). Last but not least, the pulsar's characteristic
age, as calculated from its spin parameters, is two orders
of magnitude higher than the remnant's. Therefore, these
discrepancies need to be reconciled before an association can
be claimed.
No $\gamma$-ray nebular emission was detected down to a flux of 
2.99 $\times$ 10$^{-12}$ erg/cm$^2$s (Ackermann et al. 2010).

Only one {\it XMM-Newton} observation has been performed, obs. id 0037990101, starting on 2002, November 21 at 23:25:44.52 UT,
for a total exposure of 16.5 ks.
The PN camera was operating in Small Window mode while the MOS cameras were operating in the Full Frame mode.
A thin optical filter was used for the PN camera while a medium filter for the MOS cameras.
No screening for soft proton flare events was done owing to the goodness of the observation.
A source detection was performed using both the SAS tools
and XIMAGE: the X-ray source best fit position is 20:43:43.50 +27:40:56.00 (5$"$ error radius).
No hint of diffuse emission is present in the observation.
we chose a 20$"$ radius circular region for the pulsar spectrum and 
a circular source-free region away from the source for the background.
Due to the low statistic in the XMM observation, we added the two MOS spectra using
mathpha tool and, similarly, the response
matrix and effective area files using addarf and addrmf.
we obtained a total of 180 and 107 pulsar counts from the PN and MOS cameras
(background contributions of 19.8\% and 21.4\%).
we used the C-statistic approach implemented in XSPEC.
The best fitting model is a simple powerlaw
(reduced chisquare value $\chi^2_{red}$ = 0.97, 50 dof)
with  a photon index $\Gamma$ = 2.98$_{-0.29}^{+0.44}$ ,
absorbed by a column N$_H$ = $<$ 3.62 $\times$ 10$^{20}$ cm$^{-2}$.
A simple blackbody model is not statistically acceptable while
the add of a thermal model to the simple powerlaw gives no statistically
significative improvement.
Assuming the best fit model, the 0.3-10 keV unabsorbed pulsar flux is
2.18$_{-1.13}^{+0.30}$ $\times$ 10$^{-14}$ erg/cm$^2$ s.
Using a distance
of 1.8 kpc, the pulsar luminosity is L$_{1.8kpc}^{nt}$ = 8.47$_{-4.39}^{+1.17}$ $\times$ 10$^{30}$ erg/s.

\begin{figure}
\centering
\includegraphics[angle=0,scale=.50]{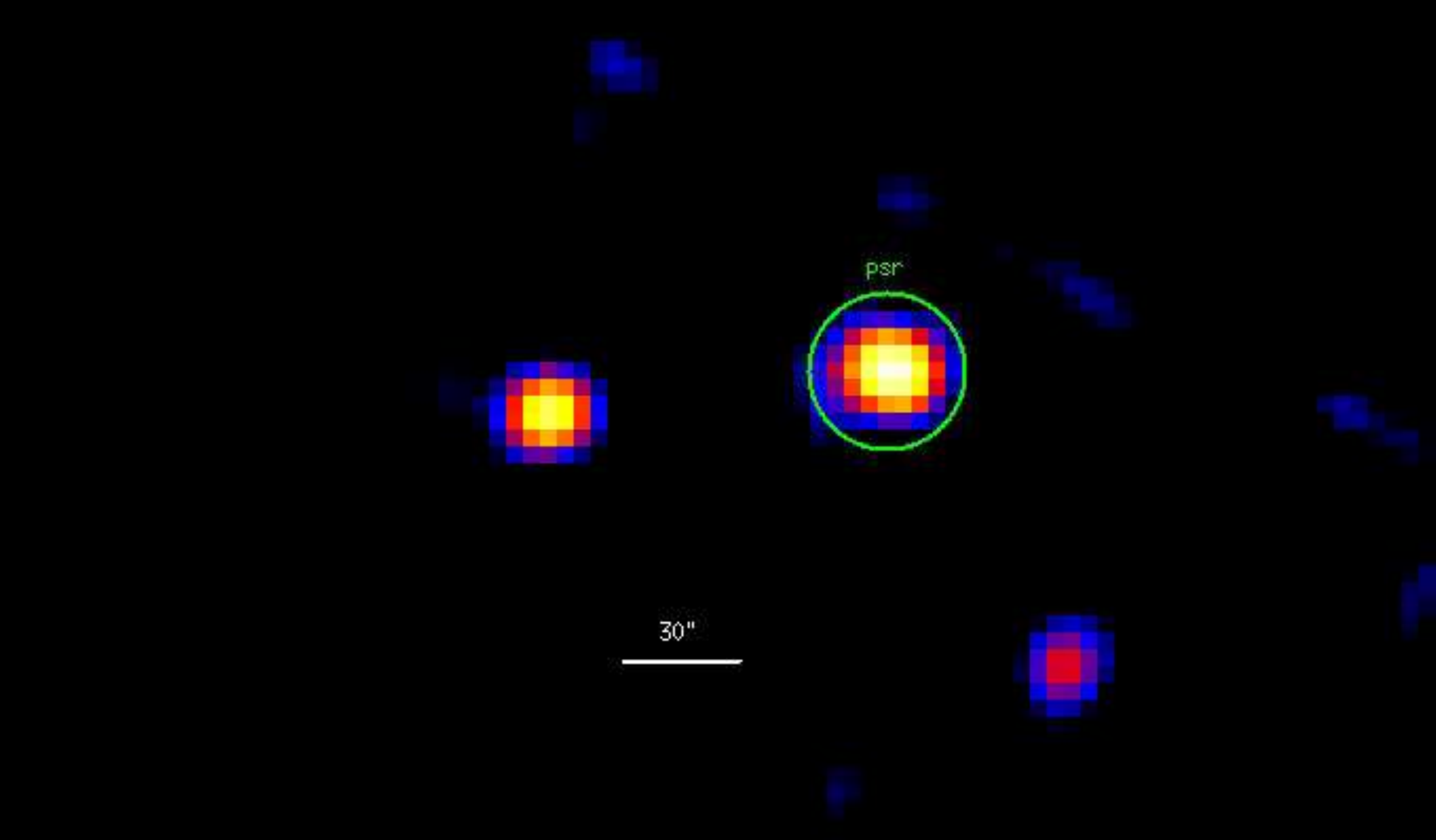}
\caption{PSR J2043+2740 0.3-10 keV {\it XMM-Newton} Imaging. The PN and the two MOS images have been added. 
The green circle marks the pulsar region used in the analysis.
\label{J2043-im}}
\end{figure}

\begin{figure}
\centering
\includegraphics[angle=0,scale=.50]{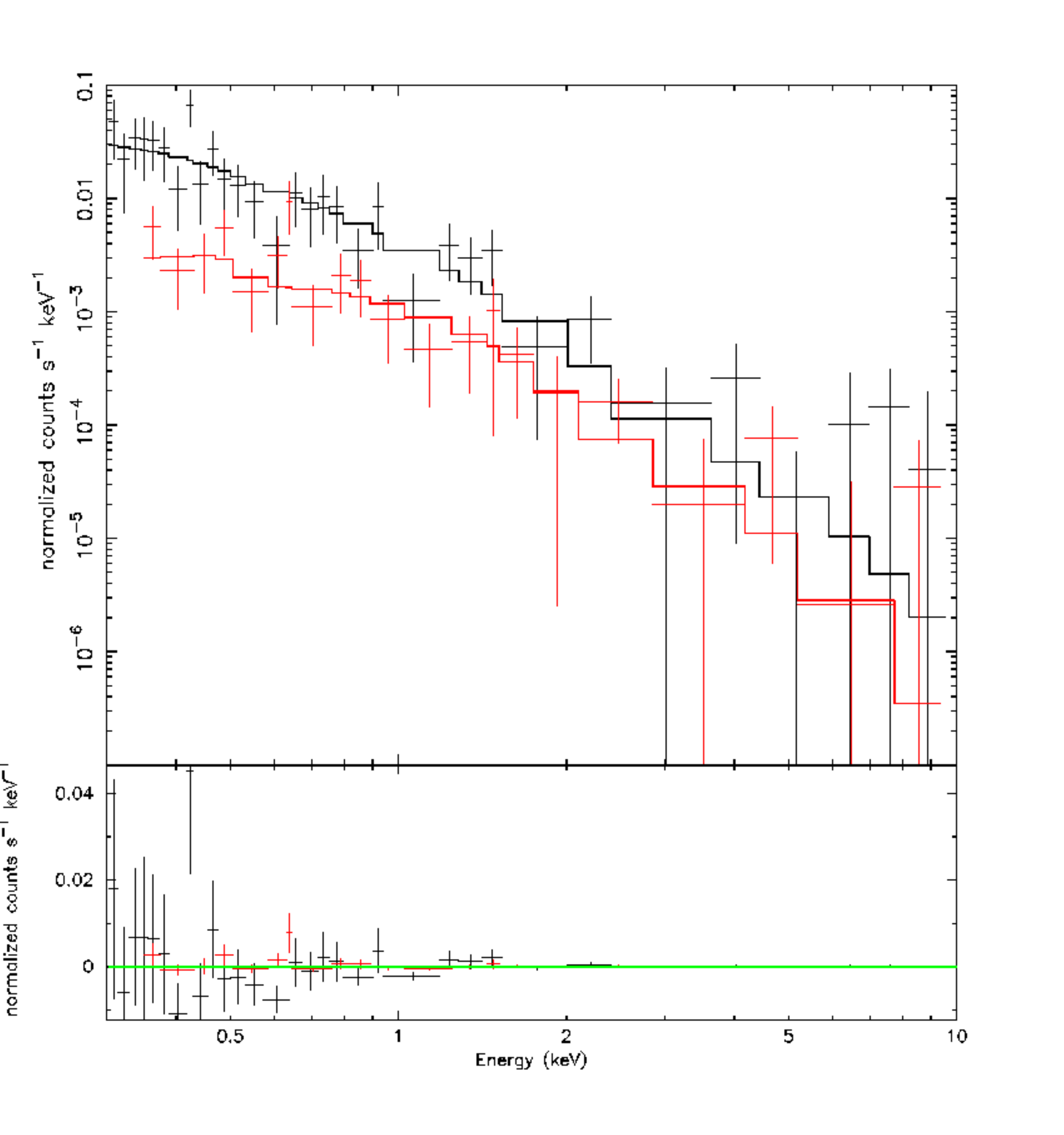}
\caption{PSR J2043+2740 {\it XMM-Newton} Spectrum (see text for details).
Residuals are shown in the lower panel.
\label{J2043-sp}}
\end{figure}

\clearpage

{\bf J2055+25 (Obama) - type 2 RQP} % rifatta con la cstat

Pulsations from J2055+25 were detected by {\it Fermi} using the
blind search technique (Saz Parkinson et al. 2010).
Such a pulsar was informally named $"$Obama$"$ for it was discovered right after
the election of the new US president.
A $\gamma$-ray nebular emission was detected with a flux of 
(1.76 $\pm$ 0.33) $\times$ 10$^{-11}$ erg/cm$^2$s (Ackermann et al. 2010).
The pseudo-distance of the object based on $\gamma$-ray data (Saz Parkinson et al. (2010))
is $\sim$ 0.4 kpc.

Two {\it XMM-Newton} observations were performed in order to find the X-ray counterpart
of the {\it Fermi} source:\\
- obs. id 0605470401 started on 2009, October 26 at 23:55:03 UT, for a total
exposure of 17.8 ks;\\
- obs. id 0605470901 started on 2010, April 21 at 14:22:51 UT, for a total exposure of 12.2 ks.\\
Both the PN and MOS cameras were operating in the Full Frame mode and a thin optical filter was used.
First, an accurate
screening for soft proton flare events was done in the {\it XMM-Newton} observations obtaining a resulting total
exposure of 12.9 ks from the first observation. Due to the very high proton flares contribute,
the second observation was rejected.
A source detection was performed using both the SAS tools
and XIMAGE: the X-ray source best fit position is 20:55:48.99 +25:39:58.78 (5$"$ error radius).
No hint of diffuse emission is present in the observation.
we chose a 20$"$ radius circular region for the pulsar spectrum and 
a circular source-free region away from the source for the background.
Due to the low statistic in the XMM observation, we added the two MOS spectra using
mathpha tool and, similarly, the response
matrix and effective area files using addarf and addrmf.
we obtained a total of 136 and 75 pulsar counts from the PN and MOS cameras
(background contributions of 13.5\% and 20.2\%).
we used the C-statistic approach implemented in XSPEC.
The best fitting model is a simple powerlaw
(reduced chisquare value $\chi^2_{red}$ = 0.98, 38 dof)
with  a photon index $\Gamma$ = 2.04$_{-0.52}^{+0.66}$ ,
absorbed by a column N$_H$ = 1.51$_{-1.17}^{+1.42}$ $\times$ 10$^{21}$ cm$^{-2}$.
Such an high value of the absorbing column, similar to the galactic one ($\sim$ 1.08 $\times$ 10$^{21}$ cm$^{-2}$
using WebTools), can suggest the presence of a thermal component. Anyway,
the add of a blackbody model to the simple powerlaw gives no statistically
significative improvement. A simple thermal model is not statistically acceptable.
Assuming the best fit model, the 0.3-10 keV unabsorbed pulsar flux is
4.33$_{-2.77}^{+1.21}$ $\times$ 10$^{-14}$ erg/cm$^2$ s.
Using a distance
of 0.4 kpc, the pulsar luminosity is L$_{0.4kpc}^{nt}$ = 8.31$_{-5.32}^{+2.32}$ $\times$ 10$^{29}$ erg/s.

\begin{figure}
\centering
\includegraphics[angle=0,scale=.50]{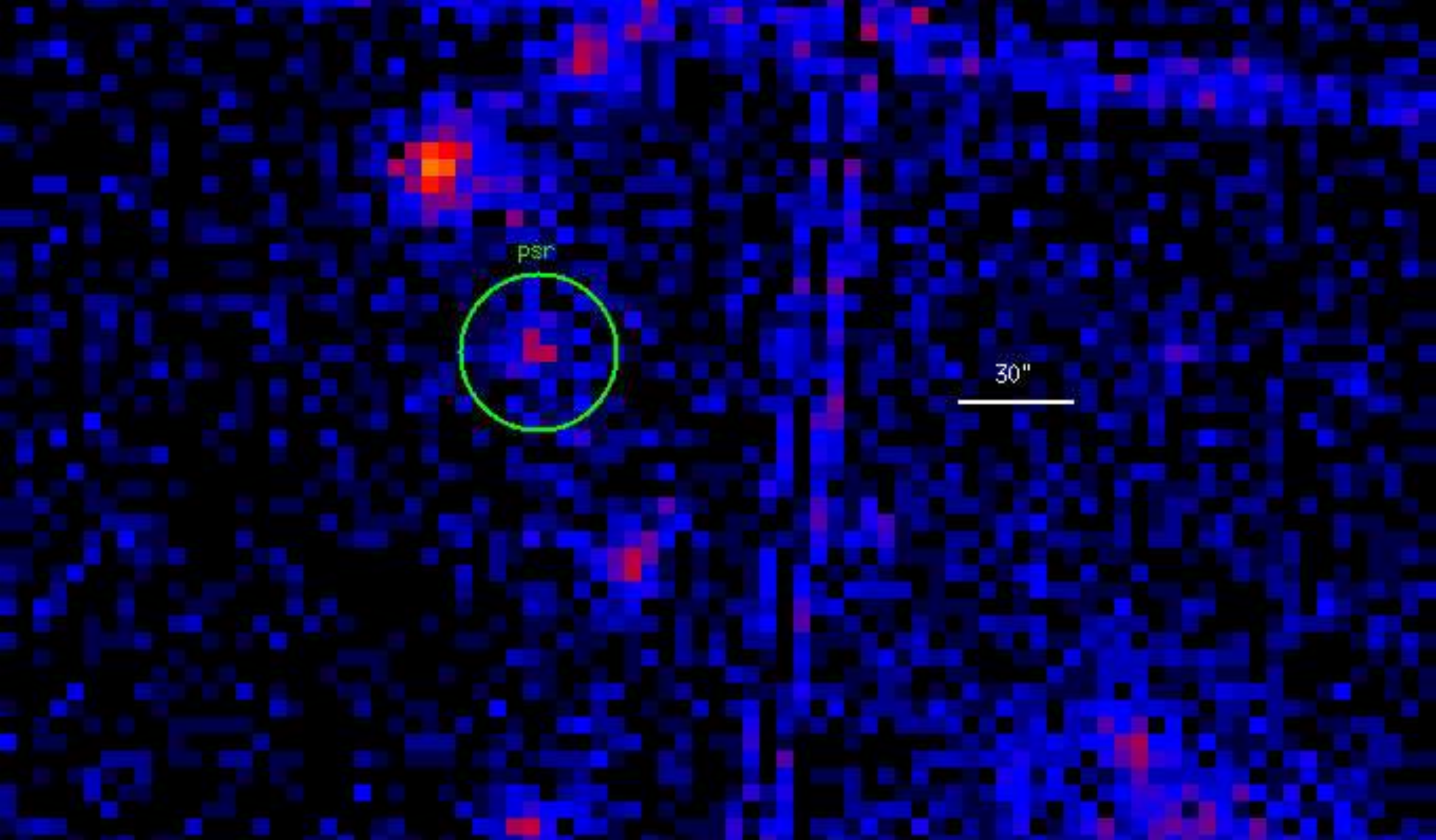}
\caption{PSR J2055+25 0.3-10 keV {\it XMM-Newton} Imaging. The PN and the two MOS images have been added. 
The green circle marks the pulsar region used in the analysis.
\label{J2055-im}}
\end{figure}

\begin{figure}
\centering
\includegraphics[angle=0,scale=.50]{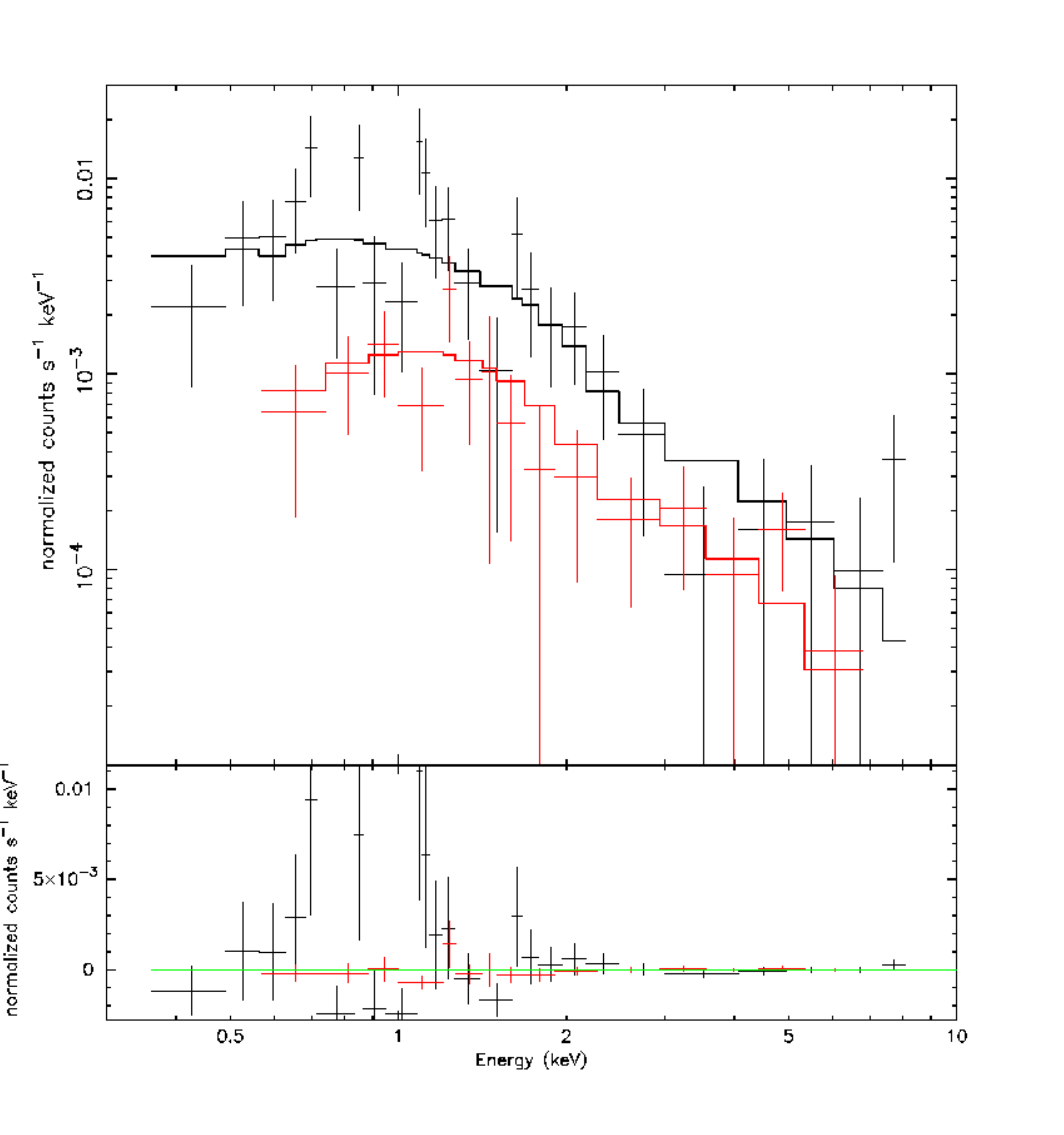}
\caption{PSR J2055+25 {\it XMM-Newton} Spectrum (see text for details).
Residuals are shown in the lower panel.
\label{J2055-sp}}
\end{figure}

\clearpage

{\bf J2124-3358 - type 2 RL MSP} %rifatta l'analisi

% Hui & Becker 2006
PSR J2124-3358 was discovered
during the Parkes 436 MHz survey of the southern sky (Bailes
et al. 1997). X-ray emission
from PSR J2124-3358 was reported by
Becker \& Trumper (1998, 1999) in ROSAT HRI data
Gaensler et al. (2002) discovered an H$\alpha$-emitting bow shock
nebula around it. This bow shock is very broad
and highly asymmetric about the direction of the pulsar's
proper motion. The asymmetric shape might be caused by a significant
density gradient in the ISM, bulk flow of ambient gas
and/or anisotropies in the pulsar's relativistic wind (Gaensler
et al. 2002).
A $\gamma$-ray nebular emission was detected by {\it Fermi} with a flux of 
(2.18 $\pm$ 0.44) $\times$ 10$^{-11}$ erg/cm$^2$s (Ackermann et al. 2010).
The distance of the pulsar based on the parallax measure is 0.25$_{-0.08}^{+0.25}$ kpc (Hotan et al. 2006).

\begin{figure}
\centering
\includegraphics[angle=0,scale=.30]{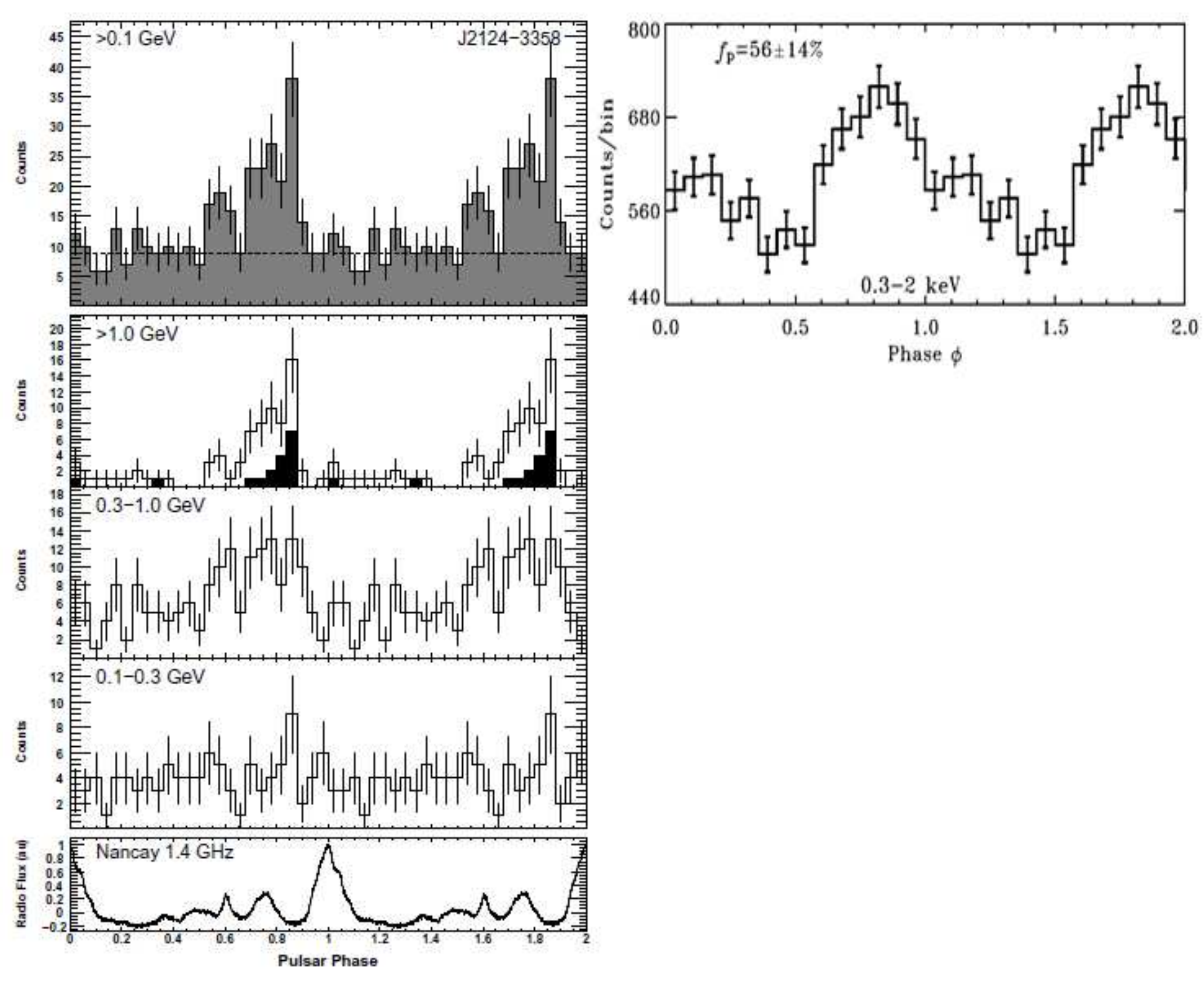}
\caption{PSR J2124-3358 Lightcurve.{\it Left: Fermi} $\gamma$-ray lightcurve folded with Radio
(Abdo et al. catalogue). {\it Right: XMM-Newton} pulse profile extracted from the EPIC-PN data
in the 0.3-2 keV range, with estimated value of the intrinsic source pulsed fraction f$_p$.
See Zavlin 2006 for details.
\label{J2124-lc}}
\end{figure}

\begin{figure}
\centering
\includegraphics[angle=0,scale=.30]{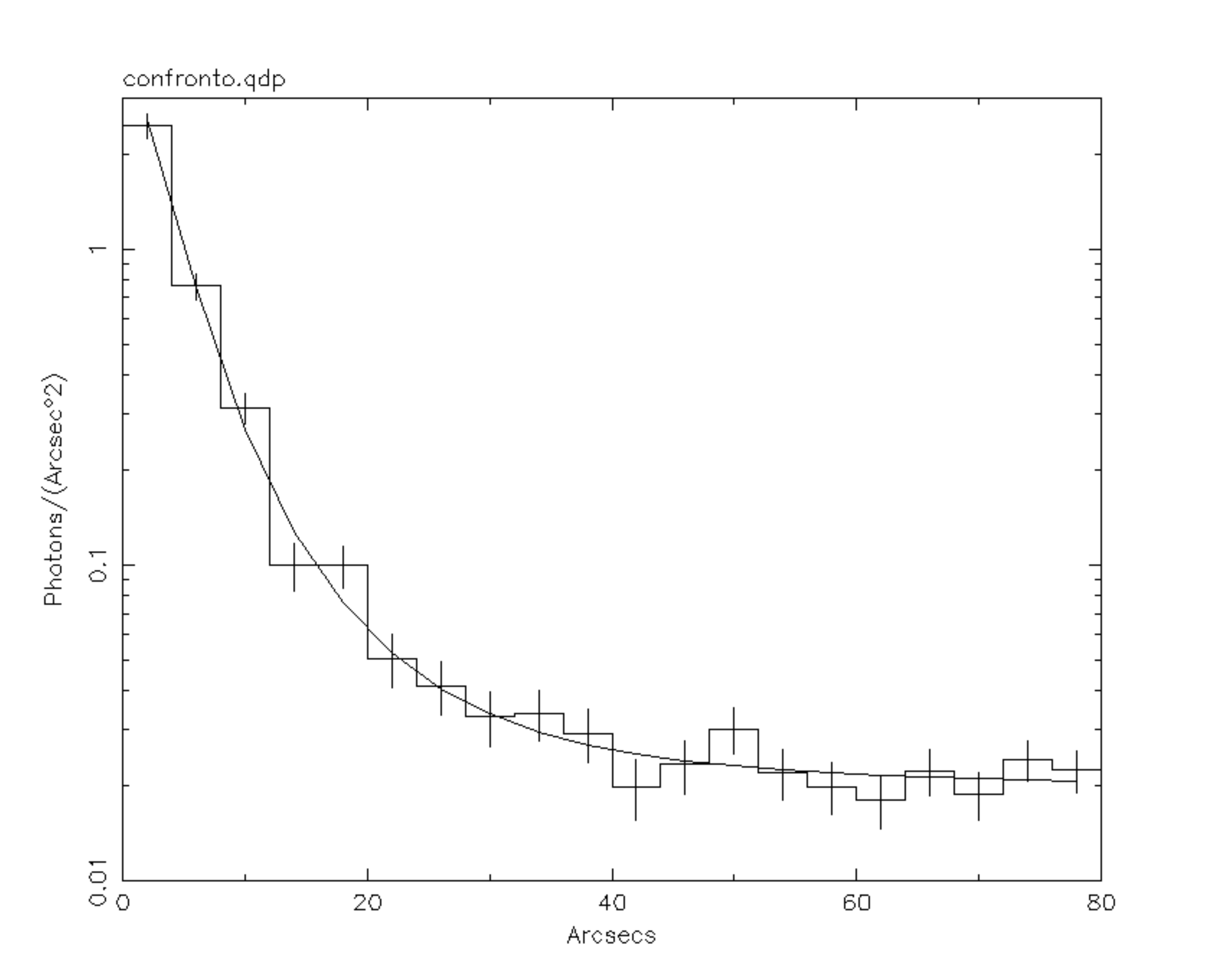}
\caption{PSR J2124-3358 {\it XMM-Newton} PN radial profile (0.3-10 keV energy range).
This was fitted by using the XMM calibration document CAL-TN-0052. No excess has been found.
\label{J2124-psf}}
\end{figure}

Two different X-ray observations of J2124-3358 were performed:\\
- obs. id 5585, {\it Chandra} ACIS-S very faint mode, start time 2004, December 19 at 23:17:10 UT, exposure 30.2 ks;\\
- obs. id 0112320601, {\it XMM-Newton} observation, start time 2002, April 14 at 22:28:17 UT, exposure 67.9 ks.\\
The PN camera of the EPIC
instrument was operated in Fast Timing mode, while the MOS detectors were set in Full frame mode.
For the PN camera a thin optical filter was used,
while for the MOS cameras a medium one.
First, an accurate
screening for soft proton flare events was done in the {\it XMM-Newton} observations obtaining a resulting total
exposure of 59.6 ks.
The X-ray source best fit position (obtained by using the celldetect
tool inside the Ciao distribution) is 21:24:43.85 -33:58:44.61 (1$"$ error radius).
A very faint jet-like nebular emission is apparent in the {\it Chandra} observation with a total length
of $\sim$ 30$"$. We searched for diffuse emission in the immediate
surroundings of the pulsar in the {\it XMM-Newton} MOS cameras data. The resulting graph was fitted
using the prescriptions of the XMM calibration document CAL-TN-0052 in order to find the
Point Spread Function and to find any possible excess of counts (see Figure \ref{J2124-psf}). No X-ray nebular emission was detected.

\begin{figure}
\centering
\includegraphics[angle=0,scale=.40]{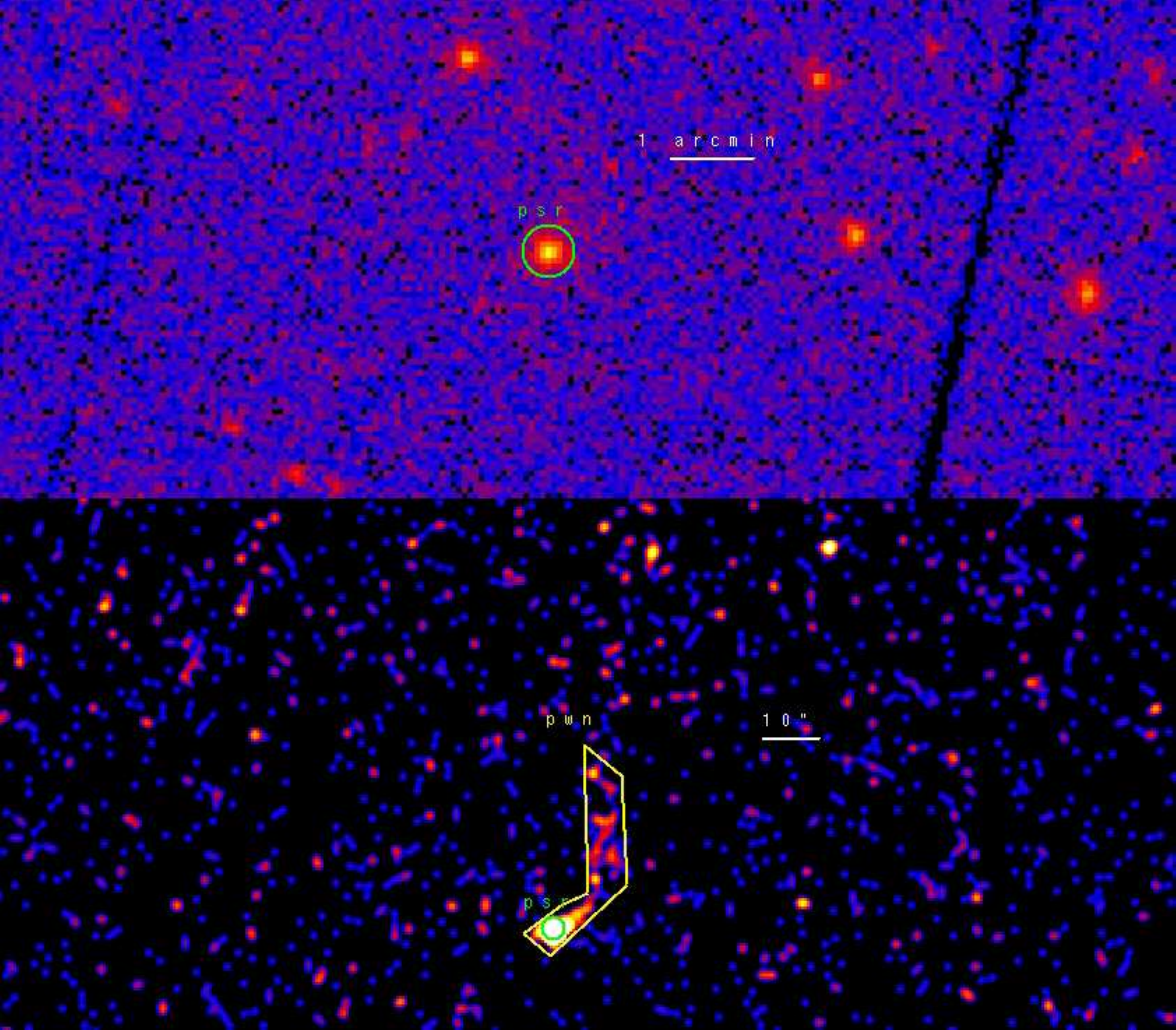}
\caption{{\it Upper Panel:} PSR J2124-3358 0.3-10 keV {\it XMM-Newton} EPIC Imaging. The PN and the two MOS images have been added.
The green circle marks the source region used in the analysis.
{\it Lower Panel:} PSR J2124-3358 0.3-10 keV {\it Chandra} Imaging. The image has been smoothed with a Gaussian
with Kernel radius of $2"$. The green circle marks the pulsar region while the yellow
polygon the nebular region used in the analysis.
\label{J2124-im}}
\end{figure}

For the {\it Chandra} observation, we chose a 2$"$ radius circular region for the
pulsar spectrum, an ad-hoc region for the nebular emission (see Figure \ref{J2124-im})
and a circular source-free region away from the source for the background.
For the {\it XMM-Newton} observation, we chose a 20$"$ radius circular region around
the pulsar while the background was extracted from a circular source-free region away from the source.
we obtained a total of 483, 15730, 810 and 834 pulsar counts respectively from the {\it Chandra}, XMM PN, MOS1 and 2 cameras
(background contributions of 0.1\%, 12.0\%, 9.4\% and 8.2\%). we also obtained
74 nebular counts from the {\it Chandra} observation (background contribution of 18.3\%).
Only for the nebular spectrum, we used the C-statistic
approach implemented in XSPEC.
For this reason, the nebular spectrum was fitted separately using the best fit N$_H$
value obtained from the pulsar spectra fit.
Due to its faintness and spatial extension, it cannot significantly contribute to the {\it XMM-Newton}
spectrum that is considered to be emitted only from the pulsar itself.
The best fitting model is a combination of a powerlaw and a blackbody
(probability of obtaining the data if the model is correct 
- p-value - of 0.42, 105 dof)
absorbed by a column N$_H$ = 2.76$_{-2.76}^{+4.87}$ $\times$ 10$^{20}$ cm$^{-2}$.
The powerlaw component has a photon index $\Gamma$ = 2.89$_{-0.35}^{+0.45}$;
the thermal component has a temperature of T = 3.11$_{-0.35}^{+0.37}$ $\times$ 10$^6$ K.
The blackbody radius R = 18.6$_{-8.8}^{+12.2}$ m determined from the
model parameters suggests that the emission is from a hot spot.
A simple blackbody model is not acceptable.
A simple powerlaw model is statistically acceptable but
an f-test performed comparing
a simple powerlaw with the composite spectrum gives a
chance probability of 7.2 $\times$ 10$^{-4}$, pointing
to a significative improvement by adding the blackbody component.
The nebular emission has a photon index $\Gamma$ = 1.90$_{-0.43}^{+0.47}$.
Assuming the best fit model, the 0.3-10 keV unabsorbed non-thermal pulsar flux is
6.68$_{-3.44}^{+1.50}$ $\times$ 10$^{-14}$, the thermal flux is 
2.91$_{-1.50}^{+0.66}$ $\times$ 10$^{-14}$ and the nebular flux is 
1.40$_{-0.69}^{+0.94}$ $\times$ 10$^{-14}$ erg/cm$^2$ s.
Using a distance
of 0.25 kpc, the luminosities are L$_{0.25kpc}^{nt}$ = 5.01$_{-2.58}^{+1.13}$ $\times$ 10$^{29}$,
L$_{0.25kpc}^{bol}$ = 2.18$_{-1.12}^{+0.49}$ $\times$ 10$^{29}$ and L$_{0.25kpc}^{pwn}$ = 1.05$_{-0.52}^{+0.71}$ $\times$ 10$^{29}$ erg/s.

\begin{figure}
\centering
\includegraphics[angle=0,scale=.50]{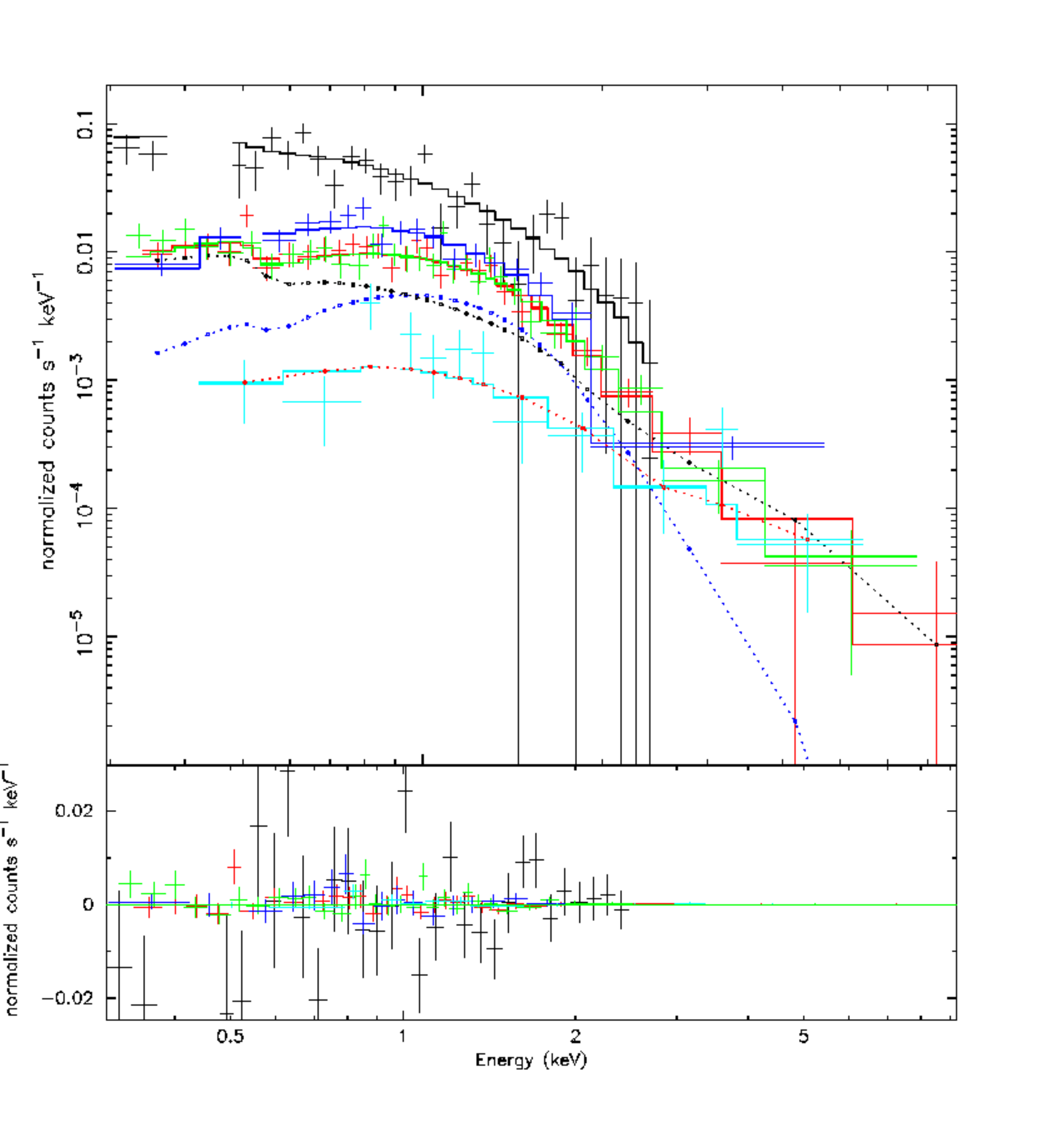}
\caption{PSR J2124-3358 Spectrum. Different colors mark all the different dataset used (see text for details).
Blue points mark the powerlaw component while black points the thermal component of the pulsar spectrum.
Red points mark the nebular spectrum.
Residuals are shown in the lower panel.
\label{J2124-sp}}
\end{figure}

\clearpage

{\bf J2214+3002 - type 2 RL MSP} % Nuova! osservazione in arrivo a giugno

After the {\it Fermi} detection, 25 constant and unassociated {\it Fermi} sources were observed by using 
the prime focus receiver at the GBT centered at 820MHz with 200MHz of bandwidth (Ransom et al. 2010).
The millisecond radio pulsar J2214+3002 was found during this campaign . It is a so-called
$"$black-widow$"$ system with a very low-mass companion ($\sim$ 0.02M$_s$) and likely timing irregularities, similar
to pulsars B1957+20 (Fruchter et al. 1988), J2051-0827
(Stappers et al. 1996), and J0610-2100 (Burgay et al.
2006). While we currently have no evidence for
radio eclipses from the pulsar (at least at frequencies
$>=$ 1.4GHz), its formation was likely similar to that of
those other systems (e.g. King et al. 2005). Its distance coming
from radio dispersion measurements is $\sim$ 1.5 kpc.

There is only a {\it Chandra} ACIS-I (faint mode) X-ray observation of J2214+3002 (obs. id 11788) for
a total exposure of 19.6 ks.
The X-ray source best fit position (obtained by using the celldetect
tool inside the Ciao distribution) is 22:14:38.85 +30:00:38.20 (1.5$"$ error radius).
The source low statistic prevents any search for diffuse emission.
We obtained 77 pulsar counts (background contribution of 0.4\%).
We used the C-statistic approach implemented in XSPEC.
The best fitting model is a simple powerlaw (reduced chisquare $\chi^2_{red}$ = 1.7, 11 dof)
with a photon index  $\Gamma$ = 3.32$_{-0.46}^{+1.01}$,
absorbed by a column N$_H$ = $<$ 2.13 $\times$ 10$^{21}$ cm$^{-2}$.
A simple blackbody model is not statistically acceptable while a thermal plus nonthermal
model cannot be studied due to the low statistic.
Assuming the best fit model, the 0.3-10 keV unabsorbed pulsar flux is
7.43 $\pm$ 0.25 $\times$ 10$^{-14}$ erg/cm$^2$ s.
Using a distance
of 1.5 kpc, the luminosities are L$_{1.5kpc}^{nt}$ = (2.00 $\pm$ 0.07) $\times$ 10$^{31}$ erg/s.

\begin{figure}
\centering
\includegraphics[angle=0,scale=.40]{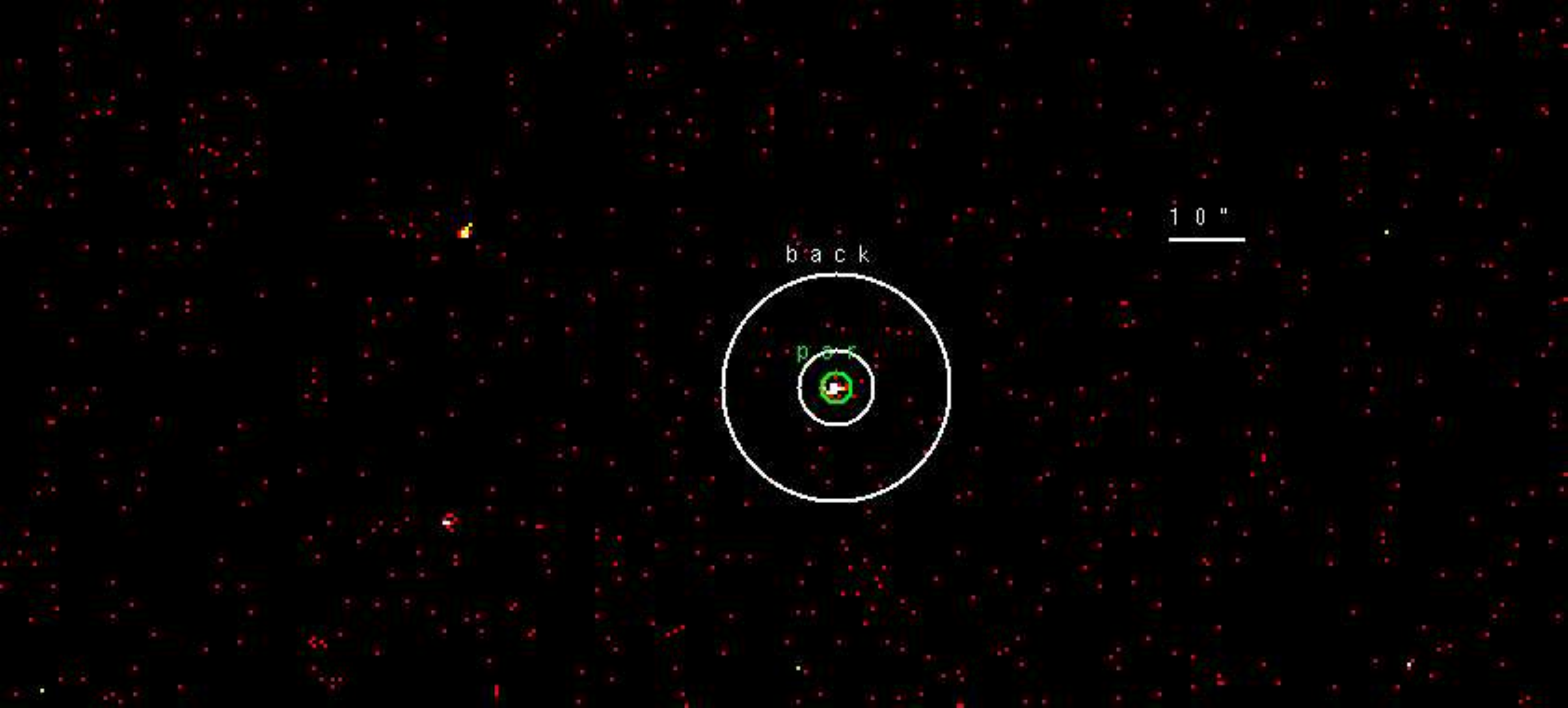}
\caption{PSR J2214+3002 0.3-10 keV {\it Chandra} Imaging.
The green circle marks the pulsar while the white annulus the background region used in the analysis.
\label{J2214-im}}
\end{figure}

\begin{figure}
\centering
\includegraphics[angle=0,scale=.50]{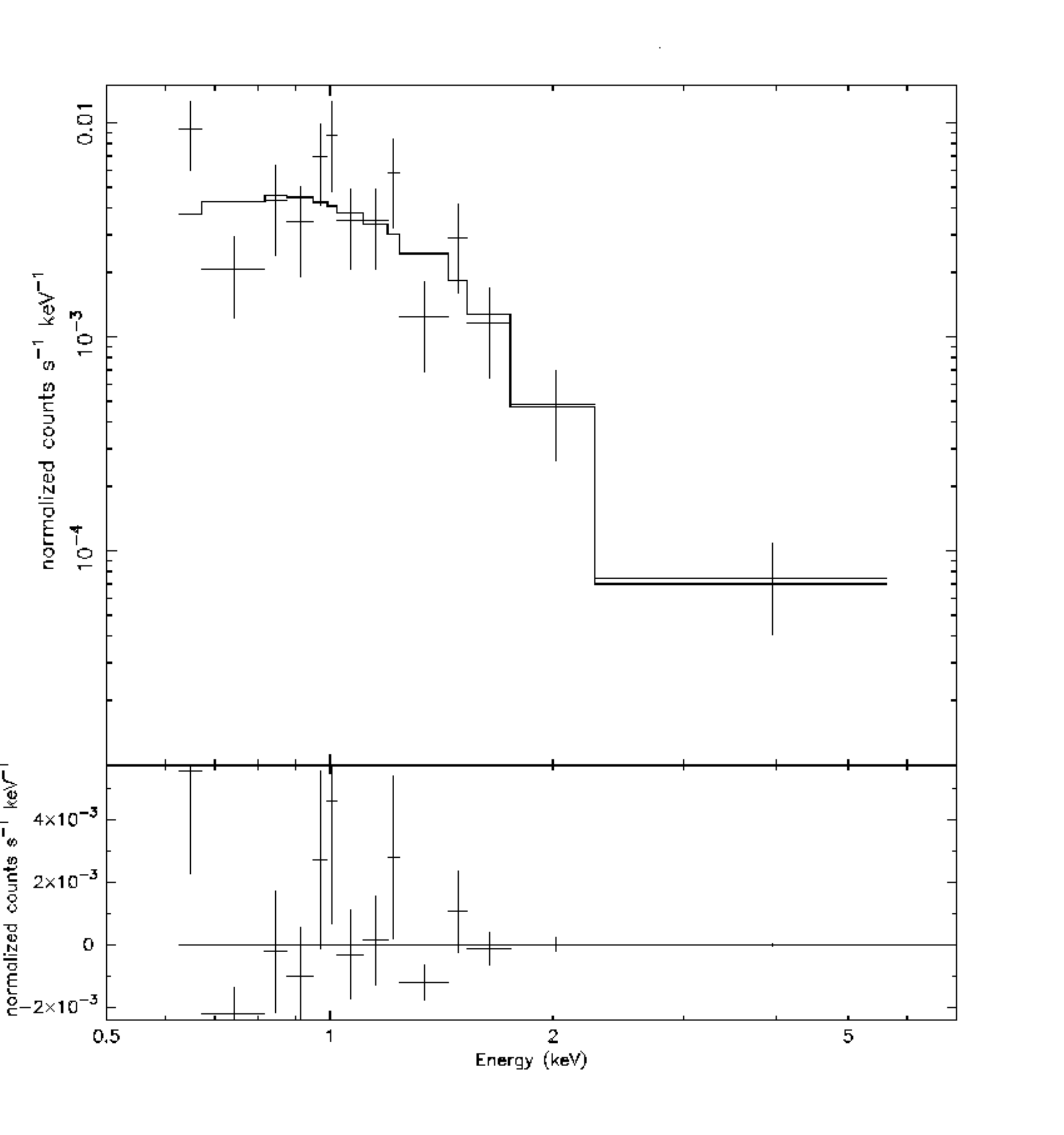}
\caption{PSR J2214+3002 {\it Chandra} Spectrum (see text for details).
Residuals are shown in the lower panel.
\label{J2214-sp}}
\end{figure}

\clearpage

{\bf J2229+6114 - type 2 RLP}

% Abdo et al. 2009
PSR J2229+6114 is located
within the error box of the EGRET source
3EG J2227+6122 (Hartman et al. 1999). Detected
as a compact X-ray source by ROSAT and ASCA
observations of the EGRET error box, it was later
discovered to be a radio and X-ray pulsar with a period
of P = 51.6 ms (Halpern et al. 2001b). The radio pulse
profile shows a single sharp peak, while the X-ray light
curve at 0.8 - 10 keV consists of two peaks, separated
by $\Delta\phi$ = 0.5.
The pulsar is as young as the Vela pulsar (characteristic
age $\tau_c$ = 10 kyr), as energetic ( $\dot{E}$ = 2.2 $\times$ 10$^{37}$ erg s$^{-1}$),
and is evidently the energy source of the $"$Boomerang$"$
arc-shaped PWN G106.65+2.96, suggested to be part of
the supernova remnant (SNR) G106.3+2.7 discovered
by Joncas \& Higgs (1990). Recently, the PWN has been
detected at TeV energies by {\it MILAGRO}. Studies of the radial velocities of both neutral
hydrogen and molecular material place the system at
$\sim$ 800 pc (Kothes et al. 2001), while Halpern et al.
(2001a) suggest a distance of 3 kpc estimated from its
X-ray absorption. The pulsar DM, used in conjunction
with the NE2001 model, yields a distance of 7.5 kpc ,
significantly above all other estimates. An HI distance
determination for the pulsar yields between 2.5 and 6.6
kpc (Johnston et al. 1996). Kothes et al. 2001 found the distance
to be between 0.8 and 6.5 kpc (see also Abdo et al.(2009c)).

\begin{figure}
\centering
\includegraphics[angle=0,scale=.30]{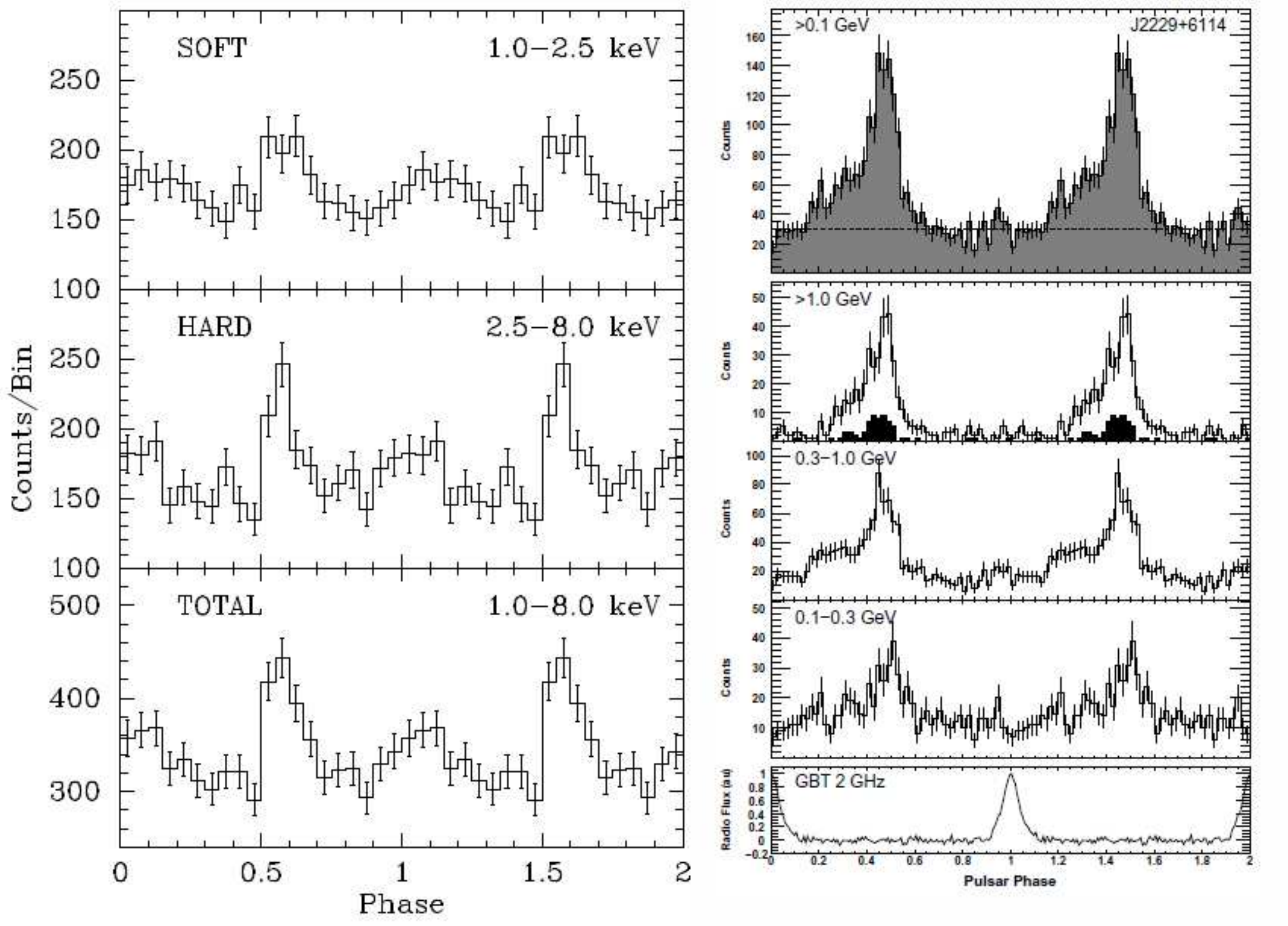}
\caption{PSR J2229+6114 Lightcurve.{\it Left: } X-ray pulse profile
from the {\it ASCA} GIS. See Halpern et al. 2001 for details.
{\it Right: Fermi} $\gamma$-ray lightcurve folded with Radio
(Abdo et al. catalogue). 
\label{J2229-lc}}
\end{figure}

Three different X-ray observations of J2229+6114 were performed:\\
- obs. id 1948, {\it Chandra} ACIS-I faint mode, start time 2001, February 14 at 04:51:12 UT, exposure 18.0 ks;\\
- obs. id 2787, {\it Chandra} ACIS-I very faint mode, start time 2002, March 15 at 12:25:01 UT, exposure 95.2 ks;\\
- obs. id 0012660101, {\it XMM-Newton} observation, start time 2002, June 15 at 09:51:20 UT, exposure 28.5 ks.\\
The PN camera of the EPIC instrument was operated in Small Window mode. The MOS cameras were operating in
Fast Uncompressed mode, so that they were not used in order to obtain the spectrum of the pulsar.
For the PN camera a thin optical filter was used.
First, an accurate
screening for soft proton flare events was done in the {\it XMM-Newton} observations obtaining a resulting total
exposure of 16.7 ks.
The X-ray source best fit position (obtained by using the celldetect
tool inside the Ciao distribution) is 22:29:05.27 +61:14:09.14 (1$"$ error radius).
A bright nebular emission is apparent in the {\it Chandra} observation with a radius of $\sim$ 20$"$
while a fainter emission is present until $\sim$ 1$'$.
For the {\it Chandra} observation, we chose a 2$"$ radius circular region for the
pulsar spectrum and a 1$'$ radius circular region for the nebula; all the pointlike
sources were excluded from the nebular extraction region. The
background was extracted from a circular source-free region away from the source.
For the {\it XMM-Newton} observation, we chose a 1$'$ radius circular region around
the pulsar in order to take in account both the pulsar and the nebula;
the background was extracted from a circular source-free region away from the source.
we added the two observations' {\it Chandra} spectra using
the mathpha tool and, similarly, the response
matrix and effective area files using addarf and addrmf.
we obtained a total of 2518 pulsar and 7198 nebular counts from the {\it Chandra} data
(background contributions of 0.1\% and 20.4\%); we also obtained 3182 counts
from the XMM data (background contribution of 34.0\%).
The best fitting model is a simple powerlaw (probability of obtaining the data if the model is correct 
- p-value - of 0.02, 261 dof fitting both the pulsar and nebular spectra)
with a photon index  $\Gamma$ = 1.22$_{-0.07}^{+0.06}$,
absorbed by a column N$_H$ = 4.09$_{-0.44}^{+0.48}$ $\times$ 10$^{21}$ cm$^{-2}$.
A simple blackbody model is not statistically acceptable while an 
f-test performed comparing
a simple powerlaw with a composite powerlaw plus blackbody spectrum gives the quite high
chance probability of 8.2 $\times$ 10$^{-2}$.
The nebular emission has a photon index $\Gamma$ = 1.31$_{-0.04}^{+0.06}$.
Assuming the best fit model, the 0.3-10 keV unabsorbed pulsar flux is
4.93 $\pm$ 0.49 $\times$ 10$^{-13}$ and the nebular flux is 
1.14$_{-0.10}^{+0.08}$ $\times$ 10$^{-12}$ erg/cm$^2$ s.
Using a distance
of 3.65 kpc, the luminosities are L$_{3.65kpc}^{nt}$ = 7.88 $\pm$ 0.78 $\times$ 10$^{32}$
and L$_{3.65kpc}^{pwn}$ = 1.82$_{-0.16}^{+0.13}$ $\times$ 10$^{33}$ erg/s.

\begin{figure}
\centering
\includegraphics[angle=0,scale=.50]{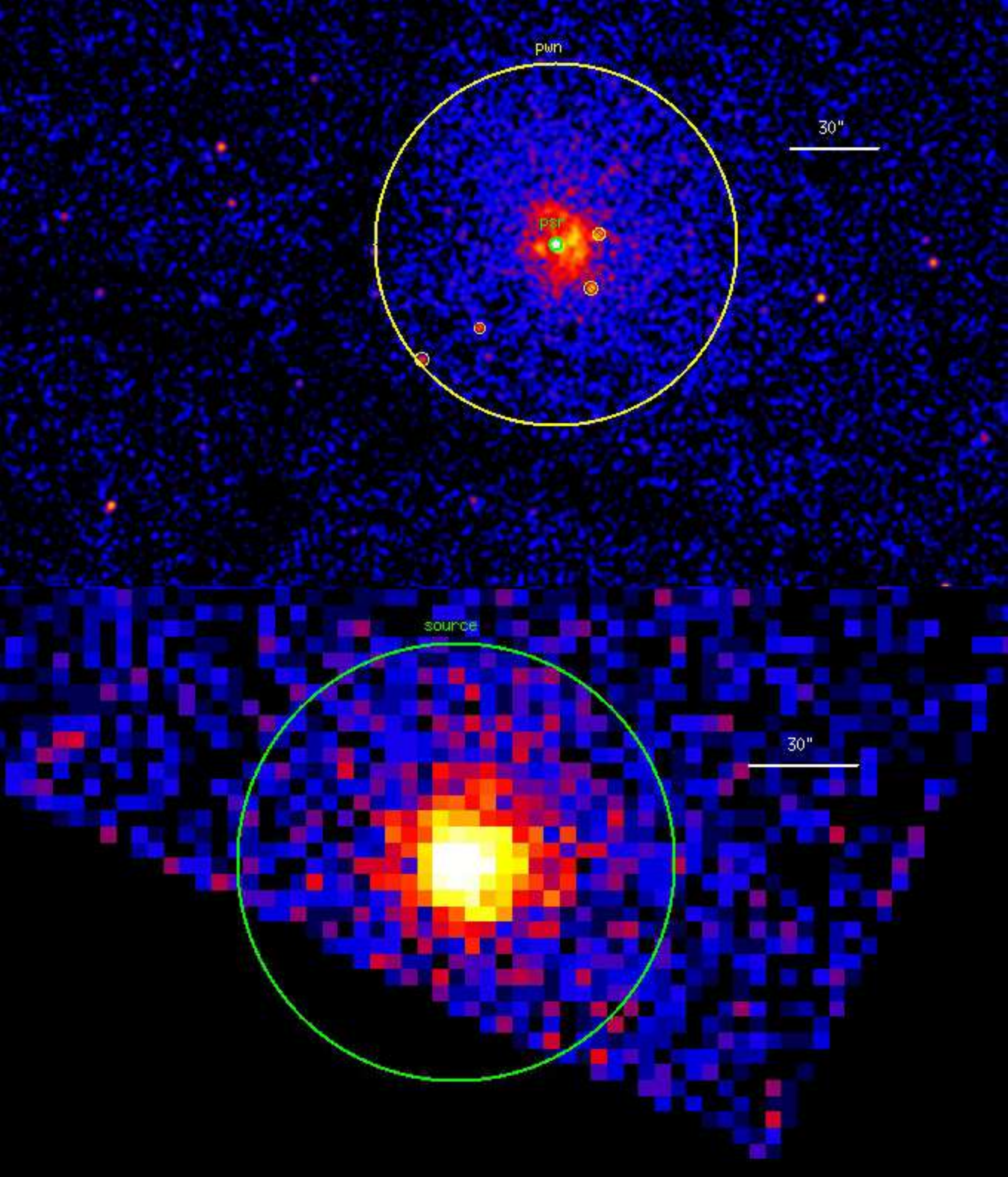}
\caption{{\it Upper Panel:} PSR J2229+6114 0.3-10 keV {\it Chandra} Imaging. The image has been smoothed with a Gaussian
with Kernel radius of $2"$. The green circle marks the pulsar region while the yellow
polygon the nebular region used in the analysis.
{\it Lower Panel:} PSR J2229+6114 0.3-10 keV {\it XMM-Newton} EPIC Imaging. The PN and the two MOS images have been added.
The green circle marks the source region used in the analysis.
\label{J2229-im}}
\end{figure}

\begin{figure}
\centering
\includegraphics[angle=0,scale=.50]{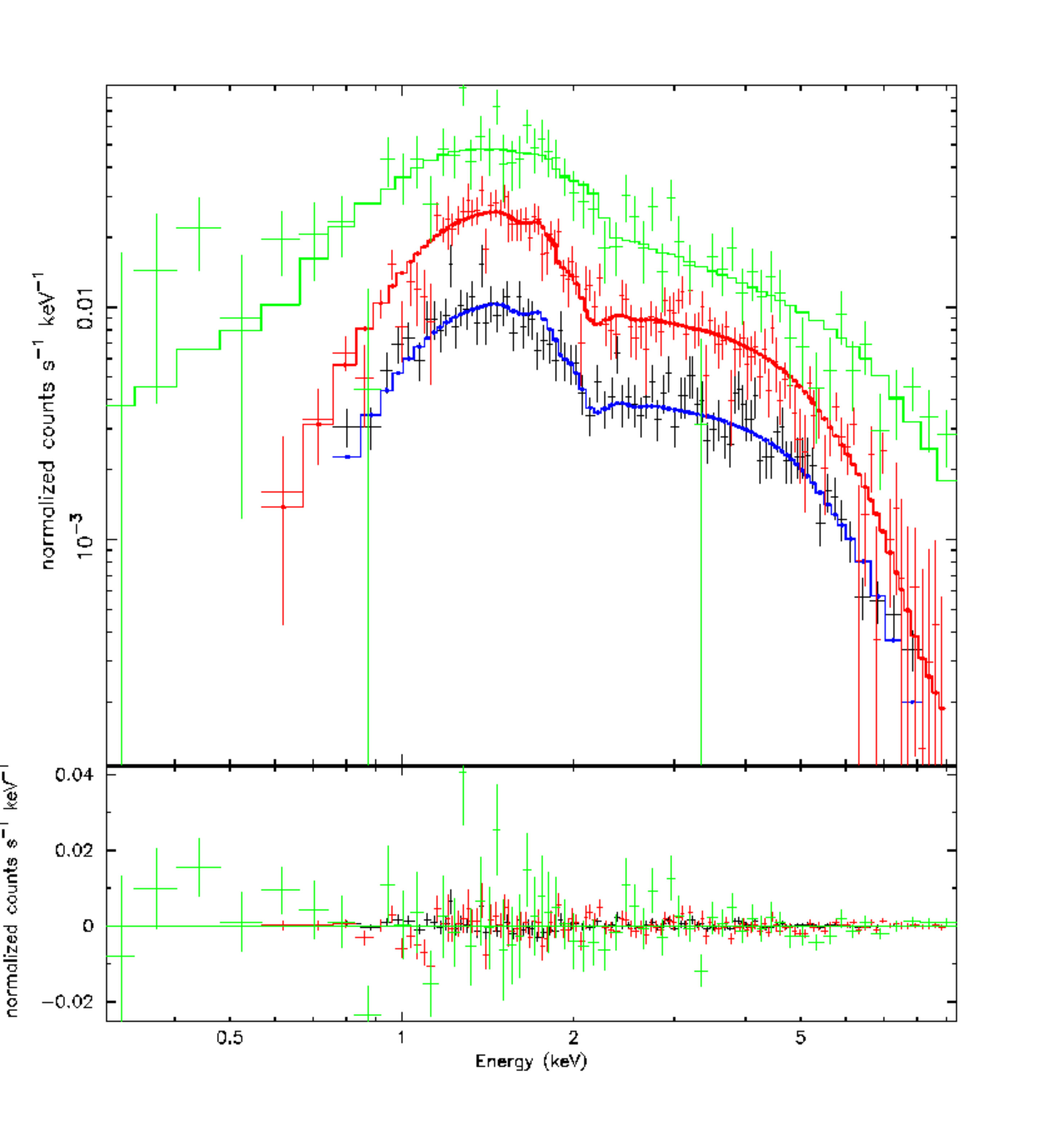}
\caption{PSR J2229+6114 Spectrum. Different colors mark all the different dataset used (see text for details).
Black points mark the pulsar spectrum while the red ones the nebular spectrum.
Residuals are shown in the lower panel.
\label{J2229-sp}}
\end{figure}

\clearpage

{\bf J2238+59 - type 0 RQP} % osservazione in arrivo

J2238+59 was one of the fist pulsars discovered using the
blind search technique (Abdo et al. 2009).
No $\gamma$-ray nebular emission was detected down to a flux of 
1.65 $\times$ 10$^{-10}$ erg/cm$^2$s (Ackermann et al. 2010).
The pseudo-distance of the object based on $\gamma$-ray data (Saz Parkinson et al. (2010))
is $\sim$ 2.1 kpc.

After the {\it Fermi} detection, we asked for a {\it SWIFT} observation
of the $\gamma$-ray error box (obs id. 00031398001, 7.83 ks exposure).
After the data reduction, no X-ray source were found inside
the {\it Fermi} error box.
For a distance of 2.1 kpc we found a
rough absorption column value of 7 $\times$ 10$^{21}$ cm$^{-2}$
and using a simple powerlaw spectrum
for PSR+PWN with $\Gamma$ = 2 and a signal-to-noise of 3,
we obtained an upper limit non-thermal unabsorbed flux of 4.49 $\times$ 10$^{-13}$ erg/cm$^2$ s,
that translates in an upper limit luminosity L$_{2.1kpc}^{nt}$ = 1.09 $\times$ 10$^{32}$ erg/s.

{\bf J2240+5832 - type 0 RLP} % Nuova!

J2240+5832 is an isolated radio pulsar found by the Pulsar Search Consortium.
Radio dispersion measurements found the distance to be $\sim$ 7.7 kpc.

The only X-ray observation pointing the radio position of the pulsar
is the {\it SWIFT} obs. id 00058431001, with a 5.72 ks exposure.
After the data reduction, no X-ray source were found at the radio position.
For a distance of 7.7 kpc we found a
rough absorption column value of 7 $\times$ 10$^{21}$ cm$^{-2}$
and using a simple powerlaw spectrum
for PSR+PWN with $\Gamma$ = 2 and a signal-to-noise of 3,
we obtained an upper limit non-thermal unabsorbed flux of 4.60 $\times$ 10$^{-13}$ erg/cm$^2$ s,
that translates in an upper limit luminosity L$_{7.7kpc}^{nt}$ = 3.27 $\times$ 10$^{33}$ erg/s.

{\bf J2241-5236 - type 2 RL MSP} % Nuova!

J2240+5832 is a millisecond radio pulsar in a binary system, found by the Pulsar Search Consortium
using the Parkes radio telescope.
Radio dispersion measurements found the distance to be $\sim$ 0.5 kpc (Keith et al. 2011).

There is only a {\it Chandra} ACIS-I (faint mode) X-ray observation of J2241-5236 (obs. id 11789) for
a total exposure of 19.9 ks.
The X-ray source best fit position (obtained by using the celldetect
tool inside the Ciao distribution) is 22:41:42.01 -52:36:36.21 (1.3$"$ error radius).
The source low statistic prevents any search for diffuse emission.
we obtained 76 pulsar counts (background contribution of 0.2\%).
we used the C-statistic approach implemented in XSPEC.
The best fitting model is a simple powerlaw (reduced chisquare $\chi^2_{red}$ = 1.7, 12 dof)
with a photon index  $\Gamma$ = 2.59$_{-0.42}^{+0.95}$,
absorbed by a column N$_H$ = $<$ 2.48 $\times$ 10$^{21}$ cm$^{-2}$.
A simple blackbody model is not statistically acceptable while a thermal plus nonthermal
model cannot be studied due to the low statistic.
Assuming the best fit model, the 0.3-10 keV unabsorbed pulsar flux is
5.22 $\pm$ 0.72 $\times$ 10$^{-14}$ erg/cm$^2$ s.
Using a distance
of 0.5 kpc its luminosity is L$_{0.5kpc}^{nt}$ = 1.57 $\pm$ 0.22 $\times$ 10$^{30}$ erg/s.

\begin{figure}
\centering
\includegraphics[angle=0,scale=.40]{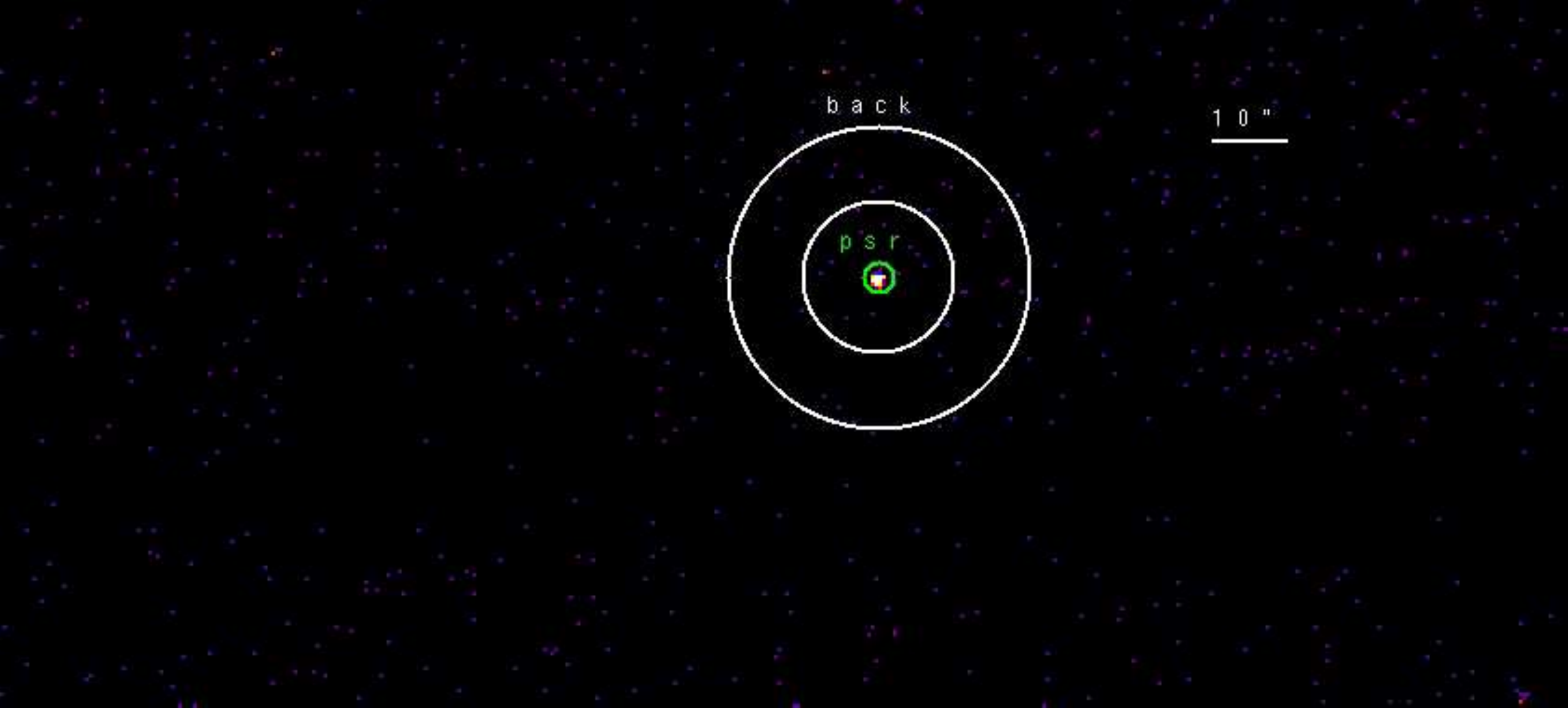}
\caption{PSR J2241-5236 0.3-10 keV {\it Chandra} Imaging.
The green circle marks the pulsar while the white annulus the background region used in the analysis.
\label{J2241-im}}
\end{figure}

\begin{figure}
\centering
\includegraphics[angle=0,scale=.50]{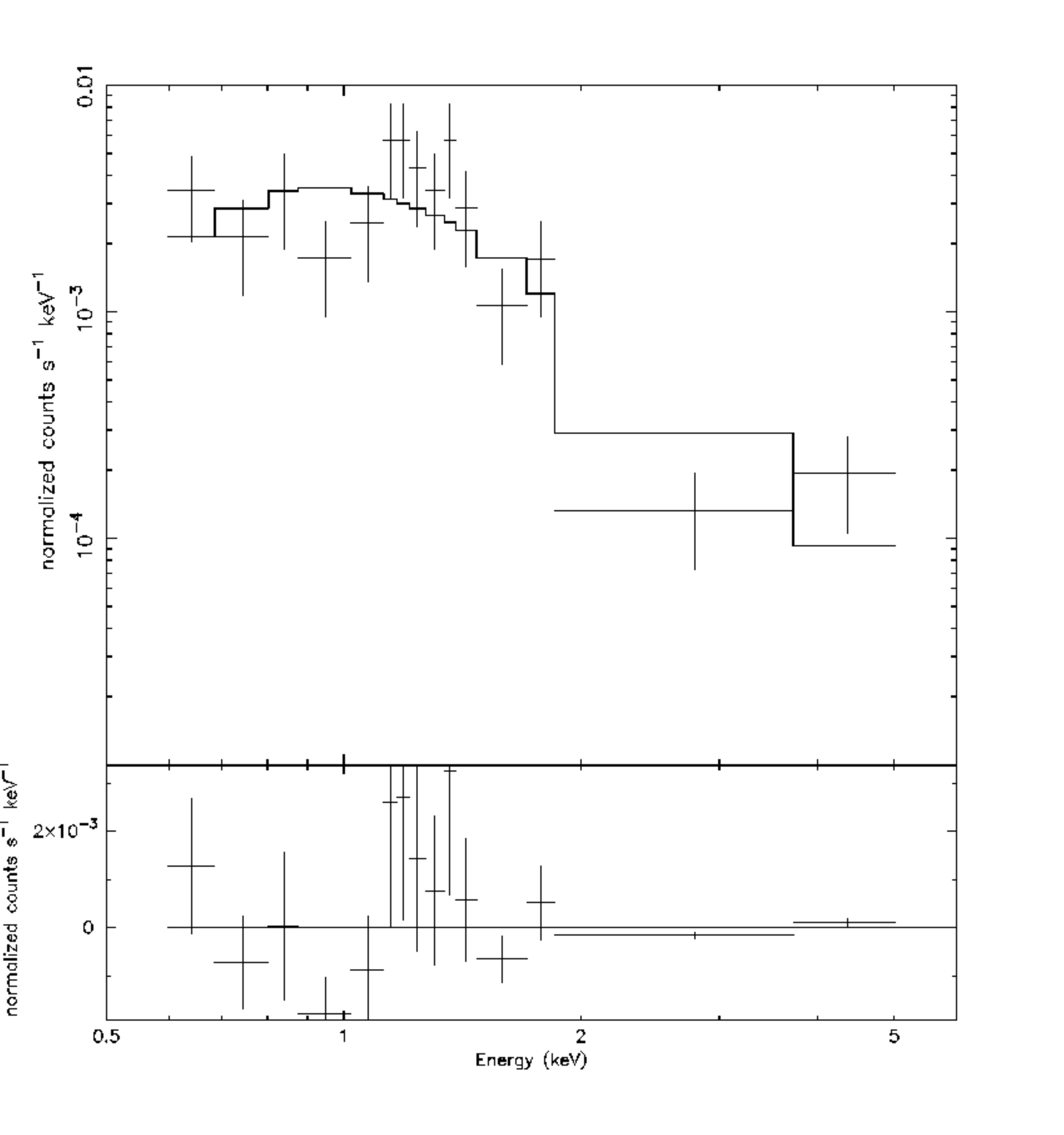}
\caption{PSR J2241-5236 {\it Chandra} Spectrum (see text for details).
Residuals are shown in the lower panel.
\label{J2241-sp}}
\end{figure}

\clearpage

{\bf J2302+4442 - type 2 RL MSP} % Nuova!

J2302+4442 is a millisecond pulsar in a binary system
found by the Pulsar Search Consortium.
Radio dispersion measurements found the distance to be $\sim$ 1.2 kpc (Cognard et al. 2011).

After the detection of the {\it Fermi} source, an {\it XMM-Newton} observation
was asked in order to find its X-ray counterpart:\\
- obs. id 0605470501, starting on 2009, December 25 at 20:18:27 UT, exposure 21.4 ks.\\
Both the PN and MOS cameras were operating in the Full Frame mode; a thin optical filter was used
for the PN camera while a medium filter was used for the MOS cameras.
First, an accurate
screening for soft proton flare events was done in the {\it XMM-Newton} observations obtaining a resulting
exposure of 16.1 ks.
A source detection was performed using both the SAS tools
and XIMAGE: the X-ray source best fit position is 23:02:46.5 +44:42:21.5 (5$"$ error radius).
No hint of diffuse emission is present in the observation.
we chose a 20$"$ radius circular region for the pulsar spectrum and 
a circular source-free region away from the source for the background.
Due to the low statistic in the XMM observation, we added the two MOS spectra using
mathpha tool and, similarly, the response
matrix and effective area files using addarf and addrmf.
we obtained a total of 277 and 127 pulsar counts from the PN and MOS cameras
(background contributions of 17.5\% and 13.5\%).
The best fitting model is a simple powerlaw
((probability of obtaining the data if the model is correct 
- p-value - of 0.19, 13 dof fitting both the pulsar and nebular spectra)
with  a photon index $\Gamma$ = 2.91$_{-0.33}^{+0.46}$ ,
absorbed by a column N$_H$ = 1.30$_{-0.52}^{+0.91}$ $\times$ 10$^{21}$ cm$^{-2}$,
in agreement with the galactic one ($\sim$ 1.31 $\times$ 10$^{21}$ cm$^{-2}$
using WebTools).
A simple blackbody model is not statistically accepted while a composite model
gives no statistical improvement.
Assuming the best fit model, the 0.3-10 keV unabsorbed pulsar flux is
6.81$_{-3.83}^{+1.40}$ $\times$ 10$^{-14}$ erg/cm$^2$ s.
Using a distance
of 1.2 kpc, the pulsar luminosity is L$_{1.2kpc}^{nt}$ = 1.18$_{-0.66}^{+0.24}$ $\times$ 10$^{31}$ erg/s.
\begin{figure}
\centering
\includegraphics[angle=0,scale=.30]{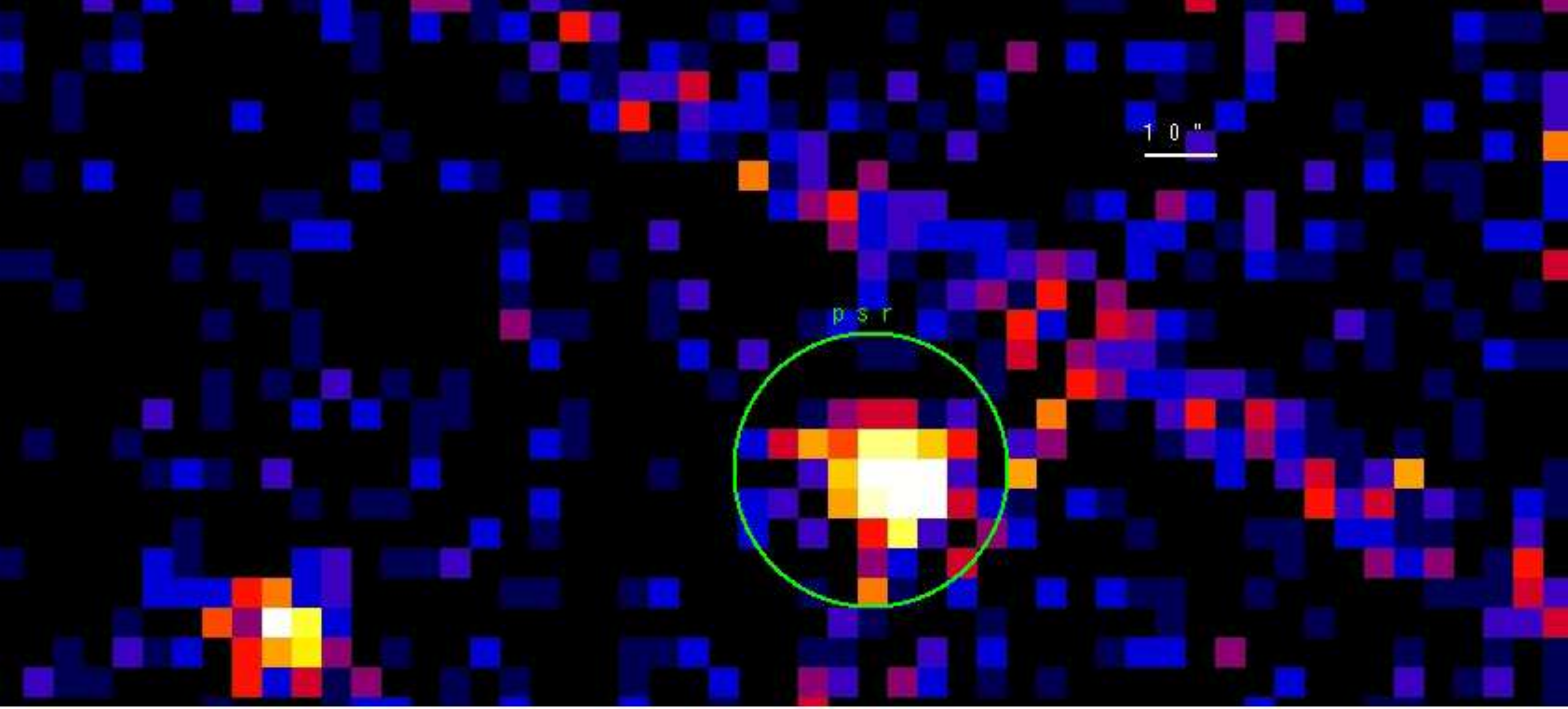}
\caption{PSR J2302+4442 0.3-10 keV {\it XMM-Newton} EPIC Imaging. The PN and the two MOS images have been added. 
The green circle marks the pulsar region used in the analysis.
\label{J2302-im}}
\end{figure}

\begin{figure}
\centering
\includegraphics[angle=0,scale=.40]{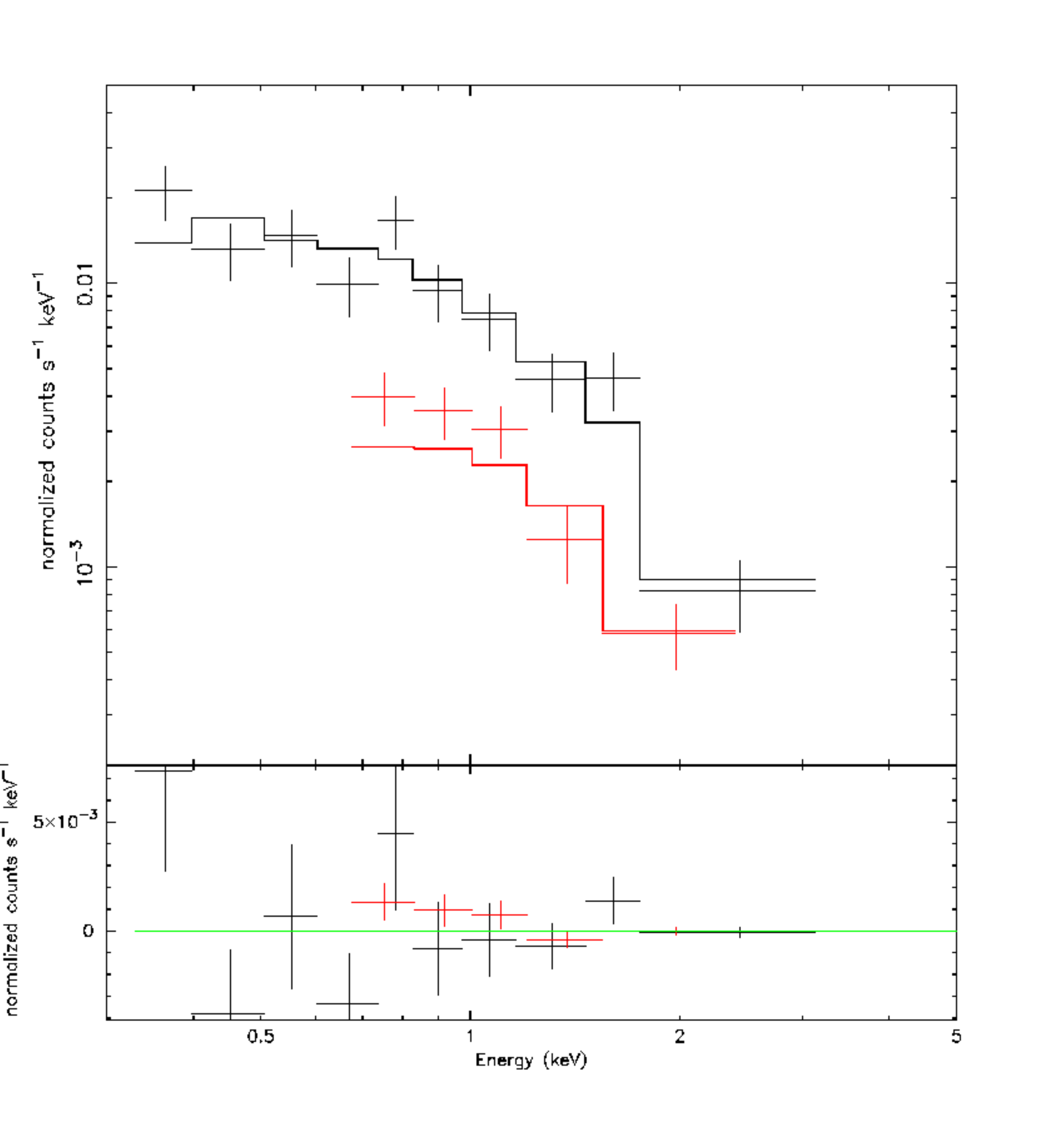}
\caption{PSR J2302+4442 {\it XMM-Newton} Spectrum (see text for details).
Residuals are shown in the lower panel.
\label{J2302-sp}}
\end{figure}

\clearpage

\appendix

\chapter{New observations of Radio-quiet pulsars}

In order to make this thesis' work, we obtained a number of X-ray observations of the newly discovered RQ {\it Fermi} pulsars.
Here we report the accepted proposals we made for CTA-1 and MORLA pulsars plus a number of {\it SWIFT} TOO we obtained to explore
the X-ray field of previously unobserved {\it Fermi} pulsars.

\section{Searching for X-ray pulsations from CTA-1}

{\bf Searching for X-ray pulsations from the
radio quiet Gamma-ray pulsar within CTA-1 plerion}

P.A. Caraveo, A. De Luca, M. Marelli, G.F. Bignami, G. Kanbach, Saz Parkinson, P. 
on behalf of the {\it Fermi}/LAT collaboration

Accepted in Cycle 8 {\it XMM-Newton} Call for Proposal.

{\bf Abstract}\\
During its activation phase, the Fermi Large Area Telescope has
discovered the time signature of a radio quiet neutron star
coincident with RX J0007.0+7302, the source at the center of the
young SNR CTA-1. The inferred timing parameters point to a
Vela-like neutron star with an age comparable to that estimated
for the SNR. Existing XMM-Newton observations of RX J0007.0+7302
show a tantalizing, although far from compelling, evidence for
pulsation. In view of the obvious interest of this newly
discovered radio quiet neutron star, here we ask for an orbit-long
XMM-Newton observation of the CTA-1 central source. With a full orbit
it will be possible to study
its X-ray timing behaviour, taking advantage of the contemporary
timing parameters provided by the Fermi LAT telescope.

\clearpage

\section{Searching for the X-ray counterpart of Morla}

{\bf A DEEP CHANDRA/NOAO INVESTIGATION TO IDENTIFY THE
COUNTERPART OF AN OLD PULSAR DISCOVERED IN GAMMA-RAYS}

A. De Luca, P.A. Caraveo, M. Marelli, R. Mignani , P. Saz Parkinson
on behalf of the {\it Fermi}/LAT collaboration

Accepted in Cycle 11 {\it Chandra} Call for Proposal.

{\bf Abstract}\\
After 5 months of all-sky scanning, the {\it Fermi} Large Area Telescope (LAT) has discovered 15 new
gamma-ray pulsars using a blind search algorithm. The discovery of a population of radio-quiet
(or - at least - radio-faint), bright gamma-ray pulsars has far reaching implications for $\gamma$-ray
source population studies, as well as for our overall understanding of pulsar physics. We
propose a multiwavelength project within the frame of {\it Chandra}/NOAO joint observations, aimed
at identifying the X-ray counterpart of one of the most exciting members of the newly discovered
pulsar sample: a bright gamma-ray pulsar with timing properties pointing to an old neutron star
with a very low rotational energy loss. Possibly very close to us, such pulsar is a factor 10 less
energetic - and significantly older - than Geminga.

\clearpage

\section{{\it SWIFT} TOOs}

As soon as a new radio-quiet pulsar was detected by {\it Fermi}, we analyzed all the available X-ray observations.
If the pulsar field was uncovered, we asked for a 5-10 ks {\it SWIFT} target of opportunity observation. In three years we
obtained {\it SWIFT} observations for all the following pulsars for a total of 117 ks:\\
J1741-2050 (09/10/2008), J1028-5819 (19/11/2008), J0357+2311 (21/11/2008-20/01/2009), J1459-6056 (25/02/2009), J1732-31 (03/03/2009),
J1958+2841 (10/03/2009), J1813-12 (19/03/2009), J1907+06 (30/03/2009), J2238+59 (09/04/2009), J1846+0919 (06/08/2009),
J1957+50 (01/09/2009), J1429-5911 (22/10/2009), J1044-5737 (23/11/2009), J1954+28 (02/12/2009), J0734-1557 (31/08/2010).\\
Moreover, we asked for a survey (proposal id. 5080075 inside cycle 5 GI {\it SWIFT} program)
of 12 galactic unidentified {\it Fermi} objects of the first three months catalogue (Abdo et al. 2009b) with a low $\gamma$-ray variability.
At now, six of them have been detected as pulsars (J0633+0632, J1732-31, J1741-2050, J1813-12, J1958+2841 and J2238+59).

Using these observations, at now we have found confirmed X-ray counterparts of five pulsars, 3 RQP (J0633+0632, J1813-12 and J1958+2841)
and 2 RLP (J1028-5819 and J1741-2050).

\clearpage

\chapter{The High-Energy behaviour of Fermi pulsars}

{\bf A MULTIWAVELENGTH STUDY ON THE HIGH-ENERGY BEHAVIOUR OF THE FERMI/LAT PULSARS}

Published in Astrophysical Journal - Marelli, M., De Luca, A. \& Caraveo P.A., 2011, ApJ 733 82.

\section{ABSTRACT}
Using archival as well as freshly acquired data, we assess the X-ray behaviour of the {\it Fermi}/Large Area Telescope
$\gamma$-ray pulsars listed in the First {\it Fermi} source catalogue After revisiting the relationships between the pulsars' rotational
energy losses and their X-ray and $\gamma$-ray luminosities, we focus on the distance-independent $\gamma$-to-X-ray flux ratios.
When plotting our F$_{\gamma}$/F$_X$ values as a function of the pulsars' rotational energy losses, one immediately sees that
pulsars with similar energetics have F$_{\gamma}$/F$_X$ spanning three decades. Such spread, most probably stemming from
vastly different geometrical configurations of the X-ray and $\gamma$-ray emitting regions, defies any straightforward
interpretation of the plot. Indeed, while energetic pulsars do have low F$_{\gamma}$/F$_X$ values, little can be said for the bulk
of the {\it Fermi} neutron stars. Dividing our pulsar sample into radio-loud and radio-quiet subsamples, we find that, on
average, radio-quiet pulsars do have higher values of F$_{\gamma}$/F$_X$, implying an intrinsic faintness of their X-ray emission
and/or a different geometrical configuration. Moreover, despite the large spread mentioned above, statistical tests
show a lower scatter in the radio-quiet data set with respect to the radio-loud one, pointing to a somewhat more
constrained geometry for the radio-quiet objects with respect to the radio-loud ones.

\section{INTRODUCTION}
The vast majority of the 1800 rotation-powered pulsars known
to date (Manchester et al. 2005) were discovered by radio
telescopes. While only few pulsars have also been seen in the
optical band (see, e.g., Mignani 2009, 2010), the contribution of
{\it Chandra} and {\it XMM-Newton} telescopes increased the number of
X-ray counterparts of radio pulsars bringing the grand total to
$\sim$100 (see, e.g., Becker 2009). Such high-energy emission can
yield crucial information on the pulsar physics, disentangling
thermal components from non-thermal ones, and tracing the
presence of pulsar wind nebulae (PWNe).

{\it Chandra}'s exceptional spatial resolution made it possible to
discriminate clearly the PWN and the pulsar (PSR) contributions
while {\it XMM-Newton}'s high spectral resolution and throughput
unveiled the multiple spectral components which characterize
pulsars (see, e.g., Possenti et al. 2002). Although the X-ray
non-thermal power-law index seems somehow related to the
gamma-ray spectrum (see, e.g., Kaspi et al. 2004), extrapolating
the X-ray data underpredicts the $\gamma$-ray flux by at least one order
of magnitude (see, e.g., Abdo et al. 2010c).

Until the launch of {\it Fermi}, only seven pulsars were seen
in high-energy gamma rays (Thompson 2008), and only one
of them, Geminga, was not detected by radio telescopes. The
{\it Fermi}/Large Area Telescope (LAT) dramatically changed such
scenario establishing radio-quiet pulsars as a major family of
$\gamma$-ray emitting neutron stars. After one year of all-sky monitoring
{\it Fermi}/LAT has detected 54 gamma-ray pulsars, 22 of which
are radio quiet (Abdo et al. 2010a; Saz Parkinson et al. 2010;
Camilo et al. 2009). Throughout this paper, we shall classify
as radio quiet all the pulsars detected by {\it Fermi} through blind
searches (Abdo et al. 2010a; Saz Parkinson et al. 2010) but
not seen in radio in spite of dedicated deep searches. Containing
a sizeable fraction of radio-quiet pulsars, the {\it Fermi} sample
provides, for the first time, the possibility to compare the
phenomenology of radio-loud and radio-quiet neutron stars assessing
their similarities and their differences (if any).

While our work rests on the {\it Fermi} data analysis and results
(Abdo et al. 2010a; Saz Parkinson et al. 2010) for the X-ray side
we had to first build an homogeneous data set relying both on
archival sources and on fresh observations.

In the following, we will address the relationship between the
classical pulsar parameters, such as age and overall energetics
$\dot{E}$, and their X-ray and $\gamma$-ray yields. While the evolution of
the X-ray and $\gamma$-ray luminosities as a function of $\dot{E}$ and
the characteristic age $\tau_c$ have been already discussed, we
will concentrate on the ratio between the X-ray and $\gamma$-ray
luminosities, thus overcoming the distance conundrum which
has hampered the studies discussed so far in the literature. We
note that F$_{\gamma}$/F$_X$ parameter probes both pulsar efficiencies at
different wavelengths and distribution of the emitting regions
in the pulsar magnetosphere. Thus, such a distance-independent
approach does magnify the role of both geometry and geography
in determining the high-energy emission from pulsars.

\section{DATA ANALYSIS}
\subsection{$\gamma$-Ray Analysis}

We consider all the pulsars listed in the First Year Catalogue of
{\it Fermi} $\gamma$-ray sources (Abdo et al. 2010b) which contains the $\gamma$-
ray pulsars listed in the First {\it Fermi} pulsar catalogue (Abdo et al.
2010a) as well as the new blind-search pulsars found by Saz
Parkinson et al. (2010). Our sample comprises 54 pulsars:\\
1. twenty-nine detected using radio ephemerides and\\
2. twenty-five found through blind searches; of these three
were later found to have also a radio emission and, as such,
they were added to the radio emitting ones.

Thus, our sample of $\gamma$-ray emitting neutron stars consists of
32 radio pulsars and 22 radio-quiet pulsars. Here, we summarize
the main characteristics of the analysis performed in the two
articles.

The pulsar spectra were fitted with an exponential cutoff
power-law model of the form

\
$dN/dE = KE_{GeV}^{-\gamma}exp(-E/E_{cutoff})$
\

1 GeV has been chosen to define the normalization factor
because it is the energy at which the relative uncertainty on the
differential flux is minimal.

The spectral analysis was performed taking into account the
contribution of all the neighboring sources (up to 17$^{\circ}$) and the
diffuse emission. Sources at more than 3$^{\circ}$ from any pulsars
were assigned fixed spectra, taken from the all-sky analysis.
$\gamma$-rays with E$>$100MeV have been used and the contamination
produced by cosmic-ray interactions in the Earth's atmosphere
was avoided by selecting a zenith angle $>$105$^{\circ}$.

At first, all events have been used in order to obtain a phase-averaged
spectrum for each pulsar. Next the data have been
split into on-pulse and off-pulse samples. The off-pulse sample
has been described with a simple power law while, for the on-pulse
emission, an exponentially cutoff power law has been
used, with the off-pulse emission (scaled to the on-pulse phase
interval) added to the model. Such an approach is adopted in
order to avoid a possible PWN contamination to the pulsar
spectrum.

For completeness, we included in our sample also the four
radio pulsars listed in the fourth IBIS/ISGRI catalogue (Bird et al.
2010) but, so far, not seen by {\it Fermi}. Searching in the one
year {\it Fermi} catalogue (Abdo et al. 2010b), we found a potential
counterpart for PSR J0540-6919 but the lack of a pulsation
prevents us to associate the IBIS/ISGRI pulsar with the {\it Fermi}
source. We therefore used the 1FGL flux as an upper limit.
The three remaining IBIS pulsars happen to be located near the
galactic center, where the intense radiation from the disk of our
Galaxy hampers the detection of $\gamma$-ray sources. We used the
sensitivity map taken from Abdo et al. (2010b) to evaluate the
{\it Fermi} flux upper limit.

\subsection{X-Ray Data}

The X-ray coverage of the {\it Fermi}/LAT pulsars is uneven
since the majority of the newly discovered radio-quiet PSRs
have never been the target of a deep X-ray observation, while
for other well-known $\gamma$-ray pulsars - such as Crab, Vela, and
Geminga - one can rely on a lot of observations. To account for
such an uneven coverage, we classify the X-ray spectra on the
basis of the public X-ray data available, thus assigning\\
1. label $"$0$"$ to pulsars with no confirmed X-ray counterparts
(or without a non-thermal spectral component);\\
2. label $"$1$"$ to pulsars with a confirmed counterpart but too
few photons to assess its spectral shape; and\\
3. label $"$2$"$ to pulsars with a confirmed counterpart for which
the data quality allows for the analysis of both the pulsar
and the nebula (if present).\\

An $"$ad hoc$"$ analysis was performed for seven pulsars for
which the standard analysis could not be applied (e.g., owing
to the very high thermal component of Vela or to the closeness
of J1418-6058 to an active galactic nucleus, AGN). Table \ref{art-tab-2}
provides details on such pulsars.

We consider an X-ray counterpart to be confirmed if\\
1. X-ray pulsation has been detected;\\
2. X-ray and radio coordinates coincide; and\\
3. X-ray source position has been validated through the blind search
algorithm developed by the {\it Fermi} collaboration
(Abdo et al. 2009; Ray et al. 2010).

If none of these conditions apply, $\gamma$-ray pulsar is labeled as $"$0$"$.

According to our classification scheme, we have 14 type-0,
7 type-1, and 37 type-2 pulsars. In total 44 $\gamma$-ray neutron stars,
31 radio-loud, and 13 radio-quiet have an X-ray counterpart.

Since the X-ray observation database is continuously growing,
the results available in literature encompass only fractions
of the X-ray data now available. Moreover, they have been obtained
with different versions of the standard analysis software
or using different techniques to account for the PWN contribution.
Thus, with the exception of the well-known and bright
X-ray pulsars, such as Crab or Vela, we re-analyzed all the X-ray
data publicly available following an homogeneous procedure.
If only a small fraction of the data are publicly available, we
quoted results from a literature search.

In order to assess the X-ray spectra of {\it Fermi} pulsars, we
used photons with energy 0.3 keV $<$ E $<$ 10 keV collected by
{\it Chandra}/ACIS (Garmire et al. 2003), {\it XMM-Newton} (Struder
et al. 2001; Turner et al. 2001), and {\it SWIFT}/XRT (Burrows
et al. 2005). We selected all the public observations (as of 2010
April) that overlap the error box of {\it Fermi} pulsars or the radio
coordinates.

We neglected all {\it Chandra}/HRC observations owing to the
lack of energy resolution of the instrument. To analyze {\it Chandra}
data, we used the {\it Chandra} Interactive Analysis of Observation
software (CIAO ver. 4.1.2). The {\it Chandra} point-spread function
(PSF) depends on the off-axis angle: we used for all the point
sources an extraction area around the pulsar that contains 90\%
of the events. For instance, for on-axis sources we selected all
the photons inside a 20$"$ radius circle, while we extracted photons
from the inner part of PWNs (excluding the 20$"$ radius circle of
the point source) in order to assess the nebular spectra: such
extended regions vary from pulsar to pulsar as a function of the
nebula dimension and flux.

We analyzed all the {\it XMM-Newton} data (both from PN and
MOS1/2 detectors) with the {\it XMM-Newton} Science Analysis
Software (SASv8.0). The raw observation data files (ODFs)
were processed using standard pipeline tasks (epproc for PN,
emproc for MOS data); we used only photons with event pattern
0-4 for the PN detector and 0-12 for the MOS1/2 detectors.
When necessary, an accurate screening for soft proton flare
events was done, following the prescription by De Luca \&
Molendi (2004).

If, in addition to {\it XMM-Newton} data, deep {\it Chandra} data
were also available, we made an {\it XMM-Newton} spectrum of
the entire PSR+PWN and used the {\it Chandra} higher resolution
to disentangle the two contributions. When only {\it XMM-Newton}
data were available, the point source was analyzed by selecting
all the photons inside a 20$"$ radius circle while the whole PWN
(with the exception of the 20$"$ radius circle of the point source)
was used in order to assess the nebular spectrum.

We analyzed all the {\it SWIFT}/XRT data with HEASOFT
version 6.5 selecting all the photons inside a 20$"$ radius circle. If
multiple data sets collected by the same instruments were found,
spectra, response, and effective area files for each data set were
added by using the mathpha, addarf, and addrmf HEASOFT
tools.

All the spectra have been studied with XSPEC v.12
(Arnaud 1996) choosing, whenever possible, the same background
regions for all the different observations of each pulsar.
All the data were rebinned in order to have at least
25 counts per channel, as requested for the validity of $\chi^2$
statistic.

The XMM-{\it Chandra} cross-calibration studies (Stuhlinger
et al. 2008) report only minor changes in flux ($<$10\%) between
the two instruments. When both XMM and {\it Chandra} data were
available, a constant has been introduced to account for such
uncertainty. Conversely, when the data were collected only by
one instrument, a systematic error was introduced. All the PSRs
and PWNs have been fitted with absorbed power laws; when
statistically needed, a blackbody component has been added
to the pulsar spectrum. Since PWNs typically show a powerlaw
spectrum with a photon index which steepens moderately
as a function of the distance from the PSR (Gaensler \& Slane
2006), we used only the inner part of each PWN. Absorption
along the line of sight has been obtained through the fitting procedure
but for the cases with very low statistic for which we
used information derived from observations taken in different
bands.

\subsection{X-ray Analysis}

For pulsars with a good X-ray coverage, we carried out the
following steps.

If only {\it XMM-Newton} public observations were available,
we tried to take into account the PWN contribution. First, we
searched the literature for any evidence of the presence of a PWN
and, if nothing was found, we analyzed the data to search for
extended emission. If no evidence for the presence of a PWN was
found, we used PN and MOS1/2 data in a simultaneous spectral
fit. On the other hand, if a PWN was present, its contribution
was evaluated on a case by case basis. If the statistic was good
enough, we studied simultaneously the inner region, containing
both PSR and PWN, and the extended source region surrounding
it. The inner region data were described by two absorbed (PWN
and PSR) power laws, while the outer one by a single (PWN)
power law. The N$_H$ and the PWN photon index values were the
same in the two (inner and outer) data sets.

When public {\it Chandra} data were available, we evaluated
separately PSR and PWN (if any) in a similar way.

If both {\it Chandra} and XMM public data were available,
we exploited {\it Chandra} space resolution to evaluate the PWN
contribution by\\
1. obtaining two different spectra of the inner region (a),
encompassing both PSR and PWN and of the outer region
(b) encompassing only the PWN;\\
2. extracting a total XMM spectrum (c) containing both PSR
and PWN: this is the only way to take into account the
XMM's larger PSF;\\
3. fitting simultaneously (a)-(c) with two absorbed power
laws and eventually (if statistically significant) an absorbed
blackbody, using the same N$_H$; a constant multiplicative
was also introduced in order to account for a possible
discrepancy between {\it Chandra} and XMM calibrations; and\\
4. forcing to zero the normalization(s) of the PSR model(s)
in the {\it Chandra} outer region and freeing the other normalizations
in the {\it Chandra} data sets; fixing the XMM PSR
normalization(s) at the inner {\it Chandra} data set one and the
XMM PWN normalization at the inner+outer normalizations
of the {\it Chandra} PWN.

Only for few well-known pulsars, or pulsars for which the data
set is not yet entirely public, we used results taken from the
literature (see Table \ref{art-tab-2}). Where necessary, we used XSPEC in
order to obtain the flux in the 0.3-10 keV energy range and to
evaluate the unabsorbed flux.

For pulsars with a confirmed counterpart but too few photons
to discriminate the spectral shape, we evaluated a hypothetical
unabsorbed flux by assuming a single power-law spectrum
with a photon index of 2 to describe PSR+PWN. We also
assumed that the PWN and PSR thermal contributions are
30\% of the entire source flux (a sort of mean value of all the
considered type-2 pulsars). To evaluate the absorbing column,
we need a distance value which can come either from the radio
dispersion or - for radio-quiet pulsars - from the following
pseudo-distance reported in Saz Parkinson et al. (2010):\\
$d=0.51\dot{E_{34}}^{1/4}/F_{\gamma,10}^{1/2}$ kpc \\
where $\dot{E}=\dot{E}_{34}\times10^{34}erg/s$ and $F_{\gamma}=F_{\gamma,10}\times10^{-10}erg/cm^2s$
and the beam correction factor f$_{\gamma}$ is assumed to be 1 (Watters et al. 2009) for all pulsars.

Then, the HEASARC WebTools was used to find the galactic column density
(N$_H$) in the direction of the pulsar; with the distance information,
we could rescale the column density value of the pulsar.
We found the source count rate by using the XIMAGE
task (Giommi et al. 1992). Then, we used the WebPimms
tool inside the WebTools package to evaluate the source unabsorbed
flux. Such a value has to be then corrected to account
for the PWN and PSR thermal contributions. We are aware
that each pulsar can have a different photon index, as well
as thermal and PWN contributions, so we used these mean
values only as a first approximation. All the low-quality pulsars
(type 1) will be treated separately and all the considerations
in this paper will be based only on high-quality objects
(type 2).

For pulsars without a confirmed counterpart, we evaluated the
X-ray unabsorbed flux upper limit assuming a single power-law
spectrum with a photon index of 2 to describe PSR+PWN and
using a signal-to-noise ratio of 3.

The column density has been evaluated as above. Under the
previous hypotheses, we used the signal-to-noise definition in
order to compute the upper limit to the absorbed flux of the
X-ray counterpart. Next we used XSPEC to find the unabsorbed
upper limit flux.

On the basis of our X-ray analysis, we define a subsample of
{\it Fermi} $\gamma$-ray pulsars for which we have, at once, reliable X-ray
data (type-2 pulsars) and satisfactory distance estimates such
as parallax, radio dispersion measurement, column density estimate,
supernova remnant (SNR) association. Such a subsample
contains 24 radio emitting neutron stars and 5 radio-quiet ones.
The low number of radio quiet is to be ascribed to lack of high quality
X-ray data. Only one of the IBIS pulsars has a clear
distance estimate. Moreover, we have four additional radio-quiet
pulsars with reliable X-ray data but without a satisfactory
distance estimate.

In Tables \ref{art-tab-2} and \ref{art-tab-3}, we reported the $\gamma$-ray and X-ray
parameters of the 54 {\it Fermi} first year pulsars. We also included
the four hard X-ray pulsars taken from the $"$Fourth IBIS/ISGRI
soft gamma-ray survey catalogue$"$ (Bird et al. 2010). We use $\dot{E}=4\pi^2I\dot{P}/P$, $\tau_C=P/2\dot{P}$ and 
$B_{lc}=3.3\times10^{19}(P\dot{P})^{1/2}\times(10km)/(R_{lc}^3)$, where $R_{lc}=cP/2\pi$, P is the pulsar spin period, $\dot{P}$ its derivative 
and the standard value for moment of inertia of the neutron star I=$10^{45}g/cm^2$ (see e.g. Steiner et al. 2010).
Using the P and $\dot{P}$ values taken from Abdo et al. (2009c), Saz Parkinson et al. (2010), we computed
the values reported in Table \ref{art-tab-1}. Most of the distance values are taken from Abdo et al. (2009c), Saz Parkinson et al. (2010) (see Table \ref{art-tab-1}).

\section{DISCUSSION}
\subsection{Study of the X-Ray Luminosity}
\label{art-xdata}

The X-ray luminosity, L$_X$, is correlated with the pulsar spindown
luminosity $\dot{E}$. The scaling was first noted by Seward
\& Wang (1988) who used Einstein data of 22 pulsars - most
of them just upper limits - to derive a linear relation between
logF$^X_{0.2-4keV}$ and log $\dot{E}$. Later, Becker \& Trumper (1997) investigated
a sample of 27 pulsars using ROSAT, yielding the simple
scaling L$_X^{0.1-2.4keV}\simeq10^{-3}\dot{E}$. The uncertainty due to soft
X-ray absorption translates into very high flux errors; moreover
it was very hard to discriminate between the thermal and
power-law spectral components. A reanalysis was performed by
Possenti et al. (2002), who studied in the 2-10 keV band a sample
of 39 pulsars observed by several X-ray telescopes. However,
they could not separate the PWN from the pulsar contribution.
Moreover, they conservatively adopted, for most of the pulsars,
an uncertainty of 40\% on the distance values. A better comparison
with our data can be done with the results by Kargaltsev \&
Pavlov (2008), who recently used high-resolution {\it Chandra} data
in order to disentangle the PWN and pulsar fluxes. Focusing
just on {\it Chandra} data, and rejecting XMM observations, they
obtain a poor spectral characterization which translates in high
errors on fluxes. They also adopted an uncertainty of 40\% on the
distance values for most pulsars. Despite the big uncertainties,
mainly due to poor distance estimates, all these data sets show
that the L$_X$ versus $\dot{E}$ relation is quite scattered. The high values
of the $\chi^2_{red}$ seem to exclude a simple statistical effect.

We are now facing a different panorama, since our ability to
evaluate pulsars' distances has improved (Abdo et al. 2010a;
Saz Parkinson et al. 2010) and we are now much better in
discriminating pulsar emission from its nebula. The use of
XMM data makes it possible to build good quality spectra
allowing to disentangle the non-thermal from the thermal
contribution, when present. In particular, we can study the
newly discovered radio-quiet pulsar population and compare
them with the $"$classical$"$ radio-loud pulsars. We investigate the
relations between the X-ray and $\gamma$-ray luminosities and pulsar
parameters, making use of the data collected in Tables \ref{art-tab-1}, \ref{art-tab-2} and \ref{art-tab-3}.

Using the 29 {\it Fermi} type-2 pulsars with a clear distance
estimate and with a well-constrained X-ray spectrum, the
weighted least-square fit yields
\begin{equation}
log_{10}L^X_{29}=(1.11_{-0.30}^{+0.21})+(1.04\pm0.09)log_{10}\dot{E}_{34}
\end{equation}
where $\dot{E}=\dot{E}_{34}\times10^{34}erg/s$ and $L_X=L^X_{29}\times10^{29}erg/s$. All the uncertainties are at 90\% confidence level.
We can evaluate
the goodness of this fit using the reduced $\chi^2$ value $\chi^2_{red}$
= 3.7; a double-linear fit does not significantly change the value of $\chi^2_{red}$.
A more precise way to evaluate the dispersion of the data set
around the fitted curve is the parameter
$W^2=(1/n)\sum_{i=1->n}(y_{oss}^i-y_{fit}^i)^2$\\
where $y_{oss}^i$ is the actual i$^{th}$ value of the dataset (in our case $log_{10}L^X_{29}$) and $y_{fit}^i$ the expected one. A lesser spread in the dataset
translate into a lower value of $W^2$. We obtain $W^2=0.436$ for the $L_x-\dot{E}$ relationship.
Such high values of both $W^2$ and $\chi_{red}^2$ are an indication of an important scattering of the $L_X$ values around the fitted relation.\\

Our results are in agreement with Possenti et al. (2002) and
Kargaltsev \& Pavlov (2008).

\subsection{Study of the $\gamma$-Ray Luminosity}
\label{art-gdata}

The gamma-ray luminosity, L$_{\gamma}$, is correlated with the pulsar spin-down luminosity $\dot{E}$.
Such a trend is expected in many theoretical models (see e.g. Zhang et al. 2004, Muslimov\&Harding 2003)
and it's shortly discussed in the {\it Fermi} LAT catalogue of gamma-ray pulsars (Abdo et al. 2009c).

Selecting the same subsample of {\it Fermi} pulsar used in the previous chapter to assess the relation
between $L_{\gamma}$ and $\dot{E}$, we found that a linear fit:
\begin{equation}
log_{10}L_{32}^{\gamma}=(0.45_{-0.17}^{+0.50})+(0.88\pm0.07)log_{10}\dot{E}_{34}
\end{equation}
yields an high value of $\chi^2_{red}=4.2$.\\
Inspection of the distribution of residuals lead us to try a double-linear relationship, 
which yields:
\begin{eqnarray}
log_{10}L_{32}^{\gamma}=(2.45\pm0.76)+(0.20_{-0.31}^{+0.27})log_{10}\dot{E}_{34} & , & \dot{E}>E_{crit}\\
log_{10}L_{32}^{\gamma}=(0.52\pm0.18)+(1.43_{-0.23}^{+0.31})log_{10}\dot{E}_{34} & , & \dot{E}<E_{crit}
\end{eqnarray}
with $E_{crit}=3.72_{-3.44}^{+3.55}\times10^{35}erg/s$ and $\chi^2_{red}=2.2$. An f-test shows
that the probability for a chance $\chi^2$ improvement is 0.00011.
Such a result is in agreement with the data reported in Abdo et al. (2009c) for the entire dataset
of {\it Fermi} $\gamma$-ray pulsars.
Indeed, the $\chi^2_{red}$ obtained for the double linear fit is better than that obtained for the 
$L_X$-$\dot{E}$ relationship. We obtain $W^2=0.344$ for the double linear $L_{\gamma}-\dot{E}$ relationship. 
Both the $\chi^2_{red}$ and $W^2$ are in agreement with a little higher scatter in the $L_X-\dot{E}$ graph.
A difference between the X-ray and $\gamma$-ray emission geometries - that translates in different values of 
f$_{\gamma}$ and f$_X$ - could explain such a behaviour.

The existence of an $\dot{E}_{crit}$ has been posited from the theoretical point for different
pulsar emission models.
Revisiting the outer-gap model for pulsars with $\tau<10^7$ yrs and assuming initial conditions as well as
pulsars' birth rates, Zhang et al. (2004) found a sharp boundary, due to the saturation of the gap size, for $L_{\gamma}=\dot{E}$.
They obtain the following distribution of pulsars' $\gamma$-ray luminosities:
\begin{eqnarray}
log_{10}L_{\gamma}=log_{10}\dot{E}+const. & , &  \dot{E}<\dot{E}_{crit}\\
log_{10}L_{\gamma}\sim0.30log_{10}\dot{E}+const. & , & \dot{E}>\dot{E}_{crit}
\end{eqnarray}
By assuming the fractional gap size from Zhang\&Cheng(1997), they obtain $\dot{E}_{crit}=1.5\times10^{34}P^{1/3}erg/s$.
While Equation 4 is similar to our double linear fit (Equation 3), the $\dot{E}_{crit}$ they obtain
seems to be lower than our best fit value.\\
On the other hand, in slot-gap models (Muslimov\&Harding 2003), the break occurs
at about $10^{35}erg/s$, when the gap is limited by screening of the acceleration field by pairs.\\
We can see from Figure \ref{art-fig-2} that radio-quiet pulsars have higher luminosities than the radio-loud ones, for similar values of
$\dot{E}$. As in the $L_X-\dot{E}$ fit, we can't however discriminate between the two population due to the big errors stemming from distance estimate.

\subsection{Study of the $\gamma$-to-X-Ray Luminosity Ratio}

At variance with the X-ray and gamma-ray luminosities,
the ratio between the X-ray and gamma-ray luminosities is
independent from pulsars' distances. This makes it possible
to significatively reduce the error bars leading to more precise
indications on the pulsars' emission mechanisms.

Figure \ref{art-fig-3} reports the histogram of the F$_{\gamma}$/F$_X$ values using
only type-2 (high-quality X-ray data) pulsars. The radio-loud
pulsars have $<F_{\gamma}/F_X>\sim800$ while the radio-quiet population
has $<F_{\gamma}/F_X>\sim4800$. Applying the Kolmogorov-Smirnov
(K-S) test to type-2 pulsars' F$_{\gamma}$/F$_X$ values, we obtained that the
probability for the two data sets to belong to the same population
is 0.0016. By using all the pulsars with a confirmed X-ray
counterpart (i.e., including also type-1 objects), this probability
increases to 0.00757. We can conclude, with a 3$\sigma$ confidence
level, that the radio-quiet and radio-loud data sets we used are
somewhat different.

{\bf A Distance-independent Spread in F$_{\gamma}$/F$_X$}
\label{3p3}

\begin{figure}
\centering
\includegraphics[angle=0,scale=.40]{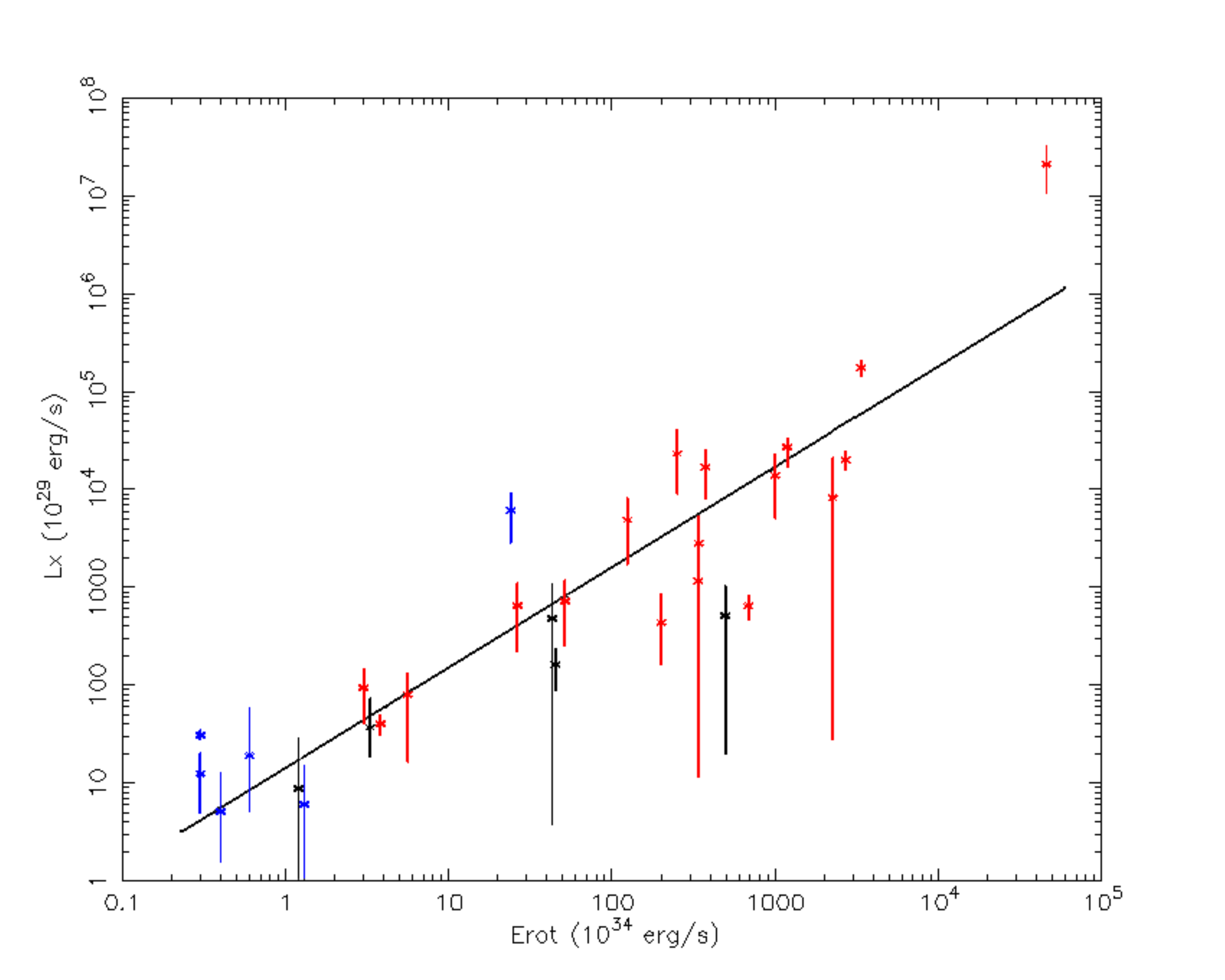}
\caption{$\dot{E}$-L$_X$ diagram for all pulsars classified as type 2 and with a clear distance estimation, assuming f$_X$=1 (see Equation 5). Black: radio-quiet pulsars; red: radio-loud pulsars; blue: millisecond pulsars. The linear best fit of the logs of the two quantities is shown. \label{art-fig-1}}
\end{figure}

\begin{figure}
\centering
\includegraphics[angle=0,scale=.40]{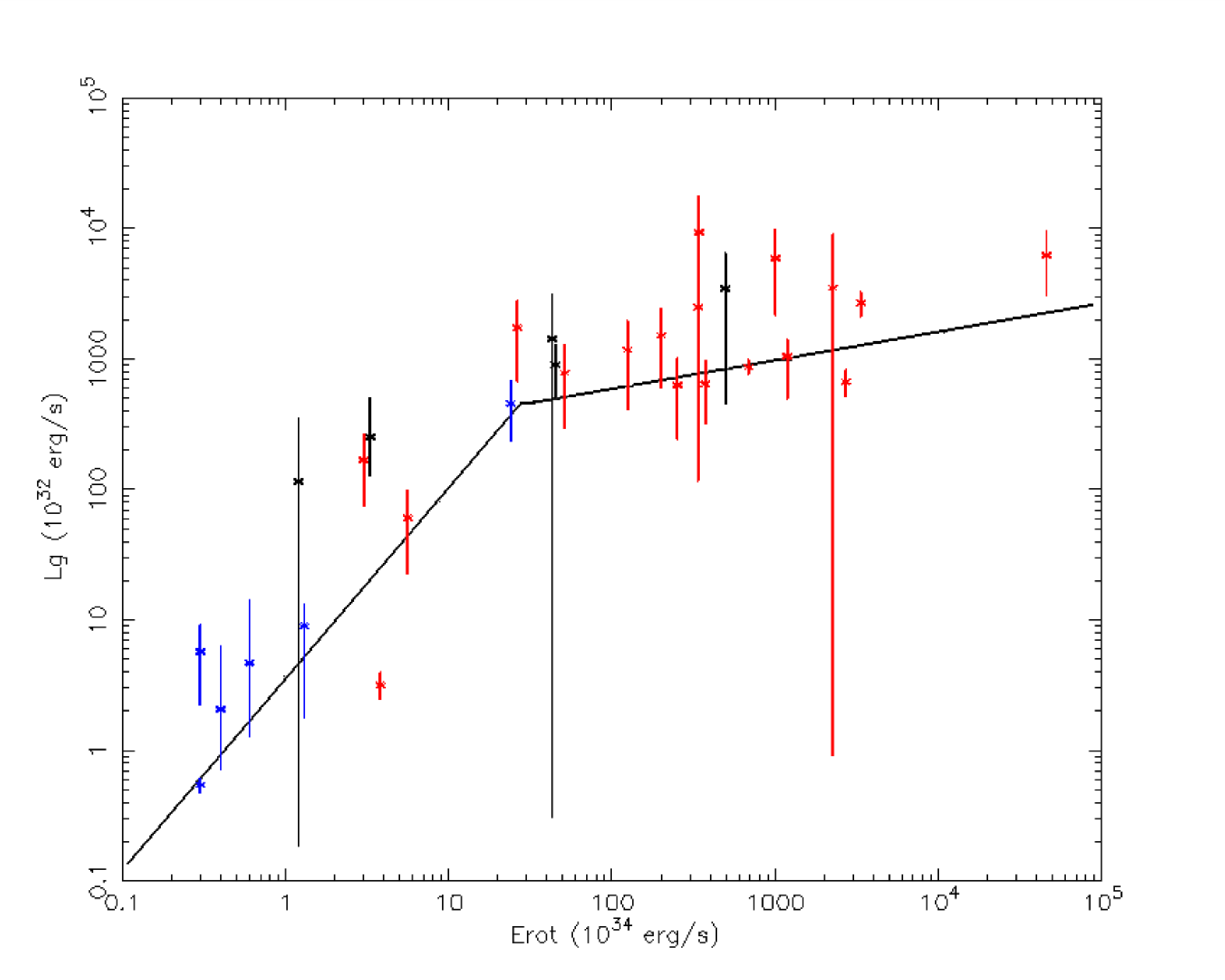}
\caption{$\dot{E}$-L$_{\gamma}$ diagram for all pulsars classified as type 2 and with a clear distance estimation, assuming f$_{\gamma}$=1 (see Equation 5). Black: radio-quiet pulsars; red: radio-loud pulsars; blue: millisecond pulsars. The double linear best fit of the logs of the two quantities is shown. \label{art-fig-2}}
\end{figure}

Figure \ref{art-fig-4} shows F$_{\gamma}$/F$_X$ as a function of $\dot{E}$ for our entire
sample of $\gamma$-ray-emitting neutron stars while in Figure \ref{art-fig-5}
only the pulsars with $"$high-quality$"$ X-ray data have been
selected. Even neglecting the upper and lower limits (shown
as triangles) as well as the low-quality points (see Figure \ref{art-fig-5}),
one immediately notices the scatter on the F$_{\gamma}$/F$_X$ parameter
values for a given value of $\dot{E}$. Such an apparent spread cannot
obviously be ascribed to a low statistic. An inspection of
Figure\ref{art-fig-4}  makes it clear that a linear fit cannot satisfactorily
describe the data. In a sense, this finding should not come
as a surprise since Figure \ref{art-fig-4} is a combination of Figures \ref{art-fig-1}
and \ref{art-fig-2}, and we have seen that Figure \ref{art-fig-2} requires a double-linear
fit. However, combining the results of our previous fits
we obtain the dashed line in Figure \ref{art-fig-4},
clearly a very poor description of the data. For $\dot{E}\sim<5\times10^{36}$
the F$_{\gamma}$/F$_X$ values scatter around a mean value of $\sim$1000 with
a spread of a factor of about 100. For higher $\dot{E}$ the values of
F$_{\gamma}$/F$_X$ seem to decrease drastically to an average value of $\sim$50,
reaching the Crab with F$_{\gamma}$/F$_X$ $\sim$ 0.1.

The spread in the F$_{\gamma}$/F$_X$ values for pulsars with similar $\dot{E}$ is
obviously unrelated to distance uncertainties. Such a scatter can
be due to geometrical effects. For both X-ray and $\gamma$-ray energy
bands,
\begin{equation}
L_{\gamma,X}=4\pi f_{\gamma,X}F_{obs}D^2
\end{equation}
where f$_X$ and f$_{\gamma}$ account for the X and $\gamma$ beaming geometries (which may or may not be related).
If the pulse profile observed along the line-of-sight at $\zeta$ (where $\zeta_E$ is the Earth line-of-sight) for a pulsar with
magnetic inclination $\alpha$ is $F(\alpha,\zeta,\phi)$, where $\phi$ is the pulse phase, than we can write:
\begin{equation}
f=f(\alpha,\zeta_E)=\frac{\int\int F(\alpha,\zeta,\phi)sin(\zeta)d\zeta d\phi}{2\int F(\alpha,\zeta_E,\phi)d\phi}
\end{equation}
where f depends only on the viewing angle and the magnetic
inclination of the pulsar. With a high value of this correction
coefficient, the emission is disfavored. Obviously F$_{\gamma}$/F$_X$ =
L$_{\gamma}$/L$_X$ $\times$ f$_X$/f$_{\gamma}$. Different f$_X$/f$_{\gamma}$ values for different pulsars
can explain the scattering seen in the F$_{\gamma}$/F$_X$ - $\dot{E}$ relationship.

Watters et al. (2009) assume a nearly uniform emission
efficiency while Zhang et al. (2004) compute a significant
variation in the emission efficiency as a function of the geometry
of pulsars. In both cases, geometry plays an important role
through magnetic field inclination as well as through viewing
angle.

The very important scatter found for F$_{\gamma}$/F$_X$ values is obviously
due to the different geometrical configurations which
determine the emission at different wavelength of each pulsar.
While geometry is clearly playing an equally important role in
determining pulsar luminosities, the F$_{\gamma}$/F$_X$ plot makes its effect
easier to appreciate.

The dashed line in Figures \ref{art-fig-4} and \ref{art-fig-5} is the combination of the
best fits of L$_{\gamma}$ - $\dot{E}$ and L$_X$ - $\dot{E}$ relationship, considering f$_{\gamma}$ = 1
and f$_X$ = 1 so that it represents the hypothetical value of F$_{\gamma}$/F$_X$
that each pulsar would have if f$_{\gamma}$ = f$_X$: all the pulsars with a
value of F$_{\gamma}$/F$_X$ below the line have f$_X$ $<$ f$_{\gamma}$. We have seen in
Section ~\ref{3p3} that the radio-quiet data set shows a higher mean
value of F$_{\gamma}$/F$_X$. This is clearly visible in Figure \ref{art-fig-5} where all
the radio-quiet points are above the expected values (dashed
line), so that all the radio-quiet pulsars should have f$_X$ $>$ f$_{\gamma}$.
Moreover, the radio-quiet data set shows a lower scatter with
respect to the radio-loud one pointing to more uniform values of
f$_{\gamma}$/f$_X$ for the radio-quiet pulsars. A similar viewing angle or a
similar magnetic inclination for all the radio-quiet pulsars could
explain such a behaviour.

Figure \ref{art-fig-6} shows the F$_{\gamma}$/F$_X$ behaviour as a function of the
characteristic pulsar age. In view of the uncertainty of this
parameter, we have also built a similar plot using $"$real$"$ pulsar
age, as derived from the associated SNRs (see Figure \ref{art-fig-6}).
Similarly to the $\dot{E}$ relationship, for $\tau$ $<$ 104 yr, F$_{\gamma}$/F$_X$ values
increase with age (both the characteristic and real ones), while
for $\tau$ $>$ 104 yr the behaviour becomes more complex.

\subsection{Study of the Selection Effects}

There are two main selections we have done in order to
obtain our sample of pulsars with both good $\gamma$-ray and X-ray
spectra (type 2). First, the two populations of radio-quiet and
radio-loud pulsars are unveiled with different techniques: using
the same data set, pulsars with known rotational ephemerides
have a detection threshold lower than pulsars found through
blind period searches. In the First {\it Fermi}/LAT pulsar catalogue
(Abdo et al. 2010a), the faintest gamma-ray-selected pulsar has
a flux $\sim$ 3 $\times$ higher than the faintest radio-selected one. Second,
we chose only pulsars with a good X-ray coverage. Such a
coverage depends on many factors (including the policy of
X-ray observatories) that cannot be modeled.

Our aim is to understand if these two selections influenced in
different ways the two populations of pulsars we are studying:
if this was the case, the results obtained would have been distorted.
The $\gamma$-ray selection is discussed at length in the {\it Fermi}/
LAT pulsar catalogue (Abdo et al. 2010a). Since the radio-quiet
population obviously has a detection threshold higher than the
radio-loud, we could avoid such bias by selecting all the pulsars
with a flux higher than the radio-quiet detection threshold (6 $\times$
10$^{-8}$ photon cm$^{-2}$ s$^{-1}$). Only five radio-loud type-2 pulsars
are excluded (J0437-4715, J0613+1036, J0751+1807,
J2043+2740, and J2124-3358) with F$_{\gamma}$/F$_X$ values ranging
from 87 to 1464. We performed our analysis on
such a reduced sample and the results do not change
significantly.

We can, therefore, exclude the presence of an important bias
due to the $\gamma$-ray selection on type-2 pulsars.

\begin{figure}
\centering
\includegraphics[angle=0,scale=.40]{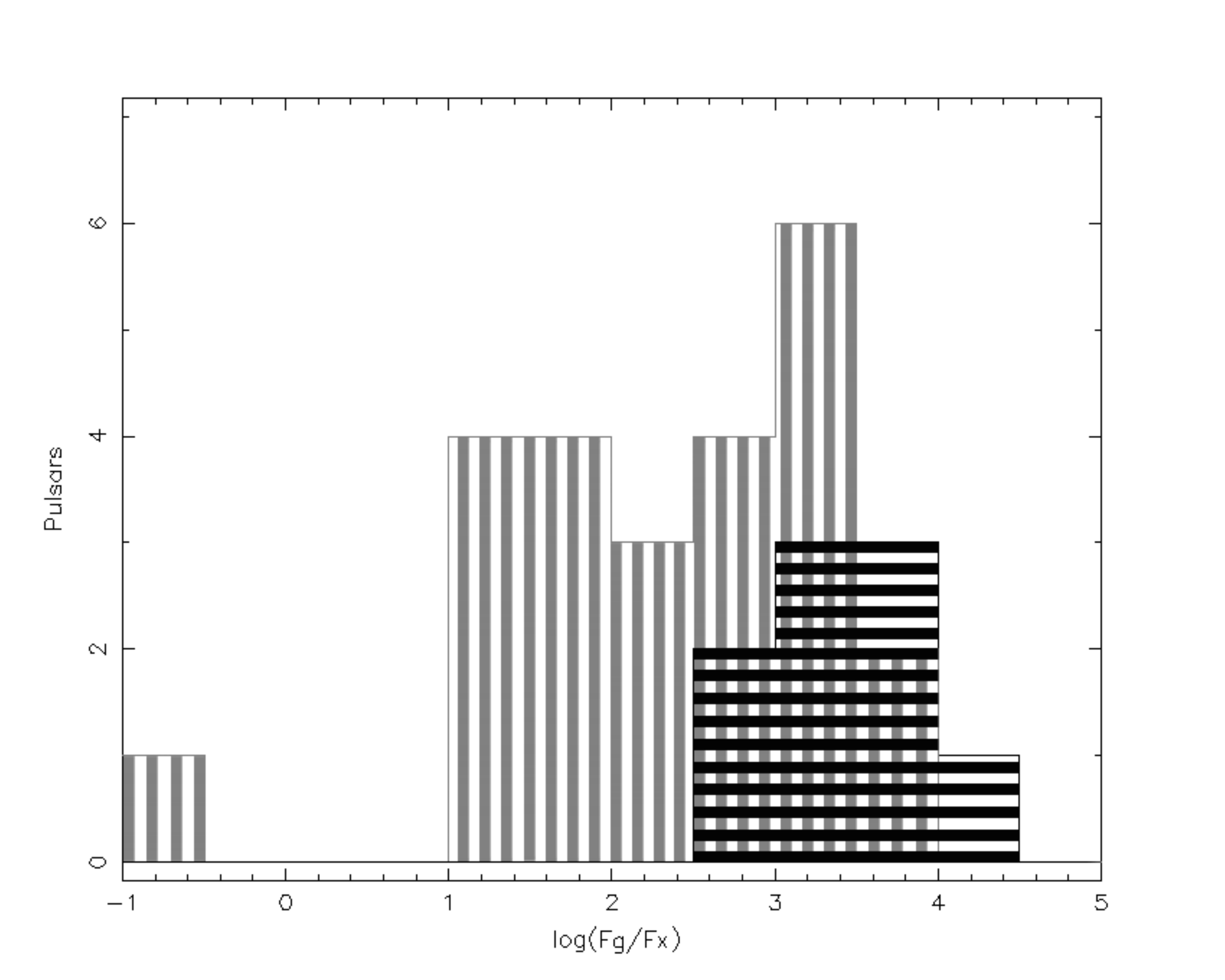}
\caption{log(F$_{\gamma}$/F$_X$) histogram. The step is 0.5; the radio-loud (and millisecond) pulsars are indicated in grey and the radio-quiet ones in black. Only high confidence pulsars (type 2) have been used for a total of 24 radio-loud and 9 radio-quiet pulsars. \label{art-fig-3}}
\end{figure}

\begin{figure}
\centering
\includegraphics[angle=0,scale=.40]{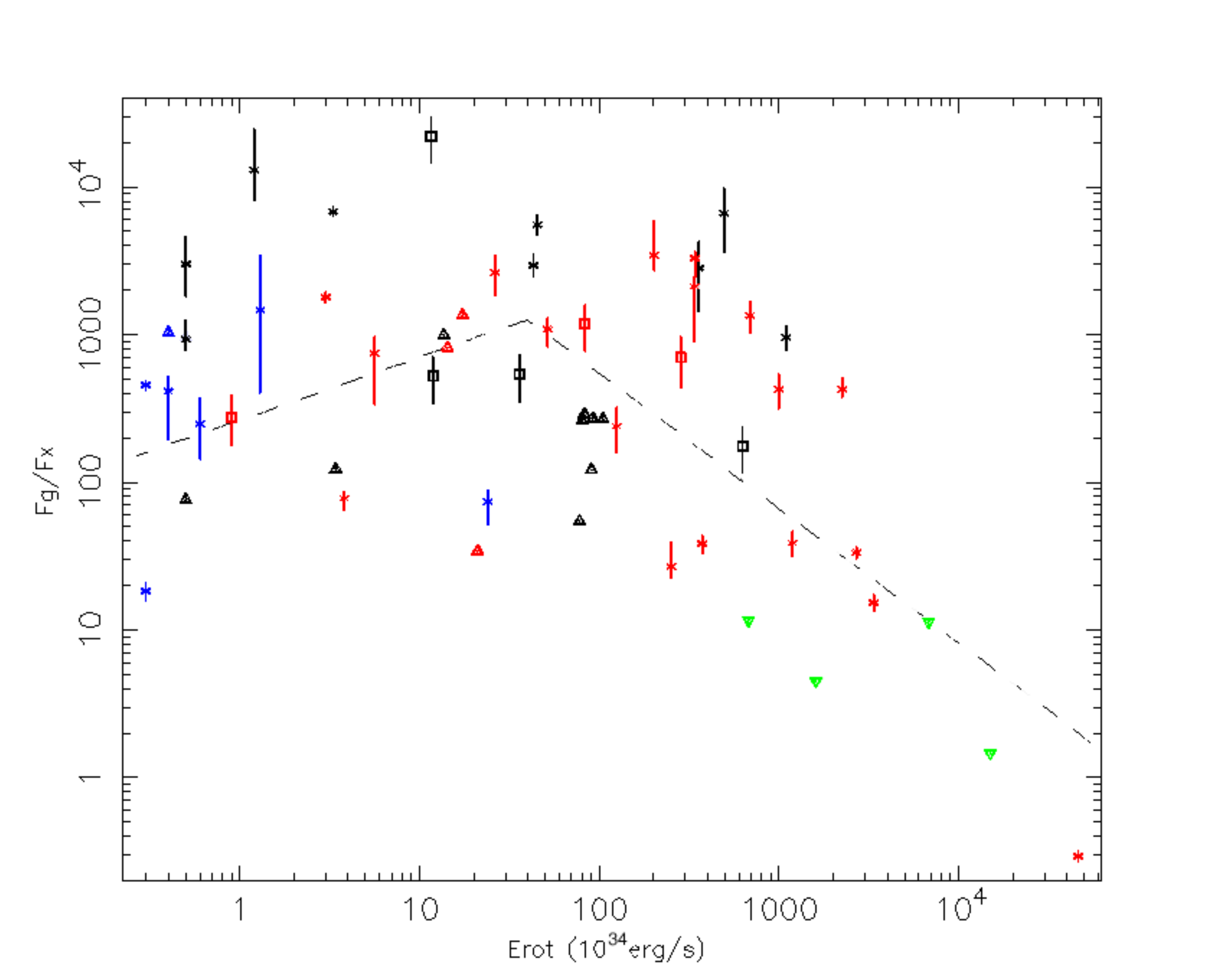}
\caption{$\dot{E}$-F$_{\gamma}$/F$_X$ diagram. Green: IBIS pulsars; black: radio-quiet pulsars; red: radio-loud pulsars; blue: millisecond pulsars. The triangles are upper and lower limits, the squares indicate pulsars with a type 1 X-ray spectrum (see Table \ref{art-tab-2}) and the stars pulsars with a high quality X-ray spectrum The dotted line is the combination of the best fitting functions obtained for Figure \ref{art-fig-1} and \ref{art-fig-2} with the geometrical correction factor set to 1 for both the X and $\gamma$-ray bands. \label{art-fig-4}}
\end{figure}

\begin{figure}
\centering
\includegraphics[angle=0,scale=.20]{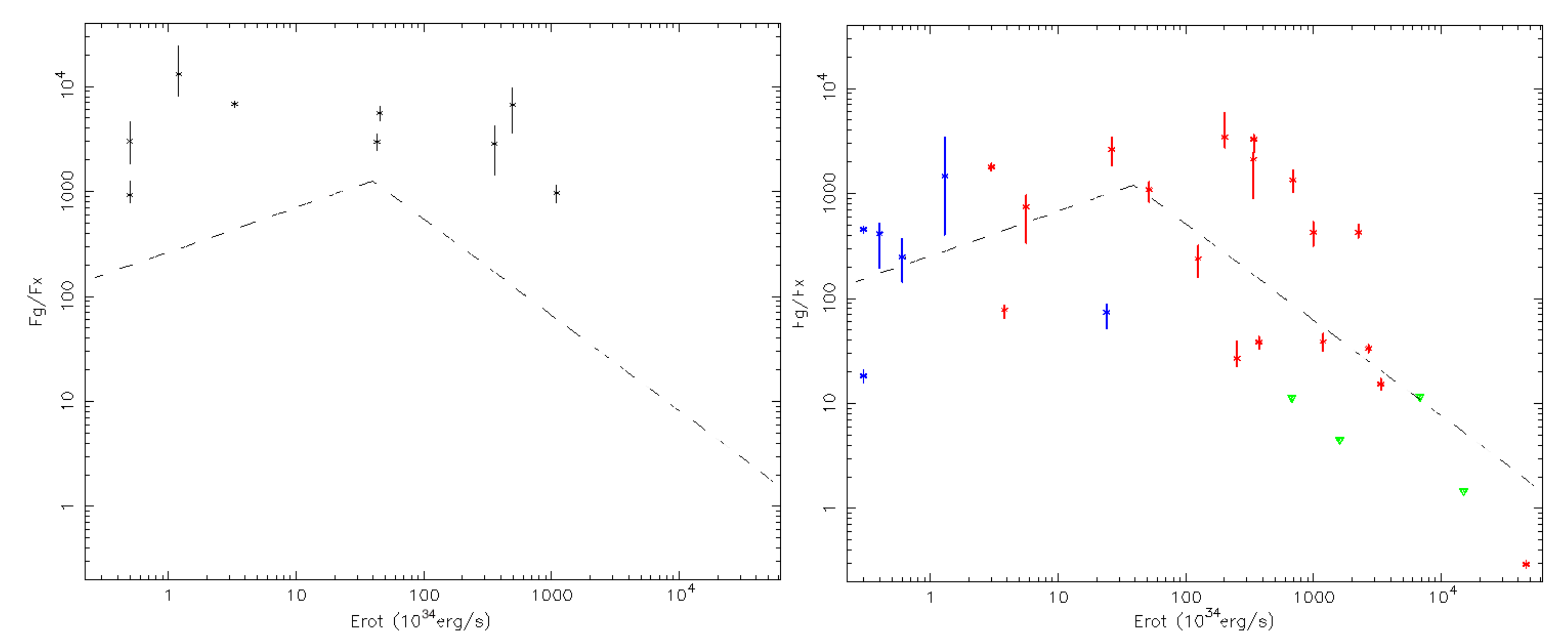}
\caption{$\dot{E}$-F$_{\gamma}$/F$_X$ diagram for high confidence pulsars only (type 2). Green: IBIS pulsars; black: radio-quiet pulsars; red: radio-loud pulsars; blue: millisecond pulsars. The triangles are upper limits. \label{art-fig-5}}
\end{figure}

\begin{figure}
\centering
\includegraphics[angle=0,scale=.20]{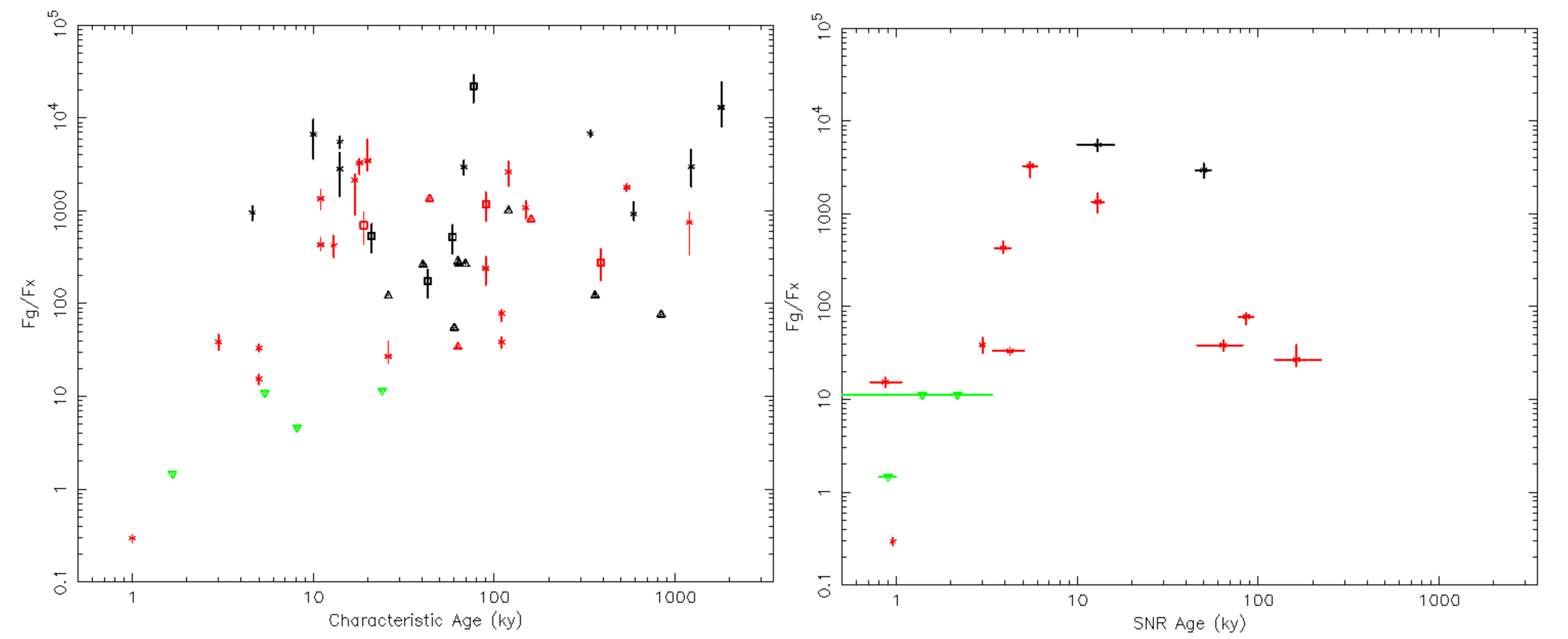}
\caption{Left: Characteristic Age-F$_{\gamma}$/F$_X$ diagram. Right: SNR Age-F$_{\gamma}$/F$_X$ diagram. Green: IBIS pulsars; black: radio-quiet pulsars; red: radio-loud pulsars. Triangles are upper limits, squares are pulsars with a type 1 X-ray spectrum while stars are pulsars with a type 2 X-ray spectrum. \label{art-fig-6}}
\end{figure}

In order to roughly evaluate the selection affecting the
X-ray observations, we used the method developed by Schmidt
(1968) to compare the current radio-quiet and radio-loud samples'
spatial distributions, following the method also used in
Abdo et al. (2010a). For each object with an available distance
estimate, we computed the maximum distance still allowing
detection from $D_{max}=D_{est}(F_{\gamma}/F_{min})^{1/2}$, where $D_{est}$ comes
from Table \ref{art-tab-1}, the photon flux and F$_{min}$ are taken from Abdo
et al. (2010a) and Saz Parkinson et al. (2010). We limited D$_{max}$
to 15 kpc, and compared V, the volume enclosed within the
estimated source distance, to that enclosed within the maximum
distance, V$_{max}$, for a galactic disk with radius 10 kpc and
thickness 1 kpc (as in Abdo et al. 2010a). The inferred values
of $<V/V_{max}>$ are 0.462, 0.424, 0.443, and 0.516 for the entire
gamma-ray pulsars' data set, the radio-quiet pulsars, millisecond
pulsars, and the radio-loud pulsars. These are quite close
to the expected value of 0.5 even if $<V/V_{max}>^{rq}$ is lower than
$<V/V_{max}>^{rl}$. If we use only type-2 pulsars we obtain 0.395,
0.335, 0.462, and 0.419. These lower values of $<V/V_{max}>$ indicate
that we have a good X-ray coverage only for close-by -
or very bright - pulsars, not a surprising result. By using the
X-ray-counterpart data set, both the radio-loud and radio-quiet
$<V/V_{max}>$ values appear lower of about 0.1: this seems to indicate
that we used the same selection criteria for the two populations
and we minimized the selection effects in the histogram
of Figure \ref{art-fig-3}.

We can conclude that the $\gamma$-ray selection introduced no
changes in the two populations, while the X-ray selection
excluded objects both faint and/or far away; any distortion,
if present, is not overwhelming.

Only if deep future X-ray observations centered on radio-quiet
{\it Fermi} pulsars fail to unveil lower values of F$_{\gamma}$/F$_X$, it will
be possible to be sure that radio-quiet pulsars have a different
geometry (or a different emission mechanism) than radio-loud
ones.

\section{CONCLUSIONS}

The discovery of a number of radio-quiet pulsars comparable
to that of radio-loud ones together with the study of their X-ray
counterparts made it possible, for the first time, to address their
behaviour using a distance-independent parameter such as the
ratio of their fluxes at X-ray and $\gamma$-ray wavelengths.

First, we reproduced the well-known relationship between
the neutron star luminosities and their rotational energy losses.
Next, selecting only the {\it Fermi} pulsars with good X-ray data, we
computed the ratio between the $\gamma$-ray and X-ray fluxes
and studied its dependence on the overall rotational energy loss
as well as on the neutron star age.

Much to our surprise, the distance-independent F$_{\gamma}$/F$_X$ values
computed for pulsars of similar age and energetic differ by
up to three orders of magnitude, pointing to important (yet
poorly understood) differences both in position and height of
the regions emitting at X-ray and $\gamma$-ray wavelengths within the
pulsars magnetospheres. Selection effects cannot account for
the spread in the F$_{\gamma}$/F$_X$ relationship and any further distortion,
if present, is not overwhelming.

In spite of the highly scattered values, a decreasing trend is
seen when considering young and energetic pulsars. Moreover,
radio-quiet pulsars are characterized by higher values of the
F$_{\gamma}$/F$_X$ parameter ($<F_{\gamma}/F_X>_{rl}\sim800$ and $<F_{\gamma}/F_X>_{rq}\sim4800$),
so that a K-S test points to a probability of 0.0016 for them to
belong to the same population as the radio-loud ones. While
it would be hard to believe that radio-loud and radio-quiet
pulsars belong to two different neutron star populations, the
K-S test probably points to different geometrical configurations
(possibly coupled with viewing angles) that characterize radio-loud
and radio-quiet pulsars. Indeed the radio-quiet population
we analyzed is less scattered than the radio-loud one, pointing
to a more uniform viewing or magnetic geometry of radio-quiet
pulsars.

Our work is just a starting point, based on the first harvest of
gamma-ray pulsars. The observational panorama will quickly
evolve. The gamma-ray pulsar list will certainly grow and
this will trigger more X-ray observations, improving both in
quantity and quality the database of the neutron stars detected in
X-rays and $\gamma$-rays to be used to compute our multiwavelength,
distance-independent parameter. However, to fully exploit the
information packed in F$_{\gamma}$/F$_X$, a complete three-dimensional
modeling of the pulsar magnetosphere is needed to account for
the different locations and heights of the emitting regions at
work at different energies. Such modeling could provide the
clue to account for the spread we have observed for the ratios
between $\gamma$-ray and X-ray fluxes as well as for the systematically
higher values measured for radio-quiet pulsars.

\setlength{\LTleft}{-1pt}
\begin{footnotesize}
\begin{landscape}
\begin{center}
\begin{longtable}{cccccccccc}
PSR Name & P$^a$ & $\dot{P}^a$ & $\tau_c$ & $\tau_{snr}^b$ & B$_{lc}$ & d$^a$ & $\dot{E}$ & Type$^e$ & PWN$^f$ \\
 & (ms) & $(10^{-15})$ & (ky) & (ky) & (kG) & (kpc) & ($10^{34}$erg/s) & & \\
\endhead
J0007+7303 & 316 & 361 & 14 & 13 & 3.1 & 1.4$\pm$0.3 & 45.2 & g & Y\\
J0030+0451 & 4.9 & $10^{-5}$ & 7.7$\times10^{6}$ & - & 17.8 & 0.30$\pm$0.09 & 0.3 & m & N\\
J0205+6449 & 65.7 & 194 & 5 & 4.25$\pm$0.85 & 115.9 & 2.9$\pm$0.3 & 2700 & r & Y\\
J0218+4232 & 2.3 & 7.7$\times10^{-5}$ & 5$\times10^{5}$ & - & 313.1 & 3.25$\pm$0.75 & 24 & m & N\\
J0248+6021 & 217 & 55.1 & 63 & - & 3.1 & 5.5$\pm3.5$ & 21 & r & ?\\
J0357+32  & 444 & 12 & 590 & - & 0.2 & 0.5$^c$ & 0.5 & g & Y\\
J0437-4715 & 5.8 & 1.4$\times10^{-5}$ & 6.6$\times10^{6}$ & - & 13.7 & 0.1563$\pm$0.0013 & 0.3 & m & N\\
J0534+2200 & 33.1 & 423 & 1.0 & 0.955 & 950 & 2.0$\pm$0.5 & 46100 & r & Y\\
J0540-6919 & 50.5 & 480 & 1.67 & 0.9$\pm$0.1 & 364 & 50$^d$ & 15000 & i & Y\\
J0613-0200 & 3.1 & 9$\times10^{-6}$ & 5.3$\times10^{6}$ & - & 54.3 & 0.48$^{+0.19}_{-0.11}$ & 1.3 & m & N\\
J0631+1036 & 288 & 105 & 44 & - & 2.1 & 2.185$\pm$1.440 & 17.3 & r & ?\\
J0633+0632 & 297 & 79.5 & 59 & - & 1.7 & 1.1$^c$ & 11.9 & g & ?\\
J0633+1746 & 237 & 11 & 340 & - & 1.1 & 0.250$_{-0.062}^{+0.12}$ & 3.3 & g & N\\
J0659+1414 & 385 & 55 & 110 & 86$\pm$8 & 0.7 & 0.288$_{-0.027}^{+0.033}$ & 3.8 & r & N\\
J0742-2822 & 167 & 16.8 & 160 & - & 3.3 & $2.07_{-1.07}^{+1.38}$ & 14.3 & r & ?\\
J0751+1807 & 3.5 & 6.2$\times10^{-6}$ & 8$\times10^{6}$ & - & 32.3 & 0.6$_{-0.2}^{+0.6}$ & 0.6 & m & N\\
J0835-4510 & 89.3 & 124 & 11 & 13$\pm$1 & 43.4 & 0.287$_{-0.017}^{+0.019}$ & 688 & r & Y\\
J1023-5746 & 111 & 384 & 4.6 & - & 44 & 2.4$^c$ & 1095 & g & ?\\
J1028-5819 & 91.4 & 16.1 & 90 & - & 14.6 & 2.33$\pm$0.70 & 83.2 & r & Y\\
J1044-5737 & 139 & 54.6 & 40.3 & - & 9.5 & 1.5$^c$ & 80.3 & g & ?\\
J1048-5832 & 124 & 96.3 & 20 & - & 16.8 & 2.71$\pm$0.81 & 201 & r & Y\\
J1057-5226 & 197 & 5.8 & 540 & - & 1.3 & 0.72$\pm$0.20 & 3.0 & r & N\\
J1124-5916 & 135 & 747 & 3 & 2.99$\pm$0.06 & 37.3 & 4.8$_{-1.2}^{+0.7}$ & 1190 & r & Y\\
J1413-6205 & 110 & 27.7 & 62.9 & - & 12.3 & 1.4$^c$ & 82.7 & g & ?\\
J1418-6058 & 111 & 170 & 10 & - & 29.4 & 3.5$\pm$1.5 & 495 & g & Y\\
J1420-6048 & 68.2 & 83.2 & 13 & - & 69.1 & 5.6$\pm$1.7 & 1000 & r & N\\
J1429-5911 & 116 & 30.5 & 60.2 & - & 11.3 & 1.6$^c$ & 77.5 & g & ?\\
J1459-60 & 103 & 25.5 & 64 & - & 13.6 & 1.5$^c$ & 91.9 & g & ?\\
J1509-5850 & 88.9 & 9.2 & 150 & - & 11.8 & 2.6$\pm$0.8 & 51.5 & r & N\\
J1614-2230 & 3.2 & 4$\times10^{-6}$ &  1.2$\times10^{6}$ & - & 33.7 & 1.27$\pm$0.39 & 0.5 & m & N\\
J1617-5055 & 69 & 135 & 8.13 & - & 86.6 & 6.5$\pm$0.4$^d$ & 1600 & i & Y\\
J1709-4429 & 102 & 93 & 18 & 5.5$\pm$0.5 & 26.4 & 2.5$\pm$1.1 & 341 & r & Y\\
J1718-3825 & 74.7 & 13.2 & 90 & - & 21.9 & 3.82$\pm$1.15 & 125 & r & Y\\
J1732-31 & 197 & 26.1 & 120 & - & 2.7 & 0.6$^c$ & 13.6 & g & ?\\
J1741-2054 & 414 & 16.9 & 390 & - & 0.3 & 0.38$\pm$0.11 & 0.9 & r & ?\\
J1744-1134 & 4.1 & 7$\times10^{-6}$ & 9$\times10^{6}$ & - & 24 & 0.357$_{-0.035}^{+0.043}$ & 0.4 & m & N\\
J1747-2958 & 98.8 & 61.3 & 26 & 163$_{-39}^{+60}$ & 23.5 & 2.0$\pm$0.6 & 251 & r & Y\\
J1809-2332 & 147 & 34.4 & 68 & 50$\pm$5 & 6.5 & 1.7$\pm$1.0 & 43 & g & Y\\
J1811-1926 & 62 & 41 & 24 & 2.18$\pm$1.22 & 64 & 7$\pm$2$^d$ & 678 & i & Y\\ 
J1813-1246 & 48.1 & 17.6 & 43 & - & 76.2 & 2.0$^c$ & 626 & g & ?\\
J1813-1749 & 44.7 & 150 & 5.4 & 1.3925$\pm$1.1075 & 272 & 4.70$\pm$0.47$^d$ & 680 & i & Y\\
J1826-1256 & 110 & 121 & 14 & - & 25.2 & 1.2$^c$ & 358 & g & Y\\
J1833-1034 & 61.9 & 202 & 5 & 0.87$_{-0.15}^{+0.20}$ & 137.3 & 4.7$\pm$0.4 & 3370 & r & Y\\
J1836+5925 & 173 & 1.5 & 1800 & - & 0.9 & 0.4$_{-0.15}^{+0.4 d}$ & 1.2 & g & N\\
J1846+0919 & 226 & 9.93 & 360 & - & 1.2 & 1.2$^c$ & 3.4 & g & ?\\
J1907+06 & 107 & 87.3 & 19 & - & 23.2 & 1.3$^c$ & 284 & r & ?\\
J1952+3252 & 39.5 & 5.8 & 110 & 64.0$\pm$18 & 71.6 & 2.0$\pm$0.5 & 374 & r & Y\\
J1954+2836 & 92.7 & 21.2 & 69.5 & - & 16.4 & 1.7$^c$ & 105 & g & ?\\
J1957+5036 & 375 & 7.08 & 838 & - & 0.3 & 0.9$^c$ & 0.5 & g & ?\\
J1958+2841 & 290 & 222 & 21 & - & 3.0 & 1.4$^c$ & 35.8 & g & ?\\
J2021+3651 & 104 & 95.6 & 17 & - & 26 & 2.1$_{-1.0}^{+2.1}$ & 338 & r & Y\\
J2021+4026 & 265 & 54.8 & 77 & - & 1.9 & 1.5$\pm$0.45 & 11.6 & g & ?\\
J2032+4127 & 143 & 19.6 & 120 & - & 5.3 & 3.60$\pm$1.08 & 26.3 & r & ?\\
J2043+2740 & 96.1 & 1.3 & 1200 & - & 3.6 & 1.80$\pm$0.54 & 5.6 & r & N\\
J2055+25 & 320 & 4.08 & 1227 & - & 0.3 & 0.4$^c$ & 0.5 & g & ?\\
J2124-3358 & 4.9 & 1.2$\times10^{-5}$ & 6$\times10^{5}$ & - & 18.8 & 0.25$_{-0.08}^{+0.25}$ & 0.4 & m & N\\
J2229+6114 & 51.6 & 78.3 & 11 & 3.90$\pm$0.39 & 134.5 & 3.65$\pm$2.85 & 2250 & r & Y\\
J2238+59 & 163 & 98.6 & 26 & - & 8.6 & 2.1$^c$ & 90.3 & g & ?\\
\caption{\\
{\bf General Characteristics of {\it Fermi} pulsars.}\\
a : P, $\dot{P}$ and most of the values of the distance are taken from Abdo et al. (2009c), Saz Parkinson et al. (2010).\\
b : Age derived from the associated SNR. Respectively taken from Slane et al.(2004), Gotthelf et al.(2007), Rudie et al.(2008), Hwang et al.(2001), Thorsett et al.(2003), Gorenstein et al.(1974), Winkler et al.(2009), Bock\&Gvaramadze(2002), Hales et al.(2009), Roberts\&Brogan(2008), Tam\&Roberts(2008), Brogan et al.(2005), Bietenholz\&Bartel(2008), Migliazzo et al.(2008), Kothes et al.(2006).\\
c : These distances are taken from Saz Parkinson et al. (2010) and are obtained under the assumption of a beam correction factor f$_{\gamma}$ = 1 for the gamma-ray emission cone of all pulsars. In this way one obtains:\\
$d=0.51\dot{E_{34}}^{1/4}/F_{\gamma,10}^{1/2}$ kpc\\
where $\dot{E}=\dot{E}_{34}\times10^{34}erg/s$ and $F_{\gamma}=F_{\gamma,10}\times10^{-10}erg/cm^2s$.
See also Saz Parkinson et al. (2010).\\
d : Respectively taken from Campana et al.(2008), Kaspi et al.(1998), Kaspi et al.(2001), Gotthelf\&Halpern(2009), Halpern et al.(2007).\\
e : g = radio-quiet pulsars ; r = radio-loud pulsars ; m = millisecond pulsars ; i = pulsars detected by INTEGRAL/IBIS but not yet by {\it Fermi} (see Bird et al. 2009).\\
f : Only bright PWNs have been considered (with F$^{pwn}_x$ $>$ 1/5 F$^{psr}_x$). The presence or the absence of a bright PWN has been valued by re-analyzing the X-ray data (except for the X-ray analyses taken from literature, see table \ref{art-tab-2}).\\
\label{art-tab-1}}
\end{longtable}

\begin{longtable}{cccccccccc}
PSR Name & X$^a$ & Inst$^b$ & F$_X^{nt}$ & F$_X^{tot}$ & N$_H$ & $\gamma_X$ & kT & R$_{BB}$ & Eff$_X^d$\\
 & & & ($10^{-13}erg/cm^{2}s$) & ($10^{-13}erg/cm^{2}s$) & ($10^{20}cm^{-3}$) & & (keV) & (km) & \\
\endhead
J0007+7303 & 2 & X/C & 0.686$\pm$0.100 & 0.841$\pm$0.098 & 16.6$_{-7.6}^{+8.9}$ & 1.30$\pm$0.18 & 0.102$_{-0.018}^{+0.032}$ & 0.64$_{-0.20}^{+0.88}$ & 2.84$\times10^{-5}$\\
J0030+0451 & 2 & X & 1.16$\pm$0.02 & 2.8$\pm$0.1 & 0.244$_{-0.244}^{+7.470}$ & 2.8$_{-0.4}^{+0.5}$ & 0.194$_{-0.021}^{+0.015}$ & 0.6$_{-0.1}^{+0.225}$ & 3.32$\times10^{-4}$\\
J0205+6449 & 2 & C & 19.9$\pm$0.5 & 19.9$\pm$0.5 & 40.2$\pm$0.11 & 1.82$\pm$0.03 & - & - & 5.92$\times10^{-5}$\\
J0218+4232 & 2 & L$^f$ & 4.87$_{-1.28}^{+0.57}$ & 4.87$_{-1.28}^{+0.57}$ & 7.6$\pm$4.3 & 1.19$\pm$0.12 & - & - & 2.05$\times10^{-3}$\\
J0248+6021 & 0 & S & $<$9.00 & $<$9.00 & 80$^c$ & 2 & - & - & -\\
J0357+32 & 2 & C & 0.64$_{-0.06}^{+0.09}$ & 0.64$_{-0.06}^{+0.09}$ & 8.0$\pm$4.0 & 2.53$\pm$0.25 & - & - & -\\
J0437-4715 & 2 & X/C & 10.1$_{-0.6}^{0.8}$ & 14.3$_{-0.7}^{0.9}$ & 4.4 $\pm$ 1.7 & 3.17 $\pm$ 0.13 & 0.228$_{-0.003}^{+0.006}$ & 0.060$_{-0.008}^{+0.009}$ & 8.23$\times10^{-5}$\\
J0534+2200 & 2 & L$^f$ & 44300$\pm$1000 & 44300$\pm$1000 & 34.5$\pm$0.2 & 1.63$\pm$0.09 & - & - & 3.67$\times10^{-3}$\\
J0540-6919 & 2 & L$^f$ & 568$\pm$6 & 568$\pm$6 & 37$\pm$1 & 1.98$\pm$0.02 & - & - & 1.13$\times10^{-1}$\\
J0613-0200 & 2* & X & 0.221$_{-0.158}^{+0.297}$ & 0.221$_{-0.158}^{+0.297}$ & 1$^e$ & 2.7$\pm$0.4 & - & - & 3.74$\times10^{-5}$\\
J0631+1036 & 0 & X & $<$0.225 & $<$0.225 & 20$^c$ & 2 & - & - & -\\
J0633+0632 & 1 & S & 1.53$\pm$0.51 & 1.53$\pm$0.51 & 20$^c$ & 2 & - & - & -\\
J0633+1746 & 2 & L$^f$ & 4.97$_{-0.27}^{+0.09}$ & 12.6$_{-0.7}^{+0.2}$ & 1.07$^e$ & 1.7$\pm$0.1 & 0.190$\pm$0.030 & 0.04$\pm$0.01 & 8.99$\times10^{-5}$\\
J0659+1414 & 2 & L$^f$ & 4.06$_{-0.59}^{+0.03}$ & 168$_{-24}^{+1}$ & 4.3$\pm$0.2 & 2.1$\pm$0.3 & 0.125$\pm$0.003 & 1.80$\pm$0.15 & 8.46$\times10^{-5}$\\
J0742-2822 & 0 & X & $<$0.225 & $<$0.225 & 20$^c$ & 2 & - & - & -\\
J0751+1807 & 2 & L$^f$ & 0.44$_{-0.13}^{+0.18}$ & 0.44$_{-0.13}^{+0.18}$ & 4$^e$ & 1.59$\pm$0.30 & - & - & 2.52$\times10^{-4}$\\
J0835-4510 & 2* & L$^f$ & 65.1$\pm$15.7 & 281$\pm$67 & 2.2$\pm$0.5 & 2.7$\pm$0.6 & 0.129$\pm$0.007 & 2.5$\pm$0.3 & 9.78$\times10^{-6}$\\
J1023-5746 & 2* & C & 1.61$\pm$0.27 & 1.61$\pm$0.27 & 115$_{-41}^{+47}$ & 1.15$_{-0.22}^{+0.24}$ & - & - & -\\
J1028-5819 & 1 & S & 1.5$\pm$0.5 & 1.5$\pm$0.5 & 50$^c$ & 2 & - & - & -\\
J1044-5737 & 0 & S & $<$3.93 & $<$3.93 & 50$^c$ & 2 & - & - & -\\
J1048-5832 & 2* & C+X & 0.50$_{-0.10}^{+0.35}$ & 0.50$_{-0.10}^{+0.35}$ & $90_{-20}^{+40}$ & 2.4$\pm$0.5 & - & - & 1.74$\times10^{-5}$\\
J1057-5226 & 2 & C+X & 1.51$_{-0.13}^{+0.02}$ & 24.5$_{-2.5}^{+0.3}$ & 2.7$\pm$0.2 & 1.7$\pm$0.1 & 0.179$\pm$0.006 & 0.46$\pm$0.06 & 2.49$\times10^{-4}$\\
J1124-5916 & 2 & C & 9.78$_{-1.03}^{+1.18}$ & 10.90$_{-1.26}^{+1.32}$ & 30.0$_{-4.8}^{+2.8}$ & 1.54$_{-0.17}^{+0.09}$ & 0.426$_{-0.018}^{+0.034}$ & 0.274$_{-0.077}^{+0.089}$ & 2.27$\times10^{-4}$\\
J1413-6205 & 0 & S & $<$4.9 & $<$4.9 & 40$^c$ & 2 & - & - & -\\
J1418-6058 & 2 & C+X & 0.353$\pm$0.154 & 0.353$\pm$0.154 & 233$_{-106}^{+134}$ & 1.85$_{-0.56}^{+0.83}$ & - & - & 1.05$\times10^{-5}$\\
J1420-6048 & 2* & X & 1.6$\pm$0.7 & 1.6$\pm$0.7 & 202$_{-106}^{+161}$ & 0.84$_{-0.37}^{+0.55}$ & - & - & 1.11$\times10^{-4}$\\
J1429-5911 & 0 & S & $<$16.9 & $<$16.9 & 80$^c$ & 2 & - & - & -\\
J1459-60 & 0 & S & $<$3.93 & $<$3.93 & 100$^c$ & 2 & - & - & -\\
J1509-5850 & 2 & C+X & 0.891$_{-0.186}^{+0.132}$ & 0.891$_{-0.186}^{+0.132}$ & 80$^e$ & 1.31$\pm$0.15 & - & - & 1.12$\times10^{-4}$\\
J1614-2230 & 0 & C+X & $<$0.286 & 0.286$_{-0.086}^{+0.015}$ & 2.9$_{-2.9}^{+4.3}$ & 2 & 0.236$\pm$0.024 & 0.92$_{-0.35}^{+0.73}$ & -\\
J1617-5055 & 2 & L$^f$ & 64.2$\pm$0.3 & 64.2$\pm$0.3 & 345$\pm$14 & 1.14$\pm$0.06 & - & - & 2.03$\times10^{-3}$\\
J1709-4429 & 2 & C+X & 3.78$_{-0.94}^{+0.37}$ & 9.04$_{-2.25}^{+0.87}$ & 45.6$_{-2.9}^{+4.4}$ & 1.88$\pm$0.21 & 0.166$\pm$0.012 & 4.3$_{-0.86}^{+1.72}$ & 6.62$\times10^{-5}$\\
J1718-3825 & 2 & X & 2.80$\pm$0.67 & 2.80$\pm$0.67 & 70$^e$ & 1.4$\pm$0.2 & - & - & 3.12$\times10^{-4}$\\
J1732-31 & 0 & S & $<$2.42 & $<$2.42 & 50$^c$ & 2 & - & - & -\\
J1741-2054$^{g}$ & 1 & S & 4.64$_{-1.63}^{+1.84}$ & 4.64$_{-1.63}^{+1.84}$ & 0$^e$ & 2.10$_{-0.28}^{+0.50}$ & - & - & 9.93$\times10^{-4}$\\
J1744-1134 & 0 & C & $<$0.272 & 0.272$\pm$0.020 & 12$_{-12}^{+42}$ & 2 & 0.272$_{-0.098}^{+0.094}$ & 0.132$_{-0.120}^{+1.600}$ & -\\
J1747-2958 & 2* & C+X & 48.7$_{-6.0}^{+21.3}$ & 48.7$_{-6.0}^{+21.3}$ & 256$_{-6}^{+9}$ & 1.51$_{-0.44}^{+0.12}$ & - & - & 7.41$\times10^{-4}$\\
J1809-2332 & 2 & C+X & 1.40$_{-0.23}^{+0.25}$ & 3.14$_{-0.53}^{+0.57}$ & 61$_{-8}^{+9}$ & 1.85$_{-0.36}^{+1.89}$ & 0.190$\pm$0.025 & 1.54$_{-0.44}^{+1.26}$ & 8.98$\times10^{-5}$\\
J1811-1926 & 2 & C & 26.6$_{-3.7}^{+2.3}$ & 26.6$_{-3.7}^{+2.3}$ & 175$_{-12}^{+11}$ & 0.91$_{-0.08}^{+0.09}$ & - & - & 1.18$\times10^{-3}$\\
J1813-1246 & 1 & S & 9.675$\pm$3.225 & 9.675$\pm$3.225 & 100$^c$ & 2 & - & - & 1.13$\times10^{-3}$\\
J1813-1749 & 2 & C & 24.4$\pm$11.5 & 24.4$\pm$11.5 & 840$_{-373}^{+433}$ & 1.3$\pm$0.3 & - & - & -\\
J1826-1256 & 2 & C & 1.18$\pm$0.58 & 1.18$\pm$0.58 & 100$^e$ & 0.63$_{-0.63}^{+0.90}$ & - & - & -\\
J1833-1034 & 2 & X+C & 66.3$\pm$2.0 & 66.3$\pm$2.0 & 230$^e$ & 1.51$\pm$0.07 & - & - & 4.15$\times10^{-4}$\\
J1836+5925 & 2 & X+C & 0.459$_{-0.174}^{+0.403}$ & 0.570$_{-0.216}^{+0.500}$ & $<$0.792 & 1.56$_{-0.73}^{+0.51}$ & 0.056$_{-0.009}^{+0.012}$ & 4.47$_{-1.31}^{+3.03}$ & 5.84$\times10^{-5}$\\
J1846+0919 & 0 & S & $<$2.92 & $<$2.92 & 20$^c$ & 2 & - & - & -\\
J1907+06 & 1 & C & 3.93$\pm$1.45 & 3.93$\pm$1.45 & 398$_{-375}^{+468}$ & 3.16$_{-2.28}^{+2.76}$ & - & - & -\\
J1952+3252 & 2 & L$^f$ & 35.0$\pm$4.4 & 38.0$\pm$3.0 & 30$\pm$1 & 1.63$_{-0.05}^{+0.03}$ & 0.13$\pm$0.02 & 2.2$_{-0.8}^{+1.4}$ & 3.57$\times10^{-4}$\\
J1954+2836 & 0 & S & $<$3.65 & $<$3.65 & 50$^c$ & 2 & - & - & -\\
J1957+5036 & 0 & S & $<$2.98 & $<$2.98 & 10$^c$ & 2 & - & - & -\\
J1958+2841 & 1 & S & 1.57$\pm$0.53 & 1.57$\pm$0.53 & 40$^c$ & 2 & - & - & -\\
J2021+3651 & 2 & C+X & 2.21$_{-1.27}^{+0.35}$ & 6.01$_{-3.44}^{+0.96}$ & 65.5$\pm$6.0 & 2$\pm$0.5 & 0.140$_{-0.018}^{+0.023}$ & 4.94$\pm$1.40 & 2.75$\times10^{-5}$\\
J2021+4026 & 1 & C & 0.443$\pm$0.148 & 0.443$\pm$0.148 & 40$^c$ & 2 & - & - & 1.03$\times10^{-4}$\\
J2032+4127 & 2* & C+X & 0.423$\pm$0.118 & 0.423$\pm$0.118 & 38.7$_{-38.7}^{+75.6}$ & 1.87$_{-0.76}^{+0.96}$ & - & - & 1.99$\times10^{-4}$\\
J2043+2740 & 2 & X & 0.208$_{-0.208}^{+0.480}$ & 0.208$_{-1.08}^{+0.48}$ & $<$20 & 3.1$\pm$0.4 & - & - & 1.44$\times10^{-4}$\\
J2055+25 & 2 & X & 0.382$_{-0.148}^{+0.197}$ & 0.382$_{-0.148}^{+0.197}$ & 7.3$_{-7.3}^{+10.4}$ & 2.2$_{-0.6}^{+0.5}$ & - & - & 1.79$\times10^{-3}$\\
J2124-3358 & 2 & X & 0.668$_{-0.344}^{+0.150}$ & 0.959$_{-0.494}^{+0.216}$ & 2.76$_{-2.76}^{+4.87}$ & 2.89$_{-0.35}^{+0.45}$ & 0.268$_{-0.032}^{+0.034}$ & 0.019$_{-0.009}^{+0.012}$ & 1.25$\times10^{-4}$\\
J2229+6114 & 2 & C+X & 51.3$_{-5.8}^{+9.3}$ & 51.3$_{-5.8}^{+9.3}$ & 30$_{-4}^{+9}$ & 1.01$_{-0.12}^{+0.06}$ & - & - & 2.90$\times10^{-4}$\\
J2238+59 & 0 & S & $<$4.49 & $<$4.49 & 70$^c$ & 2 & - & - & -\\
\caption{
{\bf X-ray Spectra of the Pulsars}\\
The fluxes are unabsorbed and here the non-thermal and total fluxes are shown. The model used is an absorbed powerlaw plus blackbody, where statistically necessary. The only exceptions are PSR J0437-4715 (double PC plus powerlaw), J0633+1746 and J0659+1414 (double BB plus powerlaw): here only the most relevant thermal component is reported. All the errors are at a 90\% confidence level.\\
a : This parameter shows the confidence of the X-ray spectrum of each pulsar, based on the available X-ray data. An asterisk mark the pulsars for which ad ad-hoc analysis was necessary. See section ~\ref{art-xdata}.\\
b : C = {\it Chandra}/ACIS ; X = XMM/PN+MOS ; S = {\it SWIFT}/XRT ; L = literature. Only public data have been used (at December 2010).\\
c : here, the column density has been fixed by using the galactic value in the pulsar direction obtained by Webtools and scaling it for the distance (see Table \ref{art-tab-1}).\\
d : the beam correction factor f$_X$ is assumed to be 1, which can result in an efficiency $>$ 1. See Watters et al. (2009). Here the errors are not reported.\\
e : The statistic is very low so that it was necessary to freeze the column density parameter; the values have been evaluated by using WebTools.\\
f : Respectively taken from Webb et al.(2004a), Kargaltsev\&Pavlov(2008), Campana et al.(2008), De Luca et al. (2005), De Luca et al. (2005), Webb et al.(2004b), Mori et al.(2004), Kargaltsev et al.(2009), Li et al.(2005).\\
g : The spectrum is well fitted also by a single blackbody.\\
\label{art-tab-2}}
\end{longtable}

\begin{longtable}{cccccccc}
PSR Name & F$_R$ & F$_{\gamma}$ & $\gamma_{\gamma}$ & Cutoff$_G$ & Eff$_{\gamma}^a$ & F$_{\gamma}$/F$_{X nt}$ & F$_{\gamma}$/F$_{X tot}$\\
 & (mJy) & ($10^{-10}erg/cm^{2}s$)& & (GeV) & & & \\ 
\endhead
J0007+7303 & $<$0.006$^c$ & 3.82$\pm$0.11 & 1.38$\pm$0.04 & 4.6$\pm$0.4 & 0.2 & 5570$\pm$827 & 4544$\pm$546\\
J0030+0451 & 0.6 & 0.527$\pm$0.035 & 1.22$\pm$0.16 & 1.8$\pm$0.4 & 0.17 & 454$\pm$31 & 188$\pm$14\\
J0205+6449 & 0.04 & 0.665$\pm$0.054 & 2.09$\pm$0.14 & 3.5$\pm$1.4 & 0.0025 & 33.4$\pm$2.9 & 33.4$\pm$2.9\\
J0218+4232 & 0.9 & 0.362$\pm$0.053 & 2.02$\pm$0.23 & 5.1$\pm$4.2 & 0.2 & 74.3$_{-22.4}^{+13.9}$ & 74.3$_{-22.4}^{+13.9}$\\
J0248+6021 & 9 & 0.308$\pm$0.058 & 1.15$\pm$0.49 & 1.4$\pm$0.6 & 0.735 & $>$34.2 & $>$34.2\\
J0357+32 & $<$0.043$^c$ & 0.639$\pm$0.037 & 1.29$\pm$0.18 & 0.9$\pm$0.2 & 5.23 & 1000$_{-150}^{+200}$ & 1000$_{-150}^{+200}$\\
J0437-4715 & 140 & 0.186$\pm$0.022 & 1.74$\pm$0.32 & 1.3$\pm$0.7 & 0.02 & 18.4$\pm$2.5 & 13.0$\pm$1.7\\
J0534+2200 & 14 & 13.07$\pm$1.12 & 1.97$\pm$0.06 & 5.8$\pm$1.2 & 0.001 & 0.295$\pm$0.026 & 0.295$\pm$0.026\\
J0540-6919 & 0.024 & $<$0.833 & 2 & - & - & $<$1.47 & $<$1.47\\
J0613-0200 & 1.4 & 0.324$\pm$0.035 & 1.38$\pm$0.24 & 2.7$\pm$1.0 & 0.07 & $1464_{-1059}^{+1974}$ & $1464_{-1059}^{+1974}$\\
J0631+1036 & 0.8 & 0.304$\pm$0.051 & 1.38$\pm$0.35 & 3.6$\pm$1.8 & 0.14 & $>$1350 & $>$1350\\
J0633+0632 & $<$0.003$^c$ & 0.801$\pm$0.064 & 1.29$\pm$0.18 & 2.2$\pm$0.6 & 1.4 & 524$\pm$179 & 524$\pm$179\\
J0633+1746 & $<$1 & 33.85$\pm$0.29 & 1.08$\pm$0.02 & 1.90$\pm$0.05 & 0.78 & 6812$_{-375}^{+136}$ & 2687$_{-151}^{+48}$\\
J0659+1414 & 3.7 & 0.317$\pm$0.030 & 2.37$\pm$0.42 & 0.7$\pm$0.5 & 0.01 & 78.1$_{-13.6}^{+7.5}$ & 1.89$_{-0.32}^{+0.18}$\\
J0742-2822 & 15 & 0.183$\pm$0.035 & 1.76$\pm$0.40 & 2.0$\pm$1.4 & 0.07 & $>$812 & $>$812\\
J0751+1807 & 3.2 & 0.109$\pm$0.032 & 1.56$\pm$0.58 & 3.0$\pm$4.3 & 0.08 & 248$_{-103}^{+125}$ & 248$_{-103}^{+125}$\\
J0835-4510 & 1100 & 88.06$\pm$0.45 & 1.57$\pm$0.01 & 3.2$\pm$0.06 & 0.01 & 1353$\pm$326 & 313$\pm$75\\
J1023-5746 & $<$0.031 & 1.55$\pm$0.10 & 1.47$\pm$0.14 & 1.6$\pm$0.3 & 0.12 & 963$\pm$173 & 963$\pm$173\\
J1028-5819 & 0.36 & 1.77$\pm$0.12 & 1.25$\pm$0.17 & 1.9$\pm$0.5 & 0.14 & 1182$\pm$403 & 1182$\pm$403\\
J1044-5737 & $<$0.021 & 1.03$\pm$0.07 & 1.60$\pm$0.12 & 2.5$\pm$0.5 & 0.45 & $>$262 & $>$262\\
J1048-5832 & 6.5 & 1.73$\pm$0.11 & 1.31$\pm$0.15 & 2.0$\pm$0.4 & 0.08 & 3451$_{-725}^{+2426}$ & 3451$_{-725}^{+2426}$\\
J1057-5226 & 11 & 2.72$\pm$0.08 & 1.06$\pm$0.08 & 1.3$\pm$0.1 & 0.56 & 1804$_{-164}^{+59}$ & 111$_{-12}^{+4}$\\
J1124-5916 & 0.08 & 0.380$\pm$0.058 & 1.43$\pm$0.33 & 1.7$\pm$0.7 & 0.01 & 38.9$\pm$7.4 & 34.9$\pm$6.7\\
J1413-6205 & $<$0.025 & 1.29$\pm$0.10 & 1.32.$\pm$0.16 & 2.6$\pm$0.6 & 0.43 & $>$287 & $>$287\\
J1418-6058 & $<$0.03$^c$ & 2.36$\pm$0.32 & 1.32$\pm$0.20 & 1.9$\pm$0.4 & 0.08 & 6672$\pm$3049 & 6672$\pm$3049\\
J1420-6048 & 0.9 & 1.59$\pm$0.29 & 1.73$\pm$0.20 & 2.7$\pm$1.0 & 0.06 & 426$\pm$112 & 426$\pm$112\\
J1429-5911 & $<$0.022 & 0.926$\pm$0.081 & 1.93$\pm$0.14 & 3.3$\pm$1.0 & 0.45 & $>$55.0 & $>$55.0\\
J1459-60 & $<$0.038$^c$ & 1.06$\pm$0.10 & 1.83$\pm$0.20 & 2.7$\pm$1.1 & 0.52 & $>$269 & $>$269\\
J1509-5850 & 0.15 & 0.969$\pm$0.101 & 1.36$\pm$0.23 & 3.5$\pm$1.1 & 0.15 & 1088$_{-254}^{+197}$ & 1088$_{-254}^{+197}$\\
J1614-2230 & - & 0.274$\pm$0.042 & 1.34$\pm$0.36 & 2.4$\pm$1.0 & 1.03 & $>$958 & 958$_{-434}^{+196}$\\
J1617-5055 & - & $<$0.3 & 2 & - & - & $<$4.5 & $<$4.5\\
J1709-4429 & 7.3 & 12.42$\pm$0.22 & 1.70$\pm$0.03 & 4.9$\pm$0.4 & 0.33 & 3285$_{-819}^{+327}$ & 1374$_{-343}^{+134}$\\
J1718-3825 & 1.3 & 0.673$\pm$0.160 & 1.26$\pm$0.62 & 1.3$\pm$0.6 & 0.09 & 240$\pm$81 & 240$\pm$81\\
J1732-31 & $<$0.008$^c$ & 2.42$\pm$0.12 & 1.27$\pm$0.12 & 2.2$\pm$0.3 & 1.33 & $>$1000 & $>$1000\\
J1741-2054 & 0.156$^c$ & 1.28$\pm$0.07 & 1.39$\pm$0.14 & 1.2$\pm$0.2 & 0.24 & 277$_{-98}^{+111}$ & 277$_{-98}^{+111}$\\
J1744-1134 & 3 & 0.280$\pm$0.046 & 1.02$\pm$0.59 & 0.7$\pm$0.4 & 0.1 & $>$1030 & 1030$\pm$187\\
J1747-2958 & 0.25 & 1.31$\pm$0.14 & 1.11$\pm$0.28 & 1.0$\pm$0.2 & 0.02 & 26.9$_{-4.3}^{+12.1}$ & 26.9$_{-4.3}^{+12.1}$\\
J1809-2332 & $<$0.026$^c$ & 4.13$\pm$0.13 & 1.52$\pm$0.06 & 2.9$\pm$0.3 & 0.33 & 2951$_{-494}^{+535}$ & 1316$_{-226}^{+242}$\\
J1811-1926 & - & $<$0.3 & 2 & - & - & $<$11.25 & $<$11.25\\
J1813-1246 & $<$0.028$^c$ & 1.69$\pm$0.11 & 1.83$\pm$0.12 & 2.9$\pm$0.8 & 0.20 & 175$\pm$59 & 175$\pm$59\\
J1813-1749 & - & $<$0.3 & 2 & - & - & $<$11.25 & $<$11.25\\
J1826-1256 & $<$0.044$^c$ & 3.34$\pm$0.15 & 1.49$\pm$0.09 & 2.4$\pm$0.3 & 20.7 & 2834$\pm$1398 & 2834$\pm$1398\\
J1833-1034 & 0.07 & 1.02$\pm$0.12 & 2.24$\pm$0.15 & 7.7$\pm$4.8 & 0.01 & 15.3$\pm$1.9 & 15.3$\pm$1.9\\
J1836+5925 & $<$0.01$^c$ & 6.00$\pm$0.11 & 1.35$\pm$0.03 & 2.3$\pm$0.1 & 2 & 13065$_{-4958}^{+11474}$ & 10520$_{-3991}^{+9231}$\\
J1846+0919 & $<$0.004 & 0.358$\pm$0.035 & 1.60$\pm$0.19 & 4.1$\pm$1.5 & 2.1 & $>$123 & $>$123\\
J1907+06 & 0.0034$^b$ & 2.75$\pm$0.13 & 1.84$\pm$0.08 & 4.6$\pm$1.0 & 0.30 & 700$\pm$260 & 700$\pm$260\\
J1952+3252 & 1 & 1.34$\pm$0.07 & 1.75$\pm$0.10 & 4.5$\pm$1.2 & 0.02 & 38.3$\pm$5.3 & 35.2$\pm$3.4\\
J1954+2836 & $<$0.004 & 0.975$\pm$0.068 & 1.55$\pm$0.14 & 2.9$\pm$0.7 & 0.39 & $>$267 & $>$267\\
J1957+5036 & $<$0.025 & 0.227$\pm$0.020 & 1.12$\pm$0.28 & 0.9$\pm$0.2 & 5.6 & $>$76.2 & $>$76.2\\
J1958+2841 & $<$0.005$^c$ & 0.846$\pm$0.069 & 0.77$\pm$0.26 & 1.2$\pm$0.2 & 1.6 & 539$\pm$187 & 539$\pm$187\\
J2021+3651 & 0.1 & 4.70$\pm$0.15 & 1.65$\pm$0.06 & 2.6$\pm$0.3 & 0.07 & 2129$_{-1225}^{+343}$ & 783$_{-449}^{+127}$\\
J2021+4026 & $<$0.011$^c$ & 9.77$\pm$0.18 & 1.79$\pm$0.03 & 3.0$\pm$0.2 & 2.2 & 22061$\pm$7381 & 22061$\pm$7381\\
J2032+4127 & 0.05$^c$ & 1.12$\pm$0.12 & 0.68$\pm$0.38 & 2.1$\pm$0.6 & 0.64 & 2636$\pm$790 & 2636$\pm$790\\
J2043+2740 & 7 & 0.155$\pm$0.027 & 1.07$\pm$0.55 & 0.8$\pm$0.3 & 0.09 & 747$_{-409}^{+217}$ & 747$_{-409}^{+217}$\\
J2055+25 & $<$0.106 & 1.15$\pm$0.07 & 0.71$\pm$0.19 & 1.0$\pm$0.2 & 5.4 & 3010$_{-1181}^{+1563}$ & 3010$_{-1181}^{+1563}$\\
J2124-3358 & 1.6 & 0.276$\pm$0.035 & 1.05$\pm$0.28 & 2.7$\pm$1.0 & 0.05 & 413$_{-219}^{+107}$ & 288$_{-153}^{+74}$\\
J2229+6114 & 0.25 & 2.20$\pm$0.08 & 1.74$\pm$0.07 & 3.0$\pm$0.5 & 0.025 & 42.9$_{-5.1}^{+7.9}$ & 42.9$_{-5.1}^{+7.9}$\\
J2238+59 & $<$0.007$^c$ & 0.545$\pm$0.059 & 1.00$\pm$0.36 & 1.0$\pm$0.3 & 0.52 & $>$121 & $>$121\\
\caption{{\bf $\gamma$-ray spectra of the pulsars.}\\
A broken powerlaw spectral shape is assumed for all the pulsars and the values are taken from Abdo et al. (2009c), Saz Parkinson et al. 2010. The gamma-ray flux is above 100 GeV. The 4 sources with an upper limit flux are taken from Bird et al. (2009) (see section ~\ref{art-gdata}). The radio flux densities (at 1400MHz) are taken from Abdo et al. (2009c), Saz Parkinson et al. 2010. All the errors are at a 90\% confidence level.\\
a : f${_\gamma}$ is assumed to be 1, which can result in an efficiency $>$ 1. See Watters et al. 2009. Here the errors are not reported.\\
b : taken from Abdo et al.(2010e).\\
c : taken from Ray et al.(2011).
\label{art-tab-3}}
\end{longtable}
\end{center}
\end{landscape}
\end{footnotesize}

\begin{footnotesize}

\end{footnotesize}


\begin{thebibliography}{99}
\bibitem[Abdo et al.(2007)]{Abd07} Abdo, A.A., et al., 2007, ApJ, 664, 91
\bibitem[Abdo et al.(2008)]{Abd08} Abdo, A.A., et al., 2008, Science 322, 1218
\bibitem[Abdo et al.(2009a)]{Abd09c} Abdo, A.A. et al. 2009, Sci, 325, 840A
\bibitem[Abdo et al.(2009b)]{Abd09a} Abdo, A.A. et al., 2009, Sci, 325, 848
\bibitem[Abdo et al.(2009c)]{Abd09b}  Abdo, A.A. et al., 2009, ApJS 183, 46
\bibitem[Abdo et al.(2009d)]{Abd09d}  Abdo, A.A. et al., 2009, ApJ, 699, 102
\bibitem[Abdo et al.(2010a)]{Abd09} Abdo, A.A et al., 2010, ApJ, 187, 460A
\bibitem[Abdo et al.(2010b)]{Abd10b} Abdo, A.A. et al., 2010 ApJ, 188, 405A
\bibitem[Abdo et al.(2010c)]{Abd10d} Abdo, A.A. et al., 2010 ApJ, 712, 1209A
\bibitem[Abdo et al.(2010d)]{Abd10e} Abdo, A.A. et al., 2010 ApJ, 711, 64A
\bibitem[Abdo et al.(2010e)]{Abd10f} Abdo, A.A. et al., 2010 ApJ, 713, 154
\bibitem[Abdo et al.(2011)]{Abd11} Abdo, A.A. et al., 2011, in preparation
\bibitem[Abdo et al.(2012)]{Abd12} Abdo, A.A. et al., 2012, in preparation (the 2nd pulsar catalogue)
\bibitem[Ackermann et al.(2011)]{Ack11} Ackermann, M. et al., 2011, ApJ, 726, 81
\bibitem[Aharonian et al.(2005)]{Aha05} Aharonian, F. et al., 2005, A\&A, 435, 17
\bibitem[Aharonian et al.(2006)]{Aha06} Aharonian, F. et al., 2007, A\&A, 456, 245
\bibitem[Aharonian et al.(2007)]{Aha07} Aharonian, F. et al., 2007, A\&A, 472, 489
\bibitem[Aharonian et al.(2009)]{Aha09} Aharonian, F. et al., 2009, A\&A, 499, 723
\bibitem[Alpar et al.(1984)]{Alp84} Alpar, M.A. et al., 1984, ApJ, 276, 325
\bibitem[Arnaud(1996)]{Arn96} Arnaud, K. A., 1996, in Astronomical Data Analysis software Systems V, eds. G. Jacoby \& J. Barnes (ASP Conf. Ser. 101), 17
\bibitem[Arnett\&Bowers(1977)]{Arn77} Arnett, W.D. \& Bowers, R.L., 1977, ApJS, 33, 415
\bibitem[Atwood et al.(2006)]{Atw06} Atwood, W.W. et al., 2006, ApJ, 652, 49
\bibitem[Bailes et al.(1997)]{Bai97} Bailes, M. et al. 1997, ApJ, 481, 386
\bibitem[Bandiera(2008)]{Ban08} Bandiera, R., 2008, A\&A 490, L3
\bibitem[Bassa et al.(2003)]{Bas03} Bassa, C.G. et al., 2003, A\&A, 403, 1067
\bibitem[Backer\&Sallmen(1982)]{Bac82} Backer, D.C. \& Sallmen, S.T., 1982, Nature, 300, 615
\bibitem[Becker et al.(1982)]{Bec82} Becker, W.E., et al., 1982, ApJ, 255, 557
\bibitem[Becker et al.(1995)]{Bec95} Becker, W.E., et al., 1995, A\&A, 298, 528
\bibitem[Becker et al.(1996)]{Bec96} Becker, W.E., et al., 1996, MNRAS, 282, 33
\bibitem[Becker et al.(2000)]{Bec00} Becker, W.E., et al., 2000, ApJ 545, 1015
\bibitem[Becker et al.(2004)]{Bec04} Becker, W.E., et al., 2004, ApJ 615, 908
\bibitem[Becker et al.(2005)]{Bec05} Becker, W.E., et al., 2005, ApJ 633, 367
\bibitem[Becker et al.(2006)]{Bec06} Becker, W., et al., 2006, ApJ 645, 1421
\bibitem[Becker(2009)]{Bec09} Becker, W., 2009, in Astrophysics and Space Science Library, Vol. 357, Astrophysics and Space Science Library, ed W. Becker
\bibitem[Becker\&Aschenbach(2002)]{Bec02} Becker, W. \& Aschenbach, B., 2002, nsps.conf.64B
\bibitem[Becker\&Pavlov(2002)]{Bec02b} Becker, W. \& Pavlov, G.G., 2002, astro.ph, 8356B
\bibitem[Becker\&Trumper(1997)]{Bec97} Becker, W. \& Trumper, J., 1997, A\&A, 326, 682
\bibitem[Becker\&Trumper(1998)]{Bec98} Becker, W. \& Trumper, J., 1998, IAU circ, 6829
\bibitem[Becker\&Trumper(1999)]{Bec99} Becker, W. \& Trumper, J., 1999, A\&A, 341, 803
\bibitem[Becker \& Hui(2005)]{becker05} Becker, W.E. \& Hui, C.Y., 2005, APJ 633, 367
\bibitem[Bell et al.(1993)]{Bel93} Bell, J.F. et al., 1993, Nature, 364, 603
\bibitem[Bell et al.(1995)]{Bel95} Bell, J.F. et al., 1995, ApJ, 440, 81
\bibitem[Bertsch et al.(1992)]{Ber92} Bertsch, D.L. et al., 1992, Nature, 357, 306
\bibitem[Bietenholz\&Bartel(2006)]{Bie06} Bietenholz, M.F.\& Bartel, N., 2006, cosp 36, 3143
\bibitem[Bietenholz\&Bartel(2008)]{Bie08a} Bietenholz, M.F.\& Bartel, N., 2008, MNRAS, 386, 1411B
\bibitem[Biggs et al.(1989)]{Big89} Biggs, J.D. et al., 1989, tns. conf. 157
\bibitem[Biggs et al.(1989)]{Big89b} Biggs, J.D. et al., 1989, ESASP, 296, 293
\bibitem[Biggs et al.(1990)]{Big90} Biggs, J.D. et al., 1990, IAUC, 4988, 2
\bibitem[Bignami et al.(1983)]{Big83} Bignami, G.F. et al., 1983, ApJ, 272, 9
\bibitem[Bignami et al.(1988)]{Big88} Bignami, G.F. et al., 1988, A\&A, 202, 1
\bibitem[Bignami et al.(1993)]{Big93} Bignami, G.F. et al., 1993, Nature, 361, 704
\bibitem[Bignami\&Caraveo(1996)]{Big96} Bignami, G.F. \& Caraveo, P.A., 1996, ARA\&A 34, 331
\bibitem[Bird et al.(2009)]{Bir09} Bird, A.J. et al., 2009, arXiv:0910.1704
\bibitem[Blair et al.(2005)]{Bla05} Blair, W.P. et al., 2005, AJ, 129, 2268
\bibitem[Blair et al.(2009)]{Bla09} Blair, W.P. et al., 2009, ApJ, 692, 335
\bibitem[Bock\&Gvaramadze(2002)]{Bock02a} Bock, D.C.-J. \& Gvaramadze, V.V., 2002, A\&A, 394, 533B
\bibitem[Bogdanov\&Grindley(2009)]{Bog09} Bogdanov, S. \& Grindley, J.E., 2009, ApJ, 703, 1557
\bibitem[Braje et al.(2002)]{Bra02} Braje, T.M. et al., 2002, ApJ, 565, 91
\bibitem[Braun et al.(1986)]{Bra86} Braun, R.G. et al., 1986, A\&A, 162, 259
\bibitem[Brinkmann\&Oegleman(1987)]{Bri87} Brinkmann, W. \& Oegleman, H., 1987, A\&A, 182, 71
\bibitem[Brisken et al.(2002)]{Bri02} Brisken, W.F. et al., 2002, ApJ, 571, 906
\bibitem[Brogan et al.(2005)]{Bro05a} Brogan, C.L. et al., 2005, ApJ, 629L, 105B
\bibitem[Bucciantini(2002)]{Buc02} Bucciantini, N., 2002, A\&A 387, 1066
\bibitem[Bucciantini et al.(2005)]{Buc05} Bucciantini, N., et al., 2005, A\&A 443, 519
\bibitem[Burgay et al.(2006)]{Bur06} Burgay, M. et al., 2006, MNRAS, 368, 283
\bibitem[Burgay et al.(2006b)]{Bur06b} Burgay, M. et al., 2006, MNRAS, 372, 410
\bibitem[Burrows et al.(2005)]{Bur05} Burrows, D. N. et al., 2005, Space Sci. Rev., 120, 165
\bibitem[Camilo et al.(2000)]{Cam00} Camilo, F. et al., 2000, ASPC, 202, 3
\bibitem[Camilo et al.(2001)]{Cam01} Camilo, F. et al., 2001, ApJ, 557, 51
\bibitem[Camilo et al.(2002)]{Cam02} Camilo, F. et al., 2002, ApJ, 571, 41
\bibitem[Camilo et al.(2004)]{Cam04} Camilo, F. et al., 2004, ApJ, 611, 25
\bibitem[Camilo et al.(2006)]{Cam06} Camilo, F. et al., 2006, ApJ, 637, 456
\bibitem[Camilo et al.(2009)]{Cam09} Camilo, F. et al., 2009, ApJ, 705, 1C
\bibitem[Campana et al.(2008)]{Cam08} Campana R. et al., 2008, MNRAS, 389, 691C
\bibitem[Caraveo et al.(1996)]{Car96} Caraveo, P.A., et al., 1996, ApJ, 461, 91
\bibitem[Caraveo et al.(2001)]{Car01} Caraveo, P.A. et al., 2001, ApJ, 561, 930
\bibitem[Caraveo et al.(2003)]{Car03} Caraveo, P.A. et al., 2003, Science 305, 376
\bibitem[Caraveo et al.(2004)]{caraveo04} Caraveo, P.A., et al., 2004, Science, 305, 376
\bibitem[Caraveo(2009)]{Car09} Caraveo, P.A., 2009, arXiv, 0912, 4857
\bibitem[Caraveo(2010)]{Car10a} Caraveo, P.A., 2010, in  High Time Resolution Astrophysics IV, to be published in PoS, arXiv:1009.2421
\bibitem[Caraveo et al.(2010)]{Car10b} Caraveo, P.A., et al., 2010, ApJL  in press, arXiv:1010.4167
\bibitem[Castelletti et al.(2003)]{Cas03} Castelletti, G. et al., 2003, AJ, 126, 2114
\bibitem[Caswell et al.(2004)]{Cas04} Caswell, J.L. et al., 2004, MNRAS, 352, 1405
\bibitem[Caswell\&Goss(1970)]{Cas70} Caswell, J.L. \& Goss, W.M., 1970, ApJ, 7, 141
\bibitem[Chandler et al.(2001)]{Cha01} Chandler, A.M. et al., 2001, ApJ, 680, 620
\bibitem[Cheng et al.(2006)]{Che06} Cheng, K.S. et al., 2006, ApJ, 641, 427
\bibitem[Cheng\&Helfand(1983)]{Che83} Cheng, A.F. \& Helfand, D.J., 1983, ApJ, 271, 271
\bibitem[Cheng\&Rudeman(1980)]{Che80} Cheng, A.F. \& Rudeman, M.A., 1980, ApJ, 235, 576
\bibitem[Chin\&Salpeter(1964)]{Chi64} Chin, H.Y. \& Salpeter, E.E., 1964, PhRvl, 12, 413
\bibitem[Cognard et al.(1996)]{Cog96} Cognard, L. et al. 1996, ApJ, 457, 81
\bibitem[Cognard et al.(2011)]{Cog11} Cognard, L. et al. 2011, ApJ, 732, 47
\bibitem[Condon et al.(1998)]{Con98} Condon, J.J., et al., 1998, AJ 115, 1693
\bibitem[Cordes\&Lazio(2002)]{Cor02} Cordes, J.M. \& Lazio, T.J.W., 2002, astro-ph/0207156
\bibitem[Cordova et al.(1989)]{Cor89} Cordova, F.A. et al., 1989, ApJ, 345, 451
\bibitem[Cranford et al.(2001)]{Cra01} Cranford, F. et al., 2001, ApJ, 554, 152
\bibitem[Crawford\&Tiffany(2007)]{Cra07} Crawford, F. \& Tiffany, C.L., 2007, AJ, 134, 1231
\bibitem[Cusumano et al.(2003)]{Cus03} Cusumano, G. et al., 2003, A\&A, 410, 9
\bibitem[D'Amico et al.(2001)]{DAm01} D'Amico, N. et al., 2001, ApJ, 552, 45
\bibitem[De Jager et al.(1996)]{Dej96} De Jager, O.C., et al., 1996, ApJ 457, 253
\bibitem[De Jager \& Djannati-Ata\"i(2008)]{Dej08} De Jager, O.C. \&  Djannati-Ata\"i, A., 2008, to appear in Springer Lecture Notes on Neutron
Stars and Pulsars: 40 years after their discovery, eds. W. Becker, arXiv:0803.0116
\bibitem[De Luca\&Molendi(2004)]{DeL04} De Luca, A. \& Molendi, S. 2004, ApJ, 419, 837D
\bibitem[De Luca et al.(2005)]{DeL05} De Luca, A. et al., 2005, ApJ, 623, 1051
\bibitem[De Luca et al.(2006)]{DeL06} De Luca, A. et al., 2006, A\&A 445, L9
\bibitem[Deller et al.(2008)]{Del08} Deller, A.T. et al., 2008, ApJ, 685, 67
\bibitem[Dickey \& Lockman(1990)]{Dic90} Dickey, J.M. \& Lockman, F.J., 1990, ARA\&A 28, 215
\bibitem[Dodson\&Golap(2002)]{Dod02} Dodson, R. \& Golap, K., 2002, MNRAS, 334, 1
\bibitem[Dodson et al.(2003)]{Dod03} Dodson, R. et al., 2003, ApJ, 596, 1137
\bibitem[Dwarakanath\&Shankar(1990)]{Dwa90} Dwarakanath, K. \& Shankar, U.N., 1990, JApA, 11, 323
\bibitem[Elston et al.(2003)]{Els03} Elston, R., et al., 2003, Proc.SPIE 4841, 1611
\bibitem[Facondi et al.(1973)]{Fac73} Facondi, S.R. et al., 1973, A\&A, 27, 67
\bibitem[Fesen et al.(1988)]{Fes88} Fesen, R.A. et al., 1988, Nature, 334, 229
\bibitem[Finley et al.(1992)]{Fin92} Finley, J.P. et al., 1992, ApJ, 394, 21
\bibitem[Finley et al.(1998)]{Fin98} Finley, J.P. et al., 1998, ApJ, 493, 884
\bibitem[Foster et al.(1997)]{Fos97} Foster, R.S. et al., 1997, AAS, 19111110F
\bibitem[Foster\&Wolszczan(1993)]{Fos93} Foster, R.S. \& Wolszczan, A., 1993, AAS, 183, 3707
\bibitem[Frail et al.(1994)]{Fra94} Frail, D.A. et al., 1994, ApJ, 437, 781
\bibitem[Frail\&Kulkarni(1991)]{Fra91} Frail, D.A. \& Kulkarni, S.R., 1991, Nature, 352, 785
\bibitem[Fruchter et al.(1988a)]{Fru88a} Fruchter, A.S., 1988, Nature, 333, 237
\bibitem[Fruchter et al.(1988b)]{Fru88b} Fruchter, A.S., 1988, Nature, 334, 686
\bibitem[Gaensler et al.(1999)]{Gae99} Gaensler, B.M., et al., 1999, MNRAS, 305, 742
\bibitem[Gaensler et al.(2002)]{Gae02} Gaensler, B.M., et al., 2002, ApJ, 580, 137
\bibitem[Gaensler et al.(2004)]{Gae04} Gaensler, B.M., et al., 2004, ApJ 616, 383
\bibitem[Gaensler\&Slane(2006)]{Gae06} Gaensler, B.M. \& Slane, P.O., 2006, ARA\&A 44, 17
\bibitem[Garmire et al.(2003)]{Gar03} Garmire, G.G. et al., 2003, SPIE 4851, 28
\bibitem[Giacani et al.(2001)]{Gia01} Giacani, E.B. et al., 2001, AJ, 121, 3133
\bibitem[Giommi et al.(1992)]{Gio92} Giommi, P. et al. 1992, ASPC, 25, 100G
\bibitem[Gold(1967)]{Gol67} Gold, T., 1968, Nature, 218, 731
\bibitem[Golden et al.(2005)]{Gol05} Golden, A. et al. 2005, ApJ, 635, 153
\bibitem[Goldreich \& Julian(1969)]{Gol69} Goldreich, P. \& Julian, W.H., 1969, ApJ 157, 869
\bibitem[Gonzalez et al.(2004)]{Gon04} Gonzalez, M. et al., 2004, AAS, 20510205G
\bibitem[Gonzalez et al.(2005)]{Gon05} Gonzalez, M. et al., 2005, ApJ, 630, 489
\bibitem[Gonzalez et al.(2006)]{Gon06} Gonzalez, M. et al., 2006, ApJ, 652, 569
\bibitem[Gonzalez \& Safi-Harb(2003)]{Gon03} Gonzalez, M. \& Safi-Harb, S., 2003, ApJ, 591, 143
\bibitem[Gorenstein et al.(1974)]{Gor74a} Gorenstein, P. et al., 1974, ApJ, 192, 661G
\bibitem[Gotthelf et al.(2002)]{Got02} Gotthelf, E.V. et al., 2002, ApJ, 567, 125
\bibitem[Gotthelf et al.(2007)]{Got07a} Gotthelf, E.V. et al., 2007, ApJ, 654, 267G
\bibitem[Gotthelf\&Halpern(2009)]{Got09} Gotthelf, E.V. \& Halpern, J.P., ApJ, 700L, 158G
\bibitem[Gouiffes(1998)]{Gou98} Gouiffes, C., 1998, nspt.conf 363G
\bibitem[Green\&Gull(1982)]{Gre82} Green, D.A. \& Gull, S.F., 1982, Nature, 299, 606
\bibitem[Green et al.(2004)]{Gre04} Green, P.J. et al., 2004, ApJS 150, 43
\bibitem[Greiveldinger et al.(1996)]{Gre96} Greiveldinger, C. et al., 1996, ApJ, 465, 35
\bibitem[Grenier et al.(1988)]{Gre88} Grenier, I.A. et al., 1988, A\&A, 204, 117
\bibitem[Groth(1975)]{Gro75} Groth, E.J., 1975, ApJ, 200, 278
\bibitem[Gullahorn(1977)]{Gul77} Gullahorn, G.E., 1977, AJ, 82, 309
\bibitem[Gupta et al.(2005)]{Gup05} Gupta, Y. et al., 2005, Current Science, 89, 853
\bibitem[Hales et al.(2009)]{Hal09a} Hales, C.A. et al., 2009, ApJ, 706 1316H
\bibitem[Halpern et al.(2001)]{Hal01} Halpern, J.P. et al., 2001, ApJ, 547, 323
\bibitem[Halpern et al.(2001b)]{Hal01b} Halpern, J.P. et al., 2001, ApJ, 688, 33
\bibitem[Halpern et al.(2002)]{Hal02} Halpern, J.P. et al., 2002, ApJ, 573, 41
\bibitem[Halpern et al.(2004)]{Hal04} Halpern, J.P. et al., 2004, ApJ 612, 398
\bibitem[Halpern et al.(2007)]{Hal07} Halpern, J.P. et al., 2007, ApJ, 668, 1154
\bibitem[Halpern\&Holt(1992)]{Hal92} Halpern, J.P. \& Holt, S.S., 1992, Nature, 357, 222
\bibitem[Halpern\&Ruderman(1993)]{Hal93b} Halpern, J.P. \& Ruderman, M., 1993, ApJ, 415, 286
\bibitem[Harding \& Muslimov(2001)]{Har01} Harding, A.K. \& Muslimov, A.G., 2001, ApJ 556, 987
\bibitem[Harding \& Muslimov(2002)]{Har02} Harding, A.K. \& Muslimov, A.G., 2002, ApJ 568, 862
\bibitem[Harding et al.(2002)]{Har02b} Harding, A.K. et al., 2002, ApJ, 576, 376
\bibitem[Harris \& Krawczynski(2006)]{Har06} Harris, D.E. \& Krawczynski, H., 2006, ARA\&A 44, 463
\bibitem[Hartman et al.(1999)]{Har99} Hartman, R.C. et al., 1999, yCat 21230079H
\bibitem[Hartman et al.(1999b)]{Har99b} Hartman, R.C. et al., 1999, ApJS, 123, 79
\bibitem[Hales et al.(1993)]{Hal93} Hales, S.E.G. et al., 1993, MNRAS, 263, 25
\bibitem[Helfand et al.(1980)]{Hel80} Helfand, D.J. et al. 1980, Nature, 283, 337
\bibitem[Helfand et al.(2001)]{Hel01} Helfand, D.J. et al. 2001, ApJ, 556, 380
\bibitem[Hessels et al.(2004)]{Hes04} Hessels, J.W.T., et al., 2004, IAUS, 218, 131
\bibitem[Hester(2000)]{Hes00} Hester, J., 2000, BAAS Abstr., 32, 1542
\bibitem[Hewish et al.(1968)]{Hew68} Hewish, A. et al., 1968, Nature, 217, 709
\bibitem[Hinton et al.(2007)]{Hin07} Hinton, J.A. et al., 2007, A\&A, 476, 25
\bibitem[Hobbs et al.(2005)]{Hob05} Hobbs, G. et al. 2005, MNRAS, 360, 974
\bibitem[Hotan et al.(2006)]{Hot06} Hotan, A.W. et al., 2006, MNRAS, 369, 1502
\bibitem[Huang\&Becker(2006)]{Hua06} Huang, H.H. \& Becker, W., 2006, IAUJD, 2, 56
\bibitem[Huges et al.(2001)]{Hug01} Huges, J.P. et al., 2001, ApJ, 559, 153
\bibitem[Hui \& Becker(2006)]{Hui06} Hui C.Y. \& Becker, W.E., 2006, A\&A A\&A, 448, 13
\bibitem[Hui \& Becker(2007)]{Hui07} Hui C.Y. \& Becker, W.E., 2007, A\&A 467, 1209
\bibitem[Humphrey et al.(2009)]{Hum09} Humphrey, P.J., et al., 2009, ApJ 693, 822
\bibitem[Hwang et al.(2001)]{Hwa01a} Hwang, U. et al., 2001, ApJ, 560, 742H
\bibitem[Jacoby et al.(1998)]{Jac98} Jacoby, G.H., et al., 1998, Proc. SPIE, 3355, 721
\bibitem[Jacoby et al.(2006)]{Jac06} Jacoby, G.H., et al., 2006, ApJ, 640, 183
\bibitem[Jacoby et al.(2007)]{Jac07} Jacoby, G.H., et al., 2007, ApJ, 656, 408
\bibitem[Jackson et al.(2005)]{jackson05} Jackson, M.S., et al., 2005, ApJ 633, 1114
\bibitem[Johnson et al.(1995)]{Joh95} Johnson, S.P. et al., 1995, A\&A, 293, 795 
\bibitem[Johnson \& Romani(2003)]{Joh03} Johnson, S.P. \& Romani, R.W., 2003, ApJ, 590, 95 
\bibitem[Johnson \& Wang(2010)]{Joh10} Johnson, S.P. \& Wang, Q.D., 2010, MNRAS 408, 1216 
\bibitem[Johnston et al.(1992)]{Joh92} Johnston, S. et al., 1992, MNRAS, 255, 401 
\bibitem[Johnston et al.(1993)]{Joh93} Johnston, S. et al., 1993, Nature, 361, 613 
\bibitem[Johnston et al.(1995)]{Joh95b} Johnston, S. et al., 1995, MNRAS, 274, 43 
\bibitem[Joncas\&Higgs(1990))]{Jon90} Joncas, G. \& Higgs, L.A., 1990, A\&AS, 82, 113 
\bibitem[Kalberla et al.(2005)]{Kal05} Kalberla, P.M.W., et al., 2005, A\&A 440, 775
\bibitem[Kanbach et al.(1980)]{Kan80} Kanbach, G. et al., 1980, A\&A, 90, 163
\bibitem[Kanbach et al.(1994)]{Kan94} Kanbach, G. et al., 1994, A\&A, 289, 855
\bibitem[Kargaltsev et al.(2006)]{Kar06} Kargaltsev, O., et al., 2006, ApJ 636, 406
\bibitem[Kargaltsev et al.(2006b)]{Kar06b} Kargaltsev, O., et al., 2006, A\&AS HEAD meeting, 9
\bibitem[Kargaltsev\&Pavlov(2007)]{Kar07} Kargaltsev O. \& Pavlov, G.G., 2007, Ap\&SS, 308, 287K
\bibitem[Kargaltsev\&Pavlov(2008)]{Kar08} Kargaltsev O. \& Pavlov, G.G., 2008, in AIP Conf. Proc. 983, 171
\bibitem[Kargaltsev \& Pavlov(2008b)]{Kar08b} Kargaltsev, O. \& Pavlov, G.G., 2008, ApJ 684, 542
\bibitem[Kargaltsev et al.(2009)]{Kar09} Kargaltsev, O. et al., 2009, ApJ, 690, 891
\bibitem[Kaspi et al.(1998)]{Kas98} Kaspi, V.M. et al., 1998, ApJ, 503L, 161K
\bibitem[Kaspi et al.(2001)]{Kas01} Kaspi, V.M. et al., 2001, ApJ, 560, 371K
\bibitem[Kaspi et al.(2004)]{Kas04} Kaspi, V.M. et al., 2004, astro.ph. 2136K
\bibitem[Kaspi et al.(2006)]{kaspi06} Kaspi, V., et al., 2006, in $"$Compact Stellar X-ray Sources$"$,
eds. Lewin, W. and van der Klis, M., Cambridge University Press, p. 279 
\bibitem[Keith et al.(2008)]{Kei08} Keith, M.J. et al. 2008, 389, 1881
\bibitem[Keith et al.(2011)]{Kei11} Keith, M.J. et al. 2011, MNRAS, 414, 1292
\bibitem[Kennea et al.(2002)]{Ken02} Kennea, J. et al., 2002, astro.ph 2055
\bibitem[King et al.(2005)]{Kin05} King, A.R. et al., 2005, MNRAS, 358, 1501
\bibitem[Kinkhabwala\&Thorsett(2000)]{Kin00} Kinkhabwala, A. \& Thorsett, S.E., 2000, ApJ, 554, 316
\bibitem[Koo et al.(1990)]{Koo90} Koo, B.C. et al., 1990, ApJ, 364, 178
\bibitem[Koribalski et al.(1995)]{Kor95} Koribalski, B. et al., 1995, ApJ, 441, 756
\bibitem[Kothes et al.(2001)]{Kot01} Kothes, R. et al., 2001, ApJ, 560, 236
\bibitem[Kothes et al.(2006)]{Kot06} Kothes, R. et al., 2006, ApJ, 238, 225
\bibitem[Kramer et al.(2003)]{Kra03} Kramer, M. et al., 2003, MNRAS, 342, 1299
\bibitem[Kramer et al.(2003b)]{Kra03b} Kramer, M. et al., 2003, yCat 73421299K
\bibitem[Kuiper et al.(1999)]{Kui99} Kuiper, L. et al., 1999, A\&A, 351, 119
\bibitem[Kuiper et al.(2004)]{Kui04} Kuiper, L. et al., 2004, AdSpR, 33, 507
\bibitem[Kulkarni \& Hester(1988)]{Kul88} Kulkarni, S.R. \& Hester, J.J., 1988, Nature, 335, 801
\bibitem[Kulkarni et al.(1988)]{Kul88b} Kulkarni, S.R., et al., 1988, Nature, 331, 50
\bibitem[Kulkers et al.(2003)]{Kul03} Kulkers, E. et al., 2003, A\&A, 399, 663
\bibitem[Lamb\&Macomb(1997)]{Lam97} Lamb, R.C. \& Macomb, D.J., 1997, ApJ, 488, 872
\bibitem[Large et al.(1968)]{Lar68} Large, M.I. et al., 1968, Nature, 220, 753
\bibitem[Lasker et al.(2008)]{Las08} Lasker, B.M., et al., 2008, AJ 136, 735
\bibitem[Lattanzi et al.(1997)]{Lat97} Lattanzi, M.G., et al., 1997, A\&A 318, 997 
\bibitem[Li et al.(2005)]{Li05} Li, X.H. et al., 2005, ApJ, 628, 931
\bibitem[Lin et al.(1992)]{Lin92} Lin, Y.C. et al., 1992, IAU Circ., 5676, 2
\bibitem[Livingstone et al.(2005)]{Liv05} Livingstone, M.A. et al., 1982, ApJ, 619, 1046
\bibitem[Lohsen(1981)]{Loh81} Lohsen, E.H.G., 1981, A\&AS, 44, 1
\bibitem[Lommen et al.(2000)]{Lom00} Lommen, A.N. et al., 2000, ApJ, 545, 1007
\bibitem[Lommen et al.(2006)]{Lom06} Lommen, A.N. et al., 2006, ApJ, 642, 1012
\bibitem[Lundgren et al.(1993)]{Lun93} Lundgren, S.C. et al., 1993, AAS, 183, 3703
\bibitem[Lundgren et al.(1995)]{Lun95} Lundgren, S.C. et al., 1995, ApJ, 453, 419
\bibitem[Lundgren et al.(1995b)]{Lun95b} Lundgren, S.C. et al., 1995, ApJ, 453, 433
\bibitem[Lynds(1962)]{Lyn62} Lynds, B.T., 1962, ApJS, 7, 1
\bibitem[Lyne et al.(1988)]{Lyn88} Lyne, A.G. et al., 1988, MNRAS, 233, 667
\bibitem[Lyne\&Lorimer(1994)]{Lyn94} Lyne, A.G. \& Lorimer, D.R., 1994, Nature, 369, 127
\bibitem[Lyne\&Graham-Smith(1999)]{Lyn99} Lyne, A.G. \& Graham-Smith, F., 1998, CAS, 31L
\bibitem[Mainieri et al.(2002)]{Mai02} Mainieri, V. et al., 2002, A\&A 393, 425
\bibitem[Manchester et al.(1982)]{Man82} Manchester, R.N. et al., 1982, ApJ, 262, 45
\bibitem[Manchester et al.(1991)]{Man91} Manchester, R.N. et al., 1991, Nature, 253, 7
\bibitem[Manchester et al.(2001)]{Man01} Manchester, R.N. et al., 2001, MNRAS, 328, 17
\bibitem[Manchester et al.(2005)]{Man05} Manchester, R.N. et al., 2005, AJ, 129, 1993
\bibitem[Manchester et al.(2005b)]{Man05b} Manchester, R.N. et al., 2005, yCat, 7245, 0M
\bibitem[Manzali et al.(2007)]{Man07} Manzali, A., et al., 2007, ApJ 669, 470
\bibitem[Mason et al.(1998)]{Mas98} Mason, B.D., et al., 1998, AJ, 116, 2975.
\bibitem[Matheson\&Safi-Harb(2010)]{Mat10} Matheson, H. \& Safi-Harb, S., 2010, 724, 572 
\bibitem[McAdam et al.(1993)]{McA93} McAdam, W.B. et al. 1993, Nature, 361, 516 
\bibitem[McGowan et al.(2004)]{Mcg04} McGowan, K.E., et al., 2004, ApJ, 600, 343 
\bibitem[McGowan et al.(2006)]{Mcg06} McGowan, K.E., et al., 2006, ApJ  647, 1300 
\bibitem[Michel(1991)]{Mic91} Michel, F.C., 1991, PASP, 103770M
\bibitem[Migliazzo et al.(2002)]{Mig02} Migliazzo, J.M, et al., 2002, ApJ, 567, 141
\bibitem[Migliazzo et al.(2008)]{Mig08a} Migliazzo, J.M, et al., 2008, arXiv:astro-ph/0202063v1
\bibitem[Mignani (2008)]{Mig08} Mignani, Roberto P., 2009, arXiv0912.2931M
\bibitem[Mignani (2010)]{Mig10} Mignani, Roberto P., 2010, ihea.book, 47M
\bibitem[Mignani et al.(2011)]{Mig11} Mignani, Roberto P., 2011, in preparation
\bibitem[Mignani et al.(2010)]{Mig10b} Mignani, R.P., et al., 2010, ApJ 720, 1635
\bibitem[Mineo et al.(2002)]{Min02} Mineo, T. et al., 2002, A\&A, 392, 181
\bibitem[Mirabal et al.(2000)]{Mir00} Mirabal, N. et al., 2000, ApJ, 541, 180
\bibitem[Mirabal\&Halpern(2001)]{Mir01} Mirabal, N. \& Halpern, J.P., 2001, ApJ, 547, 137
\bibitem[Misanovic et al.(2008)]{Mis08} Misanovic, Z., et al., 2008, ApJ 685, 1129
\bibitem[Monet et al.(2003)]{Mon03} Monet, D.G., et al., 2003, AJ 125, 984
\bibitem[Moon et al.(2004)]{Moo04} Moon, D.S., et al. 2004, ApJ, 610, 33
\bibitem[Mori et al.(2004)]{Mor04} Mori, K. et al., 2004, AdSpR, 33, 503M
\bibitem[Murray et al.(2002)]{Mur02} Murray, S.S. et al., 2002, ApJ, 654, 267
\bibitem[Muslimov\&Harding(2003)]{Mus03} Muslimov, A.G. \& Harding, A.K., 2003, ApJ, 588, 430
\bibitem[Navano et al.(1995)]{Nav95} Navano, J. et al., 1995, ApJ, 455, 55
\bibitem[Ng et al.(2005)]{Ng05} Ng, C.Y. et al., 2005, ApJ, 627, 904
\bibitem[Nicastro et al.(2002)]{Nic02} Nicastro, L. et al., 2002, Proceedings of the Seminar on NSs, pulsars and SNRs
\bibitem[Nicastro et al.(2003)]{Nic03} Nicastro, L. et al., 2003, A\&A, 413, 1065
\bibitem[Nice et al.(2005)]{Nic05} Nice, D.J. et al., 2005, ApJ, 634, 1242
\bibitem[Nolan et al.(1996)]{Nol96} Nolan, P.L. et al., 1996, ApJ, 459, 100
\bibitem[Noutsos\&Guillermot(2010)]{Nou10} Noutsos, A. \& Guillermot, L., 2010, tsra conf., 119
\bibitem[Novara et al.(2006)]{Nov06} Novara, G., et al., 2006, A\&A 448, 93
\bibitem[O'Brien et al.(2008)]{Obr08} O'Brien, J.T. et al., 2008, MNRAS, 388, 1
\bibitem[Oegelman et al.(1993)]{Oeg93} Oegelman, H. et al., 1993, Nature, 361, 154
\bibitem[Oegelman\&Finley(1993b)]{Oeg93b} Oegelman, H. \& Finley, J.P., 1993, ApJ, 413, 31
\bibitem[Oegelman et al.(1995)]{Oeg95} Oegelman, H. et al., 1995, ASPC, 72, 309
\bibitem[Pacini(1967)]{Pac67} Pacini, F., 1967, Nature, 216, 567
\bibitem[Page et al.(2009)]{Pag09} Page, D., et al., 2009, ApJ 707, 1131
\bibitem[Pavlov et al.(2001)]{Pav01} Pavlov, G.G. et al., 2001, ApJ, 552, 129
\bibitem[Pavlov et al.(2003)]{Pav03} Pavlov, G.G. et al., 2003, ApJ, 591, 1157
\bibitem[Pellizzoni et al.(2008)]{pellizzoni08} Pellizzoni, A., et al., 2008, ApJ 679, 664
\bibitem[Pellizzoni et al.(2009)]{Pel09} Pellizzoni, A., et al., 2009, ApJ 695, 115
\bibitem[Pineault et al.(1993)]{pineault93} Pineault, S., et al., 1993, AJ 105, 1060 
\bibitem[Pivovaroff et al.(2000)]{Piv00} Pivovaroff, M.J. et al., 2000, ApJ, 535, 379 
\bibitem[Pivovaroff et al.(2001)]{Piv01} Pivovaroff, M.J. et al., 2001, ApJ, 554, 161 
\bibitem[Possenti et al.(1996)]{Pos96} Possenti, A. et al., 1996, A\&A, 313, 565
\bibitem[Possenti et al.(2002)]{Pos02} Possenti, A. et al., 2002, A\&A, 387, 993
\bibitem[Radhakraishnan et al.(2000)]{Rad} Radhakraishnan, V. et al., 2000, Nature, 221, 443
\bibitem[Rappaport et al.(1995)]{Rap95} Rappaport, S. et al., 1995, MNRAS, 273, 731
\bibitem[Ransom et al.(2002)]{Ran02} Ransom, S.M. et al., 2002, AJ, 124, 1788
\bibitem[Ransom et al.(2010)]{Ran10} Ransom, S.M. et al., 2010, AAS, 21545312R
\bibitem[Ray et al.(1996)]{Ray96} Ray, P.S. et al., 1996, ApJ, 470, 1103
\bibitem[Ray et al.(2011)]{Ray10} Ray, P.S. et al., 2011, ApJS, 194, 17
\bibitem[Ray \& Saz Parkinson(2010)]{Ray10b} Ray, P.S., \& Saz Parkinson, P.M., 2010, Proceedings of ICREA Workshop on The High-Energy Emission from Pulsars
and their Systems, in press, arXiv:1007.2183
\bibitem[Reimer et al.(2000)]{Rei00} Reimer, O. et al., 2000, in American Institute of Physics Conference Series, vol 510, ed. M.L. McConnell \& J.M. Ryan, 489-493
\bibitem[Reimer et al.(2001)]{Rei01} Reimer, O. et al., 2001, MNRAS, 324, 772
\bibitem[Roberts et al.(1993)]{Rob93} Roberts, D.A. et al., 1993, A\&A, 274, 427
\bibitem[Roberts et al.(1999)]{Rob99} Roberts, M.S.E. et al., 1999, ApJ, 515, 712
\bibitem[Roberts et al.(2001)]{Rob01} Roberts, M.S.E. et al., 2001, ApJS, 133, 451
\bibitem[Roberts et al.(2001b)]{Rob01b} Roberts, M.S.E. et al., 2001, ApJS, 561, 187
\bibitem[Roberts et al.(2002)]{Rob02} Roberts, M.S.E. et al., 2002, ApJ, 577, 19
\bibitem[Roberts\&Brogan(2008)]{Rob08a} Roberts, M.S.E.\& Brogan, C.L., 2008, ApJ, 681, 320R
\bibitem[Romani et al.(2005)]{Rom05} Romani, R.W. et al., 2005, ApJ, 627, 383
\bibitem[Romani et al.(2010)]{Rom10} Romani, R.W. et al., 2010, ApJ, 724, 908
\bibitem[Romani\&Johnston(2001)]{Rom01} Romani, R.W. \& Johnston, S., 2001, ApJ, 557, 93
\bibitem[Rudie et al.(2008)]{Rud08a} Rudie, G.C. et al., 2008, MNRAS, 384, 1200R
\bibitem[Ruiz\&May(1986)]{Rui86} Ruiz, M.T. \& May, J., 1986, ApJ, 309, 667
\bibitem[Russell et al.(1990)]{Rus90} Russell, J.L., et al., 1990, AJ 99, 2059
\bibitem[Safi-Harb\&Kumar(2008)]{Saf08} Safi-Harb, S. \& Kumar, H.S., 2008, arXiv, 0805.3807
\bibitem[Sallman\&Bacher(1995)]{Sal95} Sallman, S. \& Bacher, D.C., 1995, in ASP. conf. ser. 72
\bibitem[Sankrit\&Blair(2002)]{Sak02} Sankrit, R., \& Blair, W.P., 2002, ApJ, 565, 297
\bibitem[Sanwal et al. (2002)]{San02} Sanwal, D. et al. 2002, ASPC, 271, 353
\bibitem[Saz Parkinson et al.(2010)]{Saz10} Saz Parkinson, P.M., et al., 2010, ApJ, 725, 571
\bibitem[Schmidt(1968)]{Sch68} Schmidt, M., 1968, ApJ, 151, 393S
\bibitem[Seward\&Harnden(1982)]{Sew82} Seward, F.D. \& Harnden, F.R.Jr, 1982, ApJ, 256, 45
\bibitem[Seward\&Wang(1988)]{Sew88} Seward, F.D. \& Wang, Z.R., 1988, ApJ, 332, 199
\bibitem[Seward et al.(1995)]{Sew95} Seward, F.D. et al., 1995, ApJ 453, 284
\bibitem[Sivakoff et al. (2011)]{Siv11} Sivakoff et al., 2011, in preparation
\bibitem[Skrutskie et al.(2006)]{Skr6} Skrutskie, M.F., et al., 2006, AJ 131, 1163
\bibitem[Slane et al.(1997)]{Sla97} Slane, P. et al., 1997, ApJ, 485, 221
\bibitem[Slane et al.(2004)]{Sla04a} Slane, P. et al., 2004, ApJ, 601, 1045S
\bibitem[Stappers et al.(1996)]{Sta96} Stappers, B.W. et al., 1996, ApJ, 465, 199
\bibitem[Stappers et al.(1999)]{Sta99} Stappers, B.W. et al., 1999, MNRAS, 308, 609
\bibitem[Stark et al.(1992)]{Sta92} Strk, A.A. et al., 1992, ApJS, 79, 77
\bibitem[Steiner et al.(2010)]{Ste10} Steiner, A.W. et al., 2010, ApJ, 722, 33
\bibitem[Strom\&Stappers(2000)]{Str00} Strom, R.G. \& Stappers, B.W., 2000, in ASP Conf. Ser. 202, 509
\bibitem[Struder et al.(2001)]{Str01} Struder, L. et al., 2001, A\&A, 365, L18
\bibitem[Stuhlinger et al.(2008)]{xmm} Stuhlinger, M. et al., 2008, XMM-SOC-CAL-TN-0052
\bibitem[Takahashi et al.(2001)]{Tak01} Takahashi, M. et al., 2001, ApJ, 554, 316
\bibitem[Tam\&Roberts(2008)]{Tam08a} Tam, C. \& Roberts, M.S.E., 2008, arXiv:astro-ph/0310586v1
\bibitem[Taylor\&Cordes(1993)]{Tay93} Taylor, J.H. \& Cordes, J.M., 1993, ApJ, 411, 674
\bibitem[Thompson et al.(1975)]{Tom75} Thompson, D.J. et al., 1975, ApJ, 200, 79
\bibitem[Thompson(2008)]{Tom08} Thompson, D.J., 2008, Reports on Progress in Physics, 71, 116901
\bibitem[Thorsett et al.(1994)]{Tho94} Thorsett, S.E. et al., 1994, IAU circular, 6012
\bibitem[Thorsett et al.(2002)]{Tho02} Thorsett, S.E. et al., 2002, ApJ, 573, 111
\bibitem[Thorsett et al.(2003)]{Tho03a} Thorsett, S.E. et al., 2003, ApJ, 592L, 71T
\bibitem[Tian\&Leahy(2008)]{Tia08} Tian, W.W. \& Leahy, D.A., 2008, MNRAS, 391, 54
\bibitem[Torii et al.(1998)]{Tor98} Torii, K. et al., 1998, in IAU Symp 188, The Hot Universe, ed. K. Koyama et al., 258
\bibitem[Torii et al.(2001)]{Tor01} Torii, K. et al., 2001, ApJ, 551, 151
\bibitem[Toscano et al.(1999)]{Tos99} Toscano, M. et al., 1999, ApJ, 523, 171
\bibitem[Totani et al.(2002)]{Tot02} Totani, T. et al., 2002, PASJ, 54, 45
\bibitem[Trepl et al.(2010)]{Tre10} Trepl, L. et al., 2010, MNRAS, 405, 1339
\bibitem[Trussoni et al.(1996)]{Tru96} Trussoni, E. et al., 1996, A\&A, 306, 581
\bibitem[Tsuruta(1998)]{Tsu98} Tsuruta, S., 1988, PhR, 292, 1
\bibitem[Tsuruta et al.(2009)]{Tsu09} Tsuruta, S., et al., 2009, ApJ 691, 621
\bibitem[Turner et al.(2001)]{Tur01} Turner, M.J.L. et al. 2001, A\&A, 365, L27
\bibitem[Wallace et al.(1977)]{Wal77} Wallace, P.T. et al., 1977, Nature, 266, 692
\bibitem[Watters et al.(2009)]{Wat09} Watters, K.P. et al., 2009, ApJ, 695, 1289
\bibitem[Webb et al.(2004a)]{Web04a} Webb, N.A. et al., 2004, A\&A, 417, 181
\bibitem[Webb et al.(2004b)]{Web04b} Webb, N.A. et al., 2004, A\&A, 419, 269
\bibitem[Weiler\&Panagia(1978)]{Wei78} Weiler, K.W. \& Panagia, N., 1978, A\&A, 70, 419
\bibitem[Weltevrede et al.(2009)]{Wel09} Weltevrede, P. et al. 2009, arXiv, 0911.3063
\bibitem[Winkler et al.(2009)]{Win09a} Winkler, P.F. et al., 2009, ApJ, 692, 1489W
\bibitem[Woosley(1987)]{Woo87} Woosley, S.E., 1987, IAUS, 125, 255
\bibitem[Word(2008)]{Wor08} Word, J.E., 2008, in American Institute of Physics Conference Series, vol. 1085, ed. F.A. Aharonian et al., 301-303
\bibitem[Van der Swaluw(2003)]{Van03} Van der Swaluw, E., 2003, A\&A 404, 939
\bibitem[Van Etten et al.(2008)]{VEt08} Van Etten, A. et al. 2008, ApJ, 680, 141
\bibitem[Van Etten\&Romani(2010)]{VEt10} Van Etten, A. \& Romani, R.W., 2010, ApJ, 711, 1168
\bibitem[Van Paradijs et al.(1988)]{VPa88} Van Paradijs, J. et al., 1988, Nature, 334, 684
\bibitem[Yadigaroglu\&Romani(1997)]{Yad97} Yadigaroglu, I.A. \& Romani, R.W., 1997, ApJ, 476, 347
\bibitem[Yusef-Zadeh\&Bally(1987)]{Yus87} Yusef-Zadeh, F. \& Bally, J., 1987, Nature, 330, 455
\bibitem[Zavlin et al.(2002)]{Zav02} Zavlin, V.V., et al., 2002, ApJ 569, 894
\bibitem[Zavlin et al.(2006)]{Zav06} Zavlin, V.V., et al., 2006, ApJ 638, 951
\bibitem[Zavlin\&Pavlov(2002)]{Zav02b} Zavlin, V.V., \& Pavlov, G.G., 2002, nsps conf, 263
\bibitem[Zavlin\&Pavlov(2004)]{Zav04} Zavlin, V.V., \& Pavlov, G.G., 2004, ApJ 616, 452
\bibitem[Zhang\&Cheng(1997)]{Zha97} Zhang, L. \& Cheng, K.S., 1997, ApJ, 487, 370
\bibitem[Zhang et al.(2004)]{Zha04} Zhang, L. et al., 2004, ApJ, 604, 317Z
\bibitem[Zhang et al.(2005)]{Zha05} Zhang, B., et al, 2005, ApJ 624, L109
\bibitem[Zhang\&Harding(2000)]{Zha00} Zhang, B., \& Harding, A.K., 2000, ApJ, 532, 1150
\bibitem[Zepka et al.(1996)]{Zep96} Zepka, A. et al., 1996, ApJ, 456, 305
\bibitem[Zieger et al.(2008)]{Zie08} Zieger, B.R. et al., 2008, ApJ, 674, 271
\bibitem[Ziegler et al.(2008)]{Zie08a} Ziegler, M. et al., 2007, ApJ, 664, 91
\end{thebibliography}
\end{document}